\documentclass[12pt,screen,twoside,pagebackref]{ibtesis}

\usepackage[utf8]{inputenc}
\usepackage{verbatim}
\usepackage{ibextra}
\usepackage{bm}
\usepackage{float}
\usepackage{caption}
\usepackage{subcaption}
\usepackage{grffile}
\usepackage{hyperref}
\usepackage{color}

\hypersetup{
   linkcolor=blue,          
    citecolor=red,        
    filecolor=green,      
    urlcolor=magenta           
}

\usepackage{appendix}
\usepackage{amssymb}
\usepackage{amsmath}

\usepackage{amsthm}
\usepackage{cancel}
\usepackage{stmaryrd}
\usepackage{setspace}
\usepackage{wasysym}
\usepackage{esint}
\usepackage{enumitem}

\let\oldFootnote\footnote
\newcommand\nextToken\relax

\renewcommand\footnote[1]{%
    \oldFootnote{#1}\futurelet\nextToken\isFootnote}

\newcommand\isFootnote{%
    \ifx\footnote\nextToken\textsuperscript{,}\fi}

 \makeatletter

\DeclareTextSymbolDefault{\textquotedbl}{T1}

\theoremstyle{plain}
\newtheorem{thm}{\protect\theoremname}[chapter]
\theoremstyle{remark}
\newtheorem{rem}[thm]{\protect\remarkname}
\theoremstyle{definition}
\newtheorem{defn}[thm]{\protect\definitionname}
\theoremstyle{plain}
\newtheorem{lem}[thm]{\protect\lemmaname}
\theoremstyle{definition}
\newtheorem{example}[thm]{\protect\examplename}
\theoremstyle{plain}
\newtheorem{cor}[thm]{\protect\corollaryname}
\theoremstyle{plain}
\newtheorem{prop}[thm]{\protect\propositionname}
\theoremstyle{plain}
\newtheorem{fact}[thm]{\protect\factname}
\theoremstyle{plain}
\newtheorem{assumption}[thm]{\protect\assumptionname}
\theoremstyle{remark}
\newtheorem{notation}[thm]{\protect\notationname}

\usepackage[font=small,labelfont=bf]{caption}

\makeatother

\addto\captionsamerican{\renewcommand{\assumptionname}{Assumption}}
\addto\captionsamerican{\renewcommand{\corollaryname}{Corollary}}
\addto\captionsamerican{\renewcommand{\definitionname}{Definition}}
\addto\captionsamerican{\renewcommand{\examplename}{Example}}
\addto\captionsamerican{\renewcommand{\factname}{Fact}}
\addto\captionsamerican{\renewcommand{\lemmaname}{Lemma}}
\addto\captionsamerican{\renewcommand{\notationname}{Notation}}
\addto\captionsamerican{\renewcommand{\propositionname}{Proposition}}
\addto\captionsamerican{\renewcommand{\remarkname}{Remark}}
\addto\captionsamerican{\renewcommand{\theoremname}{Theorem}}
\addto\captionsenglish{\renewcommand{\assumptionname}{Assumption}}
\addto\captionsenglish{\renewcommand{\corollaryname}{Corollary}}
\addto\captionsenglish{\renewcommand{\definitionname}{Definition}}
\addto\captionsenglish{\renewcommand{\examplename}{Example}}
\addto\captionsenglish{\renewcommand{\factname}{Fact}}
\addto\captionsenglish{\renewcommand{\lemmaname}{Lemma}}
\addto\captionsenglish{\renewcommand{\notationname}{Notation}}
\addto\captionsenglish{\renewcommand{\propositionname}{Proposition}}
\addto\captionsenglish{\renewcommand{\remarkname}{Remark}}
\addto\captionsenglish{\renewcommand{\theoremname}{Theorem}}
\providecommand{\assumptionname}{Assumption}
\providecommand{\corollaryname}{Corollary}
\providecommand{\definitionname}{Definition}
\providecommand{\examplename}{Example}
\providecommand{\factname}{Fact}
\providecommand{\lemmaname}{Lemma}
\providecommand{\notationname}{Notation}
\providecommand{\propositionname}{Proposition}
\providecommand{\remarkname}{Remark}
\providecommand{\theoremname}{Theorem}

\addto\captionsamerican{}
\addto\captionsenglish{}

\makeatletter
\newcommand{\customlabel}[2]{\protected@write  \@auxout{}{\string \newlabel {#1}{{#2}{\thepage}{#2}{#1} {} } }   \hypertarget{#1}{} }
\makeatother

\title{Aspects of Entanglement Entropy in Algebraic Quantum Field
Theory}
\author{M.Sc. Diego Esteban Pontello}
\director{Dr. Horacio Casini}
\carrera{Thesis for the Degree of Doctor of Philosophy}
\grado{Ph.D. candidate}
\laboratorio{Grupo de Part\'iculas y Campos --  Centro At\'omico Bariloche}
\jurado{Dr.~Mauricio~Leston (IAFE-UBA) \\ 
Dr.~Francisco~D.~Mazzitelli (CAB/IB)\\
Ph.D.~Robert~C.~Myers (Perimeter Institute)\\
Dr.~Gonzalo~Torroba (CAB/IB)}
\palabrasclave{Teoría algebraica de Campos, Álgebras de von Neumann, Entropía de entrelazamiento, Hamiltonianos modulares, Sectores de superselección}
\keywords{Algebraic QFT, von Neumann algebras, Entanglement entropy, Modular Hamiltonians, Superselection sectors}
\date{December 13, 2019}

\begin{document}

\raggedbottom
\setlength{\parskip}{3pt}


\begin{preliminary}


\dedicatoria{
\textit{“329 pages,} \\
\textit{not great, not terrible."} \\
Comrade Anatoly Dyatlov\\
}

\selectlanguage{english}
\begin{abstract}%
\hspace{0.6 cm}In this thesis, we study aspects of entanglement theory of quantum
field theories from an algebraic point of view. The main motivation
is to gain insights about the general structure of the entanglement
in QFT, towards a definition of an entropic version of QFT. In the
opposite direction, we are also interested in exploring any consequence
of the entanglement in algebraic QFT. This may help us to reveal unknown
features of QFT, with the final aim of finding a dynamical principle
which allows us to construct non-trivial and rigorous models of QFT.
The algebraic approach is the natural framework to define and study
entanglement in QFT, and hence, to pose the above inquiries.

After a self-contained review of algebraic QFT and quantum information
theory in operator algebras, we focus on our results. We compute,
in a mathematically rigorous way, exact solutions of entanglement
measures and modular Hamiltonians for specific QFT models, using algebraic
tools from modular theory of von Neumann algebras. These calculations
show explicitly non-local features of modular Hamiltonians and help
us to solve ambiguities that arise in other non-rigorous computations.
We also study aspects of entanglement entropy in theories having superselection
sectors coming from global symmetries. We follow the algebraic perspective
of Doplicher, Haag, and Roberts. In this way, we find an entropic
order parameter that ``measures''  the size of the symmetry group, which
is made out of a difference of two mutual informations. Moreover,
we identify the main operators that take account of such a difference,
and we obtain a new quantum information quantity, the entropic certainty
relation, involving algebras containing such operators. This certainty
relation keeps an intrinsic connection with subfactor theory of von
Neumann algebras.
\end{abstract}


\tableofcontents                

\begin{abreviaturas}

  a.e.: almost everywhere
  
  BF: Buchholz-Fredenhagen  
  
  BW: Bisognano-Wichmann
  
  CFT: conformal field theory
  
  CFTs: conformal field theories 
  
  DHR: Doplicher-Haag-Roberts  
  
  EE: entanglement entropy
  
  EEs: entanglement entropies
  
  l.h.s.: left-hand side
 
  MI: mutual information
  
  MIs:  mutual informations 
  
  QIT: quantum information theory
  
  QFT: quantum field theory
  
  QFTs: quantum field theories
  
  qubit: quantum bit
                             
  RE: relative entropy
 
  REs: relative entropies

  r.h.s.: right-hand side  
  
  RS: Reeh-Schlieder
  
  SS: superselection sectors
  
  vN: von Neumann
    
\end{abreviaturas}




\end{preliminary}


\renewcommand\chaptername{Chapter}
\selectlanguage{english}

\chapter{Introduction\label{INTRO}}

Quantum field theory (QFT) is the physical theory that emerged when
physicists tried to construct a quantum theory that would be compatible
with special relativity.\footnote{Along this thesis, the term QFT refers to a local Poincaré covariant
quantum field theory. We exclude from our analysis non-relativistic
or non-local quantum field theories. } In other words, it is the combination of two physical concepts: quantum
mechanics (QM) + special relativity. In the usual approach to QFT,
both concepts are unified in the notion of a quantum field. On one
hand, a quantum field is an operator acting on some physical Hilbert
space (quantum physics), whereas on the other hand, it satisfies locality
and transformation properties compatible with the symmetries of the
spacetime (special relativity). Algebraic quantum field theory (AQFT)
settles as an alternative way to unify these two concepts. More appropriately,
the term “algebraic” in AQFT refers to the quantum physics part.

Algebraic quantum theory could be thought of as an extension of the
usual approach to quantum mechanics \cite{Bogolyubov,haag}. In the
later, one first specified a Hilbert space $\mathcal{H}$ and recognized
the set of pure states as the unit rays in $\mathcal{H}$. Convex
combinations of pure states represent mixed states, and they are in
one-to-one correspondence with the statistical operators (density
matrices) in the Hilbert space. Any self-adjoint operator in $\mathcal{H}$
represents an (ideal) observable, and we are able, in principle, to
construct any operator of the theory. However, in any real situation,
we do not have at our disposal all the operators of theory. This may
happen because of experimental constraints. For example, in our laboratory,
we usually do not have the equipment to perform any measurement with
arbitrary precision. 

Whatever the circumstances may be, the mathematical way to describe
this situation is limiting the operator content of the theory as a
subset of all operators in the Hilbert space. Following physical motivation,
we can argue that such a subset is indeed an algebra (a subalgebra
of the algebra of all operators). More precisely, the quantum system
is generally described by a $C^{*}$ or von Neumann (vN) algebra $\mathfrak{A}$,
where the algebra of all (bounded) operators in some Hilbert space
is a particular example. Such an algebra is usually called the algebra
of observables. The elements of $\mathfrak{A}$ are the operators
of the theory, and the self-adjoint ones represent observables. In
this approach there is no prior reference to a Hilbert space, the
algebra is considered as an abstract object, and the states are just
linear functionals on the algebra. The number obtained by evaluating
an observable on a state has to be interpreted as the expectation
value and/or transition probability. Hilbert space arises when one
considers a representation of the algebra. Contrary to what happens
in the usual approach to QM, there exists many unitarily inequivalent
representations of the observable algebra. One of the successes of
the algebraic approach was to show that all the inequivalent representations
represent, in the end, the same physical situation. Or in other words,
every physical phenomenon describable in one representation could
be also be described in any other representation. That means that
the physical information is encoded in the algebra of observables
itself. However, the choice of a particular representation has very
computational importance. A physical process may have a simple description
in a given representation whereas its description in other representations
could be highly cumbersome. This is in fact what happens in QFT. The
algebra of QFT admits many unitarily inequivalent representations.
Among them, the Fock representation is very accurate to describe free
fields but not for interacting fields. When one uses the Fock representation
to make computations about interacting fields, we obtain divergent
quantities everywhere. This is nothing more than the consequence of
Haag's theorem that asserts that the interaction representation does
not exist in QFT. This problem could be bypassed by introducing a
cutoff and renormalizing the theory, to obtain analytical results
that are in high accordance with physical experiments. However, from
the conceptual and mathematical standpoint, this renormalization program
is far from being satisfactory.

In the usual approach to QFT, the theory is described in a Hilbert
space $\mathcal{H}$ and the operator content is described by field
operators $\phi\left(x\right)$. The local physics around a finite
region $\mathcal{O}$ of the spacetime is described by the field operators
$\phi\left(x\right)$ with $x\in\mathcal{O}$. It is well-known that
all such operators commute with the field operators $\phi\left(y\right)$
whenever $y$ is spacelike separated with all points in O. That means,
that the operators “available” within the region $\mathcal{O}$ are
far from being the set of all operators in the Hilbert space $\mathcal{H}$.
Then, they must be described by a proper subset $\mathfrak{A}\left(\mathcal{O}\right)$
of operators, depending on the region $\mathcal{O}$. It is for this
reason that the algebraic approach to quantum theory naturally fits
for QFT. Furthermore, the algebraic approach to quantum theory typically
arises interest when one wants to study QFT from an axiomatic point
of view. The strong mathematical and conceptual basis of the theory
of operators algebras allows us to establish a solid mathematical
scenario for QFT, from which general statements could be proven. From
this perspective, we can remark the algebraic proofs of the CPT and
spin-statistics theorems, and the famous connection between superselection
sectors and internal (global and local) symmetries. On the other hand,
it has been very difficult to exhibit concrete models satisfying the
axioms beyond the free theories, $d=2$ CFTs, and some superrenormalizable
theories.

The theory of von Neumann algebras could be thought of as a non-commutative
generalization of the usual classical probability theory of Kolmogorov.
More concretely, it can be shown that any classical probability space
could be regarded as an Abelian (commutative) von Neumann algebra.
The origin and development of telecommunications, classical computation,
and data processing in the last century gave place to a strong interest
in the study of statistical and information measures in (classical)
probability spaces. It was Shannon who first interpreted the concept
of entropy as information, giving place to the origin of classical
information theory. Afterward, the concept of entropy in classical
information theory is named Shannon entropy. Almost simultaneously,
the theoretical proposal of the quantum computer by R. Feynman and
P. Benioff gave birth quantum information theory (QIT). Like its classical
counterpart, the mathematical formulation of QIT is described using
“quantum” probability spaces, i.e. von Neumann algebras. It is for
that reason that QIT theory matches perfectly with the algebraic approach
to quantum physics. From the physical point of view, the central difference
between classical and quantum physics is that quantum systems exhibit
a distinct kind of statistical correlations (called generically entanglement)
that are absent in classical systems. This significant difference
has been used to show (at least theoretically) the potential ability
of quantum computing devices to solve problems that classical computers
practically cannot, leading to the notion of quantum supremacy in
quantum computing. These correlations, due to the entanglement, can
be captured with the help of the von Neumann entropy, the quantum
version of the Shannon entropy, and for this reason, it is usually
called entanglement entropy (EE).

However, the interest in studying statistical measures in quantum
(or classical) systems does not only rely on information or computation
theories. From a physical standpoint, such an interest relies on the
study of bulk or macroscopic quantities of systems governing a large
number of individual degrees of freedom. This explains the fact that
QIT (and the theory of vN algebras) has much more to do with QFT (or
many-body systems) than with the usual Schrodinger's mechanics theory.
In condensed matter physics, the entanglement entropy has been used
as an order parameter to distinguish phases that could not be identified
by symmetry \cite{Kitaev:2005dm,Levin:2006zz}, and also to localize
and characterize critical points in phase transitions \cite{Calabrese:2009qy,Holzhey:1994we,Vidal:2002rm,Fradkin:2006mb}.
In quantum gravity, more specifically in the context of holography,
there is a deep connection between entanglement entropy and geometry,
materialized in the Ryu-Takayanagi formula \cite{Ryu:2006bv}. This
formula generalizes the usual expression for the black hole entropy,
and it has helped us to better specify the holographic dictionary
in AdS/CFT \cite{Jafferis:2015del,Faulkner:2013ana,Aharony:1999ti,Maldacena:1997re}.
In particular, the continuity and emergence of spacetime could be
explained in terms of the entanglement of the boundary degrees of
freedom \cite{VanRaamsdonk:2018zws}. In QFT, the EE of the vacuum
state holds the distinct feature that it grows as the area of the
region \cite{Wolf07,Srednicki93,Eisert08}. The first insight of this
area law was observed while trying to explain the statistical origin
of the entropy of black holes as an EE \cite{Bombelli:1986rw,Susskind:1994sm,Solodukhin:2011gn}.
From the algebraic perspective, given a vN algebra representing some
physical system and a state on such an algebra, there always exists
a natural dynamics on the system, depending on the state, even if
the original system has no preferred dynamics \cite{haag,Bratteli1}.
This evolution is called modular flow, and its Hermitian generator
is called modular Hamiltonian. Of course, the conceptual similarity
of this feature with gravitational physics is almost evident, and
it has been exploited in the AdS/CFT scenario to improve the holographic
dictionary \cite{Faulkner:2017vdd,Faulkner:2018faa}. Moreover, the
modular Hamiltonian is also used in the definition of the relative
entropy on general vN algebras \cite{Araki_entropy,Araki77}.

Needless to say, the interest of EE in high energy physics is beyond
quantum gravity. In fact, in QFT, it has an interest in its own right.
The EE has been exploited to give ``entropic'' proofs of the renormalization
group irreversibility theorems in $d=2,3,$ and $4$ \cite{Casini:2004bw,Casini:2006es,Myers:2010xs,Casini:2012ei,Casini:2016fgb,Casini:2017vbe,Casini:2018kzx}.
There also exists a deep connection between spacetime symmetries and
modular evolutions for some particular types of regions. This is the
famous theorem of Bisognano-Wichmann, which asserts that, for any
QFT, the modular dynamics for wedge regions and the vacuum state coincides
with the one-parameter of boost transformations that leave the wedge
region invariant \cite{bisognano}. For CFTs, this could be generalized
for double cones. In that case, the modular evolution is given by
the one-parameter group of conformal transformations that leave the
double cone invariant \cite{Hislop81,chm}. However, for other types
of regions, this feature is lost, and even more, the modular evolution
is no longer local \cite{reduced_density,Arias:2016nip}. There have
been important efforts to understand these non-local features \cite{Arias:2016nip,Arias17},
and in particular, to compute modular Hamiltonians and entanglement
measures for concrete models and more general regions \cite{reduced_density,scalar_nuestro,Blanco:2019xwi}.

There is another formal aspect that relates QIT with QFT. The Wightman
axiomatic approach of QFT has an equivalent description in terms of
vacuum correlators \cite{Streater}. As it is well-known, the Wightman
axiomatic approach of QFT has an equivalent description in terms of
vacuum correlators. The knowledge of all such correlators allows us
to reconstruct the theory uniquely. In other words, a complete set
of field vacuum expectation values, satisfying a certain set of axioms,
gives place to a unique QFT (up to unitary equivalence) satisfying
the Garding-Wightman axioms. In AQFT, the physical content of the
theory is no-longer codified in a point-like localized field $\phi\left(x\right)$,
but in a collection of algebras $\mathfrak{A}\left(\mathcal{O}\right)$,
assigned to spacetime regions $\mathcal{O}$, satisfying a set of
axioms that take into account the algebraic and local properties of
the fields. The analogous description of quantum fields in terms of
correlators would be a map that assigns to each local algebra (or
equivalently to each spacetime region) a number that ``quantifies''
the collective behavior of the degrees of freedom inside the region
$\mathcal{O}$. A natural candidate is the EE of such regions. However,
the EE is UV divergent in QFT, and hence, it needs a cutoff to be
defined, making this quantity ambiguous in the continuum QFT. This
problem could be solved invoking the mutual information, which is
finite and well-defined quantity for any pair of disjoint regions.
We can imagine that the knowledge of all mutual information $I\left(\mathcal{O}_{1},\mathcal{O}_{2}\right)$
for all pairs of disjoint regions $\mathcal{O}_{1}$ and $\mathcal{O}_{2}$
in the vacuum state should be enough to reconstruct the underlying
QFT model uniquely. This would result in an entropic formulation of
QFT, strongly supporting the idea that “information” is a very fundamental
concept in (quantum) physics. Towards this aim, we need to reach deeper
knowledge about the behavior of EE and entanglement measures in QFT.

The aim of this thesis is to make some contributions along the lines
discussed in the last two paragraphs. In particular, we study exact
and mathematical rigorous computations of entanglement measures and
modular Hamiltonians in specific models of QFT from the algebraic
perspective. We also study formal aspects of the entanglement theory
in AQFT. The main results of this work are the following.
\begin{enumerate}
\item To begin with, we compute the relative modular Hamiltonian and the
relative entropy for coherent states and the Rindler wedge for the
theory of a free scalar \cite{Casini:2019qst}. Previous non-rigorous calculations, including
path integral methods and computations from the lattice, give a result
for such a relative entropy, which involves integrals of expectation
values of the energy-momentum stress tensor along the considered region.
However, the non-uniqueness of the stress tensor leads to an ambiguity
in the modular Hamiltonian, which manifests in the usual formula for
the relative entropy of coherent states. Then, the main motivation
for the computation of such a relative entropy is to solve this puzzle.
\item In the same lines, we compute the vacuum modular Hamiltonian and the
vacuum mutual information for a two-interval region and the theory
of a free chiral current \cite{scalar_nuestro}. As in the case of a free chiral fermion,
the modular Hamiltonian turns out to be non-local, but in this case,
the non-local term is given by a smooth kernel. Contrary to the free
chiral fermion, this model shows a failure of duality for two intervals
that translates into a loss of a symmetry property for the mutual
information usually associated with modular invariance. Moreover,
we find explicitly the operators responsible for the failure of the
duality condition. These operators commute with all the operators
of the two intervals, but they do not belong to the algebra of the
complement region of the two intervals. Such operators naturally appear
when the theory has non-trivial superselection sectors (SS).
\item The above observation motivated us to study the consequences of the
superselection structure in the EE from a general perspective \cite{Casini:2019kex}. Following
the algebraic approach to sector theory, we find a novel connection
between EE and internal global symmetries. In this way, we find that
the mutual information between spacelike regions encodes information
about the superselection structure of the theory. More precisely,
we find an entropic order parameter, constructed by a difference of
two mutual informations, that measures the size of the global symmetry
group, which accounts for the superselection structure of the theory.
Moreover, we identify the main operators that take account of such
a difference, and we find an entropic certainty relation involving
relative entropies on algebras containing such operators. We argue
that this entropic certainty relation may have a deep connection with
the theory of vN subfactors, and it may quantify the algebraic $\mu$-index
of the inclusion of algebras \cite{Longo:1989tt,Longo:1990zp}. We
complement this analysis by applying the mentioned entropic order
parameter to different physical situations, such as finite and Lie
symmetry groups, compactified scalars, regions with different topologies,
excited states, and spontaneous symmetry breaking.
\end{enumerate}
We expect that all these contributions will help to better understand
the possible correspondence between AQFT and QIT.

The content of the thesis is organized as follows. In chapter \ref{AQFT},
we review the algebraic approach to quantum field theory. To separate
the algebraic features from the ones related to locality and covariance,
we divided the chapter into two sections. The first one is dedicated
to algebraic quantum theory alone, and in the second one, we explain
how the content of the first section is used to describe QFT from
an algebraic perspective. In chapter \ref{INFO}, we review the theory
of quantum information in general operator algebras, with particular
emphasizes on entanglement and information measures. For a better
discussion, we treat first the case of finite quantum systems (finite-dimensional
algebras) and then the case of general quantum systems (infinite-dimensional
algebras). In chapter \ref{EE_QFT}, we combine the ideas developed
in the two previous chapters to review the basic aspects of EE in
QFT. We put special emphasizes on how to describe entanglement in
the algebraic framework, using tools that are well-defined in the
continuum QFT. The reader who is familiar with these concepts may
skip these chapters and go directly to chapter \ref{RE_CS}, where
we study the relative entropy for coherent states for the theory of
a free scalar field. The computation of such a relative entropy is
performed in a complete algebraic and mathematically rigorous way,
and it helps us to solve the discussed ambiguities on the modular
Hamiltonian, which arise in computations using non-rigorous methods.
In chapter \ref{CURRENT}, we compute explicitly the vacuum modular
Hamiltonian and the mutual information for two intervals for the theory
of a chiral free scalar field. We determine explicitly the eigenvectors
that diagonalized the correlator kernel using a novel method of finding
complex analytic functions obeying specific boundary conditions. We
first also apply this method to rederive the vacuum modular Hamiltonian
for a free chiral fermion in multiple intervals, previously studied
in \cite{reduced_density}. In chapter \ref{EE_SS}, we study aspects
of entanglement entropy in theories having superselection sectors.
We focus on theories with sectors arising from global symmetries following
the algebraic approach of Doplicher, Haag, and Roberts \cite{Doplicher:1969tk,Doplicher:1969kp,Doplicher:1971wk,Doplicher:1972kr,Doplicher:1973at,Doplicher:1990pn}.
We end in chapter \ref{CONCL} with a summary and the conclusions of the work.


\renewcommand\chaptername{Chapter} 
\selectlanguage{english}

\chapter{Algebraic approach to QFT\label{AQFT}}

The main purpose of this chapter is to define the concept of an algebraic
quantum field theory (AQFT). As we explain in the introduction, QFT
is the combination of two physical concepts: quantum mechanics and
special relativity. The term ``algebraic'' in AQFT refers to the
first of such concepts. In fact, the first section of the chapter
is devoted to explaining what an algebraic quantum theory is, and
in which sense it generalizes the usual approach to quantum mechanics.
We first define the concept of $C^{*}$-algebra, usually called algebra
of observables, which is the mathematical object used to describe
quantum systems. After we introduce the notion of states and representations,
we proceed to study general consequences of the theory. Among them,
we remark the concept of physical equivalence between representations
and a theorem that asserts that all representations are physically
equivalent. In other words, it means that the physical content of
the theory is encoded in the algebra of observables itself, rather
than in the representation used for calculations. We also study the
structure of the space of (pure) states of the theory, which naturally
decomposes into coherent subsets or sectors. Consequently, we show
a one-to-one correspondence between irreducible representations of
the algebra of observables and sectors of pure states. At the end
of the first section, we introduce the notion of von Neumann algebra,
which is a mathematical concept closely related to $C^{*}$-algebra.

In the second section of this chapter, we turn to study AQFT itself.
First of all we state the axioms of AQFT, followed by the concept
of vacuum states and vacuum representations. This last concepts allow
us to state the famous theorem of Reeh-Schlieder. After all the formal
aspects are explained, we give some comments about the relation between
this approach to QFT and the usual axiomatic approach of Garding and
Wightman. Then, we explain how AQFT incorporates the notion of fermionic
operators, which are excluded in our first definition of QFT. The
last part of this second section is dedicated to the study of the
lattice structure of AQFT, where the concept of duality for local
algebras plays a salient role. 

\section{Algebraic quantum theory\label{c1-sec-aqft}}

In the usual approach to quantum theory, the mathematical description
of the system uses a Hilbert space $\mathcal{H}$ and the set of its
bounded linear operators $\mathcal{B}\left(\mathcal{H}\right)$.\footnote{\label{c1_foot_BH}A bounded operator $A\in\mathcal{B}\left(\mathcal{H}\right)$
is a linear map $A:\mathcal{H}\rightarrow\mathcal{H}$ for which exists
a constant $K_{A}\geq0$ such that $\left\Vert A\left|\Psi\right\rangle \right\Vert \leq K_{A}\left\Vert \left|\Psi\right\rangle \right\Vert $
for all $\left|\Psi\right\rangle \in\mathcal{H}$. This last formula
allows to define the norm of such an operator as $\left\Vert A\right\Vert :=\mathrm{sup}\left\{ \left\Vert A\left|\Psi\right\rangle \right\Vert \,:\,\left\Vert \left|\Psi\right\rangle \right\Vert =1\right\} $.
The outstanding property about bounded operators is that they are
well-defined over the whole Hilbert space, wheres an unbounded operator
$X$ has as a domain a proper subset of $\mathcal{D}\left(X\right)\subsetneq\mathcal{H}$,
depending on $X$. This implies that the composition (product) of
two bounded operators is always well-defined, whereas the composition
of unbounded operators is not in general well-defined because of problems
concerning domains. For further details see \cite{ReedSimon}. } The set of all observables coincides with the subset $\mathcal{B}_{sa}\left(\mathcal{H}\right)\subset\mathcal{B}\left(\mathcal{H}\right)$
of all self-adjoint operators, and the set of states coincides with
the space of density matrices, i.e. 
\begin{equation}
\mathfrak{S}\left(\mathcal{H}\right):=\left\{ \rho\in\mathcal{B}\left(\mathcal{H}\right)\,:\,\rho\geq0\textrm{ and }\mathrm{Tr}\left(\rho\right)=1\right\} \,.
\end{equation}
In fact, any density matrix $\rho$ gives place to a linear functional
$\omega_{\rho}:\mathcal{B}\left(\mathcal{H}\right)\rightarrow\mathbb{C}$
through the relation
\begin{equation}
\omega_{\rho}\left(A\right):=\mathrm{Tr}\left(\rho A\right)\,,\quad\textrm{for all }\mathrm{A\in\mathcal{B}\left(\mathcal{H}\right)}\,.\label{c1_exp_value}
\end{equation}
The physical interpretation of \eqref{c1_exp_value} has two equivalent
meanings. Suppose that we are able to perform a physical experiment
where we measure the physical observable represented by $A\in\mathcal{B}_{sa}\left(\mathcal{H}\right)$
in the physical sate represented by $\rho\in\mathfrak{S}\left(\mathcal{H}\right)$. 
\begin{enumerate}
\item Equation \eqref{c1_exp_value} corresponds to the expectation value
of the physical observable represented by $A$ in the physical state
represented by $\rho$. such an expectation value coincides with the
average of the results of the measurement experiment after the experiment
is repeated $N$ times, in the limit when $N$ is large enough.
\item Due to the spectral theorem, any self-adjoint operator $A$ can be
uniquely decomposed as 
\begin{equation}
A=\int_{\mathbb{R}}\lambda\,dE_{A}\left(\lambda\right)\,,\label{c1_spec_dec}
\end{equation}
where $E_{A}\left(\Delta\right)$ are their spectral measures corresponding
to the measurable set $\Delta\subset\mathbb{R}$. Then, the expectation
value 
\begin{equation}
p_{A}\left(\Delta\right):=\omega_{\rho}\left(E_{A}\left(\Delta\right)\right)=\mathrm{Tr}\left(\rho E_{A}\left(\Delta\right)\right)\,,
\end{equation}
represents the probability that the measured value lies in the set
$\Delta$.
\end{enumerate}
\begin{rem}
We do not loose any generality considering only bounded observables.
Indeed, any unbounded self-adjoint $X$ operator can be decomposed
as in \eqref{c1_spec_dec}, and its spectral measures $E_{X}\left(\Delta\right)\in\mathcal{B}\left(\mathcal{H}\right)$
whenever $\Delta\subset\mathbb{R}$ bounded. In the practice, when
we measure any observable associated with an unbounded operator $X$
(e.g. the energy, which is associated with the Hamiltonian), we employ
a measurement device, which has some ``finite measurement range''
(e.g. the interval between minimum and maximum value of energy that
the instrument is capable to detect). Measurement devices with infinite
measurement range do not exist.\footnote{In the theory, we used them as idealized observables.}
According to \eqref{c1_spec_dec}, this means that, in practice we
measure the observable associated with the bounded operator
\begin{equation}
X=\int_{x_{min}}^{x_{MAX}}\lambda\,dE_{A}\left(\lambda\right)\,.
\end{equation}
\end{rem}

\begin{rem}
The correspondence principle, i.e. the identification of physical
observables and physical states with the mathematical objects in $\mathcal{B}_{sa}\left(\mathcal{H}\right)$
and $\mathfrak{S}\left(\mathcal{H}\right)$, is a very difficult theoretical-experimental
task that has to be solved for any particular physical situation.
A general solution to this ``correspondence problem'' has no been
achieved, and it is one of the open problems in the foundations of
quantum mechanics. This problem is beyond the purpose of this thesis.
Here, we always start with a mathematical set up as above (or with
any of its generalizations as we argue below), which may describe
some physical system.
\end{rem}

Consider the physical situation when we do not have all the observable
operators, i.e. all the set $\mathcal{B}_{sa}\left(\mathcal{H}\right)$,
to our disposal. How does the mathematical description of the physical
system change? The algebraic approach to quantum theory borns with
the purpose to give a formal description to this situation. In fact,
the solution is quite simple: the observables are described by the
self-adjoint elements of some subset $\mathfrak{A}\subset\mathcal{B}\left(\mathcal{H}\right)$,
and the states are just a specific subset of the space of linear functionals
over $\mathfrak{A}$. From the mathematical point of view, such a
subset $\mathfrak{A}$ is described by a $C^{*}$-algebra, which is
the starting point for the next subsection. 
\begin{rem}
The situation explained above covers what usually happens in QFT.
There, the local physics around a finite region $\mathcal{O}$ of
the spacetime, is described by field operators $\phi\left(x\right)$
with $x\in\mathcal{O}$. It is well-known that all such operators
commute with the (non-trivial) field operators $\phi\left(y\right)$
whenever $y$ is spacelike separated with all points in $\mathcal{O}$.
That means that the operators ``available'' within the region $\mathcal{O}$
are far from being the set of all operators in the Hilbert space $\mathcal{H}$.
They must be described by a proper subset $\mathfrak{A}\left(\mathcal{O}\right)\subsetneq\mathcal{B}\left(\mathcal{H}\right)$.
It is for this reason that the algebraic approach to quantum theory
naturally fits for QFT.
\end{rem}

\subsection{$C^{*}$-algebras}

It is well-known that the set of all bounded operators $\mathcal{B}\left(\mathcal{H}\right)$
has a natural structure of algebra: it is closed under linear combinations
and multiplication (composition of operators). However, the subset
$\mathcal{B}_{sa}\left(\mathcal{H}\right)$ is not an algebra because
the product of two self-adjoint operators is not, in general, a self-adjoint
operator. Furthermore, the algebra generated by $\mathcal{B}_{sa}\left(\mathcal{H}\right)$
is indeed $\mathcal{B}\left(\mathcal{H}\right)$, because any operator
$A\in\mathcal{B}\left(\mathcal{H}\right)$ is a linear combination
of two self-adjoint operators: $A=\frac{A+A^{*}}{2}+i\frac{A-A^{*}}{2i}$
. For all that, in the usual approach to quantum theory, it is better
to consider $\mathcal{B}\left(\mathcal{H}\right)$ as the primary
mathematical object describing the operator content of the theory,
and then, the observables are defined as the self-adjoint elements
of $\mathcal{B}\left(\mathcal{H}\right)$. 

As we explained above, in the case where not all operators are available,
the theory is described by a subset of $\mathfrak{A}\subset\mathcal{B}\left(\mathcal{H}\right)$.
From the operational point of view we expect that, if two operators
$A,B\in\mathcal{B}\left(\mathcal{H}\right)$ are available for measurements/operations
on the system, i.e. $A,B\in\mathfrak{A}$, then, any linear combination
($\lambda_{1}A+\lambda_{2}B$ with $\lambda_{1},\lambda_{2}\in\mathbb{C}$)
and the product ($AB$) of such operators, also belong to $\mathfrak{A}$.
These imply that the subset $\mathfrak{A}$ should be a subalgebra
of $\mathcal{B}\left(\mathcal{H}\right)$. This is one of our first
conclusions: a general quantum system is described by a subalgebra
$\mathfrak{A}$ of $\mathcal{B}\left(\mathcal{H}\right)$, or in a
more abstract sense, a general quantum system is described by an algebra
$\mathfrak{A}$, which is not in general isomorphic\footnote{See definition \ref{c1_iso_def}.}
to the algebra $\mathcal{B}\left(\mathcal{H}\right)$ of all operators
in a Hilbert space.

The set $\mathcal{B}\left(\mathcal{H}\right)$ has further algebraic
and topological properties, which makes it a $C^{*}$-algebra. These
properties are expected to be satisfied for any of its subalgebras,
which describe physical systems. Such properties motivate the following
definitions.
\begin{defn}
Let $\mathfrak{A}$ be a complex (abstract\footnote{The term abstract is placed there to emphasize that the algebra may
not be a concrete subalgebra $\mathfrak{A}\subset\mathcal{B}\left(\mathcal{H}\right)$
of operators acting on some Hilbert space $\mathcal{H}$. }) algebra.\footnote{All the algebras considered along this thesis are \textit{unital}
algebras, which means that exists $\mathbf{1}\in\mathfrak{A}$ such
that $\mathbf{1}\cdot A=A\cdot\mathbf{1}=A$ for all $A\in\mathfrak{A}$.} An antilinear map $*:\mathfrak{A}\rightarrow\mathfrak{A}\:/\:A\mapsto A^{*}$
is called an \textit{involution} if $A=\left(A^{*}\right)^{*}$ and
$\left(AB\right)^{*}=B^{*}A^{*}$ for all $A,B\in\mathfrak{A}$. An
algebra with an involution is called a $*$\textit{-algebra} or \textit{involution
algebra}.
\end{defn}

\begin{defn}
Let $\mathfrak{A}$ be a complex (abstract) algebra. A map $\left\Vert \cdot\right\Vert :\mathfrak{A}\rightarrow\mathbb{R}_{\geq0}\:/\:A\mapsto\left\Vert A\right\Vert $
is called a \textit{norm} if it satisfies 
\end{defn}

\begin{enumerate}
\item $\left\Vert A\right\Vert =0$ if and only if $A=0$ ,
\item $\left\Vert \lambda A\right\Vert =\left|\lambda\right|\left\Vert A\right\Vert $
,
\item $\left\Vert A+B\right\Vert \leq\left\Vert A\right\Vert +\left\Vert B\right\Vert $
,
\item $\left\Vert AB\right\Vert \leq\left\Vert A\right\Vert \left\Vert B\right\Vert $
,
\end{enumerate}
for all $A,B\in\mathfrak{A}$ and all $\lambda\in\mathbb{C}$. An
algebra with a norm is called a\textit{ normed algebra}.
\begin{defn}
\label{c1_uniftopo}Let $\mathfrak{A}$ be a normed algebra. The topology
arising from the norm is called the \textit{uniform} or \textit{norm
topology}. A normed algebra complete in the uniform topology is called
a \textit{Banach algebra}.
\end{defn}

\begin{defn}
\label{c1_cstar}Let $\mathfrak{A}$ be an involution Banach algebra.
$\mathfrak{A}$ is said to be a $C^{*}$\textit{-algebra} if the norm
satisfies $\left\Vert A^{*}A\right\Vert =\left\Vert A\right\Vert ^{2}$
for all $A\in\mathfrak{A}$ ($C^{*}$-property).
\end{defn}

All the properties described in the previous definition are satisfied
by $\mathcal{B}\left(\mathcal{H}\right)$, which give place to the
following lemma.
\begin{lem}
For any Hilbert space $\mathcal{H}$, the set of all bounded linear
operators $\mathcal{B}\left(\mathcal{H}\right)$ is a $C^{*}$-algebra.
The norm is given by the operator norm\footnote{See footnote \ref{c1_foot_BH}.}
and the involution is given by taking adjoints.
\end{lem}

\begin{proof}
See \cite{Bratteli1}.
\end{proof}
The above lemma gives an abstract characterization of the set of operators
in the usual approach to quantum theory. The algebraic approach to
quantum theory generalizes the usual approach just claiming that the
mathematical description of a quantum system is given by some $C^{*}$-algebra
$\mathfrak{A}$, where the case $\mathfrak{A}=\mathcal{B}\left(\mathcal{H}\right)$
is just a particular case. In this context, $\mathfrak{A}$ is called
the \textit{algebra of observables} and any element $A\in\mathfrak{A}$
is called an \textit{operator}.

This generalization allows us to include in the theory the description
of more general and sophisticated systems. As we argue above, this
more general algebras are needed to describe the local physics in
QFT in a rigorous mathematical way. Of course, the operators are such
one part of the description of the theory, the other part corresponds
to the states. We postponed the explanation of how the states are
described in this algebraic setting for the following subsection.

Now, we borrow some definitions from the usual theory of operator
algebras.
\begin{defn}
Let $\mathfrak{A}$ be a $C^{*}$-algebra. An element $A\in\mathfrak{A}$
is said to be \textit{self-adjoint} if $A=A^{*}$. An element $U\in\mathfrak{A}$
is said to be \textit{unitary} if $U^{*}=U^{-1}$. An element $B\in\mathfrak{A}$
is said to be \textit{positive} if exists $A\in\mathfrak{A}$ such
that $B=A^{*}A$. An element $P\in\mathfrak{A}$ is said to be (orthogonal\footnote{Throughout this thesis, all projectors are orthogonal.})
\textit{projector} if $P=P^{2}=P^{*}$. 
\end{defn}

With all these definitions, we recognize that the observables of the
theory are described by self-adjoints operators in $\mathfrak{A}$.
On the other hand, subalgebras of $C^{*}$-algebras are characterized
by the following lemma.
\begin{defn}
Let $\mathfrak{A}$ be a $C^{*}$-algebra, and let $\mathfrak{B}\subset\mathfrak{A}$
a $*$-subalgebra in the pure algebraic sense (closed by linear combinations,
products and involution) and closed in the uniform topology. Then,
we say that $\mathfrak{B}$ is a \textit{$C^{*}$-subalgebra} of $\mathfrak{A}$,
and of course, it is a $C^{*}$-algebra on its own.
\end{defn}

Among the class of all $C^{*}$-algebras, we have the subclass of
\textit{concrete} $C^{*}$-algebras, i.e. the subclass of all $C^{*}$-subalgebras
of the form $\mathfrak{A}\subset\mathcal{B}\left(\mathcal{H}\right)$
for some Hilbert space $\mathcal{H}$. The term concrete refers to
the fact that the elements of $\mathfrak{A}$ are indeed operators
acting on some Hilbert space $\mathcal{H}$. At first glance, it seems
that the class of concrete $C^{*}$-algebra is just a proper subclass
of the class of all $C^{*}$-algebras. However, there is a theorem
that asserts that any abstract $C^{*}$-algebra can be viewed as $C^{*}$-subalgebra
of $\mathcal{B}\left(\mathcal{H}\right)$ for some Hilbert space $\mathcal{H}$.
This is summarized in the following.
\begin{lem}
Let $\mathfrak{A}_{1}$ and $\mathfrak{A}_{2}$ be $C^{*}$-algebras.
Any morphism\footnote{A morphism $\Phi:\mathfrak{A}_{1}\rightarrow\mathfrak{A}_{2}$ between
$*$-algebras is any linear map satisfying $\Phi\left(AB\right)=\Phi\left(A\right)\Phi\left(B\right)$
and $\Phi\left(A^{*}\right)=\Phi\left(A\right)^{*}$ .} $\Phi:\mathfrak{A}_{1}\rightarrow\mathfrak{A}_{2}$ is continuous
(in the uniform topology), i.e. it is an homomorphism.
\end{lem}

\begin{proof}
See \cite{Bratteli1}.
\end{proof}
\begin{defn}
\label{c1_iso_def}Let $\mathfrak{A}_{1}$ and $\mathfrak{A}_{2}$
be $C^{*}$-algebras. A bijective homomorphism $\Phi:\mathfrak{A}_{1}\rightarrow\mathfrak{A}_{2}$
is called a $C^{*}$\textit{-isomorphism}. Moreover, given any $\mathfrak{A}_{1}$,
$\mathfrak{A}_{2}$ $C^{*}$-algebras, $\mathfrak{A}_{1}$ and $\mathfrak{A}_{2}$
are said to be \textit{$C^{*}$-isomorphic} (and denoted by $\mathfrak{A}_{1}\simeq\mathfrak{A}_{2}$)
if there exists a $C^{*}$-isomorphism $\Phi:\mathfrak{A}_{1}\rightarrow\mathfrak{A}_{2}$.
\end{defn}

\begin{thm}
\label{c1_iso}Let $\mathfrak{A}$ be a $C^{*}$-algebra. Then, there
exists\footnote{Possibly non-separable.} a Hilbert space $\mathcal{H}$
and a $C^{*}$-subalgebra $\mathfrak{B}\subset\mathcal{B}\left(\mathcal{H}\right)$
such that $\mathfrak{A}\simeq\mathfrak{B}$.
\end{thm}

\begin{proof}
See \cite{Bratteli1}.
\end{proof}
The above lemma shows that we do not lose any generality assuming
that our quantum system is described by a concrete $C^{*}$-algebra.
However, the description of the quantum system without reference to
any Hilbert space, allows us to study general properties of the algebra
of observables that are intrinsic to the algebra itself. The isomorphism
of lemma \ref{c1_iso} is just a representation of the abstract algebra
in some Hilbert space. There are many representations of it, which
are $C^{*}$-isomorphic but not unitarily equivalent. In the end,
all representations describe the same physics. But the choice of one
particular representation is a matter of convenience related to the
physical process one is interested to study. A given physical process
may admit a simple description in some representation, but, at the
same time, the description of the same process may be highly complex
in a different representation. We discuss all these issues in the
following subsection.

In the algebraic approach, the dynamics of the system is implemented
as follows.
\begin{defn}
Let $\mathfrak{A}$ be a $C^{*}$-algebra of observables. A \textit{dynamics}
for $\mathfrak{A}$ is a one-parameter family of automorphisms $t\in\mathbb{R}\mapsto\alpha_{t}\in\mathrm{Aut}\left(\mathfrak{A}\right)$.
The pair $\left(\mathfrak{A},\alpha_{t}\right)$ is called a \textit{$C^{*}$-dynamical
system}.
\end{defn}

Automorphisms are the algebraic version of unitary evolution. A particular
kind of \textit{$C^{*}$}-dymanical systems are the \textit{closed
systems}, which is described by a one-parameter group of automorphisms,
i.e. $\alpha_{t_{1}}\circ\alpha_{t_{2}}=\alpha_{t_{1}+t_{2}}$ for
all $t_{1},t_{2}\in\mathbb{R}$.
\begin{example}
\label{c1_weyl_alg}(CCR and Weyl algebra) In Schrodinger's mechanics,
the operator content of the theory is encoded in two basic operators:
position and momentum. The \textit{Canonical Commutation Relations
algebra} (\textit{CCR algebra}) for $N$ degrees of freedom, is defined
as the unital $*$-algebra generated by the objects $q_{1},\ldots,q_{n},p_{1},\ldots,p_{n}$
modulo the relations

\begin{align}
 & \left[q_{j},p_{k}\right]=i\delta_{jk}\mathbf{1}\,,\quad\left[q_{j},q_{k}\right]=\left[p_{j},p_{k}\right]=0\,,\label{c1_ccr_un1}\\
 & q_{k}=q_{k}^{*}\textrm{ and }p_{k}=p_{k}^{*}\,.\label{c1_ccr_un_2}
\end{align}
Here, we want to emphasize that we are considering the CCR algebra
as an abstract algebra, independently of whether it admits a representation
as a subalgebra of operators acting on some Hilbert space. We postpone
such a problem for section \ref{c1_subsec_rep}. Unfortunately, it
can be shown, that the CCR is not a $C^{*}$-algebra. In other words,
it is impossible to equip the CCR algebra with a norm in order that
all its elements have a finite norm and they satisfy all the algebraic
relations.\footnote{In the context of representations, there is no representation of CCR
algebra as bounded operators in a Hilbert space. If one tries to represent
it, one finds that at least one of the elements $q_{k}$ or $p_{k}$
(for all $k$), must be represented by an unbounded operator.} The canonical way to bypass this issue is to transform such self-adjoint
elements into unitaries. For that, we defined the $n$-parameter groups
of unitaries
\begin{equation}
W_{q}\left(\bar{a}\right):=\mathrm{e}^{i\bar{q}\cdot\bar{a}}\textrm{ and }W_{p}\left(\bar{b}\right):=\mathrm{e}^{i\bar{p}\cdot\bar{b}}\,,\label{c1_ccr_untobun}
\end{equation}
where $\bar{a},\bar{b}\in\mathbb{R}^{n}$. Formally, relations (\ref{c1_ccr_un1}-\ref{c1_ccr_un_2})
are converted into
\begin{align}
 & W_{p}\left(\bar{b}\right)W_{q}\left(\bar{a}\right)=\mathrm{e}^{i\bar{a}\cdot\bar{b}}W_{q}\left(\bar{a}\right)W_{p}\left(\bar{b}\right)\,,\label{c1_ccr2}\\
 & W_{q}\left(\bar{a}_{1}\right)W_{q}\left(\bar{a}_{2}\right)=W_{q}\left(\bar{a}_{1}+\bar{a}_{2}\right)\,,\;W_{q}\left(\bar{a}_{1}\right)W_{q}\left(\bar{a}_{2}\right)=W_{q}\left(\bar{a}_{1}+\bar{a}_{2}\right)\,,\\
 & W_{q}\left(\bar{a}\right)^{*}=W_{p}\left(-\bar{a}\right)\textrm{ and }W_{p}\left(\bar{b}\right)^{*}=W_{p}\left(-\bar{b}\right)\,.\label{c1_ccr3}
\end{align}
Any $N$-parameter family of unitaries $W_{q}\left(\bar{a}\right),W_{p}\left(\bar{b}\right)$
satisfying (\ref{c1_ccr2}-\ref{c1_ccr3}) defines a \textit{Weyl
algebra }$\mathcal{W}_{n}$. The Weyl algebra encodes all the same
information contained in the CCR algebra, but in a way that involves
unitary (bounded) operators. It can be shown that the Weyl algebra
$\mathcal{W}_{N}$ can be equipped with a (unique) norm $\left\Vert \cdot\right\Vert $
in order to convert it into a $C^{*}$-algebra \cite{Brattelli2}.
The Weyl algebra can be easily generalized to infinitely many degrees
of freedom. Given a real Hilbert space $\mathfrak{H}$, the Weyl algebra
$\mathcal{W}_{\mathfrak{H}}$, associated with $\mathfrak{H}$, is
defined as the algebra generated by the unitary elements $W_{\varphi}\left(f\right),W_{\pi}\left(g\right)$
with $f,g\in\mathfrak{H}$, modulo the relations
\begin{align}
 & W_{\pi}\left(g\right)W_{\varphi}\left(f\right)=\mathrm{e}^{i\left\langle f,g\right\rangle _{\mathfrak{H}}}W_{\varphi}\left(f\right)W_{\pi}\left(g\right)\,,\label{c1_ccr_inf_ini}\\
 & W_{\varphi}\left(f_{1}\right)W_{\varphi}\left(f_{2}\right)=W_{\varphi}\left(f_{1}+f_{2}\right)\,,\;W_{\pi}\left(g_{1}\right)W_{\pi}\left(g_{2}\right)=W_{\pi}\left(g_{1}+g_{2}\right)\,,\\
 & W_{\varphi}\left(f\right)^{*}=W_{\varphi}\left(-f\right)\,,\;W_{\pi}\left(g\right)^{*}=W_{\pi}\left(-g\right)\,.\label{c1_ccr_inf_fin}
\end{align}
As in the finite case, the Weyl algebra $\mathcal{W}_{\mathfrak{H}}$
can be equipped with a (unique) $C^{*}$-norm \cite{Brattelli2}.
When $\mathfrak{H}=L^{2}\left(\mathbb{R}^{d},\mathbb{R}\right)$,
\eqref{c1_ccr_untobun} suggests that we may heuristically write
\begin{equation}
W_{\varphi}\left(f\right)=\mathrm{e}^{i\int\varphi\left(x\right)f\left(x\right)\,d^{d}x}\textrm{ and }W_{\pi}\left(g\right)=\mathrm{e}^{i\int\pi\left(x\right)g\left(x\right)\,d^{d}x}\,.
\end{equation}
Performing formal manipulations to relations (\ref{c1_ccr_inf_ini}-\ref{c1_ccr_inf_fin}),
we may arrive to
\begin{align}
 & \left[\varphi\left(x\right),\pi\left(y\right)\right]=i\delta^{\left(d\right)}\left(x-y\right)\mathbf{1}\,,\quad\left[\varphi\left(x\right),\varphi\left(y\right)\right]=\left[\pi\left(x\right),\pi\left(y\right)\right]=0\,,\\
 & \varphi\left(x\right)=\varphi\left(x\right)^{*}\textrm{ and }\pi\left(x\right)=\pi\left(x\right)^{*}\,.
\end{align}
\end{example}

\subsection{States}

Given a $C^{*}$-algebra $\mathfrak{A}$, expression \eqref{c1_exp_value}
suggests that a state has to be described by a linear functional on
the algebra $\mathfrak{A}$. We denote the dual space of $\mathfrak{A}$
by $\mathfrak{A}^{*}$, i.e. $\mathfrak{A}^{*}$ is the linear space
of continuous\footnote{For technical reasons, we only consider functionals that are uniform
continuous. } linear functionals over $\mathfrak{A}$. We define the norm of any
functional $\phi\in\mathfrak{A}^{*}$ as
\begin{equation}
\left\Vert \phi\right\Vert :=sup\left\{ \left|\phi\left(A\right)\right|\,:\,\left\Vert A\right\Vert =1\right\} \,.\label{c1_dualnorm}
\end{equation}
In fact, the space $\mathfrak{A}^{*}$ equipped with such a norm is
a Banach space. The functionals of particular interest are defined
as follows.
\begin{defn}
Let $\mathfrak{A}$ be a $C^{*}$-algebra. A linear functional $\omega\in\mathfrak{A}^{*}$
is defined to be \textit{positive} if $\omega\left(A^{*}A\right)\geq0$
for all $A\in\mathfrak{A}$, and it is defined to be \textit{normalized}
if $\left\Vert \omega\right\Vert =1$. Any positive and normalized
linear functional is called a \textit{state}, and the set of all states
is denoted by $\mathfrak{S}\left(\mathfrak{A}\right)$. Any strictly
positive state, i.e. $\omega\left(A^{*}A\right)>0$ for all $A\in\mathfrak{A}$,
is called a \textit{faithful} state.
\end{defn}

\begin{rem}
In the above definition, the condition of a state to be a continuous
functional could be ignored. This is because, any positive linear
functionals over $C^{*}$-algebra is continuous. 
\end{rem}

\begin{rem}
Any positive functional $\phi\in\mathfrak{A}^{*}$ is automatically
\textit{hermitian}, i.e. $\phi\left(A\right)\in\mathbb{R}$ for all
$A\in\mathfrak{A}$ self-adjoint.
\end{rem}

\begin{rem}
Positive functionals $\phi\in\mathfrak{A}^{*}$ also satisfy $\left\Vert \phi\right\Vert =\phi\left(\mathbf{1}\right)$.
Therefore, the normalization property of a state $\omega\in\mathfrak{S}\left(\mathfrak{A}\right)$
is equivalent to $\omega\left(\mathbf{1}\right)=1.$
\end{rem}

The interpretation of states is the same as we explained in section
\ref{c1-sec-aqft}. Given a state $\omega\in\mathfrak{S}\left(\mathfrak{A}\right)$
and a self-adjoint operator $A\in\mathfrak{A}$, the real number $\omega\left(A\right)$
represents the expectation value of the physical observable represented
by $A$ in the physical state represented by $\omega$. Similarly,
given any projector $P$ representing some logic proposition (e.g.
``the result of measuring the physical observable represented by
the self-adjoint operator $A$ in the physical state represented by
$\omega$ lies in the measurable set $\Delta\subset\mathbb{R}$''),
$\omega\left(P\right)$ corresponds to the probability that such a
proposition is true, and henceforth, $\omega\left(\mathbf{1}-P\right)$
corresponds to the probability that such a proposition is false.

If $\alpha$ is an automorphism of the $C^{*}$-algebra $\mathfrak{A}$
and $\omega\in\mathfrak{S}\left(\mathfrak{A}\right)$ a state, then
$\omega\circ\alpha$ defined by $\left(\omega\circ\alpha\right)\left(A\right):=\omega\left(\alpha\left(A\right)\right)$
is also an state. When $\mathfrak{B}\subset\mathfrak{A}$ a $C^{*}$-subalgebra,
we can define the state $\left.\omega\right|_{\mathfrak{B}}\in\mathfrak{S}\left(\mathfrak{B}\right)$,
called \textit{the restriction of $\omega$ to $\mathfrak{B}$}, just
by restricting the expectation values to the subalgebra, i.e. $\left.\omega\right|_{\mathfrak{B}}\left(B\right):=\omega\left(B\right)$
for all $B\in\mathfrak{B}$. When there is no ambiguity, we denote
$\left.\omega\right|_{\mathfrak{B}}$ simply as $\omega$.

In the usual approach to quantum theory, it is well-known that any
convex combination of density matrices is a density matrix, and pure
states are given just by unit rays in the Hilbert space. The algebraic
counterpart of such facts is summarized in the following lemma and
definition.
\begin{lem}
\label{c1_convex}Let $\mathfrak{A}$ be a $C^{*}$-algebra. The set
of states $\mathfrak{S}\left(\mathfrak{A}\right)$ is a convex set,
i.e. $\lambda\omega_{1}+\left(1-\lambda\right)\omega_{2}\in\mathfrak{S}\left(\mathfrak{A}\right)$
whenever $\omega_{1},\omega_{2}\in\mathfrak{S}\left(\mathfrak{A}\right)$
and $0\leq\lambda\leq1$. 
\end{lem}

\begin{proof}
See \cite{Bratteli1}.
\end{proof}
\begin{defn}
\label{c1_purestates}Any extremal element of $\mathfrak{S}\left(\mathfrak{A}\right)$
(i.e. any state which cannot be written as a non-trivial convex combination
of any other two states) is a \textit{pure state}. Any non-pure state
is called a \textit{mixed state}. The set of pure states is denoted
by $\mathfrak{S}_{p}\left(\mathfrak{A}\right)$.
\end{defn}

\begin{rem}
The notions of pure and mixed states have no relation to the notion
of subalgebras. Indeed, if $\mathfrak{B}\subset\mathfrak{A}$ is a
$C^{*}$-subalgebra of $\mathfrak{A}$ and $\omega\in\mathfrak{S}\left(\mathfrak{A}\right)$
is a state, then $\left.\omega\right|_{\mathfrak{B}}\in\mathfrak{S}\left(\mathfrak{B}\right)$
could be pure or mixed independently of whether $\omega\in\mathfrak{S}\left(\mathfrak{A}\right)$
is pure or mixed.
\end{rem}

\begin{example}
\label{c1_exa_states}For $\mathcal{B}\left(\mathcal{H}\right)$,
the set of pure states is given by 
\begin{equation}
\mathfrak{S}_{p}\left(\mathcal{B}\left(\mathcal{H}\right)\right):=\left\{ \omega\left(\cdot\right):=\mathrm{Tr}\left(\left|\Psi\right\rangle \left\langle \Psi\right|\cdot\right)\textrm{ with }\left|\Psi\right\rangle \in\mathcal{H}\textrm{ and }\left\Vert \left|\Psi\right\rangle \right\Vert =1\right\} \,,
\end{equation}
i.e. it is in one-to-one correspondence with the set of unit rays
in the Hilbert space $\mathcal{H}$.
\end{example}

Definition \ref{c1_purestates} gives a precise definition of what
a pure state is in the algebraic setting. The question which automatically
arises if is such states exists, i.e. if the set $\mathfrak{S}_{p}\left(\mathfrak{A}\right)$
is non-empty. In order to answer that question, we have to introduce
the weak$^{*}$-topology in the set $\mathfrak{A}^{*}$. For any finite
set of operators $A_{1},\ldots,A_{n}\in\mathfrak{A}$, we can define
a seminorm $\sigma_{A_{1},\cdots,A_{n}}$ on $\mathfrak{A}^{*}$ by
\begin{equation}
\sigma_{A_{1},\cdots,A_{n}}\left(\phi\right)=\mathrm{sup}\left\{ \left|\phi\left(A_{k}\right)\right|\,:\,k=1,\ldots,n\right\} \,.
\end{equation}
The weak$^{*}$-topology in $\mathfrak{A}^{*}$ is defined as the
topology generated by the open neighborhoods
\begin{equation}
\mathcal{U}\left(\phi;A_{1},\ldots,A_{n};\epsilon\right)=\left\{ \phi'\in\mathfrak{A}^{*}\,:\,\sigma_{A_{1},\cdots,A_{n}}\left(\phi-\phi'\right)<\epsilon\right\} \,,\label{c1_neigh}
\end{equation}
for all $\phi\in\mathfrak{A}^{*}$, $A_{1},\ldots,A_{n}\in\mathfrak{A}$
and $\varepsilon>0$. In particular, a sequence $\phi_{k}\in\mathfrak{A}^{*}$
($k\in\mathbb{N}$) converges to $\phi\in\mathfrak{A}^{*}$ in the
weak$^{*}$-topology iff 
\begin{equation}
\phi_{k}\left(A\right)\underset{k\rightarrow\infty}{\longrightarrow}\phi\left(A\right)\,,\quad\textrm{for all }A\in\mathfrak{A}\,.
\end{equation}

\begin{lem}
Let $\mathfrak{A}$ be a $C^{*}$-algebra.\footnote{For this lemma, the algebra must be unital, i.e. it contains an identity
operator.} Then the set of pure states $\mathfrak{S}_{p}\left(\mathfrak{A}\right)$
is non-empty, and any state in $\mathfrak{S}\left(\mathfrak{A}\right)$
is the weak$^{*}$-limit of some sequence of finite convex combinations
of pure states.
\end{lem}

\begin{proof}
See \cite{haag}.
\end{proof}
In other words, the set of pure states $\mathfrak{S}_{p}\left(\mathfrak{A}\right)$
``generates'' the whole set of states $\mathfrak{S}\left(\mathfrak{A}\right)$,
if we allow infinite convex combinations converging in the weak$^{*}$-topology.

The weak$^{*}$-topology has also a physical meaning. Suppose that
we prepare our physical system in some physical state represented
by the state $\omega\in\mathfrak{S}\left(\mathfrak{A}\right)$. Since
in physics, we can never perform infinitely many experiments and since
each experiment has a limited accuracy we can, by monitoring the state
$\omega$, never determine more than some weak$^{*}$-neighborhood
in $\mathfrak{S}\left(\mathfrak{A}\right)$ of the form \eqref{c1_neigh}.
In other words, the state $\omega$ is ``physically indistinguishable''
from any other state belonging to such a weak$^{*}$-neighborhood,
or equivalently, any other state belonging to such a weak$^{*}$-neighborhood
may represent the same physical state than $\omega$. We will return
to this issue when we discuss the notion of physical equivalence between
representations.

\subsection{Representations\label{c1_subsec_rep}}

In general, one chooses an appropriate representation of the abstract
algebra $\mathfrak{A}$, as bounded operators in some Hilbert space,
to better describe the physical system.
\begin{defn}
Let $\mathfrak{A}$ be a $C^{*}$-algebra. A \textit{representation}
$\pi$ is a morphism $\pi:\mathfrak{A}\rightarrow\mathcal{B}\left(\mathcal{H}_{\pi}\right)$
where $\mathcal{H}_{\pi}$ is some Hilbert space depending on the
representation. $\pi$ is called the \textit{physical representation}
and $\mathcal{H}_{\pi}$ the \textit{physical Hilbert space}.
\end{defn}

The following lemma asserts that any representation is continuous,
and the image of algebra of observables $\mathfrak{A}$ under any
representation is indeed a concrete $C^{*}$-algebra.
\begin{lem}
Let $\mathfrak{A}$ be a $C^{*}$-algebra and $\pi:\mathfrak{A}\rightarrow\mathcal{B}\left(\mathcal{H}_{\pi}\right)$
a representation. Then, $\pi$ is an homomorphism and $\pi\left(\mathfrak{A}\right)$
is a $C^{*}$-subalgebra of $\mathcal{B}\left(\mathcal{H}_{\pi}\right)$.
\end{lem}

\begin{proof}
See \cite{Bratteli1}.
\end{proof}
If our physical system has finitely many coordinates and momenta which
satisfy the canonical commutation relations (example \ref{c1_weyl_alg}),
the choice of representation is uniquely determined (up to unitary
equivalence) by the Stone-von Neumann theorem.
\begin{thm}
(Stone-von Neumann) Let $\mathcal{W}_{n}$ the Weyl algebra of example
\ref{c1_weyl_alg}. Then there exists a unique (up to unitary equivalence)
irreducible representation $\pi:\mathcal{W}_{n}\rightarrow L^{2}\left(\mathbb{R}^{n}\right)$,
called the Schrodinger's representation. It is given by
\begin{equation}
\pi\left(W_{q}\left(\bar{a}\right)\right)\psi\left(\bar{x}\right)=\mathrm{e}^{i\bar{x}\cdot\bar{a}}\psi\left(\bar{x}\right)\textrm{ and }\pi\left(W_{p}\left(\bar{b}\right)\right)\psi\left(\bar{x}\right)=\psi\left(\bar{x}+\bar{b}\right)\,.
\end{equation}
If we define $\pi\left(W_{q}\left(\bar{a}\right)\right):=\mathrm{e}^{i\bar{q}\cdot\bar{a}}$
and $\pi\left(W_{p}\left(\bar{b}\right)\right):=\mathrm{e}^{i\bar{p}\cdot\bar{b}}$,
then we obtain the well-known expressions
\begin{equation}
q_{j}\psi\left(\bar{x}\right)=x_{j}\psi\left(\bar{x}\right)\textrm{ and }p_{j}\psi\left(\bar{x}\right)=-i\frac{\partial}{\partial x_{j}}\psi\left(\bar{x}\right)\,.
\end{equation}
Furthermore, any reducible representation is a copy of Schrodinger's
representation.
\end{thm}

\begin{proof}
See \cite{Brattelli2}.
\end{proof}
However, if the system has infinitely many degrees of freedom, as
in QFT, the Stone-von Neumann theorem is no longer applicable, and
in such cases, there are many unitarily inequivalent representations
of the canonical commutation relations \cite{Bogolyubov}. Therefore,
for such systems the physical representation $\pi$ should be chosen
carefully, depending on the particular dynamics of the system at hand;
for instance, the Fock representation cannot be used for interacting
fields. Without loss of generality, we may always assume that the
physical representation $\pi$ is \textit{faithful}, i.e. that $\pi$
is injective. The reason for this is as follows. Suppose that it would
have been possible to physically describe a quantum system by using
a non-faithful representation $\pi$ of the algebra of observables
$\mathfrak{A}$. Then the representation $\pi$ defines a faithful
representation $\tilde{\pi}$ of the quotient $\tilde{\mathfrak{A}}:=\mathfrak{A}/\mathrm{ker}\left(\pi\right)$,
and we could just as well have started with this quotient algebra
(as the algebra of observables) from the beginning.

Once a representation is specified, we can identify a preferred subset
of the set of all states.
\begin{defn}
Let $\mathfrak{A}$ be a $C^{*}$-algebra and $\pi:\mathfrak{A}\rightarrow\mathcal{B}\left(\mathcal{H}_{\pi}\right)$
a representation. A state $\omega\in\mathfrak{S}\left(\mathfrak{A}\right)$
is said to be \textit{normal with respect to the representation} $\pi$
or $\pi$\textit{-normal} if there exists a density matrix $\rho_{\omega}$
such that $\omega\left(\cdot\right)=\mathrm{Tr}\left(\rho_{\omega}\pi\left(\cdot\right)\right)$.
The set of all $\pi$\textit{-}normal states is a convex subset of
the set of all sates $\mathfrak{S}\left(\mathfrak{A}\right)$. It
is called the \textit{folium of }$\pi$ and it is denoted by $\mathfrak{S}^{\left(\pi\right)}\left(\mathfrak{A}\right)$.
Furthermore, any normal state of the form $\omega\left(\cdot\right)=\mathrm{Tr}\left(\left|\Psi\right\rangle \left\langle \Psi\right|\pi\left(\cdot\right)\right)$,
with unit vector $\left|\Psi\right\rangle \in\mathcal{H}_{\pi}$,
is called a \textit{vector state of the representation} $\pi$.
\end{defn}

The set of all $\pi$\textit{-}normal states is the subset of the
set of all states that, in some way, have a simpler description in
the representation $\pi$. In the rest of this subsection, we explain
that the choice of the physical representation is a sort of mathematical
convenience. In other words, if we want to study some physical processes
involving a particular state, the best idea is to pick up a representation
where such a state is represented by a density matrix, or even better,
is represented by a vector state.\footnote{It is always possible. See theorem \eqref{c1_gns}.}
Indeed, it can be defined a notion of ``physical equivalence'' between
representations, that relates representations whose set of normal
states can not be distinguished performing finitely many experiments
with limited accuracy. In the end, it can be proven, that all faithful
representations of the algebra of observables are physically equivalent.

The following theorem asserts that any general state of $\mathfrak{S}\left(\mathfrak{A}\right)$
can be approximated by $\pi$\textit{-}normal states.
\begin{lem}
The set of normal states $\mathfrak{S}^{\left(\pi\right)}\left(\mathfrak{A}\right)$
of any faithful representation $\pi$ is weakly$^{*}$-dense in the
set of all states $\mathfrak{S}\left(\mathfrak{A}\right)$.
\end{lem}

\begin{proof}
See \cite{haag}.
\end{proof}
The following definition concerns two different notions of equivalence
between representations.
\begin{defn}
Let $\mathfrak{A}$ be a $C^{*}$-algebra, and $\pi_{1}:\mathfrak{A}\rightarrow\mathcal{B}\left(\mathcal{H}_{\pi_{1}}\right)$
and $\pi_{2}:\mathfrak{A}\rightarrow\mathcal{B}\left(\mathcal{H}_{\pi_{1}}\right)$
two representations. We say that
\end{defn}

\begin{itemize}
\item $\pi_{1}$ is \textit{unitarily equivalent} to $\pi_{2}$ ($\pi_{1}\cong\pi_{2}$)
if exists a unitary operator $U:\mathcal{H}_{\pi_{1}}\rightarrow\mathcal{H}_{\pi_{2}}$
such that $U\pi_{1}\left(A\right)U^{*}=\pi_{2}\left(A\right)$ for
all $A\in\mathfrak{A}$,
\item and $\pi_{1}$ is \textit{phenomenologically} \textit{equivalent}\footnote{In the literature, \textit{quasi-equivalent} is used as a synonym
of phenomenological equivalent.} to $\pi_{2}$ if $\mathfrak{S}^{\left(\pi_{1}\right)}\left(\mathfrak{A}\right)=\mathfrak{S}^{\left(\pi_{2}\right)}\left(\mathfrak{A}\right)$.
\end{itemize}
The notion of unitary equivalence says that both representations are
essentially equal from the mathematical point of view, wheres the
notion of phenomenological equivalence asserts that both representations
are good for the study of the same subset of states. Of course, any
two unitarily equivalent representations are also phenomenological
equivalent, but the converse is false. The third notion of equivalence
that we introduce is physical equivalence.
\begin{defn}
Let $\mathfrak{A}$ be a $C^{*}$-algebra and $\pi_{1}:\mathfrak{A}\rightarrow\mathcal{B}\left(\mathcal{H}_{\pi_{1}}\right)$
and $\pi_{2}:\mathfrak{A}\rightarrow\mathcal{B}\left(\mathcal{H}_{\pi_{1}}\right)$
two representations. We say that $\pi_{1}$ is \textit{physically
equivalent} to $\pi_{2}$ if for every state $\omega_{1}\in\mathfrak{S}^{\left(\pi_{1}\right)}\left(\mathfrak{A}\right)$
and every weak$^{*}$-neighborhood $\mathcal{U}\left(\omega_{1};A_{1},\ldots,A_{n};\epsilon\right)$
of $\omega_{1}$ theres exists a state $\omega_{2}\in\mathfrak{S}^{\left(\pi_{2}\right)}\left(\mathfrak{A}\right)$
such that $\omega_{2}\in\mathcal{U}\left(\omega_{1};A_{1},\ldots,A_{n};\epsilon\right)$.
\end{defn}

\begin{rem}
The definition of physical equivalence does not look like symmetric
between $\pi_{1}$ and $\pi_{2}$, however, it can be shown that it
is indeed an equivalence relation.
\end{rem}

The physical significance of the above definition comes from the following
mental experiment. One starts with a physical state\footnote{Here $\Omega$ does not represent any object in any mathematical space.
It is just a way to call the real physical state prepared in our laboratory.} $\Omega$ that is represented by the element $\omega_{1}\in\mathfrak{S}^{\left(\pi_{1}\right)}\left(\mathfrak{A}\right)$
belonging to the folium of some representation $\pi_{1}$. But, how
could we ensure that $\omega_{1}$ indeed represents the physical
state $\Omega$? What we do in a real experiment, is to take a collection
of observables represented by selfadjoint operators $A_{1},\ldots,A_{n}\in\mathfrak{A}$
and measure such observables in our physical state $\Omega$, to obtain
the results $a_{1},\ldots,a_{n}\in\mathbb{R}$.\footnote{We have to repeat the experiment many times in order to pursue all
such measurements.} Since we can only measure a finite number of observables, and since
each observation has a finite precision, the claim that our state
$\omega_{1}$ represents the physical state $\Omega$, actually means
that
\begin{equation}
\left|\omega_{1}\left(A_{k}\right)-a_{k}\right|<\epsilon_{k}\,,\quad\textrm{for all }k=1,\ldots,n\,,\label{c1_measure}
\end{equation}
where $\epsilon_{k}>0$ represents the accuracy of such measurements.
Equation \eqref{c1_measure} tells us that any real experiment does
not determine the mathematical state $\omega_{1}$ uniquely, but instead
a weak$^{*}$-neighborhood of it. Hence, for theoretical computations,
any other mathematical state $\omega_{2}$ belonging to such a weak$^{*}$-neighborhood
represents the same physical state $\Omega$. Two representations
$\pi_{1}$ and $\pi_{2}$ are physical equivalent if the folium of
$\pi_{1}$ is indistinguishable from the folium of $\pi_{2}$ after
performing any real experiment. The foregoing discussion culminates
in the following theorem.
\begin{thm}
Any two faithful representations of the algebra of observables are
physically equivalent.
\end{thm}

\begin{proof}
See \cite{Haag63}.
\end{proof}
In other words, all faithful representations describe the same physics.
The selection of a particular representation is a matter of convenience
without physical implications. If one wants to study some process
related to some particular state $\omega\in\mathfrak{S}\left(\mathfrak{A}\right)$,
it is better to choose a representation where such a state is normal,
i.e. is described by a density matrix. As a general conclusion, the
physics is encoded in the algebra of observables $\mathfrak{A}$ itself.\footnote{And of course, in the particular chosen state.}

\subsection{Relation between states and representations\label{c1-sec-rep-states}}

The following theorem not only ensures that representations already
exist, but also that any state $\omega\in\mathfrak{S}\left(\mathfrak{A}\right)$
can be viewed as a vector state in some representation.\footnote{This is the algebraic abstract version of purification, which is well-known
for finite dimensional algebras or full operators algebras $\mathcal{B}\left(\mathcal{H}\right)$.}
\begin{thm}
\label{c1_gns}(GNS construction) Let $\mathfrak{A}$ be a $C^{*}$-algebra
and $\omega\in\mathfrak{S}\left(\mathfrak{A}\right)$ a state. There
exists a unique (up to unitarily equivalence) representation $\pi_{\omega}:\mathfrak{A}\rightarrow\mathcal{B}\left(\mathcal{H}_{\omega}\right)$
with cyclic\footnote{Cyclic vector means that $\pi_{\omega}\left(A\right)\left|\Omega\right\rangle $
is dense in $\mathcal{H}_{\omega}$.} vector $\left|\Omega\right\rangle \in\mathcal{H}_{\omega}$ such
that
\begin{equation}
\omega\left(A\right)=\left\langle \Omega\right|\pi_{\omega}\left(A\right)\left|\Omega\right\rangle \quad\textrm{for all }A\in\mathfrak{A}\,.
\end{equation}
The above representation is usually called as the GNS-representation
of the state $\omega$.
\end{thm}

\begin{proof}
Here we give an sketch of the proof. For a complete mathematical proof
see \cite{Bratteli1}. The idea is to start defining the pre-Hilbert
space $\mathcal{H}_{0}\equiv\mathfrak{A}$ as a set, with inner product
given by $\left\langle A\mid B\right\rangle _{0}=\omega\left(A^{*}B\right)$.
Then we define the set\footnote{$\mathcal{N}$ is a left ideal of the algebra $\mathfrak{A}$.}
of null vectors $\mathcal{N}:=\left\{ A\in\mathcal{H}_{0}\,:\,\omega\left(A^{*}A\right)=0\right\} $.
The Hilbert space $\mathcal{H}_{\omega}$ is defined as the completation
of the quotient $\mathcal{H}_{0}/\mathcal{N}$, where the completation
is in the norm derived from $\left\langle \cdot\mid\cdot\right\rangle _{0}$.
A dense subset of vectors in $\mathcal{H}_{\omega}$ is given by the
equivalence class of elements in $\mathcal{H}_{0}=\mathfrak{A}$,
which are simply denoted by $\left|A\right\rangle $ with $A\in\mathfrak{A}$.
The inner product in $\mathcal{H}_{\omega}$ is the one derived from
$\mathcal{H}_{0}$, i.e. $\left\langle A\mid B\right\rangle =\omega\left(A^{*}B\right)$.
The representation is defined by $\pi_{\omega}\left(A\right)\left|B\right\rangle =\left|AB\right\rangle $
over the elements in $\mathcal{H}_{0}/\mathcal{N}$ and extended by
continuity to the whole $\mathcal{H}_{\omega}$. The unit vector $\left|\Omega\right\rangle $
is represented (up to a phase) by $\left|\mathbf{1}\right\rangle $,
where $\mathbf{1}\in\mathfrak{A}$ is the identity operator.
\end{proof}
The above theorem gives us the first relation between states and representations.
The following theorem shows that the notions of purity of the state
and irreducibility of its GNS-representation are intimately related.
\begin{thm}
Let $\mathfrak{A}$ a $C^{*}$-algebra, $\omega\in\mathfrak{S}\left(\mathfrak{A}\right)$
a state and $\pi_{\omega}:\mathfrak{A}\rightarrow\mathcal{B}\left(\mathcal{H}_{\omega}\right)$
its GNS-representation. Then $\pi_{\omega}$ is irreducible if and
only if $\omega$ is pure.\footnote{A representation $\pi:\mathfrak{A}\rightarrow\mathcal{B}\left(\mathcal{H}\right)$
is said to be \textit{irreducible} if $\mathcal{H}$ has no proper
closed $\pi(\mathfrak{A})$-invariant subspaces. According to Schur's
lemma, $\pi$ is irreducible if and only if $\pi(\mathfrak{A})'=\left\{ \lambda\cdot\mathbf{1}\right\} $.}
\end{thm}

\begin{proof}
See \cite{Bratteli1}.
\end{proof}
There is a second relation between states and representations, based
on the fact that the set of pure states $\mathfrak{S}_{p}\left(\mathfrak{A}\right)$
is decomposed as a disjoint union of subsets called coherent sets
or sectors. For that, we define the \textit{transition probability}
$\omega_{1}\cdot\omega_{2}$ between two pure states $\omega_{1},\omega_{2}\in\mathfrak{S}_{p}\left(\mathfrak{A}\right)$
as
\begin{equation}
\omega_{1}\cdot\omega_{2}=1-\frac{1}{4}\left\Vert \omega_{1}-\omega_{2}\right\Vert \,,\label{c1_trans_prob}
\end{equation}
where the norm is the one defined in \eqref{c1_dualnorm}.\footnote{Equation \eqref{c1_trans_prob} coincides with the usual expression
$\omega_{1}\cdot\omega_{2}=\left|\left\langle \Psi_{1}\mid\Psi_{2}\right\rangle \right|^{2}$
when $\mathfrak{A}=\mathcal{B}\left(\mathcal{H}\right)$ and the pure
states are defined by unit vectors $\left|\Psi_{i}\right\rangle \in\mathcal{H}$,
i.e. $\omega_{i}\left(\cdot\right)=\mathrm{Tr}\left(\left|\Psi_{i}\right\rangle \left\langle \Psi_{i}\right|\cdot\right)$.} Because $0\leq\left\Vert \omega_{1}-\omega_{2}\right\Vert \leq\left\Vert \omega_{1}\right\Vert +\left\Vert \omega_{2}\right\Vert $,
it is clear that $\omega_{1}\cdot\omega_{2}\in\left[0,1\right]$,
and it follows from the positive-definiteness of $\left\Vert \cdot\right\Vert $
that $\omega_{1}\cdot\omega_{2}=1$ if and only if $\omega_{1}=\omega_{2}$.
When $\omega_{1}\cdot\omega_{2}=0$, we say that the states $\omega_{1}$
and $\omega_{2}$ are \textit{orthogonal}. Two subsets $\mathcal{S}_{1},\mathcal{S}_{2}\subset\mathfrak{S}_{p}\left(\mathfrak{A}\right)$
are called \textit{mutually orthogonal} if $\omega_{1}\cdot\omega_{2}=0$
for all $\omega_{1}\in\mathcal{S}_{1}$ and $\omega_{2}\in\mathcal{S}_{2}$,
and it is denoted by $\mathcal{S}_{1}\bot\mathcal{S}_{2}$. This motivates
the following definition.
\begin{defn}
A non-empty subset $\mathcal{S}\subset\mathfrak{S}_{p}\left(\mathfrak{A}\right)$
is called \textit{indecomposable} iff it cannot be written as the
union of two non-empty, mutually disjoint and mutually orthogonal
subsets, i.e. $\cancel{\exists}\mathcal{S}_{1},\mathcal{S}_{2}\subset\mathfrak{S}_{p}\left(\mathfrak{A}\right)$
such that $\mathcal{S}_{1},\mathcal{S}_{2}\neq\emptyset$, $\mathcal{S}_{1}\cap\mathcal{S}_{2}=\emptyset$,
$\mathcal{S}_{1}\bot\mathcal{S}_{2}$ and $\mathcal{S}=\mathcal{S}_{1}\cup\mathcal{S}_{2}$. 
\end{defn}

\begin{proof}
See \cite{Bogolyubov}.
\end{proof}
With the help of the above definition, we define the following relation
on $\mathfrak{S}_{p}\left(\mathfrak{A}\right)$.
\begin{defn}
Let $\omega_{1},\omega_{2}\in\mathfrak{S}_{p}\left(\mathfrak{A}\right)$.
We define the relation $\omega_{1}\sim\omega_{2}$ if there exists
an indecomposable set $\mathcal{S}\subset\mathfrak{S}_{p}\left(\mathfrak{A}\right)$
such that $\omega_{1},\omega_{2}\in\mathcal{S}$.
\end{defn}

\begin{lem}
The relation $\omega_{1}\sim\omega_{2}$ on $\mathfrak{S}_{p}\left(\mathfrak{A}\right)$
is an equivalence relation. Then the set of pure states can be uniquely
partitioned into non-empty sets (mutually disjoint and mutually orthogonal),
which are precisely the equivalence classes in $\mathfrak{S}_{p}\left(\mathfrak{A}\right)$.
Furthermore, each equivalence class is an indecomposable set.
\end{lem}

\begin{proof}
See \cite{Bogolyubov}.
\end{proof}
Each equivalence class in above lemma is called a \textit{(superselection)
sector}. We usually write $\mathfrak{S}_{p}\left(\mathfrak{A}\right)=\bigcup_{\alpha\in\mathcal{I}}^{\left(d\right)}\mathcal{C}_{\alpha}$,
where $\left\{ \mathcal{C}_{\alpha}\right\} _{\alpha\in\mathcal{I}}$
are the sectors and $\left(d\right)$ denotes that union is along
non-empty, mutually disjoint ($\mathcal{C}_{\alpha}\cap\mathcal{C}_{\beta}=\emptyset$
if $\alpha\neq\beta$) and mutually orthogonal ($\mathcal{C}_{\alpha}\bot\mathcal{C}_{\beta}=\emptyset$
if $\alpha\neq\beta$) sets. The following example helps us to better
understand the above definitions.
\begin{example}
Let $\mathcal{H}_{1},\mathcal{H}_{2}$ be Hilbert spaces and define
$\mathfrak{A}:=\mathcal{B}\left(\mathcal{H}_{1}\right)\oplus\mathcal{B}\left(\mathcal{H}_{2}\right)$.
Any operator $A\in\mathfrak{A}$ can be uniquely represented as $A=\left(A_{1},A_{2}\right)$,
with $A_{k}\in\mathcal{B}\left(\mathcal{H}_{k}\right)$ for $k=1,2$.
The algebraic operations in $\mathfrak{A}$ are component-wise, and
within each component are just the usual operations in $\mathcal{B}\left(\mathcal{H}_{k}\right)$.
Following definition \ref{c1_purestates} we find that the set of
pure states $\mathfrak{S}_{p}\left(\mathfrak{A}\right)$ is in one-to-one
correspondence with the subset of vectors $\mathcal{H}_{1}\cup\mathcal{H}_{2}$,
where $\cup$ denotes the union of sets, and not the direct sum. In
fact, any state given by a vector of the form $\alpha_{1}\left|\Psi_{1}\right\rangle +\alpha_{2}\left|\Psi_{2}\right\rangle $,
with $\left|\Psi_{k}\right\rangle \in\mathcal{H}_{k}$ and $\alpha_{k}\neq0$,
is a mixed state for $\mathfrak{A}$. It correspond to the state $\left|\alpha_{1}\right|^{2}\omega_{1}+\left|\alpha_{2}\right|^{2}\omega_{2}$,
where $\omega_{i}\left(\cdot\right)=\mathrm{Tr}\left(\left|\Psi_{i}\right\rangle \left\langle \Psi_{i}\right|\cdot\right)$.
The set of pure states $\mathfrak{S}_{p}\left(\mathfrak{A}\right)$
is not indecomposable since the subsets of pure states represented
by $\mathcal{H}_{1}$ and $\mathcal{H}_{2}$ gives a non-trivial decomposition
of $\mathfrak{S}_{p}\left(\mathfrak{A}\right)$. Furthermore, the
subsets of pure states represented by $\mathcal{H}_{1}$ and $\mathcal{H}_{2}$
are themselves indecomposable. Hence, the sectors are in one-to-one
correspondence with the subspaces $\left\{ \mathcal{H}_{1},\mathcal{H}_{2}\right\} $.
\end{example}

There is a close relation between sectors and irreducible representations.
It is summarized in the following theorem.
\begin{thm}
Each superselection sector of $\mathfrak{S}_{p}\left(\mathfrak{A}\right)$
coincides with the set of vector states in some irreducible representation.
Conversely, the set of vector states of an irreducible representation
coincides with some superselection sector of $\mathfrak{S}_{p}\left(\mathfrak{A}\right)$.
Moreover, sets of vector states of unitarily equivalent representations
corresponds to the same sector.
\end{thm}

\begin{proof}
See \cite{Bogolyubov}.
\end{proof}
The above theorem gives the famous one-to-one correspondence between
sectors of pure states and irreducible representations. Since any
other state of $\mathfrak{S}\left(\mathfrak{A}\right)$ is a weak$^{*}$-limit
of convex combinations of pure states, then for computational purposes,
it is enough to consider only unitarily inequivalent irreducible representations.

\subsection{von Neumann algebras}

Among the class of concrete $C^{*}$-algebras,\footnote{Remember that due to theorem \ref{c1_iso}, we do not loose generality
if we only consider concrete $C^{*}$-algebras. } there is the subclass of von Neumann algebras. To properly define
it, we have to introduce the \textit{weak operator topology} in $\mathcal{B}\left(\mathcal{H}\right)$.
It is defined by the set of seminorms $\varrho_{\Psi,\Phi}:\mathcal{B}\left(\mathcal{H}\right)\rightarrow\mathbb{R}_{\geq0}$,
\begin{equation}
\varrho_{\Psi,\Phi}\left(A\right):=\left|\left\langle \Psi\right|A\left|\Phi\right\rangle \right|\,,\label{c1_weak_op_topo_1}
\end{equation}
where $\left|\Psi\right\rangle ,\left|\Phi\right\rangle \in\mathcal{H}$.
In particular, a sequence $A_{k}\in\mathcal{B}\left(\mathcal{H}\right)$
($k\in\mathbb{N}$) converges to $A\in\mathcal{B}\left(\mathcal{H}\right)$
in the weak operator topology iff 
\begin{equation}
\left\langle \Psi\right|A_{k}\left|\Phi\right\rangle \underset{k\rightarrow\infty}{\longrightarrow}\left\langle \Psi\right|A\left|\Phi\right\rangle \,,\quad\textrm{for all }\left|\Psi\right\rangle ,\left|\Phi\right\rangle \in\mathcal{H}\,.\label{c1_weak_op_topo_2}
\end{equation}

\begin{defn}
\label{c1_vN_algebra}Any $*$-subalgebra $\mathcal{A}\subset\mathcal{B}\left(\mathcal{H}\right)$
that contains the identity operator $\mathbf{1}$ and is closed in
the weak operator topology is called a \textit{von Neumann (vN) algebra}.
\end{defn}

Since the weak operator topology is weaker than the uniform topology,
we have that any vN algebra $\mathcal{\mathcal{A}}\subset\mathcal{B}\left(\mathcal{H}\right)$
is also a concrete $C^{*}$-algebra. The converse is false.\footnote{For finite-dimensional algebras, both notions coincide.}
In fact, given any $C^{*}$-algebra $\mathfrak{A}\subset\mathcal{B}\left(\mathcal{H}\right)$
we can define the \textit{vN algebra generated} by $\mathfrak{A}$
simply as its closure in the weak operator topology, i.e. including
in the algebra $\mathfrak{A}$ the limit points of weak convergent
sequences. 
\begin{rem}
Besides the uniform topology (see definition \ref{c1_uniftopo}) and
the weak operator topology (see equations (\ref{c1_weak_op_topo_1}-\ref{c1_weak_op_topo_2}))
there are other five natural topologies in $\mathcal{B}\left(\mathcal{H}\right)$:
$\sigma$-weak, $\sigma$-weak$^{*}$, strong, $\sigma$-strong, $\sigma$-strong$^{*}$.
The above seven topologies are inequivalent over $\mathcal{B}\left(\mathcal{H}\right)$,
but a $*$-subalgebra $\mathcal{\mathcal{A}}\subset\mathcal{B}\left(\mathcal{H}\right)$
is closed in the weak operator topology iff is closed in any of the
other five topologies (excluding the uniform topology). In other words,
in definition \ref{c1_vN_algebra}, we can replace the closeness in
the weak operator topology with any of the other five topologies (except
the uniform topology). See \cite{Bratteli1}.
\end{rem}

\begin{rem}
vN algebras could be defined in an abstract sense without referring
to any Hilbert space. Such objects are known as $W^{*}$-algebras.
Fortunately, there is a theorem that asserts that any $W^{*}$-algebra
is equivalent to a vN subalgebra of $\mathcal{B}\left(\mathcal{H}\right)$.
See \cite{sakai}.
\end{rem}

The famous von Neumann bicommutant theorem gives a pure algebraic
characterization for vN algebras. To state it, we have first to introduce
the following definition.
\begin{defn}
Let $\mathcal{S}\subset\mathcal{B}\left(\mathcal{H}\right)$. We define
its \textit{commutant} as 
\begin{equation}
\mathcal{S}':=\left\{ A\in\mathcal{B}\left(\mathcal{H}\right)\,:\,\left[A,B\right]=0\textrm{ for all }B\in\mathcal{S}\right\} \,,
\end{equation}
where $\left[A,B\right]=AB-BA$ denotes the \textit{commutator}. The
\textit{bicommutant} of $\mathcal{S}$ is simply defined as $\mathcal{S}'':=\left(\mathcal{S}'\right)'$.
Of course, we have that $\mathcal{S}\subset\mathcal{S}''$.
\end{defn}

\begin{thm}
(von Neumann bicommutant theorem) Let $\mathcal{S}\subset\mathcal{B}\left(\mathcal{H}\right)$
be a self-adjoint set.\footnote{$\mathcal{S}\subset\mathcal{B}\left(\mathcal{H}\right)$ is called
a self-adjoint set if $A\in\mathcal{S}\Rightarrow A^{*}\in\mathcal{S}$.} Then $\mathcal{S}'$ is a vN algebra and $\mathcal{S}''$ is the
smallest vN algebra containing $\mathcal{S}$. Furthermore, $\left(\mathcal{S}''\right)'=\mathcal{S}'$.
In particular, if $\mathfrak{A}\subset\mathcal{B}\left(\mathcal{H}\right)$
is a $C^{*}$-algebra, then $\mathfrak{A}''$ is the closure of $\mathfrak{A}$
in the weak operator topology, i.e. $\mathfrak{A}''$ is the vN algebra
generated by $\mathfrak{A}$.
\end{thm}
\begin{proof}
See \cite{Bratteli1}.
\end{proof}

\begin{cor}
$\mathcal{B}\left(\mathcal{H}\right)$ and $\mathbf{1}:=\left\{ \lambda\mathbf{1}\,:\,\lambda\in\mathbb{C}\right\} $
are vN algebras.
\end{cor}

The above theorem is very useful for practical purposes. Suppose that
we have a quantum system represented by $\mathcal{B}\left(\mathcal{H}\right)$,
but we have at our disposal some subset of observables represented
by the self-adjoint operators $\left\{ A_{1},A_{2},\ldots\right\} $.
Then, $\left\{ A_{1},A_{2},\ldots\right\} ''$ is the smallest vN
algebra generated by such a set of operators. 

Now we introduce some useful definitions.
\begin{defn}
The \textit{center} of a vN algebra $\mathcal{A}\subset\mathcal{B}\left(\mathcal{H}\right)$
is defined as $\mathcal{Z}\left(\mathcal{A}\right):=\mathcal{A}\cap\mathcal{A}'$,
which is also a vN algebra. If the center is trivial, i.e. $\mathcal{Z}\left(\mathcal{A}\right)=\mathbf{1}$,
we say that $\mathcal{A}$ is a \textit{factor}.
\end{defn}

\begin{defn}
Let $\mathcal{A}_{1},\mathcal{A}_{2}\subset\mathcal{B}\left(\mathcal{H}\right)$
be vN algebras. If $\mathcal{A}_{1}\subset\mathcal{A}_{2}$ we say
that $\mathcal{A}_{1}$ is a \textit{vN subalgebra} of $\mathcal{A}_{2}$.
In particular, $\mathcal{A}_{1}$ and $\mathcal{A}_{2}$ are vN subalgebras
of $\mathcal{B}\left(\mathcal{H}\right)$.
\end{defn}

\begin{defn}
Let $\mathcal{A}\subset\mathcal{B}\left(\mathcal{H}\right)$ be a
vN algebra and $\left|\Psi\right\rangle \in\mathcal{H}$ a vector.
We say that $\left|\Psi\right\rangle $ is \textit{cyclic} for $\mathcal{A}$
if $\mathcal{A}\left|\Psi\right\rangle $ is dense in $\mathcal{H}$,
and we say that $\left|\Psi\right\rangle $ is \textit{separating}
for $\mathcal{A}$ if $\cancel{\exists}A\in\mathcal{A}$ such that
$A\left|\Psi\right\rangle =0$. Any vector which is cyclic and separating
is called \textit{standard}.
\end{defn}

\begin{lem}
Let $\mathcal{A}\subset\mathcal{B}\left(\mathcal{H}\right)$ be a
vN algebra and $\left|\Psi\right\rangle \in\mathcal{H}$ a vector.
Then, $\left|\Psi\right\rangle $ is cyclic (rep. separating) for
$\mathcal{A}$ if and only if $\left|\Psi\right\rangle $ is separating
(rep. cyclic) for $\mathcal{A}'$.
\end{lem}
\begin{proof}
See \cite{Bratteli1}.
\end{proof}
vN algebras can be thought of as abstract algebras represented in
some Hilbert space. In fact, any representation $\pi:\mathfrak{A}\rightarrow\mathcal{B}\left(\mathcal{H}_{\pi}\right)$
of some $C^{*}$-algebra $\mathfrak{A}$, gives place naturally to
the vN algebra $\pi\left(\mathfrak{A}\right)''$. A vN algebra has
naturally a preferred set of normal states, i.e. all the states represented
by density matrices in the Hilbert space where the vN algebra acts.

Despite that there is a very clear mathematical distinction between
$C^{*}$-algebras and vN algebras, there is no general consensus if
the algebras of observables of quantum systems may be described by
uniform or weakly closed algebras. Both type of algebras are used
in the literature, and sometimes simultaneously in the description
of a system. Below we give some partial arguments in favor of vN algebras
with respect to  $C^{*}$-algebras. We also do not have to forget
that vN algebras are also $C^{*}$-algebras, and hence, all the discussion
we have done about $C^{*}$-algebras also applies to vN algebras.
\begin{enumerate}
\item When we construct examples of algebras of observables (in particular
in algebraic quantum field theory), we first define some set of observables
through concrete mathematical expressions, and then, we define the
algebra of observables as the closure of such a set. From the technical
point of view, it is easier to take a weak closure than the uniform
closure. It is because the weak closure is equivalent to the double
commutant, which is an algebraic operation.
\item In quantum information theory, most of the information/statistical/entanglement
measures have easier definitions for vN algebras rather than for $C^{*}$-algebras.
In fact, some measures over $C^{*}$-algebras are simply defined in
term of the vN algebra generated of some particular representation
(e.g. the GNS-representation of some state).
\item In QFT, in the DHR theory of superselection sectors (see chapter \ref{EE_SS}),
the local algebras corresponding to bounded regions are defined to
be vN algebras in order to have a simpler description of the structure
of superselection sectors. However, the algebras associated with unbounded
regions are defined to be $C^{*}$-algebras but not vN algebras.
\item In QFT, it is well-known that operators belonging to spacelike separated
regions commute. In fact, for any spacetime region $\mathcal{O}_{1}$
we have an associated algebra of observables $\mathfrak{A}\left(\mathcal{O}_{1}\right)$.\footnote{We make such a correspondence clearer in the following section.}
Suppose for a moment that all local algebras are subalgebras of a
concrete algebra $\mathfrak{A}\subset\mathcal{B}\left(\mathcal{H}\right)$.
Given any region $\mathcal{O}_{2}$ spacelike separated to $\mathcal{O}_{1}$,
we must have $\mathfrak{A}\left(\mathcal{O}_{2}\right)\subset\mathfrak{A}\left(\mathcal{O}_{1}\right)'$.
$\mathfrak{A}\left(\mathcal{O}_{1}\right)$ may not be a vN algebra,
but $\mathfrak{A}\left(\mathcal{O}_{1}\right)'$ is always a vN algebra.
vN algebras arise naturally in QFT when we impose the principle of
microcausality.
\end{enumerate}
Finally, we want to remark that the set of vN subfactors of $\mathcal{B}\left(\mathcal{H}\right)$
has a natural structure of orthocomplemented lattice.\footnote{See appendix \ref{APP_LATTICE} for the definition of orthocomplemented
lattice.}
\begin{prop}
\label{c1_lat_vn}The set of vN subfactors of $\mathcal{B}\left(\mathcal{H}\right)$
has a natural structure of orthocomplemented lattice (ordered by inclusion)
given by
\begin{align}
 & \left\{ \lambda\mathbf{1}\right\} \subset\mathcal{A}\subset\mathcal{B}\left(\mathcal{H}\right)\,,\\
 & \mathcal{A}_{1}\vee\mathcal{A}_{2}=\left(\mathcal{A}_{1}\cup\mathcal{A}_{2}\right)''\,,\\
 & \mathcal{A}_{1}\land\mathcal{A}_{2}=\mathcal{A}_{1}\cap\mathcal{A}_{2}\,,
\end{align}
where $\mathcal{A},\mathcal{A}_{1},\mathcal{A}_{2}\subset\mathcal{B}\left(\mathcal{H}\right)$
are any vN subalgebras and $\cup$ (resp. $\cap$) denote the union
(resp. intersection) of sets. The complementation in the lattice is
given by the commutant $\mathcal{A}\rightarrow\mathcal{A}'$.
\end{prop}

\begin{proof}
See \cite{haag}.
\end{proof}
Such a structure plays an important role in the correspondence between
algebras and regions in QFT since the set of causally complete regions
of the spacetime has also a natural structure of orthocomplemented
lattice. We will extend this discussion in subsection \ref{c1_subsec_lattice}.

\section{Algebraic quantum field theory}

In the algebraic approach to quantum field theory (AQFT), we associate
for each region of the spacetime a $C^{*}$ or vN algebra which encodes
the algebraic relations between the quantum fields. Such an assignment
must satisfy a set of axioms that encode the physical conditions in
the algebraic framework. Unless the specific set of axioms considered
could depend on the underlying theory (especially on the spacetime
considered), the minimal assumptions we list below are very standard
for the treatment of QFT on Minkowski spacetime. 

\subsection{Axioms of AQFT\label{c1-sec_aqft}}

To start, we define the $d$-dimensional Minkowski spacetime as the
smooth manifold $\mathbb{R}^{d}$ with Lorentzian metric $\eta_{\mu\nu}=\mathrm{diag}\left(+1,-1,\ldots,-1\right)$.
The (proper orthochronous) Poincaré group is denoted by $\mathcal{P}_{+}^{\uparrow}$,
and a general element $g\in\mathcal{P}_{+}^{\uparrow}$ is denoted
by $g=\left(\Lambda,a\right)$, where $\Lambda\in\mathcal{L}_{+}^{\uparrow}$
is a (proper orthochronous) Lorentz matrix and $a\in\mathbb{R}^{d}$
is a $d$-vector which implements the spacetime translations. $\mathcal{P}_{+}^{\uparrow}$
acts geometrically on a spacetime region $\mathcal{O}\subset\mathbb{R}^{d}$
as
\begin{equation}
g\mathcal{O}:=\left\{ \Lambda x+a\,:\,x\in\mathcal{O}\right\} \,.\label{c1_poin_region}
\end{equation}
Among the class of all spacetime regions, we work with the subclass
of causally complete regions.
\begin{defn}
\label{c1-def-k}Given an $\mathcal{O}\subset\mathbb{R}^{d}$, we
define its (open) \textit{spacelike complement} $\mathcal{O}'$ as
the set of all points in $\mathbb{R}^{d}$ that are spacelike separated
with all points in $\mathcal{O}$, i.e.
\begin{equation}
\mathcal{O}':=\mathrm{int}\left\{ x\in\mathbb{R}^{d}\::\:\left(x-y\right)^{2}<0,\;\forall y\in\mathcal{O}\right\} \,,
\end{equation}
and its \textit{causal completion} as $\mathcal{O}'':=\left(\mathcal{O}'\right)'$.
It is always true that $\left(\mathcal{O}''\right)'=\mathcal{O}'$.
We say that a region $\mathcal{O}$ is \textit{causally complete}
if $\mathcal{O}''=\mathcal{O}$.\textit{}\footnote{\textit{A causally complete region is always an open set of }$\mathbb{R}^{d}$.}
The set of all causally complete regions is denoted by $\mathcal{K}$.
\end{defn}

A useful way to construct causally complete regions is as follows.
First, define a (smooth) Cauchy surface $\Sigma\subset\mathbb{R}^{d}$
and take any relative open subset $\mathcal{C}\subset\Sigma$. Then,
its Cauchy development\textit{}\footnote{Given any set $\mathcal{S}\subset\mathbb{R}^{d}$ , its \textit{future
}(resp.\textit{ past})\textit{ Cauchy development} $D^{+}\left(\mathcal{S}\right)$
(resp. $D^{-}\left(\mathcal{S}\right)$) is defined as the set of
all spacetime points $x\in\mathbb{R}^{d}$ for which every past (resp.
future) directed inextendible causal curve through $x$ intersects
$\mathcal{S}$ at least once. The \textit{Cauchy development} $D\left(\mathcal{S}\right)$
is the union of the future and past Cauchy developments.} $D\left(\mathcal{C}\right)$ is a causally complete region. Furthermore,
any causally complete region can be constructed in that way. We also
have that $\mathcal{P}_{+}^{\uparrow}$ maps the set of causally complete
regions onto the set of causally complete regions, i.e. $g\left(\mathcal{K}\right)=\mathcal{K}$
for all $g\in\mathcal{P}_{+}^{\uparrow}$. 

There is a particular kind of causally complete regions, known as
double cones.
\begin{defn}
We call a \textit{double cone} to any non-empty open region $\mathcal{O}\subset\mathbb{R}^{d}$
of the Minkowski spacetime defined by the intersection of the future open
null cone of some point $x\in\mathbb{R}^{d}$ with the past open null
cone of another point $y\in\mathbb{R}^{d}$.\footnote{In particular, $y$ must be in the timelike future of $x$ in order
to have $\mathcal{O}\neq\emptyset$.}
\end{defn}

\begin{rem}
Every double cone is a bounded, connected and simply connected region.
Moreover, $\mathcal{P}_{+}^{\uparrow}$ maps the set of double cones
onto the set of double cones.
\end{rem}

Before we list the axioms of AQFT, we define two different notions
of spacelike separated regions.
\begin{defn}
We say that $\mathcal{O}_{1},\mathcal{O}{}_{2}\in\mathcal{K}$ are
\textit{spacelike separated} if $\mathcal{O}_{1}\subset\mathcal{O}'_{2}$,
and \textit{strictly spacelike separated} if $\overline{\mathcal{O}}_{1}\subset\mathcal{O}'_{2}$,
where $\overline{\mathcal{O}}_{1}$ denotes the closure of $\mathcal{O}{}_{1}$.
Spacelike and strictly spacelike separated regions are respectively
denoted by $\mathcal{O}_{1}\text{\Large\ensuremath{\times}}\mathcal{O}_{2}$
and $\mathcal{O}_{1}\text{\Large\ensuremath{\times\!\negmedspace\!\times}}\mathcal{O}_{2}$.
strictly spacelike separated regions are spacelike separated regions
such that there is a finite non-zero separation distance between them.
\end{defn}

\begin{defn}
\label{c1_def_aqft}An \textit{algebraic quantum field theory (AQFT)}
is defined by a $C^{*}$-algebra $\mathfrak{A}$, called the\textit{
algebra of observables}, and an assignment to every causally complete
region $\mathcal{O}\in\mathcal{\mathcal{K}}$ a $C^{*}$-subalgebra
$\mathfrak{A}\left(\mathcal{O}\right)\subset\mathfrak{A}$, i.e.
\begin{eqnarray}
\mathcal{O}\in\mathcal{\mathcal{K}} & \mapsto & \mathfrak{A}\left(\mathcal{O}\right)\subset\mathfrak{A}\,,
\end{eqnarray}
which are called the \textit{local algebra}s \textit{(of observables}).
This collection of local algebras must satisfy: 
\begin{enumerate}
\item \textit{(generating property}) $\mathfrak{A=\overline{\bigcup_{\mathcal{O}\in\mathcal{K}}\mathfrak{A}\left(\mathcal{O}\right)}}^{\left\Vert .\right\Vert }$.
\item \textit{(isotony}) For any pair of regions $\mathcal{O}_{1}\subset\mathcal{O}_{2}$,
then $\mathfrak{A}\left(\mathcal{O}_{1}\right)\subset\mathfrak{A}\left(\mathcal{O}_{2}\right)$. 
\item \textit{\label{c1-defAQFT-locality}(locality)} For any pair of regions
$\mathcal{O}_{1}\text{\Large\ensuremath{\times}}\mathcal{O}_{2}$,
then $\left[\mathfrak{A}\left(\mathcal{O}_{1}\right),\mathfrak{A}\left(\mathcal{O}_{2}\right)\right]=\left\{ 0\right\} $.
\item \textit{(Poincaré covariance}) There is a uniform continuous linear
representation $\alpha:\mathcal{P}_{+}^{\uparrow}\rightarrow\mathrm{Aut}\left(\mathfrak{A}\right)$,
such that $\alpha_{g}\left(\mathfrak{A}\left(\mathcal{O}\right)\right)=\mathfrak{A}\left(g\mathcal{O}\right)$
for all $\mathcal{O}\in\mathcal{K}$ and all $g\in\mathcal{P}_{+}^{\uparrow}$.
\end{enumerate}
Any collection $\mathfrak{A}\left(\mathcal{O}\right)$ of $C^{*}$-algebras
satisfying the above axioms is called a \textit{net of local algebras}
\textit{(of observables}).\footnote{Mathematically, due to axiom 1, the collection of local algebras form
a net indexed by the elements of $\mathcal{K}$. The elements of $\mathcal{K}$
form a directed set when it is ordered by the usual set inclusion.}
\end{defn}
\begin{rem}
In the usual approach to QFT, the theory is defined through a collection
of field operators $\phi\left(x\right)$ acting on some Hilbert space
$\mathcal{H}$. From a heuristic point of view, such field operators
form a $*$-algebra: we are allowed to take linear combinations of
field operators, the product in the algebra is given by the composition
and the involution is defined by taking adjoints. One is tempted to
define the local algebras just as 
\begin{equation}
\mathfrak{A}\left(\mathcal{O}\right):=\left\langle \left\{ \phi\left(x\right)\,:\,x\in\mathcal{O}\right\} \right\rangle \,,\label{c1_bad_alegbras}
\end{equation}
where $\left\langle \cdot\right\rangle $ means the algebra generated.
Morally that is correct, but from the mathematical point of view,
the field operators $\phi\left(x\right)$ are not operators in the
usual sense (linear transformations acting on the Hilbert space).
They are very singular objects known as (unbounded) operator-valued
distributions. Luckily, such a technical problem can be bypassed,
and a net of $C^{*}$-algebras can be properly defined from such a
collection of fields (see subsection \ref{c1_subsec_aqft_vs_gw}).
Despite these technical issues, we want to emphasize that, relation
\eqref{c1_bad_alegbras} is an informal, but physically correct, way
to define the local algebras.
\end{rem}

There is an extra property that states that any local algebra may
be generated by algebras of smaller regions.
\begin{enumerate}
\setcounter{enumi}{4}
\item \label{aqft_add}\textit{(additivity) }For any $\mathcal{O}\in\mathcal{K}$
and any of its open coverings $\mathcal{O}=\bigcup_{\alpha}\mathcal{O}_{\alpha}$
with $\mathcal{O}_{\alpha}\in\mathcal{K}$, then $\mathfrak{A\left(\mathcal{O}\right)=\overline{\bigcup_{\alpha}\mathfrak{A}\left(\mathcal{O}_{\alpha}\right)}}^{\left\Vert .\right\Vert }$.
\end{enumerate}
This property may hold in physical models, but it is excluded from
the axioms because there are consistent models of AQFT where additivity
fails. We will discuss more about this fact in chapter \ref{EE_SS},
when we study theories having non-trivial superselection sectors.

One of the challenges of AQFT, is to find non-trivial models satisfying
the axioms 1. to 4. above. However, this task is very difficult, and
the number of known examples is very limited. Between them we want
to emphasize the free scalar fields (see chapter \ref{RE_CS} and
\cite{Hollands17}), free Dirac fields (see \cite{Hollands17})
and chiral CFTs (see chapter \ref{CURRENT} and \cite{Carpi07,Hollands17}).

\subsection{Vacuum states and vacuum representations}

In the usual approach to QFT, it is postulated the existence of a
distinguished state, the vacuum state, which has the following characteristic
properties: it is pure and Poincaré invariant. This motivates the
following definition.
\begin{defn}
\label{c1_def_vacuum}Let $\mathfrak{A}\left(\mathcal{O}\right)$
be a net of local algebras and $\omega_{0}\in\mathfrak{S}\left(\mathfrak{A}\right)$
a state. We say that $\omega_{0}$ is a \textit{vacuum state} if it
satisfies the above three conditions:
\end{defn}

\begin{itemize}
\item $\omega_{0}$ is pure, i.e. $\omega_{0}\in\mathfrak{S}_{p}\left(\mathfrak{A}\right)$.
\item $\omega_{0}$ invariant under all $\alpha_{g}$, i.e. $\omega_{0}\circ\alpha_{g}=\omega_{0}$
for all $g\in\mathcal{P}_{+}^{\uparrow}$.
\item In the GNS-representation $\pi_{0}:\mathfrak{A}\rightarrow\mathcal{B}\left(\mathcal{H}_{0}\right)$
of $\omega_{0}$, the linear representation $\alpha:\mathcal{P}_{+}^{\uparrow}\rightarrow\mathrm{Aut}\left(\mathfrak{A}\right)$
is implemented by a positive energy unitary representation of $U:\mathcal{P}_{+}^{\uparrow}\rightarrow\mathcal{B}\left(\mathcal{H}\right)$
in the sense that $U\left(g\right)\pi\left(A\right)U\left(g\right)^{*}=\pi\left(\alpha_{g}\left(A\right)\right)$
for all $A\in\mathfrak{A}$ and all $g\in\mathcal{P}_{+}^{\uparrow}$.
Positive energy means that the representation is strongly continuous
and the infinitesimal generators $P^{\mu}$ of the translation subgroup
(i.e. $U\left(0,a\right)=\mathrm{e}^{iP^{\mu}a_{\mu}}$) have their
joint spectrum included in the \textit{closed forward light cone}
$\overline{V}_{+}:=\left\{ p\in\mathbb{R}^{d}\,:\,p\cdot p>0\textrm{ and }p^{0}>0\right\} $.
Such a representation is called a \textit{vacuum representation}.
\end{itemize}
\begin{prop}
In the context of the previous definition, the GNS-vector, which is
denoted by $\left|0\right\rangle \in\mathcal{H}_{0}$, is called the
vacuum vector, and it satisfies $U\left(g\right)\left|0\right\rangle =\left|0\right\rangle $
for all $g\in\mathcal{P}_{+}^{\uparrow}$.
\end{prop}

\begin{rem}
In the context of definition \ref{c1_def_vacuum}, a vacuum representation
is always irreducible since the vacuum state is pure.
\end{rem}

The question that arises is: given a net of local algebras satisfying
the axioms of the definition \ref{c1_def_aqft}, could one guarantee
the existence and uniqueness of the vacuum state? Up to our knowledge,
this question is an open problem. However, some insights on such problem
can be found in \cite{haag,Halvorson06}. On the other hand, all known
interesting AQFT examples have a unique pure vacuum state. In this
thesis, the existence and uniqueness of the vacuum are considered
as extra postulates.

Once we have constructed the GNS-representation of the vacuum state,
we can define the net of vN algebras
\begin{eqnarray}
\mathcal{O}\in\mathcal{K} & \mapsto & \mathcal{A}\left(\mathcal{O}\right):=\pi_{0}\left(\mathfrak{A}\left(\mathcal{O}\right)\right)''\subset\mathcal{B}\left(\mathcal{H}_{0}\right)\,.
\end{eqnarray}
It can be shown that such a net of vN algebras satisfies all the axioms
of the definition \ref{c1_def_aqft}. Moreover, under our assumptions,
the algebra $\mathcal{A}\left(\mathbb{R}^{d}\right)=\mathcal{B}\left(\mathcal{H}_{0}\right)$
has a unique vacuum state given by the GNS-vector $\left|0\right\rangle $.
For many purposes, it is very useful to consider this representation
as being the defining net of our AQFT. In other words, the local algebras
$\mathfrak{A}\left(\mathcal{O}\right)$ are indeed concrete vN subalgebras
of $\mathcal{B}\left(\mathcal{H}_{0}\right)$, such that the Poincaré
automorphisms are unitarily implemented by positive energy unitary
representation, and such that there is a unique Poincaré invariant
vector $\left|0\right\rangle $ representing the vacuum state. For
example, when one wants to construct a concrete model of an AQFT satisfying
the axioms above, it is, in general, easier to construct a net of
vN algebras $\mathcal{O}\rightarrow\mathcal{A}\left(\mathcal{O}\right)$
acting on a Hilbert space.

Finally, we want to state the famous Reeh-Schlieder theorem. In order
to do that, we have to introduce a weaker notion of additivity.
\begin{defn}
\label{c1-weak_add}Let $\mathfrak{A}\left(\mathcal{O}\right)$ be
a net of local algebras. We say that the vacuum representation $\pi_{0}:\mathfrak{A}\rightarrow\mathcal{B}\left(\mathcal{H}_{0}\right)$
satisfies \textit{weak additivity} if
\begin{equation}
\mathcal{B}\left(\mathcal{H}_{0}\right)=\left(\bigcup_{x\in\mathbb{R}^{d}}\pi_{0}\left(\mathfrak{A}\left(\mathcal{O}+x\right)\right)\right)''\,,
\end{equation}
for all regions $\mathcal{O}\in\mathcal{K}$.
\end{defn}

We emphasize that weak additivity holds in any interesting known model,
and it is expected to hold in general. For the purpose of this thesis,
weak additivity is considered as an extra axiom. Finally, we state
the Reeh-Schlieder theorem.
\begin{thm}
\label{c1_rs}(Reeh-Schlieder) Let $\mathfrak{A}\left(\mathcal{O}\right)$
be a net of local algebras which satisfies weak additivity. Then the
vacuum vector $\left|0\right\rangle $ is cyclic for any algebra $\pi_{0}\left(\mathfrak{A}\left(\mathcal{O}\right)\right)''$
of any non-empty region $\mathcal{O}\in\mathcal{K}$, and it is also
separating for any algebra $\pi_{0}\left(\mathfrak{A}\left(\mathcal{O}\right)\right)''$
of any region $\mathcal{\mathcal{O}\in\mathcal{K}}$, whenever $\mathcal{O}'\neq\emptyset$. 
\end{thm}

\begin{proof}
See \cite{araki}.
\end{proof}
\begin{defn}
Motivated in the previous theorem, we say that a causally complete
region $\mathcal{O}\in\mathcal{K}$ is \textit{standard} whenever
$\mathcal{O}$ and $\mathcal{O}'$ are non-empty.
\end{defn}

\subsection{AQFT vs. Garding-Wightman axioms\label{c1_subsec_aqft_vs_gw}}

It is a very interesting exercise to try to relate the algebraic approach
to other known approaches to QFT. A different axiomatic framework
of QFT is the Garding-Wightman approach \cite{Streater}. There, it
is postulated the existence of point-like quantum fields $\phi\left(x\right)$,
which are operator valued distributions in some Hilbert space $\mathcal{H}$.\footnote{Along this section, we just consider a single real scalar field.}
Operator valued distribution means that
\begin{equation}
\phi\left(f\right):=\int_{\mathbb{R}^{d}}\phi\left(x\right)f\left(x\right)\,d^{d}x\,,
\end{equation}
is a well defined (unbounded) operator once it has been smeared by
a test function $f\in\mathcal{S}\left(\mathbb{R}^{d}\right)$.\footnote{$\mathcal{S}\left(\mathbb{R}^{d}\right)$ denotes the Schwartz space
of test functions (see for example \cite{Streater}). } Since both approaches are mathematically well-defined, we can ask
ourselves if they are equivalent (in some sense). This question can
be split into two: 1) given a QFT theory satisfying the Garding-Wightman
axioms, is it possible to construct an AQFT which encodes the same
physical information? And conversely 2) given an AQFT, is it possible
to construct a QFT satisfying the Garding-Wightman axioms?

The first question is well-understood, and the answer is affirmative
if we supplement the Garding-Wightman axioms with some extra technical
assumptions.\footnote{For the real scalar field, the assumption is that the operators $\phi\left(f\right)$
are self-adjoint whenever $f$ is a real function. See \cite{ReedSimon}
for the distinction between self-adjoint and hermitian operators. } We sketch here such a construction for the real scalar field. It
can be done in many different ways, but all of them giving the same
result \cite{horuzhy}. First of all, one could be tempted to define
the local algebras for a region $\mathcal{O}\in\mathcal{K}$ as
\begin{equation}
\mathfrak{A}_{U}\left(\mathcal{O}\right):=\left\langle \left\{ \phi\left(f\right)\,:\,\mathrm{supp}\left(f\right)\subset\mathcal{O}\right\} \right\rangle \,,\label{c1_gq_aqft_mal}
\end{equation}
where $\left\langle \cdot\right\rangle $ means the algebra generated.
However, the operators $\phi\left(f\right)$ are unbounded, and hence,
they cannot belong to the local algebras, which have to be defined
including only bounded operators. Furthermore, an algebra containing
unbounded operators can never be closed in order to obtain a $C^{*}$
or a vN algebra. To bypass this issue, we have to turn from unbounded
to bounded operators. Here, we sketch three different ways.
\begin{enumerate}
\item Since the operators $\phi\left(f\right)$ are self-adjoint (when $f$
is a real function), due to the spectral theorem, they can be decomposed
as
\begin{equation}
\phi\left(f\right)=\int_{\mathbb{R}}\lambda dE_{\phi\left(f\right)}\left(\lambda\right)\,,
\end{equation}
where $E_{\phi\left(f\right)}\left(\Delta\right)$, with $\Delta\subset\mathbb{R}$
a measurable set, are its spectral measures. For bounded measurable
sets $\Delta\subset\mathbb{R}$, $E_{\phi\left(f\right)}\left(\Delta\right)$
are bounded operators. Then, the local (vN) algebras are defined as
\[
\mathfrak{A}_{1}\left(\mathcal{O}\right):=\left\{ E_{\phi\left(f\right)}\left(\Delta\right)\,:\,f\textrm{ real, }\mathrm{sup}\left(f\right)\subset\mathcal{O}\textrm{ and }\Delta\subset\mathbb{R}\textrm{ bounded measurable}\right\} ''\,.
\]
\item Since the operators $\phi\left(f\right)$ are self-adjoint (when $f$
a is real function), due to Stone's theorem we have that the operators
$W\left(f\right):=\mathrm{e}^{i\phi\left(f\right)}$ are unitaries
and hence bounded \cite{ReedSimon}. Then, the local (vN) algebras
are defined as
\begin{equation}
\mathfrak{A}_{2}\left(\mathcal{O}\right):=\left\{ W\left(f\right)\,:f\textrm{ real, }\,\mathrm{sup}\left(f\right)\subset\mathcal{O}\right\} ''\,.
\end{equation}
\item We first define the $*$-algebra of (unbounded) operators $\mathfrak{A}_{U}\left(\mathcal{O}\right)$
as in equation \eqref{c1_gq_aqft_mal}. Then, we define its \textit{weak
commutant} as\footnote{$\mathcal{D}\subset\mathcal{H}$ denotes the common invariant dense
domain of all quantum fields \cite{Streater}.}
\begin{equation}
\mathfrak{A}\left(\mathcal{O}\right)^{w}:=\left\{ A\in\mathcal{B}\left(\mathcal{H}\right):\left\langle X^{*}\Psi\right|A\left|\Phi\right\rangle \! = \! \left\langle \Psi\right|AX\left|\Phi\right\rangle ,\,\forall X\in\mathfrak{A}_{U}\left(\mathcal{O}\right)\textrm{, }\forall\left|\Psi\right\rangle ,\left|\Phi\right\rangle \! \in  \mathcal{D}\right\} .
\end{equation}
And finally, we define the local (vN) algebras as $\mathfrak{A}_{3}\left(\mathcal{O}\right):=\left(\mathfrak{A}\left(\mathcal{O}\right)^{w}\right)'$.
\end{enumerate}
It can be shown that $\mathfrak{A}_{1}\left(\mathcal{O}\right)=\mathfrak{A}_{2}\left(\mathcal{O}\right)=\mathfrak{A}_{3}\left(\mathcal{O}\right)=:\mathfrak{A}\left(\mathcal{O}\right)$,
and the net $\mathfrak{A}\left(\mathcal{O}\right)$ satisfies all
the axioms of the definition \ref{c1_def_aqft}, regarding the quantum
field $\phi\left(x\right)$ satisfies the Garding-Wightman axioms
\cite{horuzhy}.

Question 2) is much harder, and at first sight, we must argue that
it has a negative answer. In fact, the axioms of the definition \ref{c1_def_aqft}
do not guarantee the existence of a vacuum state. If we supplement
such axioms with the existence and uniqueness of the vacuum state, we
should have a priori all the ``physical'' ingredients to perform
such a construction. However, it is a very difficult task to construct
a single point-like quantum field $\phi\left(x\right)$, with the
property that $\phi\left(f\right)$ ``generates'' all the operators
of all local algebras $\mathfrak{A}\left(\mathcal{O}\right)$, when
$f$ runs along the set of all test functions. Given some extra technical
assumptions, point-like objects $\phi\left(x\right)$, having some
of the properties we expect for a quantum field, can be found out
in the algebra $\mathfrak{A}$ \cite{Fredenhagen81}. However, many
details about the properties of such field operators remain unclear,
e.g., the question of whether the fields possess finite spin, or the
existence of product expansions \cite{Bostelmann04}. On the other
hand, there is some consensus among the algebraic community, that
the point-like quantum fields are just ``parameterizations'' or
``basis'' for the abstract AQFT's. We argue in favor of such a statement
with the help of some examples.
\begin{itemize}
\item The theory of the free massive scalar field $\phi\left(x\right)$
may look different than the theory of its derivatives $\partial_{\mu}\phi\left(x\right)$.
But due to the equation of motion, we can reconstruct $\phi=\frac{1}{m^{2}}\boxempty\phi$
from $\partial_{\mu}\phi$ using algebraic relations.
\item Less trivial, the free massive Dirac fermion in $3+1$ dimensions
has four field components. In the chiral basis, the algebra generated
by these four field components is the same as the algebra generated
by the $2$-dimensional spinor components of definite chirality. This
is because the other two components of the field are related to the
former due to the equation of motion.
\item Still less trivial, there is a duality that relates the free Maxwell
in $2+1$ dimensions with the theory of the derivatives of a free
massless scalar field in the same spacetime dimension. Such a duality
is given by $\partial_{\mu}\phi=\varepsilon_{\mu\nu\sigma}F^{\nu\sigma}$.
Then, the local algebras obtained from both theories must be equivalent.
\end{itemize}
These examples show that there may exist many different ways to ``parametrize''
the same AQFT using different sets of quantum fields. To find a complete
and non-redundant set of quantum fields describing some AQFT looks
very similar to the problem of finding a basis for an abstract (infinite-dimensional)
vector space.

\subsection{Fermions in AQFT\label{c1-sec-fermions}}

AQFT was born as a theory of local observables, i.e. the theory is
defined to contain only observable quantities. As we have emphasized
in the axioms, operators belonging to spacelike separated regions
always commute. This is in contradiction with the usual approach of
QFT, where the existence of fermionic fields that anticommute at spacelike
distances is allowed. If we want to introduce fermionic operators,
we have to relax the locality axiom (axiom \ref{c1-defAQFT-locality}
in definition \ref{c1_def_aqft}), and we must introduce the notion
of graded locality and graded net of local algebras.

For this purpose, let us assume that the local algebras are concrete
$C^{*}$-algebras $\mathfrak{F}\left(\mathcal{O}\right)$ acting on
some Hilbert space $\mathcal{H}$, i.e. we have an assignment
\begin{eqnarray}
\mathcal{O}\in\mathcal{K} & \mapsto & \mathfrak{F}\left(\mathcal{O}\right)\subset\mathcal{B}\left(\mathcal{H}\right)\,.
\end{eqnarray}
The global algebra is denoted by $\mathfrak{F}$. We first relax condition
4. in definition \ref{c1_def_aqft}, imposing that the symmetry group
is the universal covering $\mathcal{\tilde{P}}_{+}^{\uparrow}$ of
the Poincaré group $\mathcal{P}_{+}^{\uparrow}$. A general element
$\widetilde{g}\in\widetilde{\mathcal{P}}_{+}^{\uparrow}$ is denoted
by $\widetilde{g}=(\widetilde{\Lambda},a)$, where $\tilde{\Lambda}\in\widetilde{\mathcal{L}}_{+}^{\uparrow}$
and $a\in\mathbb{R}^{d}$. We denote by $g\in\mathcal{P}_{+}^{\uparrow}$
the image of $\widetilde{g}\in\widetilde{\mathcal{P}}_{+}^{\uparrow}$
under the covering map. We assume that the Poincaré symmetry is unitarily
implemented in the Hilbert space $\mathcal{H}$, i.e. there exists
a positive energy unitary representation $\tilde{g}\in\mathcal{\tilde{P}}_{+}^{\uparrow}\mapsto U\left(\tilde{g}\right)\in\mathcal{B}\left(\mathcal{H}\right)$
such that $U\left(\tilde{g}\right)\mathfrak{F}\left(\mathcal{O}\right)U\left(\tilde{g}\right)^{*}=\mathfrak{F}\left(g\mathcal{O}\right)$.
We also assume that there is a unique (up to a phase) unit vector
$\left|0\right\rangle \in\mathcal{H}$ such that $U\left(\tilde{g}\right)\left|0\right\rangle =\left|0\right\rangle $
for all $\widetilde{g}\in\widetilde{\mathcal{P}}_{+}^{\uparrow}$.
\begin{defn}
A \textit{$\mathbb{Z}_{2}$-grading} in the net of local algebra $\mathfrak{F}\left(\mathcal{O}\right)$
is defined by an operator $\Gamma\in\mathcal{B}\left(\mathcal{H}\right)$
such that $\Gamma=\Gamma^{-1}=\Gamma^{*}$, $\Gamma\left|0\right\rangle =\left|0\right\rangle $,
$\Gamma\mathfrak{F}\left(\mathcal{O}\right)\Gamma=\mathfrak{F}\left(\mathcal{O}\right)$
for all $\mathcal{O}\in\mathcal{K}$ and $\left[\Gamma,U\left(\tilde{g}\right)\right]=0$
for all $\widetilde{g}\in\widetilde{\mathcal{P}}_{+}^{\uparrow}$.
An operator $A\in\mathfrak{F}$ such that $\Gamma A\Gamma=\pm A$
is called \textit{homogeneous}, indeed a \textit{Bose} or \textit{Fermi}
operator according to the $\pm$ alternative, or simply \textit{even}
or \textit{odd} operator. We shall say that the degree $\partial_{A}$
of the homogeneous operator $A\in\mathfrak{F}$ is $0$ in the Bose
case and $1$ in the Fermi case. A net of local algebras with a $\mathbb{Z}_{2}$-grading
is called a \textit{$\mathbb{Z}_{2}$-graded net of local algebras}.
\end{defn}

Given a grading we have the following facts.
\begin{prop}
Let $\mathfrak{F}\left(\mathcal{O}\right)$ be a $\mathbb{Z}_{2}$-graded
net of local algebras. Then
\end{prop}

\begin{enumerate}
\item \textit{Any operator $A\in\mathfrak{F}$ can be uniquely decomposed
as a sum of a Bose operator and a Fermi operator. In fact, $A=A_{+}+A_{-}$
with $\Gamma A_{\pm}\Gamma=\pm A_{\pm}$, where $A_{\pm}:=\frac{A\pm\Gamma A\Gamma}{2}$. }
\item \textit{If $A=A_{+}+A_{-}\in\mathfrak{F}\left(\mathcal{O}\right)$,
then $A_{+},A_{-}\in\mathfrak{F}\left(\mathcal{O}\right).$}
\end{enumerate}
Once we have defined the grading and we have identified the Bose and
Fermi operators, we introduce the graded commutator.
\begin{defn}
Let $\mathfrak{F}\left(\mathcal{O}\right)$ be a $\mathbb{Z}_{2}$-graded
net of local algebras. We define the \textit{graded commutator} as
follows: for $A,B\in\mathfrak{F}$ homogeneous we set
\begin{equation}
\left[A,B\right]_{\Gamma}:=AB-\left(-1\right)^{\partial_{A}\cdot\partial_{B}}BA\,,
\end{equation}
and we extend it to more general operators by linearity.
\end{defn}

With the help of the graded commutator we could state the axiom of
graded locality which replace the axiom \ref{c1-defAQFT-locality}
in definition \ref{c1_def_aqft}.
\begin{defn}
Let $\mathfrak{F}\left(\mathcal{O}\right)$ be a $\mathbb{Z}_{2}$-graded
net of local algebras. We say that the local algebra satisfies \textit{graded
locality} if for any pair of regions $\mathcal{O}_{1}\text{\Large\ensuremath{\times}}\mathcal{O}_{2}$,
we have that $\left[\mathfrak{F}\left(\mathcal{O}_{1}\right),\mathfrak{F}\left(\mathcal{O}_{2}\right)\right]_{\Gamma}=\left\{ 0\right\} $.
\end{defn}

The above notion is equivalent to the notion of twisted locality very
often used in the literature.
\begin{defn}
\label{c1-def-Z}Let $\mathfrak{F}\left(\mathcal{O}\right)$ be a
$\mathbb{Z}_{2}$-graded net of local algebras and let $Z:=\frac{1-i\Gamma}{1-i}$.
We defined the \textit{twisted commutant} of a local algebra as \foreignlanguage{english}{${\mathfrak{F}\left(\mathcal{O}\right)^{t}}\sp{\prime}:=Z\mathfrak{F}\left(\mathcal{O}\right)'Z^{*}$}.
We say that the net satisfies \textit{twisted locality} if for any
pair of regions $\mathcal{O}_{1}\text{\Large\ensuremath{\times}}\mathcal{O}_{2}$,
we have that $\mathfrak{F}\left(\mathcal{O}_{1}\right)\subset{\mathfrak{F}\left(\mathcal{O}_{2}\right)^{t}}\sp{\prime}$.
\end{defn}

\begin{lem}
Let $\mathfrak{F}\left(\mathcal{O}\right)$ be a $\mathbb{Z}_{2}$-graded
net of local algebras. It satisfies graded locality if and only if
it satisfies twisted locality.
\end{lem}

\begin{proof}
See \cite{Longo_xu}.
\end{proof}
We can summarize the content of this section in the following definition,
which are the axioms of a net of local algebras of fields (in the
vacuum representation).
\begin{defn}
\label{c1_def_fields}A \textit{$\mathbb{Z}_{2}$-graded AQFT} is
defined by a concrete $C^{*}$-algebra $\mathfrak{F}\subset\mathcal{B}\left(\mathcal{H}\right)$,
called the\textit{ algebra of fields}, and an assignment to every
causally complete region $\mathcal{O}\in\mathcal{\mathcal{K}}$ a
$C^{*}$-subalgebra $\mathfrak{F}\left(\mathcal{O}\right)\subset\mathfrak{F}$,
i.e.
\begin{eqnarray}
\mathcal{O}\in\mathcal{\mathcal{K}} & \mapsto & \mathfrak{F}\left(\mathcal{O}\right)\subset\mathfrak{F}\,.
\end{eqnarray}
which are called the \textit{local algebra}s \textit{(of fields}).
This collection of local algebras must satisfy: 
\end{defn}
\begin{enumerate}
\item \textit{(generating property}) $\mathfrak{F=\overline{\bigcup_{\mathcal{O}\in\mathcal{K}}\mathfrak{F}\left(\mathcal{O}\right)}}^{\left\Vert .\right\Vert }$.
\item \textit{(isotony}) For any pair of regions $\mathcal{O}_{1}\subset\mathcal{O}_{2}$,
then $\mathfrak{F}\left(\mathcal{O}_{1}\right)\subset\mathfrak{F}\left(\mathcal{O}_{2}\right)$. 
\item \textit{(causality}) There exists a $\mathbb{Z}_{2}$-grading $\Gamma$
such that for any pair of regions $\mathcal{O}_{1}\text{\Large\ensuremath{\times}}\mathcal{O}_{2}$,
we have that $\mathfrak{F}\left(\mathcal{O}_{1}\right)\subset{\mathfrak{F}\left(\mathcal{O}_{2}\right)^{t}}\sp{\prime}$.
\item \textit{(Poincaré covariance}) There is a positive energy unitary
representation $\tilde{g}\in\mathcal{\tilde{P}}_{+}^{\uparrow}\mapsto U\left(\tilde{g}\right)\in\mathcal{B}\left(\mathcal{H}\right)$
such that $U\left(\tilde{g}\right)\mathfrak{F}\left(\mathcal{O}\right)U\left(\tilde{g}\right)^{*}=\mathfrak{F}\left(g\mathcal{O}\right)$
for all $\mathcal{O}\in\mathcal{K}$ and $\widetilde{g}\in\widetilde{\mathcal{P}}_{+}^{\uparrow}$,
and such that it commutes with the grading, i.e. $\left[\Gamma,U\left(\tilde{g}\right)\right]=0$
for all $\widetilde{g}\in\widetilde{\mathcal{P}}_{+}^{\uparrow}$.
\item \textit{(vacuum}) There is a unique (up to a phase) unit vector $\left|0\right\rangle \in\mathcal{H}$,
such that $U\left(\tilde{g}\right)\left|0\right\rangle =\left|0\right\rangle $
for all $\widetilde{g}\in\widetilde{\mathcal{P}}_{+}^{\uparrow}$.
\item \textit{(irreducibility}) $\mathfrak{F}'=\left\{ \lambda\cdot\mathbf{1}\right\} $.
\end{enumerate}
Any collection $\mathfrak{F}\left(\mathcal{O}\right)$ of $C^{*}$-algebras
satisfying the above axioms is called a\textit{ net of local algebras}
\textit{of fields}.
\begin{rem}
The above net is called algebra of fields or net of fields, in order
to be distinguished from the algebra of observables or net of observables.
Indeed, the term algebra of observables is reserved for a net where
all its operators commute at spacelike distance, whereas the term
algebra of fields emphasizes the fact that there are some operators,
the ``fermionic fields'', which anticommute at spacelike distance.
\end{rem}

\begin{rem}
The above definition reduces to the axioms of a net of local algebras
of observables in the vacuum representation (definitions \ref{c1_def_aqft}
and \ref{c1_def_vacuum}) in the case when the grading is trivial,
i.e. $\Gamma=\mathbf{1}$. In this sense, a net of fields is a generalization
of a net of observables.
\end{rem}

\begin{rem}
Under the assumption of weak additivity, Reeh-Schlieder theorem \ref{c1_def_fields}
also holds for the net of fields.
\end{rem}

Given any net of fields $\mathfrak{F}\left(\mathcal{O}\right)$, we
can define its observable subnet just as $\mathfrak{A}\left(\mathcal{O}\right):=\left\{ A\in\mathfrak{F}\left(\mathcal{O}\right)\,:\,\Gamma A\Gamma=A\right\} $.
The observable subnet satisfies all the axioms of definitions \ref{c1_def_aqft}
and \ref{c1_def_vacuum}. It seems to be that the observable subnet
$\mathfrak{A}\left(\mathcal{O}\right)$ has less information than
the net of fields $\mathfrak{F}\left(\mathcal{O}\right)$. However,
one of the big success of the algebraic approach to QFT was to prove
that all the physical relevant information is encoded in the observable
algebra itself and that the field algebra $\mathfrak{F}\left(\mathcal{O}\right)$
could indeed be reconstructed from $\mathfrak{A}\left(\mathcal{O}\right)$.
This is the famous Doplicher-Roberts reconstruction theorem \cite{Doplicher:1969kp,Doplicher:1990pn}.
We will further discuss this topic in chapter \ref{EE_SS}.

\subsection{Lattice structure of AQFT\label{c1_subsec_lattice}}

According to proposition \ref{c1_lat_vn}, the set of vN subfactors
of $\mathcal{B}\left(\mathcal{H}\right)$ form an orthocomplemented
lattice. The following proposition shows that the class of causally
complete regions also form an orthocomplemented lattice.
\begin{prop}
The set of causally complete regions $\mathcal{K}$ has a natural
structure of orthocomplemented lattice (ordered by inclusion) given
by
\begin{align}
 & \emptyset\subset\mathcal{O}\subset\mathbb{R}^{d}\,,\\
 & \mathcal{O}_{1}\vee\mathcal{O}_{2}:=\left(\mathcal{O}_{1}\cup\mathcal{O}_{2}\right)''\,,\\
 & \mathcal{O}_{1}\land\mathcal{O}_{2}:=\mathcal{O}_{1}\cap\mathcal{O}_{2}\,,
\end{align}
where $\mathcal{O},\mathcal{O}_{1},\mathcal{O}_{2}\in\mathcal{K}$
and $\cup$ (resp. $\cap$) denote the union (resp. intersection)
of sets. The complementation in the lattice is given by the causal
complement $\mathcal{O}\mapsto\mathcal{O}'$.
\end{prop}

\begin{figure}[htb!]
\centering
\includegraphics[width=15cm]{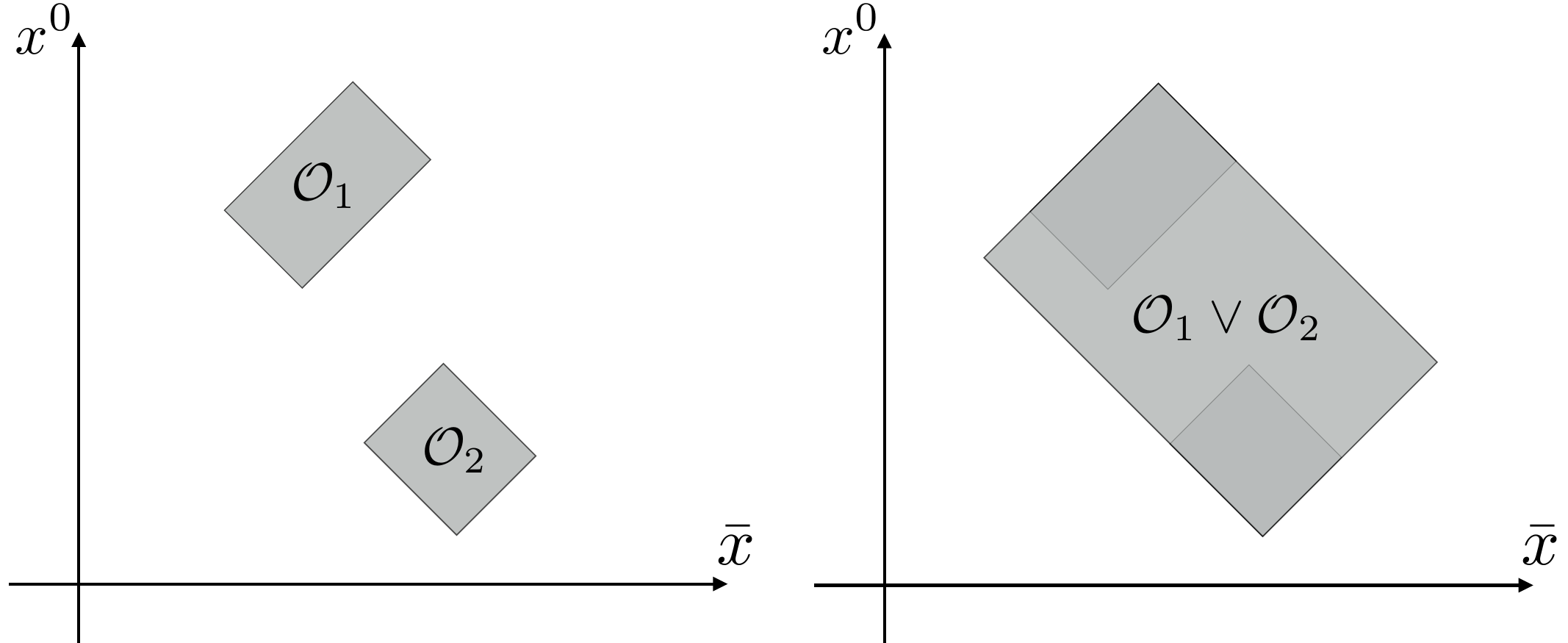}\caption{\label{c1_fig_union_regions}The supremum $\mathcal{O}_{1}\vee\mathcal{O}_{2}$
of two causally complete regions $\mathcal{O}_{1},\mathcal{O}_{2}\in\mathcal{K}$.}
\end{figure}

We remark that the region $\mathcal{O}_{1}\vee\mathcal{O}_{2}$ is
in general much bigger than $\mathcal{O}_{1}\cup\mathcal{O}_{2}$
(see figure \ref{c1_fig_union_regions}). However, for strictly spacelike
separated regions $\mathcal{O}_{1}\text{\Large\ensuremath{\times\!\negmedspace\!\times}}\mathcal{O}_{2}$
we have $\mathcal{O}_{1}\vee\mathcal{O}_{2}=\mathcal{O}_{1}\cup\mathcal{O}_{2}$.
This is not longer true for non-strict spacelike separated regions.
For example we have that $\mathcal{O}\lor\mathcal{O}'=\mathbb{R}^{d}$.

Thought it is not the most general scenario, we expect that a ``complete''
QFTs satisfied a stronger version of the axioms of AQFT.
\begin{defn}[\textbf{tentative}]
\label{c1_conj}A \textit{complete QFT} in the vacuum sector $\mathcal{H}_{0}$
is given by a lattice homomorphism\footnote{See appendix \ref{APP_LATTICE} for the definition of lattice homomorphism.}
$\mathcal{A}$ from the lattice of causally complete regions $\mathcal{K}$
into the lattice of vN subfactors of $\mathcal{B}\left(\mathcal{H}_{0}\right)$.
In other words, is given by an assignment 
\begin{eqnarray}
\mathcal{O}\in\mathcal{K} & \mapsto & \mathcal{A}\left(\mathcal{O}\right)\subset\mathcal{B}\left(\mathcal{H}_{0}\right)\,,\label{c1-lat-ass}
\end{eqnarray}
 satisfying
\begin{enumerate}
\item \label{c1-conj-irr} $\mathcal{A}\left(\emptyset\right)=\mathbf{1}$
and $\mathcal{A}\left(\mathbb{R}^{d}\right)=\mathcal{B}\left(\mathcal{H}_{0}\right)$
,
\item \label{c1-conj-add} $\mathcal{A}\left(\mathcal{O}_{1}\vee\mathcal{O}_{2}\right)=\mathcal{A}\left(\mathcal{O}_{1}\right)\vee\mathcal{A}\left(\mathcal{O}_{2}\right)$
,
\item \label{c1-conj-inter} $\mathcal{A}\left(\mathcal{O}_{1}\wedge\mathcal{O}_{2}\right)=\mathcal{A}\left(\mathcal{O}_{1}\right)\wedge\mathcal{A}\left(\mathcal{O}_{2}\right)$
,
\item \label{c1-conj-duality} $\mathcal{A}\left(\mathcal{O}'\right)=\mathcal{A}\left(\mathcal{O}\right)'$
,
\end{enumerate}
for all $\mathcal{O},\mathcal{O}_{1},\mathcal{O}_{2}\in\mathcal{K}$. 
\end{defn}

First of all, we are assuming that the local algebras of QFT are factors.
At first glance, there is no fundamental reason to do that. However,
under general circumstances it can be shown that the center $\mathcal{Z}(\mathcal{O})$
of a local algebra $\mathcal{A}(\mathcal{O})$ is a subalgebra of
the center $\mathcal{Z}$ of the global algebra $\mathcal{A}$ \cite{horuzhy}.
Since the vacuum representation is irreducible (assumption \ref{c1-conj-irr}
above), we have that $\mathcal{Z}$ is trivial, and hence all local
algebras must be factors.

\begin{figure}[h]
\centering
\includegraphics[width=15cm]{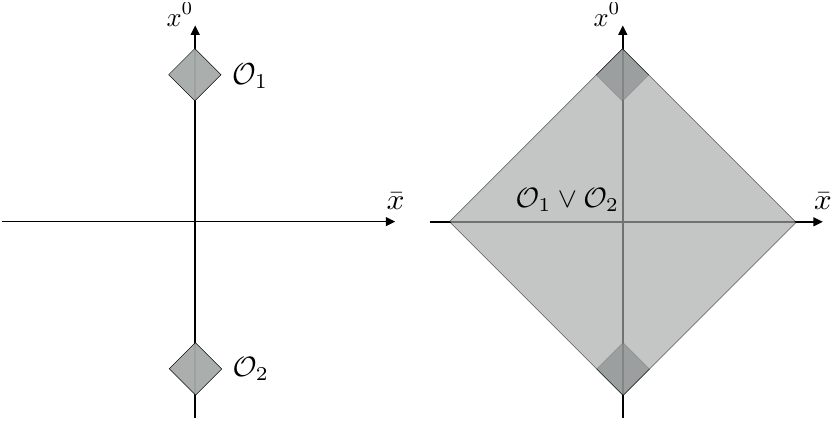}\caption{\label{c1_fig_no_add}Additivity (assumption 2 in conjecture \ref{c1_conj})
is not usually expected to hold for this timelike separated regions.}
\end{figure}

\begin{figure}[h]
\centering
\includegraphics[width=15cm]{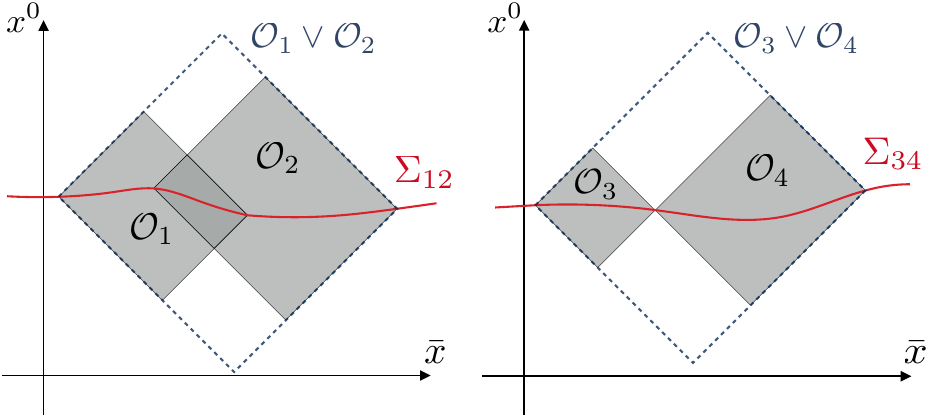}\caption{\label{c1_fig_si_add}Additivity (assumption 2 in conjecture \ref{c1_conj})
it is expected to hold for $\mathcal{\mathcal{O}}_{1}$ and $\mathcal{\mathcal{O}}_{2}$
despite they are not strictly spacelike separated. This is because
both these regions are the Cauchy development of spacelike regions
lying in a common Cauchy surface $\Sigma_{12}$. The same holds for
the regions $\mathcal{\mathcal{O}}_{3}$ and $\mathcal{\mathcal{O}}_{4}$.}
\end{figure}

The additivity property (assumption \ref{c1-conj-add}) is a stronger
version than the one stated in section \ref{c1-sec_aqft}. This is
because, as we explained above, the region $\mathcal{O}_{1}\vee\mathcal{O}_{2}$
is in general much bigger than $\mathcal{O}_{1}\cup\mathcal{O}_{2}$.
Moreover, the additivity property stated in this way is too much strong
that we do not expect it holds for many physically reasonable QFT
models. For example, consider two small double cones $\mathcal{O}_{1}$
and $\mathcal{O}_{2}$ arranged along the time axis $x^{0}$ with
vertices $\{\left(T-\varepsilon,\bar{0}\right),\left(T+\varepsilon,\bar{0}\right)\}$
and $\{\left(-T-\varepsilon,\bar{0}\right),\left(-T+\varepsilon,\bar{0}\right)\}$
respectively, with $T,\varepsilon>0$ and $\frac{T}{\varepsilon}\gg1$.
Then, the region $\mathcal{O}_{1}\lor\mathcal{O}_{2}$ is the double
cone with vertices $\{\left(-T-\varepsilon,\bar{0}\right),\left(T+\varepsilon,\bar{0}\right)\}$
(see figure \ref{c1_fig_no_add}). In this case, additivity means
that the algebra generated by those two small double cones $\mathcal{O}_{1}$
and $\mathcal{O}_{2}$ coincides with the algebra of the big double
cone $\mathcal{O}_{1}\lor\mathcal{O}_{2}$. Such a behavior is not
expected to be satisfied in all QFT models, but only for some sufficiently
complete models. On the other hand, additivity is expected to hold
in general for strictly spacelike separated regions, since in that
case $\mathcal{O}_{1}\vee\mathcal{O}_{2}=\mathcal{O}_{1}\cup\mathcal{O}_{2}$.
Moreover, we expect that additivity also holds for a bigger class
of regions, like the ones we exhibit in figure \ref{c1_fig_si_add}.
These pairs of regions are characterized for being the Cauchy development
of spacelike regions lying in a common Cauchy surface. In lattice
language, this is equivalent to the fact that $\mathcal{O}_{1}$ and
$\mathcal{O}_{2}$ commute,\footnote{See appendix \ref{APP_LATTICE}.}
i.e.
\begin{equation}
\mathcal{O}_{1}=(\mathcal{O}_{1}\wedge\mathcal{O}_{2})\vee(\mathcal{O}_{1}\wedge\mathcal{O}'_{2})\,.\label{c1-comm-sets}
\end{equation}
Furthermore, $\mathcal{O}_{1}$ and $\mathcal{O}_{2}$ commute if
and only if $\mathcal{O}_{1}\vee\mathcal{O}_{2}=\mathrm{int}\left(D\left(\overline{\mathcal{O}_{1}\cup\mathcal{O}_{2}}\right)\right)$.

According to the discussion above, assumption \ref{c1-conj-add} above
has to be weakened by
\begin{enumerate}
\item[2'.] \customlabel{c1-conj-add-bis}{2'}$\mathcal{A}\left(\mathcal{O}_{1}\lor\mathcal{O}_{2}\right)=\mathcal{A}\left(\mathcal{O}_{1}\right)\vee\mathcal{A}\left(\mathcal{O}_{2}\right)$
whenever $\mathcal{O}_{1}$ and $\mathcal{O}_{2}$ commute.
\end{enumerate}
In particular, additivity holds for complementary regions, i.e. $\mathcal{A}\left(\mathcal{O}\right)\lor\mathcal{A}\left(\mathcal{O}'\right)=\mathcal{A}\left(\mathbb{R}^{d}\right)=\mathcal{B}\left(\mathcal{H}_{0}\right)$.
The price that we pay for replacing \ref{c1-conj-add} by \ref{c1-conj-add-bis} is that
the map \eqref{c1-lat-ass} is no longer a lattice homomorphism. We
also have to remark that not every QFT satisfies assumption \ref{c1-conj-add-bis}.

Intersection property (assumption \ref{c1-conj-inter}) has no counterpart
in our previous definition \ref{c1-sec_aqft}. As additivity property
(assumptions \ref{c1-conj-add} or \ref{c1-conj-add-bis}), this property is not expected
to hold for any QFT, but only for sufficiently complete models. Moreover,
in an specific model, such properties may hold for some, but not all,
pairs of causally complete regions. For example, in chapter \ref{RE_CS}
we exhibit a specific model where additivity holds for any pair of
regions as the ones of figure \ref{c1_fig_si_add}. However, the same
model does not satisfy additivity for strictly spacelike separated
regions $\mathcal{O}_{1}\text{\Large\ensuremath{\times\!\negmedspace\!\times}}\mathcal{O}_{2}$.

Assumption \ref{c1-conj-duality} above is a stronger version of the
causality axiom \ref{c1-defAQFT-locality} in definition \ref{c1_def_aqft},
which means $\mathcal{A}\left(\mathcal{O}'\right)\subset\mathcal{A}\left(\mathcal{O}\right)'$.
In fact, assumption \ref{c1-conj-duality} is a kind of maximality
condition for the local algebras, since $\mathcal{A}\left(\mathcal{O}\right)'$
is the largest algebra that can be attached to region $\mathcal{O}'$
without spoiling causality. When assumption \ref{c1-conj-duality}
holds for a region $\mathcal{O}$ we say that the net algebra satisfies
\textit{duality} for region the $\mathcal{O}$. There are theories
which satisfy duality for a large class of regions (e.g. the real
scalar field, see chapter \ref{RE_CS}), whereas there other theories
which satisfy duality just for regions that are topologically double
cones (see chapters \ref{CURRENT} and \ref{EE_SS}). One may be tempted
to enlarge the local algebras defining $\mathcal{A}^{d}\left(\mathcal{O}\right):=\mathcal{A}\left(\mathcal{O}'\right)'$.
The problem is when we replace simultaneously
\begin{eqnarray}
\mathcal{A}\left(\mathcal{O}\right)\!\!\! & \rightarrow & \!\!\!\mathcal{A}^{d}\left(\mathcal{O}\right)=\mathcal{A}\left(\mathcal{O}'\right)'\,,\\
\mathcal{A}\left(\mathcal{O}'\right)\!\!\! & \rightarrow & \!\!\!\mathcal{A}^{d}\left(\mathcal{O}'\right)=\mathcal{A}\left(\mathcal{O}\right)'\,,
\end{eqnarray}
it could be the case that $\left[\mathcal{A}^{d}\left(\mathcal{O}\right),\mathcal{A}^{d}\left(\mathcal{O}'\right)\right]\neq\left\{ 0\right\} $,
because we have enlarged simultaneously both algebras. This motivates
the following definition.
\begin{defn}
\label{c1-def-ess-dual}Let $\mathcal{A}\left(\mathcal{O}\right)$
be a net of vN algebras. We say that $\mathcal{A}\left(\mathcal{O}\right)$
satisfies \textit{essential duality }if $\mathcal{A}^{d}\left(\mathcal{O}\right)$
satisfies causality. If this happens, then $\mathcal{A}^{d}\left(\mathcal{O}\right)$
satisfies duality for all $\mathcal{O}\in\mathcal{K}$.
\end{defn}

When a net satisfies essential duality, i.e. the local algebras can
be enlarged to obtain a net satisfying duality. However, it may happen
that additivity and/or intersection property fail in the enlarged
theory, even when in the original theory they are satisfied. In other
words, there may be a balance between additivity and duality: non
of them can be imposed without spoiling the other. A typical case
is when the theory has non-trivial superselection sectors. We will
further discuss this issue in chapter \ref{EE_SS}. On the other hand,
essential duality may hold for some, but not all, regions. For example,
in the case of spontaneous symmetry breaking, it happens that the
algebras attached to double cones do not satisfy duality but they
do satisfy essential duality. Moreover, essential duality for double
cones is a very fundamental property that might well be required of
any ‘reasonable’ QFT \cite{Halvorson06}.

Using the De Morgan laws,\footnote{See appendix \ref{APP_LATTICE}.}
we can formulate a different kinds of relations between additivity,
intersection property and duality. For example, if unrestricted duality
holds, then unrestricted additivity is equivalent to unrestricted
intersection property. Also, we have that duality for a non-connected
region with a finite number of components follows from duality for
connected regions, unrestricted additivity and unrestricted intersection
property.

Definition \ref{c1_conj} just involves the algebraic relations between
the local algebras of observables. We have to supplement them with
a postulate concerning Poincaré covariance. Indeed, the lattice of
causally complete regions and the lattice of vN subfactors of $\mathcal{B}\left(\mathcal{H}_{0}\right)$
are both $\mathcal{P}_{+}^{\uparrow}$-covariant lattices in the following
sense.
\begin{defn}
Let $L$ be an orthocomplemented lattice and $G$ a group. We say
that $L$ is a\textit{ $G$-covariant lattice} if there exists a representation
$\alpha:G\rightarrow\mathrm{Aut}\left(L\right)$ of $G$ into the
lattice automorphism of $L$.\footnote{See appendix \ref{APP_LATTICE} for the definition of lattice automorphism.}
\end{defn}

\begin{prop}
Let $\mathcal{K}$ be the lattice of causally complete regions. Then
$\mathcal{K}$ is a $\mathcal{P}_{+}^{\uparrow}$-covariant lattice
when we consider the representation \eqref{c1_poin_region} of $\mathcal{P}_{+}^{\uparrow}$
into $\mathrm{Aut}\left(\mathcal{K}\right)$.
\end{prop}

\begin{prop}
Let $g \mapsto U\left(g\right)$ be a unitary representation of
$\mathcal{P}_{+}^{\uparrow}$ on $\mathcal{H}_{0}$. Then it defines
a representation $\alpha$ of $\mathcal{P}_{+}^{\uparrow}$ into the
lattice of vN subalgebras of $\mathcal{B}\left(\mathcal{H}_{0}\right)$
by $\alpha_{g}\left(\mathcal{A}\right):=U\left(g\right)\mathcal{A}U\left(g\right)^{*}$.
Then, the lattice of vN subalgebras of $\mathcal{B}\left(\mathcal{H}_{0}\right)$
is $\mathcal{P}_{+}^{\uparrow}$-covariant when we consider such a
representation.
\end{prop}

\begin{defn}
Let $G$ be a group and $L_{1},L_{2}$ be $G-$covariant lattices.
A lattice homomorphism $\Phi:L_{1}\rightarrow L_{2}$ is called \textit{$G$-covariant
homomorphism} if $\Phi\circ\alpha_{g}^{\left(L_{1}\right)}=\alpha_{g}^{\left(L_{2}\right)}\circ\Phi$
for all $g\in G$.
\end{defn}

Finally, the assumptions \ref{c1-conj-irr} to \ref{c1-conj-duality} of conjecture \ref{c1_conj}, have
to be complemented with the following assumption:
\begin{enumerate}
\setcounter{enumi}{4}
\item \label{c1-conj-cov} There is a positive energy unitary representation $g\in\mathcal{P}_{+}^{\uparrow} \mapsto U\left(g\right)\in\mathcal{B}\left(\mathcal{H}_{0}\right)$
such that the lattice homomorphism $\mathcal{O}\in\mathcal{K} \mapsto \mathcal{A}\left(\mathcal{O}\right)\subset\mathcal{B}\left(\mathcal{H}_{0}\right)$
is $\mathcal{P}_{+}^{\uparrow}$-covariant. There is also a unique
(up to a phase) $\mathcal{P}_{+}^{\uparrow}$-invariant unit vector
$\left|0\right\rangle \in\mathcal{H}_{0}$.
\end{enumerate}
The lattice structure of conjecture \ref{c1_conj} supplemented with
assumption \ref{c1-conj-cov}, gives an elegant correspondence between geometrical
aspects (causally complete regions) and algebraic aspects (vN algebras)
of QFT. However, this structure is too strong\footnote{In particular, specially due to assumption \ref{c1-conj-add}.}
that it would not hold for some concrete models with physical interest. 

On the other hand, the lattice structure could be (partially) restored
if we work with causally complete regions within a common Cauchy surface.
In this way, we fix a Cauchy surface $\Sigma\subset\mathbb{R}^{d}$
and we consider the set $\mathcal{K}_{\Sigma}$ of all relative open
sets of $\Sigma$. Then we have the following proposition.
\begin{prop}
\label{c1-latt-cauchy}Given any Cauchy surface $\Sigma\subset\mathbb{R}^{d}$,
the set $\mathcal{K}_{\Sigma}$ of relative open sets of $\Sigma$
has a natural structure of orthocomplemented lattice (ordered by inclusion)
given by
\begin{align}
 & \emptyset\subset\mathcal{\mathcal{C}}\subset\Sigma\,,\\
 & \mathcal{\mathcal{C}}_{1}\vee\mathcal{\mathcal{C}}_{2}:=\mathrm{Int}\left(\overline{\mathcal{C}_{1}\cup\mathcal{C}_{2}}\right)\,,\\
 & \mathcal{C}_{1}\land\mathcal{C}_{2}:=\mathcal{C}_{1}\cap\mathcal{C}_{2}\,,
\end{align}
where $\mathcal{C},\mathcal{C}_{1},\mathcal{C}_{2}\in\mathcal{K}$
and $\cup$ (resp. $\cap$) denote the union (resp. intersection)
of sets. The complementation in the lattice is given by the set complement
$\mathcal{C}':=\Sigma-\overline{\mathcal{C}}$.
\end{prop}

\begin{defn}
In the context of the above definition, the set $\mathcal{C}'$ is
usually called the \textit{space complement} of $\mathcal{C}$.
\end{defn}

\begin{rem}
The orthocomplemented lattice $\mathcal{K}_{\Sigma}$ of the previous
definition is also a \textit{Boolean lattice}, i.e. every two elements
of $\mathcal{K}_{\Sigma}$ commute.\footnote{See appendix \ref{APP_LATTICE} for the definition of Boolean lattice
automorphism.}
\end{rem}

Given $\mathcal{C}\in\mathcal{K}_{\Sigma}$, we have that $D\left(\mathcal{C}\right)\in\mathcal{K}$.
Then we could define the net of vN factors
\begin{eqnarray}
\mathcal{C}\in\mathcal{K}_{\Sigma} & \mapsto & \mathcal{A}_{\Sigma}\left(\mathcal{C}\right):=\mathcal{A}\left(D\left(\mathcal{C}\right)\right)\subset\mathcal{B}\left(\mathcal{H}_{0}\right)\,.\label{c1-latt-net-ft}
\end{eqnarray}
Assumptions \ref{c1-conj-irr}, \ref{c1-conj-add-bis}, \ref{c1-conj-inter} and \ref{c1-conj-duality}
of definition \ref{c1_conj} applied for the restricted set of double
cones $\left\{ D\left(\mathcal{C}\right)\,:\,\mathcal{C}\in\mathcal{K}_{\Sigma}\right\} $
gives place to a lattice homomorphism according to the following lemma.
\begin{lem}
\label{c1_conj-2}Let $\mathcal{O}\in\mathcal{K}\mapsto\mathcal{A}\left(\mathcal{O}\right)\subset\mathcal{B}\left(\mathcal{H}_{0}\right)$
be net of vN factors satisfying the assumptions \ref{c1-conj-irr},
\ref{c1-conj-add-bis}, \ref{c1-conj-inter} and \ref{c1-conj-duality} of definition
\ref{c1_conj}, and let $\Sigma\subset\mathbb{R}^{d}$ be a Cauchy
surface. Then, the net
\begin{eqnarray}
\mathcal{C}\in\mathcal{K}_{\Sigma} & \mapsto & \mathcal{A}_{\Sigma}\left(\mathcal{C}\right)\subset\mathcal{B}\left(\mathcal{H}_{0}\right)\,,\label{c1-lat-ass-1}
\end{eqnarray}
defined through \ref{c1-latt-net-ft} is a lattice homomorphism from
the Boolean lattice of open regions $\mathcal{K}_{\Sigma}$ into the
lattice of vN subfactors of $\mathcal{B}\left(\mathcal{H}_{0}\right)$.
In other words, the map \eqref{c1-lat-ass-1} satisfies
\begin{enumerate}
\item \textit{$\mathcal{A}_{\Sigma}\left(\emptyset\right)=\mathbf{1}$ and
$\mathcal{A}_{\Sigma}\left(\Sigma\right)=\mathcal{B}\left(\mathcal{H}_{0}\right)$
,}
\item \textit{$\mathcal{A}_{\Sigma}\left(\mathcal{C}_{1}\vee\mathcal{C}_{2}\right)=\mathcal{A}_{\Sigma}\left(\mathcal{C}_{1}\right)\vee\mathcal{A}_{\Sigma}\left(\mathcal{C}_{2}\right)$
,}
\item \textit{$\mathcal{A}_{\Sigma}\left(\mathcal{C}_{1}\wedge\mathcal{C}_{2}\right)=\mathcal{A}_{\Sigma}\left(\mathcal{C}_{1}\right)\wedge\mathcal{A}_{\Sigma}\left(\mathcal{C}_{2}\right)$
,}
\item \textit{$\mathcal{A}_{\Sigma}\left(\mathcal{C}'\right)=\mathcal{A}_{\Sigma}\left(\mathcal{C}\right)'$
,}
\end{enumerate}
\textit{for all $\mathcal{C},\mathcal{C}_{1},\mathcal{C}_{2}\in\mathcal{K}_{\Sigma}$. }
\end{lem}

Since every Boolean sublattice $\mathcal{K}_{S}\subset\mathcal{K}$
is included in $\mathcal{K}_{\Sigma}$ for some Cauchy surface $\Sigma$,
we can say that \eqref{c1-lat-ass} is a lattice homomorphism for
every Boolean sublattice of $\mathcal{K}$.\footnote{See appendix \ref{APP_LATTICE} for the definition of sublattice.}

To end, we again remark that the axioms of definition \ref{c1_conj},
or its weaker version summarized in lemma \ref{c1_conj-2} may not
hold for any QFT model. For example, the free massive scalar field
satisfies properties \ref{c1-conj-irr}, \ref{c1-conj-add-bis}, and \ref{c1-conj-inter}-\ref{c1-conj-cov}
of definition \ref{c1_conj}, but it does not satisfy \ref{c1-conj-add}
(see chapter \ref{RE_CS}). In \cite{haag} there is a heuristic argument
on why the lattice structure of lemma \ref{c1_conj-2} should hold
for quantum electrodynamics. In any case, it is a very interesting
job to find models satisfying all the assumptions of definition \ref{c1_conj}.


\renewcommand\chaptername{Chapter}
\selectlanguage{english}

\chapter{Quantum information theory in operator algebras\label{INFO}}

In this chapter, we study the general aspects of quantum information
theory in operator algebras. The main goal is to define and study
many quantum information measures that we will use frequently in this
thesis. A quantum information measure is defined to be any function
over the space of states of a quantum system that describes the statistical
properties of the system as a whole. Quantum information measures
play a central role in the description and quantification of entanglement.
As we discuss in the introduction, entanglement is a property of bipartite
quantum systems, and it is manifested by non-classical correlations
between its subsystems. It is a purely quantum phenomenon and can
be viewed as an experimental resource, which can be exploited to perform
quantum computations, teleportations, or other often counter-intuitive
experiments. The organization of the chapter is as follows. First,
we introduce the notion of a quantum channel, which has to be thought
of as any “allowed” experimental operation over a quantum system which
is restricted according to physical considerations. Subsequently,
we study and introduce many quantum information measures and their
properties. Such measures are usually monotonic under quantum channels
which, roughly speaking, mean that the correlations along the system
cannot be increased, on average, applying such a type of operations.
For convenience, we separate the study of information measures into
two sections: measures for finite and general quantum systems. We
outline the central role of the relative entropy as a fundamental
information quantity for general quantum systems and its connection
with the modular theory of vN algebras. Reaching the end of the chapter,
we study a particular kind of quantum channels known as conditional
expectations, which have an important relation with relative entropy.
To conclude, we introduce the concept of entanglement on a bipartite
system. We study its structure and the ways to quantify it (entanglement
measures).

In this chapter all the algebras used are $C^{*}$-algebras. We will
explicitly emphasize when a $C^{*}$-algebra is also a vN algebra.

\section{Quantum channels\label{c2_sec_qc}}

The notion of quantum channel is a very useful concept in the study
of information and entanglement measures in quantum information theory.
The idea is to describe all the physical transformations $\mathcal{F}:\mathfrak{A}_{2}\rightarrow\mathfrak{A}_{1}$
between two quantum system, i.e. all the mathematical transformations
that are allowed from physical considerations.
\begin{defn}
A linear map $\mathcal{F}:\mathfrak{A}_{2}\rightarrow\mathfrak{A}_{1}$
between two $C^{*}$-algberas is called \textit{positive} if $\mathcal{F}\left(A\right)\geq0$
whenever $A\geq0$, $\mathcal{F}$ is called \textit{completely positive
(cp)} if $\mathcal{F}\otimes id_{n}:\mathfrak{A}_{1}\otimes M_{n}\left(\mathbb{C}\right)\rightarrow\mathfrak{A}_{2}\otimes M_{n}\left(\mathbb{C}\right)$
is positive as a map, where $id_{n}:M_{n}\left(\mathbb{C}\right)\rightarrow M_{n}\left(\mathbb{C}\right)$
is the identity map, i.e $id_{n}\left(B\right)=B$ for all $n\in\mathbb{N}$
and all $B\in M_{n}\left(\mathbb{C}\right)$.\footnote{$M_{n}\left(\mathbb{C}\right):=\mathbb{C}^{d_{n}\times d_{n}}$ denotes the algebra of all square matrices of size $n\times n$, which is
indeed a finite dimensional $C^{*}$-algebra.} A (completely) positive map is called \textit{normalized} if $\mathcal{F}\left(\mathbf{1}\right)=\mathbf{1}$. 
\end{defn}
A normalized positive map gives place to a map $\mathcal{F}^{*}:\mathfrak{A}_{1}^{*}\rightarrow\mathfrak{A}_{2}^{*}$
that transforms a state $\omega$ on $\mathfrak{A}_{1}$ into a state
$\mathcal{F}^{*}\omega$ on $\mathfrak{A}_{2}$ by de formula
\begin{equation}
\left(\mathcal{F}^{*}\omega\right)\left(A\right)=\omega\left(\mathcal{F}\left(A\right)\right)\,.
\end{equation}
In other words, $\mathcal{F}^{*}:\mathfrak{S}\left(\mathfrak{A}_{1}\right)\rightarrow\mathfrak{S}\left(\mathfrak{A}_{2}\right)$.
Conversely, any map from the states on $\mathfrak{A}_{1}$ to states
on $\mathfrak{A}_{2}$ arises from a normalized positive linear map
in this way. It is for this reason, that any normalized positive map
is called a \textit{state transformation}. Complete positivity is
motivated by quantum information theory when it is allowed to couple
our system with an ancilla, evolve the system plus the ancilla as
a closed system, and finally trace out the ancilla. A cp normalized
map is the mathematical characterization of a \textit{quantum channel}.

In addition to a quantum channel, one could perform measurements and
post-select a sub-ensemble according to the results. For a vN measurement,
given by projections $P_{k}\in\mathfrak{A}_{1}$ with $\sum_{k}P_{k}=\mathbf{1}$,
we note that the maps $\mathcal{F}_{k}:\mathfrak{A}_{2}\rightarrow\mathfrak{A}_{1}$
given by $A\mapsto P_{k}AP_{k}$ are cp but not normalized because
$0\leq\mathcal{F}_{k}\left(\mathbf{1}\right)=P_{k}\leq\mathbf{1}$.
Performing the measurement on a state $\omega$ we obtain the new
state 
\begin{equation}
\omega_{k}:=\frac{1}{p_{k}}\mathcal{F}_{k}^{*}\omega\,,
\end{equation}
with probability $p_{k}:=\omega\left(P_{k}\right)$ when $p_{k}\geq0$.
A combination of quantum channels and measurements is called a \textit{quantum
operation} \cite{Lindblad}. It is described by a family $\mathcal{F}_{k}:\mathfrak{A}_{2}\rightarrow\mathfrak{A}_{1}$
of cp maps such that $\sum_{k}\mathcal{F}_{k}\left(\mathbf{1}\right)=\mathbf{1}$.
The composition of quantum channels (resp. quantum operations) is
also a quantum channel (resp. quantum operation). There are many examples
of quantum channels \cite{Hollands17}, between them we want to remark
the automorphisms and conditional expectations (section \ref{c2_sec_ce}).

In the case of vN algebras, we only consider normal maps as in the
following sense.
\begin{defn}
A linear map $\mathcal{F}:\mathfrak{A}_{2}\rightarrow\mathfrak{A}_{1}$
is said to be \textit{normal if} $\mathcal{F}^{*}\omega$ is a normal
state on $\mathfrak{A}_{2}$ state for any normal state $\omega$
on $\mathfrak{A}_{1}$.
\end{defn}

\section{Entropies for finite quantum systems\label{c2_sec_finite}}

A \textit{finite quantum system} is any system described by a finite
dimensional algebra. Since for finite algebras, the concepts of $C^{*}$ and vN algebra are equivalent, throughout this section we call them
simply algebras. Any finite algebra $\mathfrak{A}$ is isomorphic
to a direct sum of full matrix algebras
\begin{equation}
\mathfrak{A}\simeq\bigoplus_{n=1}^{N}M_{d_{n}}\left(\mathbb{C}\right)\,.\label{c2_can_rep}
\end{equation}
Such an algebra acts naturally in the Hilbert space $\mathcal{H}:=\bigoplus_{n=1}^{N}\mathbb{C}^{d_{n}}$.
We remark that the algebra $\mathfrak{A}$ as an abstract object,
could be represented in many different ways, which are all $C^{*}$-isomorphic
but not unitarily equivalent. Any other faithful representation of
the algebra $\mathfrak{A}$ is of the form
\begin{equation}
\mathfrak{A}\simeq\bigoplus_{n=1}^{N}M_{d_{n}}\left(\mathbb{C}\right)\otimes\mathbf{1}_{f_{n}}\,,\label{c2_any_rep}
\end{equation}
where $\mathbf{1}_{f_{n}}$ is the (trivial) algebra of scalar matrices
of size $f_{n}\times f_{n}$, and $f_{n}\in\mathbb{N}$ are arbitrary
dimensions.\footnote{Expression \eqref{c2_any_rep} reduces \eqref{c2_can_rep} when $f_{n}=1$
for all $n=1,\ldots,N$. } It is not difficult to see that \eqref{c2_can_rep} and \eqref{c2_any_rep}
are isomorphic as algebras, but they are not (in general) unitarily
equivalent because they act in finite dimensional vector spaces of
different dimensions. 

Among all the representations of the algebra $\mathfrak{A}$, the
representation \eqref{c2_can_rep} is unique (up to unitary equivalence).
It is called the \textit{canonical representation}. The important
fact of such a representation arises when one wants to compute some
quantum information quantity, e.g. vN entropy, relative entropy, etc.
Such quantities depend, in general, on a collection of algebras $\mathfrak{A}_{1},\ldots,\mathfrak{A}_{k}$
and a collection of states $\mathcal{\omega}_{1},\ldots,\mathcal{\omega}_{l}$
on such algebras, and they are expected to depend only on them, and
not on the matrix representation used to write down the expressions.
However, as we see below, such information quantities are defined
through explicit formulas that only hold when they are used in the
canonical representation. In other words, when one uses a formula
to compute any information quantity, one always has to use the canonical
representation.

On the other hand, the representation matters when we consider an
inclusion of algebras, i.e. when the algebra $\mathfrak{A}$ is a
subalgebra of a bigger algebra. For example, the algebra \eqref{c2_can_rep}
is a subalgebra of $M_{d}\left(\mathbb{C}\right)$ with $d:=\sum_{n=1}^{N}d_{n}$,
whereas the algebra \eqref{c2_any_rep} is a subalgebra of $M_{\tilde{d}}\left(\mathbb{C}\right)$
with $\tilde{d}:=\sum_{n=1}^{N}d_{n}f_{n}$. The set of all subalgebras
of $M_{\tilde{d}}\left(\mathbb{C}\right)$ is bigger than the set
of all subalgebras $M_{d}\left(\mathbb{C}\right)$. In particular,
the commutants $\mathfrak{A}'$ of \eqref{c2_can_rep} and \eqref{c2_any_rep}
(considered them as vN algebras embedded in the bigger algebras $M_{d}\left(\mathbb{C}\right)$
and $M_{\tilde{d}}\left(\mathbb{C}\right)$) are respectively
\begin{equation}
\bigoplus_{n=1}^{N}\mathbf{1}_{d_{n}}\textrm{ and }\bigoplus_{n=1}^{N}\mathbf{1}_{d_{n}}\otimes M_{f_{n}}\left(\mathbb{C}\right)\,,
\end{equation}
which are not $C^{*}$-isomorphic, and hence their canonical representations
are very different. This matters in the discussion about entanglement
in quantum systems. There, not only the subsystem considered matters,
i.e. $\mathfrak{A}$, but also the ``complement'' of our subsystem
with respect we want to quantify the entanglement, i.e. $\mathfrak{A}'$. 

For an algebra $\mathfrak{A}$ in its canonical representation \eqref{c2_can_rep},
we can define its \textit{trace} as
\begin{equation}
\mathrm{Tr}{}_{\mathfrak{A}}\left(A_{1}\oplus\cdots\oplus A_{N}\right)=\sum_{n=1}^{N}\mathrm{Tr}{}_{d_{n}}\left(A_{n}\right)\,,
\end{equation}
where $\mathrm{Tr}{}_{d_{n}}$ is the usual trace in $M_{d_{n}}\left(\mathbb{C}\right)$.
Any trace will be denoted simply as $\mathrm{Tr}$ when there is no
confusion about the algebra involved. The \textit{canonical dimension}
of $\mathfrak{A}$ is defined as $d_{\mathfrak{A}}:=\sum_{n=1}^{N}d_{n}$,
i.e. the dimension of the Hilbert space where the canonical representation
acts. It is a general fact that for any state $\omega$ on $\mathfrak{A}$,
there exists a unique density matrix $\rho_{\omega}\in\mathfrak{A}$
(in its canonical representation) such that
\begin{equation}
\omega\left(A\right)=\mathrm{Tr}{}_{\mathfrak{A}}\left(\rho_{\omega}A\right)\,,\quad\forall A\in\mathfrak{A}\,,\label{c2_dens_mat}
\end{equation}
which is called the \textit{statistical operator} of $\omega$. Conversely,
any statistical operator $\rho\in\mathfrak{A}$ gives place to a unique
state through the relation \eqref{c2_dens_mat}. In other words, the
relation between states and statistical operators is one-to-one. The
state corresponding to the statistical operator $\frac{\mathbf{1}}{d_{\mathfrak{A}}}$
is called the \textit{tracial state} or \textit{maximally mixed state},
and it is denoted by $\tau_{\mathfrak{A}}$ or $\tau$. Any statistical
operator $\rho\in\mathfrak{A}$ can be uniquely decomposed as
\begin{equation}
\rho=\bigoplus_{n=1}^{N}p_{n}\rho_{n}\,,
\end{equation}
where $\rho_{n}$ are statistical operators on $M_{d_{n}}\left(\mathbb{C}\right)$
and $p_{1},\ldots,p_{N}\geq0$ such that $\sum_{n=1}^{N}p_{n}=1$.

It is important to emphasize that the trace could be thought of as
an abstract object, depending only on $\mathfrak{A}$, but to compute
it, we must use the canonical representation. Whereas, a statistical
operator is an object that only make sense in the canonical representation.
For any other representation, many different density matrices are
satisfying a relation like \eqref{c2_dens_mat}. As a matter of fact,
given two subalgebras $\mathfrak{A}_{1},\mathfrak{A}_{2}\subset M_{d}\left(\mathbb{C}\right)$
and a state $\omega$ on $M_{d}\left(\mathbb{C}\right)$, we can consider
the states $\omega_{1}:=\left.\omega\right|_{\mathfrak{A}_{1}}$ and
$\omega_{2}:=\left.\omega\right|_{\mathfrak{A}_{2}}$, which are the
restrictions of $\omega$ to $\mathfrak{A}_{1}$ and $\mathfrak{A}_{2}$.
The statistical operators $\rho_{\omega_{1}}$ and $\rho_{\omega_{2}}$
of such states are (in general) matrices with different number of
blocks and sizes.

Finally, a short comment about quantum channels in the finite dimensional
case. Even in this case, the general structure of cp normalized maps
is still poorly understood. However, in the simple case when $\mathfrak{A}_{j}:=M_{d_{j}}\left(\mathbb{C}\right)$
and $\mathcal{F}:M_{d_{2}}\left(\mathbb{C}\right)\rightarrow M_{d_{1}}\left(\mathbb{C}\right)$,
we have the usual representation in term of Krauss operators \cite{petz_easy}
\begin{equation}
\mathcal{F}\left(A\right)=\sum_{j=1}^{n}V_{j}^{\dagger}AV_{j}\,,\;\textrm{with }V_{j}\in\mathbb{C}^{d_{2}\times d_{1}}\textrm{ and }\sum_{j=1}^{n}V_{j}^{\dagger}V_{j}=\mathbf{1}_{d_{1}}\,.
\end{equation}

\subsection{von Neumann entropy}

Throughout this section, $\mathfrak{A}$ is a finite dimensional algebra,
$\omega\in\mathfrak{S}\left(\mathfrak{A}\right)$ a state, and $\rho_{\omega}$
is its corresponding statistical operator (in the canonical representation).
\begin{defn}
The \textit{von Neumann entropy} of the state $\omega$ on the algebra
$\mathfrak{A}$ is defined as
\begin{equation}
S_{\mathfrak{A}}\left(\omega\right):=-\mathrm{Tr}{}_{\mathfrak{A}}\left(\rho_{\omega}\log\rho_{\omega}\right)=-\sum_{j=1}^{n}\lambda_{j}\log\left(\lambda_{j}\right)\,,\label{c2_vn_ent}
\end{equation}
where $\lambda_{j}\geq0$ are the eigenvalues of $\rho_{\omega}$.
It will be denoted as $S\left(\omega\right)$ whenever there is no
confusion about the algebra involved.
\end{defn}

The vN entropy is the quantum counterpart of the Shannon entropy.
The Shannon entropy of a classical probability distribution $\left(p_{1},\ldots,p_{n}\right)$
is defined as
\begin{equation}
H\left(p_{1},\ldots,p_{n}\right):=-\sum_{j=1}^{n}p_{j}\log\left(p_{j}\right)\,.\label{c2_shannon}
\end{equation}
It was Shannon who first interpreted the above quantity as information.
He showed that, up to a constant factor, \eqref{c2_shannon} is the
only function of the probabilities $p_{1},\ldots,p_{n}$ that satisfies
some natural postulates for an information quantity \cite{ohya-petz}.
The quantity \eqref{c2_shannon} may be thought of as uncertainty
because we are unable to predict exactly the outcomes of an experiment
having different outcomes with probabilities $p_{1},\ldots,p_{n}$.
On the other hand, once we have performed the experiment, the amount
of information we gained is expressed by \eqref{c2_shannon}. Hence,
we shall regard uncertainty and information as complementary concepts.
Shannon entropy appears in classical information theory as a measure
of the theoretical optimal capacity to compress a classical message
(alphabet) without loosing information \cite{cover}. As a counterpart,
vN entropy has the same meaning in quantum information theory \cite{nielsen}.
However, the original interpretation of \eqref{c2_vn_ent} was thermodynamical.
Von Neumann used a gedanken experiment about the process of a separation
of a gas formed by two different species. Computing the change on
the entropy using the thermodynamical entropy, he arrived at a formula
for the entropy under the mixing of states. Then, assuming that the
entropy for pure states vanishes, he finally obtained \eqref{c2_vn_ent}.
Nowadays, it is known that expression \eqref{c2_vn_ent} could be
deduced from some postulates in an axiomatic way \cite{ohya-petz}.

The following proposition summarizes the most relevant properties
of the vN entropy.
\begin{prop}
\label{c2_vn_prop}The von Neumann entropy satisfies the following
properties \rm{\cite{ohya-petz}}.
\begin{enumerate}
\item \textit{\label{c2-vne-p}(positivity) $S\left(\omega\right)\geq0$
and $S\left(\omega\right)=0$ if and only if $\omega$ is pure.}
\item \textit{(boundedness) $S\left(\omega\right)\leq\log\left(d_{\mathfrak{A}}\right)$
where $d_{\mathfrak{A}}$ is the canonical dimension of $\mathfrak{A}$,
and $S\left(\omega\right)=\log\left(d_{\mathfrak{A}}\right)$ if and
only if $\omega=\tau_{\mathfrak{A}}$.}
\item \textit{(continuity) The map $\omega\mapsto S\left(\omega\right)$
is continuous in the space of states.}
\item \textit{\label{c2-vne-i}(invariance) If $\alpha$ is an automorphism
of $\mathfrak{A}$, then $S\left(\omega\right)=S\left(\omega\circ\alpha\right)$.}
\item \textit{\label{c2-vne-m}(mixing of disjoint states) If $\omega_{1},\ldots,\omega_{n}$
are pairwise disjoint states}\footnote{Two states $\omega_{1},\omega_{2}$ are said to be disjoint
states if their corresponding density $\rho_{1},\rho_{2}$ matrices
has disjoint supports. The support of a density matrix is the orthogonal
complement of its kernel, or equivalently, is the span of its eigenvectors
corresponding to non-zero eigenvalues.}\textit{ and $\sum_{j=1}^{n}p_{j}=1$, then}
\begin{equation}
S\left(\sum_{j=1}^{n}p_{j}\omega_{j}\right)=\sum_{j=1}^{n}p_{j}S\left(\omega_{j}\right)+H\left(p_{1},\ldots,p_{n}\right) \, .
\end{equation}
\item \textit{\label{c2-vne-c}(concavity) If $\sum_{j=1}^{n}p_{j}=1$, then}
\begin{equation}
\sum_{j=1}^{n}p_{j}S\left(\omega_{j}\right)\leq S\left(\sum_{j=1}^{n}p_{j}\omega_{j}\right)\leq\sum_{j=1}^{n}p_{j}S\left(\omega_{j}\right)+H\left(p_{1},\ldots,p_{n}\right)\,.
\end{equation}
\end{enumerate}
\end{prop}

Now we state the properties of the vN entropy under tensor products
of algebras. For that, it is useful to introduce the following definition,
which will be used repeatedly along this thesis.
\begin{defn}
\label{c2_def_prd_state}Given any two states $\omega_{1},\omega_{2}$
on the algebras $\mathfrak{A}_{1},\mathfrak{A}_{2}$, we define the
\textit{product state} $\omega_{1}\otimes\omega_{2}$ in $\mathfrak{A}_{1}\otimes\mathfrak{A}_{2}$
as 
\begin{equation}
\omega_{1}\otimes\omega_{2}\left(A_{1}\otimes A_{2}\right):=\omega_{1}\left(A_{1}\right)\omega_{2}\left(A_{2}\right)\,,\label{c2-prod_state_def}
\end{equation}
where $A_{i}\in\mathfrak{A}_{i}$ and \eqref{c2-prod_state_def} is
extended by linearity to more general operators. 
\end{defn}

When $\mathfrak{A}_{1},\mathfrak{A}_{2}$ are finite dimensional algebras,
it is not difficult to see that if $\rho_{1},\rho_{2}$ are the statistical
operators of $\omega_{1},\omega_{2}$ then $\rho_{1}\otimes\rho_{2}$
is the statistical operator of $\omega_{1}\otimes\omega_{2}$.\footnote{The canonical representation of $\mathfrak{A}_{1}\otimes\mathfrak{A}_{2}$
is (up to unitary equivalence) the tensor product of the canonical
representations of $\mathfrak{A}_{1}$ and $\mathfrak{A}_{2}$. } 
\begin{prop}
The von Neumann entropy satisfies the following properties \rm{\cite{ohya-petz}}.
\begin{enumerate}
\item \textit{(additivity) If $\omega_{i}\in\mathfrak{S}\left(\mathfrak{A}_{i}\right)$,
then $S\left(\omega_{1}\otimes\omega_{2}\right)=S\left(\omega_{1}\right)+S\left(\omega_{2}\right)$.}
\item \textit{(subadditivity) If $\omega_{12}\in\mathfrak{S}\left(\mathfrak{A}_{1}\otimes\mathfrak{A}_{2}\right)$
and $\omega_{i}:=\left.\omega_{12}\right|_{\mathfrak{A}_{i}}$, then
$S\left(\omega_{12}\right)\leq S\left(\omega_{1}\right)+S\left(\omega_{2}\right)$.}
\item \textit{(strong subadditivity) If $\omega_{123}\in\mathfrak{S}\left(\mathfrak{A}_{1}\otimes\mathfrak{A}_{2}\otimes\mathfrak{A}_{3}\right)$,
$\omega_{2}:=\left.\omega_{123}\right|_{\mathfrak{A}_{2}}$, $\omega_{12}:=\left.\omega_{123}\right|_{\mathfrak{A}_{1}\otimes\mathfrak{A}_{2}}$
and $\omega_{3}:=\left.\omega_{123}\right|_{\mathfrak{A}_{2}\otimes\mathfrak{A}_{3}}$,
then
\begin{equation}
S\left(\omega_{123}\right)+S\left(\omega_{2}\right)\leq S\left(\omega_{12}\right)+S\left(\omega_{23}\right)\,.\label{c2_str_sub}
\end{equation}
}
\end{enumerate}
\end{prop}

Now, we state a lemma about the equality of the vN entropy for complementary
algebras.
\begin{lem}
\label{c2_vn_dual}Let $\mathfrak{A}\subset M_{d}\left(\mathbb{C}\right)$
be a finite quantum subsystem of $M_{d}\left(\mathbb{C}\right)$ and
$\omega$ a pure state on $M_{d}\left(\mathbb{C}\right)$.\footnote{The set of pure states of a full matrix algebra $M_{d}\left(\mathbb{C}\right)=\mathbb{C}^{d\times d}$
is in one-to-one correspondence with the set of unit rays in the Hilbert
space $\mathbb{C}^{d}$. See example \ref{c1_exa_states}. } Then $S_{\mathfrak{A}}\left(\omega\right)=S_{\mathfrak{A}'}\left(\omega\right)$.
\end{lem}

Finally, let $\mathcal{F}:\mathfrak{A}_{2}\rightarrow\mathfrak{A}_{1}$
be a quantum channel and $\omega\in\mathfrak{S}\left(\mathfrak{A}_{1}\right)$.
It is not always true that the vN entropy is monotonic under quantum
channels in the following sense\footnote{For example, consider $\mathfrak{A}_{1}:=M_{d}\left(\mathbb{C}\right)\oplus M_{d}\left(\mathbb{C}\right)$,
$\mathfrak{A}_{2}:=M_{d}\left(\mathbb{C}\right)$ and a quantum channel
$\mathcal{F}:\mathfrak{A}_{2}\rightarrow\mathfrak{A}_{1}$ given by
$\mathcal{F}\left(A\right)=A\oplus A$. Any state $\omega$ on $\mathfrak{A}_{1}$
is represented by a statistical operator of the form $\rho_{\omega}=p_{1}\rho_{1}\oplus p_{2}\rho_{2}$
where $\rho_{i}$ are statistical operators in $M_{d}\left(\mathbb{C}\right)$,
$p_{i}\geq0$ and $p_{1}+p_{2}=1$. If we define $\omega_{i}$ as
the states on $M_{d}\left(\mathbb{C}\right)$ associated with the
density matrices $\rho_{i}$, definition \eqref{c2_vn_ent} gives
$S_{\mathfrak{A}_{1}}\left(\omega\right)=p_{1}S_{\mathfrak{A}_{2}}\left(\omega_{1}\right)+p_{2}S_{\mathfrak{A}_{2}}\left(\omega_{2}\right)+H\left(p_{1},p_{2}\right)$.
On the other hand, it is not difficult to see that, the state $\mathcal{F}^{*}\omega\in\mathfrak{S}\left(\mathfrak{A}_{2}\right)$
is $\mathcal{F}^{*}\omega=p_{1}\omega_{1}+p_{2}\omega_{2}$, which
it has vN entropy $S_{\mathfrak{A}_{2}}\left(\mathcal{F}^{*}\omega\right)=S_{\mathfrak{A}_{2}}\left(p_{1}\omega_{1}+p_{2}\omega_{2}\right)$.
Then, the concavity of the vN entropy gives $S_{\mathfrak{A}_{2}}\left(\mathcal{F}^{*}\omega\right)\leq S_{\mathfrak{A}_{1}}\left(\omega\right)$.
Moreover, it can always be chosen a state such as the strict inequality
holds, which contradicts \eqref{c2_qc_ent}.}
\begin{equation}
S_{\mathfrak{A}_{1}}\left(\omega\right)\leq S_{\mathfrak{A}_{2}}\left(\mathcal{F}^{*}\omega\right)\,.\label{c2_qc_ent}
\end{equation}
However, the relation \eqref{c2_qc_ent} holds for a subset of quantum
channels, called \textit{double stochastic maps} \cite{ohya-petz}.
They are characterized by the extra property: $\mathrm{Tr}{}_{\mathfrak{A}_{1}}\left(\mathcal{F}\left(A\right)\right)=\mathrm{Tr}{}_{\mathfrak{A}_{2}}\left(A\right)$
for all $A\in\mathfrak{A}_{2}$.\footnote{The existence of a double stochastic map $\mathcal{F}:\mathfrak{A}_{2}\rightarrow\mathfrak{A}_{1}$,
implies that $d_{\mathfrak{A}_{1}}=d_{\mathfrak{A}_{2}}$. } In particular, any quantum channel $\mathcal{F}:M_{n}\left(\mathbb{C}\right)\rightarrow M_{n}\left(\mathbb{C}\right)$
is double stochastic.

\subsection{Rényi entropy}

A concept close to von Neumann entropy is the Rényi entropy. 
\begin{defn}
The \textit{Rényi entropy of order $\alpha$} (\textit{$\alpha>0$}
and \textit{$\alpha\neq1$}) of the state $\omega$ on the algebra
$\mathfrak{A}$ is defined as
\begin{equation}
S_{\alpha,\mathfrak{A}}\left(\omega\right):=\frac{1}{1-\alpha}\log\mathrm{Tr}{}_{\mathfrak{A}}\left(\rho_{\omega}^{\alpha}\right)=\frac{1}{1-\alpha}\log\left(\sum_{j=1}^{n}\lambda_{j}^{\alpha}\right)\,,\label{c2-renyi-def}
\end{equation}
where $\lambda_{j}\geq0$ are the eigenvalues of $\rho_{\omega}$.
It will be denoted as $S_{\alpha}\left(\omega\right)$ whenever there
is no confusion about the algebra involved.
\end{defn}

The Rényi entropy satisfies the properties \ref{c2-vne-p} to \ref{c2-vne-i}
of proposition \ref{c2_vn_prop}, but it does not satisfy \ref{c2-vne-m}
and \ref{c2-vne-c}. It also satisfies additivity, but it does not
satisfy subadditivity nor strong subadditivity. The Rényi entropy
also satisfies the duality of lemma \ref{c2_vn_dual}. In the limit
when $\alpha\rightarrow1$, the Rényi entropy becomes the vN entropy
\begin{equation}
\lim_{\alpha\rightarrow1}S_{\alpha}\left(\omega\right)=S\left(\omega\right)\,.
\end{equation}
This characteristic feature has been largely exploited in QFT since
the replica trick gives an explicit expression for the Rényi entropy
(for the vacuum state) in terms of a path integral.

\subsection{Relative entropy}\label{c2_sec_re_fin}

Throughout this section, $\mathfrak{A}$ is a finite dimensional algebra,
$\omega,\phi\in\mathfrak{S}\left(\mathfrak{A}\right)$ two states,
and $\rho_{\omega},\rho_{\phi}$ their corresponding statistical operators
(in the canonical representation).
\begin{defn}
The \textit{relative entropy} between the states $\omega,\phi$ on
the algebra $\mathfrak{A}$ is defined as
\begin{equation}
S_{\mathfrak{A}}\left(\phi\mid\omega\right):=\mathrm{Tr}{}_{\mathfrak{A}}\left(\rho_{\phi}\left(\log\rho_{\phi}-\log\rho_{\omega}\right)\right)\,.\label{c2_re_def}
\end{equation}
It will be denoted as $S\left(\phi\mid\omega\right)$ whenever there
is no confusion about the algebra involved. 
\end{defn}

The relative entropy is a measure of distinguishability between two
states since it is always positive and it is zero only when both states
are equal. The following proposition summarizes the most important
properties of the relative entropy.
\begin{prop}
\label{c2_re_prop}The relative entropy satisfies the following properties
\rm{\cite{ohya-petz,jain}}.
\begin{enumerate}
\item \textit{\label{c2-e-re-p}(positivity) $S\left(\phi\mid\omega\right)\geq0$
and $S\left(\phi\mid\omega\right)=0$ if and only if $\phi=\omega$.}
\item \textit{(unboundedness) $S\left(\phi\mid\omega\right)=+\infty$ whenever
$\mathrm{supp}\left(\rho_{\phi}\right)\not\subset\mathrm{supp}\left(\rho_{\omega}\right)$.}
\item \textit{\label{c2-e-re-c}(continuity) The map $\left(\phi,\omega\right)\mapsto S\left(\phi\mid\omega\right)$
is continuous in the space of states, whenever it is finite.}
\item \textit{\label{c2-e-re-i}(invariance) If $\alpha$ is an automorphism
of $\mathfrak{A}$, then $S\left(\phi\mid\omega\right)=S\left(\phi\circ\alpha\mid\omega\circ\alpha\right)$.}
\item \label{c2-e-re-co}\textit{(convexity) If $\sum_{n=1}^{N}p_{n}=1$, then}
\begin{equation}
\sum_{n=1}^{N}p_{n}S\left(\phi_{n}\mid\omega\right)\leq S\left(\sum_{n=1}^{N}p_{n}\phi_{n}\mid\omega_{n}\right)+H\left(p_{1},\ldots,p_{n}\right) \, .
\end{equation}
\item \textit{(joint-convexity) If $\sum_{n=1}^{N}p_{n}=1$, then}
\begin{equation}
S\left(\sum_{n=1}^{N}p_{n}\phi_{n}\mid\sum_{n=1}^{N}p_{n}\omega_{n}\right)\leq\sum_{n=1}^{N}p_{n}S\left(\phi_{n}\mid\omega_{n}\right) \, .
\end{equation}
\item \textit{\label{c2-2-re-m}(monotonicity) If $\mathfrak{B}\subset\mathfrak{A}$
is a subalgebra, then $S_{\mathfrak{B}}\left(\phi\mid\omega\right)\leq S_{\mathfrak{A}}\left(\phi\mid\omega\right)$.}
\item \textit{\label{c2-e-re-tp}(tensor products) If $\mathfrak{A}=\mathfrak{A}_{1}\otimes\mathfrak{A}_{2}$,
$\omega\in\mathfrak{S}\left(\mathfrak{A}\right)$, $\omega_{j}:=\left.\omega\right|_{\mathfrak{A}_{j}}$
and $\phi_{j}\in\mathfrak{S}\left(\mathfrak{A}_{j}\right)$ ($j=1,2$), then} 
\begin{equation}
S_{\mathfrak{A}}\left(\omega\mid\phi_{1}\otimes\phi_{2}\right)=S_{\mathfrak{A}_{1}}\left(\omega_{1}\mid\phi_{1}\right)+S_{\mathfrak{A}_{2}}\left(\omega_{2}\mid\phi_{2}\right)+S_{\mathfrak{A}}\left(\omega\mid\omega_{1}\otimes\omega_{2}\right) \, .
\end{equation}
\item \textit{(quantum channels) If $\mathcal{F}:\mathfrak{A}_{2}\rightarrow\mathfrak{A}_{1}$
is a quantum channel, then}
\begin{equation}
S_{\mathfrak{A}_{2}}\left(\mathcal{F}^{*}\phi\mid\mathcal{F}^{*}\omega\right)\leq S_{\mathfrak{A}_{1}}\left(\phi\mid\omega\right) \, .
\end{equation}
\end{enumerate}
\end{prop}
As we claimed above, property \ref{c2-e-re-p} of proposition \ref{c2_re_prop}
says that relative entropy is a measure of distinguishability between
two states. However, it is not a distance in the pure mathematical
sense since it is not, in general, symmetric. In \eqref{c2_re_def},
it is useful to think $\omega$ as a known reference state against
the unknown state $\phi$ is going to be compared. If $\omega$ is
pure, the relative entropy is always infinite, unless $\phi=\omega$.
This is because, in such a case, if $\rho_{\omega}=\left|\Omega\right\rangle \left\langle \Omega\right|$,
then the projector $P:=1-\left|\Omega\right\rangle \left\langle \Omega\right|$
has probability $\phi\left(P\right)\neq0$. Therefore, if we measure
such a projector in the state $\phi$ and we can repeat such an experiment
as many times as we want, we can infer, with certainty, that $\phi\neq\omega$.

Relative entropy has also the following operational interpretation.
Given the known reference state $\omega$, the probability $p$ of
mistaken $\phi$ with $\omega$ after $n$ judiciously chosen experimental
measurements declines with $n$ as $\sim\mathrm{e}^{-nS\left(\phi\mid\omega\right)}$.
In this sense, relative entropy is an experimentally accessible quantity
(quantum Stein lemma \cite{petz_easy,Vedral}).

Among all the properties of proposition \ref{c2_re_prop}, we want
to remark \ref{c2-2-re-m} and \ref{c2-e-re-tp}. The monotonicity
says that the two states are, in general, less indistinguishable if
we restrict our system, i.e. $\mathfrak{A}$, to a subsystem, i.e.
$\mathfrak{B}$. This is because in the subsystem there are less operators
available, and hence, the number of experiments we can perform in
order distinguish them decreases. In the same line, the quantum channel
property asserts that we can never improve the capacity of distinguishing
the two states performing physical operations. Moreover, any physical
transformation of the states, in general, changes both states in a
way that they become less distinguishable.

After a simple manipulation, expression \eqref{c2_re_def} can be
rewritten in the following way
\begin{equation}
S\left(\phi\mid\omega\right)=\Delta\left\langle K_{\omega}\right\rangle -\Delta S\,,\label{c2_dh-ds}
\end{equation}
where
\begin{equation}
\Delta S=S\left(\phi\right)-S\left(\omega\right)\,,\quad\textrm{and }\quad\Delta\left\langle K_{\omega}\right\rangle =\phi\left(K_{\omega}\right)-\omega\left(K_{\omega}\right)\,.\label{c2_dh_and_ds}
\end{equation}
The first equation in \eqref{c2_dh_and_ds} is a difference of the
vN entropies in the two states, whereas the second equation has to
be interpreted as the difference of the expectation values of the
operator $K_{\omega}:=-\log\left(\rho_{\omega}\right)$. The operator
$K_{\omega}$ is known as the ``modular Hamiltonian'' of the state
$\omega$, and it plays an important role in modular theory of vN
algebras and the generalization of the relative entropy to general
quantum systems (see sections \eqref{c2_sec-mod_th} and \eqref{c2_sec-re_araki}
below). Relation \eqref{c2_dh-ds} has been much used in QFT, because
in many applications, it is easier to compute separately each term
on the r.h.s. of \eqref{c2_dh-ds} rather than the relative entropy
(l.h.s.) itself.

\subsection{Mutual information}

Another quantity of great importance in quantum information theory
is the mutual information. It is defined for a state $\omega$ on
a bipartite system $\mathfrak{A}_{1}\otimes\mathfrak{A}_{2}$.
\begin{defn}
The \textit{mutual information (MI)} of the state $\omega$ between
$\mathfrak{A}_{1}$ and $\mathfrak{A}_{2}$ is defined as
\begin{equation}
I\left(\mathfrak{A}_{1},\mathfrak{A}_{2};\omega\right):=S_{\mathfrak{A}_{1}\otimes\mathfrak{A}_{2}}\left(\omega\mid\omega_{1}\otimes\omega_{2}\right)\,,\label{c2_mi_def}
\end{equation}
where $\omega_{i}:=\left.\omega\right|_{\mathfrak{A}_{i}}$ are the
restrictions of $\omega$ to $\mathfrak{A}_{i}$. It will be denoted
by $I\left(\mathfrak{A}_{1},\mathfrak{A}_{2}\right)$ or $I\left(\omega\right)$
whenever there is no confusion about the state or the algebras involved.
\end{defn}

The mutual information is a measure of the information contained in
the bipartite system $\mathfrak{A}_{1}\otimes\mathfrak{A}_{2}$, which
is not contained in each subsystem $\mathfrak{A}_{i}$ individually.
If we interpret vN entropy as ``information'', the interpretation
of the MI as a ``shared information'' can be easily seen from the
following relation
\begin{equation}
I\left(\mathfrak{A}_{1},\mathfrak{A}_{2};\omega\right)=S_{\mathfrak{A}_{1}}\left(\omega_{1}\right)+S_{\mathfrak{A}_{2}}\left(\omega_{2}\right)-S_{\mathfrak{A}_{1}\otimes\mathfrak{A}_{2}}\left(\omega\right)\,,\label{c2-mi_from_vn}
\end{equation}
which holds for any finite quantum system. All the properties of the
mutual information come from the ones of the relative entropy. The
most salient are summarized in the following proposition.
\begin{prop}
\label{c2_mi_prop}The mutual information satisfies the following
properties.
\begin{enumerate}
\item \textit{(positivity) $I\left(\omega\right)\geq0$ and $I\left(\omega\right)=0$
if and only if $\omega=\omega_{1}\otimes\omega_{2}$.}
\item \textit{(boundedness) $I\left(\omega\right)\leq\log\left(d_{\mathfrak{A}_{1}}\right)+\log\left(d_{\mathfrak{A}_{2}}\right)$
where $d_{\mathfrak{A}_{i}}$ is the canonical dimension of $\mathfrak{A}_{i}$,
and $I\left(\omega\right)=\log\left(d_{\mathfrak{A}_{1}}\right)+\log\left(d_{\mathfrak{A}_{2}}\right)$
if and only if $\omega$ is pure and $\omega_{i}=\tau_{\mathfrak{A}_{i}}$.}
\item \textit{\label{c2-mi-con}(continuity) The map $\omega\mapsto I\left(\omega\right)$
is continuous in the space of states.}
\item \textit{(invariance) If $\alpha_{i}$ are automorphism of $\mathfrak{A}_{i}$
and $\alpha:=\alpha_{1}\otimes\alpha_{2}$, then $I\left(\omega\right)=I\left(\omega\circ\alpha\right)$.}
\item \textit{(monotonicity) if $\mathfrak{B}_{i}\subset\mathfrak{A}_{i}$
are subalgebras, then $I\left(\mathfrak{B}_{1},\mathfrak{B}_{2}\right)\leq I\left(\mathfrak{A}_{1},\mathfrak{A}_{2}\right)$.}
\item \textit{(quantum channels) If $\mathcal{F}_{j}:\mathfrak{A}_{j,B}\rightarrow\mathfrak{A}_{j,A}$
($j=1,2$) are quantum channels, and $\mathcal{F}_{\otimes}:\mathfrak{A}_{1,B}\otimes\mathfrak{A}_{2,B}\rightarrow\mathfrak{A}_{1,A}\otimes\mathfrak{A}_{2,A}$,
then $I\left(\mathfrak{A}_{1,B},\mathfrak{A}_{2,B};\mathcal{F}_{\otimes}^{*}\omega\right)\leq I\left(\mathfrak{A}_{1,A},\mathfrak{A}_{2,A};\omega\right)$.}
\end{enumerate}
\end{prop}
The mutual information can be considered as a measure of the correlations
between both subsystems. In fact, the following inequality holds
\begin{equation}
I\left(\mathfrak{A}_{1},\mathfrak{A}_{2};\omega\right)\geq\frac{1}{2}\left(\frac{\omega\left(A_{1}\otimes A_{2}\right)-\omega\left(A_{1}\right)\omega\left(A_{2}\right)}{\left\Vert A_{1}\right\Vert \left\Vert A_{2}\right\Vert }\right)^{2}\,,\label{c2_mi_corr}
\end{equation}
for all $A_{j}\in\mathfrak{A}_{j}$ \cite{Wolf07}. This means that
the mutual information measures all the possible correlations between
the two subsystems $\mathfrak{A}_{1}$ and $\mathfrak{A}_{2}$ for
the state $\omega$ since it bounds all the connected correlators
which measure such correlations, whatever they are quantum or classic.

\section{Entropies for general quantum systems\label{c2-Sec_entro_gener}}

In this section, we generalize all the formulas of the previous section
for general (not necessarily finite) quantum systems. We use the term
\textit{general quantum system} to refer to any quantum system, which
is described by a finite or an infinite dimensional algebra. The reason
we do not reserve this section just for ``infinite'' quantum systems,
is because all the formulas developed along this section also apply
for finite quantum systems, and in that case, they coincide with the
formulas stated in the previous section.

All the formulas of the previous section rely on the existence of
a trace functional $\mathrm{Tr}$ on finite dimensional algebras.
However, for a general infinite dimensional algebra, such a trace
does not exist.\footnote{It is beyond the scope of this thesis to explain what a trace functional
means and why it can not exist for general infinite dimensional algebras
\cite{Bratteli1,haag}.} There are some exceptions to the above statement. Among them, there
are the type I factors, which are vN algebras isomorphic to $\mathcal{B}\left(\mathcal{H}\right)$
where $\mathcal{H}$ is some separable Hilbert space. For such algebras,
the trace exists and is uniquely defined (up to a constant) as the
usual trace in the Hilbert space. Then, given a normal state $\omega$
on $\mathcal{B}\left(\mathcal{H}\right)$ represented by the density
matrix $\rho_{\omega}$, the expressions of the previous section apply.
However, it can be shown, under general physical assumptions, that
local algebras of QFT are not of this kind, but they are type III
vN algebras \cite{Buchholz86}. Unfortunately, such algebras do not
admit any trace functional. Therefore, we need to address this problem
from a different perspective. This perspective is the modular theory
of vN algebras.

We start this section reviewing the modular theory of vN algebras,
from which the relative entropy is defined. Mutual information is
then defined using relative entropy. At the end of this section, we
discuss why the vN entropy admits no definition for general quantum
systems.

\subsection{Modular theory\label{c2_sec-mod_th}}

A vN algebra $\mathcal{A}\subset\mathcal{B}\left(\mathcal{H}\right)$
is said to be in\textit{ standard form} if it has a standard vector
$\left|\Psi\right\rangle \in\mathcal{H}$. It is an important fact,
that any vN algebra is isomorphic to a unique (up to unitary equivalence)
vN algebra in standard form. Such an algebra is called its \textit{standard
representation} \cite{Haagerup}. Any normal state $\omega$ over
a vN algebra in standard form has a \textit{vector representative}
$\left|\Omega\right\rangle \in\mathcal{H}$
\begin{equation}
\omega\left(A\right)=\left\langle \Omega\right|A\left|\Omega\right\rangle \,,\quad\forall A\in\mathcal{A}\,.
\end{equation}
Thus, the set of normal states is already provided by the vector states
and we do not have to resort to density matrices. Of course the correspondence
$\omega\rightarrow\left|\Omega\right\rangle $ is one to many.\footnote{If $\mathcal{H}_{\Omega}=\overline{\mathcal{\mathcal{A}}\left|\Omega\right\rangle }$
is the cyclic subspace of $\left|\Omega\right\rangle $ and $V\in\mathcal{A}'$
is a partial isometry with initial subspace $\mathcal{H}_{\Omega}$,
then the vector $V\left|\Omega\right\rangle $ is also a vector representative
of $\omega$. Furthermore, any other vector representative is of this
form.} Furthermore, when the normal state is also faithful, every vector
representative is separating and there exists at least one representative
which is also cyclic \cite{haag}. 

\subsubsection{Modular Hamiltonian and modular flow}

Throughout this section, $\mathcal{A}\subset\mathcal{B}\left(\mathcal{H}\right)$
is a vN algebra and $\left|\Omega\right\rangle \in\mathcal{H}$ is
a standard vector.
\begin{fact}
There exists a unique (generally unbounded) closed antilinear operator
$S_{\Omega}$ such that
\begin{equation}
S_{\Omega}\,A\left|\Omega\right\rangle =A^{*}\left|\Omega\right\rangle \,,\quad\forall A\in\mathcal{A}\,.\label{c2-mod_op_def}
\end{equation}
\end{fact}
\begin{proof}
See \cite{Bratteli1}.
\end{proof}

\begin{defn}
\label{c2-def_mod}In the context of the above lemma, $S_{\Omega}$
is called the \textit{modular involution} associated with the pair
$\left\{ \mathcal{A},\left|\Omega\right\rangle \right\} $. Let $S_{\Omega}=J_{\Omega}\Delta_{\Omega}^{\frac{1}{2}}$
be its polar decomposition. Then, the (generally unbounded) positive
self-adjoint $\Delta_{\Omega}$ is called the \textit{modular operator}
and the antiunitary operator $J_{\Omega}$ is called the \textit{modular
conjugation}, the self-adjoint operator $K_{\Omega}:=-\log\left(\Delta_{\Omega}\right)$
is called \textit{modular Hamiltonian}, and the (strongly continuous)
one-parameter group of unitaries $\Delta_{\Omega}^{it}$ is called
the \textit{modular group}.
\end{defn}

\begin{rem}
\label{c2_remark_mod}We remark that the modular Hamiltonian of the
previous definition is not the same modular Hamiltonian we have introduced
at the end of section \ref{c2_sec_re_fin}. Roughly speaking, the
later is like the ``half'' part of the former one. The modular Hamiltonian
of the end of section \ref{c2_sec_re_fin} is only well-defined in
the finite dimensional case, and it is an operator which belongs to
the algebra considered. For such a reason, it is sometimes called
the ``inner'' modular Hamiltonian. On the other hand, the modular
Hamiltonian of definition \ref{c2-def_mod} is always well-defined,
but it does not belong to the algebra $\mathcal{A}$. Because of this,
it is sometimes called the ``full'' modular Hamiltonian. The relation
between these two objects will be clearer in example \ref{c2_ex_finit_mod}
below.
\end{rem}

The following is the famous theorem of Tomita-Takesaki.
\begin{thm}
\label{c2-thm-tt}(Tomita-Takesaki) The modular conjugation $J_{\Omega}$
and the modular group $\Delta_{\Omega}^{it}$ associated with $\left\{ \mathcal{A},\left|\Omega\right\rangle \right\} $
act as
\begin{eqnarray}
 & J_{\Omega}\,\mathcal{A}\,J_{\Omega}=\mathcal{A}'\,,\\
 & \Delta_{\Omega}^{it}\,\mathcal{A}\,\Delta_{\Omega}^{-it}=\mathcal{A}\quad\mathrm{and}\quad\Delta_{\Omega}^{it}\,\mathcal{A}'\,\Delta_{\Omega}^{-it}=\mathcal{A}'\,,
\end{eqnarray}
for all $t\in\mathbb{R}$. 
\end{thm}
\begin{proof}
See \cite{Bratteli1}.
\end{proof}

The above lemma states that the modular group $\Delta_{\Omega}^{it}$
induces a one-parameter group of automorphisms
\begin{equation}
\sigma_{t}^{\Omega}\left(A\right):=\Delta_{\Omega}^{it}A\Delta_{\Omega}^{-it}\,,\quad A\in\mathcal{A},\,t\in\mathbb{R},\label{c2-mod-auto}
\end{equation}
on the algebra $\mathcal{A}$, which is called the \textit{modular
flow}. 

The modular flow is a kind of dynamics or evolution in the algebra
$\mathcal{A}$ that is completely defined by a given state. It is
usually called the \textit{modular evolution}. However, it can be
shown that the modular group $\Delta_{\Omega}^{it}$ does not belong
to $\mathcal{A}$ neither $\mathcal{A}'$. Furthermore, it also can
be shown that, in general, the modular flow is not \textit{inner},
i.e. $\cancel{\exists}U\left(t\right)\in\mathcal{A}$ such that $\sigma_{t}^{\Omega}\left(A\right)=U\left(t\right)AU\left(t\right)^{*}$
for all $A\in\mathcal{A}$ and $t\in\mathbb{R}$. This means that
the modular dynamics, which is uniquely determined by the state, involves
operators which are outside the algebra.
\begin{rem}
The parameter $t$ in \ref{c2-mod-auto} is usually called \textit{modular
time}. The theorem \ref{c2-thm-tt} means that any given state gives
place to a natural notion of dynamics and hence a natural notion of
time, both state-dependent. Without being rigorous, it suggests a
kind of parallelism between modular theory and general relativity.
In general relativity, the notion of time, which is encoded in the
spacetime metric, is completely determined by the configuration of
matter and energy along the spacetime, i.e. the state of the system.
Moreover, modular theory is described in terms of operator algebras,
i.e. quantum physics. This suggests that it could be a hidden connection
between modular theory and quantum gravity. This is a proposal that
is being explored nowadays. 
\end{rem}

\begin{example}
\label{c2_ex_finit_mod}We show here how these definitions are realized
in the finite dimensional case. Any finite dimensional vN factor in
standard form is of the form $\mathcal{A}:=M_{n}\left(\mathbb{C}\right)\otimes\mathbf{1}_{n}$.
This algebra acts on the Hilbert space $\mathcal{H}:=\mathbb{C}^{n}\otimes\mathbb{C}^{n}$
and it has a commutant $\mathcal{A}'=\mathbf{1}_{n}\otimes M_{n}\left(\mathbb{C}\right)$.
A state $\omega$ on $\mathcal{A}$ can be uniquely represented by
a statistical operator $\rho_{\omega}\in M_{n}\left(\mathbb{C}\right)$
such that $\omega\left(A\otimes\mathbf{1}\right)=\mathrm{Tr}_{n}\left(\rho_{\omega}A\right)$.
Let $\rho_{\omega}=\sum_{i=1}^{n}\lambda_{i}\left|i\right\rangle \left\langle i\right|$
be its the spectral decomposition. The state $\omega$ is faithful
if and only if $\lambda_{i}>0$ for all $i=1,\ldots,n$. In such a
case, the vector $\left|\Omega\right\rangle =\sum_{i=1}^{n}\sqrt{p_{i}}\left|i\right\rangle \otimes\left|i\right\rangle $
is a standard vector representative of $\omega$ in $\mathcal{A}$.
A straightforward computation shows that
\begin{align}
S_{\Omega}\left(\left|i\right\rangle \otimes\left|j\right\rangle \right) & =\sqrt{\frac{\lambda_{i}}{\lambda_{j}}}\left|i\right\rangle \otimes\left|j\right\rangle \,, &  & \negthickspace\negthickspace\negthickspace\negthickspace\negthickspace\negthickspace\negthickspace\negthickspace\negthickspace\textrm{and extended antilinearly to }\mathcal{H}\,,\\
\Delta_{\Omega}\left(\left|i\right\rangle \otimes\left|j\right\rangle \right) & =\frac{\lambda_{i}}{\lambda_{j}}\left|i\right\rangle \otimes\left|j\right\rangle \,, &  & \negthickspace\negthickspace\negthickspace\negthickspace\negthickspace\negthickspace\negthickspace\negthickspace\negthickspace\textrm{and extended linearly to }\mathcal{H}\,,\label{c2_ex_delta}\\
J_{\Omega}\left(\left|i\right\rangle \otimes\left|j\right\rangle \right) & =\left|j\right\rangle \otimes\left|i\right\rangle \,, &  & \negthickspace\negthickspace\negthickspace\negthickspace\negthickspace\negthickspace\negthickspace\negthickspace\negthickspace\textrm{and extended antilinearly to }\mathcal{H}\,.
\end{align}
Moreover, the modular operator \eqref{c2_ex_delta} can be rewritten
as $\Delta_{\Omega}=\rho_{\omega}\otimes\rho_{\omega}^{-1}$. Then
we have that
\begin{eqnarray}
\Delta_{\Omega}^{it} & \!\!\!=\!\!\! & \rho_{\omega}^{it}\otimes\rho_{\omega}^{-it}\,,\\
K_{\Omega} & \!\!\!=\!\!\! & -\log\left(\rho_{\omega}\right)\otimes\mathbf{1}+\mathbf{1}\otimes\log\left(\rho_{\omega}\right)\,,\\
\sigma_{t}^{\Omega}\left(A\otimes\mathbf{1}\right) & \!\!\!=\!\!\! & \rho_{\omega}^{it}A\rho_{\omega}^{-it}\otimes\mathbf{1}\,.
\end{eqnarray}
\end{example}

\begin{rem}
The above example shows explicitly the relation between the ``full''
modular Hamiltonian $K_{\Omega}=-\log\left(\rho_{\omega}\right)\otimes\mathbf{1}+\mathbf{1}\otimes\log\left(\rho_{\omega}\right)$
and ``inner'' modular Hamiltonian $K_{\omega}=-\log\left(\rho_{\omega}\right)$.
As we said in \ref{c2_remark_mod}, the ``inner'' modular Hamiltonian
has to be considered as half of the full modular Hamiltonian which
is contained in the algebra.
\end{rem}

Now we state the KMS-condition, which gives a sort of connection between
thermal physics and modular evolution.
\begin{defn}
\label{c2-kms_def}(KMS-condition) Let $\omega$ be a normal state
and $\alpha_{t}\in\mathrm{Aut}\left(\mathcal{A}\right)$ a one-parameter
group of automorphisms. We say that $\omega$ satisfies the \textit{KMS-condition}
at (inverse temperature) $\beta>0$ with respect to $\alpha_{t}$,
if for any $A,B\in\mathcal{A}$, there exists a continuous function
$G_{A,B}:\mathbb{R}+i\left[-\beta,0\right]\rightarrow\mathbb{C}$,
analytic on $\mathbb{R}+i\left(-\beta,0\right)$ such that
\begin{equation}
G_{A,B}\left(t\right)=\omega\left(\alpha_{t}\left(A\right)B\right)\:\textrm{ and }\:G_{A,B}\left(t-i\beta\right)=\omega\left(B\alpha_{t}\left(A\right)\right)\,,
\end{equation}
for all $t\in\mathbb{R}$.
\end{defn}

\begin{lem}
\textup{If $\omega$ satisfies the KMS-condition with respect to $\alpha_{t}$,
then $\omega\left(\alpha_{t}\left(A\right)\right)=\omega\left(A\right)$
for all $A\in\mathcal{A}$ and all $t\in\mathbb{R}$.}
\end{lem}

The KMS-condition is, within the algebraic setting, the way to characterize
thermal states according to a specific dynamics. In other words, a
thermal state (at inverse temperature $\beta$) for a given evolution
(a one-parameter group of automorphisms) is a KMS-state. The following
example helps to understand better this concept.
\begin{example}
Let be the algebra $B\left(\mathcal{H}\right)$ and a self-adjoint
operator $H$ acting $\mathcal{H}$. Let be $\beta>0$ and suppose
that the operator $\mathrm{e}^{-\beta H}$ is a trace-class operator,
i.e. $Z:=\mathrm{Tr}_{\mathcal{H}}\left(\mathrm{e}^{-\beta H}\right)<\infty$.
We may think of $H$ as the Hamiltonian of the system which gives
the usual evolution $\alpha_{t}\left(A\right)=\mathrm{e}^{iHt}A\mathrm{e}^{-iHt}$.
The state $\omega_{\beta}\left(\cdot\right):=Z^{-1}\mathrm{Tr}_{\mathcal{H}}\left(\mathrm{e}^{-\beta H}\cdot\right)$
is a thermal state at inverse temperature $\beta$. Then, following
definition \ref{c2-kms_def}, for any $A,B\in B\left(\mathcal{H}\right)$
we define the function
\begin{equation}
G_{A,B}\left(t\right):=\omega\left(\alpha_{t}\left(A\right)B\right)=Z^{-1}\mathrm{Tr}_{\mathcal{H}}\left(\mathrm{e}^{-\beta H}\mathrm{e}^{iHt}A\mathrm{e}^{-iHt}B\right)\,.
\end{equation}
A straightforward computation shows
\begin{eqnarray}
G_{A,B}\left(t-i\beta\right) & \!\!\!=\!\!\! & Z^{-1}\mathrm{Tr}_{\mathcal{H}}\left(\mathrm{e}^{-\beta H}\mathrm{e}^{iH\left(t-i\beta\right)}A\mathrm{e}^{-iH\left(t-i\beta\right)}B\right)\nonumber \\
 & \!\!\!=\!\!\! & Z^{-1}\mathrm{Tr}_{\mathcal{H}}\left(\mathrm{e}^{-\beta H}\mathrm{e}^{iHt}\mathrm{e}^{\beta H}A\mathrm{e}^{-iHt}\mathrm{e}^{-\beta H}B\right)\nonumber \\
 & \!\!\!=\!\!\! & Z^{-1}\mathrm{Tr}_{\mathcal{H}}\left(\mathrm{e}^{-H\beta}B\mathrm{e}^{iHt}A\mathrm{e}^{-iHt}\right)\nonumber \\
 & \!\!\!=\!\!\! & \omega\left(B\alpha_{t}\left(A\right)\right)\,.\label{c2_kms_ex}
\end{eqnarray}
Equation \eqref{c2_kms_ex} shows that the state $\omega_{\beta}$
is a KMS-state (at inverse temperature $\beta$) for the evolution
$\alpha_{t}$.
\end{example}

The following theorem gives a connection between modular flow and KMS-condition.
\begin{thm}
The induced state $\omega\left(\cdot\right)=\left\langle \Omega\right|\cdot\left|\Omega\right\rangle $
satisfies the KMS-condition at $\beta=-1$ with respect to the modular
flow $\sigma_{t}^{\Omega}$.
\end{thm}
\begin{proof}
See \cite{Brattelli2,Summers:2003tf}.
\end{proof}

The KMS-condition is quite restrictive as we see below.
\begin{lem}
Let $\sigma_{t}^{\Omega}$ be the modular flow associated with $\left\{ \mathcal{A},\left|\Omega\right\rangle \right\} $
and $\omega\left(\cdot\right)=\left\langle \Omega\right|\cdot\left|\Omega\right\rangle $
the induced state.
\end{lem}
\begin{itemize}
\item \textit{If $\omega$ satisfies the KMS-condition at $\beta=-1$ with
respect to some group of automorphisms $\alpha_{t}$, then $\alpha_{t}=\sigma_{t}^{\Omega}$. }
\item \textit{Now suppose that }$\mathcal{A}$ is a factor\textit{. If a
normal state $\phi$ satisfies the KMS-condition at $\beta=-1$ with
respect to $\sigma_{t}^{\Omega}$, then $\phi=\omega$.}
\end{itemize}
\begin{proof}
See \cite{Brattelli2,Summers:2003tf}.
\end{proof}

\subsubsection{Relative modular Hamiltonian and relative modular flow\label{c2-sec-rmh}}

The aim of this section is to define the relative modular operator,
which plays a central role in the definition of the relative entropy
of the next section. Throughout this section, $\mathcal{A}\subset\mathcal{B}\left(\mathcal{H}\right)$
is a vN algebra and $\left|\Omega\right\rangle ,\left|\Phi\right\rangle \in\mathcal{H}$
are two standard vectors.
\begin{fact}
There exists a unique (generally unbounded) closed antilinear operator
$S_{\Phi,\Omega}$ such that
\begin{equation}
S_{\Phi,\Omega}\,A\left|\Omega\right\rangle =A^{*}\left|\Phi\right\rangle \,,\quad\forall A\in\mathcal{A}\,.
\end{equation}
\end{fact}

\begin{proof}
See \cite{Brattelli2}.
\end{proof}
\begin{defn}
In the context of the above lemma, $S_{\Phi,\Omega}$ is called the
\textit{relative modular involution} associated with the pair $\left\{ \mathcal{A},\left|\Omega\right\rangle ,\left|\Phi\right\rangle \right\} $.
Let $S_{\Phi,\Omega}=J_{\Phi,\Omega}\Delta_{\Phi,\Omega}^{\frac{1}{2}}$
be its polar decomposition. Then, the (generally unbounded) positive
self-adjoint $\Delta_{\Phi,\Omega}$ is called the \textit{relative
modular operator} and the antiunitary operator $J_{\Phi,\Omega}$
is called the \textit{relative modular conjugation}, the self-adjoint
operator $K_{\Phi,\Omega}:=-\log\left(\Delta_{\Phi,\Omega}\right)$
is called \textit{relative modular Hamiltonian}, and the (strongly
continuous) one-parameter group of unitaries $\Delta_{\Omega}^{it}$
is called \textit{relative modular group}.
\end{defn}

The following lemma shows a connection between the modular group and
the relative modular group.
\begin{lem}
In the context of the above definition, we have that
\begin{eqnarray}
\Delta_{\Phi,\Omega}^{it}\,A\,\Delta_{\Phi,\Omega}^{-it}=\Delta_{\Phi}^{it}\,A\,\Delta_{\Phi}^{-it} &  & A\in\mathcal{A}\,,\\
\Delta_{\Phi,\Omega}^{it}\,A'\,\Delta_{\Phi,\Omega}^{-it}=\Delta_{\Omega}^{it}\,A'\,\Delta_{\Omega}^{-it} &  & A'\in\mathcal{A}'\,,
\end{eqnarray}
for all $t\in\mathbb{R}$.
\end{lem}
\begin{proof}
See \cite{haag}.
\end{proof}

As it happens with the modular group, $\Delta_{\Phi,\Omega}^{it}\notin\mathcal{A},\mathcal{A}'$.
However, due to the above lemma, the unitaries $u_{\Phi,\Omega}\left(t\right):=\Delta_{\Phi,\Omega}^{it}\Delta_{\Omega}^{-it}$
belong to $\mathcal{A}$ for all $t\in\mathbb{R}$. such a family
of unitaries $u_{\Phi,\Omega}\left(t\right)$ encodes most of the
relevant information about the relative modular group. In fact, we
have the following lemma.
\begin{lem}
\label{c2_cnr_lemma}The one-parameter family of unitaries\footnote{This one-parameter family of unitaries is not a one-parameter group.}
$u_{\Phi,\Omega}\left(t\right):=\Delta_{\Phi,\Omega}^{it}\Delta_{\Omega}^{-it}$
satisfies the following conditions.
\end{lem}
\begin{enumerate}
\item \label{crn_1} \textit{$u_{\Phi,\Omega}\left(t\right)\in\mathcal{A}$ for all $t\in\mathbb{R}$.}
\item \label{crn_2} \textit{$\sigma_{t}^{\Phi}\left(A\right)u_{\Phi,\Omega}\left(t\right)=u_{\Phi,\Omega}\left(t\right)\sigma_{t}^{\Omega}\left(A\right)$
for all $A\in\mathcal{A}$ and all $t\in\mathbb{R}$.}
\item \label{crn_3} \textit{$u_{\Phi,\Omega}\left(t+t'\right)=u_{\Phi,\Omega}\left(t\right)\sigma_{t}^{\Omega}\left(u_{\Phi,\Omega}\left(t'\right)\right)$
for all $t,t'\in\mathbb{R}$.}
\item \label{crn_4} \textit{There exists a continuous function $G:\mathbb{R}+i\left[0,1\right]\rightarrow\mathbb{C}$,
analytic on $\mathbb{R}+i(0,1)$ such that $G(t)=\left\langle \Phi \right| u_{\Phi,\Omega}(t) \left| \Phi \right\rangle $
and $G(i) = \left\langle \Omega \! \mid \! \Omega \right\rangle $.}
\end{enumerate}
\textit{Furthermore, when $\mathcal{A}$ is a factor, $u_{\Phi,\Omega}\left(t\right)$
is uniquely determined by the conditions \ref{crn_1} to \ref{crn_4} above.}

When the modular group $\Delta_{\Omega}^{it}$ of the vector $\left|\Omega\right\rangle $
is known, the knowledge of $\Delta_{\Phi,\Omega}^{it}$ is equivalent
to the knowledge of $u_{\Phi,\Omega}\left(t\right)$. However, in
practical calculations, we may expect that the operators $u_{\Phi,\Omega}\left(t\right)$
are easier to be computed because of condition \ref{crn_1} in the previous lemma.
We exploit this feature in chapter \ref{RE_CS}, when we use all these
tools to compute the relative entropy for coherent states in free
scalar theory.
\begin{example}
\label{c2_ex_finit_rel_mod}In the same context of example \ref{c2_ex_finit_mod},
now we have a second faithful state $\phi$ whose corresponding statistical
operator is denoted by $\rho_{\phi}$. As we have explained above,
such a statistical operator has no non-zero eigenvalues. A straightforward
computation shows that the relative modular operator has the expression
$\Delta_{\Phi,\Omega}^{it}=\rho_{\phi}\otimes\rho_{\omega}^{-1}$,
and hence we have that
\begin{eqnarray}
\Delta_{\Phi,\Omega}^{it} & \!\!\!=\!\!\! & \rho_{\phi}^{it}\otimes\rho_{\omega}^{-it}\,,\\
K_{\Phi,\Omega} & \!\!\!=\!\!\! & -\log\left(\rho_{\phi}\right)\otimes\mathbf{1}+\mathbf{1}\otimes\log\left(\rho_{\omega}\right)\,,\\
u_{\Phi,\Omega}\left(t\right) & \!\!\!=\!\!\! & \rho_{\phi}^{it}\rho_{\omega}^{-it}\otimes\mathbf{1}\,.
\end{eqnarray}
\end{example}

The condition \ref{crn_4} in lemma \ref{c2_cnr_lemma} is just a rewriting of
the well-known relative KMS-condition. To state such a condition,
we first introduce the following one-parameter family of automorphisms
\begin{equation}
\sigma_{t}^{\Phi,\Omega}\left(A\right):=\sigma_{t}^{\Phi}\left(A\right)u_{\Phi,\Omega}\left(t\right)=u_{\Phi,\Omega}\left(t\right)\sigma_{t}^{\Omega}\left(A\right)\,,\quad A\in\mathcal{A},\,t\in\mathbb{R},
\end{equation}
which are known as the \textit{Connes Radon-Nikodym (CNR) cocycle}. 
\begin{thm}
\label{c2-thm-kms-rel}(relative KMS-condition \cite{Bertozzini:2010us}) In the context of
this section, let $\omega\left(\cdot\right)=\left\langle \Omega\right|\cdot\left|\Omega\right\rangle $
and $\phi\left(\cdot\right)=\left\langle \Phi\right|\cdot\left|\Phi\right\rangle $
be the induced states by $\left|\Omega\right\rangle $ and $\left|\Phi\right\rangle $.
Then, for any given $A,B\in\mathcal{A}$, there exists a unique continuous
function $G_{A,B}:\mathbb{R}+i\left[0,1\right]\rightarrow\mathbb{C}$,
analytic on $\mathbb{R}+i(0,1)$ such that\textup{
\begin{equation}
G_{A,B}\left(t\right)=\phi\left(\sigma_{t}^{\Phi,\Omega}\left(A\right)B\right)\:\textrm{ and }\:G_{A,B}\left(t+i\right)=\omega\left(B\sigma_{t}^{\Phi,\Omega}\left(A\right)\right)\,.\label{c2-kms-rel}
\end{equation}
}for all $t\in\mathbb{R}$.
\end{thm}

\subsection{Relative entropy for von Neumann algebras\label{c2_sec-re_araki}}

The definition of the relative entropy for a general vN algebra is
due to Araki \cite{Araki_entropy}.
\begin{defn}
\label{c2_re_araki}Let $\mathcal{A}\subset\mathcal{B}\left(\mathcal{H}\right)$
be a vN algebra in standard form, $\omega,\phi$ two normal faithful
states, and $\left|\Omega\right\rangle ,\left|\Phi\right\rangle \in\mathcal{H}$
standard vector representatives. The \textit{relative entropy (RE)}
between the states $\omega,\phi$ in the algebra $\mathcal{A}$ is
defined using the relative modular Hamiltonian $K_{\Omega,\Phi}$
as
\begin{equation}
S_{\mathcal{A}}\left(\phi\mid\omega\right):=\left\langle \Phi\right|K_{\Omega,\Phi}\left|\Phi\right\rangle \,.\label{c2_rel_ent}
\end{equation}
It will be denoted as $S\left(\phi\mid\omega\right)$ whenever there
is no confusion about the algebra involved.
\end{defn}

\begin{rem}
When $\mathcal{A}$ is not in standard form, the relative entropy
is defined as above but computed over its standard representation
\cite{Haagerup}. When $\omega$ or $\phi$ are not faithful normal
states, i.e. $\left|\Omega\right\rangle $ or $\left|\Phi\right\rangle $
are not standard, the definition is still possible but has to be somewhat
modified \cite{ohya-petz,Araki77}.
\end{rem}

\begin{rem}
It can be shown that the r.h.s. of expression \eqref{c2_rel_ent}
is independent of the vector representatives chosen for the states
\cite{Araki_entropy}. This justifies the notation employed on the
l.h.s. of \eqref{c2_rel_ent}.
\end{rem}

\begin{rem}
When $\omega,\phi$ are not normal states, the relative entropy could
still be defined (see section \ref{c2-sec_re_cstar}).
\end{rem}

The relative entropy \eqref{c2_rel_ent} satisfies all the same properties
that in the finite dimensional case, with the exception of \ref{c2-e-re-c}
in proposition \ref{c2_re_prop}, which has to be replaced by the
following one.
\begin{enumerate}
\item[3'.] \textit{(lower-semicontinuity) If $\lim_{n\rightarrow\infty}\left\Vert \phi_{n}-\phi\right\Vert =\lim_{n\rightarrow\infty}\left\Vert \omega_{n}-\omega\right\Vert =0$,
then \\ $\liminf_{n\rightarrow\infty}S(\phi_{n}\mid\omega_{n})\geq S(\phi\mid\omega)$.}\footnote{\textit{$\liminf_{n\rightarrow\infty}x_{n}=\lim_{n\rightarrow\infty}\left(\mathrm{inf}_{m\geq n}x_{m}\right)$}
denotes the \textit{limit inferior}.}
\end{enumerate}
The following lemma is useful for computations.
\begin{lem}
\label{c2-re-fin}When the relative entropy is finite, in particular
when $\left|\Omega\right\rangle $ belongs to the domain of $K_{\Omega,\Phi}$,
the following expression holds
\begin{equation}
S\left(\phi\mid\omega\right)=i\lim_{t\rightarrow0}\frac{\left\langle \Phi\right|\Delta_{\Omega,\Phi}^{it}\left|\Phi\right\rangle -1}{t}\,.\label{c2_rel_ent_f}
\end{equation}
\end{lem}

\begin{proof}
See \cite{ohya-petz}.
\end{proof}
\begin{example}
Let be $\omega,\phi$ two faithful states on the finite quantum system
$M_{n}\left(\mathbb{C}\right)$. It is easy to see that the vN algebra
$M_{n}\left(\mathbb{C}\right)$ acting on $\mathbb{C}^{n}$ is not
in standard form. However, we can lift to its standard representation
which becomes $\mathcal{A}=M_{n}\left(\mathbb{C}\right)\otimes\mathbf{1}_{n}$
as in the examples \ref{c2_ex_finit_mod} and \ref{c2_ex_finit_rel_mod}.
Then, all the formulas described in such examples apply here. In particular,
if $\rho_{\omega},\rho_{\phi}$ are the statistical operators of $\omega,\phi$,
we have that $K_{\Omega,\Phi}=-\log\left(\rho_{\omega}\right)\otimes\mathbf{1}+\mathbf{1}\otimes\log\left(\rho_{\phi}\right)$.
Then, according to \eqref{c2_rel_ent} we have that
\begin{eqnarray}
S\left(\phi\mid\omega\right) & \!\!\!=\!\!\! & \left\langle \Phi\right|K_{\Omega,\Phi}\left|\Phi\right\rangle =\left\langle \Phi\right|-\log\left(\rho_{\omega}\right)\otimes\mathbf{1}+\mathbf{1}\otimes\log\left(\rho_{\phi}\right)\left|\Phi\right\rangle \nonumber \\
 & \!\!\!=\!\!\! & \left\langle \Phi\right|\mathbf{1}\otimes\log\left(\rho_{\phi}\right)\left|\Phi\right\rangle -\left\langle \Phi\right|\log\left(\rho_{\omega}\right)\otimes\mathbf{1}\left|\Phi\right\rangle \,.\label{c2_ex_re}
\end{eqnarray}
The second term is the expectation value of the operator $\log\left(\rho_{\omega}\right)\otimes\mathbf{1}\in M_{n}\left(\mathbb{C}\right)\otimes\mathbf{1}_{n}$
in the vector $\left|\Phi\right\rangle $. Since $\left|\Phi\right\rangle $
is a vector representative of $\phi$, then we have that
\begin{equation}
\left\langle \Phi\right|\log\left(\rho_{\omega}\right)\otimes\mathbf{1}\left|\Phi\right\rangle =\phi\left(\log\left(\rho_{\omega}\right)\right)=\mathrm{Tr}_{n}\left(\rho_{\phi}\log\left(\rho_{\omega}\right)\right)\,.\label{c2_ex_re_p1}
\end{equation}
To compute the first term of \eqref{c2_ex_re}, we have to use the
expression of example \ref{c2_ex_finit_mod}, which shows how the
vector representative $\left|\Phi\right\rangle $ is constructed in
terms of the eigenvectors and eigenvalues of the statistical operator
$\rho_{\phi}$. Once we use such a relation, the computation is straightforward
since $\mathbf{1}\otimes\log\left(\rho_{\phi}\right)$ and $\left|\Phi\right\rangle $
are diagonal on the same basis. In fact, we get
\begin{equation}
\left\langle \Phi\right|\mathbf{1}\otimes\log\left(\rho_{\phi}\right)\left|\Phi\right\rangle =\mathrm{Tr}_{n}\left(\rho_{\phi}\log\left(\rho_{\phi}\right)\right)\,.\label{c2_ex_re_p2}
\end{equation}
Replacing \eqref{c2_ex_re_p1} and \eqref{c2_ex_re_p2} into \eqref{c2_ex_re},
we arrive to
\begin{equation}
S\left(\phi\mid\omega\right)=\mathrm{Tr}_{n}\left(\rho_{\phi}\left(\log\rho_{\phi}-\log\rho_{\omega}\right)\right)\,,
\end{equation}
which coincides with the definition of relative entropy for finite
quantum systems (equation \eqref{c2_re_def}).
\end{example}

\subsection{Relative entropy for $C^{*}$-algebras\label{c2-sec_re_cstar}}

For general $C^{*}$-algebras, the relative entropy could be defined
by three different but equivalent ways. Two of them use the definition
\ref{c2_re_araki} in an appropriate representation. The third one,
Kosaki's formula, has no use of any representation and gives an expression
involving only expectation values of the states in operators of the
$C^{*}$-algebra. Of course, these three definitions are generalizations
of definition \ref{c2_re_araki}, in the sense that they give the
same result that \eqref{c2_rel_ent} when they are applied to vN algebras
and normal states. 

Throughout this section, $\omega,\phi$ are states on the $C^{*}$-algebra
$\mathfrak{A}$. The relative entropy \textit{$S_{\mathfrak{A}}\left(\phi\mid\omega\right)$}
is defined by any of the following three equivalent ways \cite{ohya-petz}.
\begin{enumerate}
\item \label{re_c_1} Using the definition \ref{c2_re_araki} for the universal enveloping
vN algebra $\mathcal{\tilde{\pi}}\left(\mathfrak{A}\right)''$, where
$\tilde{\pi}$ is the universal representation of $\mathfrak{A}$.\footnote{The enveloping representation is defined as $\tilde{\pi}:=\bigoplus_{\omega}\pi_{\omega}$
where the sum runs along all the states $\omega$ on $\mathfrak{A}$
and $\pi_{\omega}$ are their corresponding GNS-representation.} In fact, any state $\psi$ on $\mathfrak{A}$ admits a unique normal
extension to $\tilde{\psi}$. Then we set\textit{
\begin{equation}
S_{\mathfrak{A}}\left(\phi\mid\omega\right):=S_{\mathcal{\tilde{\pi}}\left(\mathfrak{A}\right)''}(\tilde{\phi}\mid\tilde{\omega})\,.
\end{equation}
}
\item \label{re_c_2} Using the definition \ref{c2_re_araki} for the algebra $\pi_{\omega}\left(\mathfrak{A}\right)''$,
where $\pi_{\omega}$ is the GNS-representation of $\omega$. Of course,
$\omega$ is a normal state for $\pi_{\omega}\left(\mathfrak{A}\right)''$,
but nothing guarantees that $\phi$ is normal state for $\pi_{\omega}\left(\mathfrak{A}\right)''$.
Then we set\textit{
\begin{equation}
S_{\mathfrak{A}}\left(\phi\mid\omega\right):=\begin{cases}
S_{\pi_{\omega}\left(\mathfrak{A}\right)''}(\phi\mid\omega) & \textrm{if }\,\phi\textrm{ is }\pi_{\omega}\textrm{-normal},\\
+\infty & \textrm{otherwise}\,.
\end{cases}
\end{equation}
}
\item \label{re_c_3} Kosaki's formula:
\begin{equation}
\hspace{-.45cm} S_{\mathfrak{A}}\left(\phi\mid\omega\right):=\sup_{n\in\mathbb{N}}\sup_{x\left(t\right)}\left\{ \log\left(n\right)-\int_{\frac{1}{n}}^{+\infty} \! \frac{dt}{t} \left[\phi\left(x^{*}\left(t\right)x\left(t\right)\right)+\frac{1}{t}\omega\left(y^{*}\left(t\right)y\left(t\right)\right)\right]\right\} ,
\end{equation}
where the second $\sup$ is over all step functions $x:\left[\frac{1}{n},+\infty\right)\rightarrow\mathfrak{A}$
such that $x\left(t\right)=\mathbf{0}$ for sufficiently large $t$
and $y\left(t\right):=\mathbf{1}-x\left(t\right)$.
\end{enumerate}
For computational purposes, expression \ref{re_c_2} is easier to handle. Expression
\ref{re_c_3} has the advantage of being independent of any representation, and
it is good to prove general statements concerning the relative entropy.
Expression \ref{re_c_1} was the original definition that was given by Araki \cite{Araki77}.
As we claimed above, all these expressions give place to the same
result, and it can be shown that they satisfy all the same properties
which are satisfied by expression \eqref{c2_rel_ent}.

\subsection{Mutual information}

As it happens for finite quantum systems, the mutual information is
defined for the case of a bipartite system $\mathcal{A}_{1}\otimes\mathcal{A}_{2}$.
It is generalized to general quantum systems though expression \eqref{c2_mi_def}.
\begin{defn}
Let $\mathcal{A}_{1},\mathcal{A}_{2}$ be two vN algebras and $\omega$
a normal faithful state in $\mathcal{A}_{1}\otimes\mathcal{A}_{2}$.
The \textit{mutual information (MI)} of the state $\omega$ between
$\mathcal{A}_{1}$ and $\mathcal{A}_{2}$ is defined as
\begin{equation}
I\left(\mathcal{A}_{1},\mathcal{A}_{2};\omega\right):=S_{\mathcal{A}_{1}\otimes\mathcal{A}_{2}}\left(\omega\mid\omega_{1}\otimes\omega_{2}\right)\,,\label{c2-def-mi-inf}
\end{equation}
where $\omega_{i}:=\left.\omega\right|_{\mathcal{A}_{i}}$ are the
restrictions of $\omega$ to $\mathcal{A}_{i}$.\footnote{The states $\omega_{1},\omega_{2}$ and $\omega_{1}\otimes\omega_{2}$
are normal whenever $\omega$ is normal.} It will be denoted by $I\left(\mathcal{A}_{1},\mathcal{A}_{2}\right)$
or $I\left(\omega\right)$ whenever there is no confusion about the
state or the algebras involved.
\end{defn}

The mutual information \eqref{c2-def-mi-inf} satisfies all the same
properties that in the finite dimensional case, with the exception
of \ref{c2-mi-con} in proposition \ref{c2_mi_prop}, which has to
be replaced by the following one.
\begin{enumerate}
\item[3'.] \textit{(lower-semicontinuity) If $\lim_{n\rightarrow\infty}\left\Vert \omega_{n}-\omega\right\Vert =0$,
then $\liminf_{n\rightarrow\infty}I\left(\omega_{n}\right)\geq I\left(\omega\right)$.}
\end{enumerate}

\subsection{Comments about von Neumann entropy}

The aim of this brief subsection is to briefly discuss why a generalization
of the vN entropy does not exist for general quantum systems. The
following are three partial arguments to support such a statement.
\begin{enumerate}
\item \label{vn_mal_1}  In the finite case, there are many mathematical relations between
the relative entropy and the vN entropy. In fact, if one assumes that
relative entropy as a more fundamental quantity, we are able to reconstruct
the vN entropy (and all its properties) entirely from the relative
entropy itself. On the other hand, we know that a rigorous definition
of the relative entropy exists for general quantum systems and it
satisfies all the same properties that the one in the finite case.
Then, it is expected that if there should exist a generalization of
the vN entropy, all the mathematical relations between RE and the
vN entropy are preserved in the general case. However, in the finite
case, the vN entropy can be expressed in term of the relative entropy
as follows
\begin{equation}
S\left(\omega\right):=\sup\left\{ \sum_{j}p_{j}S\left(\omega_{j}\mid\omega\right)\,:\,\omega=\sum_{j}p_{j}\omega_{j}\textrm{ and }\sum_{j}p_{j}=1\right\} \,.
\end{equation}
It can be shown, in many interesting examples, that such a supremum,
in general, is $+\infty$.
\item \label{vn_mal_2} The prototypal example of a general quantum system is QFT. In order
to perform calculations, the continuum QFT may be regularized in a
lattice introducing a UV cutoff $\varepsilon$. In such a way, the
infinite dimensional algebras of QFT are converted into finite ones,
and for the regularized theory, all the expressions of section \ref{c2_sec_finite}
apply. Then, the vN entropy can be computed for the regularized theory
obtaining a finite number $S_{\varepsilon}\left(\omega\right)$. However,
in all interesting examples, we have that $S_{\varepsilon}\left(\omega\right)\rightarrow+\infty$
when $\varepsilon\rightarrow0$.
\item \label{vn_mal_3} Again in QFT, given the vacuum state $\omega$ and the algebra $\mathfrak{A}\left(\mathcal{O}\right)$
attached to the spacetime region $\mathcal{O}$, we may want to compute
the vN entropy $S_{\mathfrak{A}\left(\mathcal{O}\right)}\left(\omega\right)$.
Since the vacuum state is Poincaré invariant, we must have that $S_{\mathfrak{A}\left(\mathcal{O}\right)}\left(\omega\right)$
is a Poincaré invariant quantity, i.e. $S_{\mathfrak{A}\left(g\mathcal{O}\right)}\left(\omega\right)=S_{\mathfrak{A}\left(\mathcal{O}\right)}\left(\omega\right)$
for all $g\in\mathcal{P}_{+}^{\uparrow}$. If we also require that
$S_{\mathfrak{A}\left(\mathcal{O}\right)}\left(\omega\right)$ satisfies
strong subadditivity (which can be considered as the distinguished
``quantum'' property of the vN entropy), then there is a theorem
that asserts that $S_{\mathfrak{A}\left(\mathcal{O}\right)}\left(\omega\right)$
cannot be non-negative for all $\mathcal{O}$, unless it is trivial
(see section \ref{c3-sec-ee-qft} below). This means that some of
the properties of the vN must be relaxed in order to have a well defined
and non-trivial vN entropy in QFT.
\end{enumerate}
We will return to issues \ref{vn_mal_2} and \ref{vn_mal_3} in section \ref{EE_QFT}.

\section{Conditional expectations\label{c2_sec_ce}}

Among all the normalized cp maps, there is a subclass called conditional
expectations \cite{petz_easy}, which maps an algebra $\mathfrak{A}$
onto a subalgebra $\mathfrak{B}\subset\mathfrak{A}$.
\begin{defn}
\label{c2-def-ce}Let $\mathfrak{B}\subset\mathfrak{A}$ a subalgebra
of $\mathfrak{A}$. A linear map $\varepsilon:\mathfrak{A}\rightarrow\mathfrak{B}$
is called a \textit{conditional expectation} if it is positive, unit
preserving and has the following property
\begin{equation}
\varepsilon\left(B_{1}AB_{2}\right)=B_{1}\varepsilon\left(A\right)B_{2}\,,\quad\forall A\in\mathfrak{A}\textrm{ and }\forall B_{1},B_{2}\in\mathfrak{B}\,.\label{c2_ce_def_prop}
\end{equation}
\end{defn}

\begin{rem}
Due to the positivity of $\varepsilon$, the relation $\varepsilon\left(BA\right)=B\varepsilon\left(A\right)$
for all $A\in\mathfrak{A},B\in\mathfrak{B}$ implies $\varepsilon\left(AB\right)=\varepsilon\left(A\right)B$
for all $A\in\mathfrak{A},B\in\mathfrak{B}$.
\end{rem}

\begin{rem}
It can be shown that any conditional expectation is also completely
positive.
\end{rem}

Choosing $B_{1}=A=\mathbf{1}$ in \eqref{c2_ce_def_prop} we see that
$\varepsilon$ acts identically on $\mathfrak{B}$, and hence $\varepsilon$
maps $\mathfrak{A}$ onto the whole $\mathfrak{B}$. The physical
meaning of a conditional expectation is as follows. Suppose we start
with a physical system described by $\mathfrak{A}$, but for some
reason (e.g. experimental limitations) we have at our disposal a subset
of operators $\mathfrak{B}\subset\mathfrak{A}$. The conditional expectation
$\varepsilon:\mathfrak{A}\rightarrow\mathfrak{B}$ is the mathematical
realization of such a physical procedure, which is usually called
a \textit{coarse-graining}. In fact, as we argued in chapter \ref{AQFT},
we should have no trouble in taking linear combinations and in composing
operators in the subsystem, which means that $\mathfrak{B}\subset\mathfrak{A}$
is indeed a subalgebra. The conditional expectation also has to preserve
elements in $\mathfrak{B}$ because the expectation value of any operator
$B\in\mathfrak{B}$ in any state $\omega\in\mathfrak{S}\left(\mathfrak{A}\right)$
must be preserved by the corse-graining, i.e. $\omega\left(\varepsilon\left(B\right)\right)=\omega\left(B\right)$.
The positivity property is supported by interpretational considerations
of quantum theory. The conditional expectation has to map observables
(self-adjoint operators) and logical propositions (projectors) into
observables and logical propositions. Moreover, the property \eqref{c2_ce_def_prop}
means that for the subsystem, the composition law of operators is
preserved. It is needed because, as we have claimed, it should not
be any experimental restriction to compose operators in $\mathfrak{B}$.
The linearity is imposed to preserve the linear structure of states
of spaces. Indeed, we expect that the expectation value $\omega\left(\varepsilon\left(\lambda_{1}A_{1}+\lambda_{2}A_{2}\right)\right)$
coincides with $\lambda_{1}\omega\left(\varepsilon\left(A_{1}\right)\right)+\lambda_{2}\omega\left(\varepsilon\left(A_{2}\right)\right)$
for any state $\omega$ of $\mathfrak{A}$. 

Given an algebra $\mathfrak{A}$ and a subalgebra $\mathfrak{B}\subset\mathfrak{A}$,
a conditional expectation from $\mathfrak{A}$ to $\mathfrak{B}$
may not exist and/or may not be unique \cite{Longo:1994xe}. The different
possible conditional expectations correspond to different choices
of how the operators in $\mathfrak{A}$ are projected onto $\mathfrak{B}$.
However, when $\mathfrak{A}$ is a finite dimensional algebra (hence
also $\mathfrak{B}$), there always exists a unique conditional expectation
which satisfies

\begin{equation}
\mathrm{Tr}_{\mathfrak{A}}\left(\varepsilon\left(A\right)\right)=\mathrm{Tr}_{\mathfrak{A}}\left(A\right)\,,\quad\forall A\in\mathfrak{A}\,.
\end{equation}
Such a conditional expectation is called the \textit{trace preserving
conditional expectation} of the inclusion of algebras $\mathfrak{B}\subset\mathfrak{A}$. 
\begin{example}
The prototypical example is when $\mathfrak{A}=M_{n}\left(\mathbb{C}\right)\otimes M_{m}\left(\mathbb{C}\right)$
and $\mathfrak{B}=M_{n}\left(\mathbb{C}\right)\otimes\mathbf{1}_{m}$.
The partial trace operation
\begin{equation}
\varepsilon\left(A\otimes B\right)=\mathrm{Tr}\left(B\right)\left(A\otimes\mathbf{1}\right)\,,
\end{equation}
is the (unique) trace preserving conditional expectation for such
an inclusion of algebras.
\end{example}

A conditional expectation $\varepsilon:\mathfrak{A}\rightarrow\mathfrak{B}$
can be used to lift a state $\phi$ on $\mathcal{B}$ to the state
$\phi\circ\varepsilon$ on $\mathfrak{A}$. There is also a close
relation between conditional expectations and relative entropy. For
any $\omega,\phi\in\mathfrak{S}\left(\mathfrak{A}\right)$ such that
$\phi=\phi\circ\varepsilon$, the following relation, called \textit{conditional
expectation property}, holds
\begin{equation}
S_{\mathfrak{A}}\left(\omega\mid\phi\right)-S_{\mathfrak{B}}\left(\left.\omega\right|_{\mathfrak{B}}\mid\left.\phi\right|_{\mathfrak{B}}\right)=S_{\mathfrak{A}}\left(\omega\mid\omega\circ\varepsilon\right)\,.\label{c2_ce_prop}
\end{equation}
The physical meaning of such a relation is as follows. The state $\phi=\phi\circ\varepsilon$
has to be considered a state on $\mathfrak{A}$ whose physical content
is indeed completely codified in $\mathfrak{B}$, or in other words,
$\phi$ is a state which has no more correlations in $\mathfrak{A}$
besides the correlations that are already present in $\mathfrak{B}$.
Then, the difference of distinguishing any other state $\omega$ on
$\mathfrak{A}$ with such a state $\phi$ in the algebra $\mathfrak{A}$
or in the algebra $\mathfrak{B}$ (l.h.s. of \eqref{c2_ce_prop})
is a quantity that depends only on the correlations of $\omega$ in
$\mathfrak{A}$ that are not present in $\mathfrak{B}$ (r.h.s. of
\eqref{c2_ce_prop}). In particular, if we also have that $\omega=\omega\circ\varepsilon$,
then
\begin{equation}
S_{\mathfrak{A}}\left(\omega\mid\phi\right)=S_{\mathfrak{B}}\left(\omega\mid\phi\right)\,,\label{c2-ce-igual}
\end{equation}
which means that both states are equally distinguishable in $\mathfrak{A}$
and in $\mathfrak{B}$. Relation \eqref{c2_ce_prop} plays a central
role in our study of aspects of the entanglement entropy in QFTs that
have superselection sectors (see chapter \ref{EE_SS}).

\section{Entanglement\label{c2-secc_enta}}

Along this chapter, we have studied many information measures both
for finite and general quantum systems. Now, we will see how they
are related to the concept of (quantum) entanglement. Some states
in a bipartite quantum system show a particular kind of correlations,
between the subsystems, which are possible only because of the quantum
nature of the system. Quantum entanglement refers to the collection
of such correlations, with no classical counterpart, as a whole, and
we say that a state is entangled when at least “one” of these quantum
correlations is present in such a state. In order to understand this
phenomenon with more precision, we have to invoke Bell's gedanken
experiment \cite{bell_original,bell_chsh}.

\subsection{Bell's inequality and entanglement entropy}

We start with a physical bipartite system $S$ formed by two subsystems
$S_{A}$ and $S_{B}$, each one attached to an observer, Alice and
Bob. We may think that each subsystem is just a particle, hence we
have two distinguishable particles. If the system is quantum mechanical,
it is represented by a tensor product of $C^{*}$-algebras $\mathfrak{A}:=\mathfrak{A}_{A}\otimes\mathfrak{A}_{B}$.
Initially, the system $S$ is prepared in a global state $\omega$,
and Alice and Bob perform measurements in each subsystem independently.\footnote{“Independently” means that if Alice performs any measurement in her
subsystem, this will not modify the probabilities of the outcomes
performed by Bob in his subsystem. In the classical case, this is
always true. In the quantum case, it is only true (for a general state)
whenever the operators, representing such measurements, commute. In
our case, this holds because the subsystems are in a tensor product.
Physically, it could be guaranteed whenever Alice and Bob are spacelike
separated.} 

We consider two different measurement devices on each subsystem given
by observables $\left\{ A_{1},A_{2}\right\} $ on $S_{A}$ and $\left\{ B_{1},B_{2}\right\} $
on $S_{B}$. For convenience, we label the outcomes of each measurement
with the values $A_{j},B_{j}\in\{-1,+1\}$. We are able to perform
simultaneously any pair of measurements $(A_{j},B_{k})$, but we are
not able to measure $(A_{1},A_{2})$ nor $(B_{1},B_{2})$ simultaneously.
We also assume that we can reproduce the state $\omega$ with perfect
accuracy as many times as we want and that the measurements can be
performed with perfect accuracy.

We run the experiment $4N$ times, with $N\gg1$. The first $N$ times
we measure $(A_{1},B_{1})$, the second $N$ times we measure the
$(A_{1},B_{2})$, and so on. We can estimate experimentally the expectation
values $\left\langle A_{j}B_{k}\right\rangle $, and therefore we
can estimate the expectation value of the observable
\begin{equation}
C:=A_{1}B_{1}+A_{1}B_{2}+A_{2}B_{1}-A_{2}B_{2}\,.
\end{equation}

Assume now that the bipartite system is classical. This means that
the observables $(A_{1},A_{2})$ and $(B_{1},B_{2})$ are simultaneously
measurable. According to the basic formulas of classical probability
theory, a straightforward computation gives
\begin{equation}
\left|\left\langle C\right\rangle _{\omega}\right|\leq2\:,\label{c2-bell-ine}
\end{equation}
independent of the choice of the state $\omega$ and the observables,
as long as they take values in the set $\{-1,+1\}$. Formula \eqref{c2-bell-ine}
is known as the \textit{CHSH version of Bell's inequality}.\footnote{CHSH stands for John Clauser, Michael Horne, Abner Shimony, and Richard Holt, who described it in the paper \cite{bell_chsh}.}

However, if the bipartite system is quantum mechanical the above inequality
can be violated. The prototypical example is the bipartite quantum
system of two qubits. Each subsystem is represented by the finite
dimensional algebras $\mathfrak{A}_{A}:=\mathfrak{A}_{B}:=M_{2}\left(\mathbb{C}\right)$,
and hence, the bipartite system is represented by $\mathfrak{A}:=M_{2}\left(\mathbb{C}\right)\otimes M_{2}\left(\mathbb{C}\right)$.
Each subsystem acts naturally on the Hilbert space $\mathcal{H}_{A}=\mathcal{H}_{B}=\mathbb{C}^{2}$,
and we fix an orthonormal basis $\left\{ \left|0\right\rangle ,\left|1\right\rangle \right\} $
of such a Hilbert space. In order to show explicitly the violation
of \eqref{c2-bell-ine}, we can choose observables of the form
\begin{eqnarray}
A_{j}:=\bar{\sigma}\cdot\bar{a}_{j}\otimes\mathbf{1}_{2} & \textrm{ and } & B_{j}:=\mathbf{1}_{2}\otimes\bar{\sigma}\cdot\bar{b}_{j}\,,
\end{eqnarray}
where $\bar{\sigma}:=(\sigma_{1},\sigma_{2},\sigma_{3})$ is the usual
vector formed by the Pauli matrices, and $\bar{a}_{j},\bar{b}_{j}\in\mathbb{R}^{3}$
with $|\bar{a}_{j}|=|\bar{b}_{j}|=1$ so that the eigenvalues of such
operators belong to $\left\{ -1,+1\right\} $ as we need. In particular,
taking
\begin{equation}
\bar{a}_{1}:=\left(0,0,1\right)\,,\;\bar{a}_{2}:=\left(1,0,0\right)\,,\;\bar{b}_{1}:=\dfrac{1}{\sqrt{2}}\left(1,0,1\right)\,,\;\bar{b}_{2}:=\dfrac{1}{\sqrt{2}}\left(1,0,-1\right)\,,
\end{equation}
and the pure state $\psi$ given by the vector (Bell pair)
\begin{equation}
\left|\Psi\right\rangle :=\frac{1}{\sqrt{2}}\left(\left|0\right\rangle \otimes\left|0\right\rangle +\left|1\right\rangle \otimes\left|1\right\rangle \right)\in\mathcal{H}_{A}\otimes\mathcal{H}_{B}\,,\label{c2_bell}
\end{equation}
a straightforward computation shows
\begin{equation}
\left\langle C\right\rangle _{\psi}:=\left\langle \Psi\right|A_{1}\otimes B_{1}\left|\Psi\right\rangle +\left\langle \Psi\right|A_{1}\otimes B_{2}\left|\Psi\right\rangle +\left\langle \Psi\right|A_{2}\otimes B_{1}\left|\Psi\right\rangle -\left\langle \Psi\right|A_{2}\otimes B_{2}\left|\Psi\right\rangle =2\sqrt{2}>2\,.
\end{equation}
The conclusion is that the state $\psi$ exhibits correlations between
the two subsystems that cannot be reproduced by a classical system.
This is a pure quantum phenomenon, and it is possible because of the
particular structure of the underlying state. A further computation
shows that 
\begin{equation}
\left|\left\langle C\right\rangle _{\omega}\right|\leq2\sqrt{2}\,,
\end{equation}
for any state $\omega$ and any observables having eigenvalues in
the set $\{-1,+1\}$. This means that the state $\psi$ gives a maximum
violation of the Bell's inequality. This fact is usually interpreted
stating that $\psi$ is a \textit{maximally entangled state}.

It is important to emphasize that such correlations occur even if
$\mathfrak{A}_{A}$ and $\mathfrak{A}_{B}$ are commuting algebras,
i.e. they are statistically independent. In fact, the above behavior
is only possible because of the existence of non-commuting observables
within $\mathfrak{A}_{A}$ and $\mathfrak{A}_{B}$. If instead of
the above operators, we chose commuting operators $[A_{1},A_{2}]=0$
and $[B_{1},B_{2}]=0$, we would obtain an expectation value $\bigl|\left\langle C\right\rangle _{\psi}\bigr|\leq2$,
as in the classical case. This means that the quantum character of
the system is fundamental to have entanglement. In a more general
scenario, the entanglement depends not only on the state but also
on the algebras $\mathfrak{A}_{A}$ and $\mathfrak{A}_{B}$. As long
as $\mathfrak{A}_{A}$ and $\mathfrak{A}_{B}$ are bigger, more operators
are available within each subsystem, and hence more quantum correlations
could be exhibited by the state. On the other hand, when the algebras
$\mathfrak{A}_{A}$ and $\mathfrak{A}_{B}$ are Abelian, the inequality
\eqref{c2-bell-ine} is satisfied for any state, and hence there is
no entanglement between the systems.

As we have already claimed, the form \eqref{c2_bell} of the state
$\psi$ is crucial to violate Bell's inequality. For example, the
state $\omega$, given by the statistical operator
\begin{equation}
\rho_{\omega}:=p\left|00\right\rangle \left\langle 00\right|+(1-p)\left|11\right\rangle \left\langle 11\right|\,,\quad p\in\left[0,1\right]\,,
\end{equation}
satisfies $\left|\left\langle C\right\rangle _{\omega}\right|\leq2$,
for any choice of the operators $A_{j},B_{k}$ with eigenvalues $\{-1,+1\}$.
The state $\omega$ may exhibit correlations but of the classical
type.

The violation of Bell's inequality in a real experiment implies that
the underlying system is governed by the laws of quantum mechanics.
Many experimental tests were performed obtaining successful violations
of Bell's inequality \cite{bell1,bell2,bell3,bell4,bell5,bell6,bell7,bell8,bell9,bell10,bell11,bell12,bell13,bell14,bell15}.
This pure quantum phenomenon has already caused a lot of problems
to philosophers of quantum theory \cite{epr}. However, there is an
almost general consensus that such a phenomenon does not violate Einstein's
principle of locality. The non-local correlations, due to the entanglement,
cannot be used to transmit information at spacelike distances. Moreover,
from a conceptual viewpoint, the violation of Bell's inequality rules
out the possibility of a quantum theory of local hidden variables
compatible with Poincaré invariance. However, it leaves the door open
to non-local hidden variables theories, such as De Broglie–Bohm theory,
Many Worlds Theory, etc.

Any operation (unitary evolution and/or projective measurement) performed
by Alice or Bob in each of their subsystems is called a local operation.
If we also allow them to communicate classically (e.g. by a call phone),
the operations allowed in the bipartite system are called \textit{local
operations with classical communication (LOCC)}. LOCC's have been
used experimentally to teleport a quantum state along spacelike distances
\cite{teleport}. The entanglement has been largely exploited in quantum
computation in order to show (until now in a theoretical way) that
a quantum computer can solve many computational problems in a faster
way than any classical computer does \cite{nielsen}.

The question which naturally arises is if we can better understand
the structure of the entanglement and if we can quantify it. The first
attempt to do that was using the concept of \textit{entanglement entropy
(EE)}. In the above context, it is given by the vN entropy of the
reduced states
\begin{equation}
S_{E,\mathfrak{A}}\left(\omega\right):=S_{\mathfrak{A}_{A}}\left(\left.\omega\right|_{\mathfrak{A}_{A}}\right)=S_{\mathfrak{A}_{B}}\left(\left.\omega\right|_{\mathfrak{A}_{B}}\right)\,,\label{c2_ee}
\end{equation}
where the second equality holds because the state $\omega$ is a pure
state on $\mathfrak{A}$ (lemma \ref{c2_vn_dual}). For the state
\eqref{c2_bell}, we have that $S_{E,\mathfrak{A}}\left(\psi\right)=\log\left(2\right)$,
which is the maximum value possible for this quantity \eqref{c2_ee}.
This agrees with the fact that $\psi$ is a maximally entangled state
since it gives maximum violation of Bell's inequality. The entanglement
entropy is a good entanglement measure for pure states in finite quantum
systems. However, it has (at least) two important problems: 1) it
does not give a good measure of entanglement for non-pure global states,
and 2) it is only well-defined for finite dimensional or type I vN
algebras. The second problem arises naturally in QFT, as we will see
in the following chapter. These facts force us to take alternative
ways to define EE in QFT.

In the following subsections, we explain in more detail how the example
above can be generalized to general quantum systems. We define with
precision the concepts of entangled states, separable state, and entanglement
measure.

\subsection{Entangled and separable states\label{c2_sec_ent-sep}}

A general \textit{bipartite} (quantum) system consists of two algebras
$\mathfrak{A}_{1},\mathfrak{A}_{2}$, called the \textit{subsystems},
which give place to the algebra $\mathfrak{A}=\mathfrak{A}_{1}\otimes\mathfrak{A}_{2}$,
called the \textit{system}. Any state $\omega\in\mathfrak{S}\left(\mathfrak{A}\right)$
is called a \textit{global state}, and for example, the term \textit{global
pure state} is frequently used to denote any element in $\mathfrak{S}_{p}\left(\mathfrak{A}\right)$.
Generally, the subsystems and the system are considered to be vN algebras,
and all the states to be normal. We can now introduce the distinction
between “separable states” and “entangled states”.
\begin{defn}
A global state $\omega\in\mathfrak{S}\left(\mathfrak{A}\right)$ is
said to be \textit{separable} if it can be written as 
\begin{equation}
\omega=\sum_{j=1}^{\infty}p_{j} \, \omega_{1,j}\otimes\omega_{2,j}\,,\label{c2_sep_def}
\end{equation}
for some collection of states $\omega_{i,j}\in\mathfrak{S}\left(\mathfrak{A}_{i}\right)$
and non-negative numbers $p_{j}\geq0$ such that $\sum_{j=1}^{\infty}p_{j}=1$
($i=1,2$ and $j\in\mathbb{N}$).\footnote{The sum \eqref{c2_sep_def} is considered as a uniform convergent
series.} In particular, expression \eqref{c2_sep_def} means that $\omega\left(A_{1}\otimes A_{2}\right)=\sum_{j=1}^{\infty}p_{j}\omega_{1,j}\left(A_{1}\right)\omega_{2,j}\left(A_{2}\right)$
for all $A_{i}\in\mathfrak{A}_{i}$. Any non-separable state is called
\textit{entangled}.
\end{defn}

Since any mixed state is a convex combination of pure states, then
any separable state is a convex combination of pure states. And of
course, the converse is trivially true. Then, we do not loose any
generality just considering separable states as in \eqref{c2_sep_def}
where all $\omega_{i,j}$ are pure. Depending on the global state
$\omega$, the outcomes of separate measurements on the two subsystems
$\mathfrak{A}_{1},\mathfrak{A}_{2}$ can exhibit different kinds of
correlations. When $\omega:=\omega_{1}\otimes\omega_{2}$ , there
are no correlations at all. For a general separable state, however,
there can be correlations, which are of a classical nature. Entangled
states exhibit even more general “quantum” correlations.

\subsection{Entanglement measures\label{c2_sec_ent-meas}}

The core of the entanglement measures lies in the notion of separable
operation \cite{Hollands17}. In section \ref{c2_sec_qc}, we have
introduced the set of quantum operations. Any operation is an arbitrary
combination of quantum channels (cp normalized maps) and projective
measurements. For any two given bipartite systems $\mathfrak{A}_{1,A}\otimes\mathfrak{A}_{2,A}$
and $\mathfrak{A}_{1,B}\otimes\mathfrak{A}_{2,B}$, we say that a
cp map $\mathcal{F}:\mathfrak{A}_{1,B}\otimes\mathfrak{A}_{2,B}\rightarrow\mathfrak{A}_{1,A}\otimes\mathfrak{A}_{2,A}$
is \textit{local} if there exist cp maps $\mathcal{F}_{i}:\mathfrak{A}_{i,B}\rightarrow\mathfrak{A}_{i,A}$
such that
\begin{equation}
\mathcal{F}\left(A_{1}\otimes A_{2}\right)=\mathcal{F}_{1}\left(A_{1}\right)\otimes\mathcal{F}_{2}\left(A_{2}\right)=\left(\mathcal{F}_{1}\otimes\mathcal{F}_{2}\right)\left(A_{1}\otimes A_{2}\right)\,,\label{c2_loc_cp}
\end{equation}
for all $A_{i}\in\mathfrak{A}_{i}$. More generally, a \textit{separable
operation} is by definition any map $\mathcal{F}:\mathfrak{A}_{1,B}\otimes\mathfrak{A}_{2,B}\rightarrow\mathfrak{A}_{1,A}\otimes\mathfrak{A}_{2,A}$
of the form $\mathcal{F}=\sum_{j=1}^{n}\mathcal{F}_{j}$ where each
$\mathcal{F}_{j}$ is a local cp map of the form \eqref{c2_loc_cp}
and $\mathcal{F}\left(\mathbf{1}\right)=\sum_{j=1}^{n}\mathcal{F}_{j}\left(\mathbf{1}\otimes\mathbf{1}\right)=\mathbf{1}$.
Such an operation maps a state $\omega$ of $\mathfrak{A}$, with
probability $p_{j}:=\omega\left(\mathcal{F}_{j}\left(\mathbf{\mathbf{1}\otimes\mathbf{1}}\right)\right)$,
to the state $\frac{1}{p_{j}}\mathcal{F}_{j}^{*}\omega$.

The main idea of an entanglement measure is to give a partial order
in the set of states of a bipartite system, which indicates which
states are “more entangled” than other states. Until now, given a
general bipartite system, there is no canonical or preferred way to
do that. It could be because we have not already understood the structure
of the entanglement in a final way, or perhaps, the different ways
could be attached to different experimental choices. The important
fact that they all have in common is that entanglement cannot be created
performing local operations. The following definition is not definitive,
but it collects all the properties we expect for an entanglement measure.
\begin{defn}
An entanglement measure $E$ is a function from the set of all bipartite
quantum systems $\mathfrak{A}_{1}\otimes\mathfrak{A}_{2}$ and their
corresponding states $\mathfrak{S}\left(\mathfrak{A}_{1}\otimes\mathfrak{A}_{2}\right)$
to the real numbers, i.e.
\begin{eqnarray}
\left(\mathfrak{A}_{1}\otimes\mathfrak{A}_{2},\omega\right) & \mapsto & E_{\mathfrak{A}_{1}\otimes\mathfrak{A}_{2}}\left(\omega\right)=E\left(\omega\right)\,,
\end{eqnarray}
which satisfies the following properties.
\begin{enumerate}
\item \label{ems_sym} \textit{(symmetry)} $E$ is independent of the order of the systems $\mathfrak{A}_{1}$
and $\mathfrak{A}_{2}$.
\item \label{ems_pos} \textit{(positivity)} $E\left(\omega\right)\geq0$ and $E\left(\omega\right)=0$
if and only if $\omega$ is separable.
\item \label{ems_con} \textit{(convexity)} If $\omega=\sum_{j=1}^{n}p_{j}\omega_{j}$ with $p_{j}\geq0$
and $\sum_{j=1}^{n}p_{j}=1$, then 
\begin{equation}
E\left(\omega\right)\leq\sum_{j=1}^{n}p_{j}E\left(\omega_{j}\right) \, .
\end{equation}
\item \label{ems_mon} \textit{(monotonicity under separable operations)} For any separable operation
$\mathcal{F}:\mathfrak{A}_{1,B}\otimes\mathfrak{A}_{2,B}\rightarrow\mathfrak{A}_{1,A}\otimes\mathfrak{A}_{2,A}$
of the form $\mathcal{F}=\sum_{j=1}^{n}\mathcal{F}_{j}$, then 
\begin{equation}
\sum_{j=1}^{n}p_{j}E\left(\frac{\mathcal{F}_{j}^{*}\omega}{p_{j}}\right)\leq E\left(\omega\right) \, ,
\end{equation}
where the sum runs over all $p_{j}:=\omega\left(\mathcal{F}_{j}\left(\mathbf{\mathbf{1}}\right)\right)>0$.
\item \label{ems_sup} \textit{(superadditivity)} Let $\mathfrak{A}_{1}=\mathfrak{A}_{1,A}\otimes\mathfrak{A}_{1,B}$
and $\mathfrak{A}_{2}=\mathfrak{A}_{2,A}\otimes\mathfrak{A}_{2,B}$
be two bipartite systems, and let $\omega\in\mathfrak{S}\left(\mathfrak{A}_{1}\otimes\mathfrak{A}_{2}\right)$
and let $\omega_{A}:=\left.\omega\right|_{\mathfrak{A}_{1,A}\otimes\mathfrak{A}_{2,A}}$
and $\omega_{B}:=\left.\omega\right|_{\mathfrak{A}_{1,B}\otimes\mathfrak{A}_{2,B}}$.
Then 
\begin{equation}
E_{\mathfrak{A}_{1,A}\otimes\mathfrak{A}_{2,A}}\left(\omega_{A}\right)+E_{\mathfrak{A}_{1,B}\otimes\mathfrak{A}_{2,B}}\left(\omega_{B}\right)\leq E_{\mathfrak{A}_{1}\otimes\mathfrak{A}_{2}}\left(\omega\right)\,,
\end{equation}
and equality holds when $\omega=\omega_{A}\otimes\omega_{B}$.
\end{enumerate}
\end{defn}

Property \ref{ems_con} states that entanglement cannot be increased by mixing
states, however, it can be reduced. Property \ref{ems_mon} states that no entanglement
can be won by performing separable operations. In particular if two
states $\omega,\phi$ are such that one can go from $\omega$ to $\phi$
by local operations and vice-versa, then $E\left(\omega\right)=E\left(\phi\right)$.
Property \ref{ems_sup} states the behavior of the entanglement when the subsystems
$\mathfrak{A}_{1}$ and $\mathfrak{A}_{2}$ are themselves composed
of statistically independent subsystems.

There are many examples of entanglement measures in the literature
\cite{Hollands17}. However, not all of them satisfy all the properties
above but a proper subset. In particular, we can mention the \textit{relative
entanglement entropy}
\begin{equation}
E_{R}\left(\omega\right):=\mathrm{inf}\left\{ S\left(\omega\mid\phi\right)\,:\:\phi\textrm{ is separable}\right\} \,,
\end{equation}
which satisfies properties \ref{ems_sym} to \ref{ems_mon}, and the mutual information
\begin{equation}
E_{I}\left(\omega\right):=I\left(\mathfrak{A}_{1},\mathfrak{A}_{2};\omega\right)\,,
\end{equation}
which satisfies properties \ref{ems_sym}, \ref{ems_con} and \ref{ems_mon}, but it satisfies \ref{ems_pos} only over
the subset of pure states. We always have $E_{R}\left(\omega\right)\leq E_{I}\left(\omega\right)$
\cite{Hollands17}.


\renewcommand\chaptername{Chapter}
\selectlanguage{english}

\chapter{Entanglement in QFT\label{EE_QFT}}

In this chapter, we study general aspects of entanglement in QFT.
For this purpose, we start with a general explanation about how the
previous discussion of entanglement in bipartite quantum systems fits
in AQFT. In particular, we remark that the MI for two strictly spacelike
separated regions is always well-defined in QFT. We discuss two famous
theorems concerning entanglement in QFT. The first one is the Bisognano-Wichmann
theorem, which gives a direct connection between modular evolution
and spacetime symmetries in QFT. The second one is the “entanglement
interpretation” of the Reeh-Schlieder theorem, which suggests that
the vacuum state of any QFT is entangled at any distance.

Then we study the entanglement entropy (EE) in QFT. It is a quantity
that measures the entanglement, on a given state, between the degrees
of freedom of a region $\mathcal{O}$ with the degrees of freedom
of its commutant $\mathcal{O}'$. We show that it is an ill-defined
quantity due to UV-divergences. These divergencies have their origin
in the fact that QFT is a theory with a “continuum” of degrees of
freedom. Then, to study EE in QFT we must regularize the theory placing
a short distance cutoff. In the regularized theory, we see that the
divergent behavior of the EE with the cutoff has a very rigid structure.
In fact, the understanding of such a divergence structure, allows
us to define a universal regularized EE and a minimally subtracted
EE using the MI. These entanglement measures have all the properties
expected for the EE, except for the positivity. Furthermore, they
contain all the relevant and universal information about the entanglement
between the region $\mathcal{O}$ and $\mathcal{O}'$ in the vacuum
state.

In the rest of this chapter, we show how to compute the modular Hamiltonian
and the EE for free fields using a lattice cutoff. In the presence
of such a cutoff, the infinite type III vN algebras of QFT are transformed
into type I vN algebras, and hence, all the formulas for finite quantum
systems of the previous chapter apply there. In particular, we focus
on Gaussian states. Such states are characterized by the fact that
all their information is encoded in their two-point correlators. For
these states, we obtain simple formulas for the modular Hamiltonian
and the EE. In the end, we explain how all the expressions obtained
for the lattice model have to be interpreted in the continuum limit.
All the expressions derived there are very useful for computational
purposes. We use them in chapter \ref{CURRENT} where we compute
the modular Hamiltonian and the EE for the free chiral current field.

\section{General structure of entanglement in AQFT\label{c3-gen-str}}

From the algebraic perspective, the study of the entanglement in QFT
is as follows. As we explained in chapter \ref{AQFT}, we start with
a net of local $C^{*}$-algebras $\mathfrak{A}\left(\mathcal{O}\right)$
indexed by the set of causally complete regions $\mathcal{K}$ of
the spacetime. Given two regions $\mathcal{O}_{1},\mathcal{O}_{2}\in\mathcal{K}$
we want to characterize the (quantum) correlations between the degrees
of freedom localized in $\mathcal{O}_{1}$ and $\mathcal{O}_{2}$.
According to the discussion in the previous chapter, it must be done
with the help of the algebras $\mathfrak{A}\left(\mathcal{O}_{1}\right),\mathfrak{A}\left(\mathcal{O}_{2}\right)\subset\mathfrak{A}$.
If we want to study the entanglement between these two algebras on
a state $\omega\in\mathfrak{S}\left(\mathfrak{A}\right)$, it is useful
to consider the GNS-representation $\pi_{\omega}:\mathfrak{A}\rightarrow\mathcal{B}\left(\mathcal{H}_{\omega}\right)$
of $\omega$ and define the net of vN algebras $\mathcal{A}\left(\mathcal{O}\right):=\pi_{\omega}\left(\mathfrak{A}\left(\mathcal{O}\right)\right)''\subset\mathcal{B}\left(\mathcal{H}_{\omega}\right)$
with $\mathcal{O}\in\mathcal{K}$. This net of vN algebras allows
us to study the entanglement across any two regions, not only for
the state $\omega$ but for any $\pi_{\omega}$-normal state $\phi$
(any state which is represented by a density matrix in such a representation).
Moreover, any global normal state is \textit{locally normal}, i.e.
its restriction to any local algebra $\mathcal{A}\left(\mathcal{O}\right)$
is normal. Then, if we assume that a local algebra $\mathcal{A}\left(\mathcal{O}_{1}\right)$
is in standard form, any normal state $\phi$ has a vector representative
$\left|\Phi\right\rangle \in\mathcal{H}_{\omega}$, which is also
a vector representative for any subalgebra $\mathcal{A}\left(\mathcal{O}_{2}\right)$
of any region $\mathcal{O}_{2}\subset\mathcal{O}_{1}$. This picture
fits perfectly when one considers the vacuum state $\omega_{0}$.
For the vacuum representation $\pi_{0}:\mathfrak{A}\rightarrow\mathcal{B}\left(\mathcal{H}_{0}\right)$,
due to Reeh-Schlieder theorem \ref{c1_rs}, the vacuum vector $\left|0\right\rangle $
is standard for any local algebra $\mathcal{A}\left(\mathcal{O}\right)$
whenever $\mathcal{\mathcal{O}\in\mathcal{K}}$ and $\mathcal{O}'\neq\emptyset$.
From now on, we always work in the vacuum representation, unless it
is otherwise stated.

Now given two local algebras $\mathcal{A}\left(\mathcal{O}_{1}\right),\mathcal{A}\left(\mathcal{O}_{2}\right)\subset\mathcal{B}\left(\mathcal{H}_{0}\right)$,
in order to have a well-defined entanglement structure, we need this
pair of algebras to be \textit{kinematically independent}, i.e. they
commute.\footnote{For fermion nets, we impose twisted commutativity (see section \ref{c1-sec-fermions}).}
It can be easily satisfied considering spacelike separated regions
$\mathcal{O}_{1}\text{\Large\ensuremath{\times}}\mathcal{O}_{2}$.
In particular, it includes the case when $\mathcal{O}_{2}=\mathcal{O}'_{1}$,
which is frequently used in QFT. We also assume that $\mathcal{A}\left(\mathcal{O}_{1}\right),\mathcal{A}\left(\mathcal{O}_{2}\right)$
do not have a non-trivial element in common, i.e. $\mathcal{A}\left(\mathcal{O}_{1}\right)\cap\mathcal{A}\left(\mathcal{O}_{2}\right)=\mathbf{1}$.
This is automatically satisfied whenever $\mathcal{A}\left(\mathcal{O}_{1}\right)$
or $\mathcal{A}\left(\mathcal{O}_{2}\right)$ is a factor, and this
holds under the assumption that the vacuum representation is irreducible,
i.e. $\mathcal{A}\left(\mathbb{R}^{d}\right)'=\mathbf{1}$ (see discussion
in section \ref{c1_subsec_lattice}). 

We also require that the algebras $\mathcal{A}\left(\mathcal{O}_{1}\right),\mathcal{A}\left(\mathcal{O}_{2}\right)$
are \textit{statistically independent (in the spatial product sense)}
\cite{Summers90}, i.e.
\begin{equation}
\mathcal{A}\left(\mathcal{O}_{1}\right)\vee\mathcal{A}\left(\mathcal{O}_{2}\right)\cong\mathcal{A}\left(\mathcal{O}_{1}\right)\otimes\mathcal{A}\left(\mathcal{O}_{2}\right)\,,\label{c3-sta_ind}
\end{equation}
where $\cong$ means unitarily equivalence. We remark that the vN
algebra on the l.h.s. of \eqref{c3-sta_ind} acts on $\mathcal{H}_{0}$,
whereas the algebra on the r.h.s. of \eqref{c3-sta_ind} acts on $\mathcal{H}_{0}\otimes\mathcal{H}_{0}$.
If we assume additivity for this pair of regions, we also have that
\begin{equation}
\mathcal{A}\left(\mathcal{O}_{1}\right)\vee\mathcal{A}\left(\mathcal{O}_{2}\right)=\mathcal{A}\left(\mathcal{O}_{1}\cup\mathcal{O}_{2}\right)=\mathcal{A}\left(\mathcal{O}_{1}\vee\mathcal{O}_{2}\right)\,,\label{c3-add-alg}
\end{equation}
where the last equality holds since $\mathcal{O}_{1}$ and $\mathcal{O}_{2}$
are spacelike separated. The statistical independence relation \eqref{c3-sta_ind}
is a non-trivial property which may not hold in general. It could
be derived from a stronger assumption called split-property \cite{Summers90}. 
\begin{defn}\label{def_split_property}
Let $\mathcal{A}_{1},\mathcal{A}_{2}\subset\mathcal{B}(\mathcal{H})$
be two commuting vN algebras. We say that the pair $\mathcal{A}_{1},\mathcal{A}_{2}$
satisfies the \textit{split-property} if there exists a type I factor
$\mathcal{R}$ such that $\mathcal{A}_{1}\subset\mathcal{R}\subset\mathcal{A}'_{2}$.
\end{defn}

Roughly speaking, in QFT, the split property for local algebras attached
to strictly spacelike separated regions $\mathcal{O}_{1}\text{\Large\ensuremath{\times\!\negmedspace\!\times}}\mathcal{O}_{2}$
is a condition about the number of local degrees of freedom. It is
well-known that the number of degrees of freedom of a local algebra
is infinite, but such infinity is not so ``large''. For example,
a QFT defined by a set of independent local fields $\phi_{i}\left(x\right)$
does not satisfy the split property if the number of such independent
fields is infinite. On the other hand, Buchholz has proven that free
field theories satisfy the split property for strictly separated regions
\cite{Buchholz73}. Moreover, Buchholz and Wichmann argue that the
split property for strictly separated spacelike regions is a sufficient
condition for a particle interpretation of QFT \cite{BuchholzWightman}.
For our purposes, the split-property is an extra assumption.
\begin{assumption}
The local algebras $\mathcal{A}\left(\mathcal{O}_{1}\right),\mathcal{A}\left(\mathcal{O}_{2}\right)$
of any pair of strictly separated spacelike regions $\mathcal{O}_{1}\text{\Large\ensuremath{\times\!\negmedspace\!\times}}\mathcal{O}_{2}$
satisfies the split-property. Hence, \eqref{c3-sta_ind} holds.
\end{assumption}

\begin{rem}
It is important to emphasize that the above property only holds for
algebras attached to strictly separated regions. It is a well-known
fact that, even for free fields, the algebras $\mathcal{A}\left(\mathcal{O}\right)$
and $\mathcal{A}\left(\mathcal{O}'\right)$ do not satisfy \eqref{c3-sta_ind},
and hence they do not satisfy the split-property. This means that
the entanglement structure of complementary regions is not completely
well-defined. From the mathematical point of view, this could be understood
as a reason why the entanglement entropy diverges in QFT.
\end{rem}

 \sloppy Since relation \eqref{c3-sta_ind} means unitarily equivalence, we
have that any normal $\omega$ on $\mathcal{A}\left(\mathcal{O}_{1}\vee\mathcal{O}_{2}\right)$
is mapped to a normal state $\tilde{\omega}$ on $\mathcal{A}\left(\mathcal{O}_{1}\right)\otimes\mathcal{A}\left(\mathcal{O}_{2}\right)$.
Moreover, the expectation values of both states in their corresponding
algebras are equivalent, i.e. $\omega\left(A_{1}A_{2}\right)=\tilde{\omega}\left(A_{1}\otimes A_{2}\right)$.
We will denote both states simply as $\omega$.

In conclusion, we are able to map our original system to a bipartite
quantum system. Now, we can compute any entanglement measure as we
have explained in the previous chapter. As an example, the mutual
information for any normal state $\omega$ on $\mathcal{A}\left(\mathcal{O}_{1}\vee\mathcal{O}_{2}\right)$
is
\begin{equation}
I\left(\mathcal{O}_{1},\mathcal{O}_{2};\omega\right):=I\left(\mathcal{A}\left(\mathcal{O}_{1}\right)\otimes\mathcal{A}\left(\mathcal{O}_{2}\right);\omega\right)\,,
\end{equation}
where the r.h.s. is computed using definition \ref{c2-def-mi-inf}.
It will be denoted by $I\left(\mathcal{O}_{1},\mathcal{O}_{2}\right)$
or $I\left(\omega\right)$ whenever there is no confusion about the
state or the algebras involved. Then, the mutual information is a
well-defined quantity in QFT for two strictly separated regions. Moreover,
if the underlying state is the vacuum $\omega_{0}$, it can be shown
that it is also finite. For the relative entropy, the above discussion
is even simpler. It is because, in this case, we deal with only one
algebra $\mathcal{A}\left(\mathcal{O}\right)$ and there is no tensor
product structure. Then, for any two normal states $\omega,\phi$
we define
\begin{equation}
S_{\mathcal{O}}\left(\phi\mid\omega\right):=S_{\mathcal{A}\left(\mathcal{O}\right)}\left(\phi\mid\omega\right)\,.
\end{equation}

\begin{rem}
In the case of fermionic nets, expression \eqref{c3-sta_ind} can
never hold because of the presence of fermionic operators anticommuting
at spacelike distances. However, in that case, relation \eqref{c3-sta_ind}
still holds if the tensor product is replaced by the \textit{graded
tensor product}. Then, the mutual information for fermionic nets is
also well-defined \cite{Longo_xu}.
\end{rem}

In conclusion, MI and RE are well-defined entanglement quantities
in QFT. However, as we have explained in the previous chapter, the
entanglement entropy (EE) is not well-defined since the local algebras
of QFT are type III vN algebras. At this point we could take two paths:
a) discard completely EE in QFT, or b) try to understand its divergent
structure and extract the ``finite'' relevant information about
the entanglement from such a divergent quantity. Of course, here we
follow b). In the end, we define a ``universal'' regularized EE
which works for any QFT.

\subsection{Bisognano-Wichmann theorem\label{c3-sec-bw}}

The study of modular Hamiltonians in QFT is a very intriguing problem,
both for the study of entanglement or just for merely algebraic interests.
In any QFT, we have the distinguished vacuum state $\omega_{0}$,
which is represented by the vector $\left|0\right\rangle $. Hence,
an interesting task is to compute the modular Hamiltonian for such
a vector and any local algebra $\mathcal{A}\left(\mathcal{O}\right)$.
However, this problem is far from being solved in general, except
for some particular examples, which in general, involve free field
theories. Despite this, there is a particular kind of regions, called
wedges, for which the modular flow of the vacuum vector has a ``geometrical
action''. This is the very famous theorem of Bisognano-Wichmann,
which we explain in this section.

Let first define the wedge regions. The prototypical example of a
wedge region is the \textit{right Rindler wedge}
\begin{equation}
\mathcal{W}_{R}:=\left\{ x\in\mathbb{R}^{d}\,:\,x^{1}>\left|x^{0}\right|\right\} \,.
\end{equation}
A general \textit{wedge region} is defined to be any region $\mathcal{W}$
obtained from $\mathcal{W}_{R}$ applying a general Poincaré transformation,
i.e. $\mathcal{W}=g\mathcal{W}_{R}$ with $g\in\mathcal{P}_{+}^{\uparrow}$.
Throughout this section, it is convenient to write any spacetime point
as $x=\left(x^{0},x^{1},\bar{x}_{\bot}\right)$, where $\bar{x}_{\bot}:=\left(x^{2},\ldots,x^{d-1}\right)$
are called the ``orthogonal coordinates'' with respect to the wedge
$\mathcal{W}_{R}$. Now we state the Bisognano-Wichmann theorem.
\begin{thm}
\label{c3-bw}(Bisognano-Wichmann) In the vacuum representation of
any QFT, the modular operator $\Delta_{0}$ and the modular conjugation
$J_{0}$ associated with $\left\{ \mathcal{A}\left(\mathcal{W}_{R}\right),\left|0\right\rangle \right\} $
are
\begin{equation}
J_{0}=\Theta\quad\textrm{and}\quad\Delta_{0}=\mathrm{e}^{-2\pi K_{1}}\:\textrm{,}\label{c3-bw-eq}
\end{equation}
where $\Theta$ is the CRT operator\footnote{The CRT operator is the generalization of the CPT operator to any
spacetime dimensions $d$. In $d=4$, CRT and CPT differ just by a spatial rotation.} and $K_{1}$ is the infinitesimal generator of the one-parameter
group of boost symmetries in the plane $\left(x^{0},x^{1}\right)$,
i.e.
\begin{equation}
U\left(\Lambda_{1}^{s},0\right)=\mathrm{e}^{iK_{1}s}\,,\quad\textrm{with }\Lambda_{1}^{s}:=\left(\begin{array}{ccc}
\cosh\left(s\right) & \sinh\left(s\right) & \boldsymbol{0}\\
\sinh\left(s\right) & \cosh\left(s\right) & \boldsymbol{0}\\
\boldsymbol{0} & \boldsymbol{0} & \boldsymbol{1}
\end{array}\right)\,.\label{c3-boost}
\end{equation}
Furthermore, wedge duality $\mathcal{A}\left(\mathcal{W}_{R}'\right)=\mathcal{A}\left(\mathcal{W}_{R}\right)'$
holds.
\end{thm}

Bisognano and Wichmann were the first who stated and proved this theorem
for the theory of a real scalar field \cite{bisognano}. However,
this theorem holds for any AQFT in the vacuum representation \cite{horuzhy}.
Moreover, it holds for any wedge region due to the Poincaré invariance
of the theory. In fact, for any other wedge $\mathcal{W}=g\mathcal{W}_{R}$,
the modular operator $\Delta_{\mathcal{W}}$ and the modular conjugation
$J_{\mathcal{W}}$ associated with $\left\{ \mathcal{A}\left(\mathcal{W}\right),\left|0\right\rangle \right\} $
are $J_{\mathcal{W}}=U\left(g\right)J_{0}U\left(g\right)^{*}$ and
$\Delta_{\mathcal{W}}=U\left(g\right)\Delta_{0}U\left(g\right)^{*}$.

From \eqref{c3-bw-eq} we easily get the relation $K_{0}=2\pi K_{1}$,
which could be written as
\begin{equation}
K_{0}=2\pi\int_{\mathbb{R}^{d}}d^{d-1}x\,x^{1}\,T_{00}(x)\,,\label{c3-bw-t}
\end{equation}
whenever the theory has a stress-tensor field operator. Inspired by
the discussion in remark \ref{c2_remark_mod} and example \ref{c2_ex_finit_mod},
one is tempted to write the following expression for the ``inner''
modular Hamiltonian
\begin{equation}
K_{0}^{R}=2\pi\int_{x^{1}>0}d^{d-1}x\,x^{1}\,T_{00}(x)\,,\label{c3-mh-bad}
\end{equation}
where the super-index $R$ comes from ``right'' and indicates that
it is the ``half'' of $K_{0}$ belonging to the right Wedge region.
The above formula is heuristically correct because, for any operator
$A\in\mathcal{A}\left(\mathcal{W}_{R}\right)$, the modular flow could
be written as $\sigma_{t}\left(A\right)\equiv\mathrm{e}^{-i2\pi K_{0}^{R}t}A\mathrm{e}^{i2\pi K_{0}^{R}t}$.
However, expression \eqref{c3-mh-bad} is not completely correct because
it is mathematically ill-defined. It can be shown that $K_{0}^{R}$
is not an operator,\footnote{In the sense as a linear transformation on the Hilbert space with
dense domain.} and in particular, we have that $\left\Vert K_{0}^{R}\left|0\right\rangle \right\Vert =+\infty$.
Moreover, the full modular Hamiltonian can never be decomposed as
$K_{0}=K_{0}^{R}-K_{0}^{R'}$, because the algebras $\mathcal{A}\left(\mathcal{W}_{R}\right)$
and $\mathcal{A}\left(\mathcal{W}'_{R}\right)$ are not statistically
independent in the sense of \eqref{c3-sta_ind}. However, $K_{0}^{R}$
is well-defined as a sesquilinear form: expectation values $\left\langle \Psi_{1}\right|K_{0}^{R}\left|\Psi_{2}\right\rangle $
are finite for all $\left|\Psi_{j}\right\rangle $ in a dense subset
of the Hilbert space.

Two remarkable things follow from the above theorem. The first one
is that for algebras associated with wedge regions, the modular operator
is universal, i.e. it does not depend on the theory. The second one
is that the modular flow $A\rightarrow\sigma_{t}\left(A\right)=\Delta_{0}^{it}A\Delta_{0}^{-it}$
has a geometrical action on the algebra $\mathcal{A}\left(\mathcal{W}_{R}\right)$,
in the sense that if $A\in\mathcal{A}\left(\mathcal{O}\right)$ with
$\mathcal{O}\subset\mathcal{W}_{R}$, then $\sigma_{t}\left(A\right)\in\mathcal{A}\left(U\left(\Lambda_{1}^{s},0\right)\mathcal{O}\right)$
with $s=-2\pi t$. Or equivalently, if the theory is defined by means
a field operator $\phi\left(x\right)$, then $\sigma_{t}\left(\phi\left(x\right)\right)=\phi\left(\Lambda_{1}^{-2\pi t}x\right)$
for all $x\in\mathcal{W}_{R}$ and all $t\in\mathbb{R}$. In other
words, the modular evolution is given by the boost evolution, which
is a spacetime symmetry. This gives a relation between spacetime and
modular theory. Unfortunately, such behavior does not generalize to
any other kind of region. The modular group associated with a local
algebra $\mathcal{A}\left(\mathcal{O}\right)$ of a general region
$\mathcal{O}$ other than a wedge, gives an automorphism of the algebra,
which cannot be geometrical \cite{Arias:2016nip}. There is an exception,
which is the double cones in CFTs. In those cases, the modular flow
is also universal (independent of the theory), and it is given by
the one-parameter group of conformal transformations which leaves
the double cone invariant \cite{Hislop81,chm}.

The above interesting relation between spacetime symmetries and modular
theory can be reversed. Let us start with a net of vN algebras associated
with wedges, i.e. an assignment to each wedge region $\mathcal{W}$
a vN algebra $\mathcal{W}\mapsto\mathcal{A}\left(\mathcal{W}\right)\subset\mathcal{B}\left(\mathcal{H}_{0}\right)$.
We assume that this net satisfies the three first axioms of definition
\ref{c1_def_aqft}, and we also assume that all the wedge algebras
$\mathcal{A}\left(\mathcal{W}\right)$ have a common standard vector
$\left|0\right\rangle \in\mathcal{H}$. Let us denote by $\Delta_{\mathcal{W}}^{it}$
the modular group associated with $\left\{ \mathcal{A}\left(\mathcal{W}\right),\left|0\right\rangle \right\} $.
Instead of assuming the Poincaré covariance of the net (axiom 4 of
definition \ref{c1_def_aqft}), we assume the following condition.
\begin{enumerate}
\item[4'.] \textit{(Bisognano-Wichmann property)} For any wedges $\mathcal{W}_{1},\mathcal{W}_{2}$,
we have that
\begin{equation}
\Delta_{\mathcal{W}_{1}}^{it}\mathcal{A}\left(\mathcal{W}_{2}\right)\Delta_{\mathcal{W}_{1}}^{-it}=\mathcal{A}\left(\Lambda_{\mathcal{W}_{1}}^{-2\pi t}\mathcal{W}_{2}\right)\,,
\end{equation}
 where $\Lambda_{\mathcal{W}_{1}}^{-2\pi t}$ is the one-parameter
group of Lorentz matrices leaving the wedge $\mathcal{W}_{1}$ invariant.
\end{enumerate}
This condition means that the wedge algebras transform accordingly
to the statement of theorem \ref{c3-bw}. Under the above assumptions,
it can be shown that there exists a unique unitary representation
of the Poincaré group $g\in\mathcal{P}_{+}^{\uparrow}\mapsto U\left(g\right)\in\mathcal{B}\left(\mathcal{H}_{0}\right)$
such that
\begin{eqnarray}
\Delta_{\mathcal{W}}^{it} & \!\!\!=\!\!\! & U\left(\Lambda_{\mathcal{W}}^{-2\pi t}\right)\,,\\
\mathcal{A}\left(g\mathcal{W}\right) & \!\!\!=\!\!\! & U\left(g\right)\mathcal{A}\left(\mathcal{W}\right)U\left(g\right)^{*}\,,
\end{eqnarray}
for all $g\in\mathcal{P}_{+}^{\uparrow}$, all $t\in\mathbb{R}$ and
all wedges $\mathcal{W}$ \cite{Davidson95}. In other words, the
quantum Poincaré symmetry can be reconstructed assuming the Bisognano-Wichmann
property. This leaves open the question if the Bisognano-Wichmann
property could be a ``more fundamental'' condition than the Poincaré
symmetry.

\subsection{Reeh-Schlieder theorem\label{c3-secc-rs}}

In this subsection, we show that Reeh-Schlieder theorem \ref{c1_rs}
admits an interpretation in terms of entanglement. For such a purpose,
we make use of an example concerning finite dimensional algebras.
In section \ref{c2-secc_enta}, we have introduced the bell pair
\begin{equation}
\left|\Psi\right\rangle :=\frac{1}{\sqrt{2}}\left(\left|0\right\rangle \otimes\left|0\right\rangle +\left|1\right\rangle \otimes\left|1\right\rangle \right)\,,\label{c3-bell}
\end{equation}
as a maximally entangled state for the bipartite system $\mathfrak{A}:=M_{2}\left(\mathbb{C}\right)\otimes M_{2}\left(\mathbb{C}\right)$.
such a vector has the interesting property that, acting on it with
operators belonging to one of the subsystems, we can construct any
vector of the full Hilbert space $\mathbb{C}^{2}\otimes\mathbb{C}^{2}$.
More precisely, any single qubit operator
\begin{equation}
A:=\left(\begin{array}{cc}
a_{00} & a_{01}\\
a_{10} & a_{11}
\end{array}\right)\in M_{2}\left(\mathbb{C}\right)\,,
\end{equation}
acts over the basis vectors as
\begin{eqnarray}
A\left|0\right\rangle =a_{00}\left|0\right\rangle +a_{01}\left|1\right\rangle \,, &  & A\left|1\right\rangle =a_{10}\left|0\right\rangle +a_{11}\left|1\right\rangle \,.
\end{eqnarray}
Then, acting with such an operator on \eqref{c3-bell} we have that
\begin{equation}
\left(A\otimes\mathbf{1}\right)\left|\Psi\right\rangle =\frac{1}{\sqrt{2}}\left(a_{00}\left|0\right\rangle \otimes\left|0\right\rangle +a_{01}\left|1\right\rangle \otimes\left|0\right\rangle +a_{10}\left|0\right\rangle \otimes\left|1\right\rangle +a_{11}\left|1\right\rangle \otimes\left|1\right\rangle \right)\,.\label{c3-bell_rs}
\end{equation}
In other words, choosing the operator $A$ conveniently, we can reach
any vector of the full Hilbert space $\mathbb{C}^{2}\otimes\mathbb{C}^{2}$.
This behavior is only possible because of the particular structure
of the vector \eqref{c3-bell}. Moreover, it also works for any global
vector state whose reduced state, to any of the subsystems, is faithful,
i.e. the reduced density matrices are invertible. In contrast, the
above fact does not hold for the separable vector $\left|0\right\rangle \otimes\left|0\right\rangle $.

The similarity with the vacuum vector $\left|0\right\rangle $ in
QFT is evident. Given a region $\mathcal{O}$, the full observable
algebra can be decomposed as
\begin{equation}
\mathcal{B}\left(\mathcal{H}_{0}\right)=\mathcal{A}\left(\mathcal{O}\right)\vee\mathcal{A}\left(\mathcal{O}'\right)\,.
\end{equation}
The Reeh-Schlieder theorem asserts that acting with operators $A\in\mathcal{A}\left(\mathcal{O}\right)$
on the vacuum vector $\left|0\right\rangle $ we can reach (almost\footnote{The word ``almost'' is there because $\mathcal{A}\left(\mathcal{O}\right)\left|0\right\rangle $
is not all, but a dense subset of $\mathcal{H}_{0}$.}) any vector in $\mathcal{H}_{0}$. Following the example above, we
say that the vacuum state is entangled for the pair $\left\{ \mathcal{A}\left(\mathcal{O}\right),\mathcal{A}\left(\mathcal{O}'\right)\right\} $.
Moreover, since it happens for any region $\mathcal{O}$, we say that
the vacuum vector is entangled at any distance. Against intuition,
the vacuum is a highly entangled state \cite{Verch:2004vj}.

The Reeh-Schlieder theorem may seem paradoxical at first. It implies
that by acting on the vacuum with an operator supported in a small
region $\mathcal{O}$, one can create whatever one wants. Indeed,
for any two spacelike separated regions $\mathcal{O}_{1}\text{\Large\ensuremath{\times}}\mathcal{O}_{2}$
and any operator $A_{1}\in\mathcal{A}\left(\mathcal{O}_{1}\right)$,
the vector $A_{1}\left|0\right\rangle $ can always be reproduced
by the action of an operator $A_{2}\in\mathcal{A}\left(\mathcal{O}_{2}\right)$
on $\left|0\right\rangle $, regardless how far is $\mathcal{O}_{1}$
from $\mathcal{O}_{2}$. However, it is no longer true that performing
unitary operators inside a small region $\mathcal{O}$, we can create
whatever state we want supported in a far spacelike separated region.
This is because in order that the Reeh-Schlieder theorem holds, we
have to be allowed to act on the vacuum vector with any local operator.
The Reeh-Schlieder theorem is no longer true for the restricted subset
of unitary operators.

\section{Entanglement entropy in QFT\label{c3-sec-ee-qft}}

As we have discussed in the previous section, the vacuum state $\omega_{0}$
is a ``highly'' entangled state in QFT. The aim of this section
is to quantify such entanglement between different regions of spacetime.
More concretely, we consider a causally complete region $\mathcal{O}\in\mathcal{K}$
and the local algebra $\mathcal{A}\left(\mathcal{O}\right)$. We look
for a measure of entanglement, for the vacuum state, between $\mathcal{A}\left(\mathcal{O}\right)$
and the algebra of its complementary region $\mathcal{A}\left(\mathcal{O}'\right)$.

\begin{figure}[h]
\centering
\includegraphics[width=13cm]{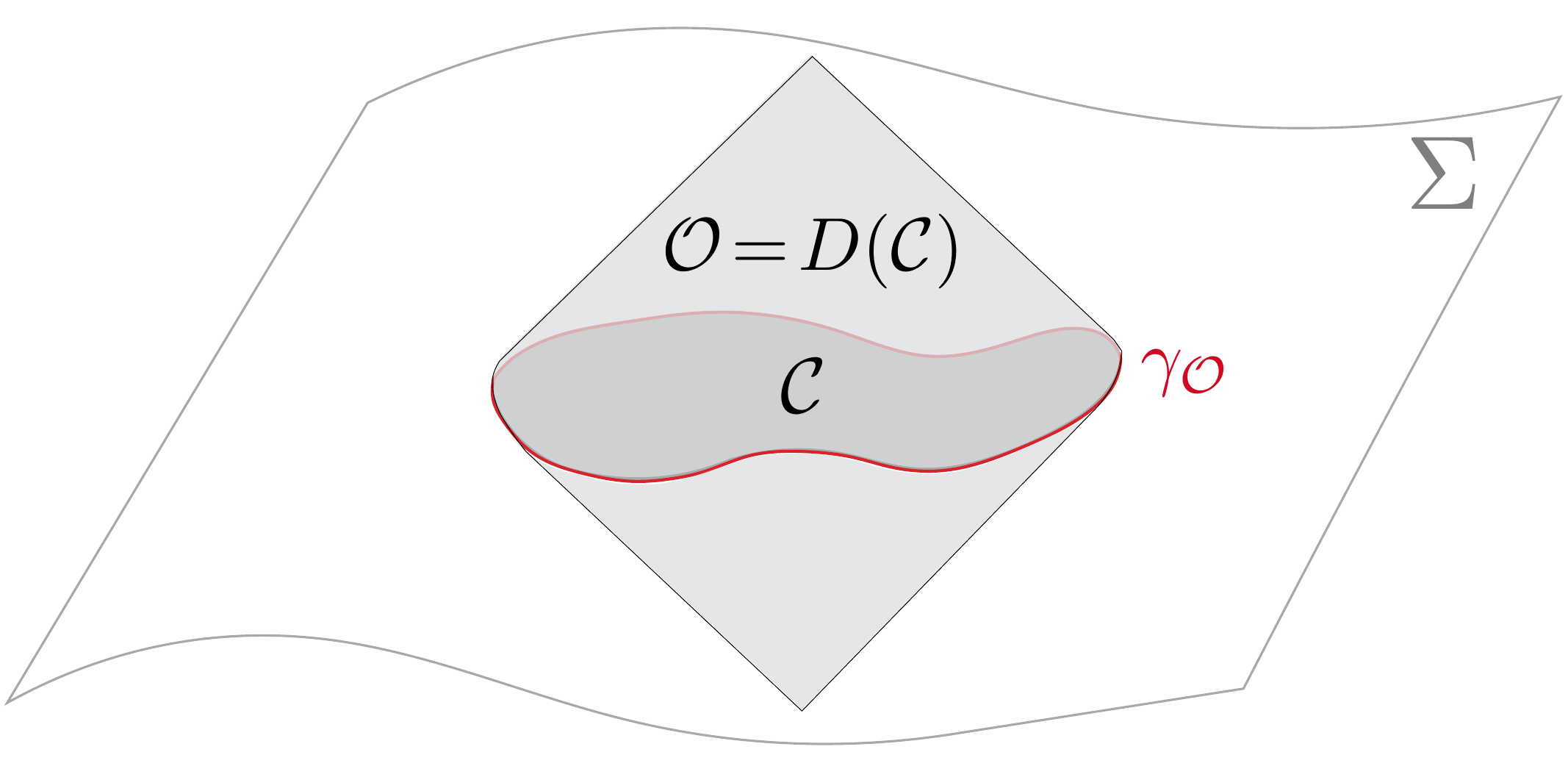}\caption{\label{c3_fig_double_cone}The causally complete region $\mathcal{O}$
is the Cauchy development of the spacelike region $\mathcal{C}$,
which lies inside the Cauchy surface $\Sigma$. The entanglement surface
$\gamma_{\mathcal{O}}$ is the boundary $\partial\mathcal{C}$ of
$\mathcal{C}$ inside $\Sigma$. For a given causally complete region
$\mathcal{O}$, there are many Cauchy surfaces and spacelike regions
giving the same configuration. However, $\gamma_{\mathcal{O}}$ is
independent of such choices.}
\end{figure}

Given a causally complete region $\mathcal{O}\in\mathcal{K}$, there
always exists a (non-unique) Cauchy surface $\Sigma$ and a spacelike
region $\mathcal{C}\subset\Sigma$, such that $\mathcal{O}$ is the
Cauchy development of $\mathcal{C}$, i.e. $\mathcal{O}=D\left(\mathcal{C}\right)$.
The (spacelike codimension $2$) boundary $\partial\mathcal{C}$ of
$\mathcal{C}$ inside $\Sigma$ is called the \textit{entanglement
surface} between $\mathcal{O}$ and $\mathcal{O}'$, and it is denoted
by $\gamma_{\mathcal{O}}$ (see figure \ref{c3_fig_double_cone}).
It is not difficult to see that the entanglement surface is independent
on the choice of $\Sigma$, and furthermore, $\gamma_{\mathcal{O}}=\gamma_{\mathcal{O}'}$
for any $\mathcal{O}\in\mathcal{K}$.

For a finite bipartite system, the entanglement entropy \eqref{c2_ee}
is a good measurement of entanglement between the two subsystems for
any pure state (section \ref{c2-secc_enta}). However, as we have
already discussed, we have the following two problems in QFT.
\begin{enumerate}
\item $\mathcal{A}\left(\mathcal{O}\right)\vee\mathcal{A}\left(\mathcal{O}'\right)\cancel{\cong}\mathcal{A}\left(\mathcal{O}\right)\otimes\mathcal{A}\left(\mathcal{O}'\right)$,
and hence, the structure explained in section \ref{c2-secc_enta}
does not apply in this case.
\item The local algebra $\mathcal{A}\left(\mathcal{O}\right)$ is a type
III vN algebra. This implies that the vN entropy of any state on such
an algebra is ill-defined (section \ref{c2-Sec_entro_gener}).
\end{enumerate}
In order to bypass such issues, we need to regularize the theory introducing
a UV cutoff $\delta$. For example, the lattice cutoff fits very well
in the algebraic setting. In this case, the local algebras of QFT
are represented by a lattice of (bosonic or fermionic) harmonic oscillators
separated by a distance $\delta>0$.\footnote{We explain this in detail in section \ref{c3-secc-lattice}.}
Then, in the regularized QFT, the regularized local algebras $\mathcal{A}_{\delta}\left(\mathcal{O}\right)$
are type I vN algebras and they satisfy
\begin{equation}
\mathcal{A}_{\delta}\left(\mathcal{O}\right)\vee\mathcal{A}_{\delta}\left(\mathcal{O}'\right)\cong\mathcal{A}\left(\mathcal{O}\right)\otimes\mathcal{A}\left(\mathcal{O}'\right)\,.
\end{equation}
Hence, we can define the regularized EE as
\begin{equation}
S_{\delta}\left(\mathcal{O}\right):=-\mathrm{Tr}_{\mathcal{A}_{\delta}\left(\mathcal{O}\right)}\left(\rho_{\mathcal{O}}\log\rho_{\mathcal{O}}\right)\,,
\end{equation}
where $\rho_{\mathcal{O}}$ is the statistical operator of the restricted
state $\left.\omega\right|_{\mathcal{A}_{\delta}\left(\mathcal{O}\right)}$.
However, in the continuum limit $\delta\rightarrow0^{+}$, we always
get $S_{\delta}\left(\mathcal{O}\right)\rightarrow+\infty$. It is
for this reason that we say that the EE is UV divergent in QFT.

The above behavior has nothing to do with the special choice of the
UV lattice cutoff. Indeed, it can be shown, using different techniques
and cutoffs, that the EE always diverges in the continuum limit $\delta\rightarrow0^{+}$,
even for free theories. Furthermore, such behavior could not be remedied
without spoiling some of the distinct properties of the EE. To be
more precise, we must remark which are these properties.
\begin{defn}
\label{c3-def-eeqft}In QFT, the EE for the vacuum state is a function
from the set of causally complete regions into the real numbers, i.e.
$\mathcal{O}\in\mathcal{K}\mapsto S\left(\mathcal{O}\right)\in\mathbb{R}$,
which satisfies the following properties.
\end{defn}

\begin{enumerate}
\item (positivity) $S\left(\mathcal{O}\right)\geq0$ for all $\mathcal{O}\in\mathcal{K}$. 
\item (strong subadditivity) $S\left(\mathcal{O}_{1}\lor\mathcal{O}_{2}\right)+S\left(\mathcal{O}_{1}\land\mathcal{O}_{2}\right)\leq S\left(\mathcal{O}_{1}\right)+S\left(\mathcal{O}_{2}\right)$
for all commuting\footnote{For the definition of commuting regions see equation \eqref{c1-comm-sets}
and definition \ref{fn_comm_sets-1} in appendix \ref{APP_LATTICE}.} regions $\mathcal{O}_{1}$ and $\mathcal{O}_{2}$.
\item \label{ee_qft_cov}(Poincaré invariance) $S\left(g\mathcal{O}\right)=S\left(\mathcal{O}\right)$
for all $g\in\mathcal{P}_{+}^{\uparrow}$.
\end{enumerate}
The first two conditions are motivated by the definition of the EE
for finite quantum systems. The third condition is stated because
the vacuum state is Poincaré invariant, and the local algebras $\mathcal{A}\left(\mathcal{O}\right)$
and $\mathcal{A}\left(g\mathcal{O}\right)$ are unitarily equivalent
because of the Poincaré covariance of the theory. However, there is
a lemma that asserts that a non-trivial EE cannot exist in QFT.
\begin{lem}
Let $\mathcal{K}_{p}\subset\mathcal{K}$ denotes the subset of causally
complete regions with polyhedral entanglement surface. Any positive,
strong subadditive and Poincaré invariant function $\mathcal{O}\in\mathcal{K}_{p}\mapsto S\left(\mathcal{O}\right)\in\mathbb{R}$
is of the form 
\begin{equation}
S\left(\mathcal{O}\right)=c_{1}+c_{2}\,\mathrm{vol}\left(\gamma_{\mathcal{O}}\right)\,,\label{c3-triv-ee}
\end{equation}
where $\mathrm{vol}\left(\gamma_{\mathcal{O}}\right)$ is the geometrical
volume of the (spacelike codimension 2) entanglement surface $\gamma_{\mathcal{O}}$
and $c_{1},c_{2}\geq0$ are independent of the region $\mathcal{O}$.
\end{lem}

\begin{proof}
See \cite{casini2004geometric}.
\end{proof}
Assuming that \eqref{c3-triv-ee} represents the EE, then we would
have for the MI 
\begin{equation}
I\left(\mathcal{O}_{1},\mathcal{O}_{2}\right)=S\left(\mathcal{O}_{1}\right)+S\left(\mathcal{O}_{2}\right)-S\left(\mathcal{O}_{1}\cup\mathcal{O}_{2}\right)=c_{1}\,,
\end{equation}
for all strictly spacelike separated regions $\mathcal{O}_{1},\mathcal{O}_{2}\in\mathcal{K}_{p}$.
Furthermore, if we assume that the theory satisfies clustering, the
correlations and hence the MI, between $\mathcal{O}_{1}$ and $\mathcal{O}_{2}$
would vanish whenever the separation between the regions goes to spacelike
infinity. This would imply that $c_{1}=0$, and therefore, $I\left(\mathcal{O}_{1},\mathcal{O}_{2}\right)=0$
for all strictly spacelike separated regions $\mathcal{O}_{1},\mathcal{O}_{2}\in\mathcal{K}_{p}$.
Moreover, given any two strictly spacelike separated regions $\mathcal{O}_{1},\mathcal{O}_{2}\in\mathcal{K}$,
there always exist two strictly spacelike separated regions $\tilde{\mathcal{O}}_{1},\tilde{\mathcal{O}}_{2}\in\mathcal{K}_{p}$
such that $\tilde{\mathcal{O}}_{j}\supset\mathcal{O}_{j}$. Then,
by monotonicity of the MI, we would have that $I\left(\mathcal{O}_{1},\mathcal{O}_{2}\right)=0$
for all strictly spacelike separated regions $\mathcal{O}_{1},\mathcal{O}_{2}\in\mathcal{K}.$

This means that \eqref{c3-triv-ee} is not a good entanglement measure
since, as we have discussed in section \ref{c3-secc-rs}, the vacuum
state is always entangled at any distance. Regularized QFTs have a
well-defined EE at the expense of giving up some of the properties
of definition \ref{c3-def-eeqft}. For example, the EE of a lattice
QFT does not satisfy property \ref{ee_qft_cov}, because Poincaré covariance is explicitly
broken by the regularization.

In conclusion, the EE is always UV divergent in QFT. However, its
divergence structure has a very rigid form. We will discuss this issue
in the following subsection.

\subsection{Divergence structure of the EE\label{c3-sec-div-ee}}

\begin{figure}[h]
\centering
\includegraphics[width=8cm]{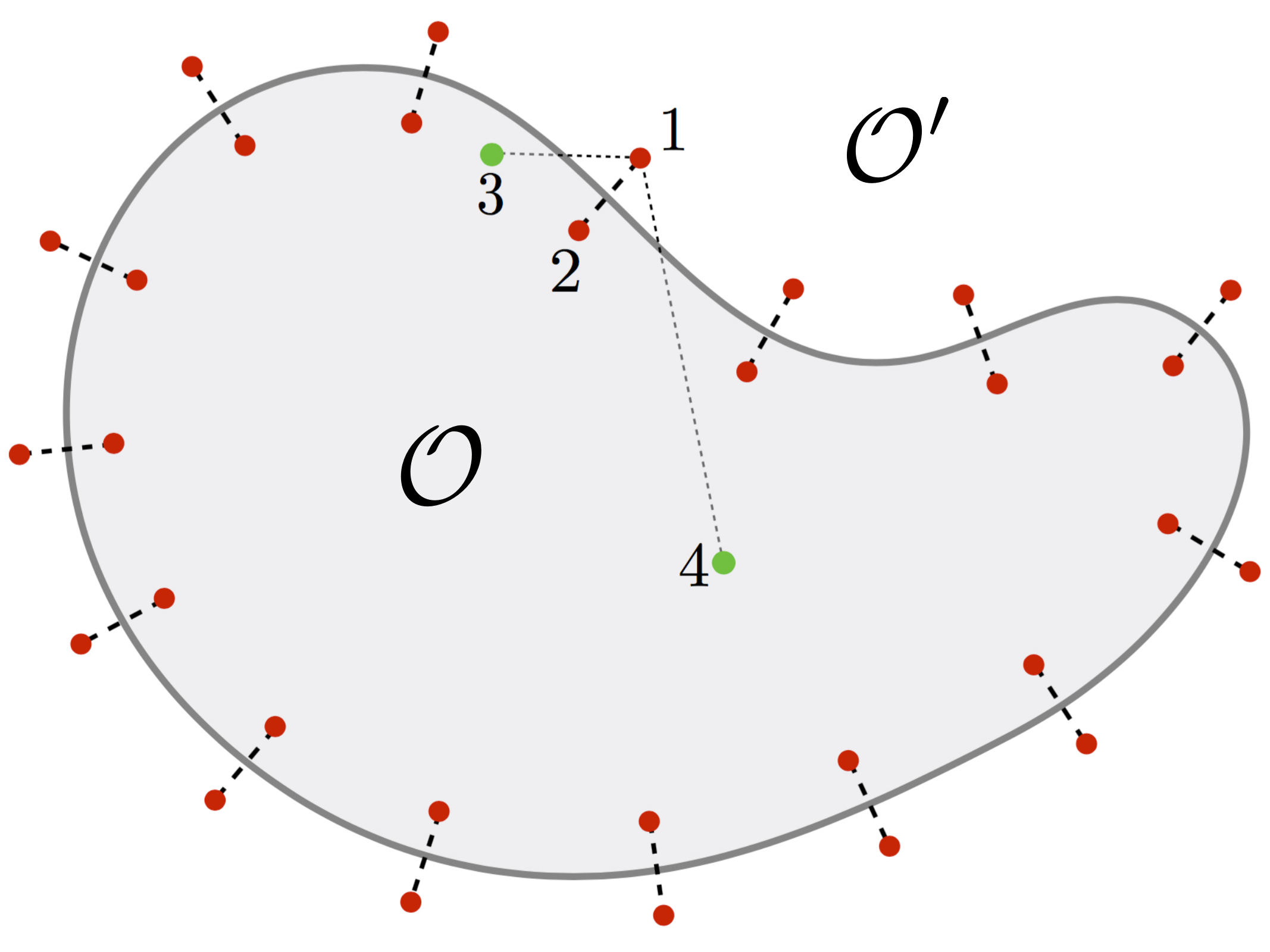}\caption{\label{c3_fig_ee_surf}In QFT, the main contribution to the EE of
the vacuum state restricted to a region $\mathcal{O}$ is given by
the entanglement between the degrees of freedom which are arbitrarily
closed and at opposite sides of the entanglement surface $\gamma_{\mathcal{O}}$.
This contribution is extensive along $\gamma_{\mathcal{O}}$. Heuristically,
because of the locality of the Hamiltonian, the entanglement between
the degree of freedom located in the point $1$ with the one located
in the point $3$ or in the point $4$, is always less than the entanglement
with the degree of freedom in the point $2$, which is the closest
degree of freedom (belonging to $\mathcal{O}$) to the one at point
$1$. As a consequence, the divergent part of the EE is local and
extensive along $\gamma_{\mathcal{O}}$.}
\end{figure}

In the presence of a UV cutoff $\delta$ with mass dimension $-1$,
the leading divergent contribution to the EE grows with the volume
of the entanglement surface \cite{Wolf07,Srednicki93,Eisert08}
\begin{equation}
S_{\delta}\left(\mathcal{O}\right)=\frac{\tilde{\mu}_{d-2}}{\delta{}^{d-2}}\mathrm{vol}\left(\gamma_{\mathcal{O}}\right)+\textrm{non-leading terms ,}\label{c3-ee-exp-a}
\end{equation}
which is commonly known as \textit{area law}. The coefficient $\tilde{\mu}_{d-2}$
is a dimensionless constant that depends on the theory and the regularization
prescription. Physically, this divergence is a consequence of the
existence, in QFT, of an infinite number of degrees of freedom which
are entangled in the vacuum state. Because the vacuum state is the
ground state of a local Hamiltonian, the degrees of freedom which
contribute more to the entanglement between $\mathcal{O}$ and its
complement $\mathcal{O}'$ are those located arbitrarily close and
at opposite sides of the entanglement surface. Due to the same locality
of the short distance correlations, the degrees of freedom which produce
the UV divergences are disposed extensively along the entanglement
surface $\gamma_{\mathcal{O}}$ (see figure \ref{c3_fig_ee_surf}).
This fact allows us to better understand the structure of divergencies
of $S_{\delta}\left(\mathcal{O}\right)$, which is generically given
by a sum of terms that are local and extensive along the entangling
surface $\gamma_{\mathcal{O}}$.

From now on, we fix a specific CFT and we take the vacuum state.
The aim is to characterize the behavior of $S_{\delta}\left(\mathcal{O}\right)$
in terms of $\mathcal{O}$ for a fixed regularization. For the moment
we just consider regions $\mathcal{O}$ with smooth entanglement surface
$\gamma_{\mathcal{O}}$. Since in a CFT there is no dimensionful parameter,
we expect the following expansion for the regularized EE \cite{Liu:2012eea}

\begin{eqnarray}
S_{\delta}\left(\mathcal{O}\right) & \!\!\!=\!\!\! & \frac{\mu_{d-2}\left(\gamma_{\mathcal{O}}\right)}{\delta^{d-2}}+\frac{\mu_{d-4}\left(\gamma_{\mathcal{O}}\right)}{\delta^{d-4}}+\frac{\mu_{d-6}\left(\gamma_{\mathcal{O}}\right)}{\delta^{d-6}}\nonumber \\
 &  & +\ldots+S_{0}\left(\mathcal{O}\right)+\textrm{ positive powers of }\delta\,.\label{c3-ee-div}
\end{eqnarray}
The above expansion is in inverse powers of the cutoff $\delta$.
The parameters $\mu_{d-2k}\left(\gamma_{\mathcal{O}}\right)$ are
local and extensive along the entanglement surface $\gamma_{\mathcal{O}}$.
Local means that they depend on the geometry of the region and in
the cutoff UV behavior close to the entanglement surface. Extensive
means that the contributions are additive along $\gamma_{\mathcal{O}}$,
and hence, they can be written as integrals along $\gamma_{\mathcal{O}}$.
The mass dimension of $\mu_{d-2k}\left(\gamma_{\mathcal{O}}\right)$
must be $2k-d$ because the EE is dimensionless. The coefficient of
leading term is $\mu_{d-2}\left(\gamma_{\mathcal{O}}\right)=\tilde{\mu}_{d-2} \, \mathrm{vol}\left(\gamma_{\mathcal{O}}\right)$ with $\tilde{\mu}_{d-2}$ dimensionless, which gives the area law
as we have explained above. For a general regularization, the parameters
$\mu_{d-2k}\left(\gamma_{\mathcal{O}}\right)$ have extensive contributions
coming from the (intrinsic and extrinsic) geometry of the entanglement
surface $\gamma_{\mathcal{O}}$, but they also have extra local and
extensive contributions constructed with parameters coming from the
regularization. These quantities are \textit{non-universal} in the
sense that they depend on the regularization scheme used.

Among the different regularization schemes, there are the so-called
``geometric regularizations''. Such regulators are characterized
by the fact that the quantities $\mu_{d-2k}\left(\gamma_{\mathcal{O}}\right)$
are purely geometrical in the sense that they can be written as

\begin{equation}
\mu_{d-2k}\left(\gamma_{\mathcal{O}}\right)=\int_{\gamma_{\mathcal{O}}}d\sigma\sqrt{h}\,R_{d-2k}\left[h_{ab},K_{ab}^{\nu}\right]\,,\label{c3-local-geo}
\end{equation}
where $R_{k}\left[h_{ab},K_{ab}\right]$ represents a generic curvature
scalar, constructed from the induced metric $h_{ab}$ and the extrinsic
curvatures $K_{ab}^{\nu}$ ($\nu=1,2$ runs over two normal directions)
of the surface $\gamma_{\mathcal{O}}$, and $\sqrt{h}d\sigma$ is
the usual volume element over the surface $\gamma_{\mathcal{O}}$.
For a geometrical cutoff, it is expected that the expansion \eqref{c3-ee-div}
is perturbative in the geometry of $\gamma_{\mathcal{O}}$. This means
that the geometrical integrands of \eqref{c3-local-geo} can only
involve positive integer powers of the curvature scalars $R_{d-2k}$.
Then, because the cutoff is the unique mass scale in the theory, only
integer powers of the cutoff can appear in the subleading terms of
\eqref{c3-ee-exp-a}, justifying the ansatz \eqref{c3-ee-div}. The
coefficients $\mu_{d-2k}\left(\gamma_{\mathcal{O}}\right)$ are also
non-universal quantities, i.e. they are regularization dependent.
However, in section \ref{c3-sec-univee}, we show a ``universal''
way to choose a geometrical regularization prescription for any QFT.
From now on, we only consider geometric regularizations. 

We remark that the expansion \eqref{c3-ee-div} does not involve terms
proportional to $\delta^{d-l}$ with $l>0$ and odd. These terms,
which could be present in an arbitrary regularization of the theory,
will involve integrals of geometric quantities of even mass dimensions.
Any such quantity must contain odd powers of the extrinsic curvatures
$K_{ab}^{\nu}$ of $\gamma_{\mathcal{O}}$, and therefore, contribute
with opposite sign for $\mathcal{O}$ and $\mathcal{O}'$. However,
in a regularization where such terms are present, the duality $S_{\delta}\left(\mathcal{O}\right)=S_{\delta}\left(\mathcal{O}'\right)$,
which is expected whenever the global state is pure, fails.\footnote{See section \ref{c3-sec-duality} for clarification.}
This is because such a contribution has opposite signs when is calculated
for $\mathcal{O}$ and $\mathcal{O}'$. Hence, whenever we write \eqref{c3-ee-div},
we are implicitly assuming a regularization scheme that preserves
the mentioned duality.

The constant term $S_{0}\left(\mathcal{O}\right)$ goes as $\sim\delta^{0}$
for odd $d$ dimensions and goes as $\sim\log\left(\delta\right)$
for even $d$ dimensions. In fact, we expect
\begin{equation}
S_{0}\left(\mathcal{O}\right)=\begin{cases}
\left(-1\right)^{\frac{d-1}{2}}s_{0}(\mathcal{O})\,, & \textrm{odd }d\,,\\
\left(-1\right)^{\frac{d-2}{2}}s_{0}(\mathcal{O})\log\left(\frac{R_{\mathcal{O}}}{\delta}\right)+\tilde{s}_{0}(\mathcal{O})\,, & \textrm{even }d\,,
\end{cases}
\end{equation}
where $R_{\mathcal{O}}$ is any length parameter of the region $\mathcal{O}$,
which scales as $R_{\mathcal{O}}\rightarrow\lambda R_{\mathcal{O}}$
for dilatations $x\rightarrow\lambda x$. The sign factors before
$s_{0}(\mathcal{O})$ are chosen for later convenience. $s_{0}(\mathcal{O})$
and $\tilde{s}_{0}(\mathcal{O})$ are non-local quantities, and they
depend only on the shape but not on the size of the region $\mathcal{O}$.
We are not usually interested in the terms with positive powers of
$\delta$ because they vanish in the limit of $\delta\rightarrow0^{+}$.

The important fact is that the term $S_{0}\left(\mathcal{O}\right)$
contains valuable information about the entanglement between $\mathcal{O}$
and $\mathcal{O}'$. In fact, under the assumptions above, $s_{0}(\mathcal{O})$
are universal quantities belonging to the continuum CFT. For example,
when $\mathcal{O}$ is the Cauchy development of a $d-1$ dimensional
sphere of radius $R$, we have that
\begin{equation}
S_{0}\left(\mathcal{O}\right)=\begin{cases}
\left(-1\right)^{\frac{d-1}{2}}F\,, & \textrm{odd }d\,,\\
\left(-1\right)^{\frac{d-2}{2}}4A\log\left(\frac{R}{\delta}\right)+\mathrm{const}\,\,, & \textrm{even }d\,,
\end{cases}
\end{equation}
where $A$ is the coefficient of the Euler density in the trace anomaly
\cite{Cardy:1988cwa} and $F$ is the constant term of the free energy
in a $d$ dimensional Euclidean sphere \cite{chm,chmy,Jafferis:2011zi}.
These coefficients are precisely the quantities that decrease under
renormalization group flow in the $c$, $F$ and $A$ theorems in
dimension $d=2,3,4$ respectively \cite{Casini:2004bw,Casini:2006es,Myers:2010xs,Casini:2012ei,Casini:2016fgb,Casini:2017vbe,Casini:2018kzx}.
For even $d$, the universal part of the term $\tilde{s}_{0}(\mathcal{O})$
is harder to be extracted, even in the case of spheres. This is because
a change in the cutoff $\delta\rightarrow\lambda\delta$, or equivalently
in the length parameter $R_{\mathcal{O}}\rightarrow\lambda^{-1}R_{\mathcal{O}}$,
leads to a change in the finite part $\tilde{s}_{0}(\mathcal{O})\rightarrow\tilde{s}_{0}(\mathcal{O})-\left(-1\right)^{\frac{d-2}{2}}s_{0}(\mathcal{O})\log\left(\lambda\right)$.
To extract its universal part, we have to invoke relative quantities.
For example, for two regions $\mathcal{O}_{1}\text{\Large\ensuremath{\times\!\negmedspace\!\times}}\mathcal{O}_{2}$ we can use the regularized mutual information
\begin{equation}
I_{\delta}(\mathcal{O}_{1},\mathcal{O}_{2}):=S_{\delta}(\mathcal{O}_{1})+S_{\delta}(\mathcal{O}_{2})-S_{\delta}(\mathcal{O}_{1}\lor\mathcal{O}_{2}) \, .
\end{equation}
In this case, the local and extensive contributions of the coefficient
$\tilde{s}_{0}(\mathcal{O}_{1})+\tilde{s}_{0}(\mathcal{O}_{2})-\tilde{s}_{0}(\mathcal{O}_{1}\lor\mathcal{O}_{2})$
cancel out and it gives a universal quantity of the continuum CFT
attached to the pair $\{\mathcal{O}_{1},\mathcal{O}_{2}\}$.

The general expression \eqref{c3-ee-div} slightly changes when the
entanglement surface is non-smooth. For example, it could be the case
that $\gamma_{\mathcal{O}}$ has a defect like a corner. The local
contributions to the entropy coming from points around such a defect
are non-longer perturbative in the local geometry since, around there,
the geometry does not look like flat geometry plus small smooth curvature
deformations. In this case, there are extra divergent contributions
attached to the defect \cite{Myers12,Bueno15,Elvang15}. 

In the case of free fields and for some simple geometries, the above
structure of divergences could be explicitly verified \cite{Casini_review}.
This has also been verified in holographic theories using the Ryu-Takayanagi
prescription.

In the case of a general QFT with non-conformal symmetry, the divergence
structure of the EE is richer. The presence of dimensionful parameters
on the theory enlarges the possible divergent terms. Now, let us consider
a general QFT that is constructed by perturbing a CFT in the UV-fixed
point by a relevant operator with mass dimension $\Delta\in(\frac{d-2}{2},d)$

\begin{equation}
S=S_{CFT}+\lambda\int O_{\Delta}\left(x\right)d^{d}x\,,
\end{equation}
and coupling constant $\lambda$ with mass dimension $d-\Delta\in(0,\frac{d+2}{2})$.
Then, the coefficients of the geometrical terms in the divergent part
of the entropy are corrected. We could have terms of the form
\begin{equation}
\mu_{d-2k}\left(\gamma_{\mathcal{O}}\right)\frac{\lambda^{n}}{\delta^{d-2k-n\left(d-\Delta\right)}}\,,\quad n\in\mathbb{N}\,,\;n\neq1\,.
\end{equation}
It means that, besides the negative integer powers of the cutoff,
there may appear new non-integer powers of the cutoff compensated
with an adequate integer powers of the coupling constant. For precise
values of $\Delta$ and $n\geq2$, these new terms could be recombined
with preexisting terms of less order in $\delta$, modifying the value
of the non-universal coefficients of $\mu_{d-2k}\left(\gamma_{\mathcal{O}}\right)$.
However, the presence of such terms cannot modify the leading divergent
term proportional to $\delta^{-d+2}$, which always grows with the
area.

As we have emphasized, the divergent structure \eqref{c3-ee-div}
is non-universal. In section \ref{c3-sec-univee}, we introduce a
general ``geometric'' way to regularize the EE, which holds for
any QFT.

\subsection{Equality of the EE for complementary regions\label{c3-sec-duality}}

Here, we give a brief comment about the equality of the EE for complementary
regions. As we explained in section \ref{c2_sec_finite}, for any
finite quantum system $\mathfrak{A}\subset M_{n}\left(\mathbb{C}\right)$
and any global pure state $\omega$ on the algebra $M_{n}\left(\mathbb{C}\right)$,
we have the following equality for the vN entropies
\begin{equation}
S\left(\mathfrak{A}\right)=S\left(\mathfrak{A}'\right)\,,\label{c3-eq-ee}
\end{equation}
where $\mathfrak{A}'$ is the algebraic commutant of $\mathfrak{A}$
inside $M_{n}\left(\mathbb{C}\right)$. In QFT, we have a similar
setup. We could ask ourselves whether for two complementary regions
$\mathcal{O}$ and $\mathcal{O}'$ and any global pure state, we would
have
\begin{equation}
S\left(\mathcal{O}\right)=S\left(\mathcal{O}'\right)\,.\label{c3-eq-ee-reg}
\end{equation}
Unfortunately, such a question is meaningless because the EE is infinite
in QFT. However, let us assume for a moment that the EE is finite
and well-defined. According to relation \eqref{c3-eq-ee}, we expect
the equality 
\begin{equation}
S\left(\mathcal{A}\left(\mathcal{O}\right)\right)=S\left(\mathcal{A}\left(\mathcal{O}\right)'\right)\,,
\end{equation}
which does not imply \eqref{c3-eq-ee-reg} unless 
\begin{equation}
\mathcal{A}\left(\mathcal{O}\right)'=\mathcal{A}\left(\mathcal{O}'\right)\,.\label{c3-duality}
\end{equation}
In other words, the equality of the EE for complementary regions (on
a global pure state) is a consequence of the duality condition \eqref{c3-duality}
for the algebra $\mathcal{A}\left(\mathcal{O}\right)$ attached to
the region $\mathcal{O}$. However, in AQFT, it is not guaranteed
that the duality \eqref{c3-duality} holds in general. In fact, it
could be the case that such equality holds for some, but not all,
regions. In chapters \ref{CURRENT} and \ref{EE_SS}, we will study
some models where the equality \eqref{c3-duality} holds for double
cones, but not for regions made out of the union of two strictly spacelike
separated double cones.

The ambiguous relation \eqref{c3-eq-ee-reg} could be unambiguously
stated for a regularized EE \eqref{c3-ee-div}. For example, in a
case of a QFT regularized by a lattice cutoff $\delta$, the regularized
EE \eqref{c3-ee-div} satisfies the equality \eqref{c3-eq-ee-reg}
whenever the regularized algebra $\mathcal{A}_{\delta}\left(\mathcal{O}\right)$
satisfies the duality condition \eqref{c3-duality}. For a QFT in
the continuum, an equivalent and complete rigorous statement of the
relation \eqref{c3-eq-ee-reg} could be stated using the relative
entropy. In fact, for finite quantum systems, we can use the relations
\eqref{c2_dh-ds} and \eqref{c2_dh_and_ds} for the algebras $\mathfrak{A}$
and $\mathfrak{A}'$, and two global pure states $\omega\left(\cdot\right)=\left\langle \Omega\right|\cdot\left|\Omega\right\rangle $
and $\phi\left(\cdot\right)=\left\langle \Phi\right|\cdot\left|\Phi\right\rangle $
\begin{align}
S_{\mathfrak{A}}\left(\phi\mid\omega\right) & =\Delta\left\langle K_{\omega,\mathfrak{A}}\right\rangle -\Delta S_{\mathfrak{A}}\,,\\
S_{\mathfrak{A}'}\left(\phi\mid\omega\right) & =\Delta\left\langle K_{\omega,\mathfrak{A}'}\right\rangle -\Delta S_{\mathfrak{A}'}\,.
\end{align}
Then, subtracting both expressions and using \eqref{c3-eq-ee}, we
have that
\begin{equation}
S_{\mathfrak{A}}\left(\phi\mid\omega\right)-S_{\mathfrak{A}'}\left(\phi\mid\omega\right)=\Delta\left\langle K_{\omega,\mathfrak{A}}-K_{\omega,\mathfrak{A}'}\right\rangle \,.\label{c3-eq-rel-1}
\end{equation}
According to remark \ref{c2_remark_mod} and example \eqref{c2_ex_finit_mod},
$K_{\omega,\mathfrak{A}}-K_{\omega,\mathfrak{A}'}$ in the above expression
is just the ``full'' modular Hamiltonian associated with $\left\{ \mathfrak{A},\left|\Omega\right\rangle \right\} $,
which will be also denoted by $K_{\omega,\mathfrak{A}}^{\mathrm{}}$.
Therefore, equation \eqref{c3-eq-rel-1} could be rewritten as
\begin{equation}
S_{\mathfrak{A}}\left(\phi\mid\omega\right)-S_{\mathfrak{A}'}\left(\phi\mid\omega\right)=\left\langle \Psi\right|K_{\omega,\mathfrak{A}}\left|\Psi\right\rangle =\phi\left(K_{\omega,\mathfrak{A}}\right)\,.\label{c3-eq-rel-2}
\end{equation}
The interesting fact of the above relation, which encodes the information
of equality \eqref{c3-eq-ee}, is that it involves relative entropies
and a ``full'' modular Hamiltonian, which are objects well-defined
in the continuum QFT. Moreover, the expression \eqref{c3-eq-rel-2}
could be directly proved using the tools of modular theory developed
in section \ref{c2_sec-mod_th}. In AQFT, relation \eqref{c3-eq-rel-2}
must be restated in the following way.
\begin{lem}
In any AQFT (in the vacuum representation) and for any two global
pure states $\omega,\phi$ on $\mathcal{B}\left(\mathcal{H}_{0}\right)$,
given by the unit vectors $\left|\Omega\right\rangle ,\left|\Phi\right\rangle $,
we have the following equality
\begin{equation}
S_{\mathcal{O}}\left(\phi\mid\omega\right)-S_{\mathcal{O}'}\left(\phi\mid\omega\right)=\phi\left(K_{\Omega}\right)\,,\label{c3-eq-rel-fin}
\end{equation}
whenever the duality relation $\mathcal{A}\left(\mathcal{O}\right)'=\mathcal{A}\left(\mathcal{O}'\right)$
holds. The operator $K_{\Omega}$ in \eqref{c3-eq-rel-fin} is the
(full) modular Hamiltonian associated with $\left\{ \mathcal{A}\left(\mathcal{O}\right),\left|\Omega\right\rangle \right\} $.
\end{lem}

\subsection{Universal regularized entanglement entropy\label{c3-sec-univee}}

As we have explained in section \ref{c3-sec-ee-qft}, EE is necessarily
divergent in QFT, and its structure of divergences is non-universal.
Then, in order to obtain universal information out of it, we are forced
to use relative entropy quantities. This is especially necessary,
in the context of the present thesis, since we are discussing subtleties
that could be hard to understand without this perspective.

We are going to use mutual information as a way to define a regularized
entanglement entropy \cite{chmy}. Or, more precisely, we are defining
entropy in QFT as a quantity derived from mutual information. MI retains
all the universal information of the EE and, as we have explained
in section \ref{c3-gen-str}, it is a well-defined quantity for strictly
separated regions, and it is always finite for the vacuum state.

The definition of EE from MI follows from the following observation.
We take a global state $\omega$ and two regions $\mathcal{O}_{1}\text{\Large\ensuremath{\times\!\negmedspace\!\times}}\mathcal{O}_{2}$.
According to \eqref{c2-mi_from_vn}, it is expected that the mutual
information $I\left(\mathcal{O}_{1},\mathcal{O}_{2}\right)$ looks
like
\begin{equation}
I\left(\mathcal{O}_{1},\mathcal{O}_{2}\right)``="S\left(\mathcal{O}_{1}\right)+S\left(\mathcal{O}_{1}\right)-S\left(\mathcal{O}_{1}\vee\mathcal{O}_{2}\right)\,.\label{c3-mi_ee}
\end{equation}
The symbol $``="$ is placed to emphasize that such equality is just
heuristic. In fact, the r.h.s. of \eqref{c3-mi_ee} is ill-defined.
The above relation must be considered as a heuristic motivation coming
from finite quantum systems. However, if we place a short distance
cutoff $\delta>0$, we are allowed to compute the bare entanglement
entropies $S_{\delta}\left(\mathcal{O}\right)$. In this case, equation
\eqref{c3-mi_ee} is no longer heuristic and could be restated as
\begin{equation}
I\left(\mathcal{O}_{1},\mathcal{O}_{2}\right)=\lim_{\delta\rightarrow0^{+}}\left[S_{\delta}\left(\mathcal{O}_{1}\right)+S_{\delta}\left(\mathcal{O}_{1}\right)-S_{\delta}\left(\mathcal{O}_{1}\vee\mathcal{O}_{2}\right)\right]\,.
\end{equation}
The above relation is independent of the chosen regularization prescription
since local ambiguities due to the cutoff are local and extensive
on the boundary, and they cancel in \eqref{c3-mi_ee} as we take the
continuum limit $\delta\rightarrow0^{+}$. 

\begin{figure}[h]
\centering
\includegraphics[width=12cm]{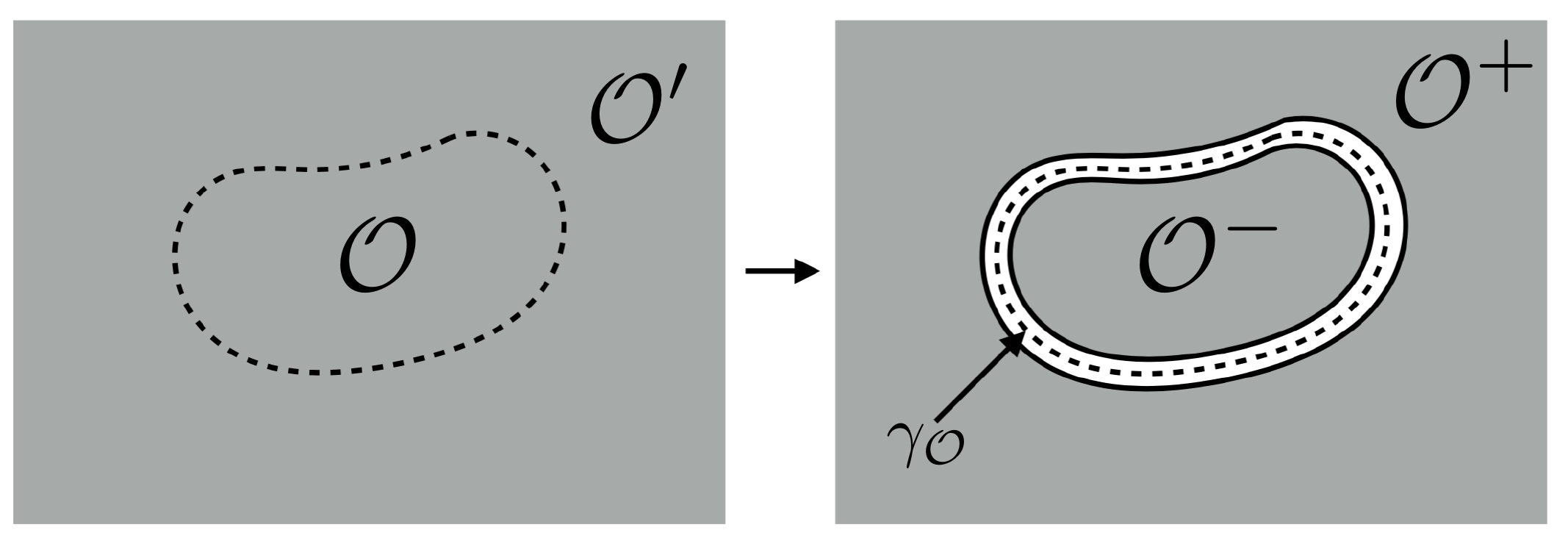}\caption{\label{c3_fig_ee_mi}Geometrical setup for the definition of the EE
through the MI.}
\end{figure}

Now we want to use $I\left(\mathcal{O}_{1},\mathcal{O}_{2}\right)$,
which is a function of two regions, to produce a quantity $S\left(\mathcal{O}\right)$
that is a function of only one region $\mathcal{O}$, and it will
be interpreted as a version of the EE suitable for any QFT. From now
on, we assume that the global state $\omega$ is pure. We also consider
the entanglement surface $\gamma_{\mathcal{O}}$ of the causally complete
region $\mathcal{O}$. We take a spatial unit vector $\eta$ exterior
to $\gamma_{\mathcal{O}}$, and a short distance scale $\epsilon>0$.\footnote{For each point $x\in\gamma_{\mathcal{O}}$, the set of spatial unit
exterior vectors forms a one-dimensional hyperbola in Minkowski spacetime. } We can construct two spatial surfaces, each on each side of $\gamma_{\mathcal{O}}$,
using the elements of the ``framing\textquotedbl{} $\left(\eta,\epsilon\right)$,
as
\begin{eqnarray}
\gamma_{\mathcal{O}^{+}}=\gamma_{\mathcal{O}}+\frac{\epsilon}{2}\eta & \textrm{ and } & \gamma_{\mathcal{O}^{-}}=\gamma_{\mathcal{O}}-\frac{\epsilon}{2}\eta\,.\label{c3-frame}
\end{eqnarray}
We call $\mathcal{O}^{-}\subset\mathcal{O}$ the causally complete
region with entanglement surface $\gamma_{\mathcal{O}^{-}}$, and
$\mathcal{O}^{+}\subset\mathcal{O}'$ the causally complete region
with entanglement surface $\gamma_{\mathcal{O}^{+}}$ (see figure
\ref{c3_fig_ee_mi}). Then, we use the mutual information $I\left(\mathcal{O}^{+},\mathcal{O}^{-}\right)$
as a regularization of the entropy. More precisely, we define the
\textit{universal regularized entanglement entropy} as
\begin{equation}
S_{reg}\left(\mathcal{O};\epsilon,\eta\right):=\frac{I\left(\mathcal{O}^{+},\mathcal{O}^{-}\right)}{2}\,,\label{c3-re_ee}
\end{equation}
where in the second equality we have used \eqref{c3-mi_ee}. Definition
\eqref{c3-mi_ee} is justified by the heuristic relation \eqref{c3-mi_ee}.
In fact, in the limit $\epsilon\rightarrow0^{+}$ we expect that the
r.h.s. of \eqref{c3-mi_ee} looks like
\begin{eqnarray}
S(\mathcal{O}^{+})+S(\mathcal{O}^{-})\!\!\! & \rightarrow & \!\!\!S(\mathcal{O}^{'})+S(\mathcal{O})=2S(\mathcal{O})\,,\label{c3-eq-reg1}\\
S(\mathcal{O}^{+}\vee\mathcal{O}^{-})\!\!\! & \rightarrow & \!\!\!S(\mathbb{R}^{d})=0\,.\label{c3-eq-reg2}
\end{eqnarray}
These last relations were motivated by the equality of the vN entropy
for complementary algebras which is always satisfied for a global
pure state on finite algebras. We remark that we do not have to assume
\eqref{c3-duality} to state \eqref{c3-eq-reg1} and \eqref{c3-eq-reg2}.
In fact, such formulas are just a guide to motivate the definition
\ref{c3-re_ee}. $S_{reg}\left(\mathcal{O};\epsilon,\eta\right)$
is a quantity that belongs to the continuum theory, and in particular,
is Poincaré invariant for the vacuum state. The symmetric framing
at both sides of $\gamma_{\mathcal{O}}$ in \eqref{c3-frame} gives
equal regularized entropy for complementary regions, i.e. $S_{reg}\left(\mathcal{O};\epsilon,\eta\right)=S_{reg}\left(\mathcal{O}';\epsilon,-\eta\right)$.
It is important to notice that the short distance $\epsilon>0$, which
plays the role of the cutoff for the EE, is indeed a geometrical quantity
belonging to the continuum QFT.

$S_{reg}$ is not a function of the region $\mathcal{O}$ alone but
depends on the framing, which includes the vector field $\eta$. As
$\epsilon\rightarrow0^{+}$ crossing all scales of the theory and
curvature scales of $\gamma_{\mathcal{O}}$, we expect that $S_{reg}$
can be written as a series of inverse integer powers in $\varepsilon$
similar to \eqref{c3-ee-div}. Since $\epsilon$ is a regulator, as
we are taking the $\epsilon\rightarrow0^{+}$ limit, we discard positive
powers of $\epsilon$. The divergent terms with negative exponents
are produced by ultra-local entanglement between degrees of freedom
arbitrarily close to both sides of $\gamma_{\mathcal{O}}$. In fact,
since the cutoff $\epsilon$ is geometrical, we expect that the coefficients
of these divergent contributions could be written as integrals of
local geometrical quantities along $\gamma_{\mathcal{O}}$ like the
one of \eqref{c3-local-geo}. This geometric integrand is constructed
with the metric, powers of the curvature of $\gamma_{\mathcal{O}}$,
and the vector field $\eta$. The dependence on $\eta$ should only
show up in the divergent terms, and we should be able to subtract
these terms to eliminate the frame dependence. Finally, we define
\begin{equation}
S\left(\mathcal{O}\right):=\lim_{\epsilon\rightarrow0^{+}}\left[S_{reg}\left(\mathcal{O};\epsilon,\eta\right)-\textrm{local divergent terms}\right]\,.
\end{equation}
In consequence, $S\left(\mathcal{O}\right)$ is finite, Poincaré invariant,
and completely defined by the continuum theory itself. It can be thought
of as a \textit{minimally subtracted entanglement entropy}. We emphasize
that it is free from ambiguities since it is uniquely and well-defined
for any QFT.

While $S\left(\mathcal{O}\right)$ does not have the property of being
positive for arbitrary regions, it should retain some other important
properties of entropy. The symmetry between complementary regions
$S\left(\mathcal{O}\right)=S\left(\mathcal{O}'\right)$ is built-in
by definition and comes from the symmetry of the mutual information
under the exchange of the two regions. $S\left(\mathcal{O}\right)$
should satisfy strong subadditivity \cite{chmy,Casini:2018kzx}. This
is important for example in the context of the entropic proof of the
irreversibility theorems.

The extensivity and locality of the divergences in $\epsilon$ of
$S_{reg}\left(\mathcal{O};\epsilon,\eta\right)$ and the strong subadditivity
property of the finite entropy $S\left(\mathcal{O}\right)$ could
be studied with complete mathematical rigor in AQFT, but this investigation
has not been done yet.\footnote{See \cite{Hollands:2018wzi} for a recent related investigation.}

\section{Free fields in the lattice\label{c3-secc-lattice}}

One convenient way to regularize QFT is introducing a lattice cutoff.
From the algebraic standpoint, this choice is the most natural, since
the regularized QFT still behaves as a sort of AQFT. In this case,
the degrees of freedom of QFT are described using a $d-1$ dimensional
square lattice $\mathcal{L}$. The square lattice $\mathcal{L}$ has
to be considered as a discretization, in cells of spacing $\delta$,
of the spacetime, over the Cauchy slice $\Sigma_{0}:=\left\{ x\in\mathbb{R}^{d}\,:\,x^{0}=0\right\} $.
In each point $j\in\mathcal{L}$, we place an operator algebra $\mathfrak{A}_{i}$,
which could be a finite dimensional algebra (in the case of fermions)
or a type I vN algebra (in the case of bosons). We define the global
algebra as the infinite tensor product $\mathfrak{A}:=\bigotimes_{j\in\mathcal{L}}\mathfrak{A}_{j}$.
For any finite subset $A\subset\mathcal{L}$, we can define the subalgebra
\begin{eqnarray}
V\subset\mathcal{L} & \mapsto & \mathfrak{A}_{A}:=\bigotimes_{j\in A}\mathfrak{A}_{j}\subset\mathfrak{A}\,.\label{c3-lat-alg}
\end{eqnarray}
Such an algebra has to be considered as the regularization of the
local algebra $\mathcal{A}\left(D\left(\mathcal{C}\right)\right)$,
where $D\left(\mathcal{C}\right)\subset\mathbb{R}^{d}$ is the Cauchy
development of any open region $\mathcal{C}\subset\Sigma_{0}$, which
contains all the lattice points $j\in A$. It is interesting to notice
that the correspondence \eqref{c3-lat-alg} satisfies the axioms 1,
2 and 3 (generating property, isotony and causality) of definition
\ref{c1_def_aqft}, but of course, it does not satisfy axiom 4 (Poincaré
invariance). For any $A_{1},A_{2}\subset\mathcal{L}$ such that $A_{1}\cap A_{2}=\emptyset$,
we have that $\mathfrak{A}_{A_{1}\cup A_{2}}:=\mathfrak{A}_{A_{1}}\otimes\mathfrak{A}_{A_{2}}$.
Furthermore, we have that
\begin{equation}
\mathfrak{A}:=\mathfrak{A}_{A}\otimes\mathfrak{A}_{A'}\,,
\end{equation}
where $A':=\mathcal{L}-A$. In other words, the regularized algebras
of complementary regions are statistically independent. In the cases
of our interest, we also have that $\mathfrak{A}'_{A}=\mathfrak{A}_{A'}$,
where $\mathfrak{A}'_{A}$ is the commutant of $\mathfrak{A}_{A}$
computed inside the global algebra $\mathfrak{A}_{\mathcal{L}}$.
Given any state on the global algebra of QFT, it could be lifted to
a global state $\omega$ in the regularized algebra $\mathfrak{A}$.
In order to calculate the EE for the local algebra $\mathfrak{A}_{A}$,
we need to compute the statistical operator $\rho_{A}$ of the restricted
state $\left.\omega\right|_{\mathfrak{A}_{A}}$.

In QFT, the algebra $\mathfrak{A}$ is usually described by a canonical
commutation (resp. anticommutation) relations algebra, usually denoted
as CCR (resp. CAR) algebra. In many situations, we consider a Gaussian
state $\omega$ on such an algebra. For example, the vacuum state
in free field theories. The complete knowledge of a Gaussian state
is entirely determined by the knowledge of its two-point correlators.
In the following, we show how to compute the statistical operator
$\rho_{A}$, and hence the modular Hamiltonian and the EE, for Gaussian
states in CCR and CAR algebras.
\begin{rem}
Throughout the following subsections, $\dagger$ denotes the Hermitian
conjugate and $*$ denotes the complex conjugate.
\end{rem}

\subsection{CAR algebra (fermions)\label{subsec-CAR-algebra-(fermions)}}

We start with the CAR algebra of $n$ degrees of freedom. Such an
algebra is defined as the free $*$-algebra generated by the operators
$\psi_{1},\ldots,\psi_{n}$ modulo the relations
\begin{equation}
\{\psi_{j},\psi_{k}^{\dagger}\}=\delta_{jk}\;\textrm{ and }\;\{\psi_{j},\psi_{k}\}=\{\psi_{j}^{\dagger},\psi_{k}^{\dagger}\}=0\,.
\end{equation}
It can be shown that the canonical representation of this algebra
is the finite dimensional algebra $\mathfrak{A}:=\bigotimes_{j=1}^{n}\mathfrak{A}_{j}$,
where each subalgebra $\mathfrak{A}_{j}:=M_{2}\left(\mathbb{C}\right)$
represents a one fermion site.\footnote{We denote by $\left(\sigma_{j}\right)_{j=1,2,3}\in M_{2}\left(\mathbb{C}\right)$
the Pauli matrices. For one degree of freedom, the fermionic operators
are represented as
\begin{equation}
\psi:=\frac{\sigma_{1}+i\sigma_{2}}{2}\in M_{2}\left(\mathbb{C}\right)\quad\textrm{and}\quad\psi^{\dagger}:=\frac{\sigma_{1}-i\sigma_{2}}{2}\in M_{2}\left(\mathbb{C}\right)\,.
\end{equation}
For $n$ degrees of freedom, the fermionic operators are represented
as
\begin{eqnarray}
\psi_{j}^{\flat}\!\!\! & := & \!\!\!\Gamma\otimes\ldots\otimes\Gamma\otimes\underset{(\textrm{site }j)}{\psi^{\flat}}\otimes\mathbf{1}\otimes\ldots\otimes\mathbf{1}\in M_{2}\left(\mathbb{C}\right)^{\otimes n}\,,\quad\flat=\cdot,\dagger\,,
\end{eqnarray}
where $\Gamma:=\sigma_{3}$.} Therefore, in this case, we have a finite quantum system.

Given any state $\omega$ on $\mathfrak{A}$, the general structure
of its two-points correlators is uniquely determined by\footnote{Here we adopt the usual convention $\left\langle A\right\rangle :=\omega\left(A\right)$.}
\begin{equation}
\langle\psi_{j}\psi_{k}^{\dagger}\rangle=:C_{jk}\,\textrm{ and }\,\langle\psi_{j}\psi_{k}\rangle=:D_{jk}\,,\label{c3-fer_gaus}
\end{equation}
where $C,D\in\mathbb{C}^{n\times n}$. The other two-points correlators
can be obtained from \eqref{c3-fer_gaus} imposing anticommutation
relations and hermiticity of $\omega$, namely
\begin{equation}
\left\langle \psi_{j}^{\dagger}\psi_{k}\right\rangle =\delta_{jk}-C_{kj}\,\textrm{ and }\,\left\langle \psi_{j}^{\dagger}\psi_{k}^{\dagger}\right\rangle =D_{kj}^{*}=\left(D^{\dagger}\right)_{jk}\,.\label{c3-fer_gaus2}
\end{equation}
Furthermore, the anticommutation relations impose $D=-D^{T}$, and
the positivity of $\omega$ imposes $C>0$ and $\mathbf{1}_{n}-C>0$.
This last fact implies that all the eigenvalues of $C$ lie in the
interval $\left[0,1\right]$. 

Let us introduce the following ``charge'' symmetry transformation.
It is a continuous (global) symmetry given by a one-parameter group
of automorphisms $\varrho_{s}$ such that
\begin{equation}
\varrho_{s}\left(\psi_{j}\right)=\mathrm{e}^{is}\psi_{j}\,\textrm{ and }\,\varrho\left(\psi_{j}^{\dagger}\right)=\mathrm{e}^{-is}\psi_{j}^{\dagger}\,.
\end{equation}
From now on, we assume that the state is invariant under $\varrho_{s}$,
i.e. $\omega=\omega\circ\varrho_{s}$.\footnote{In QFT, this happens for any state of definite charge, and in particular,
for the vacuum state. } This constrains \eqref{c3-fer_gaus} even more, imposing $D=0$.
For the structure of multipoint correlators, we now assume that the
state $\omega$ is \textit{Gaussian}, i.e. all odd-point correlators
are zero and all even-point correlators are given by the usual Wick
theorem \cite{Araki_quasifree}. Since $D=0$, we also have that all
even-point correlators with distinct numbers of fermions and anti-fermions
are zero. All non-zero multipoint correlators are obtained from the
two-point functions according to
\begin{equation}
\left\langle \psi_{j_{1}}\cdots\psi_{j_{l}}\psi_{k_{1}}^{\dagger}\cdots\psi_{k_{l}}^{\dagger}\right\rangle =\left(-1\right)^{\frac{l\left(l-1\right)}{2}}\sum_{\sigma\in S_{l}}\left(-1\right)^{\left|\sigma\right|}\prod_{m=1}^{l}\left\langle \psi_{j_{l}}\psi_{k_{\sigma\left(l\right)}}^{\dagger}\right\rangle \,,
\end{equation}
where $S_{l}$ is the permutation group of $l$ elements. It is a
well-known fact that the state $\omega$, and hence its statistical
operator $\rho$, are completely determined by the multipoint correlators
as above.

We now choose a proper subset of sites $A\subsetneq\mathcal{L}=\left\{ 1,\ldots,n\right\} $
and define the bipartite system $\mathfrak{A}=\mathfrak{A}_{A}\otimes\mathfrak{A}_{A'}$
where
\begin{equation}
\mathfrak{A}_{A}:=\bigotimes_{j\in A}\mathfrak{A}_{j}\,\textrm{ and }\,\mathfrak{A}_{A'}:=\bigotimes_{j\in A'}\mathfrak{A}_{j}\,.\label{c3-fer-bipar}
\end{equation}
To compute the entanglement entropy of the ``region'' $A$ we need
to first compute the statistical operator $\rho_{A}$ of the restricted
state $\left.\omega\right|_{\mathfrak{A}_{A}}$. For that, we use
the following ansatz
\begin{equation}
\rho_{A}:=\frac{1}{Z}\mathrm{e}^{-\sum_{j,k\in A}\psi_{j}^{\dagger}K_{jk}\psi_{k}}\,,\label{c3-fer-ans}
\end{equation}
which always works for any Gaussian state \cite{gaudin}. We have that $K\in\mathbb{C}^{n_{A}\times n_{A}}$,
$n_{A}:=\#A$, and $Z=\det(1-\mathrm{e}^{-K})>0$ in order to have
$\mathrm{Tr}_{\mathfrak{A}_{A}}\left(\rho_{A}\right)=1$. Imposing
the following relation
\begin{equation}
C_{jk}=\langle\psi_{j}\psi_{k}^{\dagger}\rangle=\frac{1}{Z}\mathrm{Tr}_{\mathfrak{A}_{A}}\left(\mathrm{e}^{-\sum_{j,k\in A}\psi_{j}^{\dagger}K_{jk}\psi_{k}}\psi_{j}\psi_{k}^{\dagger}\right)\,,
\end{equation}
a straightforward computation gives
\begin{equation}
K=-\log\left(C_{A}^{-1}-1\right) \in \mathbb{C}^{n_{A}\times n_{A}} \,,\label{c3-mh-fer}
\end{equation}
where $C_{A}\in\mathbb{C}^{n_{A}\times n_{A}}$ is just the matrix $C_{jk}$ but where the indices
are restricted to $j,k\in A$ \cite{Casini_review}. In other words,
if $\left\{ |u_{l}\rangle\right\} \subset\mathbb{C}^{n_{A}}$ is a
orthonormal basis of eigenvectors of $C_{A}$ with the corresponding
eigenvalues $\left\{ \lambda_{l}\right\} $, then we have that
\begin{equation}
K=-\sum_{l=1}^{n_{A}}\log\left(\lambda_{l}^{-1}-1\right)|u_{l}\rangle\langle u_{l}| \in\mathbb{C}^{n_{A}\times n_{A}} \,.\label{c3-fer_Km}
\end{equation}
Since the eigenvalues of $C_{A}$ are restricted to the interval $\left[0,1\right]$,
the above equation is valid whenever all $\lambda_{l}<1$. When $\lambda_{l}=1$
for some $l$, it means that the local reduced density matrix has
a zero eigenvalue, which is equivalent to the reduced state not being
faithful. In QFT, because of the Reeh-Schlieder theorem, this is forbidden
for the vacuum state and the algebra of any standard region $\mathcal{O}$.

When some eigenvalues are equal to $1$, a more careful calculation
shows that \eqref{c3-fer_Km} still holds if we replace the sum including
only eigenvalues less than $1$. The fact that all eigenvalues are
less than $1$, means that the restricted state $\left.\omega\right|_{\mathfrak{A}_{A}}$
is faithful. 

The entanglement entropy and the entanglement Rényi entropies follow from \eqref{c2_vn_ent} and \eqref{c2-renyi-def}
\begin{eqnarray}
S\left(A\right) & \!\!\!=\!\!\! & - \sum_{l=1}^{n_{A}} \left[ (1-\lambda_{l}) \log(1-\lambda_{l}) + \lambda_{l} \log(\lambda_{l}) \right] = -\mathrm{Tr}\left[\left(1-C_{A}\right)\log\left(1-C_{A}\right)+C_{A}\log C_{A}\right]\,, \nonumber \\
S_{\alpha}\left(A\right) & \!\!\!=\!\!\! & \frac{1}{1-\alpha}\sum_{l=1}^{n_{A}}\left[\log(1-\lambda_l)^\alpha + \lambda_{l}^\alpha \right]=\frac{1}{1-\alpha}\mathrm{Tr}\log\left[\left(1-C_{A}\right)^{\alpha}+C_{A}^{\alpha}\right]\,,\label{c3-fer-ee}
\end{eqnarray}
where the above traces have to be taken in the finite dimensional
algebra space $\mathbb{C}^{n_{A}\times n_{A}}$. The modular Hamiltonian
is obtained from \eqref{c3-fer-ans}
\begin{equation}
K_{A}=-\log\left(\rho_{A}\right)=\log\left(Z\right)\mathbf{1}+\sum_{j,k\in A}\psi_{j}^{\dagger}K_{jk}\psi_{k} \in \mathfrak{A}_{A} \,.\label{c3-mh-car}
\end{equation}

\subsection{CCR algebra (bosons)\label{c3-secc-bose}}

We start with the CCR algebra of $n$ degrees of freedom. Such an
algebra is defined as the free $*$-algebra generated by the operators
$\phi_{1},\ldots,\phi_{n},\pi_{1},\ldots,\pi_{n}$ modulo the relations
\begin{alignat}{1}
 & \left[\phi_{j},\pi{}_{k}\right]=i\delta_{jk}\,,\;\left[\phi_{j},\phi{}_{k}\right]=\left[\phi_{j},\phi{}_{k}\right]=0\,,\\
 & \phi_{j}=\phi_{j}^{\dagger}\,\textrm{ and }\,\pi_{j}=\pi_{j}^{\dagger}\,.
\end{alignat}
As we have explained in example \ref{c1_weyl_alg}, this algebra is
not indeed a $C^{*}$-algebra. However, we could pass to the corresponding
Weyl algebra (exponentiated version). We also have that the Weyl algebra
is not finite dimensional. In fact, its unique irreducible representation
is $\mathfrak{A}:=\mathcal{B}\left(\mathcal{H}\right)$ where $\mathcal{H}:=L^{2}\left(\mathbb{R}^{n}\right)$.
The operators act as
\begin{eqnarray}
\hspace{-5mm} \phi_{j}\Psi\left(q_{1},\ldots,q_{n}\right)=q_{j}\Psi\left(q_{1},\ldots,q_{n}\right) & \textrm{and} & \pi_{j}\Psi\left(q_{1},\ldots,q_{n}\right)=-i\frac{\partial}{\partial q_{j}}\Psi\left(q_{1},\ldots,q_{n}\right).
\end{eqnarray}
The operators $\phi_{j},\pi_{k}$ are unbounded, but we can use them
to completely characterize a given state $\omega$. Without loss of
generality, we assume that $\left\langle \phi_{j}\right\rangle =\left\langle \pi_{j}\right\rangle =0$
for all $j=1,\ldots,n$. The general structure of the two-point correlators
of $\omega$ is uniquely determined by
\begin{align}
 & \left\langle \phi_{j}\phi{}_{k}\right\rangle =:X_{jk}\,,\quad\left\langle \pi_{j}\pi{}_{k}\right\rangle =:P_{jk}\,,\label{c3-bose-state}\\
 & \left\langle \phi_{j}\pi{}_{k}\right\rangle =:\frac{i}{2}\delta_{jk}+R_{jk}\,,\label{c3-bose-state2}
\end{align}
where $X,P,R\in\mathbb{C}^{n\times n}$. We also have the following
constrains: $X,P\in\mathbb{R}^{n\times n}$ due to the hermiticity
of the state, $X,P>0$ due to the positivity of the state and $R\in\mathbb{R}^{n\times n}$
due to the commutation relations. The positivity of $\langle\left(\phi_{j}+ic_{jk}\pi{}_{k}\right)^{\dagger}\left(\phi_{j}+ic_{jk}\pi{}_{k}\right)\rangle$
for arbitrary numbers $c_{jk}\in\mathbb{R}$ implies that $XP>\frac{1}{4}$,
which means that all the eigenvalues of $XP$ lie in the interval
$\left[\frac{1}{4},+\infty\right)$. 

Let us now introduce the ``time inversion'' symmetry. It is given
by an anti-automorphism $\tau$ such that
\begin{equation}
\tau\left(\phi_{j}\right)=\phi_{j}\,\textrm{ and }\,\tau\left(\pi_{j}\right)=-\pi_{j}\,.\label{c3-bose-time}
\end{equation}
We now assume that the state is invariant under $\tau$, i.e. $\omega=\omega\circ\tau$.
This constrains \eqref{c3-fer_gaus} more, imposing $R=0$. Like in
the fermion case, we assume that the state $\omega$ is Gaussian.
Then, all odd-point correlators are zero and all even-point correlators
are given by the usual Wick theorem
\begin{equation}
\left\langle f_{j_{1}}\cdots f_{j_{2l}}\right\rangle =\frac{1}{2^{l}l!}\sum_{\sigma\in S_{2l}}\prod_{m=1}^{l}\left\langle f_{j_{2m-1}}f_{j_{\sigma\left(2m\right)}}\right\rangle \,,
\end{equation}
where $f_{j}$ are any of the field or momentum variables, and the
operators are ordered inside the expectation value so that all the
field variables are placed on the left and all the momentum variables
on the right. It can be shown that the state $\omega$ can be uniquely
determined by the collection of all multipoint correlators.

Once we have characterized the structure of the state, we compute
the entanglement entropy. For such a purpose, it is useful to notice
the following decomposition of the algebra $\mathfrak{A}$, 

\begin{equation}
\mathfrak{A}:=\bigotimes_{j=1}^{n}\mathfrak{A}_{j}\,,\;\textrm{where }\mathfrak{A}_{j}:=\mathcal{B}\left(L^{2}\left(\mathbb{R}\right)\right)\,.
\end{equation}
As in the fermion case, we choose a proper subset of sites $A\subsetneq\mathcal{L}=\left\{ 1,\ldots,n\right\} $
and define the bipartite system $\mathfrak{A}=\mathfrak{A}_{A}\otimes\mathfrak{A}_{A'}$,
according to \eqref{c3-fer-bipar}. Even though the algebras are infinite
dimensional, we can still use all the formulas for finite quantum
systems to compute any information measure.\footnote{This is because, as we discuss at the beginning of section \ref{c2-Sec_entro_gener},
these algebras are type I vN algebras.} In particular, the statistical operator $\rho_{A}$ of the restricted
state $\left.\omega\right|_{\mathfrak{A}_{A}}$ is a well-defined
density matrix.\footnote{If $\omega$ is a normal state (described by a density matrix in $\mathcal{H}$)
then $\left.\omega\right|_{\mathfrak{A}_{V}}$ must be normal. In
fact, $\rho_{V}$ is given by the usual partial trace on $\mathfrak{A}_{V'}$.} We use the following ansatz
\begin{equation}
\rho_{A}:=\frac{1}{Z}\mathrm{e}^{-\sum_{j,k\in A}\left(\pi{}_{j}N_{jk}\pi_{k}+\phi{}_{j}M_{jk}\phi_{k}\right)}\,,\label{c3-bose-ans}
\end{equation}
which always works for any Gaussian state \cite{Chung_2000,Peschel_2003}.
Similarly, as we did in the fermion case, we impose \eqref{c3-bose-state}
to the above ansatz. A straightforward calculation gives
\begin{eqnarray}
M & \!\!\!=\text{\!\!\!} & P\frac{1}{2C}\log\left(\frac{C+\frac{1}{2}}{C-\frac{1}{2}}\right)\,,\label{c3-bose-m}\\
N & \text{\!\!\!}=\text{\!\!\!} & \frac{1}{2C}\log\left(\frac{C+\frac{1}{2}}{C-\frac{1}{2}}\right)X\,,\label{c3-bose-n}
\end{eqnarray}
where $C:=\sqrt{XP}$ \cite{Casini_review}. We remark that the matrices
$X,P,C$ in the above relations are the restriction to $A$ of the
original $n\times n$ matrices, i.e. they are square matrices of dimension
$n_{A}:=\#A$. Expressions \eqref{c3-bose-m} and \eqref{c3-bose-n}
can be computed using the eigenvectors and eigenvalues of $XP$ and
$PX$.

The entanglement entropy and the entanglement Rényi entropies follow
from \eqref{c2_vn_ent} and \eqref{c2-renyi-def}
\begin{eqnarray}
S\left(A\right) & \!\!\!=\!\!\! & \mathrm{Tr}\left[\left(C+\frac{1}{2}\right)\log\left(C+\frac{1}{2}\right)-\left(C-\frac{1}{2}\right)\log\left(C-\frac{1}{2}\right)\right]\,,\\
S_{\alpha}\left(A\right) & \!\!\!=\!\!\! & \frac{1}{\alpha-1}\mathrm{Tr}\log\left[\left(C+\frac{1}{2}\right)^{\alpha}+\left(C-\frac{1}{2}\right)^{\alpha}\right]\,.
\end{eqnarray}
Finally, the modular Hamiltonian is
\begin{equation}
K_{A}=-\log\left(\rho_{A}\right)=\log\left(Z\right)\mathbf{1}+\sum_{j,k\in A}\left(\pi{}_{j}N_{jk}\pi_{k}+\phi{}_{j}M_{jk}\phi_{k}\right) \in \mathfrak{A}_{A} \,.
\end{equation}

\subsection{General CCR algebra\label{c3-sec-genCAR}}

Here we generalize the previous result to a general CCR algebra. A
general CCR algebra $\mathfrak{A}$ is defined as the free $*$-algebra generated
by the operators $f_{1},\ldots,f_{2n}$ modulo the relations
\begin{equation}
\left[f_{j},f{}_{k}\right]=:iC_{jk}\;\textrm{ and }\;f_{j}=f_{j}^{\dagger}\,,
\end{equation}
where $C\in\mathbb{R}^{2n\times 2n}$ and $C=-C^{T}$ \cite{Sorkin_ccr}.
It is useful to define the operator vector $\bar{f}:=\left(f_{1},\ldots,f_{2n}\right)^{T}$.

A general Gaussian state $\omega$ has a two-point correlator
\begin{align}
\left\langle f_{j}f_{k}\right\rangle  & =:F_{jk}\,,\label{c3-gen-ccr-state}
\end{align}
with $F\in\mathbb{C}^{n\times n}$, and $F>0$ due to the positivity
of the state. Due to the commutation relations it must be that 
\begin{equation}
C_{jk}=2\,\mathrm{Im}\left(F_{jk}\right)\,.\label{c3_genccr-cons}
\end{equation}

We first choose a non-empty subset of an even number of sites $A\subsetneq\mathcal{L}=\left\{ 1,\ldots,2n\right\} $ and define $n_{A}:=\#A/2$. In order to get a well-defined subsystem, we must assume that the matrix $C_A \in \mathbb{R}^{2 n_A\times 2 n_A} $ obtained by restricting the indices of $C_{jk}$ to $j,k\in A$ is also skew-symmetric. In other words, the subalgebra $\mathfrak{A}_{A} \subset \mathfrak{A}$ generated by $\left\{ f_{j}\right\} _{j\in A}$ defines itself a general CCR algebra.

Now, we show how to map this problem to the previous one in section \ref{c3-secc-bose}. Since $C_A$ is skew-symmetric and real, there
exist $O\in\mathbb{R}^{2 n_A\times 2 n_A}$, $O^{-1}=O^{T}$ and $D\in\mathbb{R}^{n_A\times n_A}$, $D>0$ and diagonal, such that
\begin{equation}
O C_A O^{T}=\left(\begin{array}{cc}
0 & D\\
-D & 0
\end{array}\right)\,.
\end{equation}
Defining 
\begin{equation}
Q=\left(\begin{array}{cc}
D^{-\frac{1}{2}} & 0\\
0 & D^{-\frac{1}{2}}
\end{array}\right)\,,
\end{equation}
we have that $Q\in\mathbb{R}^{2n_A\times2n_A}$, $Q>0$, and we can write
the original commutator in the canonical form, i.e.
\begin{equation}
QOC_AO^{T}Q=\left(\begin{array}{cc}
0 & 1\\
-1 & 0
\end{array}\right)\,.
\end{equation}
Accordingly, 
\begin{equation}
\bar{\Phi}=\left(\phi_{1},\ldots,\phi_{n},\pi_{1},\ldots,\pi_{n}\right)^{T}:=QO\bar{f}\,,\label{c3-genccr-new}
\end{equation}
are $2n_A$ variables, satisfying the CCR algebra of the previous section.
Implementing all the same changes to the state \eqref{c3-gen-ccr-state},
we can write
\begin{equation}
QOFO^{T}Q=:\left(\begin{array}{cc}
X & \frac{i}{2}+R\\
-\frac{i}{2}+R^{\dagger} & P
\end{array}\right)\,,\label{c3-ccr-newstate}
\end{equation}
where $X,P,R\in\mathbb{R}^{n_A\times n_A}$ and $X,P>0$ due to \eqref{c3_genccr-cons}
and the positivity of the state. Comparing this last expression with
\eqref{c3-bose-state} and \eqref{c3-bose-state2}, we conclude that
the matrix \eqref{c3-ccr-newstate} has the information about the
correlations of the state $\omega$ in the new variables \eqref{c3-genccr-new}.
As in the previous case, we assume that $R=0$. This is a consequence
of the time-inversion invariance of the state respect to the anti-automorphism
\eqref{c3-bose-time}. In term of the variables $\bar{f}$, there
is an anti-automorphism $\tau$ such that $\tau\left(f_{i}\right)=\sum_{j=1}^{2n}T_{ij}f_{j}$ with $T\in\mathbb{R}^{2n_A\times 2n_A}$ and $TCT^{\dagger}=-C$, and such that the state $\omega$ is invariant, i.e $\omega\circ\tau=\omega$.

We notice that the possibility of constructing $n_A$ pairs of canonical conjugate variables in \eqref{c3-genccr-new} is a direct consequence of assuming that the reduced correlator $C_A \in \mathbb{R}^{2n_A\times 2n_A} $ is skew-symmetric, i.e. $\mathfrak{A}_{A}$ defines itself a general CCR algebra. Furthermore, the converse is also true: if $\mathfrak{A}_{A}$ contains $n_A$ pairs of canonical conjugate variables then $C_A \in \mathbb{R}^{2n_A\times 2n_A} $ is skew-symmetric. In the concrete, the computation above is nothing but the explicit identification of these $n_A$ pairs $\{\phi_j,\pi_j\}_{j=1}^{n_A}$ of canonical conjugate variables inside $\mathfrak{A}_{A}$.

The statistical operator $\rho_{A}$ of the restricted state $\left.\omega\right|_{\mathfrak{A}_{A}}$ could be expressed in terms of the variables $\bar{\Phi}$ as in \eqref{c3-bose-ans},
where the matrix kernels $M,N$ are given by the expressions \eqref{c3-bose-m}
and \eqref{c3-bose-n}. In this case, the kernel $C_A=\sqrt{XP}$ has
to be read from the diagonal blocks of \eqref{c3-ccr-newstate}. However,
it is useful to have an equivalent expression in terms of the original
variables $\bar{f}$. For that, we define the matrix $V\in\mathbb{C}^{2n_{A}\times2n_{A}}$
\begin{equation}
V:=-iC_A^{-1}F-\frac{1}{2}=O^{T}Q\left(\begin{array}{cc}
0 & iP\\
-iX & 0
\end{array}\right)Q^{-1}O\,,\label{c3-genCCR-V}
\end{equation}
whose spectrum is of the form $\left\{ \pm\lambda_{j}\,:\,\lambda_{j}\geq\frac{1}{2}\textrm{ for }j=1,\ldots,n_{A}\right\} $
\cite{scalar_nuestro}. A straightforward computation reveals \cite{scalar_nuestro}
\begin{equation}
\rho_{A}:=\frac{1}{Z}\mathrm{e}^{-\sum_{j,k\in A}f{}_{j}K_{jk}f_{k}}\,,\label{c3-ccr-ans}
\end{equation}
where the matrix $K\in\mathbb{C}^{2n_{A}\times2n_{A}}$ is given by
\begin{equation}
K:=-\frac{i}{2}\frac{V}{\left|V\right|}\log\left(\frac{\left|V\right|+\frac{1}{2}}{\left|V\right|-\frac{1}{2}}\right)C_A^{-1}\,.\label{c3-genCCR-mat}
\end{equation}
Then, the entanglement entropy and the Rényi entanglement entropy
are \cite{scalar_nuestro}
\begin{eqnarray}
S\left(A\right) & \!\!\!=\!\!\! & \mathrm{Tr}\left[\left(V+\frac{1}{2}\right)\log\left|V+\frac{1}{2}\right|\right]\nonumber \\
 & \!\!\!=\!\!\! & \mathrm{Tr}\,\Theta\left(V\right)\left[\left(V+\frac{1}{2}\right)\log\left(V+\frac{1}{2}\right)-\left(V-\frac{1}{2}\right)\log\left(V-\frac{1}{2}\right)\right]\,,\label{c3-genCCR_ee}\\
\nonumber \\
S_{\alpha}\left(A\right) & \!\!\!=\!\!\! & \frac{1}{\alpha-1}\mathrm{Tr}\,\Theta\left(V\right)\log\left[\left(V+\frac{1}{2}\right)^{\alpha}+\left(V-\frac{1}{2}\right)^{\alpha}\right]\,,\label{c3-genCCR_renyi}
\end{eqnarray}
where $\Theta\left(V\right)$ is the orthogonal projector on the subspace
of positive eigenvalues of $V$. The modular Hamiltonian is
\begin{equation}
K_{A}=-\log\left(\rho_{A}\right)=\log\left(Z\right)\mathbf{1}+\sum_{j,k\in A}f{}_{j}K_{jk}f_{k} \in \mathfrak{A}_{A}\,.\label{c3-genCCR-mh}
\end{equation}

\subsection{Continuum limit\label{c3-sec_cont}}

Here we make a few comments on how the above expressions have to be
understood in the continuum limit. Continuum limit has to be thought
of as taking the limit of vanishing lattice spacing, i.e. $\delta\rightarrow0^{+}$.
Such a limit gives place to a divergent EE. However, it gives meaningful
expressions for the correlators kernels and the modular Hamiltonian.
Here we show this for the case of the CAR algebra. In the lattice
model, the modular Hamiltonian is given by the expression \eqref{c3-mh-car}.
It is important to emphasize that the term proportional to the identity
operator is spurious for the purpose of the modular flow,
\begin{equation}
\sigma_{t}\left(O\right)=\mathrm{e}^{iK_{A}t}O\mathrm{e}^{-iK_{A}t}=\mathrm{e}^{it\sum_{j,k\in A}\psi_{j}^{\dagger}K_{jk}\psi_{k}}O\mathrm{e}^{-it\sum_{j,k\in A}\psi_{j}^{\dagger}K_{jk}\psi_{k}}\,,\quad O\in\mathfrak{A}_{A}\,.\label{c3-con-modev}
\end{equation}
Furthermore, in QFT such a constant term is indeed infinite. Such
an infinity comes from the fact that the \eqref{c3-mh-car} is an
expression for the ``inner'' modular Hamiltonian, which, in the
continuum QFT, is a sesquilinear form rather than an operator. For
this reason, in QFT we denote the ``inner'' modular Hamiltonian
just by
\begin{equation}
K_{A}=\sum_{j,k\in A}\psi_{j}^{\dagger}K_{jk}\psi_{k}\,.\label{c3-ec-mh-con}
\end{equation}
The translation of the above equations to the continuum QFT is as
follows. First we pick a spatial region $\mathcal{C}\subset\Sigma_{0}:=\{x\in\mathbb{R}^{d}\,:\,x^{0}=0\}$.
The vacuum state $\omega_{0}$ gives place to the two-point correlator
kernel $\omega_{0}\left(\psi\left(\bar{x}\right)\psi^{\dagger}\left(\bar{y}\right)\right)=\left\langle \psi\left(\bar{x}\right)\psi^{\dagger}\left(\bar{y}\right)\right\rangle =:C\left(\bar{x}-\bar{y}\right)$
where $\bar{x},\bar{y}\in\mathcal{C}$. This correlator can be understood
as a Hermitian kernel operator acting on the ``one-particle'' Hilbert
space $L^{2}\left(\mathcal{C}\right)\otimes\mathbb{C}^{S}$ through
the relation\footnote{The space $\mathbb{C}^{S}$ appears because of the spin character
of the field operator $\psi\left(\bar{x}\right)$.}
\begin{equation}
\left(Cf\right)\left(\bar{x}\right)=\int_{\mathcal{C}}C\left(\bar{x}-\bar{y}\right)f\left(\bar{y}\right)\,d^{d-1}y\,.\label{c3-fer-corr-ker}
\end{equation}
Then, the relation \eqref{c3-mh-fer} must be understood as an operator
equation, where the action of the operators is defined through their
kernels. In fact, the Hermitian correlator kernel has a complete set
of eigenfunctions $|u_{s,k}\rangle=u_{s,k}\left(\bar{x}\right)$ satisfying\footnote{In chapter \ref{CURRENT} we show this explicitly for the chiral fermion
field. }
\begin{eqnarray}
C|u_{s,k}\rangle\!\!\! & = & \!\!\!\frac{1+\tanh\left(\pi s\right)}{2}|u_{s,k}\rangle\,,\label{c3-ferm-con-eig}\\
\int_{\mathcal{C}}u_{s,k}\left(\bar{x}\right)^{\dagger}u_{s',k'}\left(\bar{x}\right)\,d^{d-1}x\!\!\! & = & \!\!\!\delta_{k,k'}\delta\left(s-s'\right)\,,\\
\sum_{k\in\Upsilon}\int_{\mathbb{R}}u_{s,k}\left(\bar{x}\right)u_{s,k}\left(\bar{y}\right)^{\dagger}\,ds\!\!\! & = & \!\!\!\delta\left(\bar{x}-\bar{y}\right)\,,
\end{eqnarray}
where $s,s'\in\mathbb{R}$ , $\bar{x},\bar{y}\in\mathcal{C}$ and
$k\in\Upsilon$ is a possible degeneracy index.\footnote{Eventually, $k\in\Upsilon$ could be a continuous index, and hence,
the summation and Kronecker delta function above have to be replaced
by an integral and a Dirac delta function.} The above eigenfunctions depend strongly on the choice of the region $\mathcal{C}$. Using such eigenfunctions, the correlator
kernel could be expressed as
\begin{equation}
C\left(\bar{x}-\bar{y}\right)=\sum_{k\in\Upsilon}\int_{\mathbb{R}}u_{s,k}\left(\bar{x}\right)\frac{1+\tanh\left(\pi s\right)}{2}u_{s,k}\left(\bar{y}\right)^{\dagger}\,ds\,.\label{c3-corr-spec}
\end{equation}
Using expressions \eqref{c3-mh-fer} and \eqref{c3-ec-mh-con}, we
also have that
\begin{eqnarray}
K & \!\!\!=\!\!\! & \int_{\mathcal{C}\times\mathcal{C}}d^{d-1}x\,d^{d-1}y\,\psi^{\dagger}\left(\bar{x}\right)\mathcal{K}\left(\bar{x},\bar{y}\right)\psi\left(\bar{y}\right)\,,\label{c3-mh-ker}\\
\mathcal{K}\left(\bar{x},\bar{y}\right) & \!\!\!=\!\!\! & \sum_{k\in\Upsilon}\int_{\mathbb{R}}u_{s,k}\left(\bar{x}\right)\,2\pi s\,u_{s,k}\left(\bar{y}\right)^{\dagger}\,ds\,.\label{c3-mh-ker-diag}
\end{eqnarray}
It is important to remark that the modular Hamiltonian \eqref{c3-mh-ker}
has to be understood in the sense of modular evolution according to
formula \eqref{c3-con-modev}. In fact, for any operator $O$ belonging
to the algebra $\mathcal{A}\left(D\left(\mathcal{C}\right)\right)$,
the modular evolution \eqref{c3-con-modev} is given by the usual
Heisenberg equation
\begin{equation}
\partial_{t}O\left(t\right)=i\left[K,O\left(t\right)\right]\,,\label{c3-mod-ev-fer}
\end{equation}
where $O\left(t\right):=\sigma_{t}\left(O\right)$. In particular, for an operator of the form $O_{f}:=\int_{\mathcal{C}}\psi^{\dagger}\left(\bar{x}\right)f\left(\bar{x}\right)\,d^{d-1}x$ with $\mathrm{sup}\left(f\right)\subset\mathcal{C}$, equation \eqref{c3-mod-ev-fer} implies

\begin{eqnarray}
O_{f}\left(t\right) & \!\!\!=\!\!\! & \int_{\mathcal{C}}\psi^{\dagger}\left(\bar{x}\right)f\left(\bar{x};t\right)\,d^{d-1}x\,,\\
\partial_{t}f\left(\bar{x};t\right) & \!\!\!=\!\!\! & i\int_{\mathcal{C}}\mathcal{K}\left(\bar{x},\bar{y}\right)f\left(\bar{y};t\right)\,d^{d-1}y\,.
\end{eqnarray}
We define the operators 
\begin{eqnarray}
\tilde{\psi}_{s,k}^{\dagger}:=\int_{\mathcal{C}}\psi^{\dagger}\left(\bar{x}\right)u_{s,k}\left(\bar{x}\right)\,d^{d-1}x & \textrm{ and } & \tilde{\psi}_{s,k}:=\int_{\mathcal{C}}u_{s,k}\left(\bar{x}\right)^{\dagger}\psi\left(\bar{x}\right)\,d^{d-1}x\,,\label{c3-field-s}
\end{eqnarray}
which satisfy the CAR relations\foreignlanguage{english}{
\begin{equation}
\left\{ \tilde{\psi}_{s,k},\tilde{\psi}_{s',k'}^{\dagger}\right\} =\delta_{k,k'}\delta\left(s-s'\right)\,,\quad\left\{ \tilde{\psi}_{s,k},\tilde{\psi}_{s',k'}\right\} =\left\{ \tilde{\psi}_{s,k}^{\dagger},\tilde{\psi}_{s',k'}^{\dagger}\right\} =0\,.\label{c3-car-fer-mod}
\end{equation}
}Modes \eqref{c3-field-s} diagonalize the modular Hamiltonian in
the sense that they evolve under the modular flow accordingly to
\begin{eqnarray}
\sigma_{t}\left(\tilde{\psi}_{s,k}^{\dagger}\right)=\mathrm{e}^{i2\pi st}\tilde{\psi}_{s,k}^{\dagger} & \textrm{ and } & \sigma_{t}\left(\tilde{\psi}_{s,k}\right)=\mathrm{e}^{-i2\pi st}\tilde{\psi}_{s,k}\,.\label{c3-ops-ev}
\end{eqnarray}
Moreover, it is not difficult to show that the formulas \eqref{c3-ops-ev}
imply the KMS-condition for the two-point correlator
\begin{equation}
\left\langle \sigma_{t-i}\left(\tilde{\psi}_{s,k}\right)\tilde{\psi}_{s',k'}^{\dagger}\right\rangle =\left\langle \tilde{\psi}_{s',k'}^{\dagger}\sigma_{t}\left(\tilde{\psi}_{s,k}\right)\right\rangle \,,\quad\forall\bar{x},\bar{y}\in\mathcal{C},\,\forall s,s'\in\mathbb{R}\,.\label{c3-kms-fer}
\end{equation}
Since the set of operators $\{\psi_{s,k}\left(\bar{x}\right),\psi_{s,k}^{\dagger}\left(\bar{x}\right)\,:\,s\in\mathbb{R},\,k\in\Upsilon\}$
form a ``complete basis'' of the algebra of operators of the region
$\mathcal{C}$ and the multipoint correlators are determined by the
two-point correlators, we have that equations \eqref{c3-field-s}
and \eqref{c3-ops-ev} define rigorously the modular flow in the continuum
QFT. Certainly, the information about the state and the region (algebra)
in the above equations is completely codified in the eigenfunctions $u_{s,k}\left(\bar{x}\right)$. 

The same analysis could be carried out for the EE. According to \eqref{c3-fer-ee},
in the continuum QFT, we shall have
\begin{equation}
S\left(\mathcal{C}\right)=\mathrm{Tr}\left(\mathcal{S}\left(\bar{x},\bar{y}\right)\right)=\sum_{k\in\Upsilon}\int_{\mathbb{R}}ds\int_{\mathcal{C}\times\mathcal{C}}d^{d-1}x\,d^{d-1}y\,u_{s,k}\left(\bar{x}\right)^{\dagger}\mathcal{S}\left(\bar{x},\bar{y}\right)u_{s,k}\left(\bar{y}\right)\,,\label{c3-ee-fer-qft}
\end{equation}
where
\begin{eqnarray}
\mathcal{S}\left(\bar{x},\bar{y}\right) & \!\!\!=\!\!\! & \sum_{k\in\Upsilon}\int_{\mathbb{R}}ds\,u_{s,k}\left(\bar{x}\right)g\left(s\right)u_{s,k}\left(\bar{y}\right)^{\dagger}\,,\label{c3-ker-ee}\\
g\left(s\right) & \!\!\!=\!\!\! & \pi s\,\tanh(\pi s)+\log\left(\text{sech}(\pi s)\right)-\log\left(2\right)\,,
\end{eqnarray}
is the spectral resolution of $\left(-1\right)\left[\left(1-C\right)\log\left(1-C\right)+C\log C\right]$
according to \eqref{c3-corr-spec}. Performing the integral over $d^{d-1}y$
in equation \eqref{c3-ee-fer-qft} we get
\begin{equation}
S\left(\mathcal{C}\right)= \int_\mathcal{C} d^{d-1}x\, \mathcal{S}\left(\bar{x},\bar{x}\right)  =\sum_{k\in\Upsilon}\int_{\mathbb{R}}ds\,g\left(s\right)\int_{\mathcal{C}}d^{d-1}x\,u_{s,k}\left(\bar{x}\right)^{\dagger}u_{s,k}\left(\bar{x}\right)\,.\label{c3-ee-qft}
\end{equation}
If we perform the integral over $d^{d-1}x$ we obtain a $\delta\left(0\right)$
factor, giving a divergent EE. This happens because, when we have
moved to the continuum QFT, we have lost the original lattice cutoff
$\delta$. In order to regularize the entropy \eqref{c3-ee-qft},
we introduce a new geometrical cutoff which is given by changing the
integration region $\mathcal{C}$ by the ``regularized'' region
$\mathcal{C}_{\epsilon}\subset\mathcal{C}$ defined to contain all
the points of $\mathcal{C}$ that are separated by a distance greater
than $\epsilon$ from the boundary $\partial\mathcal{C}$, i.e. $\mathcal{C}_{\epsilon}:=\left\{ \bar{x}\in\mathcal{C}\,:\,\mathrm{dist}\left(\bar{x},\partial\mathcal{C}\right)>\epsilon\right\} $.
For example, for a $d=2$ QFT and a finite interval $\mathcal{C}:=\left(a,b\right)$,
we have that $\mathcal{C}_{\epsilon}=\left(a+\epsilon,b-\epsilon\right)$.
Then, we have defined the following ``geometrically'' regularized
(and finite) EE
\begin{equation}
S_{\epsilon}\left(\mathcal{C}\right)=\sum_{k\in\Upsilon}\int_{\mathbb{R}}ds\,g\left(s\right)\int_{\mathcal{C}_{\epsilon}}d^{d-1}x\,u_{s,k}\left(\bar{x}\right)^{\dagger}u_{s,k}\left(\bar{x}\right)\,.
\end{equation}
The interesting fact arises when we consider the MI for two non-intersecting
regions $\mathcal{C}_{1},\mathcal{C}_{2}\subset\Sigma_{0}$. In this
case, we have that
\begin{equation}
S_{\epsilon}\left(\mathcal{C}_{1}\right)+S_{\epsilon}\left(\mathcal{C}_{2}\right)-S_{\epsilon}\left(\mathcal{C}_{1}\cup\mathcal{C}_{2}\right)\underset{\epsilon\rightarrow0^{+}}{\longrightarrow}I\left(\mathcal{C}_{1},\mathcal{C}_{2}\right)\,.\label{c3-der-mi}
\end{equation}
This is because the divergences of each of the terms are local and
extensive along the boundaries of the regions and they cancel out.
In fact, using the kernel \eqref{c3-ker-ee} of the regions $\mathcal{C}_{1},\mathcal{C}_{2}$
and $\mathcal{C}_{1}\cup\mathcal{C}_{2}$, we can define the kernel
\begin{equation}
\mathcal{I}_{\mathcal{C}_{1},\mathcal{C}_{2}}\left(\bar{x},\bar{y}\right):=\mathcal{S}_{\mathcal{C}_{1}}\left(\bar{x},\bar{y}\right)\Theta_{\mathcal{C}_{1}}\left(\bar{y}\right)+\mathcal{S}_{\mathcal{C}_{2}}\left(\bar{x},\bar{y}\right)\Theta_{\mathcal{C}_{2}}\left(\bar{y}\right)-\mathcal{S}_{\mathcal{C}_{1}\cup\mathcal{C}_{2}}\left(\bar{x},\bar{y}\right)\,,\label{c3-ker-mi}
\end{equation}
where $\Theta_{\mathcal{C}_{j}}$ is the characteristic function of
the region $\mathcal{C}_{j}$. Equation \eqref{c3-der-mi} says that
\eqref{c3-ker-mi} is a trace-class operator in the Hilbert space $\mathfrak{H}=L^{2}\left(\mathcal{C}_{1}\cup\mathcal{C}_{2}\right)\otimes\mathbb{C}^{S}$
and that
\begin{equation}
\mathrm{Tr}_{\mathfrak{H}}\left(\mathcal{I}_{\mathcal{C}_{1},\mathcal{C}_{2}}\right)=I\left(\mathcal{C}_{1},\mathcal{C}_{2}\right)\,.
\end{equation}
To summarize, we have shown how lattice formulas (for free fields)
are still valid in the continuum QFT whenever they are interpreted
appropriately. In fact, the eigenfunctions of the correlator kernel
$C\left(\bar{x}-\bar{y}\right)$ restricted to the region $\mathcal{C}$,
contain all the information of the modular flow and the mutual information
for two non-intersecting spatial regions on free field theory.


\renewcommand\chaptername{Chapter}
\selectlanguage{english}

\chapter{Relative entropy for coherent states\label{RE_CS}}

In this chapter, we make a rigorous computation of the RE between
the vacuum state and a coherent state for a free scalar in the framework
of AQFT. We study the case of the Rindler Wedge. Previous calculations,
including path integral methods and computations from the lattice,
give a result for this RE involving integrals of expectation values
of the energy-momentum stress tensor along the considered region.
However, the stress tensor is in general non-unique. That means that,
if we start with some stress tensor, we can “improve” it by adding
a conserved term without modifying the Poincaré charges. On the other
hand, the presence of such an improving term affects the naive result
for the RE by adding a non-vanishing boundary contribution along the
entanglement surface. In other words, there is an ambiguity in the
usual formula for the RE coming from the non-uniqueness of the stress
tensor. The main motivation of this section is to solve this puzzle.
We first show that all choices of stress tensor, except the canonical
one, are not allowed by positivity and monotonicity of the RE. Then,
we fully compute the RE between the vacuum and a coherent state in
the framework of AQFT using the Araki formula and the techniques of
modular theory explained in section \ref{c2-Sec_entro_gener}. Both
results coincide and give the usual expression for the RE calculated
with the canonical stress tensor.

The chapter is organized as follows. First, in section \ref{c4-sec_ambiguitis},
we show how ambiguities can arise when one computes the RE making
naive calculations. We show, in the case of the Rindler wedge, that
such ambiguities are related to the choice of an improvement term
in the stress-tensor. Next, in section \ref{c4-sec-bounds}, we explain
how to fix such ambiguities imposing bounds which come from the positivity
and monotonicity of the RE. In the rest of the chapter, we compute
the RE using Araki formula. In section \ref{c4-sec-aqft_scalar},
we briefly review the algebraic formulation of the free real scalar
field. Because of a forthcoming necessity, we consider two different
approaches. The first one is the usual approach, where we define a
net of algebras associated with spacetime regions. The second one
consists in defining the local algebras associated with spatial sets
belonging to a common Cauchy surface. We also explain how both approaches
are related. The reader who is familiar with these concepts may skip
this section and go directly to \ref{c4-sec-re_sym}. There we study
general aspects concerning the RE for coherent states which apply
to any region, and next, in section \ref{c4-sec-re}, we explicitly
compute the proposed RE for the Rindler wedge algebra. We study separately,
the (trivial) case when the coherent state is made of a unitary operator
belonging to the wedge algebra, and the more interesting (and also
more difficult) case, when the coherent state has a non-vanishing
density along the entanglement surface. We provide a complete mathematical
rigorous proof of all the results. The proof of some theorems and
some tedious, but straightforward calculations, are placed into appendices.

This chapter is based on the article \cite{Casini:2019qst}.

\section{Ambiguities in the heuristic formula for relative entropy\label{c4-sec_ambiguitis}}

Here we show how ambiguities can arise when we try to compute entanglement
quantities making naive calculations. To be concrete, we start with
a QFT in the vacuum representation, and we want to compute the relative
entropy $S_{\mathcal{W}_{R}}\left(\phi\mid\omega_{0}\right)$ between
the vacuum state $\omega_{0}$ and any other global pure state $\phi$
for the right Rindler wedge region $\mathcal{W}_{R}$. In actual computations,
it is customary and useful to assume a cutoff model, such as a lattice,
and proceed to the computation taking the continuum limit as a final
step. In general, we expect the quantity computed belongs to the continuum
theory as far as the result does not depend on the regularization.
In the presence of a regularization, we can assume that the local
algebras $\mathcal{A}\left(\mathcal{W}_{R}\right)$ and $\mathcal{A}\left(\mathcal{W}'_{R}\right)$
are statistically independent
\begin{equation}
\mathcal{B}\left(\mathcal{H}_{0}\right)\cong\mathcal{A}\left(\mathcal{W}'_{R}\right)\otimes\mathcal{A}\left(\mathcal{W}_{R}\right)\,,
\end{equation}
and furthermore they are type I vN algebras. Then, we are able to
use all the formulas derived in section \ref{c2_sec_finite}. In particular,
if $\rho_{0}^{R}$ and $\rho_{\phi}^{R}$ are the statistical operators
of the restricted states $\left.\omega_{0}\right|_{\mathcal{A}\left(\mathcal{W}_{R}\right)}$
and $\left.\phi\right|_{\mathcal{A}\left(\mathcal{W}_{R}\right)}$,
then we can use the expressions \eqref{c2_dh-ds} and \eqref{c2_dh_and_ds},
namely
\begin{equation}
S_{\mathcal{W}_{R}}\left(\phi\mid\omega_{0}\right)=\Delta\left\langle K_{0}^{R}\right\rangle -\Delta S_{\mathcal{W}_{R}}\,,\label{c4-dh_ds}
\end{equation}
where 
\begin{eqnarray}
\Delta\left\langle K_{0}^{R}\right\rangle  & = & \phi\left(K_{0}^{R}\right)-\omega_{0}\left(K_{0}^{R}\right)\,,\\
\Delta S_{\mathcal{W}_{R}} & = & S_{\mathcal{W}_{R}}\left(\phi\right)-S_{\mathcal{W}_{R}}\left(\omega_{0}\right)\,,\label{c4-ds}
\end{eqnarray}
and $K_{0}^{R}=-\log\left(\rho_{0}^{R}\right)$ is the ``inner''
modular Hamiltonian of the vacuum vector $\left|0\right\rangle $.

One has to be very careful when using expressions like \eqref{c4-dh_ds}
in QFT. The l.h.s is a well-defined quantity in the continuum QFT,
but a mathematically rigorous definition of the continuum limit of
the two terms on the r.h.s. of \eqref{c4-dh_ds} has not been worked
out in the literature yet. In fact, the EE entropy and the ``inner''
modular Hamiltonian $K_{0}$ are not well-defined quantities of the
continuum QFT. As we have explained in section \ref{c3-sec-bw}, the
``inner'' modular Hamiltonian $K_{0}^{R}$ could be obtained cutting
the ``full'' modular Hamiltonian $K_{0}$ into two pieces
\begin{equation}
K_{0}=K_{0}^{R}-K_{0}^{L}\,,
\end{equation}
where $K_{0}^{R}$ belongs to $\mathcal{A}\left(\mathcal{W}_{R}\right)$
and $K_{0}^{L}$ belongs to $\mathcal{A}\left(\mathcal{W}_{L}\right)$.
In the case of the right Rindler wedge, it is given by \eqref{c3-mh-bad}
\begin{equation}
K_{0}^{R}=2\pi\int_{x^{1}>0}\negthickspace d^{d-1}x\,x^{1}\,T_{00}(x)\,.\label{c4-half}
\end{equation}
This is not a well-defined operator in the Hilbert space because its
fluctuation $\left\langle 0\right|\left(K_{0}^{R}\right)^{2}\left|0\right\rangle $
diverges. However, \eqref{c4-half} is a well-defined sesquilinear
form, and hence, expectation values as $\Delta\langle K_{0}^{R}\rangle$
can still be computed. Another more important issue is that cutting
the ``full'' modular Hamiltonian in two pieces generates ambiguities.
We are allowed for example to add field operators localized at the
boundary such that $K_{0}^{R}$ has still the same localization and
commutation relations with operators inside $\mathcal{W}_{R}$. Another
view of the same problem is that, hidden in expression \eqref{c4-half},
there is an ambiguity related to the non-uniqueness of the stress
tensor. For example, for the free real scalar field, starting from
the canonical stress tensor
\begin{equation}
T_{\mu\nu}^{can}=:\partial_{\mu}\varphi\partial_{\nu}\varphi-\frac{1}{2}\eta_{\mu\nu}\left(\partial_{\sigma}\varphi\partial^{\sigma}\varphi-m^{2}\varphi^{2}\right):\,,\label{c4-stress}
\end{equation}
we can add an ``improving term'' to obtain a new stress tensor
\begin{equation}
T_{\mu\nu}=T_{\mu\nu}^{can}+\frac{\lambda}{2\pi}\,(\partial_{\mu}\partial_{\nu}-g_{\mu\nu}\partial^{2}):\varphi^{2}:\,,\label{c4-stress_imp}
\end{equation}
where $\lambda$ could be, in principle, any real number. The Poincaré
generators obtained from \eqref{c4-stress_imp} equal the ones obtained
from \eqref{c4-stress}, since both expressions differ in a boundary
term which vanishes when the integration region is the whole space.
However, the expression \eqref{c4-half} for $K_{0}^{R}$ involves
an integral in a semi-infinite region, and hence, the presence of
an improving term adds a non zero extra boundary term to the result,
\begin{equation}
K_{R}\rightarrow K_{R}+\lambda\int_{x^{1}=0}\negthickspace d^{d-2}x\,:\varphi^{2}\left(x\right):\,.\label{c4-hamil_improved}
\end{equation}
This is essentially the only boundary term we can add with the correct
dimensions, and that does not require a dimensionful coefficient with
negative dimensions. This can have non-zero expectation values for
certain states and makes the definition of $\Delta\langle K_{0}^{R}\rangle$
ambiguous.

\begin{figure}[h]
\centering
\includegraphics[width=10cm]{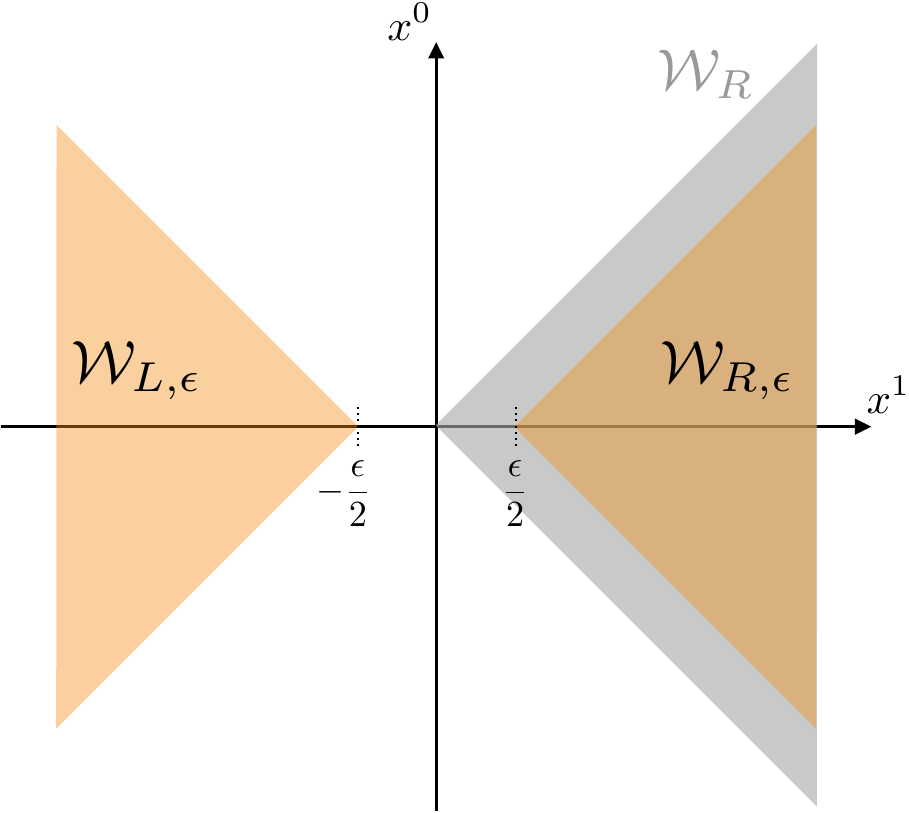}\caption{\label{c4_fig_rindler}Regularized wedge regions for the computation
of $\Delta S_{\mathcal{W}_{R}}$ according to formula \eqref{c4-ds-reg}.}
\end{figure}

Since the relative entropy is well-defined, this ambiguity must be
compensated by another one in the definition of $\Delta S_{\mathcal{W}_{R}}$
in \eqref{c4-ds}. This is the subtraction of two divergent quantities,
and again, we do not have a mathematically rigorous definition in
the continuum QFT. We can make this definition unambiguous in a natural
way by using the universal regularization of EE explained in section
\ref{c3-sec-univee}. Using such a universal prescription, one can
define
\begin{equation}
\Delta S_{\mathcal{W}_{R}}=\lim_{\epsilon\rightarrow0^{+}}\left[S_{reg}^{\phi}\left(\mathcal{W}_{R};\epsilon,\eta^{1}\right)-S_{reg}^{\omega_{0}}\left(\mathcal{W}_{R};\epsilon,\eta^{1}\right)\right]\,,\label{c4-ds-reg}
\end{equation}
where we have chosen the unit exterior vector $\eta^{1}:=\left(0,-1,\bar{0}\right)$.
In fact, according to equation \eqref{c3-re_ee}, each term of \eqref{c4-ds-reg}
is a MI between the regions $\mathcal{W}_{R,\epsilon}:=\left\{ x\in\mathbb{R}^{d}\,:\,x^{1}>\left|x^{0}\right|+\frac{\epsilon}{2}\right\} $
and $\mathcal{W}_{L,\epsilon}:=\mathcal{W}_{R,-\epsilon}'$ (see figure
\ref{c4_fig_rindler}). One expects that the local ambiguities due
to the cutoff in both terms of \eqref{c4-ds-reg} cancel out, giving
a finite $\Delta S_{\mathcal{W}_{R}}$. Defining $\Delta S_{\mathcal{W}_{R}}$
rigorously through \eqref{c4-ds-reg}, then $\Delta\langle K_{0}^{R}\rangle$
is also unambiguously defined through
\begin{equation}
\Delta\langle K_{0}^{R}\rangle=S_{\mathcal{W}_{R}}\left(\phi\mid\omega_{0}\right)+\Delta S_{\mathcal{W}_{R}}\,.
\end{equation}
Then, the question that arises is whether this definition agrees with
the expectation value of \eqref{c4-half}. In such a case, boundary
terms in this expression should be automatically fixed. In particular,
we should be able to study which value of the improvement term is
the correct one for the free real scalar field in \eqref{c4-stress_imp}.

\begin{figure}[h]
\centering
\includegraphics[width=12cm]{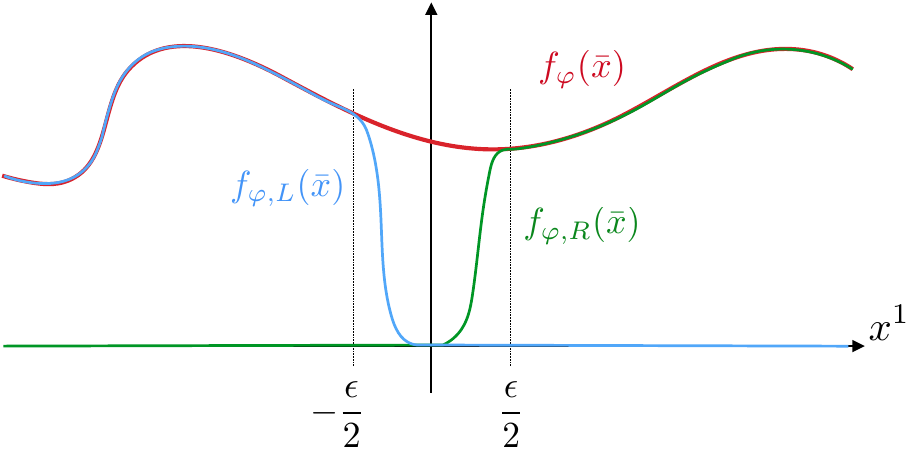}\caption{\label{c4_fig_partir_func}For the purpose of the algebra $\mathcal{A}\left(\mathcal{W}_{R,\epsilon}\cup\mathcal{W}_{L,\epsilon}\right)$,
the test function $f_{\varphi}$ can be replaced by $f_{\varphi,R}+f_{\varphi,L}$.}
\end{figure}

In order to settle this issue, we analyze the relative entropy between
a coherent state and the vacuum for a free real scalar field in the
Rindler wedge. A coherent state $\phi$ is a global pure state formed
out by acting on the vacuum vector with a unitary operator that is
the exponential of the smeared field, i.e. 
\begin{equation}
\phi\left(\cdot\right)=\left\langle \Phi\right|\cdot\left|\Phi\right\rangle \quad\textrm{ with }\quad\left|\Phi\right\rangle =\mathrm{e}^{i\int d^{d-1}x\,\left[\varphi\left(\bar{x}\right)f_{\varphi}\left(\bar{x}\right)+\pi\left(\bar{x}\right)f_{\pi}\left(\bar{x}\right)\right]}\left|0\right\rangle \,,\label{c4-coh_intro}
\end{equation}
where $\varphi\left(\bar{x}\right):=\varphi\left(0,\bar{x}\right)$,
$\pi\left(\bar{x}\right):=\partial_{0}\varphi\left(0,\bar{x}\right)$,
and $f_{\varphi},f_{\pi}\in\mathcal{S}\left(\mathbb{R}^{d-1},\mathbb{R}\right)$.\footnote{$\mathcal{S}\left(\mathbb{R}^{n},\mathbb{R}\right)$ denotes the Schwartz
space of real-valued, smooth and exponentially decreasing functions
at infinity.} For the purpose of the definition \eqref{c4-ds-reg}, we can represent
the same state $\phi$ with a different vector $|\tilde{\Phi}\rangle=U_{L}U_{R}\left|0\right\rangle $
where $U_{R},U_{L}$ are unitaries, $U_{R}\in\mathcal{A}\left(\mathcal{W}_{R}\right)$
and $U_{L}\in\mathcal{A}\left(\mathcal{W}_{L}\right)$. Indeed, we
can replace each of the smooth functions $f_{\varphi}\left(\bar{x}\right)$
and $f_{\pi}\left(\bar{x}\right)$ in \eqref{c4-coh_intro} by the
sum of two new smooth functions
\begin{equation}
f_{\varphi}\mapsto f_{\varphi,R}+f_{\varphi,L}\,,\quad f_{\pi}\mapsto f_{\pi,R}+f_{\pi,L}\,,
\end{equation}
such that $f_{\varphi,R}\left(\bar{x}\right)=0$ for $x^{1}\leq0$,
$f_{\varphi,R}\left(\bar{x}\right)=f_{\varphi}\left(\bar{x}\right)$
for $x^{1}\geq\frac{\epsilon}{2}$, $f_{\varphi,L}\left(\bar{x}\right)=0$
for $x^{1}\geq0$ and $f_{\varphi,L}\left(\bar{x}\right)=f_{\varphi}\left(\bar{x}\right)$
for $x^{1}\leq-\frac{\epsilon}{2}$ (see figure \ref{c4_fig_partir_func}).
Idem for $f_{\pi}$. Under these assumptions, the vector $|\tilde{\Phi}\rangle=U_{L}U_{R}\left|0\right\rangle $
defined by
\begin{equation}
U_{R}=\mathrm{e}^{i\int d^{d-1}x\,\left[\varphi\left(\bar{x}\right)f_{\varphi,R}\left(\bar{x}\right)+\pi\left(\bar{x}\right)f_{\pi,R}\left(\bar{x}\right)\right]}\quad\mathrm{and}\quad U_{L}=\mathrm{e}^{i\int d^{d-1}x\,\left[\varphi\left(\bar{x}\right)f_{\varphi,L}\left(\bar{x}\right)+\pi\left(\bar{x}\right)f_{\pi,L}\left(\bar{x}\right)\right]},
\end{equation}
is a vector representative of the state $\phi$ for the algebra $\mathcal{A}\left(\mathcal{W}_{R,\varepsilon}\cup\mathcal{W}_{L,\varepsilon}\right)$.
In fact, the above computation can be done because of the presence
of the finite corridor of width $\epsilon$. Moreover, we have that
the operator $U_{R}$ (resp. $U_{L}$) acts, by adjoint action, as
an automorphism of the algebra of the region $\mathcal{W}_{R,\epsilon}$
(resp. $\mathcal{W}_{L,\epsilon}$), and as the identity transformation
over the algebra of the region $\mathcal{W}_{L,\epsilon}$ (resp.
$\mathcal{W}_{R,\epsilon}$). Such automorphisms do not change the
MI, and according to definition \eqref{c4-ds-reg}, we automatically
have $\Delta S_{\mathcal{W}_{R}}=0$ for these states.

Hence, the question simplifies to see whether for coherent states
we have that
\begin{equation}
S_{\mathcal{W}_{R}}\left(\phi\mid\omega_{0}\right)=\Delta\langle K_{0}^{R}\rangle=2\pi\int_{x^{1}>0}\negthickspace d^{d-1}x\,x^{1}\,\left\langle \Phi\right|T_{00}\left(\bar{x}\right)\left|\Phi\right\rangle \,,\label{c4-cali}
\end{equation}
and which is the right improvement term of the stress tensor. Notice
that coherent states can change the expectation value of $:\varphi^{2}:$.
We answer this question in two independent ways:
\begin{enumerate}
\item Assuming that \eqref{c4-cali} is correct for some improvement, we
show, imposing bounds which come from the positivity and monotonicity
of the RE, that the only possibility is the canonical stress tensor,
i.e. $\lambda=0$ (section \ref{c4-sec-bounds}).
\item We explicitly compute the relative entropy using Araki formula, and
we show that the result \eqref{c4-cali} is correct for the canonical
stress tensor (sections \ref{c4-sec-aqft_scalar} and \ref{c4-sec-re}).
\end{enumerate}

\section{Bounds on boundary terms in the RE\label{c4-sec-bounds}}

According to the discussion above, there is an ambiguity in the expression
\eqref{c4-cali} for the RE of a coherent state coming from the different
possible choices of an improving term for the stress energy-momentum
tensor. According to \eqref{c4-hamil_improved}, the relative entropy
could be written as the usual contribution with the canonical stress
tensor plus a boundary term coming from the improving 
\begin{equation}
S\left(\phi\mid\omega_{0}\right)=\lambda\int_{x_{1}=0}\negthickspace d^{d-2}x\,\left\langle \Phi\right|\varphi^{2}\left(\bar{x}\right)\left|\Phi\right\rangle +2\pi\int_{x_{1}>0}\negthickspace d^{d-1}x\,x^{1}\left\langle \Phi\right|T_{00}^{can}\left(\bar{x}\right)\left|\Phi\right\rangle \,.\label{c4-re_lambda}
\end{equation}
In this section, we assume that this formula is correct, and we show
that the only consistent choice is $\lambda=0$.

A general coherent state can be written as in \eqref{c4-coh_intro}
with $f_{\varphi},f_{\pi}\in\mathcal{S}\left(\mathbb{R}^{d-1},\mathbb{R}\right)$.
In this case, a straightforward computation from \eqref{c4-re_lambda}
gives 
\begin{equation}
\hspace{-0.2cm} S\left(\phi\mid\omega_{0}\right)=\lambda \! \int_{x_{1}=0} \!\!\!\!\!\!\! d^{d-2}x\,f_{\pi}\left(\bar{x}\right)^{2}+2\pi \! \int_{x_{1}>0} \!\!\!\!\!\!\! d^{d-1}x\,\frac{x^{1}}{2}\left(f_{\varphi}\left(\bar{x}\right)^{2}+\left(\nabla f_{\pi}\left(\bar{x}\right)\right)^{2}+m^{2}f_{\pi}\left(\bar{x}\right)^{2}\right) .\label{eq:re_coherent}
\end{equation}
Independently of the correct value for $\lambda$, if we want that
\eqref{c4-re_lambda} and \eqref{eq:re_coherent} represent real expressions
for a RE, they must satisfy all the properties known for a RE. In
particular we will concentrate on the positivity and the monotonicity
(see proposition \ref{c2_re_prop}). Therefore, the strategy we adopt
is to choose functions $f_{\varphi}$ and $f_{\pi}$ conveniently,
and impose positivity and the monotonicity on \eqref{eq:re_coherent}
in order to bound the allowed values for $\lambda$. In fact, we will
show that from the positivity we obtain $\lambda\geq0$, and from
the monotonicity we obtain $\lambda\leq0$, an hence, it must be that
$\lambda=0$. Then, we conclude that, if we assume that \eqref{c4-cali}
is the correct result for the RE, such an expression only holds for
the canonical stress-energy-momentum tensor \eqref{c4-stress}.\footnote{Conversely, it is almost immediate to see that expression \eqref{eq:re_coherent}
is positive and monotonic for $\lambda=0$.}

Before we continue, we make two simplifications. The first one, which
is obvious, is to take $f_{\varphi}\equiv0$ and denote $f:=f_{\pi}$.
The second one is to work in $d=1+1$ dimensions. The general result
for arbitrary dimension could be easily obtained from the former.

\subsection{Lower bound from positivity}

We start with the expression

\begin{equation}
S\left(\phi\mid\omega_{0}\right)=\lambda f\left(0\right)^{2}+\pi\int_{0}^{+\infty}dx\,x\left(f'\left(x\right)^{2}+m^{2}f\left(x\right)^{2}\right)\,,
\end{equation}
where $f\in\mathcal{S}\left(\mathbb{R},\mathbb{R}\right)$. Then,
the positivity of the RE means that

\begin{equation}
0\leq\lambda f\left(0\right)^{2}+\pi\int_{0}^{+\infty}dx\,xf'\left(x\right)^{2}+\pi\,m^{2}\int_{0}^{+\infty}dx\,xf\left(x\right)^{2}\,.
\end{equation}
By scaling the function $f(x)\rightarrow f(x/\beta)$ the first two terms of the r.h.s. are constant while the last one gets
multiplies by $\beta^{2}$. Hence, we can make the last term as small
as we want and simply take $m=0$ in the following. Taking $f$ such
that $f\left(0\right)\neq0$, we get 
\begin{equation}
0\leq\lambda+\pi\frac{\int_{0}^{+\infty}dx\,xf'\left(x\right)^{2}}{f\left(0\right)^{2}}\,.\label{c4-pos_ineq_2}
\end{equation}
Now, we introduce a convenient family of real functions $\left(f_{a}\right)_{a>0}\in\mathcal{S}\left(\mathbb{R},\mathbb{R}\right)$
given by 
\begin{equation}
f_{a}\left(x\right):=\log\left(\frac{x}{L}+a\right)\mathrm{e}^{-\frac{x}{L}}\,,\quad x\geq0\,,
\end{equation}
where $L>0$ is a dimensionful fixed constant.\footnote{The functions $f_{a}$ are smoothly extended to the whole real line.
Such an extension is guaranteed by a theorem due to Seeley \cite{seeley}.\label{c4-fn-c_inf_ext}} A straightforward computation shows that the integral in equation
\eqref{c4-pos_ineq_2} behaves as

\begin{equation}
\int_{0}^{+\infty}dx\,x\,f_{a}'\left(x\right)^{2}=-\log\left(a\right)+\mathcal{O}(a^{0})\,,\quad0<a\ll1\,.\label{c4-bound1}
\end{equation}
Then, replacing \eqref{c4-bound1} into \eqref{c4-pos_ineq_2} we get 
\begin{equation}
0\leq\lambda-\pi\frac{\log\left(a\right)+\mathcal{O}(a^{0})}{\log^{2}\left(a\right)}\,.
\end{equation}
Finally, taking the limit $a\rightarrow0^{+}$ we get the desired result 
\begin{equation}
\lambda\geq0 \, . \label{c4-lower_bound}
\end{equation}

\subsection{Upper bound from monotonicity}

\begin{figure}[h]
\centering
\includegraphics[width=11cm]{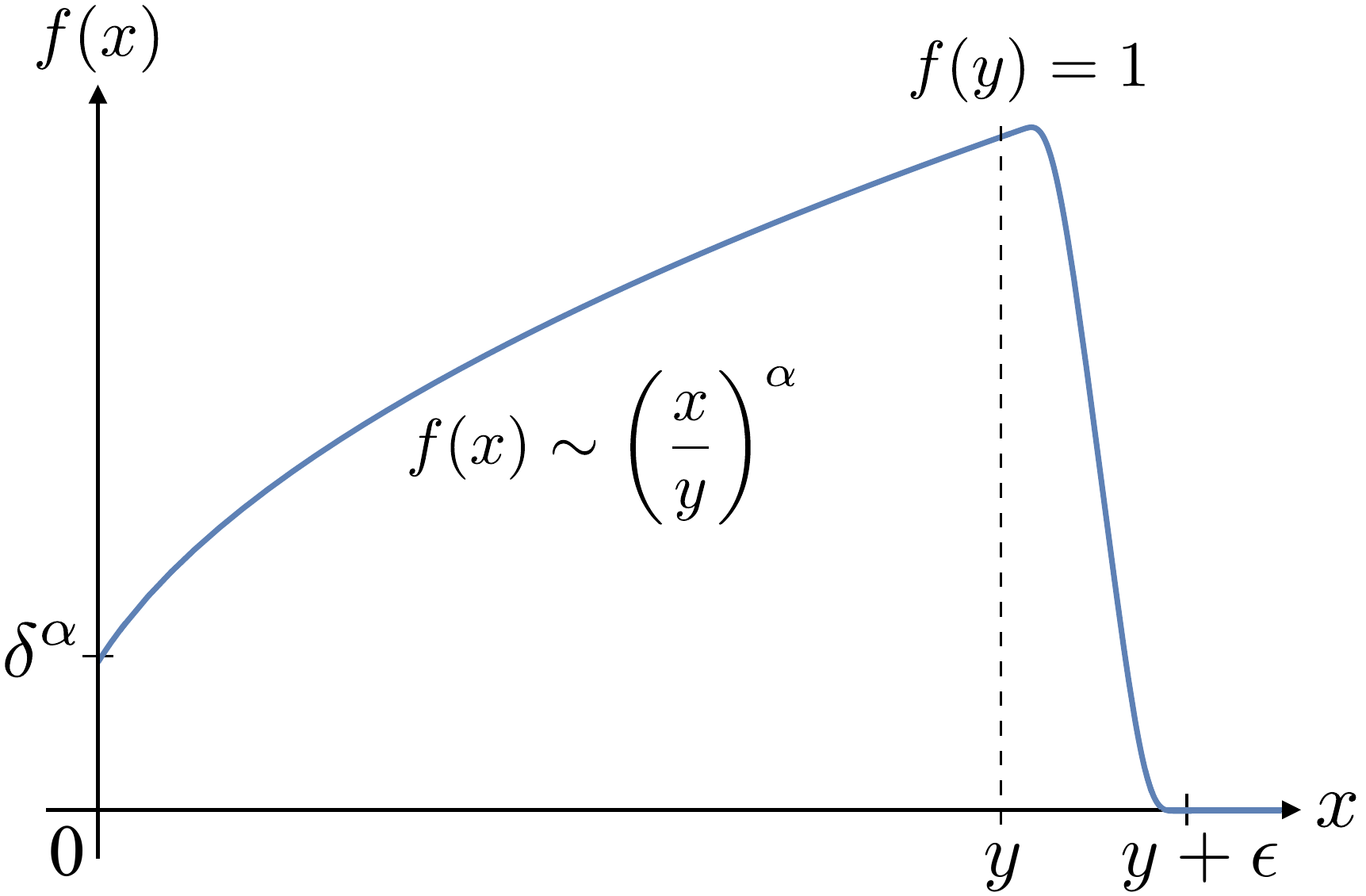}\caption{\label{c4_fig_mono}A schematic plot of the function $f=f_{\alpha,\delta,y,\epsilon}$. }
\end{figure}

Monotonicity for the case of the right wedge means 
\begin{equation}
S_{\mathcal{W}_{R,y}}\left(\phi\mid\omega_{0}\right)\geq S_{\mathcal{W}_{R,y'}}\left(\phi\mid\omega_{0}\right)\,,\quad\textrm{for any }y'\geq y\,,\label{c4-mon_re}
\end{equation}
where $\mathcal{W}_{R,y}:=\left\{ x\in\mathbb{R}^{d}\,:\,x^{1}>\left|x^{0}\right|+y\right\} $.
In fact, $\mathcal{W}_{R,y}$ is obtained applying a translation of
amount $y$, in the $x^{1}$ positive direction, to the right Rindler
wedge $\mathcal{W}_{R}$. In the following, we denote $S\left(y\right):=S_{\mathcal{W}_{R,y}}\left(\phi\mid\omega_{0}\right)$.
Then, we have the expressions 
\begin{eqnarray}
S\left(0\right)\!\!\! & = & \!\!\!\lambda f\left(0\right)^{2}+\pi\int_{0}^{+\infty}dx\,x\left(f'\left(x\right)^{2}+m^{2}f\left(x\right)^{2}\right)\,,\\
S\left(y\right)\!\!\! & = & \!\!\!\lambda f\left(y\right)^{2}+\pi\int_{y}^{+\infty}dx\,\left(x-y\right)\left(f'\left(x\right)^{2}+m^{2}f\left(x\right)^{2}\right)\,,
\end{eqnarray}
where $f\in\mathcal{S}\left(\mathbb{R},\mathbb{R}\right)$. We can
eliminate the mass terms by scaling the test function as in the previous
section. The monotonicity $S\left(0\right)\geq S\left(y\right)$ for
$y\geq0$ reads 
\begin{eqnarray}
\lambda\left(f\left(y\right)^{2}-f\left(0\right)^{2}\right) \!\!\! & \leq & \!\!\! \pi\int_{0}^{y}dx\,xf'\left(x\right)^{2}+\pi\,y\int_{y}^{+\infty}dx\,f'\left(x\right)^{2}\,.\label{c4-ineq_mon}
\end{eqnarray}
Now, we introduce a convenient family of functions parametrized by
the constants $\alpha\in\left(0,\frac{1}{2}\right),\delta\in\left(0,1\right),y>0,\epsilon>0$
and given by 
\begin{equation}
f_{\alpha,\delta,y,\epsilon}\left(x\right):=g_{\alpha,\delta,y}\left(x\right)\Theta_{y,\epsilon}\left(x\right)\,,\quad\textrm{for }x\geq0\,,\label{c4-func_mon}
\end{equation}
where 
\begin{equation}
g_{\alpha,\delta,y}\left(x\right):=\left(\frac{x}{y}\left(1-\delta\right)+\delta\right)^{\alpha}\,,
\end{equation}
and $\Theta_{y,\epsilon}$ is a smooth step function satisfying the
condition 
\begin{equation}
\Theta_{y,\epsilon}\left(x\right)=\begin{cases}
1 & x\leq y\,,\\
0 & x\geq y+\epsilon\,.
\end{cases}
\end{equation}
We introduce such a step function to ensure that $f_{\alpha,\delta,y,\epsilon}\in\mathcal{S}\left(\mathbb{R},\mathbb{R}\right)$
for any values of $\left(\alpha,\delta,y,\epsilon\right)$ in the
set specified above. The functions $f_{\alpha,\delta,y,\epsilon}$
are smoothly extended for $x\leq0$ (see footnote \ref{c4-fn-c_inf_ext}).
The behavior of this function is shown in figure \ref{c4_fig_mono}.

In particular, we use the following smooth step function
\begin{equation}
\Theta_{y,\epsilon}\left(x\right):=\left[1+\exp\left(-\frac{2\epsilon\left(x-y-\frac{\epsilon}{2}\right)}{\left(x-y-\frac{\epsilon}{2}\right)^{2}-\frac{\epsilon^{2}}{4}}\right)\right]^{-1}\,,\quad\textrm{if }y<x<y+\epsilon\,,
\end{equation}
which has the crucial property $\max_{x\in\mathbb{R}}\left|\Theta_{y,\epsilon}'\left(x\right)\right|=\frac{2}{\epsilon}$.
From now on, to lighten the notation, we do not write the cumbersome
subindices of the above functions. We evaluate the different terms
of \eqref{c4-ineq_mon} independently as

\begin{eqnarray}
f\left(y\right)^{2}-f\left(0\right)^{2}\!\!\! & = & \!\!\!1-\delta^{2\alpha}\,,
\end{eqnarray}
\begin{eqnarray}
\pi\int_{0}^{y}dx\,xf'\left(x\right)^{2}\!\!\! & = & \!\!\!\pi\frac{\alpha}{2}\left(\frac{2\alpha\delta-\delta^{2\alpha}}{1-2\alpha}+1\right)\,,
\end{eqnarray}
\begin{eqnarray}
\pi y\int_{y}^{+\infty}dx\,f'\left(x\right)^{2}\!\!\! & \leq & \!\!\!\pi y\int_{y}^{y+\epsilon}dx\left|g'\left(x\right)^{2}\Theta\left(x\right)^{2}\right|+\pi y\int_{y}^{y+\epsilon}dx\left|g\left(x\right)^{2}\Theta'\left(x\right)^{2}\right|\nonumber \\
 &  & \!\!\!+\pi y\int_{y}^{y+\epsilon}dx\left|2g'\left(x\right)g\left(x\right)\Theta\left(x\right)\Theta'\left(x\right)\right|\,.\label{c4-term_feo}
\end{eqnarray}
We deal with each term of \eqref{c4-term_feo} separately 
\begin{eqnarray}
\pi y \! \int_{y}^{y+\epsilon} \!\!\!\! dx\,\left|g'\left(x\right)^{2}\Theta\left(x\right)^{2}\right|\!\!\! & \leq & \!\!\!\pi y\! \int_{y}^{y+\epsilon} \!\!\!\! dx\,g'\left(x\right)^{2}=\frac{\pi\alpha^{2}\left(1-\delta\right)}{1-2\alpha}\left[1-\left(1+\frac{(1-\delta)\epsilon}{y}\right)^{2\alpha-1}\right]\nonumber \\
 & \underset{\epsilon\rightarrow+\infty}{\longrightarrow} & \!\!\!\frac{\pi\alpha^{2}\left(1-\delta\right)}{1-2\alpha}\,,
\end{eqnarray}
\begin{eqnarray}
\pi y\int_{y}^{y+\epsilon}dx\,\left|2g'\left(x\right)g\left(x\right)\Theta\left(x\right)\Theta'\left(x\right)\right|\!\!\! & \leq & \!\!\!\frac{2\pi y}{\epsilon}\int_{y}^{y+\epsilon}dx\,2g'\left(x\right)g\left(x\right)=\frac{2\pi y}{\epsilon}\left[g\left(y+\epsilon\right)^{2}-g\left(y\right)^{2}\right]\nonumber \\
 & = & \!\!\!\frac{2\pi y}{\epsilon}\left[\left(1+\frac{\left(1-\delta\right)\epsilon}{y}\right)^{2\alpha}-1\right]\underset{\epsilon\rightarrow+\infty}{\longrightarrow}0\,,
\end{eqnarray}
\begin{eqnarray}
\pi y\int_{y}^{y+\epsilon}dx\,\left|g\left(x\right)^{2}\Theta'\left(x\right)^{2}\right|\!\!\! & \leq & \!\!\!\frac{4\pi y}{\epsilon^{2}}\int_{y}^{y+\epsilon}dx\,g\left(x\right)^{2}\nonumber \\
 & = & \!\!\!\frac{4\pi y^{2}}{\left(1+2\alpha\right)\left(1-\delta\right)\epsilon^{2}}\left(\left(1+\frac{\left(1-\delta\right)\epsilon}{y}\right)^{2\alpha+1}-1\right) \hspace{1 cm} \nonumber \\
 & \underset{\epsilon\rightarrow+\infty}{\longrightarrow} & \!\!\!0\,,
\end{eqnarray}
where in the last steps of each computation we take the limit $\epsilon\rightarrow+\infty$.
It is valid to take this limit in the inequality since it must hold
for all $\epsilon>0$. Replacing these partial results on \eqref{c4-ineq_mon}
we arrive to
\begin{equation}
\lambda\left(1-\delta^{2\alpha}\right)\leq\pi\frac{\alpha}{2}\left(\frac{2\alpha\delta-\delta^{2\alpha}}{1-2\alpha}+1\right)+\frac{\pi\alpha^{2}\left(1-\delta\right)}{1-2\alpha}\,.
\end{equation}
Then, taking the limit $\delta\rightarrow0^{+}$ we get

\begin{equation}
\lambda\leq\pi\frac{\alpha}{2}+\pi\frac{\alpha^{2}}{1-2\alpha}\,,
\end{equation}
and finally, taking $\alpha\rightarrow0^{+}$ we arrive to the desired
result 
\begin{equation}
\lambda\leq0.\label{c4-upper_bound}
\end{equation}

\section{Algebraic theory of the free real scalar field\label{c4-sec-aqft_scalar}}

In this section, we briefly review the algebraic theory of the free
real scalar field. Concretely, we show explicitly the definition of
the net of local algebras, satisfying all the axioms of the definitions
\ref{c1_def_aqft} and \ref{c1_def_vacuum}. The material of this
section is broadly discussed in \cite{araki63,araki64,horuzhy,Guido:2008}.
For our convenience, we introduce our own notation, and we develop
some relations between the concepts, which, to our opinion, were not
very clearly discussed in these references.

\subsection{Local algebras for spacetime regions\label{c4-algebras-st}}

The algebraic theory of the free real scalar field is defined as a
net of vN algebras acting on the Fock Hilbert space $\mathcal{H}_{0}$.
In fact, this net defines the vacuum representation of the theory
according to definition \ref{c1_def_vacuum}. The Hilbert space $\mathcal{H}_{0}$
is constructed as the symmetric tensor product of the one-particle
Hilbert space. To describe it properly we introduce the following
three vector spaces.

\paragraph*{The space of test functions.}

The space of test functions is the Schwartz space $\mathcal{S}\left(\mathbb{R}^{d},\mathbb{R}\right)$
of real, smooth and exponentially decreasing functions at infinity.
This space carries naturally a representation of the restricted Poincaré
group $\mathcal{P}_{+}^{\uparrow}$ given by $f\mapsto f_{\left(\Lambda,a\right)}$,
with $f_{\left(\Lambda,a\right)}\left(x\right):=f\left(\Lambda^{-1}\left(x-a\right)\right)$
for any $\left(\Lambda,a\right)\in\mathcal{P}_{+}^{\uparrow}$.

\paragraph*{The one particle Hilbert space.}

The Hilbert space $\mathfrak{H}$ of one-particle states of mass $m>0$
and zero spin is made up of the square-integrable functions on the
mass shell hyperboloid $H_{m}:=\left\{ p\in\mathbb{R}^{d}\,:\,p^{2}=m^{2}\:,\,p^{0}>0\right\} $
with the Poincaré invariant measure $d\mu(p):=\Theta(p^{0})\delta(p^{2}-m^{2})d^{d}p$.
It can be realized as 
\begin{eqnarray}
\mathfrak{H} & \!\!\!=\!\!\! & L^{2}\left(\mathbb{R}^{d-1},\frac{d^{d-1}p}{2\omega\left(\bar{p}\right)}\right)\:\textrm{,}\\
\left\langle \boldsymbol{f}\!\mid\!\boldsymbol{g}\right\rangle _{\mathrm{1p}} & \!\!\!=\!\!\! & \int_{\mathbb{R}^{d-1}}\frac{d^{d-1}p}{2\omega\left(\bar{p}\right)}\boldsymbol{f}\left(\bar{p}\right)^{*}\boldsymbol{g}\left(\bar{p}\right)\,,
\end{eqnarray}
where $p^{0}=\sqrt{\bar{p}^{2}+m^{2}}=:\omega\left(\bar{p}\right)$.\footnote{We denote by $\left\langle \cdot\!\mid\!\cdot\right\rangle _{\mathrm{1p}}$
the inner product in the one-particle Hilbert space $\mathfrak{H}$,
whereas $\left\langle \cdot\!\mid\!\cdot\right\rangle $ denotes the
inner product in the Fock Hilbert space $\mathcal{H}_{0}$.} Such a space carries a unitary representation of $\mathcal{P}_{+}^{\uparrow}$
given by $\left(u\left(\Lambda,a\right)\boldsymbol{f}\right)\left(p\right):=\mathrm{e}^{ip\cdot a}\boldsymbol{f}\left(\Lambda^{-1}p\right)$
for any $\boldsymbol{f}\in\mathfrak{H}$ and $\left(\Lambda,a\right)\in\mathcal{P}_{+}^{\uparrow}$.

\paragraph*{The Fock Hilbert space.}

The Fock Hilbert space $\mathcal{H}_{0}$ is the direct sum of the
symmetric tensor powers of the one-particle Hilbert space $\mathfrak{H}$
\begin{equation}
\mathcal{H}_{0}:=\bigoplus_{n=0}^{\infty}\mathfrak{H}^{\otimes n,sym}\,.
\end{equation}
For each $\boldsymbol{h}\in\mathfrak{H}$, the creation and annihilation
operators $A^{*}\left(\boldsymbol{h}\right)$ and $A\left(\boldsymbol{h}\right)$
act on $\mathcal{H}$ as usual
\begin{eqnarray}
A^{*}\left(\boldsymbol{h}\right)\mathrm{sym}\left(\boldsymbol{h}_{1}\otimes\cdots\otimes\boldsymbol{h}_{n}\right) & \!\!\!:=\!\!\! & \sqrt{n+1}\,\mathrm{sym}\left(\boldsymbol{h}\otimes\boldsymbol{h}_{1}\otimes\cdots\otimes\boldsymbol{h}_{n}\right),\\
A\left(\boldsymbol{h}\right)\mathrm{sym}\left(\boldsymbol{h}_{1}\otimes\cdots\otimes\boldsymbol{h}_{n}\right) & \!\!\!:=\!\!\! & \frac{1}{\sqrt{n}}\sum_{j=1}^{n}\left\langle \boldsymbol{h}_{j}\!\mid\!\boldsymbol{h}\right\rangle _{\mathrm{1p}}\mathrm{sym}\left(\boldsymbol{h}_{1}\otimes\cdots\otimes\widehat{\boldsymbol{h}_{j}}\otimes\cdots\otimes\boldsymbol{h}_{n}\right), \hspace{.9cm} \label{aniquilacion}
\end{eqnarray}
where $\widehat{\boldsymbol{h}_{j}}$ means that the vector is excluded from the tensor product on the r.h.s. of \eqref{aniquilacion}. The Fock space naturally inherits from $\mathfrak{H}$ a unitary representation
of $\mathcal{P}_{+}^{\uparrow}$, which is denoted by $U\left(\Lambda,a\right)$.
According to that, there is a unique (up to a phase) $\mathcal{P}_{+}^{\uparrow}$-invariant
vector denoted by $\left|0\right\rangle :=\boldsymbol{1}\in\mathfrak{H}^{\otimes0}$,
which is called the \textit{vacuum vector.} For each $\boldsymbol{h}\in\mathfrak{H}$,
the normalized vector 
\begin{equation}
\mathrm{e}^{\boldsymbol{h}}:=\mathrm{e}^{-\frac{1}{2}\left\Vert \boldsymbol{h}\right\Vert _{1p}^{2}}\sum_{n=0}^{\infty}\frac{\boldsymbol{h}^{\otimes n}}{\sqrt{n!}}\in\mathcal{H}_{0}\,,
\end{equation}
is called \textit{coherent vector}, and it satisfies the relations
$\mathrm{e}^{\boldsymbol{0}}=\left|0\right\rangle $ and $\left\langle 0\!\mid\!\mathrm{e}^{\boldsymbol{h}}\right\rangle =\mathrm{e}^{-\frac{1}{2}\left\Vert \boldsymbol{h}\right\Vert _{\mathrm{1p}}^{2}}$.

It is a well-known fact that the structure of a free QFT is completely
determined by the underlying one-particle quantum theory. More concretely,
the assignment $\mathcal{O} \mapsto \mathcal{A}\left(\mathcal{O}\right)$
is defined by the composition of two different maps 
\begin{eqnarray}
\mathcal{O}\in\mathcal{K}  & \mapsto &  K\left(\mathcal{O}\right)\subset\mathfrak{H}\,,\\
K\subset\mathfrak{H}  & \mapsto &  \mathcal{A}\left(K\right)\subset\mathcal{B}\left(\mathcal{H}_{0}\right)\,,
\end{eqnarray}
which are called 1st and 2nd quantization maps respectively.\footnote{$\mathcal{K}$ denotes the set of causally complete regions of the
spacetime $\mathbb{R}^{d}$ (see definition \ref{c1-def-k}). } We will explain each map separately.

\subsubsection{First quantization map}

For any closed linear subspace $K\subset\mathfrak{H}$ we define its
\textit{symplectic complemen}t as
\begin{equation}
K':=\left\{ \boldsymbol{h}\in\mathfrak{H}\,:\, \mathrm{Im} \left\langle \boldsymbol{h}\!\mid\!\boldsymbol{k}\right\rangle _{\mathrm{1p}}=0\,,\textrm{ for all }\boldsymbol{k}\in K\right\} \,.
\end{equation}
Now, we consider the following real dense embedding $E:\,\mathcal{S}\left(\mathbb{R}^{d},\mathbb{R}\right)\rightarrow\mathfrak{H}$
\begin{equation}
\left(Ef\right)\left(\bar{p}\right):=\sqrt{2\pi}\,\hat{f}\mid_{H_{m}}\left(\bar{p}\right)=\sqrt{2\pi}\,\hat{f}\left(\omega\left(\bar{p}\right),\bar{p}\right)\:\textrm{,}
\end{equation}
where $\hat{f}\left(p\right):=\left(2\pi\right)^{-\frac{d}{2}}\int_{\mathbb{R}^{d}}f\left(x\right)\mathrm{e}^{ip\cdot x}d^{d}x$
is the usual Fourier transform. Such an embedding is Poincaré covariant,
i.e. $E\left(f_{\left(\Lambda,a\right)}\right)=u\left(\Lambda,a\right)E\left(f\right)$.
From now on, we naturally identified functions on $\mathcal{S}\left(\mathbb{R}^{d},\mathbb{R}\right)$
with vectors on $\mathfrak{H}$ through the above embedding.

The \textit{1st quantization map} is an assignment 
\begin{eqnarray}
\mathcal{O}\in \mathcal{K} & \mapsto & K\left(\mathcal{O}\right)\subset\mathfrak{H} \, ,
\end{eqnarray}
where $K\left(\mathcal{O}\right)$ is a real closed linear subspace defined as
\begin{equation}
K\left(\mathcal{O}\right):=\overline{\left\{ E\left(f\right)\,:\,f\in\mathcal{S}\left(\mathbb{R}^{d},\mathbb{R}\right)\textrm{ and }\mathrm{supp}\left(f\right)\subset\mathcal{O}\right\} }\subset\mathfrak{H}\,.
\end{equation}
It is not difficult to see that such a map satisfies the duality $K\left(\mathcal{O}'\right)=K\left(\mathcal{O}\right)'$
\cite{araki64}.

\subsubsection{Second quantization map}

We define the embedding $W:\mathfrak{H}\rightarrow\mathcal{B}\left(\mathcal{H}_{0}\right)$
\begin{equation}
W\left(\boldsymbol{h}\right):=\mathrm{e}^{i\left(A\left(\boldsymbol{h}\right)+A^{*}\left(\boldsymbol{h}\right)\right)}\:.\label{weyl_map}
\end{equation}
The operators $W\left(\boldsymbol{h}\right)$ are called \textit{Weyl
unitaries}. These operators satisfy the canonical commutation relations
(CCR) in the Segal's form \cite{araki63} 
\begin{eqnarray}
W\left(\boldsymbol{h}_{1}\right)W\left(\boldsymbol{h}_{2}\right)\!\!\! & = & \!\!\!\mathrm{e}^{-i\,\mathrm{Im}\left\langle \boldsymbol{h}_{1},\boldsymbol{h}_{2}\right\rangle _{\mathrm{1p}}}W\left(\boldsymbol{h}_{1}+\boldsymbol{h}_{2}\right)\,,\\
W\left(\boldsymbol{h}\right)^{*}\!\!\! & = & \!\!\!W\left(-\boldsymbol{h}\right)\:.
\end{eqnarray}
A Poincaré unitary $U\left(\Lambda,a\right)$ acts covariantly on
a Weyl operator according to
\begin{eqnarray}
U\left(\Lambda,a\right)W\left(\boldsymbol{h}\right)U\left(\Lambda,a\right)^{*}\!\!\! & = & \!\!\!W\left(u\left(\Lambda,a\right)\boldsymbol{h}\right)\,,\\
W\left(\boldsymbol{h}\right) \left|0\right\rangle \!\!\! & = & \!\!\!\mathrm{e}^{i\boldsymbol{h}}\:\textrm{.}
\end{eqnarray}
The \textit{2nd quantization map} is an assignment 
\begin{eqnarray}
K\subset\mathfrak{H} & \mapsto & \mathcal{A}\left(K\right)\subset\mathcal{B}\left(\mathcal{H}\right) \, ,
\end{eqnarray}
from the set of real closed linear subspaces of $\mathfrak{H}$ to
the set of vN subfactors of $\mathcal{B}\left(\mathcal{H}_{0}\right)$.
It is defined by means of
\begin{equation}
\mathcal{A}\left(K\right):=\left\{ W\left(\boldsymbol{k}\right)\,:\,\boldsymbol{k}\in K\right\} ''\subset\mathcal{B}\left(\mathcal{H}_{0}\right)\:.
\end{equation}
This map satisfies the duality $\mathcal{A}\left(K'\right)=\mathcal{A}\left(K\right)'$ \cite{araki63}.

\subsubsection{Net of local algebras\label{c4-subsec-net_local}}

According to the previous discussion, the net of local algebras $\mathcal{O}\in\mathcal{K}\rightarrow\mathcal{A}\left(\mathcal{O}\right)\subset\mathcal{B}\left(\mathcal{H}_{0}\right)$
of the free real scalar field is defined as the composition of the
1st and 2nd quantization maps, i.e. 
\begin{equation}
\mathcal{A}\left(\mathcal{O}\right):=\mathcal{A}\left(K\left(\mathcal{O}\right)\right)\:\textrm{.}\label{eq:net_spacetime}
\end{equation}
This net satisfies all the axioms of definitions \ref{c1_def_aqft}
, \ref{c1_def_vacuum}, and \ref{c1-weak_add}. Moreover, the map $\mathcal{O}\mapsto\mathcal{A}\left(\mathcal{O}\right)$
in \eqref{eq:net_spacetime} satisfies the properties \ref{c1-conj-irr}, \ref{c1-conj-add-bis}, and \ref{c1-conj-inter}-\ref{c1-conj-cov} of definition \ref{c1_conj} \cite{araki64}. For any $f\in\mathcal{S}\left(\mathbb{R}^{d},\mathbb{R}\right)$,
the field operator $\phi\left(f\right)$ is defined through the relation

\begin{equation}
W\left(E\left(f\right)\right)=\mathrm{e}^{i\phi\left(f\right)}=:W\left(f\right)\,.
\end{equation}

\subsection{Local algebras at a fixed time\label{subsec:algebras-ft}}

In this subsection, we discuss the theory of the vN algebras for the
free real scalar field at a fixed time $x^{0}=0$. Naively speaking,
they are the local algebras generated by the field operator at a fixed
time $\varphi\left(\bar{x}\right)$ and its canonical conjugate momentum
field $\pi\left(\bar{x}\right)$. This approach is very useful for
the computation of the RE in section \ref{c4-sec-re}.

We can decompose $\mathfrak{H=\mathfrak{H}_{\varphi}\oplus_{\mathbb{R}}\mathfrak{H}_{\pi}}$
into two $\mathbb{R}$-linear closed subspaces
\begin{eqnarray}
\mathfrak{H}_{\varphi}\!\!\! & := & \!\!\!\left\{ \boldsymbol{h}\in\mathfrak{H}\,:\,\boldsymbol{h}\left(\bar{p}\right)=\boldsymbol{h}\left(-\bar{p}\right)^{*}\:\textrm{a.e.}\right\} \,,\\
\mathfrak{H}_{\pi}\!\!\! & := & \!\!\!\left\{ \boldsymbol{h}\in\mathfrak{H}\,:\,\boldsymbol{h}\left(\bar{p}\right)=-\boldsymbol{h}\left(-\bar{p}\right)^{*}\:\textrm{a.e.}\right\} \,.
\end{eqnarray}
Each $\boldsymbol{h}\in\mathfrak{H}$ can be uniquely written as $\boldsymbol{h}=\boldsymbol{h}_{\varphi}+\boldsymbol{h}_{\pi}$,
where
\begin{eqnarray}
\boldsymbol{h}_{\varphi}\left(\bar{p}\right):=\frac{\boldsymbol{h}\left(\bar{p}\right)+\boldsymbol{h}\left(-\bar{p}\right)^{*}}{2} & \textrm{ and } & \boldsymbol{h}_{\pi}\left(\bar{p}\right):=\frac{\boldsymbol{h}\left(\bar{p}\right)-\boldsymbol{h}\left(-\bar{p}\right)^{*}}{2}\,.
\end{eqnarray}
We also have the following useful relations
\begin{equation}
\mathrm{Im}\left\langle \boldsymbol{h}_{\varphi}\!\mid\!\boldsymbol{h}'_{\varphi}\right\rangle _{\mathrm{1p}}=\mathrm{Im}\left\langle \boldsymbol{h}_{\pi}\!\mid\!\boldsymbol{h}'_{\pi}\right\rangle _{\mathrm{1p}}=\mathrm{Re}\left\langle \boldsymbol{h}_{\varphi}\!\mid\!\boldsymbol{h}'_{\pi}\right\rangle _{\mathrm{1p}}=0\:\textrm{,}\label{eq:rel_prod_int}
\end{equation}
for all $\boldsymbol{h}_{\varphi},\,\boldsymbol{h}'_{\varphi}\in\mathfrak{H}_{\varphi}$
and $\boldsymbol{h}_{\pi},\,\boldsymbol{h}'_{\pi}\in\mathfrak{H}_{\pi}$. 

Now, we consider the following real dense embeddings $E_{\varphi,\pi}:\,\mathcal{S}\left(\mathbb{R}^{d-1},\mathbb{R}\right)\rightarrow\mathfrak{H}_{\varphi,\pi}$
\begin{equation}
\left(E_{\varphi}f\right)\left(\bar{p}\right):=\hat{f}\left(\bar{p}\right)\quad\mathrm{and}\quad\left(E_{\pi}f\right)\left(\bar{p}\right):=i\omega\left(\bar{p}\right)\,\hat{f}\left(\bar{p}\right)\,,
\end{equation}
where $\hat{f}\left(\bar{p}\right):=\left(2\pi\right)^{-\frac{d-1}{2}}\int_{\mathbb{R}^{d-1}}f\left(\bar{x}\right)\mathrm{e}^{-i\bar{p}\cdot\bar{x}}d^{d-1}x$.
From now on, we naturally identify functions on $\mathcal{S}\left(\mathbb{R}^{d-1},\mathbb{R}\right)$
with vectors on $\mathfrak{H}_{\varphi},\mathfrak{H}_{\pi}$ through
these embeddings. The map $E_{\varphi}$ (resp. $E_{\pi}$) is actually
defined on a bigger class of test functions, namely $H^{-\frac{1}{2}}\left(\mathbb{R}^{d-1},\mathbb{R}\right)$
(resp. $H^{\frac{1}{2}}\left(\mathbb{R}^{d-1},\mathbb{R}\right)$),
i.e.
\begin{equation}
E_{\varphi}:\,H^{-\frac{1}{2}}\left(\mathbb{R}^{d-1},\mathbb{R}\right)\rightarrow\mathfrak{H}_{\varphi}\quad\mathrm{and}\quad E_{\pi}:\,H^{\frac{1}{2}}\left(\mathbb{R}^{d-1},\mathbb{R}\right)\rightarrow\mathfrak{H}_{\pi}\,,
\end{equation}
where $H^{\alpha}\left(\mathbb{R}^{d-1},\mathbb{R}\right)$ is the
real Sobolev space of order $\alpha$ (see appendix \ref{AP-sec-sobo}).
We have that $E_{\varphi}\left(H^{\frac{1}{2}}\left(\mathbb{R}^{d-1},\mathbb{R}\right)\right)=\mathfrak{H}_{\varphi}$
and $E_{\pi}\left(H^{-\frac{1}{2}}\left(\mathbb{R}^{d-1},\mathbb{R}\right)\right)=\mathfrak{H}_{\pi}$.
For any $\boldsymbol{h}_{\varphi}\in\mathfrak{H}_{\varphi}$ and $\boldsymbol{h}_{\pi}\in\mathfrak{H}_{\pi}$,
we use the map \eqref{weyl_map} to define the Weyl unitaries
\begin{equation}
W_{\varphi}\left(\boldsymbol{h}_{\varphi}\right):=W\left(\boldsymbol{h}_{\varphi}\right)\quad\textrm{and}\quad W_{\pi}\left(\boldsymbol{h}_{\pi}\right):=W\left(\boldsymbol{h}_{\pi}\right)\,,
\end{equation}
which satisfy the CCR in the Weyl form \cite{araki63}
\begin{eqnarray}
W_{\varphi}\left(\boldsymbol{h}_{\varphi}\right)W_{\pi}\left(\boldsymbol{h}_{\pi}\right)W_{\varphi}\left(\boldsymbol{h}'_{\varphi}\right)W_{\pi}\left(\boldsymbol{h}'_{\pi}\right)\!\!\! & = & \!\!\!W_{\varphi}\left(\boldsymbol{h}_{\varphi}+\boldsymbol{h}'_{\varphi}\right)W_{\pi}\left(\boldsymbol{h}_{\pi}+\boldsymbol{h}'_{\pi}\right)\mathrm{e}^{2i\,\mathrm{Im}\left\langle \boldsymbol{h}'_{\varphi}\mid\boldsymbol{h}{}_{\pi}\right\rangle _{\mathrm{1p}}}, \hspace{0.9cm}\\
W_{\varphi}\left(\boldsymbol{h}_{\varphi}\right)^{*}\!\!\! & = & \!\!\!W_{\varphi}\left(-\boldsymbol{h}_{\varphi}\right),\\
W_{\pi}\left(\boldsymbol{h}_{\pi}\right)^{*}\!\!\! & = & \!\!\!W_{\pi}\left(-\boldsymbol{h}_{\pi}\right).
\end{eqnarray}
The \textit{field operator at a fixed time} $\varphi\left(f_{\varphi}\right)$
and its \textit{canonical conjugate momentum field} $\pi\left(f_{\pi}\right)$
are defined by the formulas 
\begin{equation}
W_{\varphi}\left(E_{\varphi}\left(f_{\varphi}\right)\right)=\mathrm{e}^{i\varphi\left(f_{\varphi}\right)}=:W_{\varphi}\left(f_{\varphi}\right)\quad\mathrm{and}\quad W_{\pi}\left(E_{\pi}\left(f_{\pi}\right)\right)=:\mathrm{e}^{i\pi\left(f_{\pi}\right)}=:W_{\pi}\left(f_{\pi}\right)\,.
\end{equation}
Here again, the local algebras at a fixed time are also defined through
the 1st and 2nd quantization maps.

\subsubsection{First quantization map}

We say that $\mathcal{C}\subset\mathbb{R}^{d-1}$ is a \textit{spatially
complete} region if it is open, regular,\footnote{An open set $U\subset\mathbb{R}^{n}$ is \textit{regular} iff $U\equiv\mathrm{Int}\left(\overline{U}\right)$.}
and with regular boundary.\footnote{We say that $\mathcal{C}\subset\mathbb{R}^{d-1}$ has a \textit{regular
boundary} if $\partial\mathcal{C}\subset\mathbb{R}^{d-1}$ is a smooth
submanifold of dimension $d-2$, or several manifolds joined together
along smooth manifolds of dimension $d-3$ \cite{araki64}.} The set of all spatially complete regions is denoted by $\tilde{\mathcal{K}}_{0}$,
which is a proper subset of the set of all open regions $\mathcal{K}_{0}$.\footnote{Acording to proposition \ref{c1-latt-cauchy}, the set of all open
regions $\mathcal{K}_{0}$ is an orthocomplemented lattice. It can
be shown that $\tilde{\mathcal{K}}_{0}$ is a sublattice of $\mathcal{K}_{0}$
\cite{araki64,horuzhy}. } From now on, we only consider spatially complete regions. 

The \textit{1st quantization map}
\begin{eqnarray}
\mathcal{C}\in\tilde{\mathcal{K}}_{0} & \mapsto & \left(K_{\varphi}(\mathcal{C}),K_{\pi}(\mathcal{C})\right)\subset\left(\mathfrak{H}_{\varphi},\mathfrak{H}_{\pi}\right)\,,
\end{eqnarray}
is defined by means
\begin{eqnarray}
K_{\varphi}(\mathcal{C})\!\!\! & := & \!\!\!\overline{\left\{ E_{\varphi}\left(f\right)\,:\,f\in\mathcal{S}\left(\mathbb{R}^{d-1},\mathbb{R}\right)\textrm{ and }\mathrm{supp}\left(f\right)\subset\mathcal{C}\right\} }\subset\mathfrak{H}_{\varphi}\,,\\
K_{\varphi}(\mathcal{C})\!\!\! & := & \!\!\!\overline{\left\{ E_{\pi}\left(f\right)\,:\,f\in\mathcal{S}_{\mathbb{}}\left(\mathbb{R}^{d-1},\mathbb{R}\right)\textrm{ and }\mathrm{supp}\left(f\right)\subset\mathcal{C}\right\} }\:\subset\mathfrak{H}_{\pi}\textrm{.}
\end{eqnarray}
It can be shown that \cite{araki64}
\begin{eqnarray}
K_{\varphi}(\mathcal{C})\!\!\! & = & \!\!\!\left\{ E_{\varphi}\left(f\right)\,:\,f\in H^{-\frac{1}{2}}\left(\mathbb{R}^{d-1},\mathbb{R}\right)\textrm{ and }\mathrm{supp}\left(f\right)\subset\mathcal{\overline{C}}\:\textrm{a.e.}\right\} \,,\\
K_{\pi}(\mathcal{C})\!\!\! & = & \!\!\!\left\{ E_{\pi}\left(f\right)\,:\,f\in H^{\frac{1}{2}}\left(\mathbb{R}^{d-1},\mathbb{R}\right)\textrm{ and }\mathrm{supp}\left(f\right)\subset\mathcal{\mathcal{\overline{C}}}\:\textrm{a.e.}\right\} \:\textrm{.}
\end{eqnarray}

\subsubsection{Second quantization map}

Let $K_{\varphi}\subset\mathfrak{H}_{\varphi}$ and $K_{\pi}\subset\mathfrak{H}_{\pi}$
be $\mathbb{R}$-linear closed subspaces. The \textit{2nd quantization
map}
\begin{eqnarray}
\left(K_{\varphi},K_{\pi}\right)\subset\left(\mathfrak{H}_{\varphi},\mathfrak{H}_{\pi}\right) & \mapsto & \mathcal{A}_{0}\left(K_{\varphi},K_{\pi}\right)\subset\mathcal{B}\left(\mathcal{H}_{0}\right)\,,
\end{eqnarray}
is defined by means

\begin{equation}
\mathcal{A}_{0}\left(K_{\varphi},K_{\pi}\right):=\left\{ W_{\varphi}\left(k_{\varphi}\right)W_{\pi}\left(k_{\pi}\right)\,:\,k_{\varphi}\in K_{\varphi},\,k_{\pi}\in K_{\pi}\right\} ''\subset\mathcal{B}\left(\mathcal{H}_{0}\right)\:\textrm{.}
\end{equation}

\subsubsection{Net of local algebras at a fixed time}

The net of local algebras $\mathcal{C}\in\tilde{\mathcal{K}}_{0} \mapsto \mathcal{A}_{0}\left(\mathcal{C}\right)\subset\mathcal{B}\left(\mathcal{H}_{0}\right)$
at a fixed time is defined as the composition of the 1st and 2nd quantization
maps, i.e.
\begin{equation}
\mathcal{A}_{0}\left(\mathcal{C}\right):=\mathcal{A}_{0}\left(K_{\varphi}\left(\mathcal{C}\right),K_{\pi}\left(\mathcal{C}\right)\right)\:\textrm{.}\label{eq:net_fixed_time}
\end{equation}
The above net satisfies all the properties of proposition \ref{c1_conj-2}
\cite{araki64}. In other words, it is an homomorphism of orthocomplemented
lattices.

\subsection{Relation between the two approaches\label{subsec:Relation-between-algebras}}

In this subsection, we explain the relation existing between the two
approaches of sections \ref{c4-algebras-st} and \ref{subsec:algebras-ft}.

\paragraph{The relation between nets.}
Given a spatially complete region $\mathcal{C}\in\mathcal{\tilde{K}}_{0}$,
we denote its Cauchy development as $\mathcal{O_{C}}:=D\left(\mathcal{C}\right)\in\mathcal{K}$.
Then, the following relation holds \cite{Camassa}
\begin{equation}
K\left(\mathcal{O_{C}}\right)=K_{\varphi}\left(\mathcal{C}\right)\oplus_{\mathbb{R}}K_{\pi}\left(\mathcal{C}\right)\,,
\end{equation}
and hence, we have the following equality of vN algebras
\begin{equation}
\mathcal{A}_{0}\left(\mathcal{C}\right)=\mathcal{A}\left(\mathcal{O_{C}}\right)\,.\label{eq:rel_algebras}
\end{equation}
The relations developed along the above subsections can be summarized
in the following schematic diagram
\begin{alignat}{4}
\mathcal{O}_{\mathcal{C}} & \in\mathcal{K} & \quad\overset{E}{\longmapsto}\quad &  & K & \subset\mathfrak{H} & \quad\overset{W}{\longmapsto}\quad & \mathcal{A}\subset\mathcal{B}\left(\mathcal{H}\right)\nonumber \\
 & \,\uparrow\text{\scriptsize\ensuremath{D}} &  &  &  & \:\uparrow\oplus_{\mathbb{R}} &  & \shortparallel\\
\mathcal{C} & \in\mathcal{\tilde{K}}_{0} & \quad\overset{E_{\varphi,\pi}}{\longmapsto}\quad &  & \left(K_{\varphi},K_{\pi}\right) & \subset\mathfrak{H}_{\varphi}\oplus_{\mathbb{R}}\mathfrak{H}_{\pi} & \quad\overset{W_{\varphi,\pi}}{\longmapsto}\quad & \mathcal{A}_{0}\subset\mathcal{B}\left(\mathcal{H}\right)\:.\nonumber 
\end{alignat}

\paragraph{The relation between test functions.}

Given any $f\in\mathcal{S}\left(\mathbb{R}^{d},\mathbb{R}\right)$,
we can define
\begin{equation}
F\left(x\right):=\int_{\mathbb{R}^{d}}\Delta\left(x-y\right)\,f\left(y\right)d^{d}y\,,\label{eq:convolucion}
\end{equation}
where\footnote{In the Wightman approach to QFT, we recognize $\Delta(x-y):=-i[\phi(x),\phi(y)]$.}
\begin{equation}
\Delta\left(x\right):=-i\left(2\pi\right)^{-\left(d-1\right)}\int_{\mathbb{R}^{d}}\mathrm{e}^{-ip\cdot x}\delta\left(p^{2}-m^{2}\right)\mathrm{sgn}\left(p^{0}\right)\,d^{d}p\,.\label{c4-sf-comm}
\end{equation}
It follows that $F$ is a solution of the Klein-Gordon equation, i.e.
$\left(\boxempty+m^{2}\right)F=0$, and whereby, we can take its initial
Cauchy data at $x^{0}=0$ through
\begin{equation}
f_{\varphi}\left(\bar{x}\right):=-\frac{\partial F}{\partial x^{0}}\left(0,\bar{x}\right)\;\textrm{ and }\;f_{\pi}\left(\bar{x}\right):=F\left(0,\bar{x}\right)\:\textrm{.}\label{eq:f_fixed_time}
\end{equation}
Finally, it can be shown that $f_{\varphi},\,f_{\pi}\in\mathcal{S}\left(\mathbb{R}^{d-1},\mathbb{R}\right)$
and
\begin{equation}
E\left(f\right)=E_{\varphi}\left(f_{\varphi}\right)+E_{\pi}\left(f_{\pi}\right)\:.\label{eq:segal_vs_weyl_1p}
\end{equation}
Moreover, since $F\left(x\right)=0$ if $x\in\mathrm{supp}\left(f\right)'$,
then we have that
\begin{equation}
\mathrm{supp}\left(f\right)\subset\mathcal{O}_{\mathcal{C}}\Rightarrow\mathrm{supp}\left(f_{\varphi}\right),\,\mathrm{supp}\left(f_{\pi}\right)\subset\mathcal{C}\:\textrm{.}
\end{equation}

\paragraph{The relation between Weyl unitaries.}

For the particular case of Weyl unitaries, it follows that
\begin{equation}
W\left(f\right)=\mathrm{e}^{i\mathrm{Im}\left\langle f_{\varphi}\mid f_{\pi}\right\rangle _{\mathrm{1p}}}W_{\varphi}\left(f_{\varphi}\right)W_{\pi}\left(f_{\pi}\right)\,,\label{eq:segal_vs_weyl}
\end{equation}
where 
\begin{equation}
\mathrm{Im}\left\langle f_{\varphi},f_{\pi}\right\rangle _{\mathfrak{H}}=\frac{1}{2}\int_{\mathbb{R}^{d-1}}f_{\varphi}\left(\bar{x}\right)\,f_{\pi}\left(\bar{x}\right)d^{d-1}x\,.\label{eq:img_pe}
\end{equation}

\section{Symmetry of the RE for coherent states\label{c4-sec-re_sym}}

In this section, we study some general relations about the RE for
coherent states, which are valid for any region $\mathcal{O}\in\mathcal{K}$.
In particular, we show that the RE for coherent states is symmetric.
\begin{notation}
Let $\mathcal{A}\subset\mathcal{B}\left(\mathcal{H}\right)$ be a
vN algebra and $\left|\Omega\right\rangle ,\left|\Phi\right\rangle \in\mathcal{H}$
be standard vectors. We denote by $S_{\Phi}$ and $\Delta_{\Phi}$
the modular involution and the modular operator associated with $\left\{ \mathcal{A},\left|\Phi\right\rangle \right\} $,
and by $S_{\Phi,\Omega}$ and $\Delta_{\Phi,\Omega}$ the relative
modular involution and the relative modular operator associated with
$\left\{ \mathcal{A},\left|\Omega\right\rangle ,\left|\Phi\right\rangle \right\} $.
\end{notation}

Coherent states come from acting with a Weyl operator on the vacuum
vector. Weyl unitaries have the very important property that they
implement, by adjoint action, automorphisms for any local algebra
$\mathcal{A}\left(\mathcal{O}\right)$. More concretely, for any $\boldsymbol{h}\in\mathfrak{H}$
and any Weyl operator $W\left(f\right)\in\mathcal{A}\left(\mathcal{O}\right)$
($\mathrm{supp}\left(f\right)\subset\mathcal{O}$) we have that
\begin{equation}
W\left(\boldsymbol{h}\right)^{*}W\left(f\right)W\left(\boldsymbol{h}\right)=\mathrm{e}^{2i\mathrm{Im}\left\langle f\mid\boldsymbol{h}\right\rangle _{\mathrm{1p}}}W\left(f\right)\in\mathcal{A}\left(\mathcal{O}\right)\,,\label{eq:case_3}
\end{equation}
which implies that $W\left(\boldsymbol{h}\right)^{*}\mathcal{A}\left(\mathcal{O}\right)W\left(\boldsymbol{h}\right)=\mathcal{A}\left(\mathcal{O}\right)$.
This property has interesting implications on the RE itself. Indeed,
it implies that the relative entropy between a coherent state and
the vacuum is symmetric. To show this property, we first prove the
following lemmas.
\begin{lem}
\label{lema_auto}Let $\mathcal{A}\subset\mathcal{B}\left(\mathcal{H}\right)$
be a vN algebra, $\left|\Omega\right\rangle ,\left|\Phi\right\rangle \in\mathcal{H}_{0}$
standard vectors, and $U\in\mathcal{B}\left(\mathcal{H}\right)$ unitary
such that $U^{*}\mathcal{A}U=\mathcal{A}$. Then,
\begin{enumerate}
\item \label{lemma_tt_1} \textit{$U\left|\Omega\right\rangle $ and $U\left|\Phi\right\rangle $
are standard vectors.}
\item \label{lemma_tt_2} \textit{$S_{U\Omega}=US_{\Omega}U^{*}$ $\Rightarrow$ $\Delta_{U\Omega}=U\Delta_{\Omega}U^{\text{*}}$.}
\item \label{lemma_tt_3} \textit{$S_{U\Omega,U\Phi}=US_{\Omega,\Phi}U^{*}$ $\Rightarrow$
$\Delta_{U\Omega,U\Phi}=U\Delta_{\Omega,\Phi}U^{*}.$}
\end{enumerate}
\end{lem}
\begin{proof}
(1) $\overline{\mathcal{A}U\left|\Omega\right\rangle }=\overline{U\mathcal{A}\left|\Omega\right\rangle }=U\overline{\mathcal{A}\left|\Omega\right\rangle }=\mathcal{H}$
implies that $U\left|\Omega\right\rangle $ is cyclic. Then, $AU\left|\Omega\right\rangle =0\Leftrightarrow U^{*}AU\left|\Omega\right\rangle =0\Leftrightarrow U^{*}AU=0\Leftrightarrow A=0$
implies that $U\left|\Omega\right\rangle $ is separating. The same
holds for $U\left|\Phi\right\rangle $. (2) For any $A\in\mathcal{A}$,
we have that $\left(US_{\Omega}U^{*}\right)AU\left|\Omega\right\rangle =US_{\Omega}\left(U^{*}AU\right)\left|\Omega\right\rangle =U\left(U^{*}AU\right)^{*}\left|\Omega\right\rangle =A^{*}U\left|\Omega\right\rangle $.
Then, applying the polar decomposition, we have that $\Delta_{U\Omega}=U\Delta_{\Omega}U^{\text{*}}$.
(3) For any $A\in\mathcal{A}$, we have that $\left(US_{\Omega,\Phi}U^{*}\right)AU\left|\Phi\right\rangle =US_{\Omega,\Phi}\left(U^{*}AU\right)\left|\Phi\right\rangle =U\left(U^{*}AU\right)^{*}\left|\Omega\right\rangle =A^{*}U\left|\Omega\right\rangle \,.$
Then, $\Delta_{U\Omega,U\Phi}=U\Delta_{\Omega,\Phi}U^{*}$ follows
from applying the polar decomposition.
\end{proof}
Given a state $\omega$ on a vN algebra $\mathcal{A}\subset\mathcal{B}\left(\mathcal{H}\right)$
and a unitary $U\in\mathcal{B}\left(\mathbb{\mathcal{H}}\right)$,
we denote by $\omega_{U}$ the state defined by $\omega_{U}\left(\cdot\right):=\omega\left(U^{*}\cdot U\right)$.
\begin{lem}
\label{sim_ent_conj}Given $\mathcal{A}\subset\mathcal{B}\left(\mathcal{H}\right)$
a vN algebra in standard form, $\omega$ a faithful normal state and
$U\in\mathcal{B}\left(\mathcal{H}\right)$ unitary such that $U^{*}\mathcal{A}U=\mathcal{A}$.
Then\textup{
\begin{equation}
S_{\mathcal{A}}\left(\omega_{U}\mid\omega\right)=S_{\mathcal{A}}\left(\omega\mid\omega_{U^{*}}\right)\,.
\end{equation}
}
\end{lem}

\begin{proof}
Let $\left|\Omega\right\rangle $ be a standard vector representative
of $\omega$. Then $U\left|\Omega\right\rangle $ and $U^{*}\left|\Omega\right\rangle $
are the vector representatives of $\omega_{U}$ and $\omega_{U^{*}}$,
and they are standard vectors because of \ref{lemma_tt_1} in lemma \ref{lema_auto}.
Using \ref{lemma_tt_3} of the same lemma we have that $S_{\Omega,U\Omega}=S_{UU^{*}\Omega,U\Omega}=US_{U^{*}\Omega,\Omega}U^{*}$.
Then, $S\left(\omega_{U}\mid\omega\right)=\left\langle U\Omega\right|K_{\Omega,U\Omega}\left|U\Omega\right\rangle =\left\langle U\Omega\right|UK_{U^{*}\Omega,\Omega}U^{*}\left|U\Omega\right\rangle =\left\langle \Omega\right|K_{U^{*}\Omega,\Omega}\left|\Omega\right\rangle =S\left(\omega\mid\omega_{U^{*}}\right)$.
\end{proof}
Now, we come back to coherent states. From now on $\omega_{0}\left(\cdot\right):=\left\langle 0\right|\cdot\left|0\right\rangle $
denotes the vacuum state. Given any $f\in\mathcal{S}\left(\mathbb{R}^{d},\mathbb{R}\right)$,
we define the coherent state $\omega_{f}\left(\cdot\right):=\left\langle 0\right|W\left(f\right)^{*}\cdot\,W\left(f\right)\left|0\right\rangle $.
The Reeh-Schlieder theorem \ref{c1_rs} asserts that the vacuum vector
$\left|0\right\rangle $ is standard for any local algebra $\mathcal{A}\left(\mathcal{O}\right)$
of any standard region $\mathcal{O}\in\mathcal{K}$. Then, lemma \ref{lema_auto}
ensures the same for the coherent vector $W\left(f\right)\left|0\right\rangle $.
Then, lemma \ref{sim_ent_conj} implies
\begin{equation}
S_{\mathcal{O}}\left(\omega_{f}\mid\omega_{0}\right)=S_{\mathcal{O}}\left(\omega_{0}\mid\omega_{-f}\right)\,,\label{sr-antisym}
\end{equation}
for any coherent state $\omega_{f}$ and any standard region $\mathcal{O}\in\mathcal{K}$.

Moreover, the net algebra of the free scalar field has a global $\mathbb{Z}_{2}$-symmetry
implemented by an operator $z=z^{-1}=z^{*}$ such that\footnote{In the Lagrangian approach to QFT, this is the usual symmetry $\phi\left(x\right)\rightarrow-\phi\left(x\right)$.}
\begin{equation}
zW\left(f\right)z:=W\left(-f\right)=W\left(f\right)^{*}\,,\quad\mathrm{and}\quad z\left|0\right\rangle =\left|0\right\rangle \,.\label{z2-sym}
\end{equation}
This motivates the following lemma.
\begin{lem}
For the local algebra $\mathcal{A}\left(\mathcal{O}\right)$ of any
standard region $\mathcal{O}\in\mathcal{K}$, the RE between a coherent
state $\omega_{f}$ and the vacuum state $\omega_{0}$ satisfies 
\begin{equation}
S\left(\omega_{f}\mid\omega_{0}\right)=S\left(\omega_{-f}\mid\omega_{0}\right)\,.\label{z2-sym-sr}
\end{equation}
\end{lem}

\begin{proof}
Let $\left|0\right\rangle $, $W\left(f\right)\left|0\right\rangle $,
and $W\left(f\right)^{*}\left|0\right\rangle $ be the vector representatives
of the states $\omega_{0}$, $\omega_{f}$, and $\omega_{-f}$. Let
us denote by $S_{0,f}$ the relative modular involution associated
with $\left\{ \mathcal{A}\left(\mathcal{O}\right),W\left(f\right)\left|0\right\rangle ,\left|0\right\rangle \right\} $.
Then, employing the $\mathbb{Z}_{2}$-symmetry \eqref{z2-sym}, we
have that
\begin{equation}
\left(zS_{0,f}z\right)W\left(g\right)W\left(f\right)^{*}\left|0\right\rangle =zS_{0,f}W\left(g\right)^{*}W\left(f\right)\left|0\right\rangle =zW\left(g\right)\left|0\right\rangle =W\left(g\right)^{\text{*}}\left|0\right\rangle \,,
\end{equation}
for all $W\left(g\right)\in\mathcal{A}\left(\mathcal{O}\right)$.
Then, $S_{0,-f}=zS_{0,f}z$, and hence, $K_{0,-f}=zK_{0,f}z$. Finally,
we get $S\left(\omega_{f}\mid\omega_{0}\right)=\left\langle 0\right|K_{0,f}\left|0\right\rangle =\left\langle 0\right|K_{0,-f}\left|0\right\rangle =S\left(\omega_{-f}\mid\omega_{0}\right)$.
\end{proof}
\begin{rem}
The above lemma should apply to any scalar fiel theory with $\mathbb{Z}_{2}$-symmetry
as above, as long as it satisfies the Garding-Wightman axioms.
\end{rem}

Finally, combining \eqref{sr-antisym} and \eqref{z2-sym-sr}, we
have the following theorem concerning the symmetry for the relative
entropy between coherent states. 
\begin{thm}
\textup{\label{thm_sym_sr}}For the local algebra $\mathcal{A}\left(\mathcal{O}\right)$
of any standard region $\mathcal{O}\in\mathcal{K}$, the RE between
a coherent state $\omega_{f}$ and the vacuum state $\omega$ is symmetric,
i.e. 
\begin{equation}
S\left(\omega_{f}\mid\omega_{0}\right)=S\left(\omega_{0}\mid\omega_{f}\right)\,.\label{sym-sr}
\end{equation}
\end{thm}

To end, we have the following theorem concerning the relative entropy
between two coherent states.\footnote{This result has been found in the past using other methods. For example,
see \cite{Lashkari:2015dia} for a derivation using the replica trick
for $2d$ free CFTs.}
\begin{thm}
For the local algebra $\mathcal{A}\left(\mathcal{O}\right)$ of any
standard region $\mathcal{O}\in\mathcal{K}$, the RE between two coherent
states $\omega_{f}$ and $\omega_{g}$ satisfies
\begin{equation}
S\left(\omega_{f}\mid\omega_{g}\right)=S\left(\omega_{f-g}\mid\omega_{0}\right)\,.
\end{equation}
\end{thm}

\begin{proof}
Let $\left|0\right\rangle $, $W\left(f\right)\left|0\right\rangle $,
and $W\left(g\right)\left|0\right\rangle $ be the vector representatives
of the states $\omega_{0}$, $\omega_{f}$, and $\omega_{g}$. Let
us denote by $S_{g,f}$ the relative modular involution associated
with $\left\{ \mathcal{A}\left(\mathcal{O}\right),W\left(f\right)\left|0\right\rangle ,W\left(g\right)\left|0\right\rangle \right\} $.
Then, using \ref{lemma_tt_3} in lemma \ref{lema_auto}, we have that $W\left(g\right)^{*}S_{g,f}W\left(g\right)$
is the relative modular involution associated with $\left\{ \mathcal{A}\left(\mathcal{O}\right),W\left(g\right)^{*}W\left(f\right)\left|0\right\rangle ,\left|0\right\rangle \right\} $.
Since $W\left(g\right)^{*}W\left(f\right)\left|0\right\rangle $ is
a vector representative of $\omega_{f-g}$ , we have that $S\left(\omega_{f}\mid\omega_{g}\right)=\left\langle 0\right|W\left(f\right)^{*}S_{g,f}W\left(f\right)\left|0\right\rangle =S\left(\omega_{f-g}\mid\omega_{0}\right)$.
\end{proof}

\section{Relative entropy for the Rindler wedge\label{c4-sec-re}}

In this section, we compute the RE, in the theory of the free real
scalar field, between a coherent state and the vacuum using Araki
formula (definition \ref{c2_re_araki}). We focus on the case of the
Rindler wedge region. The aim is to show explicitly that formula \eqref{c4-cali}
holds for the canonical stress-tensor.

Let  $\mathcal{A}_{\mathcal{W}}:=\mathcal{A}\left(\mathcal{W}_{R}\right)$
be the right Rindler wedge algebra and $\mathcal{A}_{\mathcal{W}'}:=\mathcal{A}\left(\mathcal{W}'_{R}\right)$.
We define $\Sigma_{R}:=\left\{ \bar{x}\in\mathbb{R}^{d-1}\,:\,x^{1}>0\right\} $.
Then, we have that $\mathcal{W}_{R}=D\left(\Sigma_{R}\right)$ and
$\mathcal{W}{}_{L}:=\mathcal{W}'_{R}=D\left(\Sigma_{R}'\right)$,
and hence, we have that
\begin{eqnarray}
\mathcal{A}_{0}\left(\Sigma_{R}\right)=\mathcal{A}_{\mathcal{W}} & \textrm{ and } & \mathcal{A}_{0}\left(\Sigma_{R}'\right)=\mathcal{A}_{\mathcal{W}'}\,.
\end{eqnarray}
Let $\omega_{0}$ be the vacuum state and $\omega_{f}$ a coherent
state with $f\in\mathcal{S}\left(\mathbb{R}^{d},\mathbb{R}\right)$.
Let us call $\left|0\right\rangle $ and $\left|\Phi\right\rangle :=W\left(f\right)\left|0\right\rangle $
its vector representatives. In order to compute the relative entropy
$S\left(\omega_{f}\mid\omega_{0}\right)$, we need to calculate the
relative modular Hamiltonian $K_{0,\Phi}$, or $K_{\Phi,0}$ because
of theorem \ref{thm_sym_sr}. As we have explained in the last subsection,
the vectors $\left|0\right\rangle $ and $\left|\Phi\right\rangle $
are standard for $\mathcal{A}_{\mathcal{W}}$. We distinguish between
two cases 
\begin{eqnarray}
\textrm{easy case} & : & f=f_{L}+f_{R}\,,\label{eq:case_prod}\\
\textrm{hard case} & : & f\neq f_{L}+f_{R}\,,\label{eq:case_no_prod}
\end{eqnarray}
where $\mathrm{supp}\left(f_{L}\right)\in\mathcal{W}{}_{L}$ and $\mathrm{supp}\left(f_{R}\right)\in\mathcal{W}{}_{R}$.\footnote{In particular, the easy case includes the cases when $W\left(f\right)\in\mathcal{A}{}_{\mathcal{W}}$
or $W\left(f\right)\in\mathcal{A}{}_{\mathcal{W}'}$.} In the following subsections, we deal with each case \eqref{eq:case_prod}
and \eqref{eq:case_no_prod} separately. 

\subsection{Easy case: $f=f_{L}+f_{R}$\label{c4-sec-easy}}

In this case, we have that the coherent vector can be written as $W\left(f\right)=W\left(f_{L}\right)W\left(f_{R}\right)$
with $W\left(f_{L}\right)\in\mathcal{A}_{\mathcal{W}'}$ and $W\left(f_{R}\right)\in\mathcal{A}_{\mathcal{W}}$.
This case can be solved using the following lemma.
\begin{lem}
\label{lema_easy_case}Given $\mathcal{A}\subset\mathcal{B}\left(\mathcal{H}\right)$
a vN algebra, $\left|\Omega\right\rangle $ a standard vector and
$U\in\mathcal{A}$ and $U'\in\mathcal{A}'$ unitaries. Then $\left|\Phi\right\rangle =U'U\left|\Omega\right\rangle $
is standard and
\begin{equation}
S_{\Omega,\Phi}=US_{\Omega}U'^{*}\,.
\end{equation}
Then, by polar decomposition we have that $J_{\Omega,\Phi}=UJ_{\Omega}U'^{*}$,
$\Delta_{\Omega,\Phi}=U'\Delta_{\Omega}U'^{*}$ and $K_{\Omega,\Phi}=U'K_{\Omega}U'^{*}$.
\end{lem}

\begin{proof}
$\overline{\mathcal{A}\left|\Phi\right\rangle }=\overline{\mathcal{A}U'U\left|\Omega\right\rangle }=U'\overline{\mathcal{A}U\left|\Omega\right\rangle }=U'\overline{\mathcal{A}{}_{\mathcal{W}}\left|\Omega\right\rangle }=\overline{\mathcal{A}\left|\Omega\right\rangle }=\mathcal{H}$
implies $\left|\Phi\right\rangle $ is cyclic. Since the same argument
holds for $\mathcal{A}'$, then $\left|\Phi\right\rangle $ is separating
for $\mathcal{A}$. For any $A\in\mathcal{A}$, we have that $\left(US_{\Omega}U'^{*}\right)A\left|\Phi\right\rangle =US_{\Omega}U'^{*}AU'U\left|\Phi\right\rangle =U_{R}S_{\Omega}\left(AU\right)\left|\Omega\right\rangle =U\left(AU\right)^{*}\left|\Omega\right\rangle =A^{*}\left|\Omega\right\rangle $
$\Rightarrow\,S_{\Omega,\Phi}=US_{\Omega}U'^{*}$.
\end{proof}
\begin{cor}
\label{cor_re_easy}In the context of the above lemma, if $\left|\Omega\right\rangle $
and $\left|\Phi\right\rangle $ are vector representatives of the
states $\omega$ and $\phi$, then $S\left(\phi\mid\omega\right)=\left\langle \Phi\right|U'K_{\Omega}U'^{*}\left|\Phi\right\rangle =\left\langle \Omega\right|U^{*}K_{\Omega}U\left|\Omega\right\rangle $.
\end{cor}

The above corollary shows explicitly that the RE does not depend on
the unitary $U'$. This is expected because the RE is a measure of
distinguishability of the states in $\mathcal{A}$, and hence, it
has to be invariant under unitary changes of the states outside $\mathcal{A}$.

Now we apply the corollary \ref{cor_re_easy} to the case of a coherent
state, i.e. $U=W\left(f_{R}\right)$ with $\mathrm{supp}\left(f_{R}\right)\subset\mathcal{W}_{R}$.
According to Bisognano-Wichmann theorem (theorem \ref{c3-bw}), we
know that the modular Hamiltonian of the vacuum vector $\left|0\right\rangle $
is given by $K_{0}=2\pi K_{1}$, where $K_{1}$ is the infinitesimal
boost operator. Remembering that the second quantized Poincaré operator
$U\left(\Lambda_{1}^{s},0\right)=\mathrm{e}^{iK_{1}s}$, acting on
the Fock space $\mathcal{H}_{0}$, is constructed from the Poincaré
unitary operator $u\left(\Lambda_{1}^{s},0\right)=\mathrm{e}^{ik_{1}s}$,
acting on the one-particle Hilbert space $\mathfrak{H}$, then we
have that
\begin{eqnarray}
S\left(\omega_{f}\mid\omega_{0}\right)\!\!\! & = & \!\!\!\left\langle 0\right|U^{*}K_{0}U\left|0\right\rangle =2\pi\left\langle 0\right|W\left(f_{R}\right)^{*}K_{1}W\left(f_{R}\right)\left|0\right\rangle =2\pi\left\langle \mathrm{e}^{if_{R}}\right|K_{1}\left|\mathrm{e}^{if_{R}}\right\rangle \nonumber \\
 & = & \!\!\!2\pi\mathrm{e}^{-\left\Vert f_{R}\right\Vert _{\mathrm{1p}}^{2}}\sum_{n=0}^{\infty}\frac{\left(-i\right)^{n}\left(i\right)^{n}}{n!}\left\langle f_{R}^{\otimes n}\right|K_{1}\left|f_{R}^{\otimes n}\right\rangle \nonumber \\
& = & \!\!\! 2\pi\mathrm{e}^{-\left\Vert f_{R}\right\Vert _{\mathrm{1p}}^{2}}\sum_{n=0}^{\infty}\frac{n}{n!}\left\langle f_{R}^{\otimes n}\mid\left(k_{1}f_{R}\right)\otimes f_{R}^{\otimes n-1}\right\rangle \nonumber \\
 & = & \!\!\! 2\pi\mathrm{e}^{-\left\Vert f_{R}\right\Vert _{\mathrm{1p}}^{2}}\sum_{n=1}^{\infty}\frac{1}{\left(n-1\right)!}\left\langle f_{R}\right|k_{1}\left|f_{R}\right\rangle _{\mathrm{1p}}\left\langle f_{R}\mid f_{R}\right\rangle _{\mathrm{1p}}^{n-1} \nonumber \\
 & = & \!\!\! 2\pi\mathrm{e}^{-\left\Vert f_{R}\right\Vert _{\mathrm{1p}}^{2}}\left\langle f_{R}\right|k_{1}\left|f_{R}\right\rangle _{\mathrm{1p}}\mathrm{e}^{\left\Vert f_{R}\right\Vert _{\mathrm{1p}}^{2}} = 2\pi\left\langle f_{R}\right|k_{1}\left|f_{R}\right\rangle _{\mathrm{1p}}\,.\label{eq:rel_ent_easy}
\end{eqnarray}
Thus, the RE between the coherent state and the vacuum, can be expressed
in the one-particle Hilbert space $\mathfrak{H}$ in terms of the
expectation value of the boost operator $k_{1}$ in the vector $E\left(f\right)\in\mathfrak{H}$,
which generates the underlying coherent state. In the end, following
from \eqref{eq:rel_ent_easy}, we get the following theorem.
\begin{thm}
\label{thm_sr_coh_par}Let be $f_{L},f_{R}\in\mathcal{S}\left(\mathbb{R}^{d},\mathbb{R}\right)$
with $\mathrm{supp}\left(f_{L}\right)\in\mathcal{W}_{L}$, $\mathrm{supp}\left(f_{R}\right)\in\mathcal{W}_{R}$,
and $f=f_{L}+f_{R}$. Then, the relative entropy between the coherent
state $\omega_{f}$ and the vacuum $\omega_{0}$, for the algebra
of the right Rindler wedge $\mathcal{W}_{R}$ is
\begin{equation}
S\left(\omega_{f}\mid\omega_{0}\right)=2\pi\int_{x^{1}>0}\negthickspace d^{d-1}x\,x^{1}\left.\frac{1}{2}\left(\left(\frac{\partial F}{\partial x^{0}}\right)^{2}+\left|\nabla F\right|^{2}+m^{2}F^{2}\right)\right|_{x^{0}=0},\label{eq:rel_ent_easy2}
\end{equation}
where $F\left(x\right)=\int_{\mathbb{R}^{d}}\Delta\left(x-y\right)\,f\left(y\right)d^{d}y=\int_{\mathbb{R}^{d}}\Delta\left(x-y\right)\,\left[f_{L}\left(y\right)+f_{R}\left(y\right)\right]d^{d}y$
. In addition, formula \eqref{eq:rel_ent_easy2} does not depend on
the chosen function $f_{L}$ (with support in $\mathcal{W}_{L}$).
\end{thm}

\begin{proof}
A straightforward calculation explained in appendix \ref{b2-1particle},
allows us to obtain \eqref{apend_sr_easy} from \eqref{eq:rel_ent_easy}.
However, there are two differences between \eqref{apend_sr_easy}
and \eqref{eq:rel_ent_easy2} (beyond the obvious $2\pi$ in front
of the expression). The first one is that, in \eqref{apend_sr_easy}
the integral is along the whole space $\mathbb{R}^{d-1}$, and the
second one is that the function $F$ in \eqref{apend_sr_easy} is
computed using only $f_{R}$. To finally pass from \eqref{apend_sr_easy}
to \eqref{eq:rel_ent_easy2} we have to do the following two changes.
First we notice that $\mathrm{supp}\left(f_{R}\right)\subset\mathcal{W}_{R}$
implies that $\mathrm{supp}\left(\left.F\right|_{x^{0}=0}\right)\subset\Sigma_{R}$,
and hence, we can replace the integration region in \eqref{apend_sr_easy}
by $\Sigma_{R}$. In the same way, since $\mathrm{supp}\left(f_{L}\right)\subset\mathcal{W}_{L}$,
we have that the function $F_{L}\left(x\right):=\int_{\mathbb{R}^{d}}\Delta\left(x-y\right)\,f_{L}\left(y\right)d^{d}y$
vanishes along $\Sigma_{R}$, and hence, \eqref{eq:rel_ent_easy2}
holds. This also implies that \eqref{eq:rel_ent_easy2} does not depends
on $f_{L}$.
\end{proof}
As a remark, the outcome of the above theorem coincides with \eqref{c4-cali}
for the canonical stress tensor \eqref{c4-stress} as well as with
any other improvement term, since the generating function of the coherent
state identically vanishes along the entanglement surface $x^{1}=0$.

\subsection{Hard case: $f\protect\neq f_{L}+f_{R}$ \label{subsec:Hard-case}}

In this section, we assume that the function $f\in\mathcal{S}\left(\mathbb{R}^{d},\mathbb{R}\right)$
has $\mathrm{supp}\left(f\right)\not\subset\mathcal{W}_{L},\mathcal{W}_{R}$.
Moreover, we assume that $\mathrm{supp}\left(f\right)$ is compact
to avoid possible complications coming from integrals along regions
of infinite size. In the end, we are interested in the behavior of
the RE around the entanglement surface $\partial\Sigma_{R}=\left\{ \bar{x}\in\mathbb{R}^{d-1}\,:\,x^{1}=0\right\} $,
which can be captured with a compactly supported coherent state.

Before we continue, we want to remark that, in this case, the RE must
be finite. The proof is as follows. Since $\mathrm{supp}\left(f\right)$
is compact, there exists a bigger right wedge $\tilde{\mathcal{W}}_{R}\supset\mathcal{W}_{R}$
such that $W\left(f\right)\in\tilde{\mathcal{W}}_{R}$. Then the RE
between the coherent and the vacuum in the algebra $\mathcal{A}(\tilde{\mathcal{W}}_{R})$
is as the one computed in the previous section, which is finite because
the generating function $f$ is smooth. Then by monotonicity, the
RE for the original wedge $\mathcal{W}_{R}$ must be finite. In particular,
we are allowed to use expression \eqref{c2_rel_ent_f}.

The first question which arises is if we could split the unitary into
two unitaries, one belonging to the right wedge $\mathcal{W}_{R}$
and the other to the left wedge $\mathcal{W}_{L}$. In other words,
if there exist unitaries $U_{R}\in\mathcal{A}_{\mathcal{W}}$ and
$U_{L}\in\mathcal{A}_{\mathcal{W}'}$ such that $W\left(f\right)=U_{L}U_{R}$.
Unfortunately, the answer is negative for the most general interesting
case. This fact arises when we try to explicitly split $W\left(f\right)$.
To begin with, it seems natural to split the function $f$ simply
as
\begin{eqnarray}
f_{R}\left(x\right)\!\!\! & := & \!\!\!\Theta_{\mathcal{W}_{R}}\left(x\right)f\left(x\right)\,,\label{eq:partir_mal}\\
f_{L}\left(x\right)\!\!\! & := & \!\!\!\Theta_{\mathcal{W}_{L}}\left(x\right)f\left(x\right)\,,
\end{eqnarray}
where $\Theta_{\mathcal{W}_{R}}$ is the characteristic function of
the right Rindler wedge (equivalently for $\Theta_{\mathcal{W}_{L}}$).
However, it will lead to a wrong result, since $f_{R}+f_{L}\neq f$.
Moreover, if for example we start with a function $f$ supported in
the upper light-cone $V^{+}:=\left\{ x\in\mathbb{R}^{d}\,:\,x^{0}>\left|\bar{x}\right|\right\} $,
then equation \eqref{eq:partir_mal} implies that $f_{R}\equiv0$
and hence we obtain $S\left(\omega_{f}\mid\omega_{0}\right)=0$, which
is obviously the wrong result. To make a consistent splitting, we
must use the relations explained in subsection \ref{subsec:Relation-between-algebras}.
Given the spacetime function $f\in\mathcal{S}\left(\mathbb{R}^{d},\mathbb{R}\right)$,
we can construct $f_{\varphi},\,f_{\pi}\in\mathcal{S}\left(\mathbb{R}^{d-1},\mathbb{R}\right)$
satisfying the relation \eqref{eq:segal_vs_weyl}. The correct result
is to split the functions $f_{\varphi},f_{\pi}$, which are the initial
data at $x^{0}=0$ of the Klein-Gordon solution generated by $f$.
The assumption $\mathrm{supp}(f)\not\subset\mathcal{W}_{L},\mathcal{W}_{R}$
implies that $f_{\varphi}$ and $f_{\pi}$ do not vanish in any open
neighborhood of the entaglement surface $\partial\Sigma_{R}$. Now,
we would like to write
\begin{eqnarray}
f_{\varphi}=f_{\varphi,L}+f_{\varphi,R} & \mathrm{and} & f_{\pi}=f_{\pi,L}+f_{\pi,R}\:\textrm{,}
\end{eqnarray}
with $\mathrm{supp}(f_{\nu,L})\in\Sigma_{R}'$ and $\mathrm{supp}(f_{\nu,R})\in\Sigma_{R}$
($\nu=\varphi,\pi$). The right way to do this is defining
\begin{eqnarray}
f_{\nu,L}\left(\bar{x}\right):=f_{\nu}\left(\bar{x}\right)\cdot\Theta\left(-x^{1}\right) & \mathrm{and} & f_{\nu,R}\left(\bar{x}\right):=f_{\nu}\left(\bar{x}\right)\cdot\Theta\left(x^{1}\right)\,,
\end{eqnarray}
where $\Theta$ is the usual step Heaviside function. The problem
is that $f_{\nu,L}$ and $f_{\nu,R}$ are no longer smooth, and nothing
guarantees that $E_{\nu}\left(f_{\nu,R}\right)\in\mathfrak{H}_{\nu}$
(the same problem occurs for $f_{\nu,L}$). More precisely, since
$f_{\nu,R}\in L^{2}\left(\mathbb{R}^{d-1},\mathbb{R}\right)=H^{0}\left(\mathbb{R}^{d-1},\mathbb{R}\right)$,
and because of the inclusions (see appendix \ref{AP-sec-sobo})
\begin{equation}
H^{\frac{1}{2}}\left(\mathbb{R}^{d-1},\mathbb{R}\right)\subset H^{0}\left(\mathbb{R}^{d-1},\mathbb{R}\right)\subset H^{-\frac{1}{2}}\left(\mathbb{R}^{d-1},\mathbb{R}\right)\,,
\end{equation}
we have that $f_{\varphi,R}\in\mathfrak{H}_{\varphi}$ but $f_{\pi,R}\notin\mathfrak{H}_{\pi}$.
In other words, $f_{\pi,R}$ is not an appropriate smear function
for the canonical conjugate field $\pi\left(\bar{x}\right)$. This
problem does not appear because the test function is no longer smooth,
it just arises because $f_{\pi,R}$ is no longer continuous. On the
other hand, if $f_{\pi,R}$ is continuous, this problem can be solved
due to the following lemma. 
\begin{thm}
\label{par:lemma-1}Let $f\in L^{2}\left(\mathbb{R}^{n}\right)\cap C^{0}\left(\mathbb{R}^{n}\right)\cap C_{t}^{1}\left(\mathbb{R}^{n}\right)$
and $\partial_{j}f\in L^{2}\left(\mathbb{R}^{n}\right)$ for $j=1,\ldots,n$.\footnote{$C_{t}^{1}\left(\mathbb{R}^{n}\right)$ is the set of piecewise differentiable
functions. See appendix \ref{AP-sec-sobo} for a precise definition.} Then $f\in H^{1}\left(\mathbb{R}^{n}\right)$. 
\end{thm}

\begin{proof}
See appendix \ref{AP-sec-sobo}. 
\end{proof}
Then, having this in mind, the strategy we adopt below is to make
a splitting for another smear function, which, by construction, we
know  is continuous.

\subsubsection{A lemma for the relative modular group}

In this subsection, we prove a lemma that gives a general expression
for the relative modular flow, under the assumption that some operator
can be written as a product of two new operators, one belonging to
$\mathcal{A}$ and other to $\mathcal{A}'$. In the following subsection,
we prove that this assumption holds for the free real scalar field.
For simplicity and due to the symmetry relation \eqref{sym-sr}, in
the following, we work with the modular operator $\Delta_{\Phi,0}$
instead of $\Delta_{0,\Phi}$.

As a motivation, we remember from section \ref{c2_sec-mod_th}, that
contrary to the modular group $\Delta_{\Omega}^{it}$ and the relative
modular group $\Delta_{\Phi,\Omega}^{it}$, the family of operators
$u_{\Phi,\Omega}\left(t\right)=\Delta_{\Phi,\Omega}^{it}\Delta_{\Omega}^{it}$
belongs to $\mathcal{A}$ for all $t\in\mathbb{R}$ (see lemma \ref{c2_cnr_lemma}).
This suggests that the computation of $u_{\Phi,\Omega}\left(t\right)$
could involve the splitting of some test function, which would lead
to a well-defined operator. To gain some intuition, using the lemmas
\ref{lema_auto} and \ref{lema_easy_case}, we know that
\begin{equation}
u_{\Phi,\Omega}\left(t\right)=U^{*}\Delta_{\Omega}^{it}U\Delta_{\Omega}^{-it}\quad\textrm{whenever }\left|\Phi\right\rangle =U'U\left|\Omega\right\rangle \textrm{ with }U\in\mathcal{A},\,U'\in\mathcal{A}'\,.
\end{equation}
This expression motivates the following lemma.
\begin{lem}
\label{l_par_uni}Let $\mathcal{A}\subset\mathcal{B}\left(\mathcal{H}\right)$
be a vN factor, $\left|\Omega\right\rangle $ a standard vector, $U\in\mathcal{B}\left(\mathcal{H}\right)$
a unitary such that $U^{*}\mathcal{A}U=\mathcal{A}$, and $\left|\Phi\right\rangle =U\left|\Omega\right\rangle $.
If there exist families of unitaries\footnote{\textit{They are not necessarily one-parameter groups for $t\in\mathbb{R}$.}}
$V\left(t\right)\in\mathcal{A}$ and $V'\left(t\right)\in\mathcal{A}'$
such that
\begin{equation}
\begin{cases}
U^{*}\Delta_{\Omega}^{it}U\Delta_{\Omega}^{-it}=V\left(t\right)V'\left(t\right)\,,\quad\forall t\in\mathbb{R}\,,\\
V\left(0\right)=V'\left(0\right)=\mathbf{1}\,.
\end{cases}\label{eq:partir_unitario}
\end{equation}
Then, there exists a real function $\alpha:\mathbb{R}\rightarrow\mathbb{R}$
with $\alpha\left(0\right)=0$ such that
\begin{equation}
\Delta_{\Phi,\Omega}^{it}=\mathrm{e}^{-i\alpha\left(t\right)}V\left(t\right)\Delta_{\Omega}^{it}\,\textrm{.}\label{eq:rel_mod_decom_2}
\end{equation}
\end{lem}

\begin{proof}
We first see that $V\left(t\right)\Delta_{\Omega}^{it}$ acts as $\Delta_{\Phi,\Omega}^{it}$
on every $A\in\mathcal{A}$ and $A'\in\mathcal{A}'$. Indeed, we have
that
\begin{eqnarray}
\mathcal{A}\ni V\left(t\right)\Delta^{it}A\Delta^{-it}V\left(t\right)^{*}\!\!\! & = & \!\!\!V\left(t\right)V'\left(t\right)\Delta_{\Omega}^{it}A\Delta_{\Omega}^{-it}V\left(t\right)^{*}V'\left(t\right)^{*}\nonumber \\
 & = & \!\!\!U\Delta_{\Omega}^{it}U^{*}\Delta_{\Omega}^{-it}\,\Delta_{\Omega}^{it}A\Delta_{\Omega}^{-it}\,\Delta_{\Omega}^{it}U\Delta_{\Omega}^{-it}U^{*} \nonumber \\
 & = & \!\!\!U\Delta_{\Omega}^{it}U^{*}AU\Delta_{\Omega}^{-it}U^{*}=\Delta_{\Phi}^{it}A\Delta_{\Phi}^{-it}=\Delta_{\Phi,\Omega}^{it}A\Delta_{\Phi,\Omega}^{-it}\,\textrm{,} \hspace{.9cm}
\end{eqnarray}
where we have used 2 in lemma \ref{lema_auto}. Similarly, we have
that
\begin{equation}
V\left(t\right)\underset{\in\mathcal{A}{}_{\mathcal{W}'}}{\underbrace{\Delta_{\Omega}^{it}A'\Delta_{\Omega}^{-it}}}V\left(t\right)^{*}=V\left(t\right)V\left(t\right)^{*}\,\Delta_{\Omega}^{it}A'\Delta_{\Omega}^{-it}=\Delta_{\Omega}^{it}A'\Delta_{\Omega}^{-it}=\Delta_{\Phi,\Omega}^{it}A'\Delta_{\Phi,\Omega}^{-it}\,.
\end{equation}
Then, for all $B\in\mathcal{A}\cup\mathcal{A}'$ we have that
\begin{equation}
\left(V\left(t\right)\Delta_{\Omega}^{it}\right)B\left(V\left(t\right)\Delta_{\Omega}^{it}\right)^{*}=\Delta_{\Phi,\Omega}^{it}B\Delta_{\Phi,\Omega}^{-it}\:\Rightarrow\:\left[B,\left(V\left(t\right)\Delta_{\Omega}^{it}\right)^{*}\Delta_{\Phi,\Omega}^{it}\right]=0\,,
\end{equation}
and hence $\left(V\left(t\right)\Delta_{\Omega}^{it}\right)^{*}\Delta_{\Phi,\Omega}^{it}$
belongs to the center $\left(\mathcal{A}\cup\mathcal{A}'\right)'=\mathcal{A}\cap\mathcal{A}'=\left\{ \lambda\cdot\mathbf{1}\right\} $,
which is trivial since $\mathcal{A}$ is a factor. This means that
there exist a function $\lambda:\mathbb{R}\rightarrow\mathbb{C}$
such that
\begin{equation}
\Delta_{\Phi,\Omega}^{it}=\lambda\left(t\right)V\left(t\right)\Delta_{\Omega}^{it}\,.\label{eq:rel_mod_decom}
\end{equation}
Moreover, evaluating the above expression at $t=0$ we get that $\lambda\left(0\right)=1$.
Finally, since all operators in \eqref{eq:rel_mod_decom} are unitaries,
we must have that $\lambda\left(t\right)=\mathrm{e}^{-i\alpha\left(t\right)}$
for some function $\alpha:\mathbb{R}\rightarrow\mathbb{R}$ with $\alpha\left(0\right)=0$,
and then, \eqref{eq:rel_mod_decom_2} holds.
\end{proof}
Under the hypothesis of the above lemma, we can obtain the relative
modular Hamiltonian deriving \eqref{eq:rel_mod_decom_2} at $t=0$
\begin{equation}
K_{\Phi,\Omega}=i\,\left.\frac{\mathrm{d}}{\mathrm{d}t}\right|_{t=0}\Delta_{\Phi,\Omega}^{it}=i\,\left.\frac{\mathrm{d}}{\mathrm{d}t}\right|_{t=0}\mathrm{e}^{-i\alpha\left(t\right)}V\left(t\right)\Delta_{\Omega}^{it}=\alpha'\left(0\right)\mathbf{1}+i\dot{V}\left(0\right)+K_{\Omega}\,,\label{eq:rel_mod_ham_p}
\end{equation}
where the derivative in \eqref{eq:rel_mod_ham_p} has to be understood
as a limit in the strong operator topology of $\mathcal{H}$ \cite{ReedSimon}.
This formula gives a well-defined expression for the relative modular
Hamiltonian up to a constant. One way to determine such a constant
is using that $\Delta_{\Phi,\Omega}^{it}$ is a one-parameter group
of unitaries and must fulfill the concatenation equation
\begin{eqnarray}
\Delta_{\Phi,\Omega}^{it_{1}}\Delta_{\Phi,\Omega}^{it_{2}}=\Delta_{\Phi,\Omega}^{i\left(t_{1}+t_{2}\right)}\,, &  & \forall t_{1},t_{2}\in\mathbb{R}\,.\label{eq:concatenar}
\end{eqnarray}
We will discuss the computation to determine $\alpha'\left(0\right)$
in subsection \ref{subsec:determ_alpha}.

\subsubsection{Relative modular group\label{subsec:Computation-the-splitting}}

In this subsection, we show that lemma \ref{l_par_uni} applies for
the free real scalar field and coherent states. More concretely, we
have the following theorem.
\begin{thm}
\label{th_mod_rel_coh}For the algebra of the Rindler wedge $\mathcal{A}_{\mathcal{W}}$,
a Weyl unitary $U=W\left(f\right)$ with $f\in\mathcal{S}\left(\mathbb{R}^{d},\mathbb{R}\right)$,
the vacuum vector $\left|0\right\rangle $ and the vector $\left|\Phi\right\rangle =U\left|0\right\rangle $,
the hypothesis of lemma \ref{l_par_uni} holds. In particular, we
have that
\begin{equation}
\Delta_{\Phi,0}^{it}=\mathrm{e}^{i\alpha\left(s\right)}W_{\varphi}\left(g_{\varphi,R}^{s}\right)W_{\pi}\left(g_{\pi,R}^{s}\right)\Delta_{0}^{it}=\mathrm{e}^{i\alpha\left(s\right)}\mathrm{e}^{i\varphi\left(g_{\varphi,R}^{s}\right)}\mathrm{e}^{i\pi\left(g_{\pi,R}^{s}\right)}\mathrm{e}^{isK_{1}}\:\textrm{,}\label{rel_mod_flow_coh}
\end{equation}
where we have denoted $s:=-2\pi t$ and
\begin{eqnarray}
g_{\varphi,R}^{s}\left(\bar{x}\right)\!\!\! & = & \!\!\!-\frac{\partial G^{s}}{\partial x^{0}}\left(0,\bar{x}\right)\Theta\left(x^{1}\right)\,,\\
g_{\pi,R}^{s}\left(\bar{x}\right)\!\!\! & = & \!\!\!G^{s}\left(0,\bar{x}\right)\Theta\left(x^{1}\right)\,,\\
G^{s}\left(x\right)\!\!\! & = & \!\!\!\int_{\mathbb{R}^{d}}\Delta\left(x-y\right)\,\left[f\left(\Lambda_{1}^{-s}y\right)-f\left(y\right)\right]d^{d}y\,.
\end{eqnarray}
\end{thm}

\begin{proof}
According to the discussion in subsection \ref{c4-subsec-net_local},
we have that $\mathcal{A}_{\mathcal{W}}$ is a vN factor. From the
Reeh-Schlieder theorem, we have that the vacuum vector $\left|0\right\rangle $
is standard. Also, we have that any Weyl unitary satisfies $W\left(\boldsymbol{h}\right)^{*}\mathcal{A}_{\mathcal{W}}W\left(\boldsymbol{h}\right)=\mathcal{A}_{\mathcal{W}}$.
From now on, we set $s=-2\pi t$. Replacing $U=W\left(f\right)$ into
\eqref{eq:partir_unitario}, we get
\begin{eqnarray}
W\left(f\right)^{*}\Delta_{0}^{it}W\left(-f\right)\Delta_{0}^{-it}\!\!\! & = & \!\!\!W\left(-f\right)\mathrm{e}^{isK_{1}}W\left(f\right)\mathrm{e}^{-isK_{1}}=W\left(-f\right)W\left(f_{\left(\Lambda_{1}^{s},0\right)}\right)\nonumber \hspace{.9cm} \\
 & = & \!\!\!\mathrm{e}^{i\mathrm{Im}\left\langle f\mid f^{s}\right\rangle _{\mathrm{1p}}}W\left(f^{s}-f\right)\,,\label{eq:partir_unitario_2}
\end{eqnarray}
where we have defined $f^{s}:=f_{\left(\Lambda_{1}^{s},0\right)}$.
Applying the decomposition \eqref{eq:segal_vs_weyl_1p} to $g^{s}:=f^{s}-f$,
we have that
\begin{eqnarray}
 W\left(f\right)^{*}\Delta_{0}^{it}W\left(f\right)\Delta_{0}^{-it}\!\!\! & = & \!\!\! \mathrm{e}^{i\mathrm{Im}\left\langle f\mid f^{s}\right\rangle _{\mathrm{1p}}}W\left(g^{s}\right) \nonumber \\
 \!\!\! & = & \!\!\! \mathrm{e}^{i\mathrm{Im}\left\langle f\mid f^{s}\right\rangle _{\mathrm{1p}}}\mathrm{e}^{i\mathrm{Im}\langle g_{\varphi}^{s}\mid g_{\pi}^{s}\rangle_{\mathrm{1p}}}W_{\varphi}\left(g_{\varphi}^{s}\right)W_{\pi}\left(g_{\pi}^{s}\right)\,,\label{eq:partir_unitario_3}
\end{eqnarray}
with
\begin{eqnarray}
g_{\varphi}^{s}\left(\bar{x}\right)\!\!\! & := & \!\!\!-\frac{\partial G^{s}}{\partial x^{0}}\left(0,\bar{x}\right)=-\cosh\left(s\right)\frac{\partial F}{\partial x^{0}}\left(\bar{x}^{s}\right)+\sinh\left(s\right)\frac{\partial F}{\partial x^{1}}\left(\bar{x}^{s}\right)+\frac{\partial F}{\partial x^{0}}\left(0,\bar{x}\right)\,, \hspace{.7cm} \label{eq:fpi_partible}\\
g_{\pi}^{s}\left(\bar{x}\right)\!\!\! & := & \!\!\!G^{s}\left(0,\bar{x}\right)=F\left(\bar{x}^{s}\right)-F\left(0,\bar{x}\right)\,,
\end{eqnarray}
where $\bar{x}^{s}:=\left(\Lambda_{1}^{-s}x\right)_{x^{0}=0}=\left(-x^{1}\sinh\left(s\right),x^{1}\cosh\left(s\right),\bar{x}_{\bot}\right)$
and
\begin{equation}
G^{s}\left(x\right):=\int_{\mathbb{R}^{d}}\Delta\left(x-y\right)\,\left[f^{s}\left(y\right)-f\left(y\right)\right]d^{d}y\,.\label{eq:Gs}
\end{equation}
Now, we explicitly split the unitaries $W_{\varphi}\left(g_{\varphi}^{s}\right)$
and $W_{\pi}\left(g_{\pi}^{s}\right)$ in equation \eqref{eq:partir_unitario_3}
defining
\begin{eqnarray}
g_{\varphi,R}^{s}\left(\bar{x}\right):=g_{\varphi}^{s}\left(\bar{x}\right)\Theta\left(x^{1}\right) & \mathrm{and} & g_{\varphi,L}^{s}\left(\bar{x}\right):=g_{\varphi}^{s}\left(\bar{x}\right)\Theta\left(-x^{1}\right)\,,\label{eq:split_phi}\\
g_{\pi,R}^{s}\left(\bar{x}\right):=g_{\pi}^{s}\left(\bar{x}\right)\Theta\left(x^{1}\right) & \mathrm{and} & g_{\pi,L}^{s}\left(\bar{x}\right):=g_{\pi}^{s}\left(\bar{x}\right)\Theta\left(-x^{1}\right)\,,\label{eq:split_pi}
\end{eqnarray}
which evidently implies that $g_{\varphi,L}^{s}+g_{\varphi,R}^{s}=g_{\varphi}^{s}$
and $g_{\pi,L}^{s}+g_{\pi,R}^{s}=g_{\pi}^{s}$. Moreover, we have that
\begin{equation}
\!\negthickspace\negthickspace\negthickspace\negthickspace\negthickspace\begin{array}{cc}
\left.\begin{array}{c}
g_{\varphi,R}^{s},g_{\varphi,L}^{s}\in L^{2}\left(\mathbb{R}^{d-1},\mathbb{R}\right)\subset H^{-\frac{1}{2}}\left(\mathbb{R}^{d-1},\mathbb{R}\right)\\
\mathrm{supp}\left(g_{\varphi,R}^{s}\right)\subset\Sigma_{R}\textrm{ and }\mathrm{supp}\left(g_{\varphi,L}^{s}\right)\subset\Sigma{}_{R}'
\end{array}\negthickspace\right\}  & \negthickspace\negthickspace\Rightarrow\end{array}g_{\varphi,R}^{s}\in K_{\varphi}\left(\Sigma_{R}\right)\textrm{ and }g_{\varphi,L}^{s}\in K_{\varphi}\left(\Sigma{}_{R}'\right).
\end{equation}
Furthermore, we have that $g_{\pi,R}^{s},\,g_{\pi,L}^{s}$ are real-valued
functions and they satisfy the hypothesis of lemma \eqref{par:lemma-1}.
Then, we have that
\begin{equation}
\negthickspace\negthickspace\negthickspace\negthickspace\negthickspace\begin{array}{cc}
\left.\begin{array}{c}
g_{\pi,R}^{s},g_{\pi,L}^{s}\in H^{1}\left(\mathbb{R}^{d-1},\mathbb{R}\right)\subset H^{\frac{1}{2}}\left(\mathbb{R}^{d-1},\mathbb{R}\right)\\
\mathrm{supp}\left(g_{\pi,R}^{s}\right)\subset\Sigma_{R}\textrm{ and }\mathrm{supp}\left(g_{\pi,L}^{s}\right)\subset\Sigma_{R}'
\end{array}\negthickspace\right\}  & \negthickspace\negthickspace\Rightarrow\end{array}g_{\pi,R}^{s}\in K_{\pi}\left(\Sigma_{R}\right)\textrm{ and }g_{\pi,L}^{s}\in K_{\pi}\left(\Sigma_{R}'\right),
\end{equation}
which means that the splits (\ref{eq:split_phi}-\ref{eq:split_pi})
work. Coming back to \eqref{eq:partir_unitario_3}, we have that
\begin{eqnarray}
 W\!\left(f\right)\Delta_{0}^{it}W\!\left(f\right)^{*}\Delta_{0}^{-it}\!\!\!\! & = & \!\!\!\mathrm{e}^{i\mathrm{Im}\langle f\mid f^{s}\rangle_{\mathrm{1p}}}\mathrm{e}^{i\mathrm{Im}\langle g_{\varphi}^{s}\mid g_{\pi}^{s}\rangle_{\mathrm{1p}}}W_{\varphi}\left(g_{\varphi,L}^{s}+g_{\varphi,R}^{s}\right)W_{\pi}\left(g_{\pi,L}^{s}+g_{\pi,R}^{s}\right)\hspace{2cm} \nonumber \\
 & = & \!\!\!\mathrm{e}^{i\mathrm{Im}\langle f\mid f^{s}\rangle_{\mathrm{1p}}}\mathrm{e}^{i\mathrm{Im}\langle g_{\varphi}^{s}\mid g_{\pi}^{s}\rangle_{\mathrm{1p}}}W_{\varphi}\left(g_{\varphi,L}^{s}\right)W_{\varphi}\left(g_{\varphi,R}^{s}\right)W_{\pi}\left(g_{\pi,L}^{s}\right)W_{\pi}\left(g_{\pi,R}^{s}\right)\nonumber \\
 & = & \!\!\!\mathrm{e}^{i\mathrm{Im}\langle f\mid f^{s}\rangle_{\mathrm{1p}}}\mathrm{e}^{i\mathrm{Im}\langle g_{\varphi}^{s}\mid g_{\pi}^{s}\rangle_{\mathrm{1p}}}\underset{\in\mathcal{A}\mathcal{_{W}}}{\underbrace{W_{\varphi}\!\left(g_{\varphi,R}^{s}\right)\!W_{\pi}\!\left(g_{\pi,R}^{s}\right)}}\underset{\in\mathcal{A}'\mathcal{_{W}}}{\underbrace{W_{\varphi}\!\left(g_{\varphi,L}^{s}\right)\!W_{\pi}\!\left(g_{\pi,L}^{s}\right)}}.
\end{eqnarray}
Finally, replacing $V\left(t\right)=W_{\varphi}\left(g_{\varphi,R}^{s}\right)W_{\pi}\left(g_{\pi,R}^{s}\right)$
into \eqref{eq:rel_mod_decom_2}, we arrive to \eqref{rel_mod_flow_coh}.
\end{proof}
\begin{rem}
Using that the relative modular flow $\Delta_{\Phi,0}^{it}$ is strongly
continuous and that the relative entropy $S\left(\omega_{f}\mid\omega_{0}\right)$
is finite (see discussion at the beginning of section \ref{subsec:Hard-case}),
and hence the expression \eqref{c2_rel_ent_f} holds, it is not difficult
to prove that the function $t\mapsto\left\langle 0\right|\Delta_{\Phi,0}^{it}\left|0\right\rangle $
is continuously differentiable. Furthermore, taking the vacuum expectation
value on the r.h.s of \eqref{rel_mod_flow_coh}, it can be shown that
the function $\alpha\left(s\right)\in C^{1}\left(\mathbb{R}\right)$. 

To conclude, according to \eqref{eq:rel_mod_ham_p}, we have the following
expression for the relative modular Hamiltonian
\begin{equation}
K_{\Phi,0}=2\pi\left(\alpha'\left(0\right)\mathbf{1}+\varphi\left(h_{\varphi,R}\right)+\pi\left(h_{\pi,R}\right)+K_{1}\right)\,,\label{eq:explicit_rel_mod}
\end{equation}
where
\begin{eqnarray}
h_{\varphi,R}\left(\bar{x}\right)\!\!\!\! & := & \!\!\!\!\left.\frac{\mathrm{d}}{\mathrm{d}s}\right|_{s=0}g_{\varphi,R}^{s}\left(\bar{x}\right)=\left(x^{1}\frac{\partial^{2}F}{\left(\partial x^{0}\right)^{2}}\left(0,\bar{x}\right)+\frac{\partial F}{\partial x^{1}}\left(0,\bar{x}\right)\right)\cdot\Theta\left(x^{1}\right)\,,\\
h_{\pi,R}\left(\bar{x}\right)\!\!\!\! & := & \!\!\!\!\left.\frac{\mathrm{d}}{\mathrm{d}s}\right|_{s=0}g_{\pi,R}^{s}\left(\bar{x}\right)=\left(-x^{1}\frac{\partial F}{\partial x^{0}}\left(0,\bar{x}\right)\right)\cdot\Theta\left(x^{1}\right)\,.\label{eq:functions_explicit_rel_mod}
\end{eqnarray}
Using similar arguments as above, we have that $h_{\varphi,R}\in K_{\varphi}\left(\Sigma_{R}\right)$
and $h_{\pi,R}\in K_{\pi}\left(\Sigma_{R}\right)$.\footnote{An explicit computation of the strong derivative in equation \eqref{eq:explicit_rel_mod}
shows that the vacuum vector $\left|0\right\rangle $, any coherent
vector, and any vector of finite number of particles belong to the
domain of $K_{\Phi,0}$.}
\end{rem}

Before we determine the constant $\alpha'\left(0\right)$, we want
to emphasize its importance
\begin{equation}
S\left(\omega_{f}\mid\omega_{0}\right)=\left\langle 0\right|K_{\Phi,0}\left|0\right\rangle =2\pi\alpha'\left(0\right)\,.\label{srel_alpha}
\end{equation}
Thus, the constant $\alpha'\left(0\right)$ is, in fact, the desired
result for the RE. Regardless of the problem of computing the value
of $\alpha'\left(0\right)$, expressions (\ref{eq:explicit_rel_mod}-\ref{eq:functions_explicit_rel_mod})
give us an explicit exact expression for the relative modular Hamiltonian
$K_{\Phi,0}$ up to a constant. It is interesting to notice that the
difference $K_{\Phi,0}-K_{0}$ is just a linear term on the field
operators plus a constant term. We expect that this structure holds
not just for the Rindler wedge, but for any kind of region as long
as $\left|\Phi\right\rangle =W\left(f\right)\left|0\right\rangle $
is a coherent vector.

\subsubsection{Relative entropy\label{subsec:determ_alpha}}

As we have explained in equation \eqref{srel_alpha}, we need to determine
the constant $\alpha'\left(0\right)$ in order to obtain the final
result for the RE. Most of the calculations along this section are
straightforward, therefore, we present the detailed computations in
appendix \ref{subsec:appendix_3}. As in theorem \ref{th_mod_rel_coh},
throughout this section we set $s:=-2\pi t$.

We start taking the vacuum expectation value on both sides in expression
\eqref{eq:concatenar},
\begin{equation}
\left\langle 0\right|\Delta_{\Phi,0}^{it_{1}}\Delta_{\Phi,0}^{it_{2}}\left|0\right\rangle =\left\langle 0\right|\Delta_{\Phi,0}^{i\left(t_{1}+t_{2}\right)}\left|0\right\rangle \,,\label{eq:concatenar2}
\end{equation}
and we replace the expression \eqref{rel_mod_flow_coh} obtained for
the relative modular flow (see equations (\ref{apen_conca_1}-\ref{apen_conca_2})).
Applying $\left.\frac{\mathrm{d}}{\mathrm{d}s_{1}}\right|_{s_{1}=0}=-\frac{1}{2\pi}\left.\frac{\mathrm{d}}{\mathrm{d}t_{1}}\right|_{t_{1}=0}$
on both sides in \eqref{eq:concatenar2} (equations (\ref{apen_der_1}-\ref{apen_der_2}))
and matching real and imaginary parts separately, we get\footnote{Analytic properties of the relative modular flow ensures that both
sides of \eqref{eq:concatenar2} are continuous differentiable functions
on $t_{1}$ and $t_{2}$.}\footnote{The $\frac{\mathrm{d}}{\mathrm{d}s_{2}}$ in \eqref{eq:ec_dif_alpha}
appears because in some terms the dependance on $s_{1}$ is through
$s_{1}+s_{2}$. }

\begin{eqnarray}
\alpha'\left(s_{2}\right)-\frac{\mathrm{d}}{\mathrm{d}s_{2}}\mathrm{Im}\langle g_{\varphi,R}^{s_{2}}\mid g_{\pi,R}^{s_{2}}\rangle_{\mathrm{1p}}\!\!\! & = & \!\!\!\alpha'\left(0\right)-\left.\frac{\mathrm{d}}{\mathrm{d}s_{1}}\right|_{s_{1}=0}\mathrm{Im}\langle g_{R}^{s_{1}}\mid g_{R}^{s_{2}}\rangle_{\mathrm{1p}}\,,\label{eq:ec_dif_alpha}\\
\nonumber \\
\left.\frac{\mathrm{d}}{\mathrm{d}s_{1}}\right|_{s_{1}=0}\left\Vert g_{R}^{s_{1}+s_{2}}\right\Vert _{\mathrm{1p}}^{2}\!\!\! & = & \!\!\!\left.\frac{\mathrm{d}}{\mathrm{d}s_{1}}\right|_{s_{1}=0}\left\Vert g_{R}^{s_{1}}+u\left(\Lambda_{1}^{s_{1}},0\right)g_{R}^{s_{2}}\right\Vert _{1p}^{2}\,,\label{eq:obvia}
\end{eqnarray}
where $g_{R}^{s}:=E_{\varphi}\left(g_{\varphi,R}^{s}\right)+E_{\pi}\left(g_{\pi,R}^{s}\right)$.
The second equation is useless to determine $\alpha'\left(0\right)$,
then we focus in the first one, which is a differential equation for
$\alpha'\left(s\right)$ with the particularity that $\alpha'\left(0\right)$
appears on it. To solve it, let us analyze the second term on the
right-hand side in equation \eqref{eq:ec_dif_alpha}. In appendix
\ref{subsec:appendix_3}, we compute 
\begin{equation}
2\mathrm{Im}\langle g_{R}^{s_{1}}\mid g_{R}^{s_{2}}\rangle_{\mathrm{1p}}=P\left(s_{1}\right)+Q\left(s_{1},s_{2}\right)-R\left(s_{1},s_{2}\right)+\gamma\left(s_{2}\right)\,,\label{c5-gamma}
\end{equation}
where
\begin{eqnarray}
P\left(s_{1}\right)\!\!\! & := & \!\!\!\int_{x^{1}>0}\negthickspace f_{\varphi}\left(\bar{x}\right)f_{\pi}^{s_{1}}\left(\bar{x}\right)\,d^{d-1}x-\int_{x^{1}>0}\negthickspace f_{\varphi}^{s_{1}}\left(\bar{x}\right)f_{\pi}\left(\bar{x}\right)\,d^{d-1}x\,,\\
Q\left(s_{1},s_{2}\right)\!\!\! & := & \!\!\!\int_{x^{1}>0}\negthickspace f_{\varphi}^{s_{1}}\left(\bar{x}\right)f_{\pi}^{s_{2}}\left(\bar{x}\right)\,d^{d-1}x\,,\\
R\left(s_{1},s_{2}\right)\!\!\! & := & \!\!\!\int_{x^{1}>0}\negthickspace f_{\varphi}^{s_{2}}\left(\bar{x}\right)f_{\pi}^{s_{1}}\left(\bar{x}\right)\,d^{d-1}x\,\textrm{.}\label{eq:QyR}
\end{eqnarray}
The function $\gamma$ in \eqref{c5-gamma} includes all the $s_{1}$-independent
terms, which do not contribute to \eqref{eq:ec_dif_alpha}. In the
same appendix we analyze $P,\,Q,\,R$ carefully, and we obtain

\begin{equation}
\left.\frac{\mathrm{d}P}{\mathrm{d}s_{1}}\right|_{s_{1}=0}=\int_{x^{1}>0}\negthickspace d^{d-1}x\,x^{1}\left.\left(\left(\frac{\partial F}{\partial x^{0}}\right)^{2}+\left(\nabla F\right)^{2}+m^{2}F^{2}\right)\right|_{x^{0}=0}=:\boldsymbol{S}\:,\label{eq:cool_relation_1}
\end{equation}
and
\begin{equation}
\left.\frac{\mathrm{d}}{\mathrm{d}s_{1}}\left(Q-R\right)\right|_{s_{1}=0}=-\left.\frac{\mathrm{d}}{\mathrm{d}s_{2}}\left(Q-R\right)\right|_{s_{1}=0}\,.\label{eq:cool_relation_2}
\end{equation}
Turning back to \eqref{eq:ec_dif_alpha}, we can write
\begin{eqnarray}
\alpha'\left(s_{2}\right)-\frac{\mathrm{d}}{\mathrm{d}s_{2}}\mathrm{Im}\langle g_{\varphi,R}^{s_{2}}\mid g_{\pi,R}^{s_{2}}\rangle_{\mathrm{1p}}\!\!\! & = & \!\!\!\alpha'\left(0\right)-\left.\frac{\mathrm{d}}{\mathrm{d}s_{1}}\right|_{s_{1}=0} \!\!\! \mathrm{Im}\langle g_{R}^{s_{1}}\mid g_{R}^{s_{2}}\rangle_{\mathrm{1p}}\nonumber  \\
 & = & \!\!\!\alpha'\left(0\right)-\frac{1}{2}\left.\frac{\mathrm{d}}{\mathrm{d}s_{1}}\right|_{s_{1}=0} \!\!\! \left(P(s_{1})+Q\left(s_{1},s_{2}\right)-R\left(s_{1},s_{2}\right)\right)\nonumber \hspace{1cm} \\
 & = & \!\!\!\alpha'\left(0\right)-\frac{1}{2}\boldsymbol{S}+\frac{1}{2}\frac{\mathrm{d}}{\mathrm{d}s_{2}}\left(Q\left(0,s_{2}\right)-R\left(0,s_{2}\right)\right)\,.
\end{eqnarray}
Then, integrating this last equation respect to $s_{2}$, we arrive
to
\begin{equation}
\alpha\left(s_{2}\right)-\mathrm{Im}\langle g_{\varphi,R}^{s_{2}}\mid g_{\pi,R}^{s_{2}}\rangle_{\mathrm{1p}}=\alpha'\left(0\right)\,s_{2}-\frac{1}{2}\boldsymbol{S}\,s_{2}+\frac{1}{2}\left(Q\left(0,s_{2}\right)-R\left(0,s_{2}\right)\right)\,,\label{eq:alpha}
\end{equation}
where we have used that $g_{\varphi,R}^{s_{2=0}}=g_{\pi,R}^{s_{2=0}}=0\Rightarrow\mathrm{Im}\langle g_{\varphi,R}^{s_{2}=0}\mid g_{\pi,R}^{s_{2}=0}\rangle_{\mathrm{1p}}=0$,
and $Q\left(0,0\right)-R\left(0,0\right)=0$, which follows from the
definitions of $Q$ and $R$. To determine $\alpha'\left(0\right)$,
we invoke the KMS-condition stated in theorem \ref{c2-thm-kms-rel}.
Setting $A=B=\boldsymbol{1}$ in equation \eqref{c2-kms-rel}, and
simply calling $G\left(z\right)$ to the underlying function, we have
that
\begin{eqnarray}
G\left(t\right)=\left\langle 0\right|\Delta_{\Phi,0}^{it}\left|0\right\rangle  & \underset{t\rightarrow-i}{\longrightarrow} & G\left(-i\right)=\left\langle \Phi\!\mid\!\Phi\right\rangle =1\,.\label{eq:periodicidad}
\end{eqnarray}
In terms of the real variable $s=-2\pi t$, the function $G\left(s\right)$
is analytic on $\mathbb{R}+i\left(0,2\pi\right)$, and relation \eqref{eq:periodicidad}
must hold for $s\rightarrow2\pi i$. Using \eqref{rel_mod_flow_coh},
we have that
\begin{equation}
G\left(s\right)=\mathrm{e}^{i\alpha\left(s\right)}\left\langle 0\right|\mathrm{e}^{i\varphi\left(g_{\varphi,R}^{s}\right)}\mathrm{e}^{i\pi\left(g_{\pi,R}^{s}\right)}\left|0\right\rangle =\mathrm{e}^{i\alpha\left(s\right)-i\mathrm{Im}\langle g_{\varphi,R}^{s}\mid g_{\pi,R}^{s}\rangle_{\mathrm{1p}}-\frac{1}{2}\left\Vert g_{R}^{s}\right\Vert _{\mathrm{1p}}^{2}}\,,\label{eq:kms_explicita}
\end{equation}
and hence
\begin{eqnarray}
i\alpha\left(s\right)-i\mathrm{Im}\langle g_{\varphi,R}^{s}\mid g_{\pi,R}^{s}\rangle_{\mathrm{1p}}-\frac{1}{2}\left\Vert g_{R}^{s}\right\Vert _{\mathrm{1p}}^{2} & \underset{s\rightarrow2\pi i}{\longrightarrow} & i2n\pi\,,\quad n\in\mathbb{Z}\,.\label{eq:n_molesto}
\end{eqnarray}
Taking this into account, we come back to \eqref{eq:alpha} and write
\begin{equation}
i\alpha\left(s\right)-i\mathrm{Im}\langle g_{\varphi,R}^{s}\mid g_{\pi,R}^{s}\rangle_{\mathrm{1p}}-\frac{1}{2}\left\Vert g_{R}^{s}\right\Vert _{\mathrm{1p}}^{2}=i\alpha'\left(0\right)s-\frac{i}{2}\boldsymbol{S}s+\frac{i}{2}\left(Q\left(0,s\right)-R\left(0,s\right)\right)-\frac{1}{2}\left\Vert g_{R}^{s}\right\Vert _{\mathrm{1p}}^{2}.\label{eq: tomar_limite}
\end{equation}
Before we take the limit $s\rightarrow2\pi i$ in the above expression, we may notice that
$\bar{x}^{s}=\left(-x^{1}\sinh\left(s\right),x^{1}\cosh\left(s\right),\bar{x}_{\bot}\right)\underset{s\rightarrow2\pi i}{\longrightarrow}\left(0,\bar{x}\right)$,
which informally suggests that
\begin{eqnarray}
g_{R}^{s}\underset{s\rightarrow2\pi i}{\longrightarrow}0\!\!\! & \Longrightarrow & \!\!\!\left\Vert g_{R}^{s}\right\Vert _{\mathrm{1p}}^{2}\underset{s\rightarrow2\pi i}{\longrightarrow}0\,, \label{eq: is_zero_dos}\\
f_{\nu}^{s}\underset{s\rightarrow2\pi i}{\longrightarrow}f_{\nu}\!\!\! & \Longrightarrow & \!\!\!Q\left(0,s\right)-R\left(0,s\right)\underset{s\rightarrow2\pi i}{\longrightarrow}0\,,\quad\textrm{where }\nu=\varphi,\pi\,.\label{eq: is_zero}
\end{eqnarray}
More appropriately, in appendix \ref{subsec:Analytic-continuation},
we prove that the function
\begin{equation}
N\left(s\right):=\frac{i}{2}\left(Q\left(0,s\right)-R\left(0,s\right)\right)-\frac{1}{2}\left\Vert g_{R}^{s}\right\Vert _{\mathrm{1p}}^{2}\,,\quad s\in\mathbb{R}\,,
\end{equation}
can be analytically continued on the strip $\mathbb{R}+i\left(0,2\pi\right)$
and that $\lim_{s\rightarrow2\pi i}N\left(s\right)=0$. Then, taking
the limit $s\rightarrow2\pi i$ in \eqref{eq: tomar_limite}, we obtain
\begin{equation}
i2n\pi=-\alpha'\left(0\right)\,2\pi+\frac{1}{2}\boldsymbol{S}\,2\pi\,.
\end{equation}
Since $\alpha'\left(0\right),\boldsymbol{S}\in\mathbb{R}$, then,
it must be that $n=0$. Thereby, we finally get $\alpha'\left(0\right)=\frac{1}{2}\boldsymbol{S}$. 

These computation can be summarized in the following theorem, which
generalizes the theorem \ref{thm_sr_coh_par}.
\begin{thm}
Let $f\in\mathcal{S}\left(\mathbb{R}^{d},\mathbb{R}\right)$ with
$\mathrm{supp}\left(f\right)$ compact. Then, the relative entropy
between the coherent state $\omega_{f}$ and the vacuum state $\omega_{0}$
for the algebra of the right Rindler wedge $\mathcal{W}_{R}$ is
\begin{equation}
S\left(\omega_{f}\mid\omega_{0}\right)=2\pi\int_{x^{1}>0}\negthickspace d^{d-1}x\,x^{1}\left.\frac{1}{2}\left(\left(\frac{\partial F}{\partial x^{0}}\right)^{2}+\left|\nabla F\right|^{2}+m^{2}F^{2}\right)\right|_{x^{0}=0},\label{sr_coh_final}
\end{equation}
where $F\left(x\right)=\int_{\mathbb{R}^{d}}\Delta\left(x-y\right)\,f\left(y\right)d^{d}y$
. In addition, formula \eqref{sr_coh_final} only depends on the behavior
of $f$ in $\mathbb{R}^{d}-\mathcal{W}_{R}'$.
\end{thm}

In conclusion, we have shown that formula \eqref{c4-cali} holds for
the canonical stress-tensor \eqref{c4-stress}.

\section{Conclusions of the chapter}

In this chapter, we proposed to compute the RE between a coherent
state and the vacuum for the algebra of the Rindler wedge in the theory
of a free real scalar field. We adopted a complete algebraic perspective,
using Araki formula for RE and all the tools of modular theory introduced
in section \ref{c2_sec-mod_th}. The motivation to do that has two
faces. On one hand, there was not in the literature many examples
of mathematically rigorous computations of the RE using techniques
of AQFT. It was very instructive for us to understand the way that
all the theoretical material about modular theory of vN algebras is
realized in a concrete model of QFT. On the other hand, as we have
discussed at the beginning of this chapter, the use of non-rigorous
methods leads to an ambiguous formula for the RE. Such ambiguities,
which come from possible improving terms in the energy stress tensor,
can be captured with the help of coherent states.

When the support of the coherent state is contained inside the Rindler
wedge, the computation of the RE is simple as we have explained in
section \ref{c4-sec-re}. This result was already known \cite{nima}.
However, it is useless to solve the mentioned ambiguity since, in
that case, the boundary term, in the heuristic formula of the RE,
trivially vanishes independently of the chosen improving for the stress
tensor.

On the other hand, the more interesting case, when the coherent state
has non-vanishing density along the entanglement surface, is much
harder to compute. To begin with, we showed that the result of the
previous easy case does not apply here, and hence, we must use more
sophisticated tools. The conclusion was that the operator $u_{\Phi,\Omega}\left(t\right):=\Delta_{\Phi,0}^{it}\Delta_{0}^{-it}$,
which we have introduced in section \ref{c2_sec-mod_th}, is easier
to be computed than the modular group $\Delta_{\Phi,0}^{it}$ itself.
The main reason is that the group $u_{\Phi,\Omega}\left(t\right)\in\mathcal{A}$,
even though the modular group does not. In this way, we determined
unambiguously the group $u_{\Phi,0}\left(t\right)$ imposing all its
defining properties (see lemma \ref{c2_cnr_lemma}). In particular,
we showed that our candidate $u_{\Phi,0}\left(t\right)$ satisfies
the KMS-condition. In the end, the expression \eqref{c4-cali} for
the RE holds for the canonical stress tensor. This was already expected
since, as we have shown in section \ref{c4-sec-bounds}, the appearance
of any non-trivial improving term to the canonical stress tensor leads
to an expression for the RE that does not satisfy positivity nor monotonicity.


\renewcommand\chaptername{Chapter}
\selectlanguage{english}

\chapter{Entropy and modular Hamiltonian for the chiral current\label{CURRENT}}

There has been much progress in understanding local statistical properties
of the vacuum reduced to a region in a general $d=2$ CFT. The Rényi
entropies for integer index $n$ have been explicitly computed for
several models of interest \cite{Ruggiero:2018hyl,Coser:2015dvp,Alba:2011fu,Calabrese:2009ez,Calabrese:2010he,Dupic:2017hpb}.
However, analytic continuation to $n\rightarrow1$ to obtain the entanglement
entropy has shown to be a difficult task. The computation of the vacuum
modular Hamiltonians for two intervals has also proved elusive \cite{CardyTonni}.

In this chapter, we calculate the analytic form of the vacuum modular
Hamiltonian for a two-interval region and the algebra of the free
chiral current. This theory may be identified with the chiral derivatives
$J(x)=\partial\phi(x)$ of the free masless scalar field $\phi\left(x\right)$
in $d=2$. We also explicitly compute the MI between these intervals.
This model shows a failure of the duality condition (assumption \ref{c1-conj-duality}
in definition \ref{c1_conj}) for the algebras assigned to two intervals. This
fact is translated into a loss of a symmetry property for the MI usually
associated with modular invariance \cite{Cardy:2017qhl}. Contrary
to the case of the free massless fermion, the modular Hamiltonian
turns out to be completely non-local. The calculation is done by diagonalizing
the reduced density matrix by computing the eigensystem of a correlator
kernel operator, as we have explained in section \ref{c3-secc-lattice}.
The eigenvectors are obtained by a novel method that involves solving
an equivalent problem for a holomorphic function in the complex plane,
where multiplicative boundary conditions are imposed on the intervals.
Using the same technique, we also rederive the free chiral fermion
modular Hamiltonian in a more transparent way than the one used in
\cite{reduced_density}. 

We start in section \ref{c5-sec-chiralCFT} giving a brief summary
of chiral CFTs. Among all the chiral CFTs, we remark the two models
which we study along this chapter: the free fermion and the free current.
In section \ref{c5-sec-ferm}, we explain in detail how to compute
the eigenvectors of the correlator kernel with the novel method mentioned
above. In this way, we rederive the modular Hamiltonian for the $n$-interval
case. There we also introduce the main ideas to be used, in the more
complicated case, of the chiral current. However, the treatment of
the scalar case is self-contained and a reader not interested in the
discussion of the fermion field can start directly in section \ref{escalar}.
There we describe the algebra of the free chiral current and the relevant
kernels. In subsection \ref{unintervalo}, we show, for the case of
a single interval, the diagonalization of the correlator kernel, and
we explicitly compute the entropy and the modular Hamiltonian. The
case of two intervals is dealt with in subsection \ref{dosintervalos}.
The treatment is similar to the one of a single interval, but with
some extra technical difficulties. We compare the results with the
numerical calculations done in subsection \ref{numerico}. Finally,
in section \ref{conclusions}, we end with the conclusions.

\section{Chirals CFTs\label{c5-sec-chiralCFT}}

It is a very well-know fact that the stress-energy tensor of a CFT
in $1+1$ dimensions decouples into chiralities \cite{Mack88}
\begin{equation}
T\left(x^{0},x^{1}\right)=T_{+}\left(x^{0}+x^{1}\right)+T_{-}\left(x^{0}-x^{1}\right) \, .
\end{equation}
Each component $T_{\pm}$ of the stress tensor could be thought of
as living in a one-dimensional QFT over the null line $x_{\pm}:=x^{0}\pm x^{1}$.
In fact, this behavior is quite general and allows us to consider
one-dimensional CFTs, called chiral CFTs, as primary objects, and
then construct any $d=2$ CFT as direct sum of tensor products of
chiral CFTs (see \cite{FewsterHollands} and discussion around equation
\eqref{c5-cft-tensor}). Roughly speaking, a chiral CFT is an AQFT
as the one described in chapter \ref{AQFT}, but with two main differences: 
\begin{enumerate}
\item The spacetime is the null-line, which is identified with $\mathbb{R}$,
and the local algebras are attached to open regions $A\subset\mathbb{R}$.
\item The theory is covariant under the Möbius group, which is formed by
the global conformal transformations on the null-line.
\end{enumerate}
A chiral CFT could be alternatively described along the unit circle
$S^{1}:=\left\{ z\in\mathbb{C}\,:\,\left|z\right|=1\right\} $, which
is the one-point compactification of the null-line. The map which
transforms bijectively the unit circle $S^{1}$ onto the null line
(+ the point of infinity) is called \textsl{Cayley transform}
\begin{equation}
z\in S^{1}\mapsto C\left(z\right):=i\frac{1-z}{1+z}\in\mathbb{R}\cup\left\{ \infty\right\} \,.\label{c5-cayley}
\end{equation}
According to this map, any quantum field or local algebra in the circle
$S^{1}$ can be unambiguously mapped to a quantum field or a local
algebra in the line. This alternative is very useful to perform calculations.

The Möbius group $\mathsf{M\ddot{o}b}$ could be defined equivalently
as:
\begin{enumerate}
\item[a. ] The set of all transformations on the circle $S^{1}$ of the form
\begin{equation}
z\mapsto\frac{\alpha z+\beta}{\beta^{*}z+\alpha^{*}}\,,
\end{equation}
where $\alpha,\beta\in\mathbb{C}$ and $\left|\alpha\right|^{2}-\left|\beta\right|^{2}=1$.
In this way, we see that $\mathsf{M\ddot{o}b}\cong SU(1,1)/\left\{ \mathbf{1},-\mathbf{1}\right\} $.
\item[b. ] The set of all transformation of the null-line $\mathbb{R}$ of the
form 
\begin{equation}
x\mapsto\frac{ax+b}{cx+d}\,,\label{c5-mob}
\end{equation}
where $a,b,c,d\in\mathbb{R}$ and $ad-bc=1$. In this way, we see
that $\mathsf{M\ddot{o}b}\cong PSL(2,\mathbb{R})$.
\end{enumerate}
Of course, both definitions are mapped one into the other by means
of the Cayley transformation \eqref{c5-cayley}. The Möbius group
is generated by three one-parameter subgroups: $R_{\phi}$ ($\phi\in\mathbb{R}$)
corresponding to rotations on the unit circle, and $T_{s}$ ($s\in\mathbb{R}$)
and $D_{\lambda}$ ($\lambda>0$) corresponding respectively to translations
and dilations on the null-line. For any $g\in\mathsf{M\ddot{o}b}$
and any open region $A\subset\mathbb{R}$, we denote by $gA$ the
transformed open set accordingly to \eqref{c5-mob}. For completeness,
we give the definition of a chiral CFT.
\begin{defn}
\label{c5-def-chiral}An \textsl{algebraic chiral CFT} (in the vacuum
representation) is an assignment to every open set $A\subset\mathbb{R}$
a vN subalgebra $\mathcal{A}\left(A\right)\subset\mathcal{\mathcal{B}}\left(\mathcal{H}_{0}\right)$,
i.e.
\begin{eqnarray}
A\subset\mathbb{R} & \mapsto & \mathcal{A}\left(A\right)\subset\mathcal{\mathcal{B}}\left(\mathcal{H}_{0}\right)\,,
\end{eqnarray}
satisfying the following axioms.
\begin{enumerate}
\item \textit{(generating property}) $\mathcal{B}\left(\mathcal{H}_{0}\right)=\bigvee_{A\subset\mathbb{R}}\mathcal{A}\left(A\right)$.
\item \textit{(isotony}) For any open sets $A_{1}\subset A_{2}$, then $\mathcal{A}\left(A_{1}\right)\subset\mathcal{A}\left(A_{2}\right)$. 
\item \textit{(locality)} For any open set $A$, then $\mathcal{A}\left(A'\right)\subset\mathcal{A}\left(A\right)'$,
where $A':=\mathbb{R}-\overline{A}$.\footnote{In the case of fermions nets, we have to relax this axiom imposing
twisted locality respect to a $\mathbb{Z}_{2}$-grading (see section
\ref{c1-sec-fermions}).}
\item \textit{(Möbius covariance}) There is a unitary (strong continuous)
representation $g\in\mathsf{M\ddot{o}b} \mapsto U\left(g\right)\in\mathcal{\mathcal{B}}\left(\mathcal{H}_{0}\right)$,
such that $\mathcal{A}\left(gA\right)=U\left(g\right)\mathcal{A}\left(A\right)U\left(g\right)^{*}$
for all $A\subset\mathbb{R}$ and all $g\in\mathsf{M\ddot{o}b}$.
\item \textit{(positive energy)} The self-adjoint generator $H$ of the
one-parameter translation subgroup $s\in\mathbb{R}\mapsto U\left(T_{s}\right)$
is a positive operator.\footnote{An equivalent requirement is that the self-adjoint generator of the
rotation subgroup $\phi\in\mathbb{R}\mapsto U\left(R_{\phi}\right)$
is positive \cite{FewsterHollands}.} $H$ is called the \textit{chiral Hamiltonian}.
\item \textit{(vacuum)} There exists a unique (up to a phase) $\mathsf{M\ddot{o}b}$-invariant
vector $\left|0\right\rangle \in\mathcal{H}_{0}$, which is called
the vacuum vector.
\end{enumerate}
\end{defn}
\begin{rem}
It could be given an equivalent definition in terms of a net of vN
algebras on the unit circle $S^{1}$.
\end{rem}

\begin{figure}[h]
\centering
\includegraphics[width=10cm]{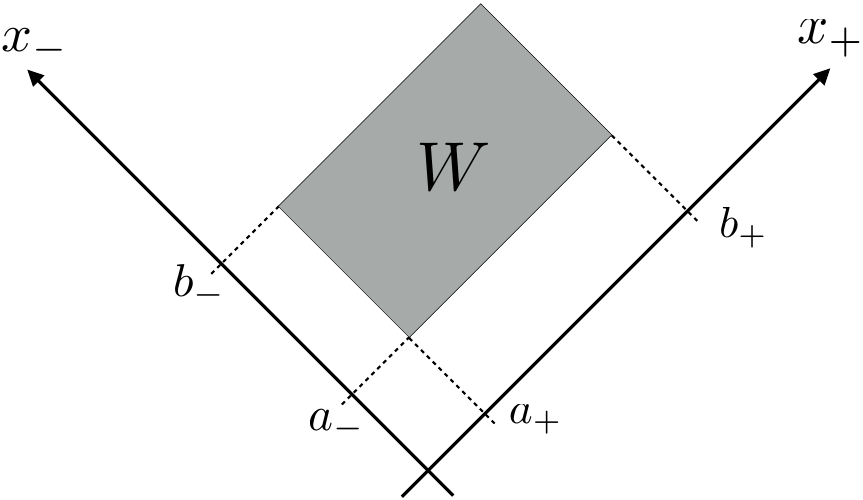}\caption{\foreignlanguage{english}{\label{c5-fig-cono_cft} The double cone $W\subset\mathbb{R}^{2}$
can be parametrized by\foreignlanguage{american}{ two one-dimensional
intervals $A_{\pm}=\left(a_{\pm},b_{\pm}\right)\in\mathcal{N}_{\pm}$}. }}
\end{figure}

The construction of a two-dimensional chiral CFT in Minkowski space using
two chiral CFTs is as follows. Let us take the basis $\left\{ e_{+},e_{-}\right\} \subset\mathbb{R}^{2}$,
where $e_{\pm}:=e^{0}\pm e^{1}$, and define the null rays $\mathcal{N}_{\pm}:=\left\{ \lambda e_{\pm}\,:\,\lambda\in\mathbb{R}\right\} $.
Then, each point $x\in\mathbb{R}^{2}$ can be uniquely decomposed
as $x=x_{+}e_{+}+x_{-}e_{-}$, and in particular, each double cone
$W\subset\mathbb{R}^{2}$ can be parametrized by two one-dimensional
intervals $A_{\pm}:=\left(a_{\pm},b_{\pm}\right)\in\mathcal{N}_{\pm}$
(see figure \ref{c5-fig-cono_cft}). Then, given two chiral CFTs 
\begin{equation}
A_{\pm}\subset\mathbb{R}\mapsto\mathcal{A}_{\pm}\left(A_{\pm}\right)\subset\mathcal{B}\left(\mathcal{H}_{\pm}\right)\,,
\end{equation}
we can define the net of vN algebras
\begin{eqnarray}
W\textrm{ (double cone)} & \mapsto & \mathcal{A}_{2d}\left(W\right):=\mathcal{A}_{+}\left(A_{+}\right)\otimes\mathcal{A}_{-}\left(A_{-}\right)\subset\mathcal{\mathcal{B}}\left(\mathcal{H}_{+}\right)\otimes\mathcal{\mathcal{B}}\left(\mathcal{H}_{-}\right)\,.\label{c5-cft-tensor}
\end{eqnarray}
Such a net satisfies all the axioms of definitions \ref{c1_def_aqft}
and \ref{c1_def_vacuum}. Furthermore, it is covariant with respect
to two copies of the Möbius group $\mathsf{M\ddot{o}b}\times\mathsf{M\ddot{o}b}\varsupsetneq\mathcal{P}_{+}^{\uparrow}$.
In particular, the vacuum vector of the $d=2$ CFT is just the tensor
product of the vacuum vectors of each chiral CFT, i.e. $\left|0\right\rangle =\left|0\right\rangle _{+}\otimes\left|0\right\rangle _{-}$.
General causally complete regions are handled in a similar fashion.

It is important to remark that the tensor product structure \eqref{c5-cft-tensor}
is not the most general possibility. In fact, in a generally non-chiral $d=2$ CFT, the algebra of a causally complete region $\mathcal{A}_{2d}\left(\mathcal{O}\right)$
may contain further operators than the ones belonging $\mathcal{A}_{+}\left(A_{+}\right)\otimes\mathcal{A}_{-}\left(A_{-}\right)$.
At the end of this chapter, in section \ref{failure}, we show how
these features appear in the case of the free massless scalar field.\footnote{Moreover, in the most general case, the Hilbert space of a two-dimensional
CFT could be defined as a direct sum of tensor products of the form
\begin{equation}
\mathcal{H}:=\bigoplus_{k=0}^{K}\mathcal{H}_{+,k}\otimes\mathcal{H}_{-,k}\,,\label{c5-fn-vira}
\end{equation}
where $K\in\mathbb{N}_{0}\cup\{\infty\}$ and the Hilbert spaces $\mathcal{H}_{\pm,k}$
carry unitaries highest weight representations of Virasoro algebra
$(c_{\pm,k},h_{\pm,k})$, where $c_{\pm,k}$ are the corresponding
central charges. Local algebras for two-dimensional causally complete
regions could be defined in similar fashion as in \eqref{c5-cft-tensor},
but regarding the richer structure coming from \eqref{c5-fn-vira}
(see \cite{FewsterHollands}).}

Whatever the case, here we are interested in one chiral CFT. Once
we have identified the chiral CFTs as an special kind of AQFTs, we
could pursue the study of the entanglement in such models. In particular,
here we study the modular Hamiltonian and the EE for two specific
models: the free chiral fermion (section \ref{c5-sec-ferm}) and the
free chiral current (section \ref{escalar}).

\section{The free chiral fermion\label{c5-sec-ferm}}

The free chiral fermion is usually defined as a quantum field $\psi\left(x\right)$
satisfying the anticommutation relations $\{\psi(x),\psi^{\dagger}(y)\}=\delta(x-y)$.
The vacuum state is a Gaussian state with two-point correlator\footnote{Changing the sign of the imaginary part in expression \eqref{corrC}
corresponds to changing chirality.}
\begin{equation}
C(x-y):=\langle0|\psi(x)\psi^{\dagger}(y)|0\rangle=\frac{1}{2}\delta(x-y)+\frac{i}{2\pi}\frac{1}{x-y}\,,\label{corrC}
\end{equation}
where the distribution on the left hand side is understood as the
principal value regularization. \eqref{corrC} is a projector when
it acts, as a kernel, on the full line. On a region $A$, it is an
Hermitian operator with continuous eigenvalues in the range $(0,1)$
\cite{reduced_density}. From the algebraic standpoint, given a test
function $f\in\mathcal{S}\left(\mathbb{R},\mathbb{R}\right)$, the
operator 
\begin{equation}
\psi\left(f\right):=\int_{\mathbb{R}}\psi\left(x\right)f\left(x\right)\,dx\,,
\end{equation}
is a well-defined operator acting on the fermionic Fock Hilbert space
$\mathcal{H}_{0}$. This Hilbert space is defined as the anti-symmetric
tensor product of the one-particle Hilbert space of massless particles
of helicity $\frac{1}{2}$. The one-particle Hilbert space carries
a unitary representation of $\mathsf{M\ddot{o}b}$, which naturally
extends to a unitary representation of $\mathsf{M\ddot{o}b}$ on $\mathcal{H}_{0}$.
For any open set $A\subset\mathbb{R}$, the local algebras are defined
as
\begin{equation}
\mathcal{A}_{\mathrm{f}}\left(A\right):=\left\{ \psi\left(f\right)\,:\,\mathrm{supp}\left(f\right)\subset A\right\} ''\,.
\end{equation}
It is not difficult to see that such a collection of algebras satisfies
all the axioms of the definition \ref{c5-def-chiral}.

Now we consider a region $A:=(a_{1},b_{1})\cup(a_{2},b_{2})\cup\cdots\cup(a_{n},b_{n})$
formed by $n$ disjoint intervals ($a_{j}<b_{j}<a_{j+1}$ for all
$j=1,\ldots,n$). According to the discussion in section \ref{c3-secc-lattice},
to obtain the modular Hamiltonian we need to solve the spectrum of
the correlator \eqref{corrC} reduced to the region $A$, i.e. $C_{A}(x,y):=C(x,y)|_{x,y\in A}$.
Then, the modular Hamiltonian is given by the equations \eqref{c3-mh-fer}
and \eqref{c3-mh-car}, which means
\begin{equation}
K=\int_{A\times A}dx\,dy\,\psi^{\dagger}(x)\mathcal{K}(x,y)\psi(y)\,,\label{citric}
\end{equation}
where the kernel $\mathcal{K}$ is 
\begin{equation}
\mathcal{K}=-\log(C_{A}^{-1}-1)\,.\label{hache}
\end{equation}
This last equation is understood as an operator equation, where the
action of the operators is defined through their kernels.

A complete description of the reduced density matrix of a massless
fermion field for multi-interval regions was given in \cite{reduced_density}.
This was achieved by diagonalizing the correlator kernel in the region,
using previous results in the literature about singular kernels of
the Cauchy type \cite{Muskhelishvili}. In the following subsection,
we make this diagonalization more transparent by mapping the problem
of integral equations in one dimension to one of partial differential
equations in two dimensions following \cite{Arias17}. The form of
the eigenvectors as well as its main properties are easily derived
using this trick.

\subsection{An equivalent problem in the complex plane\label{cp}}

In this section, we will relate the original problem of solving the
spectrum of $C_{A}$ as a kernel in $A$, to a new problem about a
function in the complex plane. For such a purpose, we think the $n$
intervals $A$ as included in the real axis of the complex plane.
For each $\lambda\in\mathbb{R}$, we consider the following problem
about a function $S(z)$ in the complex plane

\begin{eqnarray}
 &  & S(z)\text{ analytic in }\mathbb{C}-\bar{A}\,,\label{hol_A}\\
 &  & S^{+}(x_{1})=\lim_{x_{2}\rightarrow0^{+}}S(x_{1}+ix_{2})=\lambda\lim_{x_{2}\rightarrow0^{-}}S(x_{1}+ix_{2})=\lambda\,S^{-}(x_{1})\,,\qquad x_{1}\in A, \hspace{1cm} \label{disc}\\
 &  & \lim_{z\rightarrow\infty}\left|z\,S(z)\right|<\infty\,,\label{bc1}\\
 &  & \lim_{z\rightarrow\partial A}l_{z,\partial A}\,S(z)\rightarrow0\,,\label{bc}
\end{eqnarray}
where $l_{z,\partial A}$ is the distance from $z$ to the boundary
$\partial A$ (formed by $2n$ disjoint points). Thus, $S(z)$ has
a cut over $A$ with multiplicative boundary conditions. Consider
now the following complex integral
\begin{equation}
\oint dz_{2}\,\frac{1}{z_{2}-z_{1}}S(z_{2})\,,\quad z_{1}\in\mathbb{C}-\bar{A}\,,\label{circ_int}
\end{equation}
where we choose an integration contour that encircles both $A$ and
$z_{1}$ in the positive (anti-clockwise) direction. This integral
vanishes because of \eqref{bc1}. Writing it as two separated contributions
from the pole at $z_{1}$ and the integration around the cut $A$,
we get 
\begin{equation}
S(z_{1})=\frac{1}{2\pi i}\int_{A}dy\,\frac{1}{y-z_{1}}(S^{+}(y)-S^{-}(y))=\frac{1-\lambda^{-1}}{2\pi i}\int_{A}dy\,\frac{1}{y-z_{1}}S^{+}(y)\,,\label{todosla}
\end{equation}
where we have used the boundary condition \eqref{disc}. We remark
there are no contributions from the endpoints of the intervals due
to \eqref{bc}. This equation gives the value for $S(z)$ on any point
$z\in\mathbb{C}-\bar{A}$ from its values at the cut $A$. Taking
the limit $z_{1}\rightarrow x\in A$ from above, and using 
\begin{equation}
\lim_{y\rightarrow0^{+}}\frac{1}{x+iy}=\frac{1}{x}-i\pi\delta(x)\,,\label{plem}
\end{equation}
we get 
\begin{equation}
\int_{A}dy\,C_{A}(x-y)S^{+}(y)=\frac{\lambda}{\lambda-1}S^{+}(x)\,,\label{coco}
\end{equation}
which means that the boundary value of $S(z)$ plays the role of an
eigenvector with eigenvalue $\lambda(\lambda-1)^{-1}$ for the correlator
kernel on $A$. Since the spectrum of $C_{A}(x-y)$ is restricted
to $(0,1)$ (see \cite{Casini_review}), we have that $\lambda\in(-\infty,0)$.
For later convenience, we write 
\begin{equation}
\lambda:=-\mathrm{e}^{2\pi s}\,,\hspace{0.5cm}s\in\mathbb{R}\,.\label{lami}
\end{equation}

Conversely, suppose we have a solution $S^{+}(x)$ of the eq. \eqref{coco}
for some $\lambda\in\mathbb{R}_{\leq0}$ with appropriate boundary
conditions on the endpoints of the intervals as in \eqref{bc}.\footnote{This is precisely the boundary condition of the eigenvectors for the
vacuum state \cite{Casini_review}.} Then, equation \eqref{todosla} gives a complex-valued function $S(z)$
satisfying all the properties (\ref{hol_A}-\ref{bc}). For the boundary
condition \eqref{disc}, the function $S(z)$, defined in this way,
has the boundary value $S^{+}(x)$ at the upper side of the cut. For
the lower side of the cut, we have to use 
\begin{equation}
\lim_{y\rightarrow0^{+}}\frac{1}{x-iy}=\frac{1}{x}+i\pi\delta(x)\,
\end{equation}
instead of \eqref{plem}, to get the right value $S^{-}(x)=-e^{-2\pi s}S^{+}(x)$.

In conclusion, the solutions of the problem in the complex plane (\ref{hol_A}-\ref{bc})
are in one-to-one correspondence with the eigenvectors of the correlator
kernel \eqref{corrC}.

\subsection{Multiplicity and normalization of the eigenvectors\label{sec12}}

Because of conditions \eqref{disc} and \eqref{bc}, the function
$S(z)$ must have the following asymptotic behavior when $z\rightarrow\partial A$,
\begin{eqnarray}
S(z)\!\!\! & \sim & \!\!\!V_{a_{j}}\,(a_{j}-z)^{-1/2+is}\,,\label{expa}\\
S(z)\!\!\! & \sim & \!\!\!V_{b_{j}}\,(z-b_{j})^{-1/2-is}\,,\label{expa2}
\end{eqnarray}
where $V_{a_{j}}$, $V_{b_{j}}\in\mathbb{C}$ are constants. The symbol
$\sim$ means that the difference between the left and right sides
in expressions \eqref{expa} and \eqref{expa2} are analytic functions
on $\mathbb{C}-\bar{A}$ with finite limit when $z\rightarrow\partial A$.
Below, we will show that the constants $V_{a_{j}}$ and $V_{b_{j}}$
uniquely determine the solution. In order to see this, for each $s\in\mathbb{R}$,
we define the Green function $G(z,w)$ for the problem (\ref{hol_A}-\ref{bc}),
i.e.
\begin{eqnarray}
 &  & G(z,w)\,\textrm{ analytic on }\,w\in\mathbb{C}-\{w\in\bar{A}\,\textrm{or}\,\,w=z\}\,,\\
 &  & G(z,w)\sim(z-w)^{-1}\hspace{0.5cm}\text{when }w\sim z\,,\\
 &  & \lim_{x_{2}\rightarrow0^{+}}G(z,x_{1}+ix_{2})=-\mathrm{e}^{-2\pi s}\lim_{x_{2}\rightarrow0^{-}}G(z,x_{1}+ix_{2})\,,\hspace{1cm}x_{1}\in A\,,\label{jumpG}
\end{eqnarray}
and in addition $G(z,w)$ satisfies the two boundary conditions \eqref{bc1}
and \eqref{bc} as a function of $w$ for each $z\in\mathbb{C}$ fixed.\footnote{We explicitly change the sign of $s$ in \eqref{jumpG} respect to
\eqref{disc}.} For $w\rightarrow\partial A$, then we have an expansions analogous
to \eqref{expa}, 
\begin{eqnarray}
G(z,w)\!\!\! & \sim & \!\!\!U_{a_{j}}(z)\,(a_{j}-w)^{-1/2-is}\,,\label{expaG}\\
G(z,w)\!\!\! & \sim & \!\!\!U_{b_{j}}(z)\,(w-b_{j})^{-1/2+is}\,.
\end{eqnarray}
Then, the combination $G(z,w)S(w)$ does not have any jump singularity
at $A$ as a function of $w$. On the other hand, it has only simple
poles at $\partial A$ and at $z$, but it does not have a singularity
at infinity. Since the sum of all its residues must vanish, we have
that
\begin{equation}
S(z)=\sum_{j=1}^{n}\left(V_{a_{j}}U_{a_{j}}(z)-V_{b_{j}}U_{b_{j}}(z)\right)\,.\label{C2subspace}
\end{equation}

This shows there are at most $2n$ linearly independent solutions
to the problem (\ref{hol_A}-\ref{bc}) for fixed $s$, and they can
be viewed simply as elements of $\mathbb{C}^{2n}$. It also shows
that any solution which is bounded on $\partial A$ (i.e. $V_{a_{j}}=V_{b_{j}}=0$)
must vanish.

Now, we will show that the degeneracy of the space of solutions for
each $s$ fixed is at most $n$. Let us take two solutions $S_{1}(z)$
and $S_{2}(z)$ corresponding to the same value $s$. The function
$\tilde{S}_{1}(z)=(S_{1}(z^{*}))^{*}$ is a solution with parameter
$-s$ instead of $s$. The function $\tilde{S}_{1}(z)S_{2}(z)$ does
not have a cut, only poles at $\partial A$. The sum of residues must
vanish and we get 
\begin{equation}
\sum_{j=1}^{n}\left((V_{a_{j}}^{1})^{*}V_{a_{j}}^{2}-(V_{b_{j}}^{1})^{*}V_{b_{j}}^{2}\right)=0\,,\label{espacio}
\end{equation}
where $V_{a_{j}}^{k}$, $V_{b_{j}}^{k}$ are the coefficients corresponding
to $S_{k}$ ($k=1,2$). This means that any two solutions of (\ref{hol_A}-\ref{bc})
for the same $s$ must be orthogonal according to the canonical pseudo-inner
product of $\mathbb{C}^{n,n}$, which includes the case when the two
solutions are the same. The argument to justify why the space of solutions
is at most $n$ is as follows. Suppose that the $s$-valued subspace
of solutions is spanned by $\{S_{1},\dots S_{2n}\}$, where each $S_{k}$
is of the form \eqref{C2subspace}. Then, after a diagonalization
procedure,\footnote{Equivalent to the Gauss-Jordan algorithm used to diagonalize a finite
dimensional matrix.} we can get a new set of solutions $\{\tilde{S}_{1},\dots\tilde{S}_{2n}\}$
which spans the same subspace but with the property that $V_{a_{j}}^{k}=0$
for all $j=1,\dots,n$ for all $k=n+1,\dots,2n$. Automatically, because
of \eqref{espacio}, we must have that $V_{b_{j}}^{k}=0$ for all
$j=1,\dots,n$ and for all $k=n+1,\dots,2n$, and hence, $\tilde{S}_{n+1}(z)=\dots=\tilde{S}_{2n}(z)=0$.
In conclusion, the $s$-valued subspace of solutions has dimension
at most $n$. We will show, in the next subsection, that the dimension
is exactly $n$.

Now we make a final comment about the normalization of the eigenvectors
$S^{+}(x,s)$, where we are writing explicitly the dependence of the
eigenvectors through the eigenvalues $s$. Since any two eigenvectors
$S_{1}^{+}(x,s)$ and $S_{2}^{+}(x,s')$ must be orthogonal for $s\neq s'$,
we have\footnote{The function $(S_{1}^{+}(x,s))^{*}$ is the complex conjugate of the
boundary value of $S_{1}^{+}(x,s)$, which is not the same that the
boundary value of $(S_{1}(z,s))^{*}$. These two operations do not
commute.} 
\begin{equation}
\int_{A}dx\,(S_{1}^{+}(x,s))^{*}\,S_{2}^{+}(x,s')\propto\delta(s-s')\,.\label{innerp}
\end{equation}

In order to orthonormalize the eigenvectors, we need to figure out
the proportionality constant in the above equation. For this, we note
the delta function can only come from the integration around the endpoints
of the intervals on the scalar product. Using the asymptotic expansion
of the functions near the endpoints, we arrive at\footnote{More precisely, we should write each eigenvector as $S(z)=\sum_{i=1}^{n}V_{a_{i}}\,(a_{i}-z)^{-1/2+is}+V_{b_{i}}\,(z-b_{i})^{-1/2-is}+R(z)$
where the function $R(z)$ has finite limit when $z\rightarrow\partial A$.
Then, after replacing on the l.h.s. of \eqref{innerp}, we get that
the delta Dirac contributions are of the form \eqref{cierto}.} 
\begin{equation}
\int_{A}dx\,(S_{1}^{+}(x,s))^{*}\,S_{2}^{+}(x,s')=\pi\,e^{2\pi s}\,\delta(s-s')\,\sum_{j=1}^{n}\left((V_{a_{j}}^{1})^{*}V_{a_{j}}^{2}+(V_{b_{j}}^{1})^{*}V_{b_{j}}^{2}\right)\,.\label{cierto}
\end{equation}
Note that the two terms inside the parenthesis on the r.h.s. are equal
according to \eqref{espacio}.

\subsection{Construction of the eigenvectors}

In this subsection, we will explicitly construct the eigenvectors
of the correlator $C_{A}(x-y)$ using the relation developed in subsection
\ref{cp}. Concretely, we will find the general structure of any solution
$S(z)$ of the problem (\ref{hol_A}-\ref{bc}), and through them,
we will obtain the corresponding eigenvectors. In particular, we will
show that all eigenspaces of a given eigenvalue $s$ have dimension
$n$. In this subsection, $s\in\mathbb{R}$ is fixed.

We start defining the complex valued function 
\begin{equation}
\tilde{\omega}(z):=\sum_{j=1}^{n}\log\left(\frac{z-a_{j}}{z-b_{j}}\right)\,,
\end{equation}
where $\log$ is the principal branch of the complex logarithm, which
has a branch cut for $z\in\mathbb{R}_{\leq0}$. The function $\tilde{\omega}$
is analytic everywhere on the plane except at $\bar{A}$, where it
has a jump discontinuity of the form 
\begin{equation}
\tilde{\omega}^{+}(x)-\tilde{\omega}^{-}(x)=-2\pi i\,,\hspace{1cm}x\in A\,.
\end{equation}
Therefore, the function
\begin{equation}
\mathrm{e}^{\left(is+\frac{1}{2}\right)\tilde{\omega}(z)}\,,
\end{equation}
satisfies the conditions \eqref{hol_A}, \eqref{disc} and \eqref{bc},
but it does not satisfy \eqref{bc1}. On the other hand, given any
solution $S(z)$ of (\ref{hol_A}-\ref{bc}), the function 
\begin{equation}
f(z):=S(z)\,\mathrm{e}^{-\left(is+\frac{1}{2}\right)\tilde{\omega}(z)}
\end{equation}
is analytic on $\mathbb{C}-\bar{A}$, and it is also continuous on
$A$, and hence, it is analytic on $\mathbb{C}-\partial A$.\footnote{By Schwartz reflection principle.}
Then, $f(z)$ must be some meromorphic function in the whole complex
plane, with poles located at the endpoints of the intervals and also
possibly at $\infty$.\footnote{A further analysis prevents the possibility of having essential singularities
at such points.} Since $S(z)$ satisfies \eqref{bc} and because of 
\begin{equation}
\lim_{z\rightarrow x+i0^{+}}\left|\mathrm{e}^{\left(is+\frac{1}{2}\right)\tilde{\omega}(z)}\right|=\mathrm{e}^{\pi s}\prod_{j=1}^{n}\sqrt{\left|\frac{(x-a_{j})}{(x-b_{j})}\right|}\,,\hspace{0.5cm}x\in A\,,
\end{equation}
it follows that $f(z)$ must be of the form 
\begin{equation}
f(z):=\frac{g(z)}{\prod_{j=1}^{n}(z-a_{j})}\,,
\end{equation}
where $g(z)$ is an entire analytic function. In order to satisfy
the last requirement \eqref{bc1} for $S(z)$, we have that $g(z)$
must be a polynomial function of degree at most $n-1$. Taking all
this into account, any solution $S(z)$ for the problem (\ref{hol_A}-\ref{bc})
must be of the form 
\begin{equation}
S(z)=\frac{\sum_{k=0}^{n-1}c_{k}\,z^{k}}{\prod_{j=1}^{n}(z-a_{j})}\mathrm{e}^{\left(is+\frac{1}{2}\right)\tilde{\omega}(z)}\,,\label{solut}
\end{equation}
where $c_{k}\in\mathbb{C}$ parametrize $n$ linearly independent
functions. Conversely, it is easy to see that any complex valued function
of the form \eqref{solut} is a solution of the problem (\ref{hol_A}-\ref{bc}).

Taking the limit of $z\rightarrow A$ from the upper side of the cut
on expression \eqref{solut}, we obtain the eigenvectors 
\begin{equation}
S^{+}(x)=-i(-1)^{n-l}\mathrm{e}^{\pi s}\mathrm{e}^{is\omega(x)}\frac{\sum_{k=0}^{n-1}c_{k}\,x^{k}}{\sqrt{-\prod_{j=1}^{n}(x-a_{j})(x-b_{j})}}\,,\hspace{0.6cm}\text{for }x\in(a_{l},b_{l})\,,\label{solu}
\end{equation}
where
\begin{equation}
\omega(x):=\lim_{z\rightarrow x+i0^{+}}\mathrm{Re}\,\tilde{\omega}(z)=\log\left(-\frac{\prod_{j=1}^{n}(x-a_{j})}{\prod_{j=1}^{n}(x-b_{j})}\right)\,.
\end{equation}
Therefore, there are exactly $n$ degenerate eigenfunctions for the
same $s$. This space of eigenfunctions coincides with the one obtained
in \cite{reduced_density}.

\paragraph{Scalar product\protect \\
}

Due to the degeneracy, we have some arbitrariness for the election
of the eigenvectors. Such a freedom is encoded in the polynomial $P(x):=\sum_{k=0}^{n-1}c_{k}\,x^{k}$
of equation \eqref{solut}. In subsection \ref{cob}, we fix such
a freedom in order to get an orthonormal basis of eigenvectors. In
order to do that, it is useful to have an expression for the scalar
product between two eigenvectors in terms of its corresponding polynomials.
In the rest of this subsection, we will obtain such a expression. 

In equation \eqref{cierto}, using that the scalar product of two
eigenfunctions is proportional to a delta function $\delta(s-s')$,
we obtained this scalar product in terms of the coefficients of the
expansion of the eigenvectors near the endpoints of the intervals.
Here, we will reobtain this result by explicit integration of the
product of eigenfunctions.

First we take two solutions $S_{1}^{+}(x,s)$ and $S_{2}^{+}(x,s')$
of the form \eqref{solu} corresponding to two polynomials $P_{1}(x)$
and $P_{2}(x)$. Then, we compute the scalar product
\begin{equation}
\hspace{-0.4cm} \int_{A}\!dx\,S_{1}^{+*}(x,s)\,S_{2}^{+}(x,s')=-\mathrm{e}^{\pi(s+s')} \!\! \int_{-\infty}^{+\infty} \!\!\!\!\!\! d\omega\,\mathrm{e}^{-i(s-s')\omega}\sum_{l=1}^{n}\!\frac{1}{\omega'(x_{l})}\frac{P_{1}^{*}(x_{l})P_{2}(x_{l})}{\prod_{j=1}^{n}(x_{l}-a_{j})(x_{l}-b_{j})}, \!\! \label{previ}
\end{equation}
where we have changed the integration variable to $\omega$ and the
sum in \eqref{previ} runs over the distinct solutions of the equation
$\omega(x)=\omega$, which are the $n$ simple roots of the polynomial
equation 
\begin{equation}
-\mathrm{e}^{\omega}\prod_{j=1}^{n}(x-b_{j})=\prod_{j=1}^{n}(x-a_{j})\,.\label{polr}
\end{equation}
In each interval $A_{l}=(a_{l},b_{l})$, $\omega(x)$ is a monotone
increasing function that goes from $-\infty$ at $a_{l}$ to $+\infty$
at $b_{l}$. This fact implies that there exist $n$ distinct simple
roots $x_{l}$, each one belonging to a distinct interval $A_{l}$.
In \eqref{previ}, $x_{l}$ is understood as function of $\omega$,
i.e. $x_{l}(\omega)$. In order to proceed, we will show that the
following function of $\omega$ 
\begin{equation}
K(\omega):=\sum_{l=1}^{n}\frac{1}{\omega'(x_{l})}\frac{Q_{2n-2}(x_{l})}{\prod_{j=1}^{n}(x_{l}-a_{j})(x_{l}-b_{j})}\,,\label{constant}
\end{equation}
is a constant, i.e. $K(\omega)$ is independent of $\omega$ for any
polynomial $Q_{2n-2}(x)$ of degree $2n-2$. Replacing the following
expression for $\omega'(x)$ 
\begin{equation}
\omega'(x)=\frac{\prod_{i=1}^{n}(x-b_{i})\sum_{k=1}^{n}\prod_{j\neq k}(x-a_{j})-\prod_{i=1}^{n}(x-a_{i})\sum_{k=1}^{n}\prod_{j\neq k}(x-b_{j})}{\prod_{j=1}^{n}(x-a_{j})(x-b_{j})}\,,
\end{equation}
in \eqref{constant}, we arrive at 
\begin{equation}
\hspace{-.4cm} K(\omega)=\sum_{l=1}^{n}\frac{Q_{2n-2}(x_{l})}{\prod_{i=1}^{n}(x_{l}-b_{i})\sum_{k=1}^{n}\prod_{j\neq k}(x_{l}-a_{j})-\prod_{i=1}^{n}(x_{l}-a_{i})\sum_{k=1}^{n}\prod_{j\neq k}(x_{l}-b_{j})}. \!\! \label{into}
\end{equation}
Since $\omega=-\infty$ implies $x_{l}=a_{l}$ and $\omega=+\infty$
implies $x_{l}=b_{l}$, then we have the following particular limits
\begin{eqnarray}
K(-\infty)\!\!\! & = & \!\!\!\sum_{l=1}^{n}\frac{Q_{2n-2}(a_{l})}{\prod_{i=1}^{n}(a_{l}-b_{i})\prod_{j\neq l}(a_{l}-a_{j})}\,,\label{dio}\\
K(\infty)\!\!\! & = & \!\!\!-\sum_{l=1}^{n}\frac{Q_{2n-2}(b_{l})}{\prod_{i=1}^{n}(b_{l}-a_{i})\prod_{j\neq l}(b_{l}-b_{j})}\,.\label{dio1}
\end{eqnarray}
Now, we will show that $K(\omega)=K(-\infty)$, and hence constant.
For this, from equation \eqref{polr}, we have the following polynomial
identity 
\begin{equation}
\mathrm{e}^{\omega}\prod_{j=1}^{n}(x-b_{j})+\prod_{j=1}^{n}(x-a_{j})=(\mathrm{e}^{\omega}+1)\prod_{l=1}^{n}(x-x_{l})\,.\label{yyy}
\end{equation}
Evaluating \eqref{yyy} on $x=a_{k}$ (for some $k=1,\cdots,n$) we
get 
\begin{equation}
\prod_{j=1}^{n}(a_{k}-b_{j})=(1+\mathrm{e}^{-\omega})\prod_{l=1}^{n}(a_{k}-x_{l})\,,\label{refe}
\end{equation}
and taking the derivative of \eqref{yyy} respect to $x$ and evaluating
at $x=x_{l}$ (for some $l=1,\cdots,n$), we have 
\begin{equation}
\prod_{i=1}^{n}(x_{l}-b_{i})\sum_{k=1}^{n}\prod_{j\neq k}^{n}(x_{l}-a_{j})-\prod_{i=1}^{n}(x_{l}-a_{i})\sum_{k=1}^{n}\prod_{j\neq k}^{n}(x_{l}-b_{j})=-(1+\mathrm{e}^{-\omega})\prod_{i=1}^{n}(x_{l}-a_{i})\prod_{j\neq l}^{n}(x_{l}-x_{j})\,.\label{refe2}
\end{equation}
Then, replacing \eqref{refe} in the denominator of \eqref{dio},
we get 
\begin{equation}
K(-\infty)=(1+\mathrm{e}^{-\omega})^{-1}\sum_{l=1}^{n}\frac{Q_{2n-2}(a_{l})}{\prod_{i=1}^{n}(a_{l}-x_{i})\prod_{j\neq l}(a_{l}-a_{j})}\,,
\end{equation}
and replacing \eqref{refe2} in the denominator of \eqref{into},
it follows 
\begin{equation}
K(\omega)=-(1+\mathrm{e}^{-\omega})^{-1}\sum_{l=1}^{n}\frac{Q_{2n-2}(x_{l})}{\prod_{i=1}^{n}(x_{l}-a_{i})\prod_{j\neq l}^{n}(x_{l}-x_{j})}.
\end{equation}
Hence, the expected relation $K(\omega)=K(-\infty)$ follows from
\begin{eqnarray}
(1+\mathrm{e}^{-\omega})\left[K(-\infty)-K(\omega)\right]\!\!\! & = & \!\!\!\sum_{l=1}^{n}\frac{Q_{2n-2}(x_{l})}{\prod_{i=1}^{n}(x_{l}-a_{i})\prod_{j\neq l}(x_{l}-x_{j})}\nonumber \\
 &  & \!\!\!+\sum_{l=1}^{n}\frac{Q_{2n-2}(a_{l})}{\prod_{i=1}^{n}(a_{l}-x_{i})\prod_{j\neq l}(a_{l}-a_{j})}=0\,,\label{ppp}
\end{eqnarray}
where the last equality to zero is a general fact valid for any polynomial
$Q_{2n-2}$ of degree $2n-2$: evaluating the polynomial in $2n$
arbitrary points $y_{1},\cdots,y_{2n}$ there is a linear equation
that relates the value on the first $2n-1$ points to the value on
$y_{2n}$. This equation is 
\begin{equation}
\sum_{l=1}^{2n}\frac{Q_{2n-2}(y_{l})}{\prod_{j\neq l}(y_{l}-y_{j})}=0\,.
\end{equation}
Eq. \eqref{ppp} follows by specializing on $y_{j}=a_{j}$ for $j=1,\cdots,n$,
and $y_{j}=x_{j-n}$ for $j=n+1,\cdots,2n$. 

Since $K(\omega)$ is constant, we have that $K(-\infty)=K(\infty)$,
i.e. expressions \eqref{dio} and \eqref{dio1} are the same. This,
in fact, coincides with the relation \eqref{espacio} for the coefficients
of the expansions \eqref{expa} and \eqref{expa2} of the solutions
\eqref{solut} at the endpoints of the intervals. Reading off these
coefficients from the explicit form of the solutions, the relation
\eqref{espacio} writes 
\begin{eqnarray}
\sum_{j=1}^{n}(V_{a_{j}}^{1})^{*}V_{a_{j}}^{2}\!\!\! & = & \!\!\!-\sum_{l=1}^{n}\frac{P_{1}(a_{l})^{*}P_{2}(a_{l})}{\prod_{i=1}^{n}(a_{l}-b_{i})\prod_{j\neq l}(a_{l}-a_{j})}\nonumber \\
 & = & \!\!\!\sum_{l=1}^{n}\frac{P_{1}(b_{l})^{*}P_{2}(b_{l})}{\prod_{i=1}^{n}(b_{l}-a_{i})\prod_{j\neq l}(b_{l}-b_{j})}=\sum_{j=1}^{n}(V_{b_{j}}^{1})^{*}V_{b_{j}}^{2}\,.\label{vvz}
\end{eqnarray}
 Let us turn back to the scalar product \eqref{previ}. Since the
integrand on the Fourier transform in $\omega$ is constant, we get
\begin{equation}
\int_{A}dx\,S_{1}^{+*}(x,s)\,S_{2}^{+}(x,s')=-2\pi\,e^{2\pi s}\,\delta(s-s')\sum_{l=1}^{n}\frac{P_{1}(a_{l})^{*}P_{2}(a_{l})}{\prod_{i=1}^{n}(a_{l}-b_{i})\prod_{j\neq l}(a_{l}-a_{j})}\,,\label{escalaa}
\end{equation}
which coincides with the equation \eqref{cierto} because of \eqref{vvz}.

\subsection{A complete set of eigenvectors\label{cob}}

In order to construct a basis of eigenvectors for each eigenspace
of fixed $s\in\mathbb{R}$, we choose the following subset $\{u_{s}^{k}\}_{k=1}^{n}$
of eigenfunctions 
\begin{equation}
u_{s}^{k}(x):=\frac{(-1)^{l+1}}{N_{k}}\frac{\prod_{i\neq k}(x-a_{i})}{\sqrt{-\prod_{i=1}^{n}(x-a_{i})(x-b_{i})}}\mathrm{e}^{is\omega(x)}\,,\hspace{0.7cm}x\in(a_{l},b_{l})\,,\label{solunorm}
\end{equation}
with the normalization factor\footnote{Note that the expression apparently differs from eq. (36) of \cite{reduced_density}.
However, we have 
\begin{equation}
\sum_{j=1}^{n}\frac{\prod_{l\neq k}(b_{j}-a_{l})}{(b_{j}-a_{k})\prod_{l\neq j}(b_{j}-b_{l})}=\frac{\prod_{i\neq k}(a_{i}-a_{k})}{\prod_{i=1}^{n}(b_{i}-a_{k})}\,,
\end{equation}
and then, both equations are in agreement.} 
\begin{equation}
N_{k}:=\sqrt{2\pi}\left(\frac{\prod_{i\neq k}(a_{i}-a_{k})}{\prod_{i=1}^{n}(b_{i}-a_{k})}\right)^{1/2}\,.\label{Nk}
\end{equation}
In the rest of this subsection, we will show that the set $\{u_{s}^{k}\}_{k=1}^{n}$
is orthonormal and complete.

The orthonormality of $\{u_{s}^{k}\}_{k=1}^{n}$ follows immediately
from equation \eqref{escalaa}, and hence, we have 
\begin{equation}
\int_{A}dx\,u_{s}^{k*}(x)u_{s'}^{k'}(x)=\delta_{k,k'}\,\delta(s-s')\,.
\end{equation}

The completeness is quite less obvious. The general argument of section
\ref{sec12} shows that $n$ is the maximal degeneracy and then any
$n$ linearly independent vectors should form a complete basis. This
fact can be shown explicitly as follows. Using the eigenfunctions
\eqref{solunorm}, we have that
\begin{equation}
\sum_{k=1}^{n}\int_{-\infty}^{+\infty}\!\!\!\!\! ds\,u_{s}^{k}(x)u_{s}^{k*}(y)=2\pi\,\sqrt{\prod_{i=1}^{n}\frac{(x-a_{i})(y-a_{i})}{(x-b_{i})(y-b_{i})}}\left(\sum_{k=1}^{n}\frac{1}{N_{k}^{2}(x-a_{k})(y-a_{k})}\right)\sum_{l=1}^{n}\frac{\delta(x-x_{l})}{\omega'(x_{l})}\,,\label{dels}
\end{equation}
where $x_{l}\equiv x_{l}(\omega(y))\in A_{l}$ are the $n$ roots
of the polynomial equation \eqref{polr} for $\omega=\omega(y)$. In particular, when $y\in A_{l}$ then $x_{l}\equiv y$. We use the following algebraic relation\footnote{This relation is valid for any complex numbers $a_{1},\cdots,a_{n},b_{1},\cdots,b_{n},x,y$. It can be proven using the definition \eqref{Nk} for the normalization constants $N_{k}$ and decomposing the rational function at both sides into the poles for the variable $x$.}
 
\begin{eqnarray}
\sum_{k=1}^{n}\frac{1}{N_{k}^{2}(x-a_{k})(y-a_{k})}\!\!\! & = & \!\!\!\frac{\prod_{k=1}^{n}(x-b_{k})(y-a_{k})-\prod_{k=1}^{n}(y-b_{k})(x-a_{k})}{2\pi(x-y)\prod_{k=1}^{n}(x-a_{k})(y-a_{k})}\nonumber \\
 & = & \!\!\!\frac{P(x,y)}{2\pi\prod_{k=1}^{n}(x-a_{k})(y-a_{k})}\,,\label{trick}
\end{eqnarray}
where the function
\begin{equation}
P(x,y):=\frac{\prod_{k=1}^{n}(x-b_{k})(y-a_{k})-\prod_{k=1}^{n}(y-b_{k})(x-a_{k})}{x-y}\,,\label{sx}
\end{equation}
is a polynomial in $x$ of degree $n-1$ for each fixed $y$, and
its roots are the points $x=x_{l}$ except when $x_{l}=y$. As a consequence,
the only delta function which survives in \eqref{dels} is for $x=y$,
and hence 
\begin{equation}
\sum_{k=1}^{n}\int_{-\infty}^{+\infty}ds\,u_{s}^{k}(x)u_{s}^{k*}(y)=-\frac{P(x,x)}{\prod_{k=1}^{n}(x-a_{k})(x-b_{k})}\frac{1}{\omega'(x)}\delta(x-y)\,.\label{peda}
\end{equation}
In order to get a better expression for $P(x,x)$, from \eqref{sx}
we define a new function
\begin{equation}
Q(x,y)=P(x,y)(x-y)=\prod_{k=1}^{n}(x-b_{k})(y-a_{k})-\prod_{k=1}^{n}(y-b_{k})(x-a_{k})\,,
\end{equation}
which allows us to compute 
\begin{equation}
P(x,x)=\partial_{x}Q(x,y)|_{y=x}=-\omega'(x)\prod_{k=1}^{n}(x-a_{k})(x-b_{k})\,.\label{zo}
\end{equation}
Finally, replacing \eqref{zo} into \eqref{peda} we obtain the completeness
relation
\begin{equation}
\sum_{k=1}^{n}\int_{-\infty}^{+\infty}ds\,u_{s}^{k}(x)u_{s}^{k*}(y)=\delta(x-y)\,.
\end{equation}

\subsection{Modular Hamiltonian}

In this subsection, we rederive the results of \cite{reduced_density}
about the modular Hamiltonian using the information about the spectral
decomposition of the correlator kernel $C_{A}(x-y)$ obtained in the
previous subsections. The modular flow corresponding to this modular
Hamiltonian and the entanglement entropy for several intervals have
been computed in \cite{reduced_density}. Recently, the modular Hamiltonian
has also been derived using Euclidean path integral methods in \cite{Wong:2018svs}.
In \cite{Longo_xu}, the MI between several intervals have been computed
using the Araki formula without using a cutoff to compute the EE.

From \eqref{coco} and \eqref{lami}, the correlator kernel writes
\begin{equation}
C_{A}(x-y)=\sum_{k=1}^{n}\int_{-\infty}^{+\infty}ds\,u_{s}^{k}(x)\frac{1+\tanh(\pi s)}{2}u_{s}^{k\,*}(y)\,,
\end{equation}
and using this formula and \eqref{hache}, we obtain the following
expression for the modular Hamiltonian kernel 
\begin{equation}
\mathcal{K}(x,y)=\sum_{k=1}^{n}\int_{-\infty}^{+\infty}ds\,u_{s}^{k}(x)\,2\pi s\,u_{s}^{k\,*}(y)\,.\label{H_ker}
\end{equation}
Using equation \eqref{solunorm}, we get 
\begin{equation}
\mathcal{K}(x,y)=-i2\pi\,k(x,y)\,\delta'\left(\omega\left(x\right)-\omega\left(y\right)\right)\,,\label{H_ker2}
\end{equation}
where
\begin{equation}
k(x,y)=2\pi\sqrt{\prod_{i=1}^{n}\frac{\left(x-a_{i}\right)\left(y-a_{i}\right)}{\left(x-b_{i}\right)\left(y-b_{i}\right)}}\left(\sum_{k=1}^{n}\frac{1}{N_{k}^{2}\left(x-a_{k}\right)\left(y-a_{k}\right)}\right)\,.\label{defK}
\end{equation}
The aim of the rest of this subsection is to obtain a simplified expression
for the modular Hamiltonian kernel \eqref{H_ker2}. First, we have
the following identity for the Dirac delta term 
\begin{equation}
\delta'\left(\omega\left(x\right)-\omega\left(y\right)\right)=\sum_{l=1}^{n}\delta'\left(x-x_{l}\right)\frac{1}{\omega'\left(x\right)^{2}}-\delta\left(x-x_{l}\right)\frac{\omega''\left(x\right)}{\omega'\left(x\right)^{3}}\,,\label{deltacomp}
\end{equation}
where $x_{l}\equiv x_{l}(y)\in A_{l}$ are the roots of $\omega(x)=\omega(y)$
introduced in equation \eqref{dels}. From this last equation, the
modular Hamiltonian splits into the sum of a local and a non-local
kernel
\begin{equation}
\mathcal{K}(x,y)=\mathcal{K}_{loc}(x,y)+\mathcal{K}_{noloc}(x,y)\,,
\end{equation}
where $\mathcal{K}_{loc}(x,y)$ comes from the term in \eqref{deltacomp}
whit $x_{l}(y)=y$, and $\mathcal{K}_{noloc}(x,y)$ comes from the
$n-1$ terms in \eqref{deltacomp} with $x_{l}(y)\neq y$. We discuss
these two contributions separately.

\paragraph{Local part\protect \\
}

The local part for the modular Hamiltonian kernel is 
\begin{equation}
\mathcal{K}_{loc}(x,y)=-i2\pi\,k(x,y)\,\left[\frac{1}{\omega'(x)^{2}}\delta'(x-y)-\frac{\omega''(x)}{\omega'(x)^{3}}\delta(x-y)\right]\,.\label{Hloc}
\end{equation}
In order to simplify the above expression, we need to understand \eqref{Hloc}
as a distribution acting over some smooth compactly supported test
function $\varphi(x,y)$. Integrating by parts, the derivative of
the Dirac delta is converted to 
\begin{eqnarray}
\varphi(x,y)k(x,y)\frac{1}{\omega'(x)^{2}}\delta'(x-y)\!\!\! & = & \!\!\!-\Bigg[\partial_{x}\varphi(x,y) k(x,x)\frac{1}{\omega'(x)^{2}}+\varphi(x,x)\left.\partial_{x}k(x,y)\right|_{y=x}\frac{1}{\omega'(x)^{2}}\nonumber \\
 &  & \!\!\!+ \, \varphi(x,y)k(x,x)\partial_{x}\left(\frac{1}{\omega'(x)^{2}}\right)\Bigg]\delta(x-y)\,,\label{deltap}
\end{eqnarray}
The above expression can be simplified by using the following identities
\begin{eqnarray}
k(x,x)\!\!\! & = & \!\!\!\omega'(x)\,,\label{kxx}\\
\left.\partial_{x}k(x,y)\right|_{y=x}\!\!\! & = & \!\!\!\frac{1}{2}\omega''(x)\,.\label{dkxx}
\end{eqnarray}
which follow from \eqref{defK} and the algebraic relation\footnote{Like \eqref{trick}, equation \eqref{trick2} is a pure algebraic relation
valid for any complex number $a_{k}$, $b_{k}$ and $x$. It can be
shown matching the coefficients of the poles in $x$ on both sides.} 
\begin{equation}
\sum_{k=1}^{n}\frac{1}{N_{k}^{2}}\frac{1}{(x-a_{k})}=\frac{1}{2\pi}\left(1+\mathrm{e}^{-\omega(x)}\right)\,.\label{trick2}
\end{equation}
The final steps consist on replacing \eqref{kxx} and \eqref{dkxx}
into \eqref{deltap}, and integrating by parts the term containing
$\partial_{x}\varphi(x,y)$, in order to factorize the test function.
We finally get 
\begin{equation}
\mathcal{K}_{loc}(x,y)=-i2\pi\left[\frac{1}{\omega'(x)}\delta'(x-y)+\frac{1}{2}\partial_{x}\left(\frac{1}{\omega'(x)}\right)\delta(x-y)\right]\,.\label{Hlocf}
\end{equation}
The local part of the modular Hamiltonian comes from \eqref{citric}
and \eqref{Hlocf} 
\begin{equation}
K_{loc}=2\pi\int_{A}dx\,\frac{1}{\omega'(x)}T(x)\,,
\end{equation}
where $T(x)=\frac{1}{2}\left[i\partial_{x}\psi^{\dagger}(x)\psi(x)-\psi^{\dagger}(x)i\partial_{x}\psi(x)\right]$
is the energy density operator.

\paragraph{Non-local part\protect \\
}

The non-local part of the modular Hamiltonian kernel is 
\begin{equation}
\mathcal{K}_{nonloc}(x,y)=-i2\pi\,k(x,y)\,\left[\sum_{l=1\,x_{l}\neq y}^{n}\frac{1}{\omega'(x)^{2}}\delta'(x-x_{l})-\frac{\omega''(x)}{\omega'(x)^{3}}\delta(x-x_{l})\right]\,.\label{Hnonloc}
\end{equation}
The first term can be simplified by a similar computation as we did
around equation \eqref{deltap}. Here, the situation is simpler because
$k(x_{l},y)\equiv0$ for all $x_{l}\neq y$ as we showed in \eqref{trick}.
Hence, the unique term which survives is the one proportional to de
derivative of $k(x,y)$, i.e,
\begin{equation}
\left.\partial_{x}k(x,y)\right|_{x=x_{l}}=\frac{\omega'(x_{l})}{x_{l}-y}\,.
\end{equation}
Replacing it on \eqref{Hnonloc}, we arrive to 
\begin{equation}
\mathcal{K}_{noloc}(x,y)=i2\pi\sum_{l=1,\,x_{l}\neq y}^{n}\frac{1}{(x-y)}\frac{1}{\omega'(x)}\delta\left(x-x_{l}\left(\omega\left(y\right)\right)\right)\,,\label{kernelnl}
\end{equation}
and 
\begin{eqnarray}
K_{noloc}\!\!\! & = & \!\!\!i\,2\pi\sum_{l=1,\,x_{l}\neq x}^{n}\int_{A}dx\,\psi^{\dagger}\left(x_{l}\right)\frac{1}{\left(x_{l}-x\right)}\frac{1}{\omega'\left(x_{l}\right)}\psi\left(x\right)\\
 & = & \!\!\!i\,2\pi\int_{A\times A,\,x\neq y}dx\,dy\,\psi^{\dagger}\left(x\right)\frac{\delta\left(\omega\left(x\right)-\omega\left(y\right)\right)}{x-y}\psi\left(y\right)\,.
\end{eqnarray}

\paragraph{Two intervals case\protect \\
}

For the case of two intervals $A=(a_{1},b_{1})\cup(a_{2},b_{2})$,
the modular Hamiltonian operator $K=K_{loc}+K_{noloc}$ reduces to
\begin{eqnarray}
K_{loc}\!\!\! & = & \!\!\!2\pi\int_{A}dx\,\omega'(x)^{-1}\,T\left(x\right)\,,\\
K_{noloc}\!\!\! & = & \!\!\!i2\pi\int_{A}dx\,\gamma\left(x\right)\psi^{\dagger}(x)\psi\left(\bar{x}\right)\,,
\end{eqnarray}
where
\begin{eqnarray}
\omega'(x)\!\!\! & = & \!\!\!\frac{1}{x-a_{1}}+\frac{1}{x-a_{2}}-\frac{1}{x-b_{1}}-\frac{1}{x-b_{2}}\,,\\
\nonumber \\
\gamma\left(x\right)\!\!\! & = & \!\!\!\frac{\omega'(x)^{-2}}{x\left(a_{1}+a_{2}-b_{1}-b_{2}\right)+\left(b_{1}b_{2}-a_{1}a_{2}\right)}\frac{\left(b_{1}-a_{1}\right)\left(a_{2}-b_{1}\right)\left(b_{2}-a_{1}\right)\left(b_{2}-a_{2}\right)}{\left(x-a_{1}\right)\left(x-a_{2}\right)\left(x-b_{1}\right)\left(x-b_{2}\right)}\,,\hspace{1cm}\\
\nonumber \\
\bar{x}\!\!\! & = & \!\!\!\frac{a_{1}a_{2}\left(x-b_{1}-b_{2}\right)-b_{1}b_{2}\left(x-a_{1}-a_{2}\right)}{x\left(a_{1}+a_{2}-b_{1}-b_{2}\right)+\left(b_{1}b_{2}-a_{1}a_{2}\right)}\,.\label{conj-points}
\end{eqnarray}
For two symmetric intervals $A=(-R,-r)\cup(r,R)$, $0<r<R$, we can
write a more explicit form of the modular Hamiltonian
\begin{equation}
K=2\pi\int_{A}dx\,\frac{\left(x^{2}-r^{2}\right)\left(R^{2}-x^{2}\right)}{2\left(R-r\right)\left(x^{2}+rR\right)}\,T\left(x\right)+i\pi\int_{A}dx\,\psi^{\dagger}\left(x\right)\frac{rR\left(x^{2}-r^{2}\right)\left(R^{2}-x^{2}\right)}{\left(R-r\right)x\left(x^{2}+rR\right)^{2}}\psi\left(\bar{x}\right)\,,
\end{equation}
where now $\bar{x}=-\frac{rR}{x}$.

\section{The free chiral current\label{escalar}}

The free chiral current field is defined by means a field operator
$J(x)$ in the null-line, which is identified with the chiral derivative
of a massless free scalar in $d=2$, i.e. $J(x^{+})=\partial_{+}\phi(x^{+})$.
It is usually defined as a quantum field $J\left(x\right)$ satisfying
the commutation relations
\begin{equation}
\left[J(x),J(y)\right]=i\,\delta'(x-y)=:i\,C(x-y)\,,\label{commut}
\end{equation}
and having the vacuum two-point correlator
\begin{equation}
F(x,y):=\langle J(x)J(y)\rangle=-\frac{1}{2\pi}\frac{1}{(x-y-i\,0^{+})^{2}}\,.\label{cotto}
\end{equation}
This model is Gaussian and all other multipoint correlators follow
from \eqref{cotto} according to Wick's theorem. From the algebraic
standpoint, given a test function $f\in\mathcal{S}\left(\mathbb{R},\mathbb{R}\right)$,
the operator 
\begin{equation}
J\left(f\right):=\int_{\mathbb{R}}J\left(x\right)f\left(x\right)\,dx\,,
\end{equation}
is a well-defined unbounded operator acting on the bosonic Fock Hilbert
space $\mathcal{H}_{0}$. This Hilbert space is constructed as the
symmetric tensor product of the one-particle Hilbert space of massless
particles of zero helicity. This one-particle Hilbert space carries
a unitary representation of $\mathsf{M\ddot{o}b}$, which naturally
extends to a unitary representation of $\mathsf{M\ddot{o}b}$ on $\mathcal{H}_{0}$.
The chiral Hamiltonian can be expressed as
\begin{equation}
H=\frac{1}{2}\int dx\,:J^{2}(x):\,,
\end{equation}
where $T(x):=\frac{1}{2}:J^{2}(x):$ is the energy density operator.
In order to get bounded operators, we define the Weyl unitaries $W\left(f\right):=\mathrm{e}^{iJ\left(f\right)}$,
which satisfy the usual CCR relations
\begin{equation}
W\left(f\right)W\left(g\right)=\mathrm{e}^{i\int_{\mathbb{R}}dx\,f'\left(x\right)g\left(x\right)}W\left(g\right)W\left(f\right)\,.
\end{equation}
For any open set $A\subset\mathbb{R}$, the local algebras are defined as
\begin{equation}
\mathcal{A}\left(A\right):=\left\{ \mathrm{e}^{iJ\left(f\right)}\,:\,\mathrm{supp}\left(f\right)\subset A\right\} ''\,.\label{c5-curr-net}
\end{equation}
It is not difficult to show that such a collection of algebras satisfies
all the axioms of the definition \ref{c5-def-chiral}.

It is interesting to note that the free chiral current is a subnet
of the free chiral fermion. In fact, the current fermion operator
$J_{\psi}\left(x\right):=:\psi^{\dagger}\left(x\right)\psi\left(x\right):$
satisfies the commutation relations \eqref{commut}, and has the same
vacuum expectation value \eqref{cotto}. More precisely, the net of
vN algebras 
\begin{equation}
\mathcal{B}_{\mathrm{f}}\left(A\right):=\left\{ \mathrm{e}^{i\int_{\mathbb{R}}J_{\psi}\left(x\right)f\left(x\right)}\,:\,\mathrm{supp}\left(f\right)\subset A\right\} ''\subsetneq\mathcal{A}_{\mathrm{f}}\left(A\right)
\end{equation}
is a subnet of the free chiral fermion isomorphic to the current net
\ref{c5-curr-net} \cite{Kac,Bischoff:2011mx}. 

In the following, we study the modular Hamiltonian associated with
the vacuum state and a region $A\subset\mathbb{R}$. Because of the
complexity of the problem, we restrict the attention to the case of
one or two intervals. The commutation relations \eqref{commut} define
a general CCR algebra, and hence, we can use all the formulas developed
in section \ref{c3-sec-genCAR}. Then, the modular modular Hamiltonian
is given by equations \eqref{c3-genCCR-mat}, \eqref{c3-genCCR-mh}, and
\eqref{c3-genCCR-V}, which means
\begin{eqnarray}
K\!\!\! & = & \!\!\!\int_{A\times A}dx\,dy\,J(x)\mathcal{K}(x,y)J(y)\,,\\
\mathcal{K}\!\!\! & = & \!\!\!-\frac{i}{2}\,\frac{V}{|V|}\log\left(\frac{|V|+1/2}{|V|-1/2}\right)\,C^{-1}\,,\label{hamidos}\\
V\!\!\! & = & \!\!\!-iC^{-1}F-\frac{1}{2}\,.\label{dit2}
\end{eqnarray}
We emphasize that the operators $\mathcal{K},V,C,$ and $F$ above
are operators acting as kernels in the Hilbert space $L^{2}\left(A\right)$
of square-integrable functions with support in $A$. According to
the calculations from the lattice model, the operator $V$ is not
self-adjoint, but it is diagonalizable, and its spectrum belongs to
$(-\infty,-1/2]\cup[1/2,+\infty)$. In the continuum QFT, we expect
that the operator $V$ has a complete set of (generalized) eigenvectors,
with continuous spectrum $\mathrm{spec}\left(V\right)=(-\infty,-1/2]\cup[1/2,+\infty)$.
For later convenience, we parametrize the corresponding eigenvalues
by $\frac{1}{2}\coth(\pi s)$ with $s\in\mathbb{R}$. The eigenvectors
of $V$ and $V^{\dagger}$ are denoted by
\begin{eqnarray}
V|u_{s}^{k}\rangle\!\!\! & = & \!\!\!\frac{1}{2}\coth(\pi s)|u_{s}^{k}\rangle\,,\label{dit3}\\
V^{\dagger}|v_{s}^{k}\rangle\!\!\! & = & \!\!\!\frac{1}{2}\coth(\pi s)|v_{s}^{k}\rangle\,,\label{dit1}
\end{eqnarray}
where $k\in\Upsilon$ is a possible degeneracy index. We normalize
the eigenvectors according to
\begin{equation}
\langle v_{s}^{k}|u_{s'}^{k'}\rangle=\int_{A}v_{s}^{k}\left(x\right)^{*}u_{s'}^{k'}\left(x\right)\,dx=\delta_{k,k'}\,\delta(s-s')\,.\label{manja}
\end{equation}
It is not difficult to see, from \eqref{dit2}, \eqref{dit3}, and
\eqref{dit1}, that $C|u_{s}^{k}\rangle$ is an eigenvector of $V^{\dagger}$
with eigenvalue $\frac{1}{2}\coth(\pi s)$, and hence, it is a linear
combination of the vectors $\left\{ |v_{s}^{k}\rangle\,:\,k\in\Upsilon\right\} $.
The orthogonality relation \eqref{manja} leaves us the freedom to
redefine the eigenbasis $\left\{ |u_{s}^{k}\rangle\,:\,k\in\Upsilon\right\} $
by an arbitrary matrix, and the eigenbasis $\left\{ |v_{s}^{k}\rangle\,:\,k\in\Upsilon\right\} $
by the inverse adjoint matrix. We can use this freedom to set $|v_{s}^{k}\rangle\wasypropto C\,|u_{s}^{k}\rangle$
for all $s\in\mathbb{R}$ and all $k\in\Upsilon$. The phase of
the proportionality constant is fixed to be $i\,\textrm{sign}(s)$,
as we can see from taking the scalar product of \eqref{dit3} with
$\langle v_{s'}^{k'}|$ and using \eqref{manja} and the positivity
of $F$. As a result, we can further fix the eigenvectors by taking
\begin{equation}
|v_{s}^{k}\rangle:=i\,\textrm{sign}(s)\,C|u_{s}^{k}\rangle\,.\label{cuen}
\end{equation}
In terms of these vectors, the kernel \eqref{hamidos} is written simply
as 
\begin{equation}
\mathcal{K}\left(x,y\right)=\sum_{k\in\Upsilon}\int_{-\infty}^{+\infty}ds\,u_{s}^{k}\left(x\right)^{*}\,\pi|s|\,u_{s}^{k}\left(y\right)\,.\label{hkernel}
\end{equation}
The EE and the Rényi EE are given by formulas \eqref{c3-genCCR_ee}
and \eqref{c3-genCCR_renyi}. They could be rewritten in terms of
the eigenvectors (\ref{dit3}-\ref{dit1}) as
\begin{eqnarray}
S\left(A\right)\!\!\! & = & \!\!\!\sum_{k\in\Upsilon}\int_{0}^{+\infty}\!ds\int_{A}dx\,g(s)\,u_{s}^{k}(x)\,v_{s}^{k}(x){}^{*}\,,\label{c5-curr-ee}\\
S_{\alpha}\left(A\right)\!\!\! & = & \!\!\!\sum_{k\in\Upsilon}\int_{0}^{+\infty}\!ds\int_{A}dx\,g_{\alpha}(s)\,u_{s}^{k}(x)\,v_{s}^{k}(x){}^{*}\,,\label{c5-curr-renyi}
\end{eqnarray}
where
\begin{eqnarray}
g(s)\!\!\! & := & \!\!\!\frac{1+\coth\left(\pi s\right)}{2}\log\left(\frac{\coth\left(\pi s\right)+1}{2}\right)+\frac{1-\coth\left(\pi s\right)}{2}\log\left(\frac{\coth\left(\pi s\right)-1}{2}\right), \hspace{1.2cm}\label{c5-curr-eeg}\\
g_{\alpha}(s)\!\!\! & := & \!\!\!\frac{1}{\alpha-1}\,\log\left[\left(\frac{\coth(\pi s)}{2}+\frac{1}{2}\right)^{\alpha}-\left(\frac{\coth(\pi s)}{2}-\frac{1}{2}\right)^{\alpha}\right].\label{c5-curr-renyig}
\end{eqnarray}
The integrals over the $x$ coordinate in equations \eqref{c5-curr-ee}
and \eqref{c5-curr-renyi} give delta functions $\delta(0)$, and
hence, the EE and the Rényi EE are divergent. This is just the usual
divergence in QFT due to the continuum spectrum of the modular Hamiltonian.
A convenient way to regularize these entropies is to integrate up
to a distance cutoff $\epsilon>0$ from the boundary $\partial A$
(see section \ref{c3-sec_cont}). In the two interval case, the integration
region $A$ has to be replaced by $A^{\left(\epsilon\right)}:=\left(a_{1}+\epsilon,b_{1}-\epsilon\right)\cup\left(a_{2}+\epsilon,b_{2}-\epsilon\right)$.

In sections \ref{unintervalo} and \ref{dosintervalos}, we handle
the case of one and two intervals independently.

\subsection{The single interval case\label{unintervalo}}

Let us first consider the simplest case of a single interval $A:=(a,b)$.
The inverse $(\delta')^{-1}(x,y)$ of the commutator $\delta'(x-y)$,
acting on a test function $f(x)$, has to be a linear combination
of 
\begin{equation}
\int_{a}^{b}dy\,(\delta')^{-1}(x,y)f(y)=\int_{a}^{x}dy\,f(y)+ \int_{a}^{b}dy\,l(y)\,f(y)\,,
\end{equation}
where  $l:[a,b] \in \mathbb{R} \rightarrow \mathbb{C}$ is some fixed function. This last
term is linear in $f(y)$, and since it is independent of $x$, it
is annihilated by $\delta'$. In kernel notation we have to combine\footnote{Only two of these three kernels are linearly independent.}
\begin{equation}
\Theta(x-y)\,,\hspace{0.5cm}-\Theta(y-x)\,,\hspace{0.5cm}l(y)\,.
\end{equation}
There is only one antisymmetric inverse, given by 
\begin{equation}
(\delta')^{-1}(x,y):=\frac{1}{2}\left(\Theta(x-y)-\Theta(y-x)\right)\,,\label{dit}
\end{equation}
which acts on a test function $f(x)$ as
\begin{equation}
((\delta')^{-1}f)(x)=\frac{1}{2}\left(\int_{a}^{x}dy\,f(y)-\int_{x}^{b}dy\,f(y)\right)\,.
\end{equation}
Then, we have that $\delta'\circ(\delta')^{-1}=\delta$, and $(\delta')^{-1}\circ\delta'=\delta$
on test functions that vanish on the boundary of the interval.

Hence, following (\ref{dit2}-\ref{dit1}) and \eqref{dit}, our first
task is to solve the spectrum of 
\begin{equation}
2\pi C^{-1}F=\frac{1}{(x-y-i\,0^{+})}-\frac{1}{2(b-y)}-\frac{1}{2\left(a-y\right)}\,,\label{dospi}
\end{equation}
and
\begin{equation}
2\pi FC^{-1}=\frac{1}{(x-y-i\,0^{+})}-\frac{1}{2(b-x)}-\frac{1}{2(a-x)}\,.\label{dospi2}
\end{equation}

\subsubsection{A complete set of eigenvectors}

The aim of this subsection is to find a complete set of the eigenvectors
for the kernels \eqref{dospi} and \eqref{dospi2}, which are the
eigenvectors of the relevant kernels \eqref{dit3} and \eqref{dit1}
in the single interval case. In the end, the eigenvectors are given
by \eqref{c5-cur1-u} and \eqref{up} below.

We proceed as in the case of the fermion field. We think the interval
$A$ as included in the real axis of the complex plane, and we consider
an analytic function $S(z)$, having a multiplicative boundary condition
on the interval $A$ as in \eqref{disc}, i.e.
\begin{eqnarray}
 &  & S(z)\text{ analytic in }\mathbb{C}-\bar{A}\,,\label{hol_A1}\\
 &  & S^{+}(x_{1})=\lim_{x_{2}\rightarrow0^{+}}S(x_{1}+ix_{2})=\lambda\lim_{x_{2}\rightarrow0^{-}}S(x_{1}+ix_{2})=\lambda\,S^{-}(x_{1})\,,\qquad x_{1}\in A\,. \hspace{2cm} \label{disc1}
\end{eqnarray}
We further impose boundary conditions at infinity and at the endpoints
of the interval, 
\begin{eqnarray}
 &  & \lim_{z\rightarrow\infty}\left|\,S(z)\right|<\infty\,,\label{bc11}\\
 &  & \lim_{z\rightarrow\partial A}|S(z)|<\infty\,.\label{bc22}
\end{eqnarray}
Consider now the complex integral 
\begin{equation}
\oint dz_{2}\,\left(\frac{1}{z_{1}-z_{2}}-\frac{1}{2}\frac{1}{b-z_{2}}-\frac{1}{2}\frac{1}{a-z_{2}}\right)S(z_{2})=0\label{circint1}
\end{equation}
along a closed curve in the complex plane encircling both $A$ and
$z_{1}\in\mathbb{C}-\bar{A}$. This integral is zero because the integrand
has zero residue at infinity. We can shrink the integration contour
around the point $z_{1}$ and around the cut to get 
\begin{equation}
S(z_{1})=\frac{i}{2\pi}(1-\lambda^{-1})\fint_{A}dy\,\left(\frac{1}{z_{1}-y}-\frac{1}{2}\frac{1}{b-y}-\frac{1}{2}\frac{1}{a-y}\right)S^{+}(y)\,.\label{siosi}
\end{equation}
The symbol $\fint$ for the integral means that it is regularized
at the endpoints of the interval by taking the complex integral on
a small circle around these endpoints (as implied by \eqref{circint1}),
and then take the limit of zero circle size. We will soon be more
specific on how this regularization can be expressed directly by means
the function $S^{+}(y)$.

Taking the limit of $z_{1}\rightarrow x-i0^{+}$ with $x\in A$, and
using \eqref{disc1}, \eqref{dospi}, and \eqref{dit2}, we arrive
to 
\begin{equation}
\fint_{A}dy\,V(x,y)\,S^{+}(y)=\frac{\lambda+1}{2(1-\lambda)}\,S^{+}(x)\,.\label{rumania}
\end{equation}
Since the eigenvalues of $|V|$ are in $[1/2,\infty)$, we have that
$\lambda>0$ in contrast to the case of the fermion, where $\lambda$
is negative. We then write 
\begin{equation}
\lambda:=\mathrm{e}^{-2\pi s}\,,\quad s\in\mathbb{R}\,.\label{c5-curr-lambda}
\end{equation}
The eigenvalue in \eqref{rumania} then coincides with $\frac{1}{2}\coth(\pi s)$
as in \eqref{dit3}.

Therefore, for each solution $S(z)$ of the problem (\ref{hol_A1}-\ref{bc22})
in the complex plane, we get an eigenvector of the kernel $V(x,y)$
on the interval $A$. The eigenvectors, in contrast to the case of
the fermion, are bounded at the endpoints of the interval (see equation
\eqref{bc22}). This is in accordance with boundary conditions for
scalars \cite{Casini_review}. Conversely, if we have a solution of
\eqref{rumania}, we can use it as boundary data on the interval in
\eqref{siosi}, which gives a solution $S(z)$ satisfying (\ref{hol_A1}–\ref{bc22}).
These problems are then mutually equivalent.

For the single interval $A$, we can write a solution to the above
problem as 
\begin{equation}
S(z):=\mathrm{e}^{-is\tilde{\omega}(z)}\,,\quad\tilde{\omega}(z):=\log\left(\frac{z-a}{z-b}\right)\,.\label{yy}
\end{equation}
This obeys all the conditions (\ref{hol_A1}–\ref{bc22}). To show
that \eqref{yy} is the unique solution we proceed as follows. Let
us suppose that $\tilde{S}(z)$ is another solution to the same problem.
Then, the function $\tilde{S}(z)S(z)^{-1}$ is an analytic function
on the plane except, perhaps, at the endpoints of the interval. But,
because of the boundary conditions (\ref{bc11}–\ref{bc22}), we must
have that $\tilde{S}(z)S(z)^{-1}$ is bounded at infinity and at the
endpoints of the interval. Then, we have that $\tilde{S}(z)S(z)^{-1}$
is a constant function. Therefore, \eqref{yy} is the unique solution
to the problem.

The eigenvectors are given by the boundary values of $S(z)$ on the
interval
\begin{equation}
u(x)\propto S^{+}(x)\propto\mathrm{e}^{-is\omega(x)}\,,\quad\omega(x):=\log\left(\frac{x-a}{b-x}\right)\,.
\end{equation}

Now we explain more precisely the regularization in \eqref{siosi}
and \eqref{rumania}. Frequently, we will encounter integrals on the
real line of the form 
\begin{equation}
\int_{a}^{b}dx\,f(x)\,,\quad\textrm{where }\:f(x)\sim c\frac{\mathrm{e}^{-is\log(x-a)}}{x-a}\,,\;x\rightarrow a^{+}\,.
\end{equation}
Then, this integral will have an oscillatory but bounded term $c\,\mathrm{e}^{-is\log(x-a)}/(is)$
in the lower boundary as $x\rightarrow a^{+}$, and hence, it does
not converge. The regularization used above just subtracts this oscillatory
phase, i.e.
\begin{eqnarray}
\fint_{a}^{b}dx\,f(x) \!\!\! &=& \!\!\!\lim_{\epsilon\rightarrow0^{+}}\int_{a+\epsilon}^{b}dx\,f(x)-c\,\frac{\mathrm{e}^{-is\epsilon}}{is} \nonumber \\
&=& \!\!\! \int_{a}^{b}dx\,\left(f(x)-c\,\frac{\mathrm{e}^{-is\log(x-a)}}{x-a}\right)+c\,\frac{\mathrm{e}^{-is\log(b-a)}}{(-is)}\,.\label{fint_reg}
\end{eqnarray}
If the oscillatory term appears on the upper end of the integral,
an analogous subtraction is understood. As we mentioned above, this
subtraction appears naturally when the integral comes from a limit
of a complex integral around the cut, as in the transformation from
\eqref{circint1} to \eqref{siosi}. The definition of the kernel
$V$ has to be understood with this regularization.\footnote{Note that this regularization eliminates from the bare integral an
infinitely oscillatory phase in $s$, which produces vanishing terms
in any finite calculation involving integrals over the spectrum.}

Now, we have to look at the eigenvectors of \eqref{dit1}. For this,
we can just use the relation \eqref{cuen}. However, we find instructive
to compute them directly from the kernel \eqref{dospi2}. To do that,
we take a different analytic function $S(z)$ and assume that it satisfies
the same multiplicative boundary condition \eqref{disc1}. However,
in order to obtain a solution of the eigenvector problem from the
complex integral, we are now forced to impose
\begin{eqnarray}
 &  & S(z)\sim\frac{1}{|z|^{2}}\,,\quad z\rightarrow\infty\,,\\
 &  & S(z)\textrm{ have at most pole singularities at }a\textrm{ and }b\,.
\end{eqnarray}
As above, we consider the complex integral 
\begin{equation}
\oint dz_{2}\,\left(\frac{1}{z_{1}-z_{2}}-\frac{1}{2}\frac{1}{b-z_{1}}-\frac{1}{2}\frac{1}{a-z_{1}}\right)S(z_{2})\,,
\end{equation}
along a closed curve in the complex plane encircling both $A$ and
$z_{1}\in\mathbb{C}-\bar{A}$. We shrink the integration contour around
the point $z_{1}$ and around the cut to get
\begin{equation}
S(z_{1})=\frac{i}{2\pi}(1-\lambda^{-1})\fint_{A}dy\,\left(\frac{1}{z_{1}-y}-\frac{1}{2}\frac{1}{b-z_{1}}-\frac{1}{2}\frac{1}{a-z_{1}}\right)S^{+}(y)\,.
\end{equation}
The limit $z_{1}\rightarrow x-i0^{+}$, with $x\in A$, gives 
\begin{equation}
\fint_{A}dy\,V^{\dagger}(x,y)\,S^{+}(y)=\frac{\lambda+1}{2(1-\lambda)}\,S^{+}(x)\,.
\end{equation}

The value of $\lambda$ is the same as in \eqref{c5-curr-lambda},
giving the same multiplicative boundary conditions for $S(z)$ as
for the eigenvectors of $V$. However, the boundary conditions at
infinity and at the endpoints of the interval are different. These
imply that the unique solution is 
\begin{equation}
S(z):=\mathrm{e}^{-is\tilde{w}(z)}\left(\frac{1}{z-a}-\frac{1}{z-b}\right)\,.
\end{equation}
The poles have to have opposite signs so that function decays at infinity
as $|z|^{-2}$. We recognize this function is proportional to the
derivative of \eqref{yy}, and it must be because of \eqref{cuen}.

To end, we obtain the eigenvectors
\begin{eqnarray}
u_{s}\left(x\right)\!\!\! & = & \!\!\!\frac{\mathrm{e}^{-is\omega(x)}}{\sqrt{2\pi|s|}}\,,\label{c5-cur1-u}\\
v_{s}\left(x\right)\!\!\! & = & \!\!\!i\,\textrm{sign}(s)\,u_{s}'\left(x\right)=\sqrt{\frac{|s|}{2\pi}}\,\mathrm{e}^{-is\omega(x)}\left(\frac{1}{x-a}-\frac{1}{x-b}\right)\,,\label{up}
\end{eqnarray}
which are normalized in order to satisfy \eqref{manja} and \eqref{cuen}.

\subsubsection{Modular Hamiltonian}

Replacing the formula \eqref{c5-cur1-u} for the eigenvectors into
the equation \eqref{hkernel} and after a simple integration, we get
the following expression for the modular Hamiltonian kernel 
\begin{equation}
\mathcal{K}(x,y)=\int_{-\infty}^{\infty}ds\,u_{s}(x)\,\pi|s|\,(u_{s}(y))^{*}=\pi\,(\omega'(x))^{-1}\,\delta(x-y)\,.\label{mh}
\end{equation}
Then, the modular Hamiltonian operator has the well-known form for
an interval in a CFT \cite{Hislop81,chm}
\begin{equation}
K_{A}=2\pi\int_{a}^{b}dx\,\frac{(b-x)(x-a)}{b-a}\,T(x)\,,
\end{equation}
where in the present case the energy density operator is $T\left(x\right):=\frac{1}{2}:J^{2}\left(x\right):$.

\subsubsection{Entanglement entropy}

According to \eqref{c5-curr-ee} and \eqref{c5-curr-eeg}, the EE
is 
\begin{equation}
\hspace{-.3cm} S\left(A\right)=\int_{a+\epsilon}^{b-\epsilon}\!dx\,\int_{0}^{\infty}\!ds\,g(s)\,\,u_{s}(x)(v_{s}(x))^{*}=\frac{1}{12}\int_{a+\epsilon}^{b-\epsilon}\!dx\,\omega'(x)=\frac{1}{6}\log\left(\frac{b-a}{\epsilon}\right)\,.\label{s1}
\end{equation}
This gives the expected result for a conformal model with one chiral
component of central charge $c=1$. The Rényi EE can be computed analogously
using \eqref{c5-curr-renyi} and \eqref{c5-curr-renyig}
\begin{equation}
S_{\alpha}\left(A\right)=\frac{1+\alpha}{12\,\alpha}\log\left(\frac{b-a}{\epsilon}\right)\,.
\end{equation}

\subsection{The two interval case\label{dosintervalos}}

Now, we consider a two-interval region $A:=A_{1}\cup A_{2}$, where
$A_{i}:=\left(a_{i},b_{i}\right)$. To start, we need first to know
the expression of the kernel $C^{-1}=(\delta')^{-1}$ for two intervals.
The commutator is block diagonal in each of the intervals, and we
get the same result as \eqref{dit} for each of the intervals separately
\begin{equation}
C^{-1}(x,y):=\begin{cases}
\frac{1}{2}\left(\Theta(x-y)-\Theta(y-x)\right)\,, & \textrm{if }x,y\in A_{1}\textrm{ or }x,y\in A_{2}\,,\\
0\,, & \textrm{otherwise ,}
\end{cases}
\end{equation}
or equivalently, 
\begin{equation}
C^{-1}(x,y)=\frac{1}{2}\left(\Theta(x-y)-\Theta(y-x)\right)-\frac{1}{2}\Theta(x-a_{2})\Theta(b_{1}-y)+\frac{1}{2}\Theta(y-a_{2})\Theta(b_{1}-x)\,.\label{dosdos}
\end{equation}
Notice that this last expression is antisymmetric, and its derivative
is the delta function. Then, we have that
\begin{eqnarray}
\hspace{-1cm} 2\pi C^{-1}F \!\!\!&=&\!\!\! \frac{1}{(x-y-i0^{+})} \nonumber \\
&& \!\!\! + \, \frac{1}{2}\left(\Theta_{A_{1}}(x)\left(\frac{1}{y-a_{1}}+\frac{1}{y-b_{1}}\right)+\Theta_{A_{2}}(x)\left(\frac{1}{y-a_{2}}+\frac{1}{y-b_{2}}\right)\right),\label{u1w}
\end{eqnarray}
and 
\begin{eqnarray}
\hspace{-1cm} 2\pi FC^{-1} \!\!\!&=&\!\!\! \frac{1}{(x-y-i0^{+})} \nonumber \\
&& \!\!\! -\, \frac{1}{2}\left(\left(\frac{1}{x-a_{1}}+\frac{1}{x-b_{1}}\right)\Theta_{A_{1}}(y)+\left(\frac{1}{x-a_{2}}+\frac{1}{x-b_{2}}\right)\Theta_{A_{2}}(y)\right),\label{v1w}
\end{eqnarray}
where $\Theta_{A_{j}}(x)=\Theta\left(x-a_{j}\right)\Theta\left(b_{j}-x\right)$
is the characteristic functions of the region $A_{j}$.

\subsubsection{A complete set of eigenvectors}

The aim of this subsection is to find a complete set of the eigenvectors
for the kernels \eqref{u1w} and \eqref{v1w}, which are the eigenvectors
of the relevant kernels \eqref{dit3} and \eqref{dit1} in the two
intervals case. In the end, the eigenvectors are given by \eqref{v1}
and \eqref{u2} below.

Now, we have to deal with the kernels \eqref{u1w} and \eqref{u1w}
which contain theta functions. At first glance, it might seem that
the analytic method used in previous sections does not apply here.
However, we will show how to bypass this issue. To begin with, let
us consider eigenvectors $v_{s}(x)$ of \eqref{v1w} satisfying the
following extra property 
\begin{equation}
\fint_{A_{1}}dx\,v_{s}(x)=\fint_{A_{2}}dx\,v_{s}(x)=0\,.\label{cope}
\end{equation}
For such particular eigenvectors, the second and third terms in \eqref{v1w}
vanish. In the end, we will show that \eqref{cope} is true in the general
case. Under this assumption, we have that $v_{s}(x)$ is an eigenfunction
of $(x-y-i0^{+})^{-1}$. Then we use the same ideas as for a single
interval, trying to obtain $v_{s}(x)$ as a boundary value of an analytic
function. We again look for analytic functions $S(z)$ on the complex
plane with multiplicative boundary conditions on the two intervals
$A$ as in \eqref{disc1}. The class of eigenfunctions $u_{s}(x)$
of the problem must behave near the endpoints of the intervals as
in the single interval case. That is, they should behave as pure phase
factors of the form 
\begin{equation}
u_{s}(x)\sim\begin{cases}
\mathrm{e}^{-is\log(x-a_{j})}\,, & x\rightarrow a_{j}^{+}\,,\\
\mathrm{e}^{is\log(b_{j}-x)}\,, & x\rightarrow b_{j}^{-}\,.
\end{cases}
\end{equation}
Their derivatives, the functions $v_{s}\left(x\right)$, should have
at most single poles (together with a phase factor) at the interval
endpoints. Under these conditions, the most general solution is of
the form 
\begin{equation}
S(z)\propto\mathrm{e}^{-is\tilde{\omega}(z)}\left(\frac{\alpha_{1}}{z-a_{1}}+\frac{\alpha_{2}}{z-b_{1}}+\frac{\alpha_{3}}{z-a_{2}}+\frac{\alpha_{4}}{z-b_{2}}\right)\,,\label{form}
\end{equation}
with $\alpha_{l}=\alpha_{l}(s;a_{1},b_{1},a_{2},b_{2})$ ($l=1,2,3,4$)
and
\begin{equation}
\tilde{\omega}(z):=\sum_{j=1}^{2}\log\left(\frac{z-a_{j}}{z-b_{j}}\right)\,.
\end{equation}
Integrating $S(z)$ along  contours encircling the two intervals and
a large circle at infinity, it is not difficult to see that the integral
at infinity is equal to the one over the two intervals, which vanishes
because of \eqref{cope}. Then, this function must falls as $|z|^{-2}$
to cope with \eqref{cope}, and we must impose the condition 
\begin{equation}
\alpha_{1}+\alpha_{2}+\alpha_{3}+\alpha_{4}=0\,.\label{sumacero}
\end{equation}
Calling $\vec{q}:=(a_{1},b_{1},a_{2},b_{2})$, we have from \eqref{cope}
that the coefficients $\alpha_{j}$ satisfy in addition 
\begin{eqnarray}
\sum_{j=1}^{4}\alpha_{j}I_{q_{j}}^{1}\!\!\! & = & \!\!\!0\,,\label{sumaaa}\\
\sum_{j=1}^{4}\alpha_{j}I_{q_{j}}^{2}\!\!\! & = & \!\!\!0\,,\label{rest}
\end{eqnarray}
where 
\begin{equation}
I_{q_{j}}^{l}=\fint_{a_{l}}^{b_{l}}dx\,\mathrm{e}^{-is\omega(x)}\frac{1}{x-q_{j}}\,,\quad l=1,2\,,\quad j=1,2,3,4\,,\label{inti}
\end{equation}
and 
\begin{equation}
\omega(x):=\log\left(-\frac{(x-a_{1})(x-a_{2})}{(x-b_{1})(x-b_{2})}\right)\,.
\end{equation}
Only two of the equations \eqref{sumacero}, \eqref{sumaaa}, and
\eqref{rest} are independent. This follows from the fact that 
\begin{equation}
\oint dz\,\mathrm{e}^{-is\omega(z)}\left(\frac{1}{z-q_{j}}-\frac{1}{z-q_{k}}\right)=I_{q_{j}}^{1}-I_{q_{k}}^{1}+I_{q_{j}}^{2}-I_{q_{k}}^{2}=0\,.\label{fact}
\end{equation}
This complex integral around the two cuts is zero because it is equal
to the integral at infinity, which vanishes because the integrand
falls fast enough. Therefore, we can conclude that the dimension of
the space of solutions for fixed $s$ is $2$. The same argument of
the previous section shows that these solutions give the eigenvectors
of $V^{\dagger}$ once evaluated on $A$. Conversely, any eigenvector
of $V^{\dagger}$, with at most simple poles at the end of the intervals
and satisfying \eqref{cope}, is of this form.

Now a simple solution is 
\begin{equation}
\tilde{v}_{1}(z)\propto\frac{\mathrm{d}}{\mathrm{d}z}\mathrm{e}^{-is\tilde{\omega}\left(z\right)}=-i\,s\,\mathrm{e}^{-is\tilde{\omega}\left(z\right)}\left(\frac{1}{z-a_{1}}-\frac{1}{z-b_{1}}+\frac{1}{z-a_{2}}-\frac{1}{z-b_{2}}\right)\,.
\end{equation}
In fact, this satisfies \eqref{sumacero} and \eqref{sumaaa} because
it is proportional to a derivative of the phase $\mathrm{e}^{-is\tilde{\omega}(z)}$
and hence, the integral on any of the intervals vanish with the regularization
we are using. That is, integrating this function along the intervals
we have further relations for the integrals $I_{q}^{l}$ 
\begin{eqnarray}
I_{a_{1}}^{1}-I_{b_{1}}^{1}+I_{a_{2}}^{1}-I_{b_{2}}^{1}\!\!\! & = & \!\!\!0\,,\label{dely}\\
I_{a_{1}}^{2}-I_{b_{1}}^{2}+I_{a_{2}}^{2}-I_{b_{2}}^{2}\!\!\! & = & \!\!\!0\,.\label{delyout}
\end{eqnarray}
The eigenvector $v_{1}(x)$ follows from taking the boundary limit
of $\tilde{v}_{1}(z)$ on $A$ from above. The corresponding $u_{1}\left(x\right)$
eigenfunction is an integral of this function, 
\begin{equation}
u_{1}\left(x\right)=-i\,\textrm{sign}(s)\,\left(C^{-1}\,v_{1}\right)\left(x\right)\propto\,\mathrm{e}^{-is\omega\left(x\right)}\,,\label{reluv}
\end{equation}
where in applying \eqref{dosdos} to $v_{1}\left(x\right)$, boundary
terms that are oscillatory phases are discarded, in accordance with
the regularization \eqref{fint_reg}. We can check directly that \eqref{reluv}
is an eigenfunction of \eqref{u1w} by noting that 
\begin{equation}
\fint_{A}dy\,\left(\frac{1}{y-a_{1}}+\frac{1}{y-b_{1}}\right)\,u_{1}(y)=\fint_{A}dy\,\left(\frac{1}{y-a_{2}}+\frac{1}{y-b_{2}}\right)\,u_{1}(y)\,.
\end{equation}
This follows from \eqref{fact}, \eqref{dely}, and \eqref{delyout}.
Hence, the two terms involving the characteristic functions in \eqref{u1w}
are equal, and we can eliminate these functions altogether by replacing
$\Theta_{A_{1}}(x),\Theta_{A_{2}}(x)\rightarrow\frac{1}{2}$. After
these replacements, the proof that $u_{1}\left(x\right)$ is an eigenvector
of the kernel \eqref{u1w} follows from the same steps as the one
for a single interval in the previous section, by promoting $u_{1}\left(x\right)$
to a complex function $\tilde{u}_{1}(z)\propto\mathrm{e}^{-is\tilde{\omega}(z)}$.

We choose the second solution $\tilde{v}_{2}(z)$ of the form \eqref{form},
such that its boundary value on $A$ gives a $v_{2}\left(x\right)$
eigenfunction orthogonal to $u_{1}\left(x\right)$. Collecting the
coefficients of the would-be delta functions in the scalar product
between $u_{1}\left(x\right)$ and $v_{2}\left(x\right)$, which are
generated by the integral near the intervals endpoints, we have for
$\tilde{v}_{2}\left(z\right)$ 
\begin{equation}
\alpha_{1}-\alpha_{2}+\alpha_{3}-\alpha_{4}=0\,.
\end{equation}
From this and \eqref{sumacero}, we get 
\begin{equation}
\tilde{v}_{2}\left(z\right)\propto\mathrm{e}^{-isw(z)}\left(\frac{1}{z-a_{1}}+\frac{\alpha}{z-b_{1}}-\frac{1}{z-a_{2}}-\frac{\alpha}{z-b_{2}}\right)\,.
\end{equation}
And from \eqref{sumaaa}, it follows that
\begin{equation}
\alpha(s;a_{1},b_{1},a_{2},b_{2})=-\frac{I_{a_{1}}^{1}-I_{a_{2}}^{1}}{I_{b_{1}}^{1}-I_{b_{2}}^{1}}\,.\label{espre}
\end{equation}
In order to compute $u_{2}\left(x\right)$ we use \eqref{cope}. Therefore,
using \eqref{cuen} and \eqref{v1w}, we have that
\begin{equation}
u_{2}(x)=-i\,\textrm{sign}(s)\,\left\{ \begin{array}{c}
\fint_{a_{1}}^{x}dy\,v_{2}(y)=\fint_{b_{1}}^{x}dy\,v_{2}(y)\,,\quad x\in A_{1}\\
\fint_{a_{2}}^{x}dy\,v_{2}(y)=\fint_{b_{2}}^{x}dy\,v_{2}(y)\,,\quad x\in A_{2}
\end{array}\right.\,.
\end{equation}
In order to normalize the solutions, we compute the coefficient of
the delta function in the scalar product, which can only come from
the singular part of the integrals near the endpoints of the intervals.
We have that the normalized solutions satisfying \eqref{cuen} are
\begin{eqnarray}
v_{1}(x)\!\!\! & = & \!\!\!\sqrt{\frac{|s|}{4\pi}}\,\mathrm{e}^{-is\omega(x)}\left(\frac{1}{x-a_{1}}-\frac{1}{x-b_{1}}+\frac{1}{x-a_{2}}-\frac{1}{x-b_{2}}\right)\,,\label{v1}\\
u_{1}(x)\!\!\! & = & \!\!\!\frac{1}{\sqrt{4\pi|s|}}\mathrm{e}^{-is\omega(x)}\,,\label{u1}\\
v_{2}(x)\!\!\! & = & \!\!\!\sqrt{\frac{|s|}{4\pi}}\,\mathrm{e}^{-is\omega(x)}\left(\frac{1}{x-a_{1}}+\frac{\alpha}{x-b_{1}}-\frac{1}{x-a_{2}}-\frac{\alpha}{x-b_{2}}\right)\,,\label{v2}\\
u_{2}(x)\!\!\! & = & \!\!\!\frac{-is}{\sqrt{4\pi|s|}}\,\left\{ \begin{array}{c}
\fint_{a_{1}}^{x}dy\,\mathrm{e}^{-is\omega(y)}\left(\frac{1}{y-a_{1}}+\frac{\alpha}{y-b_{1}}-\frac{1}{y-a_{2}}-\frac{\alpha}{y-b_{2}}\right)\,,\hspace{0.7cm}x\in A_{1}\\
\fint_{a_{2}}^{x}dy\,\mathrm{e}^{-is\omega(y)}\left(\frac{1}{y-a_{1}}+\frac{\alpha}{y-b_{1}}-\frac{1}{y-a_{2}}-\frac{\alpha}{y-b_{2}}\right)\,,\hspace{0.7cm}x\in A_{2}
\end{array}\right. . \hspace{.8cm} \label{u2}
\end{eqnarray}

\paragraph{Completeness of the eigenvector system\protect \\
}

Now we show the eigenvector basis (\ref{v1}-\ref{u2}) is complete.
When we explicitly constructed the eigenvectors, we only consider
solutions satisfying the equation \eqref{cope} in order to simplify
the calculation, but there was no further reason to assume that. Now,
we are able to show that any other possible eigenvector must satisfy
\eqref{cope}, and hence, there are no other eigenvectors than the
ones already obtained. This fact follows by considering the $s=0$
solutions for $u_{1}\left(x\right)$ and $u_{2}\left(x\right)$. Taking
out an irrelevant factor of $|s|^{-1/2}$, which is compensated by
the inverse factor in the eigenfunctions $v\left(x\right)$, we have
that
\begin{eqnarray}
\lim_{s\rightarrow0}\sqrt{4\pi|s|}\,u_{1,s}(x)\!\!\! & = & \!\!\!1\,,\\
\lim_{s\rightarrow0}\sqrt{4\pi|s|}\,u_{2,s}(x)\!\!\! & = & \!\!\!\Theta_{A_{1}}(x)-\Theta_{A_{2}}(x)\,.
\end{eqnarray}
The first one is proportional to a constant, which is the same along
the two intervals. The second one is proportional to two opposite
constants in the two different intervals. Hence, any third solution
$v_{3,s}\left(x\right)$ would be orthogonal to $u_{1,s=0}\left(x\right)$
and $u_{2,s=0}(x)$, and therefore must satisfy \eqref{cope}. Therefore,
there cannot be any other eigenvectors for two intervals.

\subsubsection{Dependence of the eigenvectors through the cross ratio}

The aim of this subsection is to obtain simplified expressions for
the function \eqref{espre} and the eigenfunctions (\ref{v1}–\ref{u2}),
which will be useful for the final computation of the modular Hamiltonian
and the MI. For such a purpose, we study the dependence of these expressions
with the cross-ratio 
\begin{equation}
\eta:=\frac{(b_{1}-a_{1})(b_{2}-a_{2})}{(a_{2}-a_{1})(b_{2}-b_{1})}\in(0,1)\,,\label{crossratio}
\end{equation}
which is the natural geometric parameter of the problem because of
the conformal invariance of the model. The outcome of this subsection
is that the function \eqref{espre} can be expressed as in \eqref{alphah},
in term of Hypergeometric functions, and the eigenvectors (\ref{v1}–\ref{u2})
can be expressed as in (\ref{u1cr}-\ref{v2cr}) and \eqref{uiui},
which involve Appell functions.

Let us consider a general Möbius transformation $x\mapsto x':=f(x)$
given by
\begin{equation}
x'=f(x):=\frac{ax+b}{cx+d}\,,\label{mobius}
\end{equation}
where $a,b,c,d\in\mathbb{R}$ and $ad-cb>0$. Such a transformation
leaves the cross-ratio \eqref{crossratio} invariant.

Let us first understand the dependence of the function $\alpha(s;a_{1},b_{1},a_{2},b_{2})$
with the interval endpoints. We can use \eqref{mobius} to make a
change in the integration variables of the integrals \eqref{inti},
which are involved in the definition of the function $\alpha$. After
that, a straightforward computation shows that 
\begin{equation}
\alpha(s;a_{1},b_{1},a_{2},b_{2})=\alpha(s;a'_{1},b'_{1},a'_{2},b'_{2})\,,\label{c6-alpha-inv}
\end{equation}
where the two sets of intervals endpoints are related by 
\begin{equation}
a'_{j}:=f(a_{j})\,,\,\,\,\,b'_{j}:=f(b_{j})\,,\qquad j=1,2\,.\label{epr}
\end{equation}
Since \eqref{c6-alpha-inv} holds for any general Mobiüs transformation,
we have that $\alpha$ depends on the intervals endpoints only through
the cross-ratio \eqref{crossratio}, i.e. $\alpha=\alpha(s,\eta)$.

Similarly, a direct computation for the eigenfunctions shows the following
covariance properties under the change of variable \eqref{mobius} 
\begin{eqnarray}
u_{j}(x';\bar{q}')\!\!\! & = & \!\!\!\mathrm{e}^{is\Omega(\bar{q})}\,u_{j}(x;\bar{q})\,,\label{uim}\\
v_{j}(x';\bar{q}')\!\!\! & = & \!\!\!\mathrm{e}^{is\Omega(\bar{q})}\frac{1}{f'(x)}u_{j}(x;\bar{q})\,,\label{vim}
\end{eqnarray}
where $u_{j}(x;\bar{q})$ and $v_{i}(x;\bar{q})$ are the eigenfunctions
corresponding to the problem with endpoints $\bar{q}:=(a_{1},b_{1},a_{2},b_{2})$
(idem for $\bar{q}'$), and the two sets of endpoints are related
by \eqref{epr}. The real function $\Omega(\bar{q})$ is given by
\begin{equation}
\Omega(\bar{q}):=\frac{1}{2}\log\left(\frac{f'(a_{1})f'(a_{2})}{f'(b_{1})f'(b_{2})}\right)\,.
\end{equation}
Not surprisingly, the eigenfunctions $u_{j}\left(x\right)$ transforms
as a scalar wave, whereas the eigenfunction $v_{j}\left(x\right)$
as their derivatives.

Simpler expressions are obtained when we specially take the Möbius
transformation $x\mapsto x':=f_{1}(x)$ which sends the points $(a_{1},b_{1},a_{2},b_{2})\mapsto(0,\eta,1,\infty)$,
i.e.\footnote{More carefully, we shall take the Mobiüs transformation which transforms
$(a_{1},b_{1},a_{2},b_{2})\rightarrow(0,\eta,1,\Lambda)$ with $\Lambda>1$,
and in the end, we take $\Lambda\rightarrow\infty$.} 
\begin{eqnarray}
f_{1}(x)\!\!\! & := & \!\!\!\frac{\left(b_{2}-a_{2}\right)\left(x-a_{1}\right)}{\left(a_{2}-a_{1}\right)\left(b_{2}-x\right)}\,,\label{f1}\\
f'_{1}(x)\!\!\! & = & \!\!\!\frac{\left(b_{2}-a_{1}\right)\left(b_{2}-a_{2}\right)}{\left(a_{2}-a_{1}\right)\left(b_{2}-x\right)^{2}}=\frac{1}{b_{2}-a_{1}}\left(\frac{b_{2}-a_{2}}{a_{2}-a_{1}}+2x'+\frac{a_{2}-a_{1}}{b_{2}-a_{2}}x'^{2}\right)\,.
\end{eqnarray}
With this transformation, we get from \eqref{espre} the compact formula
\begin{equation}
\alpha(s,\eta)=-\frac{_{2}F_{1}\left(1+is,-is;1;\eta\right)}{_{2}F_{1}\left(1-is,is;1;\eta\right)}\,,\label{alphah}
\end{equation}
where $_{2}F_{1}(a,b;c;x)$ is the Gaussian or ordinary hypergeometric
function. To derive this result, we used the integral representation
of such a function 
\begin{equation}
_{2}F_{1}(a,b;c;x)=\frac{\Gamma(c)}{\Gamma(b)\Gamma(c-b)}\int_{0}^{1}dt\,t^{b-1}(1-t)^{c-b-1}(1-tx)^{-a}\,,\label{hyper_def}
\end{equation}
which holds for $x<1$ and Re$(c)>$ Re$(b)>0$.\footnote{When Re$(b)=0$, which occurs in \eqref{alphah}, equation \eqref{hyper_def}
has to be understood as the $\fint$-regularization explained on \eqref{fint_reg}.
\label{foot-hreg}} Expression \eqref{alphah} shows explicitily the dependence of $\alpha$
through the cross ratio, and the fact that it is a phase factor. For
$s=0$ we have $\alpha(0,\eta)=-1$, and it reaches a value dependent
on $\eta$ for $s\rightarrow\pm\infty$ (see below). We also have
$\alpha(-s,\eta)=\alpha(s,\eta)^{*}$. For a fixed $s\neq0$, we have
that $\lim_{\eta\rightarrow0}\alpha(s,\eta)=-1$ and $\lim_{\eta\rightarrow1}\alpha(s,\eta)=1$.

Applying the transformation \eqref{f1} to the eigenvectors (\ref{v1}–\ref{u2})
and using the expressions (\ref{uim}–\ref{vim}), we arrive to\footnote{Here, we discard a global constant phase factor, which is the same
for all the eigenvectors, and therefore, it does not modify the orthonormalization
condition and the condition \eqref{cuen}.} 
\begin{eqnarray}
u_{1}(x)\!\!\! & = & \!\!\!\frac{1}{\sqrt{4\pi|s|}}\mathrm{e}^{-is\log\left(\frac{x'(1-x')}{\eta-x'}\right)}\,,\label{u1cr}\\
v_{1}(x)\!\!\! & = & \!\!\!\sqrt{\frac{|s|}{4\pi}}\,f'_{1}(x)\,\mathrm{e}^{-is\log\left(\frac{x'(1-x')}{\eta-x'}\right)}\left(\frac{1}{x'}+\frac{1}{x'-1}-\frac{1}{x'-\eta}\right)\,,\label{v1cr}\\
u_{2}(x)\!\!\! & = & \!\!\!i\frac{s}{\sqrt{4\pi|s|}}\,\left(\frac{x'}{\eta}\right)^{-is}\Bigg[\frac{1}{is}F_{1}\left(-is;is,-is;1-is;x',\frac{x'}{\eta}\right)\nonumber \\
 &  & \!\!\!+\frac{\alpha(s,\eta)}{1-is}\,\frac{x'}{\eta}\,F_{1}\left(1-is;is,1-is;2-is;x',\frac{x'}{\eta}\right) \nonumber \\
 &  & \!\!\!-\frac{x'}{1-is}F_{1}\left(1-is;1+is,-is;2-is;x',\frac{x'}{\eta}\right)\Bigg]\,,\hspace{1cm}x\in A_{1}\,,\label{ua1i}\\
v_{2}(x)\!\!\! & = & \!\!\!\sqrt{\frac{\left|s\right|}{4\pi}}\,f'_{1}(x)\,\mathrm{e}^{-is\log\left(\frac{x'(1-x')}{\eta-x'}\right)}\left(\frac{1}{x'}+\frac{\alpha(s,\eta)}{x'-\eta}-\frac{1}{x'-1}\right)\,,\label{v2cr}
\end{eqnarray}
where $F_{1}(a;\beta_{1},\beta_{1};c;z_{1},z_{2})$ in \eqref{ua1i}
is the Appell Hypergeometric function of two variables. Such a function
has the following integral representation 
\begin{equation}
F_{1}(a;\beta_{1},\beta_{1};c;z_{1},z_{2})=\frac{\Gamma(c)}{\Gamma(a)\Gamma(c-a)}\int_{0}^{1}\,t^{a-1}(1-t)^{c-a-1}(1-tz_{1})^{-\beta_{1}}(1-tz_{2})^{-\beta_{2}}\,,\label{c6-appell}
\end{equation}
for $x,y<1$ and Re$(a)>0$ and Re$(c-a)>0$.\footnote{For Re$(a)=0$, the integral \eqref{c6-appell} has to be understood
in the same way we have explained in footnote \ref{foot-hreg}.} We remark that formula \eqref{ua1i} only holds for $x\in A_{1}$.
In such a case, $x'\in(0,\eta)$, and therefore, the arguments of
the Appell's functions belong to their domains of analyticity.

To get an expression for $u_{2}\left(x\right)$ valid for $x\in A_{2}$
in terms of Appell functions, we must consider a different Mobiüs
transformation $x\mapsto\tilde{x}:=f_{2}(x)$, which sends $(a_{1},b_{1},a_{2},b_{2})\mapsto(1,\infty,0,\eta)$,
\begin{eqnarray}
f_{2}(x)\!\!\! & := & \!\!\!\frac{(b_{1}-a_{1})(x-a_{2})}{(a_{2}-a_{1})(x-b_{1})}\,,\label{f2}\\
f'_{2}(x)\!\!\! & = & \!\!\!\frac{(b_{1}-a_{1})(a_{2}-b_{1})}{(a_{2}-a_{1})(x-b_{1})^{2}}=\frac{1}{a_{2}-b_{1}}\left(\frac{b_{1}-a_{1}}{a_{2}-a_{1}}+2\tilde{x}+\frac{a_{2}-a_{1}}{b_{1}-a_{1}}\tilde{x}^{2}\right)\,.
\end{eqnarray}
Then, we obtain for $x\in A_{2}$
\begin{eqnarray}
u_{2}(x)\!\!\! & = & \!\!\!\frac{-is}{\sqrt{4\pi|s|}}\left(\frac{\tilde{x}}{\eta}\right)^{-is}\Bigg[\frac{1}{is}F_{1}\left(-is;is,-is;1-is;\tilde{x},\frac{\tilde{x}}{\eta}\right)\nonumber \\
 &  & \!\!\!+\frac{\alpha(s,\eta)}{1-is}\,\frac{\tilde{x}}{\eta}\,F_{1}\left(1-is;is,1-is;2-is;\tilde{x},\frac{\tilde{x}}{\eta}\right)\nonumber \\
 &  & \!\!\!-\frac{\tilde{x}}{1-is}F_{1}\left(1-is;1+is,-is;2-is;\tilde{x},\frac{\tilde{x}}{\eta}\right)\Bigg]\,,\label{uiui}
\end{eqnarray}
which is the same expression valid for $u_{2}\left(x\right)$ in the
first interval (up to a minus global sign), but evaluated in $\tilde{x}$
instead of $x'$. This expression indicates that for any point $x_{1}\in A_{1}$
exists a point $x_{2}\in A_{2}$ such that $u_{2}(x_{1})=-u_{2}(x_{2})$,
and viceversa. In the next subsection, we show that all the eigenvectors
are classified according to such ``parity symmetry''.

\subsubsection{Parity symmetry of the eigenvectors\label{psym} \label{parity}}

The aim of this subsection is to study the behavior of the eigenfunctions
under two different conformal transformations, The first one is the
Möbius transformation that interchange the two intervals (eq. \eqref{moebius-i}),
and the second one it the Möbius transformation that reflects each
interval into itself. The outcome of this subsections is that the
eigenfunctions satisfy the symmetry properties (\ref{u1ps}-\ref{v2ps})
and (\ref{uuu1}-\ref{vvv2}). These expressions will be used in subsection
\ref{sec_mh_2} to express the final result of the modular Hamiltonian
in a more amenable way.

Let us first in introduce the Möbius transformation $x\mapsto\bar{x}:=p(x)$
that interchange the two intervals $(a_{1},b_{1},a_{2},b_{2})\overset{p}{\mapsto}(a_{2},b_{2},a_{1},b_{1})$,
namely 
\begin{eqnarray}
\bar{x}=p(x)\!\!\! & := & \!\!\!\frac{a_{1}a_{2}(x-b_{1}-b_{2})-b_{1}b_{2}(x-a_{1}-a_{2})}{x(a_{1}+a_{2}-b_{1}-b_{2})+(b_{1}b_{2}-a_{1}a_{2})}\,,\label{moebius-i}\\
p'(x)\!\!\! & = & \!\!\!\frac{(b_{1}-a_{1})(b_{2}-a_{1})(a_{2}-b_{1})(a_{2}-b_{2})}{[x(a_{1}+a_{2}-b_{1}-b_{2})+(b_{1}b_{2}-a_{1}a_{2})]^{2}}>0\,,
\end{eqnarray}
where \eqref{moebius-i} is the same as \eqref{conj-points}, which
indicates that $\bar{x}$ is the conjugate point of the point $x$,
i.e. $\omega(\bar{x})=\omega(x)$. Specializing this transformation
on the relations (\ref{uim}-\ref{vim}), we get 
\begin{eqnarray}
u_{i}(\bar{x};a_{2},b_{2},a_{1},b_{1})\!\!\! & = & \!\!\!u_{i}(x;a_{1},b_{1},a_{2},b_{2})\,,\label{uip}\\
v_{i}(\bar{x};a_{2},b_{2},a_{1},b_{1})\!\!\! & = & \!\!\!\frac{1}{p'(x)}u_{i}(x;a_{1},b_{1},a_{2},b_{2})\,,\label{vip}
\end{eqnarray}
where in this case we have that $\Omega(a_{1},b_{1},a_{2},b_{2})=0$.
On the other hand, from (\ref{v1}-\ref{v2}), we easily see that
\begin{eqnarray}
u_{1}(x;a_{2},b_{2},a_{1},b_{1})\!\!\! & = & \!\!\!u_{1}(x;a_{1},b_{1},a_{2},b_{2})\,,\\
v_{1}(x;a_{2},b_{2},a_{1},b_{1})\!\!\! & = & \!\!\!v_{1}(x;a_{1},b_{1},a_{2},b_{2})\,,\\
v_{2}(x;a_{2},b_{2},a_{1},b_{1})\!\!\! & = & \!\!\!-v_{2}(x;a_{1},b_{1},a_{2},b_{2})\,.
\end{eqnarray}
Therefore, using additionally \eqref{ua1i} and \eqref{uiui} for
$u_{2}\left(x\right)$, we conclude we have the following parity symmetries
\begin{eqnarray}
u_{1}(\bar{x})\!\!\! & = & \!\!\!u_{1}(x)\,,\label{u1ps}\\
v_{1}(\bar{x})\!\!\! & = & \!\!\!\frac{1}{p'(x)}v_{1}(x)\,,\label{v1ps}\\
u_{2}(\bar{x})\!\!\! & = & \!\!\!-u_{2}(x)\,,\label{u2ps}\\
v_{2}(\bar{x})\!\!\! & = & \!\!\!-\frac{1}{p'(x)}v_{2}(x)\,.\label{v2ps}
\end{eqnarray}
That means that the first set of eigenfunctions is even and the second
one is odd under taking the conjugate point $\bar{x}$. We remark
that in these expressions, the same eigenfunctions for the same endpoints
$(a_{1},b_{1},a_{2},b_{2})$ appear at both sides of the expressions.

Now, we consider the Möbius transformation $x\mapsto\hat{x}:=q\left(x\right)$
that reflects each interval into itself, i.e. $\left(a_{1},b_{1},a_{2},b_{2}\right)\overset{q}{\mapsto}\left(b_{1},a_{1},b_{2},a_{2}\right)$.
It is given by
\begin{equation}
\hat{x}=q\left(x\right):=\frac{x(a_{2}b_{2}-a_{1}b_{1})-a_{1}a_{2}(b_{2}-b_{1})-b_{1}b_{2}(a_{2}-a_{1})}{x(b_{2}+a_{2}-b_{1}-a_{1})+a_{1}b_{1}-a_{2}b_{2}}\,.\label{tranfi}
\end{equation}
Under this transformation, the eigenvectors transform as
\begin{eqnarray}
u_{1,-s}\left(\hat{x}\right)\!\!\! & = & \!\!\!\mathrm{e}^{is\Omega\left(\bar{q}\right)}u_{1,s}\left(x\right)\,,\label{uuu1}\\
u_{2,-s}\left(\hat{x}\right)\!\!\! & = & \!\!\!\mathrm{e}^{is\Omega\left(\bar{q}\right)}\left(-1\right)\alpha\left(-s,\eta\right)u_{2,s}\left(x\right)\,,\label{uuu2}\\
v_{1,-s}\left(\hat{x}\right)\!\!\! & = & \!\!\!\mathrm{e}^{is\Omega\left(\bar{q}\right)}\frac{\left(-1\right)}{q'\left(x\right)}v_{1,s}\left(x\right)\,,\\
v_{2,-s}\left(\hat{x}\right)\!\!\! & = & \!\!\!\mathrm{e}^{is\Omega\left(\bar{q}\right)}\frac{\alpha\left(-s,\eta\right)}{q'\left(x\right)}v_{2,s}\left(x\right)\,,\label{vvv2}
\end{eqnarray}
where now we explicitly write the dependence of the eigenfunctions
with the parameter $s$, because the above expressions relate a eigenfunction
of eigenvalue $s$ with the one of eigenvalue $-s$. In this case,
we have a no null phase factor $\Omega\left(\bar{q}\right):=2\log\left(\frac{b_{2}-b_{1}}{a_{2}-a_{1}}\right)$.

\subsubsection{Mutual information\label{c5-subsec:Mutual-information}}

\begin{figure}[h]
\centering
\includegraphics[width=13cm]{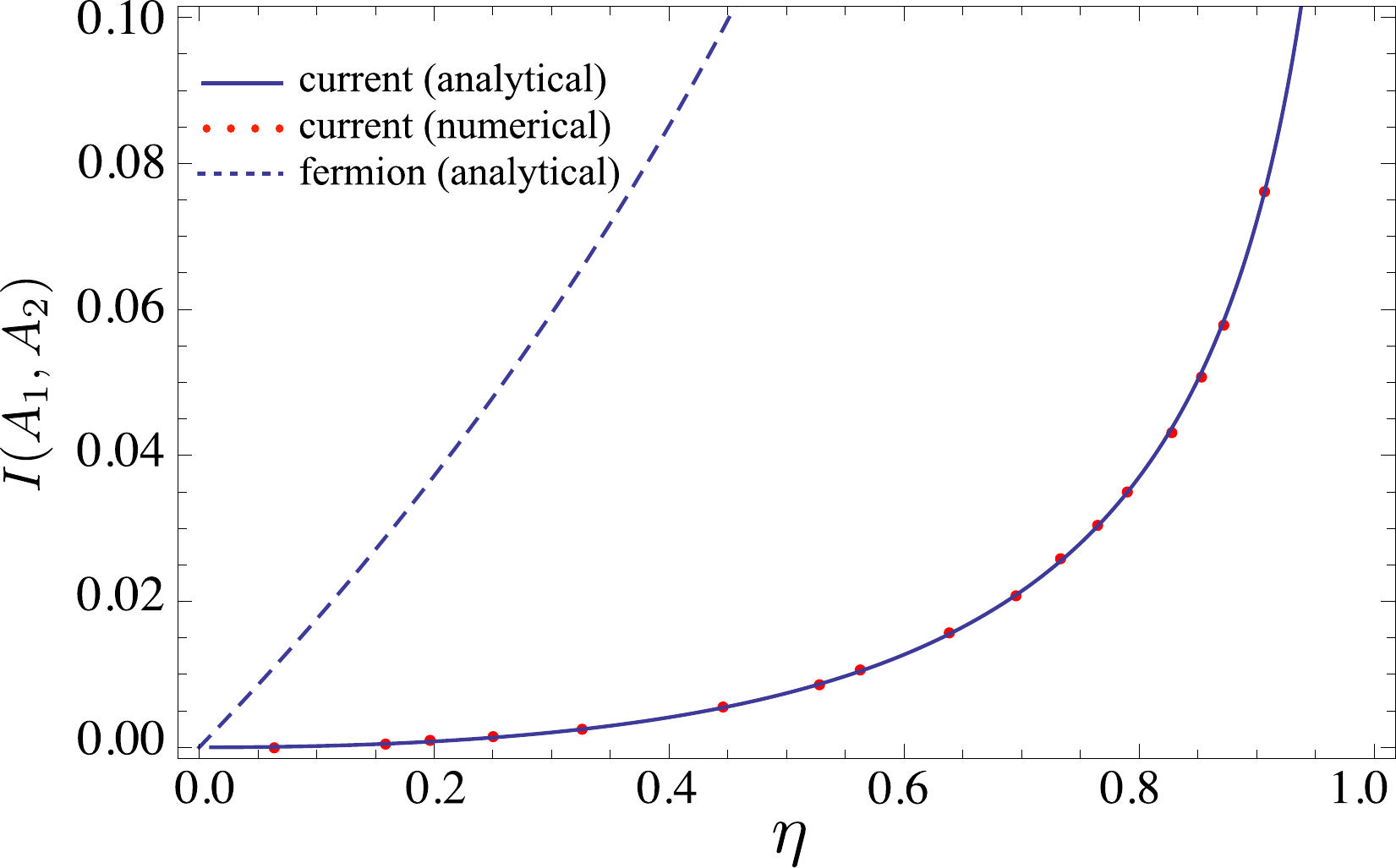}\caption{\foreignlanguage{english}{\label{figufi} The mutual information $I(A_{1},A_{2})$ as function
of the cross ratio $\eta$. The continuous solid line corresponds
to the mutual information obtained by numerical integration of the
analytic expression \eqref{mic} and \eqref{Ueta}. The red points
correspond to the simulation on the lattice, which is explained in
section \ref{numerico}. The dashed line is the free chiral fermion
mutual information $-\frac{1}{6}\log(1-\eta)$.}}
\end{figure}

\begin{figure}[h]
\centering
\includegraphics[width=13cm]{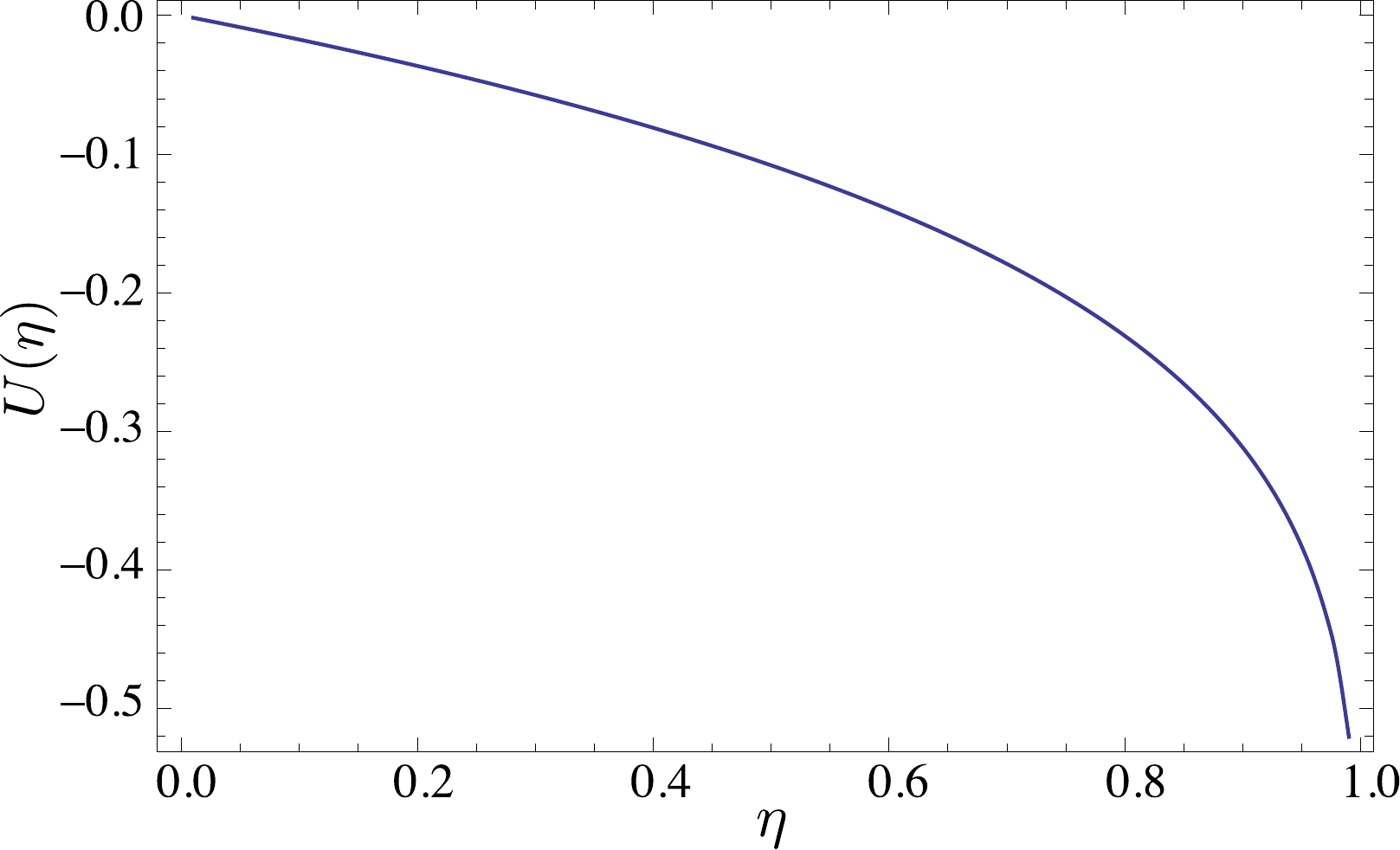}\caption{\foreignlanguage{english}{\label{figuu} The function $U(\eta)$ as a function of the cross-ratio
$\eta$. This function is negative because the free chiral current
is a subnet of the free chiral fermion, which it has $U(\eta)=0$.
It does not have the $\eta\leftrightarrow(1-\eta)$ symmetry expected
for the case where the entropy of a two-interval region is equal to
the entropy of its complement. $U(\eta)\sim-\frac{1}{2}\log(-\log(1-\eta))$
for $\eta\rightarrow1$.}}
\end{figure}

\begin{figure}[h]
\centering
\includegraphics[width=13cm]{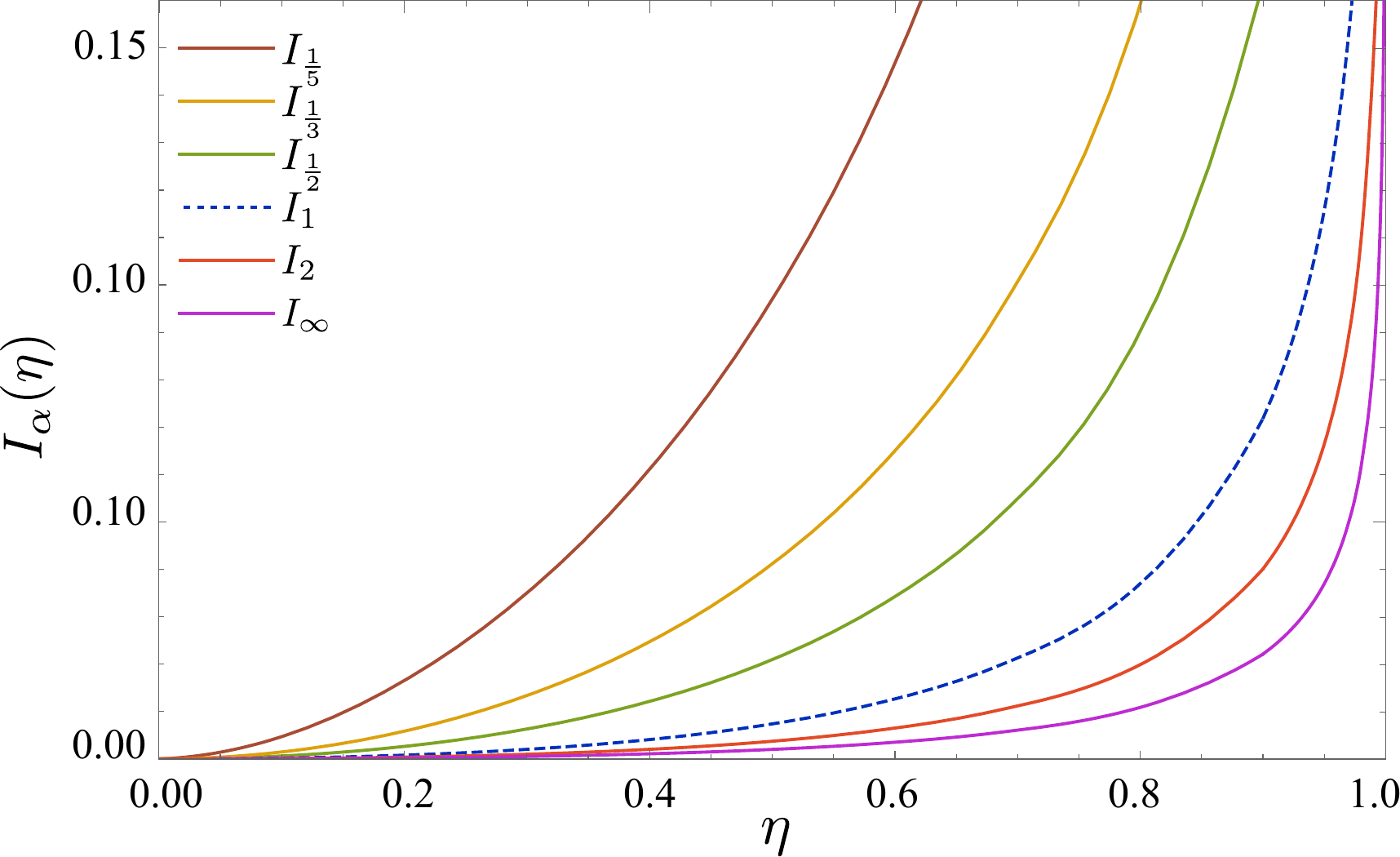}\caption{\foreignlanguage{english}{\label{rer} The Rényi mutual information $I_{\alpha}(\eta)$ as a
function of the cross-ratio $\eta$ for different values of $\alpha$. }}
\end{figure}

In this subsection, we compute the mutual information and the Rényi
mutual informations, using the expressions \eqref{c5-curr-ee} and
\eqref{c5-curr-eeg} and the eigenvectors obtained in the previous
sections. The results are given by formulas (\ref{mic}-\ref{Ueta})
and (\ref{renyi_res}-\ref{Uneta}), and showed in figures \ref{figufi},
\ref{figuu}, and \ref{rer}.

According to \eqref{c2-mi_from_vn}, the MI can be expressed as
\begin{equation}
I(A_{1},A_{2})=S(A_{1})+S(A_{2})-S(A_{1}\cup A_{2})\,,\label{mi-def}
\end{equation}
where the one interval entropies $S(A_{j})$ are obtained from \eqref{s1}
and the two-interval EE follows from \eqref{c5-curr-ee} and \eqref{c5-curr-eeg}.
This last is given by
\begin{equation}
S(A)=\sum_{k=1}^{2}\int_{0}^{+\infty}ds\int_{A^{(\epsilon)}}dx\,g(s)\,u_{k,s}(x)\,v_{k,s}(x)^{*}\,,\label{s2int}
\end{equation}
where $A^{(\epsilon)}:=A_{1}^{(\epsilon)}\cup A_{2}^{(\epsilon)}$
with $A_{j}^{(\epsilon)}:=(a_{j}+\epsilon,b_{j}-\epsilon)$ is the
regularized region. It is important to emphasize that the MI is finite
and independent of the regularization. However, each term on the r.h.s.
of equation \eqref{mi-def} is UV-divergent, and their corresponding
regularizations cannot be chosen independently. They correspond to
evaluating the integrals \eqref{s1} and \eqref{s2int} along the
regularized regions, as we have already explain, with the same cutoff
parameter $\epsilon>0$ for all the terms. In the end, we take the
limit $\epsilon\rightarrow0^{+}$ and we get the finite desired result
for the MI. 

It is convenient to express the two-interval entropy as $S(A):=S_{1}(A)+S_{2}(A)$,
where $S_{k}\left(A\right)$ denotes the term in \eqref{s2int} involving
the functions $u_{k,s}\left(x\right)$ and $v_{k,s}\left(x\right)$
($k=1,2$). Using the formulas (\ref{v1}-\ref{u1}) for the eigenvectors,
we easily obtain
\begin{eqnarray}
S_{1}(A)\!\!\! & = & \!\!\!\int_{0}^{+\infty}ds\int_{A^{(\epsilon)}}dx\,g(s)\,u_{1,s}(x)\,v_{1,s}(x)^{*}=\frac{1}{24}\int_{A^{(\epsilon)}}dx\,\omega'(x)\nonumber \\
 & = & \!\!\!\frac{1}{12}\log\left(\frac{\left(a_{2}-b_{1}\right)\left(b_{2}-a_{1}\right)}{\left(b_{2}-b_{1}\right)\left(a_{2}-a_{1}\right)}\right)+\frac{1}{12}\log\left(\frac{b_{1}-a_{1}}{\epsilon}\right)+\frac{1}{12}\log\left(\frac{b_{2}-a_{2}}{\epsilon}\right)\nonumber \hspace{.8cm} \\
 & = & \!\!\!\frac{1}{12}\log(1-\eta)+\frac{1}{2}S(A_{1})+\frac{1}{2}S(A_{2})\,.
\end{eqnarray}
Direct treatment of the integral for $S_{2}(A)$ is more complicated
due to the presence of the hypergeometric and Appell functions in
$u_{2}\left(x\right)$. We find convenient to use the following trick.
Since the integral in \eqref{s2int} is regularized, keeping $\epsilon$
fixed, we can replace 
\begin{align}
 & \int_{A^{(\epsilon)}}dx\,u_{2,s}(x)\,v_{2,s}(x)^{*}=\lim_{\delta s\rightarrow0}\int_{A^{(\epsilon)}}dx\,u_{2,s}(x)\,v_{2,s+\delta s}(x)^{*}\nonumber \\
 & =\lim_{\delta s\rightarrow0}\left[\int_{A} \! dx\,u_{2,s}(x)v_{2,s+\delta s}(x)^{*}- \! \sum_{j=1}^{2} \! \left(\int_{a_{j}}^{a_{j}+\epsilon}\!\!\!\!\!\!\! dx\,u_{2,s}(x)v_{2,s+\delta s}(x)^{*}+\!\!\int_{b_{j}-\epsilon}^{b_{j}}\!\!\!\!dx\,u_{2,s}(x)v_{2,s+\delta s}(x)^{*}\right)\right]  \nonumber \\
 & =-\lim_{\delta s\rightarrow0}\sum_{j=1}^{2}\left(\int_{a_{j}}^{a_{j}+\epsilon}\!\!dx\,u_{2,s}(x)\,v_{2,s+\delta s}(x)^{*}+\int_{b_{j}-\epsilon}^{b_{j}}\!\!dx\,u_{2,s}(x)\,v_{2,s+\delta s}(x)^{*}\right)\,,\label{tri}
\end{align}
where in the last step we have used the fact that vectors $u_{s}(x)$
and $v_{s+\delta s}(x)$ are orthogonal for $\delta s\neq0$. The
advantage of doing that is that we do not need now the precise behavior
of the eigenfunctions along the intervals but only in a small region
near the endpoint of the intervals. Then, we can just take the leading
terms of $u_{s}(x)$ and $v_{s+\delta s}(x)$ since all other subleading
terms in $\epsilon$ will disappear in the limit $\epsilon\rightarrow0^{+}$.
From (\ref{v1}–\ref{u2}), these leading terms are 
\begin{eqnarray}
u_{2,s}\left(x\right)\!\!\! & \sim & \!\!\!\begin{cases}
\frac{(-1)^{j+1}}{\sqrt{4\pi|s|}}\,\mathrm{e}^{-is\omega\left(x\right)} & \qquad\;\textrm{for}\,x\rightarrow a_{j}\,,\\
\frac{(-1)^{j}}{\sqrt{4\pi|s|}}\,\alpha(s,\eta)\,\mathrm{e}^{-is\omega\left(x\right)} & \qquad\;\textrm{for}\,x\rightarrow b_{j}\,,
\end{cases}\\
v_{2,s}\left(x\right)\!\!\! & \sim & \!\!\!\begin{cases}
(-1)^{j+1}\sqrt{\frac{|s|}{4\pi}}\,\frac{\mathrm{e}^{-is\omega\left(x\right)}}{x-a_{j}} & \textrm{for}\,x\rightarrow a_{j}\,,\\
(-1)^{j+1}\alpha(s,\eta)\sqrt{\frac{|s|}{4\pi}}\,\frac{\mathrm{e}^{-is\omega\left(x\right)}}{x-b_{j}} & \textrm{for}\,x\rightarrow b_{j}\,,
\end{cases}
\end{eqnarray}
Replacing these expressions into \eqref{tri}, we get 
\begin{equation}
S_{2}(A)=\frac{1}{12}\log(1-\eta)+\frac{1}{2}S(A_{1})+\frac{1}{2}S(A_{2})-\int_{0}^{+\infty}ds\,\frac{g(s)}{2\pi}\,i\,\alpha(s,\eta)\partial_{s}\alpha^{*}(s,\eta)\,.\label{tio}
\end{equation}
Taking into account that $-i\,\alpha(s,\eta)\partial_{s}\alpha^{*}(s,\eta)=i\partial_{s}\log(\alpha(s,\eta))$
and integrating by parts the last term in \eqref{tio}, we finally
obtain

\begin{eqnarray}
I(A_{1},A_{2})\!\!\! & = & \!\!\!-\frac{1}{6}\log(1-\eta)+U(\eta)\,,\label{mic}\\
U(\eta)\!\!\! & = & \!\!\!-\frac{i\,\pi}{2}\int_{0}^{+\infty}\!ds\,\frac{s}{\sinh^{2}(\pi s)}\,\log\left(\frac{_{2}F_{1}\left(1+is,-is;1;\eta\right)}{_{2}F_{1}\left(1-is,is;1;\eta\right)}\right)\,.\label{Ueta}
\end{eqnarray}
The the first term in \eqref{mic} coincides with the mutual information
of the free chiral fermion field \cite{Casini:2005rm,reduced_density}.
We could not express the integral in \eqref{Ueta} in terms of standard
known functions, and hence, it has to be computed numerically. The
result for $U(\eta)$ is always negative, as it must be, considering
that the free chiral current is a subnet of the free chiral fermion
algebra, and hence, its MI has to be smaller. In figure \ref{figufi},
we show a plot of the MI, while the function $U(\eta)$ is shown in
figure \ref{figuu}.

The Rényi mutual information $I_{\alpha}(\eta)=S_{\alpha}(A_{1})+S_{\alpha}(A_{2})-S_{\alpha}(A_{1}\cup A_{2})$
can be computed in the same fashion, using \eqref{c5-curr-renyi}
and \eqref{c5-curr-renyig}. A straightforward computation gives\footnote{Do not confuse the $\alpha$-index of the Rényi MI with the function
$\alpha\left(s,\eta\right)$ defined above in equation \eqref{alphah}. } 
\begin{eqnarray}
I_{\alpha}(\eta)\!\!\! & = & \!\!\!-\frac{1+\alpha}{12\,\alpha}\,\log(1-\eta)+U_{\alpha}(\eta)\,,\label{renyi_res}\\
U_{\alpha}(\eta)\!\!\! & = & \!\!\!\frac{i\,\alpha}{2(\alpha-1)}\int_{0}^{+\infty}\negthickspace\negthickspace ds\left(\coth(\pi s\alpha)-\coth(\pi s)\right)\log\left(\frac{_{2}F_{1}\left(1+is,-is;1;\eta\right)}{_{2}F_{1}\left(1-is,is;1;\eta\right)}\right). \hspace{1cm} \label{Uneta}
\end{eqnarray}
In figure \ref{rer}, we display $I_{\alpha}(\eta)$ for some values
of $\alpha$.

\paragraph{Asymptotic behavior of the mutual information\protect \\
}

Now, we study the asymptotic behavior of the MI for large and short
distances between the intervals. The $\eta\rightarrow0^{+}$ limit
corresponds to the large distance limit between the intervals. Since
the integrand of \eqref{Ueta} is analytic at $\eta=0$, a simple
Taylor expansion reveals the following asymptotic behavior 
\begin{equation}
U(\eta)=-\frac{1}{6}\eta-\frac{1}{15}\eta^{2}-\frac{13}{315}\eta^{3}+\mathcal{O}(\eta^{4})\,,\qquad\eta\rightarrow0^{+}\,,
\end{equation}
This gives
\begin{equation}
I(\eta)\sim\frac{\eta^{2}}{60}+\frac{\eta^{3}}{70}+\mathcal{O}(\eta^{4})\,,\qquad\eta\rightarrow0^{+}\,.
\end{equation}
The first term coincides with the general result for the leading term
of the large distance expansion of the mutual information for a CFT
\cite{Calabrese:2009ez,Calabrese:2010he,Cardy.esferaslejanas,Agon:2015ftl}.
For two intervals, this is given by 
\begin{equation}
I(\eta)\sim\frac{\sqrt{\pi}\Gamma(2\Delta+1)}{4^{2\Delta+1}\Gamma(2\Delta+3/2)}\eta^{2\Delta}\,,\qquad\eta\rightarrow0^{+}\,,
\end{equation}
where $\Delta$ is the lowest dimensional operator of the theory.
In the present model, this is $J(x)=\partial\phi(x)$ and has $\Delta=1$.
The fermion has $\Delta=1/2$ and a different behavior $I(\eta)\sim\frac{1}{6}\eta$
at large distances, which is quite visible in figure \ref{figufi}.

The short distance limit $\eta\rightarrow1^{-}$ is more tricky, since
the integrand of \eqref{Ueta} converges to zero in a non-uniform
way in such a limit. The main contribution to $U(\eta)$ in this limit
comes from $s\sim0$. We have to expand the Hypergeometric functions
at the numerator and denominator inside the logarithm in \eqref{Ueta}
for $\eta\rightarrow1^{-}$ and $s\sim0$ to get 
\begin{equation}
U(\eta)\sim-\frac{i\pi}{2}\int_{0}^{+\infty}\!\!ds\,\frac{s}{\sinh^{2}(\pi s)}\,\log\left(\frac{i-s\,\log(1-\eta)}{i+s\,\log(1-\eta)}\right)\sim-\frac{1}{2}\log(-\log(1-\eta))\,.
\end{equation}
This gives the expansion 
\begin{equation}
I(\eta)\sim-\frac{1}{6}\log(1-\eta)-\frac{1}{2}\log(-\log(1-\eta))\,,\qquad\eta\rightarrow1\,.
\end{equation}

\subsubsection{Modular Hamiltonian\label{sec_mh_2} }

Now, ee have now all the necessary elements to compute the modular
Hamiltonian. It is given by
\begin{equation}
K_{A}=\int_{A\times A}dx\,dy\,J(x)\mathcal{K}(x,y)J(y)\,,
\end{equation}
where, according to \eqref{mh}, the kernel is 
\begin{equation}
\mathcal{K}\left(x,y\right)=\sum_{k=1}^{2}\int_{-\infty}^{+\infty}ds\,u_{k,s}\left(x\right)\pi\left|s\right|u_{k,s}\left(y\right)^{*}\,.\label{formass}
\end{equation}
This kernel is real and symmetric.

Because the expression for the modular Hamiltonian turns out to be
quite complex, we start with some preliminaries about how we are going
to express the results. Our first simplification will be to express
the kernel in the four regions $A_{1}\times A_{1}$, $A_{1}\times A_{2}$,
$A_{2}\times A_{1}$ and $A_{2}\times A_{2}$, called respectively
$\mathcal{K}_{11}$, $\mathcal{K}_{12}$, $\mathcal{K}_{21}$ and
$\mathcal{K}_{22}$, in terms of the kernel in the first interval
alone $A_{1}\times A_{1}$, using the parity symmetry of the eigenfunctions
developd in section \ref{parity}.

Let us start by computing the contribution of the first eigenvector
$u_{1}$ to $\mathcal{K}_{11}$, 
\begin{eqnarray}
\int_{-\infty}^{+\infty}ds\,u_{1,s}\left(x\right)\pi\left|s\right|u_{1,s}\left(y\right)^{*}\!\!\! & = & \!\!\!\frac{1}{4}\int_{-\infty}^{+\infty}ds\,s\,\mathrm{e}^{is\left(\omega(x)-\omega(y)\right)}\nonumber \\
 & = & \!\!\!\frac{\pi}{2}\frac{\delta\left(x-x_{1}\left(\omega\left(y\right)\right)\right)}{\omega'\left(x\right)}=\frac{\pi}{2}\omega'(x)^{-1}\,\delta\left(x-y\right)\,,\label{dell}
\end{eqnarray}
where we set $x,y\in A_{1}$, and hence, we have summed over only
one of the roots of $\omega(x)=\omega(y)$ in the delta function.
We will find convenient to write 
\begin{equation}
\mathcal{K}_{11}(x,y)=\pi\,\omega'(x)^{-1}\delta(x-y)+\mathcal{N}(x,y)\,,\hspace{0.8cm}x,y\in A_{1}\,,\label{toje}
\end{equation}
that is, we have doubled the delta function contribution from $u_{1}\left(x\right)$
and the remaining part comes from \eqref{formass} and \eqref{dell}
\begin{equation}
\mathcal{N}(x,y):=\int_{-\infty}^{+\infty}ds\,\pi|s|\left[u_{2}(x)u_{2}(y)^{*}-u_{1}(x)u_{1}(y)^{*}\right]\,,\hspace{0.8cm}x,y\in A_{1}\,.
\end{equation}
It turns out that $\mathcal{N}(x,y)$ is a regular distribution.\footnote{It is given by a locally integrable function.}
Hence, it gives a purely non-local contribution to the modular Hamiltonian.

Now, we recall the parity symmetry of the eigenfunctions studied in
section \ref{parity} under the conformal transformation $x\mapsto\bar{x}=p(x)$
that interchanges the intervals (eq. \eqref{moebius-i}). We have
that $u_{1}(\bar{x})=u_{1}(x)$ and $u_{2}(\bar{x})=-u_{2}(x)$. These
relations give the following relations between the kernel of the modular
Hamiltonian in the different sectors 
\begin{align}
 & \mathcal{K}_{12}(x,y)=\mathcal{K}_{21}(y,x)=-\mathcal{N}(x,\bar{y})\,, &  & x\in A_{1}\,,\:y\in A_{2}\,,\\
 & \mathcal{K}_{22}(x,y)=\mathcal{K}_{11}(\bar{x},\bar{y})\,, &  & x\in A_{2}\,,\:y\in A_{2}\,.\label{tiro}
\end{align}
These relations, together with \eqref{toje}, reduce the problem to
the one of finding the form of the kernel $\mathcal{N}(x,y)$ in $A_{1}\times A_{1}$.
Unless otherwise is stated, in the following we will assume that $x,y\in A_{1}$.

A different parity symmetry (eq. \eqref{tranfi}) implies that the
kernel $\mathcal{N}(x,y)$ satisfies 
\begin{equation}
\mathcal{N}(x,y)=\mathcal{N}(\hat{x},\hat{y})\,,\hspace{0.8cm}x,y\in A_{1}\,,
\end{equation}
where the transformation $x\mapsto\hat{x}=q(x)$ is given by \eqref{tranfi}.
This follows from the corresponding symmetry of the eigenvectors \eqref{uuu1}
and \eqref{uuu2}.

Another simplification is that we can relate all two-interval cases
with cross-ratio $\eta$, between the four endpoints of the intervals,
to the case where the two-interval region is the standard one $A_{\eta}:=(0,\eta)\cup(1,\infty)$.
This is done applying the conformal transformation $x\mapsto x'=f_{1}(x)$,
given by \eqref{mobius}, on the eigenvectors, eqs. (\ref{uim}-\ref{vim}).
This simply gives 
\begin{equation}
\mathcal{N}_{A}(x,y)=\mathcal{N}_{\eta}(x',y')\,,\label{right}
\end{equation}
where we wrote explicitly the dependence on the two-interval regions.
In the following, we will call simply $\mathcal{N}(x',y')$ to the
kernel on the r.h.s. of \eqref{right}, and keep $x',y'\in(0,\eta)$.
That is, we focus on the first interval $(0,\eta)$ in the case where
the region is $A_{\eta}$.

An evaluation of $\mathcal{N}(x',y')$ requires the integration over
$s$, which turns out to be the Fourier transform of products of Appell
functions contained in $u_{2}\left(x\right)$, eq. \eqref{ua1i}.
This obscures the structure of the kernel due to the complexity of
these functions, and in particular, the analysis of the possible singular
terms. Instead, we proceed in the following way. We first write the
vectors $u_{s}(x)$ as 
\begin{equation}
u_{s}(x')=-i\,\textrm{sign}(s)\,\fint_{0}^{x'}d\tilde{x}\,v_{s}(\tilde{x})\,.
\end{equation}
Then, we make the integral in $s$ more amenable writing
\begin{equation}
\mathcal{N}(x',y')=\int_{0}^{x'}d\tilde{x}\,\int_{0}^{y'}d\tilde{y}\,\mathcal{M}(\tilde{x},\tilde{y})\,,\label{ene}
\end{equation}
where 
\begin{equation}
\mathcal{M}(\tilde{x},\tilde{y}):=\int_{-\infty}^{+\infty}ds\,\pi|s|\left[v_{2}(\tilde{x})v_{2}^{*}(\tilde{y})-v_{1}(\tilde{x})v_{1}^{*}(\tilde{y})\right]\,.\label{masas}
\end{equation}
We split the above kernel into two contributions
\begin{equation}
\mathcal{M}\left(x',y'\right)=:\mathcal{M}_{1}\left(x',y'\right)+\mathcal{M}_{2}\left(x',y'\right)\,,
\end{equation}
corresponding to $v_{1}\left(x\right)$ and $v_{2}\left(x\right)$.
Using \eqref{v1cr}, we get 
\begin{equation}
\mathcal{M}_{1}\left(x',y'\right)=\frac{\pi}{2}\tilde{\omega}'\left(x'\right)\tilde{\omega}'\left(y'\right)\delta''\left(\tilde{\omega}\left(x'\right)-\tilde{\omega}\left(y'\right)\right)\,,\label{chio0}
\end{equation}
where 
\begin{equation}
\tilde{\omega}\left(x'\right):=\log\left(\frac{x'\left(1-x'\right)}{\eta-x'}\right)\,.
\end{equation}
For the other term we use \eqref{v2cr}, and hence we get 
\begin{eqnarray}
\mathcal{M}_{2}\left(x',y'\right)\!\!\! & = & \!\!\!\int_{-\infty}^{+\infty}\!\!\!ds\,\pi\left|s\right|\frac{1}{f_{1}'\left(x\right)}\frac{1}{f_{1}'\left(x\right)}v_{2,s}\left(x\right)v_{2,s}^{*}\left(y\right)\nonumber \\
 & = & \!\!\!\frac{1}{4}\!\int_{-\infty}^{+\infty}\!\!\!\!\!\!ds\,s^{2}\mathrm{e}^{-is\left(\tilde{\omega}\left(x'\right)-\tilde{\omega}\left(y'\right)\right)}\!\left(\frac{1}{x'}+\frac{\alpha\left(s,\eta\right)}{x'-\eta}-\frac{1}{x'-1}\right)\!\!\left(\frac{1}{y'}+\frac{\alpha\left(-s,\eta\right)}{y'-\eta}-\frac{1}{y'-1}\right)\nonumber \\
 & =: & \!\!\!\mathcal{M}_{2,\cancel{\alpha}}\left(x',y'\right)+\mathcal{M}_{2,\alpha}\left(x',y'\right)\,,
\end{eqnarray}
where 
\begin{eqnarray}
\mathcal{M}_{2,\cancel{\alpha}}\left(x',y'\right)\!\!\! & := & \!\!\!-\frac{\pi}{2}\left[\left(\frac{1}{x'}-\frac{1}{x'-1}\right)\left(\frac{1}{y'}-\frac{1}{y'-1}\right)+\frac{1}{x'-\eta}\frac{1}{y'-\eta}\right]\delta''\left(z\right),\hspace{1cm}\label{chio}\\
\mathcal{M}_{2,\alpha}\left(x',y'\right)\!\!\! & := & \!\!\!\frac{\pi}{2}\left(\frac{1}{x'}-\frac{1}{x'-1}\right)\frac{1}{y'-\eta}\hat{\alpha}\left(z\right)+\frac{\pi}{2}\frac{1}{x'-\eta}\left(\frac{1}{y'}-\frac{1}{y'-1}\right)\hat{\alpha}\left(-z\right),\hspace{1cm}\label{propi}
\end{eqnarray}
where $z:=\tilde{\omega}\left(x'\right)-\tilde{\omega}\left(y'\right)$
and we have also introduced the function 
\begin{equation}
\hat{\alpha}\left(z,\eta\right):=\frac{\text{1}}{2\pi}\int_{-\infty}^{\infty}ds\,s^{2}\alpha\left(s,\eta\right)\,\mathrm{e}^{isz}\,.
\end{equation}
\eqref{chio} and \eqref{propi} are respectively the $\alpha$-independent
and $\alpha$-dependent contributions to the kernel $\mathcal{M}_{2}\left(x',y'\right)$.

This gives the kernel $\mathcal{N}(x',y')$ as a double integral over
the sum of \eqref{chio0}, \eqref{chio}, and \eqref{propi}. The
final result depends on the Fourier transform of the function $s^{2}\,\alpha\left(s,\eta\right)$,
which has to be computed numerically. This numerical computation can
be done after we have extracted the leading terms for $s\rightarrow\infty$
from $\alpha(s,\eta)$. This will also help understanding the structure
of singularities of these kernels. In the following ,we will make
a further analysis of their local and non-local parts.

\paragraph{Structure of singular terms\protect \\
}

\begin{figure}[h]
\centering
\includegraphics[width=13cm]{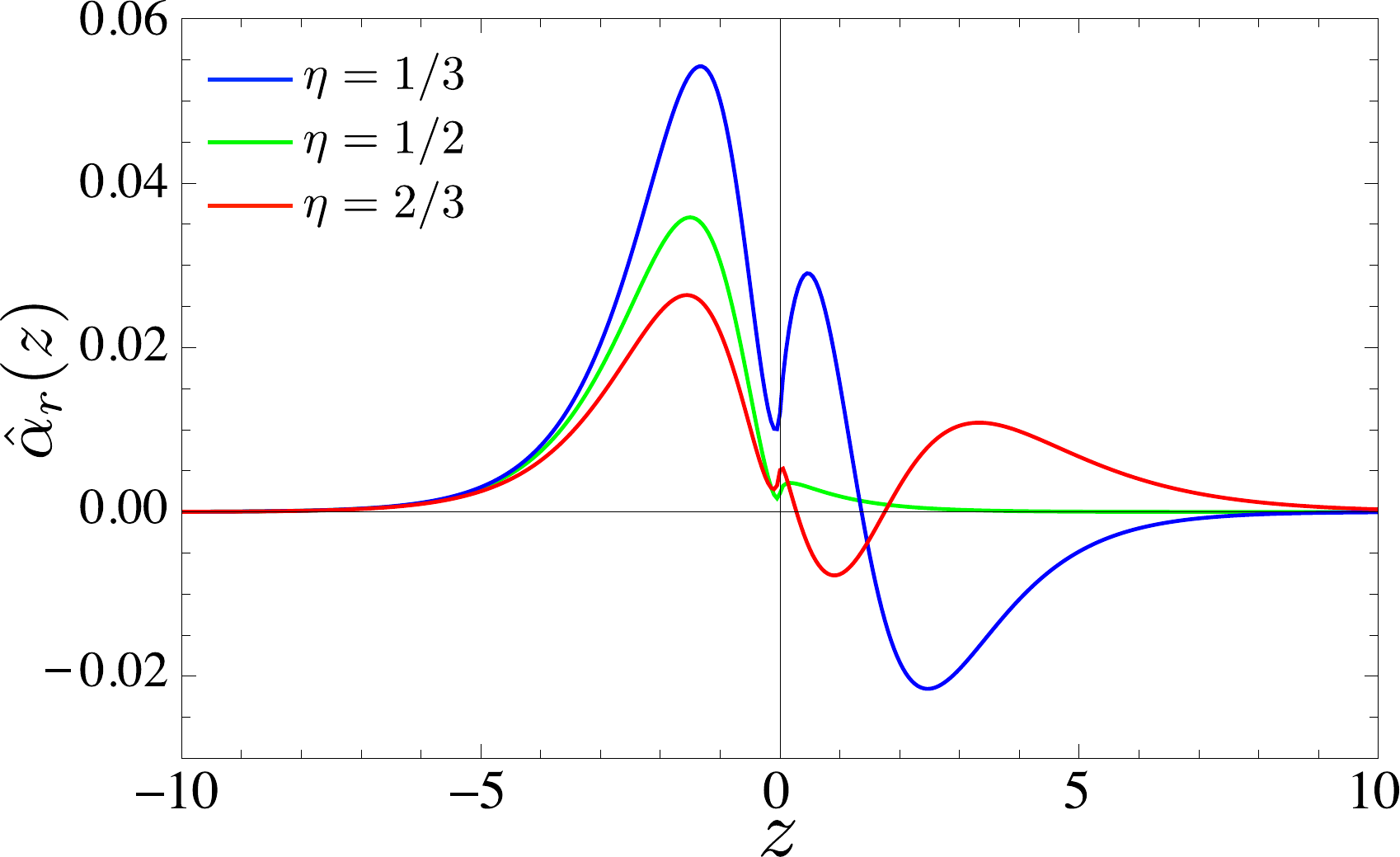}\caption{\foreignlanguage{english}{\label{fourier-alfa} The function $\hat{\alpha}_{r}(z,\eta)$ for
different values of $\eta$.}}
\end{figure}

The asymptotic behavior of the hypergeometric functions in $\alpha(s,\eta)$
for large $s$ can be computed using the integral representation \eqref{hyper_def}
and the saddle point approximation. This is straightforward. The leading
term was computed for example in \cite{Cvitkovi_2017}. Extending
this calculation to include fluctuations around the saddle point,
we get the asymptotic expansion 
\begin{equation}
\alpha\left(s,\eta\right)=\alpha_{0}+\frac{\alpha_{1}}{s}+\frac{\alpha_{2}}{s^{2}}+\frac{\alpha_{3}}{s^{3}}+{\cal O}(|s|^{-4})\,,\hspace{0.7cm}|s|\rightarrow\infty\,,\label{falfa}
\end{equation}
where 
\begin{eqnarray}
\alpha_{0}\!\!\! & = & \!\!\!\left(2\eta-1\right)+i\,2\sqrt{\eta\left(1-\eta\right)}\,\mathrm{sign}\left(s\right)\,,\\
\alpha_{1}\!\!\! & = & \!\!\!\frac{i}{2}\left(2\eta-1\right)-\sqrt{\eta\left(1-\eta\right)}\,\mathrm{sign}\left(s\right)\,,\\
\alpha_{2}\!\!\! & = & \!\!\!-\frac{i}{16}\frac{1}{\sqrt{\eta\left(1-\eta\right)}}\,\mathrm{sign}\left(s\right)\,,\\
\alpha_{3}\!\!\! & = & \!\!\!-\frac{1}{32}\frac{1}{\sqrt{\eta(1-\eta)}}\mathrm{sign}(s)+i\frac{1}{32}\frac{(2\eta-1)}{\eta(1-\eta)}\,.
\end{eqnarray}
Instead of extracting these asymptotic terms directly, we write 
\begin{equation}
\alpha\left(s,\eta\right)=\tilde{\alpha}_{0}+\frac{\tilde{\alpha}_{1}}{s}+\frac{\tilde{\alpha}_{2}}{s^{2}}+\frac{\tilde{\alpha}_{3}}{s^{3}}+\alpha_{r}\left(s,\eta\right)\,,
\end{equation}
where now 
\begin{eqnarray}
\tilde{\alpha}_{0}\!\!\! & = & \!\!\!\left(2\eta-1\right)+i2\sqrt{\eta\left(1-\eta\right)}\,\tanh\left(\frac{\pi}{2}s\right)\,,\\
\tilde{\alpha}_{1}\!\!\! & = & \!\!\!\frac{i}{2}\left(2\eta-1\right)-\sqrt{\eta\left(1-\eta\right)}\,\tanh\left(\frac{\pi}{2}s\right)\,,\\
\tilde{\alpha}_{2}\!\!\! & = & \!\!\!-\frac{i}{16}\frac{1}{\sqrt{\eta\left(1-\eta\right)}}\tanh\left(\frac{\pi}{2}s\right)\,,\\
\tilde{\alpha}_{3}\!\!\! & = & \!\!\!-\frac{1}{32}\frac{1}{\sqrt{\eta(1-\eta)}}\tanh\left(\frac{\pi s}{2}\right)+i\frac{1}{32}\frac{(2\eta-1)}{\eta(1-\eta)}\frac{s^{2}}{s^{2}+1}\,,
\end{eqnarray}
and the reminder function $\alpha_{r}\left(s,\eta\right)$ is smooth
in the parameter $s$ and $\alpha_{r}\left(s,\eta\right)\sim\frac{1}{s^{4}}$
when $|s|\rightarrow\infty$. The Fourier transform of $s^{2}\alpha(s,\eta)$
is 
\begin{eqnarray}
\hat{\alpha}\left(z,\eta\right)\!\!\! & = & \!\!\!\frac{1}{2\pi}\int_{-\infty}^{\infty}ds\,s^{2}\alpha\left(s,\eta\right)\,\mathrm{e}^{isz}=\frac{1}{2\pi}\int_{-\infty}^{\infty}ds\,s^{2}\left[\tilde{\alpha}_{0}+\frac{\tilde{\alpha}_{1}}{s}+\frac{\tilde{\alpha}_{2}}{s^{2}}+\frac{\tilde{\alpha}_{3}}{s^{3}}+\alpha_{r}\left(s,\eta\right)\right]\mathrm{e}^{isz}\nonumber \\
 & = & \!\!\!\left(1-2\eta\right)\delta''\left(z\right)+\frac{1}{2}\left(2\eta-1\right)\delta'\left(z\right)+\frac{1}{\pi}\sqrt{\eta\left(1-\eta\right)}\left(3+\cosh\left(2z\right)\right)\text{csch}^{3}\left(z\right)\nonumber \\
 &  & \!\!\!+\frac{1}{2\pi}\sqrt{\eta\left(1-\eta\right)}\sinh\left(2z\right)\,\text{csch}^{3}\left(z\right)+\frac{1}{16\pi}\frac{1}{\sqrt{\eta\left(1-\eta\right)}}\text{csch}\left(z\right)\nonumber \\
 &  & \!\!\!+\frac{1}{32\pi}\frac{1}{\sqrt{\eta(1-\eta)}}\log\left(\tanh\left|\frac{z}{2}\right|\right)+\frac{1}{64}\frac{1-2\eta}{\eta(1-\eta)}\mathrm{sign}\left(z\right)\mathrm{e}^{-\left|z\right|}+\hat{\alpha}_{r}\left(z,\eta\right)\,,
\end{eqnarray}
where $\hat{\alpha}_{r}\left(z,\eta\right)$ is the Fourier transform
of $s^{2}\alpha_{r}\left(s,\eta\right)$. This is a continuous function
vanishing exponentially fast at infinity. This real function is computed
numerically, and it is shown in figure \eqref{fourier-alfa} for some
values of $\eta$. Putting all together, we finally get 
\begin{eqnarray}
\mathcal{M}\left(x',y'\right)\!\!\! & = & \!\!\!\mathcal{M}_{1}\left(x',y'\right)+\mathcal{M}_{2}\left(x',y'\right)=\mathcal{M}_{1}\left(x',y'\right)+\mathcal{M}_{2,\alpha}\left(x',y'\right)+\mathcal{M}_{2,\cancel{\alpha}}\left(x',y'\right)\nonumber \\
 & = & \!\!\!m_{\delta''}\!\left(x',y';\eta\right)\delta''\left(z\right)+m_{\delta'}\!\left(x',y';\eta\right)\delta'\left(z\right)+m_{i}\!\left(x',y';\eta\right)+m_{r}\!\left(x',y';\eta\right),\hspace{1.1cm}
\end{eqnarray}
where
\begin{eqnarray}
m_{\delta''}\left(x',y';\eta\right)\!\!\! & := & \!\!\!\frac{\pi}{2}\tilde{\omega}'\left(x'\right)\tilde{\omega}'\left(y'\right)-\frac{\pi}{2}\left[\frac{1}{x'\left(1-x'\right)y'\left(1-y'\right)}+\frac{1}{\left(\eta-x'\right)\left(\eta-y'\right)}\right]\nonumber \\
 &  & \!\!\!+\frac{\pi}{2}\left(2\eta-1\right)\left[\frac{1}{r\left(x',y'\right)}+\frac{1}{r\left(y',x'\right)}\right],\\
\nonumber \\
m_{\delta'}\left(x',y';\eta\right)\!\!\! & := & \!\!\!-\frac{\pi}{4}\left(2\eta-1\right)\left[\frac{1}{r\left(x',y'\right)}-\frac{1}{r\left(y',x'\right)}\right],\\
\nonumber \\
m_{i}\left(x',y';\eta\right)\!\!\! & := & \!\!\!\sqrt{\eta(1-\eta)}\frac{\left[r\left(x',y'\right)-r\left(y',x'\right)\right]^{2}}{\left[r\left(x',y'\right)+r\left(y',x'\right)\right]^{3}}\nonumber \\
 &  & \!\!\!+2\sqrt{\eta(1-\eta)}\frac{r\left(x',y'\right)r\left(y',x'\right)}{\left[r\left(x',y'\right)-r\left(y',x'\right)\right]^{2}\left[r\left(x',y'\right)+r\left(y',x'\right)\right]}\nonumber \\
 &  & \!\!\!+\frac{1}{16}\frac{1}{\sqrt{\eta(1-\eta)}}\frac{1}{r\left(x',y'\right)+r\left(y',x'\right)}\nonumber \\
 &  & \!\!\!-\frac{1}{64}\frac{1}{\sqrt{\eta(1-\eta)}}\frac{r\left(x',y'\right)+r\left(y',x'\right)}{r\left(x',y'\right)r\left(y',x'\right)}\log\left|\frac{r\left(x',y'\right)-r\left(y',x'\right)}{r\left(x',y'\right)+r\left(y',x'\right)}\right|\nonumber \\
 &  & \!\!\!-\frac{\pi}{128}\frac{(2\eta-1)}{\eta(1-\eta)} \! \left[r\left(x',y'\right)-r\left(y',x'\right)\right] \!\! \left[\frac{\Theta\left(x-y\right)}{r\left(x',y'\right)^{2}}-\frac{\Theta\left(y-x\right)}{r\left(y',x'\right)^{2}}\right], \label{kirr}\\
\nonumber \\
m_{r}\left(x',y';\eta\right)\!\!\! & := & \!\!\!-\frac{\pi}{2}\frac{1}{r\left(x',y'\right)}\hat{\alpha}_{r}\left(z,\eta\right)-\frac{\pi}{2}\frac{1}{r\left(y',x'\right)}\hat{\alpha}_{r}\left(-z,\eta\right),\label{kreg}
\end{eqnarray}
and we have defined the positive polynomial function 
\begin{equation}
r\left(x',y'\right):=x'\left(1-x'\right)\left(\eta-y'\right)>0\,.
\end{equation}

The singular terms can be simplified rewriting the Dirac delta distributions
in terms of the variable $x-y$. A careful computation reveals the
relations\footnote{The reason for a term containing two derivatives of the Dirac delta
function becomes just a term proportional to $\delta\left(x'-y'\right)$
is that the function $k_{\delta''}\left(x',y';\eta\right)$ behaves
as $\left(x'-y'\right)^{2}$ for $x'\rightarrow y'$.} 
\begin{eqnarray}
m_{\delta''}\left(x',y';\eta\right)\delta''\left(\tilde{\omega}\left(x'\right)-\tilde{\omega}\left(y'\right)\right)\!\!\! & = & \!\!\!2\pi\,\eta\left(1-\eta\right)\frac{x'\left(1-x'\right)\left(\eta-x'\right)}{\left(\eta+x'^{2}-2\eta x'\right)^{3}}\delta\left(x'-y'\right)\,,\hspace{1.5cm}\label{deltauno}\\
m_{\delta'}\left(x',y';\eta\right)\delta'\left(\tilde{\omega}\left(x'\right)-\tilde{\omega}\left(y'\right)\right)\!\!\! & = & \!\!\!\frac{\pi}{4}\frac{\left(1-2\eta\right)}{\eta+x'^{2}-2\eta x'}\delta\left(x'-y'\right)\,.\hspace{1.5cm}\label{deltados}
\end{eqnarray}
Therefore, upon integration in \eqref{ene}, these terms behave as
ordinary functions, and they produce a singularity in $\mathcal{N}(x',y')$
which is just a jump in the first derivative for $x'=y'$.

In a similar way, expanding for $x'\sim y'$, the term \eqref{kirr}
reveals that it behaves as $(x'-y')^{-2}$ for $x'\rightarrow y'$.
Upon integration, this gives a $\log|x'-y'|$ singularity for $\mathcal{N}(x',y')$.
Hence, this shows that the non-local kernel $\mathcal{N}(x',y')$
is given by a regular distribution.

The behavior of the kernel $\mathcal{N}(x',y')$ on the extremes of
the interval (for example for $x'\rightarrow0$ and $y'$ fixed) can
be seen more clearly from \eqref{chio0}, \eqref{chio}, and \eqref{propi}.
The only non-local contribution comes from the Fourier transform of
$s^{2}\alpha(\eta,s)$, 
\begin{equation}
\mathcal{M}(x',y')\sim\frac{\pi}{2\,x'}\frac{1}{y'-\eta}\hat{\alpha}(\log(x'),\eta)\,,\qquad x'\rightarrow0\,.
\end{equation}
Since $\alpha(\eta,s)$ is an infinite differentiable function of
$s$, its Fourier transform fall faster than any power of the variable
$z$. Therefore $\mathcal{M}(x',y')$ is integrable on the boundary
and that is the reason we have not needed to use a regularization
in \eqref{ene}.\footnote{That is, the regularizing terms in the integral of $v_{s}\left(x\right)$
as in \eqref{fint_reg} do not contribute if we make the integral
in $s$ first.} As a result, $\mathcal{N}(x',y')$ falls to zero faster than any
power of $\log(x')^{-1}$ for $x'\rightarrow0$.

\begin{figure}[h]
\centering
\includegraphics[width=13cm]{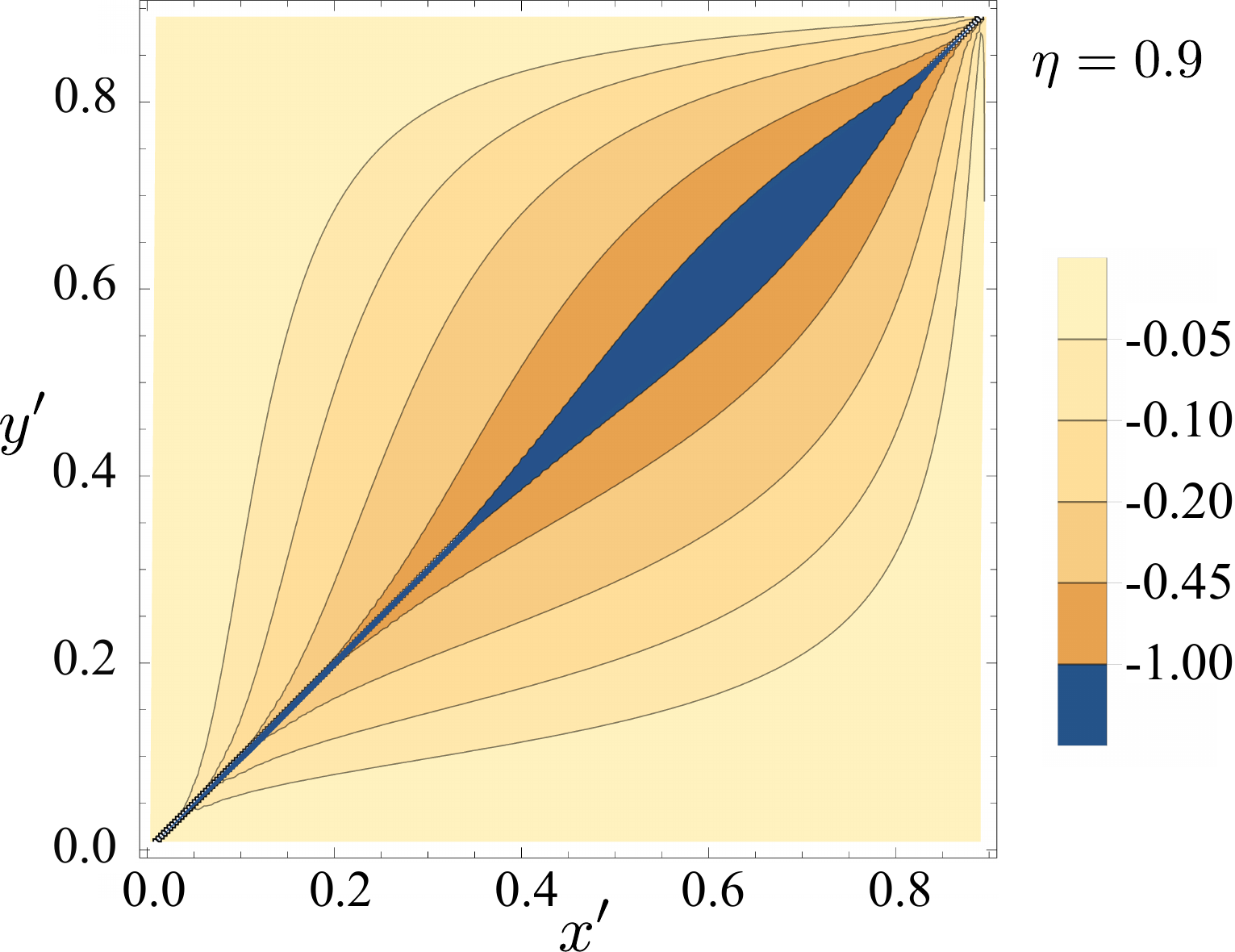}\caption{\foreignlanguage{english}{\label{contor} A contour plot of the kernel $\mathcal{N}(x',y')$
giving the non-local part of the modular Hamiltonian for two intervals
for $\eta=9/10$. $\mathcal{N}(x',y')$ has a logarithmic divergence
along the diagonal.}}
\end{figure}

A simplification in the structure of the non local kernel $\mathcal{N}(x',y')$
arises if we take into account that the integration $\int_{A_{1}'}dx'\,v_{s}^{k}(x')=0$.
Hence, using \eqref{masas} we could write \eqref{ene} as 
\begin{eqnarray}
\mathcal{N}(x',y')=-\int_{0}^{x'}d\tilde{x}\,\int_{y'}^{\eta}d\tilde{y}\,(m_{i}(\tilde{x},\tilde{y})+m_{r}(\tilde{x},\tilde{y}))\,,\hspace{0.8cm}x'<y'\,,\label{mas}\\
\mathcal{N}(x',y')=-\int_{x'}^{\eta}d\tilde{x}\,\int_{0}^{y'}d\tilde{y}\,(m_{i}(\tilde{x},\tilde{y})+m_{r}(\tilde{x},\tilde{y}))\,,\hspace{0.8cm}x'>y'\,.\label{menos}
\end{eqnarray}
In this way, we avoid crossing the $\tilde{x}=\tilde{y}$ line in
the integration, and therefore the delta functions \eqref{deltauno}
and \eqref{deltados} do not contribute. Moreover, the integrals are
now completely regular and can be done numerically since we do not
have to cross the singular points of the distributions. We checked
these expressions coincide with \eqref{ene}.\footnote{We have evaluated \eqref{ene} extracting the singular contribution
and integrating it analytically, and adding the numerical integral
of the regular parts.} A contour plot of $\mathcal{N}(x',y')$ for $\eta=9/10$ is shown
in figure \ref{contor}.

\paragraph{Summary of the modular Hamiltonian\protect \\
}

Summarizing the results, the modular Hamiltonian contains a local
part and a non-local part 
\begin{equation}
K=K_{\textrm{loc}}+K_{\textrm{noloc}}\,.
\end{equation}
The local part, generated by the delta function in \eqref{toje} and
\eqref{tiro}, gives a contribution to the modular Hamiltonian that
writes on the full region $A$ as 
\begin{equation}
K_{\textrm{loc}}=\int_{A}dx\,\pi\,(\omega'(x))^{-1}\,:J(x)^{2}:=\int_{A}dx\,2\pi\,(\omega'(x))^{-1}\,T(x)\,,
\end{equation}
where $T(x)=\frac{1}{2}:J^{2}(x):$ is the energy density. The quantity
$\beta(x):=2\pi\,(\omega'(x))^{-1}$ acts as the local inverse temperature
multiplying the energy density operator, and it dominates the limit
of relative entropy between the vacuum and energetic localized excitations
around $x$ \cite{Arias:2016nip}. This term, written in terms of
the energy density, is equal to the local term of the modular Hamiltonian
for the free chiral fermion studied in section \ref{c5-sec-ferm}.
This result coincides with general expectations for this term to be
universal across two dimensional theories \cite{Arias:2016nip,Arias17}.

The non-local part of the modular Hamiltonian is given by
\begin{eqnarray}
K_{\textrm{noloc}}\!\!\! & = & \!\!\!\int_{A_{1}\times A_{1}}\!\!\!dx\,dy\,J(x)\,\mathcal{N}(x,y)\,J(y)-\int_{A_{1}\times A_{2}}\!\!\!dx\,dy\,J(x)\,\mathcal{N}(x,\bar{y})\,J(y)\nonumber \\
 &  & \!\!\!-\int_{A_{2}\times A_{1}}\!\!\!dx\,dy\,J(x)\,\mathcal{N}(\bar{x},y)\,J(y)+\int_{A_{2}\times A_{2}}\!\!\!dx\,dy\,J(x)\,\mathcal{N}(\bar{x},\bar{y})\,J(y)\,.
\end{eqnarray}
The relevant kernel $\mathcal{N}(x,y)$ follows from \eqref{mas},
\eqref{menos}, \eqref{kirr}, and \eqref{kreg}. In contrast to the
case of the free chiral fermion, $\mathcal{N}(x,y)$ is less singular
than the local term. It is given by an integrable function, having
at most a $\log|x-y|$ singularity for $x\sim y$. Again in contrast
to the fermion case, the modular Hamiltonian is completely non local,
i.e. the kernel does not vanish identically in any open set of $A\times A$.

\paragraph{Modular flow\protect \\
}

The modular flow acts over an operator $O\in\mathcal{A}\left(A\right)$
as the unitary transformation $O(t):=\mathrm{e}^{itK}O\mathrm{e}^{-itK}$.
For an operator of the form
\begin{equation}
O(t):=\int_{A}dx\,f(x,t)\,J(x)\quad\textrm{with }\mathrm{supp}\left(f\right)\subset A\,,
\end{equation}
we have the linear flow equation 
\begin{equation}
\partial_{t}f(x,t)=-\,\beta(x)\,\partial_{x}\,f(x,t)-2\,\int_{A}dy\,\mathcal{N}(x,y)\,\partial_{y}f(y,t)\,.
\end{equation}
Then, if we start with $f(x,0)$ localized in an open small interval
inside the first interval $A_{1}$ and separated away from the boundary
of $A$ by a finite distance, the function $f(x,t)$ will spread everywhere
in both intervals $A_{1}\cup A_{2}$ for any $t\neq0$. The expectation
value $\langle O^{\dagger}(t)O(t)\rangle$ should be finite. From
the correlator \eqref{cotto} it follows that the Fourier transform
$\hat{f}(p,t):=\int_{-\infty}^{+\infty}f(x,t)\mathrm{e}^{ipx}\,dx$
should satisfy $\int_{-\infty}^{+\infty}dp\,|\hat{f}(p,\tau)|^{2}|p|<\infty$.
This does not allow for a sharp discontinuity in the test function
$f(x,t)$ at the boundary of $A$, since that would give $\hat{f}(p,t)\sim p^{-1}$.
However, a term falling like the boundary behavior of $\mathcal{N}(x,y)$
(i.e. falling to zero with $x$ as $x\rightarrow0$ faster than any
power of $\log(x)^{-1}$) can keep the test function in the space
of allowed functions.

Of course, the eigenvectors of the modular Hamiltonian kernel diagonalize
the modular flow. If we decompose the test function as
\begin{eqnarray}
f(x,t)\!\!\! & := & \!\!\!\sum_{k=1}^{2}\,\int_{-\infty}^{+\infty}ds\,\tilde{f}_{k}(s,t)\,u_{s}^{k}(x)\,,\\
\tilde{f}_{k}(s,t)\!\!\! & := & \!\!\!\int_{A}dx\,v_{s}^{k}(x)^{*}\,f(x,t)\,,
\end{eqnarray}
then, the flow equation gets diagonalized according to \eqref{ene}
\begin{equation}
\tilde{f}_{k}(s,t)=\mathrm{e}^{i2\pi st}\,\tilde{f}(s,0)\,.
\end{equation}

\subsubsection{Failure of duality for two intervals\label{failure}}

For simplicity, in this section, we work in the unit circle $S^{1}$
picture. The intervals along the null-line are mapped into the unit
circle using the inverse of the Cayley transformation \eqref{c5-cayley}.
Let us consider two disjoint intervals $A_{1}$, $A_{2}\subset S^{1}$.
The complement of $A_{1}\cup A_{2}$ is formed by two disjoint intervals,
that we call $A_{3}$, $A_{4}$. Then, we have that $(A_{1}\cup A_{2})'=A_{3}\cup A_{4}$.
If the global state is pure one usually assumes 
\begin{equation}
S(A_{1}\cup A_{2})=S(A_{3}\cup A_{4})\,,\label{syy}
\end{equation}
which is equivalent to 
\begin{equation}
I(A_{1},A_{2})=I(A_{3},A_{4})+(S(A_{1})+S(A_{2})-S(A_{3})-S(A_{4}))\,.
\end{equation}
Taking into account that the MI is a function of the cross ratio and
that the entropies for single intervals are equal to the one of complementary
intervals, we can express this relation as
\begin{equation}
I(\eta)=I(1-\eta)+\frac{1}{6}\log\left(\frac{\eta}{1-\eta}\right)\,,\label{cucas}
\end{equation}
or, equivalently \cite{Casini:2004bw} 
\begin{equation}
U(\eta)=U(1-\eta)\,.
\end{equation}
That is, the symmetry property for the entropy of complementary regions
\eqref{syy} gives the symmetry of the function $U(\eta)$. This symmetry
has also been shown as a consequence of modular invariance for two
dimensional CFTs. We have seen this symmetry is not present for the
free chiral current. This is not a problem of the continuum limit,
the same happens in a finite lattice as we will show in the next section.
As we have explained in section \ref{c3-sec-duality}, this comes
from the failure of the duality relation 
\begin{equation}
{\cal A}\left(A_{1}\cup A_{2}\right)={\cal A}\left((A_{3}\cup A_{4})'\right)\subsetneq{\cal A}\left(A_{3}\cup A_{4}\right)'\,.
\end{equation}
In fact, we can define the operator
\begin{equation}
\Phi_{12}^{\lambda}:=\mathrm{e}^{i\lambda\int_{\mathbb{R}}dx\,f(x)\,J(x)}\,,\label{dose}
\end{equation}
where $\lambda\in\mathbb{R}$ and $f(x)$ is any smooth function such
that 
\begin{equation}
f\left(x\right):=\begin{cases}
1 & x\in A_{3}\,,\\
0 & x\in A_{4}\,.
\end{cases}\label{c5-ll-func}
\end{equation}
This operator does not belong to the algebra ${\cal A}\left(A_{1}\cup A_{2}\right)$
but it belongs to ${\cal A}(A_{3}\cup A_{4})'$, since it commutes
with all the operators belonging ${\cal A}(A_{3}\cup A_{4})$ because
of the commutation relations \eqref{commut}. It is evident that any
two operators like \eqref{dose}, with the same parameter $\lambda\in\mathbb{R}$
but given by two different smear functions $f(x)$ satisfying \eqref{c5-ll-func},
will differ by an element of ${\cal A}(A_{1}\cup A_{2})$. Therefore,
the choice of $f(x)$ satisfying these properties will not change
the algebra ${\cal A}(A_{1}\cup A_{2})\vee\{\Phi_{12}^{\lambda}\}$.
In fact, we have that
\begin{equation}
{\cal A}\left((A_{1}\cup A_{2})'\right)'={\cal A}(A_{1}\cup A_{2})\vee\{\Phi_{12}^{\lambda}\,:\,\lambda\in\mathbb{R}\}\,.
\end{equation}
We call the operator $\Phi_{12}^{\lambda}$ a \textit{long-link} joining
$A_{1}$ with $A_{2}$, since being the exponential of the integral
of $\partial_{+}\phi\left(x\right)$, it is equivalent to some difference
of the field $\phi$ localized inside the two intervals. However,
in this model the field $\phi\left(x\right)$ does not actually exist,
and the only way to write this operator is through an integral of
the current as in \eqref{dose}.

There is still the possibility of considering a long link crossing
the interval $A_{4}$, rather than $A_{3}$ as in $\Phi_{12}^{\lambda}$.
However, a sum of these two long links is equivalent (modulo operators
in ${\cal A}(A_{1}\cup A_{2})$) to the integral of the current along
the whole unit circle $\oint dx\,J(x)$. This last operator commutes
with all the algebra and it is multiple of the identity operator in
the vacuum representation. Hence, the two options for the long link
are actually equivalent.

Now, in an analogously way, we can define a long link crossing $A_{1}$.
For example, we can define the operator
\begin{equation}
\Phi_{34}^{\lambda'}:=\mathrm{e}^{i\lambda'\int_{\mathbb{R}}dx\,g(x)\,J(x)}\,,
\end{equation}
with $\lambda'\in\mathbb{R}$ and $g(x)$ is any smooth function such
that 
\begin{equation}
g\left(x\right):=\begin{cases}
1 & x\in A_{1}\,,\\
0 & x\in A_{2}\,.
\end{cases}
\end{equation}
It follows from the commutation relations that
\begin{equation}
\Phi_{12}^{\lambda}\Phi_{34}^{\lambda'}=\mathrm{e}^{i\lambda\lambda'}\Phi_{34}^{\lambda'}\Phi_{12}^{\lambda}\,.
\end{equation}
Therefore, ${\cal A}\left((A_{1}\cup A_{2})'\right)'$ and ${\cal A}\left((A_{3}\cup A_{4})'\right)'$
do not commute. Hence, the theory does not satisfy essential duality
(see definition \ref{c1-def-ess-dual}), and the local algebras cannot
be enlarged (with out spoiling locality) in order to obtain a theory
satisfying the duality relation. Instead, we have that\footnote{This can be easily shown in the finite lattice model of the next subsection.}
\begin{eqnarray}
{\cal A}(A_{1}\cup A_{2})'\!\!\! & = & \!\!\!{\cal A}(A_{3}\cup A_{4})\vee\left\{ \Phi_{34}^{\lambda'}\,:\,\lambda'\in\mathbb{R}\right\} \,,\\
{\cal A}(A_{3}\cup A_{4})'\!\!\! & = & \!\!\!{\cal A}(A_{1}\cup A_{2})\vee\left\{ \Phi_{12}^{\lambda}\,:\,\lambda\in\mathbb{R}\right\} \,.
\end{eqnarray}
In order to have duality for a two-interval region in this model,
one should choose to add the long link to the algebra of one of the
pairs of intervals (e.g. $A_{1}\cup A_{2}$) and not for the complementary
one (e.g. $A_{3}\cup A_{4}$). This prescription necessarily does
not treat in an equivalent way all pairs of intervals. Another perhaps
more disturbing consequence of this choice is that the algebra ${\cal A}\left((A_{1}\cup A_{2})'\right)'$
containing the long link is not additive, meaning that it is not the
generated algebra by the algebras of the two intervals. This is because
the long link does not belong to the algebra generated by the single
intervals, i.e. 
\begin{equation}
{\cal A}(A_{3}\cup A_{4})'={\cal A}\left((A_{1}\cup A_{2})'\right)'\supsetneq{\cal A}(A_{1})\vee{\cal A}(A_{2})\,.
\end{equation}
Hence, we can have additivity at the expense of duality or viceversa,
but not both properties together. The natural choice is \eqref{c5-curr-net}
because it can be consistently assigned to any pair of interval regions,
and because of the additivity property, it is the only choice that
allows the definition of the MI.

It is worth to notice that the free chiral fermion, which is an ``extension''
of the current algebra, satisfies (twisted) duality for two intervals
and also additivity. This is why for the fermion $U(\eta)=U(1-\eta)$
since trivially $U(\eta)=0$. It is by reducing the theory to the
current algebra that we run into this particular trouble.

This failure of duality for two intervals in general chiral conformal
models has been associated with an algebraic $\mu$-index of the inclusion
of subalgebras ${\cal A}(A_{3}\cup A_{3})\subsetneq{\cal A}(A_{1}\cup A_{2})'$
\cite{Kawahigashi:1999jz}. This index ``quantifies'' the presence
of superselection sectors in the theory, and should also determine
the amount of asymmetry in the MI \cite{Longo_xu} as\footnote{The paper \cite{Longo_xu} proves this relation for subalgebras of
the free chiral fermion field, and conjectures its greater validity
for chiral CFT models.} 
\begin{equation}
U(0)-U(1)=\frac{1}{2}\log(\mu)\,.
\end{equation}
In the present model we have seen this is divergent, in accordance
with the fact that the $\mu$-index of the current algebra is infinity.
In the following chapter we study this issue in general, finding a
clear connection between EE and the superselection structure of the
theory.

\paragraph{Restoration of the duality\protect \\
}

From the point of view of a CFT in $d=2$, it might seem strange that
we could have duality violation for two intervals and hence $U(\eta)=U(1-\eta)$
fails. This property can be derived from modular invariance of the
twist operators giving place to the Rényi entropies \cite{Cardy:2017qhl}.
The reason is that duality can be restored by adequately combining
two chiral theories.

Let us look at the example of the massless limit of a free massive
scalar. The usual local algebra for a massive scalar in $d=2$ satisfies
duality and additivity for two double cones, corresponding to two
intervals in the spatial line at $t=0$ (see chapter \ref{RE_CS}).
In the massless case, the zero mode of the scalar field has divergent
mean quadratic variation and has to be removed. However, the spatial
and time derivatives of the field remain. With them we can form the
two chiral currents $\partial_{\pm}\phi$. For a single diamond, the
algebra is then equivalent to the one of two decoupled chiral currents
in single interval regions. For two diamonds however, the algebra
also contains the difference $\phi(x_{2})-\phi(x_{1})$, with $x_{1}$
and $x_{2}$ belonging to two different diamonds. We can take $x_{1},x_{2}$
on the two intervals at $t=0$, i.e. 
\begin{equation}
\phi(x_{2})-\phi(x_{1})=\int_{x_{1}}^{x_{2}}dx\,\partial_{x}\phi(0,x)=\int_{x_{1}^{+}}^{x_{2}^{+}}dx^{+}\,\partial_{+}\phi(x)-\int_{x_{1}^{-}}^{x_{2}^{-}}dx^{-}\,\partial_{-}\phi(x)\,.\label{sdf}
\end{equation}
Hence, the two diamonds contain the difference of long link operators
corresponding to the two chiral algebras. However, they do not contain
the sum of these long link operators, and therefore the chiralities
do not decouple for two diamonds. This is the reason why these algebras
for the two diamonds are compatible with the ones of the two complementary
diamonds: the chiral long link operators do not commute to each other
but their sum does, since commutators come out with opposite sign.
Thinking in terms of the field differences \eqref{sdf}, this commutation
is evident. Therefore, these algebras have the same form for all pairs
of diamonds and duality is restored. However, without the zero mode,
additivity is lost in this example. The massless limit of the MI of
two intervals is divergent as $I\sim\frac{1}{2}\log(-\log(m))$ \cite{Casini_review}.

\subsection{The free chiral current in the lattice\label{numerico} }

For the numerical calculation we put the model in a lattice. We take
the lattice Hamiltonian 
\begin{equation}
H:=\frac{1}{2}\sum_{j\in\mathbb{Z}}f_{j}^{2}\,,
\end{equation}
and the commutator 
\begin{equation}
[f_{j},f_{l}]=i(\delta_{l,j+1}-\delta_{l,j-1})=:i\,C_{jl}\,.\label{c}
\end{equation}
Let us take a periodic system and, in order for $C$ to be invertible,
we take an even number $N:=2n$ of points. The eigenvectors of the
commutator are given by phase factors 
\begin{equation}
\sum_{l=1}^{N}C_{jl}e^{ikl}=2i\sin(k)\mathrm{e}^{ikj}\,,\qquad k:=\frac{2\pi m}{N}\,,\,\,\,m=-(n-1),\cdots,n\,.
\end{equation}
From here, it follows that defining the following variables 
\begin{eqnarray}
\phi_{k}\!\!\! & := & \!\!\!\frac{1}{\sqrt{N\sin(k)}}\sum_{j=1}^{N}\cos(kj)\,f_{j}\,,\label{c5-lat-f}\\
\pi_{k}\!\!\! & := & \!\!\!\frac{1}{\sqrt{N\sin(k)}}\sum_{j=1}^{N}\sin(kj)\,f_{j}\,,\label{c5-lat-f2}
\end{eqnarray}
where $k=\frac{2\pi m}{N}\in\left(0,\pi\right)$ with $m\in(1,n-1)$,
we have that $[\phi_{k},\pi_{k'}]=i\,\delta_{k,k'}$. We have also
two other variables $\psi_{0},\,\psi_{\pi}$ that form the center
of the global algebra since they commute with all other elements,
\begin{eqnarray}
\psi_{0}:=\frac{1}{\sqrt{N}}\sum_{j=1}^{N}f_{j} & \textrm{ and } & \psi_{\pi}:=\frac{1}{\sqrt{N}}\sum_{j=1}^{N}(-1)^{j}f_{j}\,.
\end{eqnarray}
The inverse relation of (\ref{c5-lat-f}-\ref{c5-lat-f2}) is
\begin{equation}
f_{j}=\frac{2}{\sqrt{N}}\sum_{m=1}^{n-1}\sqrt{\sin(k)}\left(\cos(kj)\,\phi_{k}+\sin(kj)\pi_{k}\right)+\frac{\psi_{0}}{\sqrt{N}}+\frac{\psi_{\pi}(-1)^{j}}{\sqrt{N}}\,,
\end{equation}
where $k=\frac{2\pi m}{N}\in\left(0,\pi\right)$. The Hamiltonian
writes in these new variables as
\begin{equation}
H=\sum_{m=1}^{n-1}\sin(k)(\phi_{k}^{2}+\pi_{k}^{2})+\frac{1}{2}\psi_{0}^{2}+\frac{1}{2}\psi_{\pi}^{2}\,.\label{hmy}
\end{equation}
This gives for the vacuum state $\langle\phi_{k}^{2}\rangle=\langle\pi_{k}^{2}\rangle=\frac{1}{2}$
and $\langle\phi_{k}\pi_{k}\rangle=i/2$. The center can take any
value and we set $\psi_{0}=\psi_{\pi}=0$. Hence, we impose these
relations as a constraint. In this way, we get a pure vacuum state
and a global algebra without center. The full system has now $n-1$
degrees of freedom: $n-1$ coordinates and $n-1$ momentum variables.

We see from \eqref{hmy} that we have two sets of low energy degrees
of freedom, for $k\sim0$ and $k\sim\pi$. Thereby, the system shows
a doubling of the degrees of freedom in the continuum, analogous to
the usual fermion doubling. This is the reason we also have two commuting
operators $\psi_{0}$ and $\psi_{\pi}$.

The correlator of the original variables $F(j-l)=\langle f_{j}f_{l}\rangle$
is given by 
\[
F\left(x\right)=\begin{cases}
\frac{1}{N}\frac{\cos^{2}\left(\frac{\pi x}{2}\right)\sin\left(\frac{2\pi}{N}\right)}{\sin\left(\frac{\pi(x+1)}{N}\right)\cos\left(\frac{\pi(x-1+N/2)}{N}\right)}\,, & |x|\neq1\,\\
\frac{i}{2}C(x)\,, & |x|=1\,
\end{cases}
\]
In the limit of a large circle $N\rightarrow\infty$, we have that
\begin{equation}
F\left(x\right)=\begin{cases}
-\frac{1+(-1)^{x}}{\pi(x^{2}-1)}\,, & |x|\neq1\,\\
\frac{i}{2}C(x)\,, & |x|=1\,
\end{cases}\label{f}
\end{equation}

The entropy of a region follows from \eqref{c3-genCCR-V} and \eqref{c3-genCCR_ee}.
We first check numerically the entropy for a single interval. We calculate
the matrices \eqref{c} and \eqref{f} for intervals of length $R=10k$
with $k=1,...,20$. We fit the pairs $(R_{k},S(R_{k}))$ with $c_{0}+c_{\log}\log k+c_{-1}\frac{1}{k}+c_{-2}\frac{1}{k^{2}}$
obtaining the logarithmic coefficient $c_{\log}=1/3$ with high precision.
Notice that this coefficient is twice the expected one for the chiral
current model. This reflects the doubling on the lattice.

To calculate the MI between two intervals of length $a$ and $b$
separated by a distance $c$, that is $I(A,B)=S(A)+S(B)-S(A\cup B)$,
we need the entropies of the single intervals $S(A)$ and $S(B)$,
and the entropy of the two intervals $S(A\cup B)$. Each of these
entropies is calculated using \eqref{c3-genCCR-V} and \eqref{c3-genCCR_ee}.
In the continuum limit the MI is a function of the cross-ratio $\eta$,
where $\eta$ is defined as 
\begin{equation}
\eta:=\frac{a.b}{(a+c)(b+c)}\,,
\end{equation}
in accordance to \eqref{crossratio}. For a given cross-ratio, we
repeat the calculation for different configurations that differ one
from another just by a dilatation with parameter $k=2,4,...,20$.
We then fit the pairs $(k\,,I_{k}(\eta))$ with $c_{0}+c_{-1}\frac{1}{k}+c_{-2}\frac{1}{k^{2}}+c_{-3}\frac{1}{k^{3}}$,
and take the constant coefficient $c_{0}$ as the continuum limit
of the mutual information for the lattice model, which is twice the
chiral current model due to doubling. We then take $I(\eta)=c_{0}/2$.
We repeat the same procedure for different values of $\eta$ obtaining
the red points showed in figure \ref{figufi}.

In doing simulations for this model it is important that, if $N$
is finite, we take the total number of points even $N=2n$, and, in
order to not have a center, the subsystems need to have an even number
of points or variables, i.e, intervals of even size. This is because
half of them are coordinates and half are momenta. The complementary
subsystems automatically must have equal entropy because the global
state is pure. For example, for an interval of size $2k$ in a circle
of size $2n$, the commutant is an interval of size $2n-2k-2$, because
there are two points in the complementary region adjacent to the interval
that do not commute with the original interval. The entropies are
indeed equal. When we consider two-interval regions the commutant
algebra contains a long link as explained in the previous section.
In the lattice, it contains two long links operators because of the
doubling. More precisely, these commutant algebras with long links
for two intervals are of the form: all points in the intervals $\left(a_{1},b_{1}\right)\cup\left(a_{2},b_{2}\right)$,
and two long links given by the sums $\sum_{j}f_{j}$ and $\sum_{j}(-1)^{j}f_{j}$,
where the sums are over all the points in the open interval $(b_{1},a_{2}):=(b_{1}+1,\cdots,a_{2}-1)$.
The long links crossing the other gap between the intervals are related
to these by elements of the algebra of the intervals and the global
constraints, and hence they do not give additional operators. The
counting of degrees of freedom is as follows: for a circle of $2n$
points, if the original intervals have $2k_{1}$ and $2k_{2}$ points,
the commutant will have $2n-2k_{1}-2k_{2}-4$ points plus two long
links. This gives a total of $2n-2k_{1}-2k_{2}-2$ linearly independent
operators. This is precisely (twice) the complementary number of degrees
of freedom: $2n-2$ is twice the total number of degrees of freedom
in the lattice.

We have checked the entropies of complementary algebras of two intervals
are equal in the circle. The entropy for the two intervals with the
long links $\tilde{S}$ can also be completed to form a ``kind of
mutual information'', eliminating UV divergences in the continuum,
as 
\begin{equation}
\tilde{I}(A_{1},A_{2})=S(A_{1})+S(A_{2})-\tilde{S}(A_{1}\cup A_{2})\,.
\end{equation}
The equality of the entropies for two intervals and the one of the
complementary region including the long links 
\begin{equation}
S(A_{1}\cup A_{2})=\tilde{S}(A_{3}\cup A_{4})\,,
\end{equation}
can be completed with single interval entropies to form a relation
between the mutual informations 
\begin{equation}
\tilde{I}(\eta)=I(1-\eta)+\frac{1}{6}\log\left(\frac{\eta}{1-\eta}\right)\,.\label{illvsi}
\end{equation}
We can define the $U$ function for the entropies with the long link
\begin{equation}
\tilde{I}(\eta)=-\frac{1}{6}\log(1-\eta)+\tilde{U}(\eta)\,.
\end{equation}
Then relation \eqref{illvsi} is just the complementary relation for
the $U(\eta)$ 
\begin{equation}
\tilde{U}(1-\eta)=U(\eta)\,.
\end{equation}
These two should be symmetric and equal for a model satisfying duality
for two intervals, but this is not the case of the present model.
We have also checked numerically the relation \eqref{illvsi} in the
infinite lattice limite. For that, we calculate $\tilde{I}(\eta)$
and we note that the convergence to the continuum limit is much improved
for this case using the fitting function as $c_{0}+c_{-1/2}\frac{1}{k^{1/2}}+c_{-3/2}\frac{1}{k^{3/2}}$,
instead of using integer powers, as long as we increase the global
size $k$ of the region. The continuum limit again corresponds to
the coefficient $\tilde{I}(\eta)=\frac{c_{0}}{2}$.

\section{Conclusions of the chapter\label{conclusions}}

We obtained the vacuum modular Hamiltonian for the free chiral current
in two intervals. The modular Hamiltonian contains the usual local
term given by an integral of the energy density times a position dependent
inverse temperature. This term is identical to the free chiral fermion
one, and very probably it is universal for all theories in $d=2$.
Besides, there is a non-local term. This is given by a quadratic expression
in the current with a locally integrable kernel which does not vanish
in any open subset of $A\times A$. Hence, the modular Hamiltonian
is completely non-local in contrast to the fermion case. The MI does
not have the symmetry property \eqref{cucas}, and the origin of this
is the failure of duality for two intervals.

We treated the case of two intervals. More intervals could in principle
be treated in a similar fashion, but the expressions will depend on
a higher number of cross-ratios, and besides, the Hypergeometric and
Appell functions that parametrize the eigenvectors should be replaced
by higher dimensional Lauricella functions.

It would be interesting to understand why the free chiral fermion
modular Hamiltonian is ``quasilocal'' while the one of the chiral
current is completely non-local. The technical reason is that one
of the eigenvectors of the bosonic model has a dependence on the eigenvalue
$s$ that is not simply a phase factor $\mathrm{e}^{is\omega(x)}$.
For the fermion field and any number of intervals, or the current
field in a single interval, this same phase factor determines completely
the dependence of all eigenvectors in $s$. Perhaps a reason for the
fermion to be special is the multi-local symmetries described by Rehren
and Tedesco \cite{Rehren:2012wa}.

We have shown that the current MI is smaller than the fermion one
because the former model is a subalgebra of the later. It would be
interesting to explore other consequences of this inclusion. For example,
the difference of modular Hamiltonians $K_{\psi}-K_{J}$ between these
models should be a positive operator. We can compute the expectation
value of this difference of operators in a state generated from the
vacuum by acting with a unitary in $A$, for example, a coherent state
$\mathrm{e}^{i\int dx\,f(x)J(x)}|0\rangle$. The local contribution
vanishes in the difference of expectation values of the two modular
Hamiltonians, and we get an inequality involving exclusively the non-local
parts of $K_{\psi}$ and $K_{J}$.


\renewcommand\chaptername{Chapter}
\selectlanguage{english}

\chapter{Entanglement entropy and superselection sectors\label{EE_SS}}

As we have explained in section \ref{c1-sec-rep-states}, given an
algebra of observables $\mathcal{\mathfrak{A}}$, its space of pure
states decomposes into superselection sectors (SS). These SS are related
to unitarily inequivalent representations of $\mathcal{\mathfrak{A}}$.
In any given (non-irreducible) representation $\pi$ of $\mathfrak{A}$,
pure states belonging to the same superselection sector are related by elements of $\pi(\mathfrak{A})$. On the other hand,
there is no operator in $\pi(\mathfrak{A})$ connecting pure states belonging to different SS.

In QFT, one is usually interested in states which are constructed
from the vacuum one through local perturbations by elements of $\mathcal{\mathfrak{A}}$.
Hence, it seems plausible to restrict the attention to the vacuum GNS-representation
$\mathcal{A}:=\pi_{0}(\mathfrak{A})\subset\mathcal{B}\left(\mathcal{H}_{0}\right)$,
and dispense with the structure of the other states that the model
$\mathcal{\mathfrak{A}}$ admits. However, it is the result of a large
body of research into the superselection structure of QFT that the
SS leave a definite imprint in the relations between the different
local subalgebras assigned to regions in the model ${\cal A}$
itself. More precisely, there is a subclass of sectors representing
localizable states, such that they can be fully reconstructed from
the vacuum sector \cite{Doplicher:1969kp,Doplicher:1972kr,Doplicher:1990pn,Longo:1994xe}.
This superselection structure could be accomplished with
the help of a bigger local QFT $\mathfrak{F}\subset\mathcal{B}\left(\mathcal{H}\right)$,
called the field algebra,\footnote{Traditionally, the algebra $\mathcal{A}$ is thought to be the algebra
of local physical observables, while the charged operators in $\mathfrak{F}$
retain some locality properties but are not physically realizable
in local laboratories, e. g. an operator that can change the baryonic
number. In the theoretical setting of this thesis, we do not make
this epistemological distinction.} such that it contains copies of all the other representations corresponding
to superselection sectors of localizable states. In this way, the
Hilbert space $\mathcal{H}$ decomposes as a direct sum of subspaces,
where the observable algebra $\mathcal{A}$ acts irreducibly. In such
a decomposition, the vacuum representation $(\mathcal{A},\mathcal{H}_{0})$
appears once. The observable algebra $\mathcal{A}$ could be obtained
from $\mathfrak{F}$ as the pointwise invariant part under the action
of a compact symmetry group $G$. States belonging to the vacuum subspace
$\mathcal{H}_{0}\subset\mathcal{H}$ are invariant (neutral) with respect to
the action of $G$, whereas vector states belonging to any other subspaces
transforms non-trivially under $G$, and hence they define \textit{charged
states}. The elements of $\mathcal{A}$ cannot transform between vectors
in disjoint subspaces. However, there are charged local operators
in $\mathfrak{F}$, which allows us to create any charged state from
the vacuum subspace $\mathcal{H}_{0}$.

This superselection structure affects the relations between algebras
and regions in ${\cal A}$, either violating the property of duality
and/or additivity for some topologically non-trivial regions (see
definition \ref{c1_conj}). In this way, a theory having non-trivial
SS is considered as an incomplete model. The superselection structure
also affects the vacuum fluctuations through charge-anticharge virtual
pairs. The EE should be sensitive to these behaviors. Therefore, the main focus of this
chapter is the analysis of the consequences of the SS for the EE.

Localizable charged states remarkably come in two types, corresponding
to global or gauge symmetry charges. These correspond to two abstract
types of superselection sectors, called DHR (because of Haag, Doplicher,
and Roberts \cite{Doplicher:1969tk,Doplicher:1969kp,Doplicher:1971wk,Doplicher:1972kr,Doplicher:1973at,Doplicher:1990pn})
and BF sectors (because of Buchholz and Fredenhagen \cite{Buchholz:1981fj}) respectively. The main difference between these two cases is geometric,
global charges creating operators can be localized in a compact region, while gauge charges creating operators can be only localized in cones, allowing the Wilson line extends to infinity. Here we deal only with DHR SS.

\begin{figure}[t]
\centering
\begin{centering}
\includegraphics[width=0.4\textwidth]{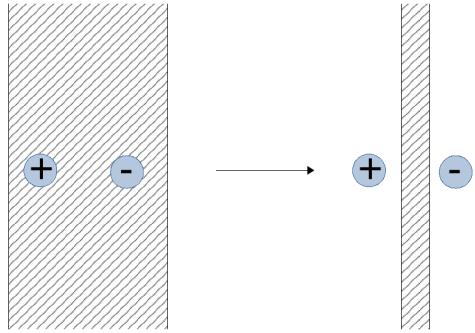} 
\par\end{centering}
\centering{}\captionsetup{width=0.9\textwidth} \caption{\label{fig6} Mutual information between two regions separated by
a strip of width $\epsilon>0$ (shaded region in the figure). For
$\epsilon$ wide enough, the typical charge-anticharge fluctuations
are not sensed by the mutual informations of both models (left panel).
When the width $\epsilon$ becomes small enough to allow for charge-anticharge
fluctuations to occur on each side of the wall with enough probability
(right panel), the MI of $\mathfrak{F}$ will take
into account these correlations while the one of the neutral model
${\cal A}$ will not.}
\end{figure}

An essential feature of the EE in QFT is that it cannot
be defined without the introduction of an ultraviolet regulator, making
this quantity inherently ambiguous through the regularization scheme
choice. As we have explained in section \ref{c3-sec-ee-qft}, this ambiguity
can be cured by computing half the MI between nearly complementary
regions, which are separated by a regulating distance $\epsilon>0$.
This is a natural quantity, taking the place of EE, that is well-defined
in the continuum model and it is used as a universal regularized EE.
On the other hand, there is also another source of ambiguities that has been discussed in the literature concerning the assignation of
local algebras to regions. In this sense, local algebras may contain a center related to ambiguities in the choice of the algebra at the boundary
of the region in a lattice model. This type of ambiguities has attracted especially attention in relation to gauge models (for the discussion
around this topic see for example \cite{Ghosh:2015iwa,Soni:2016ogt,Soni:2015yga,VanAcoleyen:2015ccp,Donnelly:2014fua,Donnelly:2011hn,Donnelly:2015hxa,Donnelly:2014gva,Camps:2018wjf}).
However, this kind of local ambiguities does not survive the continuum
limit, leaving the MI as a well-defined quantity \cite{Casini:2013rba}.
In the new scenario we are presenting here, where models with SS sectors
are considered, the analysis is enriched giving place to more interesting
consequences. In models with SS, there is more than one choice of
the algebra for topologically non-trivial regions. The MI is sensitive to these possible choices. However, these mutual informations can be reinterpreted as corresponding
to different models, with and without SS.

More concretely, the MI crucially depends on the physical regulating distance
$\epsilon$ that allows us to sense or not the presence of virtual
charge pairs according to the comparison of the size of $\epsilon$
with the typical scale $\Lambda$ of these fluctuations (see figure
\ref{fig6}). Hence, two possible results may come out in the limits
$\epsilon/\Lambda\gg1$ or $\epsilon/\Lambda\ll1$, independently
of the size $R$ of the region, whenever $R$ is much larger than both
$\epsilon$ and $\Lambda$. Then, in terms of the MI, it may seem
we still have an apparent ambiguity. One of the main results that
come from the analysis of SS is the clarification of this issue. Each
result for the mutual information corresponds to a particular choice of the algebra, where the SS have been included or not respectively. This means,
we should not interpret this as an ambiguity but as a consequence
of alternative model choices.

With this perspective, partially following previous works in the mathematical
literature \cite{Xu:2018fsv,Longo_xu}, we develop entropic order
parameters capable to sense these differences. In this way, we study models with finite or Lie group symmetry groups, and spontaneous symmetry
breaking symmetries.

While throughout this chapter we try to keep the discussion as simple
and physical as possible, with a mixed degree of mathematical rigor,
we are forced to use some specific mathematical tools to avoid making
ambiguous statements. We do not treat explicit examples where the
superselection sectors do not come from a symmetry group. This includes
models with DHR sectors in $d=2$. This would require more formal
developments but would not add any new to the general physical picture.

This chapter is structured as follows. In section \ref{algebra-regions},
we describe the problems in the relations between algebras and regions
in theories with superselection sectors, and we introduce the main elements of the theory of DHR superselection
sectors. In section \ref{entropyDHR}, we investigate the EE in the
case of DHR sectors, describing the relevant entropic order parameters and their
mutual relationships, that take the form of entropic certainty and
uncertainty relations. We explicitly compute the relevant quantities
in several cases of interest. This includes the cases of finite and
Lie symmetry groups, compactified scalars, regions with different
topologies and charge excitations. In section \ref{bounds},
we study in concrete examples, the behavior of the expectation values of
some operators (intertwiners and twists), which are the main witnesses
of the superselection sectors and play a central role in the evaluation
of the entropic quantities. We end, in section \ref{c6-conclu}, with a
summary and the conclusions.

\section{Algebras, regions and superselection sectors\label{algebra-regions}}

We are interested in some particularities of the relation between
algebras and regions that affect the EE. To start the discussion, we assume
that our observable QFT is defined as net of vN algebras 
\begin{eqnarray}
\mathcal{O}\in\mathcal{K} & \mapsto & \mathcal{A}\left(\mathcal{O}\right)\subset\mathcal{B}\left(\mathcal{H}_{0}\right)\,,
\end{eqnarray}
in the vacuum representation, satisfying all the axioms of definitions
\ref{c1_def_aqft} and \ref{c1_def_vacuum}. We also assume that the
vacuum vector $\left|0\right\rangle \in\mathcal{H}_{0}$ is the unique
Poincaré invariant vector in the Hilbert space $\mathcal{H}_{0}$.
Moreover, we assume that such a vacuum representation is unique. Weak
additivity (definition \ref{c1-weak_add}) is also assumed for the observable algebra $\mathcal{A}$.
\begin{rem}
In the following subsection, we slightly modify these assumptions in order to study the SS structure of the theory.
\end{rem}
Among the above minimal assumptions, we want to remark isotony
\begin{equation}
{\cal A}(\mathcal{O}_{1})\subset{\cal A}(\mathcal{O}_{2})\,,\quad\textrm{if }\mathcal{O}_{1}\subset\mathcal{O}_{2}\,,\label{isotonia}
\end{equation}
and causality,
\begin{equation}
{\cal A}(\mathcal{O}')\subset{\cal A}(\mathcal{O})'\,.\label{causality}
\end{equation}

We could also ask whether the observable algebra would satisfy the
stronger assumptions developed in section \ref{c1_subsec_lattice}.
More concretely, we are interested in considering the assumptions of
the weaker version of definition \ref{c1_conj}, where \ref{c1-conj-add}
is replaced by \ref{c1-conj-add-bis}, or equivalently the assumptions of definition \ref{c1_conj-2}
once a Cauchy surface is chosen. We remark here again that these assumptions
are expected to hold only for ``sufficiently complete'' theories.
However, as we will see below, some of them fail for physically significant
models.

Duality condition (\ref{c1-conj-duality} in definition \ref{c1_conj})
is an enhancement of the causality property above. When the duality
condition does not hold, it may seem rather simple to complete a given
net of algebras to have duality, just by enlarging the algebras taking
$\mathcal{A}\left(\mathcal{O}\right)\rightarrow\mathcal{A}^{d}\left(\mathcal{O}\right)=\mathcal{A}\left(\mathcal{O}'\right)'$.
However, in this process of completion, some problems might arise,
since we are enlarging the algebra of the region $\mathcal{O}$ and
its complement $\mathcal{O}'$ at the same time. In partiucular, it may happen that $\mathcal{A}^{d}\left(\mathcal{O}\right)$
and $\mathcal{A}^{d}\left(\mathcal{O}'\right)$ do not commute. Moreover,
the enlarged algebra may not satisfy additivity and/or intersection
property (\ref{c1-conj-add-bis} and \ref{c1-conj-inter} in definition \ref{c1_conj})
even though the original algebra does.

Additivity property means that algebra of a region $\mathcal{O}$
could be generated by the algebras of operators of smaller regions
included in $\mathcal{O}$. This expresses that the algebra is locally
generated. In the Garding-Wightman approach based on point-like localized
fields, this is how one would form the algebra of a region $\mathcal{O}$,
by taking arbitrary polynomials of smeared fields with support in
arbitrary small regions inside $\mathcal{O}$. Intersection property
is a sort of dual version of additivity. According to the discussion
in section \ref{c1_subsec_lattice}, if the theory satisfies duality
for any causally complete region, then, due to the De Morgan laws,
additivity holds for $\mathcal{O}_{1}$ and $\mathcal{O}_{2}$ if
and only if intersection property holds.

It is important to emphasize that when the duality condition fails, it could
be the case that it still holds for a restricted subset of causally
complete regions. For example, we will assume above that it holds
for topological trivial regions, such as double cones. However, in
a specific model, the imposition of \eqref{isotonia} and \eqref{causality}
prevents duality to hold for non-trivial topological regions.This means that the failure of the duality condition must be considered as a physical fact and not as a sort of mathematical unphysical issue due to an inadequate election of algebras. Moreover, in this case,
the different algebras $\mathcal{A}\left(\mathcal{O}\right)$ and
$\mathcal{A}^{d}\left(\mathcal{O}\right)$ give place to different
macroscopic statistical behaviors even for the vacuum state, which
are detectable with the use of entanglement measures.

Summarizing, properties \eqref{isotonia} and \eqref{causality} are
elementary axioms which always hold for any QFT model (see definition \ref{c1_def_aqft}). However, duality, additivity and intersection property are assumptions that will hold only for ``sufficiently complete''
models, and fail for some physically significant models. Here we are
interested in such failures due to the presence of non-trivial SS. We will see a simple example below.

It is important to realize that the failure of these properties does
not (necessarily) have to do with the fact that QFT is a continuum
theory with infinitely many degrees of freedom or the intricate nature
of the specific type III vN algebras. The problems we want
to address here are not related to UV physics, but they already appear in lattice models.
\begin{notation}
Throughout this chapter, we usually consider causally complete regions
which are the Cauchy development of spacelike regions over the Cauchy
surface of fixed time $x^{0}=0$. We denote such a Cauchy surface
by $\Sigma_{0}$. Moreover, we frequently use causally complete regions
$D\left(\mathcal{C}_{R}\right)$, where $\mathcal{C}_{R}$ is the
sphere of radius $R$
\begin{equation}
\mathcal{C}_{R}:=\left\{ \left(0,\bar{x}\right)\in\Sigma_{0}\,:\,\left|\bar{x}\right|<R\right\} \,.\label{c6-reg1}
\end{equation}
\end{notation}
\begin{notation}
To lighten the notation, the vacuum state is denoted by $\omega$ instead of $\omega_{0}$, as we have used before.
\end{notation}

\paragraph{Example: bosonic subnet of the free fermion field\protect \\
}

In order to introduce the main ideas, we find useful to start with
a specific simple model, where all the relevant objects can be displayed
explicitly. Once the main ideas are understood, we will discuss the
general case. Let us then focus on the following model. We take a
free Dirac field in $d$ dimensions. This theory is defined by means of the local fields 
\begin{equation}
\psi_{j}\left(x\right),\,\psi_{k}^{\dagger}\left(x\right)\,,\quad\;j,k=1,\ldots,n:=2^{\left\lfloor \frac{d}{2}\right\rfloor }\,,
\end{equation}
which satisfy the canonical anticommutation relations\footnote{\label{fn_fer_t}In the free theory, the time-zero fields $\psi_{j}\left(0,\bar{x}\right)$ and $\psi_{j}^{\dagger}\left(0,\bar{x}\right)$ are well-defined operators valued distributions in the fermionic Fock Hilbert space (see section \ref{c1_subsec_aqft_vs_gw} and \cite{Streater}). In this case, (\ref{c6-ccr-fer-1}-\ref{c6-ccr-fer-2}) imply
\begin{eqnarray}
\{\psi_{j}\left(0,\bar{x}\right),\psi_{k}\left(0,\bar{y}\right)\}\!\!\! & = & \!\!\!\{\psi_{j}^{\dagger}\left(0,\bar{x}\right),\psi_{k}^{\dagger}\left(0,\bar{y}\right)\}=0\,,\\
\{\psi_{j}\left(0,\bar{x}\right),\psi_{k}^{\dagger}\left(0,\bar{y}\right)\}\!\!\! & = & \!\!\!\delta(\bar{x}-\bar{y})\, . \label{fer_anti}
\end{eqnarray}
}
\begin{eqnarray}
\{\psi_{j}\left(x\right),\psi_{k}\left(y\right)\}\!\!\! & = & \!\!\!\{\psi_{j}^{\dagger}\left(x\right),\psi_{k}^{\dagger}\left(y\right)\}=0\,,\label{c6-ccr-fer-1}\\
\{\psi_{j}\left(x\right),\psi_{k}^{\dagger}\left(y\right)\}\!\!\! & = & \!\!\!iS_{jl}\left(x-y\right)\gamma_{lk}^{0}\,,\qquad S(x)=(i\cancel{\partial}+m)\Delta(x)\,,\label{c6-ccr-fer-2}
\end{eqnarray}
where $\cancel{\partial}:=\gamma^{\mu}\partial_{\mu}$, $\gamma^{\mu}\in\mathbb{C}^{n\times n}$
($\mu=0,\ldots,d-1$) are the Gamma matrices (in some representation),
and $\Delta(x)$ is the commutator function of the scalar field (see equation
\eqref{c4-sf-comm}). For any test function $f:=\left(f_{1},\cdots,f_{n}\right)^{T}$
with $f_{j}\in\mathcal{S}\left(\mathbb{R}^{d}\right)$, the smeared
operators are denoted by
\begin{eqnarray}
\psi\left(f\right):=\int_{\mathbb{R}^{d}}d^{d}x\,f\left(x\right)^{\dagger}\psi\left(x\right) & \textrm{ and } & \psi^{\dagger}\left(f\right):=\int_{\mathbb{R}^{d}}d^{d}x\,\psi^{\dagger}\left(y\right)f\left(x\right)\,,\label{c6-ferm-uni}
\end{eqnarray}
which are bounded operators acting on the fermionic Fock Hilbert space
$\mathcal{H}$. The local algebras for a causally complete region
$\mathcal{O}\in\mathcal{K}$ are defined as the vN algebras
\begin{equation}
\mathfrak{F}\left(\mathcal{O}\right):=\left\{ \psi\left(f_{1}\right)+\psi^{\dagger}\left(f_{2}\right)\,:\,\mathrm{supp}\left(f_{j}\right)\subset\mathcal{O}\right\} ''\,.
\end{equation}
The fermionic net satisfies the axioms \ref{c1-conj-irr}, \ref{c1-conj-add-bis}, and \ref{c1-conj-inter}-\ref{c1-conj-cov} of definition \ref{c1_conj}, including additivity and (twisted) duality \cite{Dadashyan}.\footnote{For fermionic nets, we must consider \textit{twisted duality} $\mathfrak{F}\left(\mathcal{O}'\right) = {\mathfrak{F}\left(\mathcal{O}\right)^{t}}\sp{\prime}$ (see section \ref{c1-sec-fermions}), instead of assumption \ref{c1-conj-duality} of definition \ref{c1_conj}.}

However, we are interested in the bosonic subnet of the fermion net,
i.e. the subalgebra $\mathcal{A}\subset\mathfrak{F}$ consisting of
all operators with even fermionic number. It is generated by all operators
made out of an even number of fermion fields, i.e. $\mathbf{1},\,\psi_{j}(x)\psi_{k}(y),\,\psi_{j}^{\dagger}(x)\psi_{k}^{\dagger}(y),\,\psi_{j}(x)\psi_{k}^{\dagger}(y)$, etc. All these fields have to be smeared with test functions, and we
can take arbitrary polynomials with even fermionic number. This subalgebra
acts naturally in the subspace $\mathcal{H}_{0}\subsetneq\mathcal{H}$
formed by the states with even number of (fermionic) particles. Moreover, the vacuum vector $\left|0\right\rangle \in\mathcal{H}_{0}$ is cyclic for the algebra $\mathcal{A}$ in the bosonic Hilbert space
$\mathcal{H}_{0}.$ Regardless we have defined the bosonic net as
a subnet of the fermion net, we want to emphasize that the former
exists in its own right. In fact, it could be defined without referring
to a bigger algebra nor a bigger Hilbert space. The bosonic
net $\mathcal{A}$ is called the algebra of observables, since it
is formed uniquely by "observables quantities", which commute at
spacelike distance. Remarkably, all the physics is encoded in the observable algebra itself from which the bigger algebra $\mathfrak{F}$ could be reconstructed. We will explain this fact more detailed in section \ref{c6-secc-field}.

Once we have recognized the global algebra $\mathcal{A}\subset\mathcal{B}\left(\mathcal{H}_{0}\right)$,
we are interested in the assignation of subalgebras to regions $\mathcal{O}\in\mathcal{K} \mapsto \mathcal{A}\left(\mathcal{O}\right)\subset\mathcal{B}\left(\mathcal{H}_{0}\right)$. For a double cone $W\subset\mathbb{R}^{d}$, we define
\begin{equation}
\mathcal{A}\left(W\right):=\left\{ \left[\psi\left(f_{1}\right)+\psi^{\dagger}\left(f_{2}\right)\right]\left[\psi\left(f_{3}\right)+\psi^{\dagger}\left(f_{4}\right)\right]\,:\,\mathrm{supp}\left(f_{j}\right)\subset W\right\} ''\,,\label{c6-bos-con}
\end{equation}
where the double commutant is taken inside $\mathcal{B}\left(\mathcal{H}_{0}\right)$.
At this point, it could be intriguing why we have decided to use definition
\eqref{c6-bos-con} only for double cones, and not for any causally
complete region. The reason is that for more general regions, there is
more than one alternative. For example, let us take two spacelike
separated double cones $W_{1}$ and $W_{2}$. If we define the algebra
$\mathcal{A}\left(W_{1}\vee W_{2}\right)$ using \eqref{c6-bos-con},
we have that the theory would not satisfy additivity. In fact, the
``intertwiner'' operator 
\begin{equation}
\mathcal{I}_{12}:=\psi\left(f_{1}\right)\psi\left(f_{2}\right)\,,\quad\mathrm{supp}\left(f_{j}\right)\subset W_{j}\,,\label{c6-exa-int}
\end{equation}
belongs to the global algebra $\mathcal{A}$ and would belong to the
algebra of the $W_{1} \vee W_{2}$ according to \eqref{c6-bos-con}. However, it does not belong to the additive algebra $\mathcal{A}\left(W_{1}\right)\vee\mathcal{A}\left(W_{2}\right)$.
This is because of the presence of a dual operator of $\mathcal{I}_{12}$,
called the ``twist'' operator. In order to give a simple expression
for such an operator, let us assume for a moment that $W_{1}:=D\left(\mathcal{C}_{R}\right)$.
Then, the ``twist'' operator is defined as\footnote{The reason why these operators are called ``intertwiner'' and ``twist''
will be clearer in the next section.}\footnote{A general double cone can be represented as $W_{1}:=D\left(\mathcal{C}\right)$,
where $\mathcal{C}\subset\Sigma$ is a spacelike region over some
Cauchy surface $\Sigma$. A twist operator for $W_{1}$ can be defined as
\begin{equation}
\tau:=\mathrm{e}^{i\pi\int d^{d}x\,g(x):\overline{\psi}\left(x\right)\gamma^{\mu}\psi\left(x\right):\,n_{\mu}}\,,
\end{equation}
where $n_{\mu}$ is the unit normal future pointing timelike vector to $\Sigma$ and $g(x)$ is an appropriate smearing function.}
\begin{equation}
\tau:=\mathrm{e}^{i\pi\int d^{d}x\,g\left(x^{0}\right)h\left(\bar{x}\right):\psi^{\dagger}\left(x\right)\psi\left(x\right):}\,,\label{c6-twist}
\end{equation}
where\footnote{In other words, the function $g\left(x^{0}\right)h\left(\bar{x}\right)$
must be any smooth regularization of the function $\delta\left(x^{0}\right)\Theta_{\mathcal{C}}\left(\bar{x}\right)$
with support strictly spacelike separated to $W_{2}$.}
\begin{align}
 & g\in C_{c}^{\infty}\left(\mathbb{R},\mathbb{R}\right)\:\textrm{ and }\:\int_{\mathbb{R}}g\left(t\right)\,dt=1\,,\\
 & h\in C_{c}^{\infty}\left(\mathbb{R}^{d-1},\mathbb{R}\right)\:\textrm{ and }\:h(\bar{x})=1\;\forall\bar{x}\in\mathcal{C}_{R}\,,\\
 & \mathrm{supp}(gh)\text{\Large\ensuremath{\times\negmedspace\!\!\times}}W_{2}\,.
\end{align}
The twist is a unitary operator, which belongs to the global algebra
${\cal A}$, trivially commutes with $\mathcal{A}\left(W_{2}\right)$,
and anticommutes with the fermion operators in $W_{1}$, i.e. $\tau\psi\left(f\right)\tau^{-1}=-\psi\left(f\right)$
for all smearing functions with $\mathrm{supp}\left(f\right)\subset W_{1}$. Then,
$\tau$ also commutes with $\mathcal{A}\left(W_{1}\right)$. In other
words, $\tau\in\left(\mathcal{A}\left(W_{1}\right)\cup\mathcal{A}\left(W_{2}\right)\right)'$.
However, $\tau$ does not commute with $\mathcal{I}_{12}$. In fact,
we have that
\begin{equation}
\left\{ \tau,{\cal I}_{12}\right\} =0\Rightarrow\left[\tau,{\cal I}_{12}\right]\neq0\,.\label{c6-no-comm}
\end{equation}
Because of this, we have that $\mathcal{I}_{12}$ cannot
belong to the additive algebra $\mathcal{A}\left(W_{1}\right)\vee\mathcal{A}\left(W_{2}\right)=\left(\mathcal{A}\left(W_{1}\right)\cup\mathcal{A}\left(W_{2}\right)\right)''$.

On the other hand and in order to preserve additivity, we can define the
algebra of $W_{1}\vee W_{2}$ as
\begin{equation}
\mathcal{A}\left(W_{1}\vee W_{2}\right):=\mathcal{A}\left(W_{1}\right)\vee\mathcal{A}\left(W_{2}\right)\,.
\end{equation}
We will show now that such a choice does not satisfies duality. In
fact, if we want to preserve additivity for any causally complete region, the algebra of
the of complement of the union of two double cones $\left(W_{1}\vee W_{2}\right)'$,
has to be defined according \eqref{c6-bos-con}, i.e.
\begin{equation}
\mathcal{A}\left(\left(W_{1}\vee W_{2}\right)'\right)=\bigvee_{W\subset\left(W_{1}\vee W_{2}\right)'}\mathcal{A}\left(W\right)\,,
\end{equation}
where the union runs along all the double cones $W$ included in $(W_{1}\vee W_{2})'$. It is important to emphasize that the above definition is the only possible one if we want to preserve additivity. However, this algebra
does not contain the operator \eqref{c6-twist}, because it cannot
be formed additively inside $\left(W_{1}\vee W_{2}\right)'$. Or more
precisely, the operator $\mathcal{I}_{12}$ commutes with every local
algebra $\mathcal{A}\left(W\right)$ with $W\subset\left(W_{1}\vee W_{2}\right)'$,
and hence 
\begin{equation}
\mathcal{I}_{12}\in\left(\bigcup_{W\subset\left(W_{1}\vee W_{2}\right)'}\mathcal{A}\left(W\right)\right)'\,.\label{c6-int-no}
\end{equation}
Then, we have that $\tau\notin\mathcal{A}\left(\left(W_{1}\vee W_{2}\right)'\right)$
because of \eqref{c6-no-comm} and \eqref{c6-int-no}. This implies that
\begin{equation}
\mathcal{A}\left(\left(W_{1}\vee W_{2}\right)'\right)\subsetneq\mathcal{A}\left(W_{1}\vee W_{2}\right)'=\mathcal{A}\left(\left(W_{1}\vee W_{2}\right)'\right)\vee\left\{ \tau\right\} \,,
\end{equation}
showing explicitly the failure of the duality condition for the union of two
double cones. On the other hand, for any connected region $\mathcal{O}\in\mathcal{K}$,
the additivily generated algebra  $\bigvee_{W\subset\mathcal{O}}\mathcal{A}\left(W\right)$
coincides with the one defined through \ref{c6-bos-con}. Moreover,
for such a connected region duality holds. In other words, in this
model, the ambiguity between algebras and regions, and the failure
of the duality condition are attainable only to non-connected regions.

The intersection property also conflicts with additivity since we
can take two connected causally complete regions $\mathcal{O}_{1},\mathcal{O}_{2}$, but whose intersection is the union of two disconnected regions $\mathcal{O}_{3}$ and $\mathcal{O}_{4}$, i.e. $\mathcal{O}_{1}\cap\mathcal{O}_{2}=\mathcal{O}_{3}\cup\mathcal{O}_{4}$. It not difficult to see, that an intertwiner $\mathcal{I}_{34}$ (as
the one of \eqref{c6-exa-int}) between $\mathcal{O}_{3}$ and $\mathcal{O}_{4}$
belongs to the additive algebras $\mathcal{A}(\mathcal{O}_{1})$ and
$\mathcal{A}(\mathcal{O}_{2})$, and therefore, it belongs to the intersection
of the algebras $\mathcal{A}(\mathcal{O}_{1})\wedge\mathcal{A}(\mathcal{O}_{2})$
. However, as we have discussed above, $\mathcal{I}_{34}$ does not
belong to the additive algebra $\mathcal{A}(\mathcal{O}_{3})\vee\mathcal{A}(\mathcal{O}_{4})$.
In other words, if we accept the intersection property, the intertwiner
will belong to the algebra of $\mathcal{O}_{3}\cup\mathcal{O}_{4}$, and this algebra will not be additive.

To summarize, if we start with the operators belonging to $\mathcal{A}\left(W_{1}\right)\cup\mathcal{A}\left(W_{2}\right)$,
its commutant will contain the twist operator, and then, the double commutant will not contain the intertwiner. Conversely, if we start with the algebra of two double cones plus the intertwiner $\mathcal{A}\left(W_{1}\right)\cup\mathcal{A}\left(W_{2}\right)\cup\left\{ \mathcal{I}_{12}\right\} $, its commutant will not contain the twist. In other words, denoting
by ${\cal A}_{\mathcal{O}}:={\cal A}\left(\mathcal{O}\right)$ the additive algebra for any region $\mathcal{O}\in\mathcal{K}$, then we have
\begin{eqnarray}
{\cal A}_{W_{1}\vee W_{2}}'\!\!\! & = & \!\!\!{\cal A}_{(W_{1}\vee W_{2})'}\vee\left\{ \tau\right\} \,,\\
{\cal A}_{(W_{1}\lor W_{2})'}'\!\!\! & = & \!\!\!{\cal A}_{W_{1}\lor W_{2}}\vee\left\{ {\cal I}_{12}\right\} \,.\label{c6-ferm-vio}
\end{eqnarray}
Therefore, duality for the union of two double cones (resp. the complement of two double cones)
would require that we enlarge the additive algebra of its complementary
region with the twist operator (resp. intertwiner), losing
additivity.\footnote{This is exactly what happen for the free chiral current, as we have discussed
in the previous chapter (see section \ref{failure}).} Also, this possible enlargement does not treat in the same way the
algebra of a region and the one of its commutant. Notice that we could
change the definition of the intertwiner and/or the twist by choosing
different smearing functions satisfying the stated requisites. But,
these different operators differ between them by elements of the additive
algebra.

From now on, ${\cal A}_{\mathcal{O}}$ will always denote the additive
algebra of the region $\mathcal{O}\in\mathcal{K}$, which by definition
is
\begin{equation}
\mathcal{A}\left(\mathcal{O}\right)=\bigvee_{\alpha}\mathcal{A}\left(W_{\alpha}\right)\,,\label{c6-add-ferm}
\end{equation}
where the union runs over any family of double cones such that $\mathcal{O}=\bigcup_{\alpha}W_{\alpha}$.
In a general theory, the above relation will be always assumed, i.e.
we always assume that our net of observables satisfies additivity.

Relation \eqref{c6-ferm-vio}, which ``measures'' the violation
of the duality relation for two double cones, plays a central role
in this chapter since it shows explicitly the possible ambiguities
in the assignation of algebras to regions. In fact, the difference
between the two choices of algebras is due to the existence of the intertwiner operator
${\cal I}_{12}$. The following sections aim to show that the violation
of the duality for two double cones occurs in any QFT having non-trivial
superselection sectors. This relation between the failure of the
duality condition and the existence of SS will allow us to give a general and concrete description
of the difference of the algebras ${\cal A}_{(W_{1}\lor W_{2})'}'$ and ${\cal A}_{W_{1}\lor W_{2}}$.

\begin{figure}[t]
\centering
\centering{}\includegraphics[width=10cm]{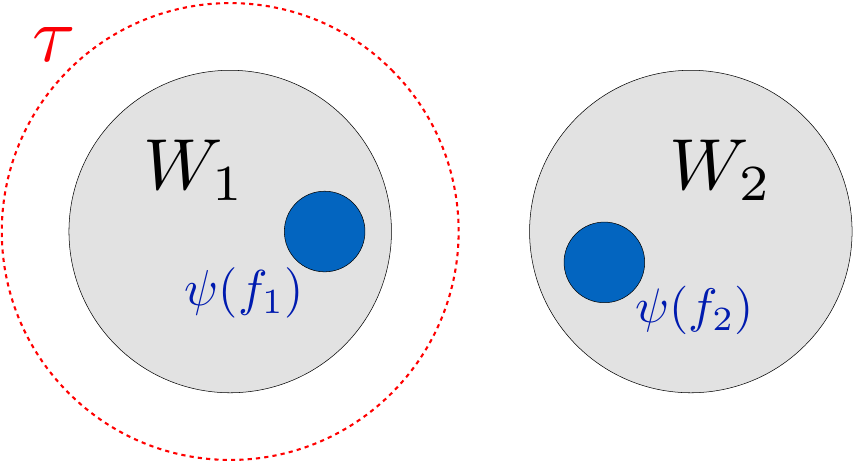} \captionsetup{width=0.9\textwidth}
\caption{\foreignlanguage{english}{\label{f3}The intertwiner $\mathcal{I}_{12}=\psi\left(f_{1}\right)\psi\left(f_{2}\right)$
commutes with the additive algebra of the exterior of the two balls
$W_{1}$ and $W_{2}$. It cannot be formed additively with the algebras
of these balls. The twist $\tau$ cannot be formed additively on the
exterior of the two balls but commutes with the algebras of the balls.
The intertwiner and the twist do not commute with each other. }}
\end{figure}

\subsection{DHR superselection sectors\label{DHR}}

Here we start with an AQFT given by a net of $C^{*}$-algebras $\mathcal{O}\in\mathcal{K} \mapsto \mathcal{A}\left(\mathcal{O}\right)\subset\mathcal{A}$.
As we have discussed in section \ref{c1-sec-rep-states}, the space
of pure states $\mathfrak{S}_{p}\left(\mathcal{A}\right)$ of the
global algebra $\mathcal{\mathcal{A}}$ decomposes into coherent subsets
called sectors. Between the collection of all sectors, there is a
particular subcollection, called DHR sectors, because of the work
of Doplicher, Haag and Roberts \cite{Doplicher:1969tk,Doplicher:1969kp,Doplicher:1971wk,Doplicher:1972kr,Doplicher:1973at,Doplicher:1990pn} (for a review see \cite{haag,Halvorson06,araki}). These sectors, whose
definition is given by criterion \ref{c6-cri-dhr} below, contain most of the states which are interesting in elementary particle physics. These states are such that they behave asymptotically like the vacuum state for
observations in far away regions of the space. It turns out that the
set of all DHR pure states is larger than the collection of vector
states in one irreducible representation (e.g. the vacuum representation),
but certainly, smaller than the set of all pure states over $\mathcal{A}$.
For our interest, we will show that the existence of non-trivial DHR
sectors is the cause of the failure of the duality for two double
cones.

Recall the connection between states and Hilbert space representations
we explained in section \ref{c1-sec-rep-states}. Each representation
has an affiliated family of normal states, which are given by the density matrices on the representation Hilbert space. The pure
states affiliated with one irreducible representation form a superselection
sector. Pure states affiliated with inequivalent irreducible representations
belongs to different sectors. Then, instead of describing ``the states
of interest'' we may thus equally well describe the ``classes
of representations of interest''.
\begin{defn}
\label{c6-cri-dhr}Let $\pi_{0}:\mathcal{A}\rightarrow\mathcal{B}\left(\mathcal{H}_{0}\right)$
be the vacuum representation of $\mathcal{A}$. A \textit{DHR representation}
is any representation $\pi:\mathcal{A}\rightarrow\mathcal{B}\left(\mathcal{H}_{\pi}\right)$
which is unitarily equivalent to the vacuum one outside any double
cone, or in other words
\begin{equation}
\left.\pi\right|_{\mathcal{A}\left(W'\right)}\cong\left.\pi_{0}\right|_{\mathcal{A}\left(W'\right)}\,,\label{c6-dhr-uni}
\end{equation}
for all double cone $W$.\footnote{The unitary operator $V_{W}:\mathcal{H}_{0}\leftrightarrow\mathcal{H}_{\pi}$
that provides the above unitarily equivalence may depends (and in general
does) on the region $W$.} Any normal state on an irreducible DHR representation is called a
\textit{DHR state}. The collection of all vector states affiliated
to an irreducible DHR representation corresponds a \textit{DHR (superselection)
sector}. 
\end{defn}

The physical meaning of the above definition may be illustrated by
the following remark \cite{Doplicher:1971wk}. Let us take a sequence of double cones
$W_{n}$ which exhaust the spacetime in the limit $n\rightarrow\infty$,
i.e. $W_{n}\subset W_{n+1}$ and $\bigcup_{n\in\mathbb{N}}W_{n}=\mathbb{R}^{d}$.
Then, $\phi\in\mathfrak{S}\left(\mathcal{A}\right)$ is a DHR state
if and only if
\begin{equation}
\left\Vert \left.\left(\phi-\omega\right)\right|_{\mathcal{A}\left(W'_{n}\right)}\right\Vert \underset{n\rightarrow\infty}{\longrightarrow}0\,,
\end{equation}
where $\omega$ is the vacuum state. In other words, the state $\phi$
describes a localized excitation respect to the vacuum in the sense
that it cannot be distinguished from it making observations in a region
far away from a bounded region.\footnote{In this case, the charge associated to the state $\phi$ has to be
a global charge, not corresponding to a gauge symmetry, since gauge
charges can be measured by the electric flux through a shell of large
radius around the charge. Hence the state $\phi$ can be distinguished
from the vacuum in the asymptotic spacelike infinity.}

Before we study the consequences of the definition \ref{c6-cri-dhr}, we
give the structural assumptions concerning the observable algebra
$\mathcal{A}$, which we need for the subsequent analysis.
\begin{assumption}
\textup{\label{c6-ass-dhr}The observable algebra is described by
a net of $C^{*}$-algebras $\mathcal{O}\in\mathcal{K}\mapsto\mathcal{A}\left(\mathcal{O}\right)\subset\mathcal{B}\left(\mathcal{H}_{0}\right)$
in the vacuum representation, satisfying all the axioms of definitions
\ref{c1_def_aqft} and \ref{c1_def_vacuum}.}\footnote{This is possible without loss of generality because of the local unitary
equivalence of all representations of interest (equation \eqref{c6-dhr-uni}).}\textup{ We also assume the following extra assumptions.}
\end{assumption}
\vspace{-.5 cm}
\begin{enumerate}
\item Type of local algebras:
\begin{enumerate}
\item Local algebras $\mathcal{A}\left(B\right)$ associated with bounded
regions $B\in\mathcal{K}$ are vN algebras, i.e. $\mathcal{A}\left(B\right)=\mathcal{A}\left(B\right)''$. 
\item Local algebras associated with unbounded regions are $C^{*}$-algebras
(norm-closed) but they are not, in general, vN algebras (weakly closed). In particular,
the global algebra $\mathcal{A}:=\mathcal{A}\left(\mathbb{R}^{d}\right)$
is a norm closed, but not a weakly closed, algebra.
\end{enumerate}
\item \label{c6-ass-add}Additivity:\footnote{This additivity property is equivalent to the one defined in \ref{c1-conj-add-bis} of
definition \ref{c1_conj}, but adjusted to the present scenario.}
\begin{enumerate}
\item For a bounded region $B\in\mathcal{K}$, we have that $\mathcal{A}\left(B\right)=\left(\bigcup_{\alpha}\mathcal{A}\left(B_{\alpha}\right)\right)''$
for any causal cover $B:=\mathrm{int}\left(D\left(\overline{\bigcup_{\alpha}B_{\alpha}}\right)\right)$.
\item For an unbounded region $\mathcal{O}\in\mathcal{K}$, we have that $\mathcal{A}\left(\mathcal{O}\right)=\overline{\bigcup_{\alpha}\mathcal{A}\left(D_{\alpha}\right)}^{\left\Vert \cdot\right\Vert }$
for any causal cover $\mathcal{O}:=\mathrm{int}\left(D\left(\overline{\bigcup_{\alpha}\mathcal{O}_{\alpha}}\right)\right)$.\footnote{This implies that $\mathcal{A}$ satisfies weak additivity in the sense of definition \ref{c1-weak_add}.}
\end{enumerate}
\item \label{c6-ass-dua}Duality for double cones:\footnote{This version of the duality is weaker than the usual one $\mathcal{A}\left(W\right)'=\mathcal{A}\left(W'\right)$
(condition \ref{c1-conj-duality} in definition \ref{c1_conj}). This
last one implicitly assumes that the algebra $\mathcal{A}\left(W'\right)$
is a vN algebra since it is the commutant of the $\mathcal{A}\left(W\right)$.
In fact, assumption \ref{c6-ass-dua} above is equivalent to $\mathcal{A}\left(W\right)'=\mathcal{A}\left(W'\right)''$.}
\begin{equation}
\mathcal{A}\left(W\right)=\mathcal{A}\left(W'\right)'\,,\quad\textrm{ for all double cones }W\,.
\end{equation}
\end{enumerate}

Let us give some comments about these assumptions. Defining the algebras
for bounded regions as vN algebras is a useful technical assumption,
whereas defining the algebras for unbounded regions just as $C^{*}$-algebras,
but not vN algebras, is more than a technical assumption. In fact,
if $W$ is a double cone, the vN algebra $\mathcal{A}\left(W'\right)''$
associated with the unbounded region $W'$ is too big enough that
it spoils the criterion \eqref{c6-dhr-uni}. In other words, it can
be shown that, in any AQFT and under vey general assumptions, there is
no representation (beyond the vacuum one) that satisfies definition
\ref{c6-cri-dhr} if we replace $\mathcal{A}\left(W'\right)$ by $\mathcal{A}\left(W'\right)''$
in \eqref{c6-dhr-uni}. Moreover, if we assume that the global algebra
$\mathcal{A}=\mathcal{A}\left(\mathbb{R}^{d}\right)$ is a vN algebra, then
we have that $\mathcal{A}''=\mathcal{B}\left(\mathcal{H}_{0}\right)$ since
the vacuum representation is irreducible (the vacuum state is pure). But,
such an algebra has only one sector since it has only one (non-equivalent)
irreducible representation, which is given by itself. We will show
how these issues appear in the example of the bosonic subnet
of the free fermion field introduced above.\footnote{Equivalently, we could define all the local algebras to be vN algebras
paying the cost of changing the relation \eqref{c6-dhr-uni} by 
\begin{equation}
\left.\pi\right|_{\bigcup_{\alpha}\mathcal{A}\left(W_{\alpha}\right)}\cong\left.\pi_{0}\right|_{\bigcup_{\alpha}\mathcal{A}\left(W_{\alpha}\right)}\,,
\end{equation}
where the union runs over all the double cones $W_{\alpha}\subset W'$.
However, the global algebra $\mathcal{A}$ cannot be a vN algebra
in order to have non-trivial SS.} It is important to emphasize, the physical fact, that the existence
of different choices for the algebra of two double cones is independent if the algebras of unbounded regions are defined to be
vN algebras or not. But, in order to relate this ambiguity, on the
choice of algebras, with the existence of SS, we must use the above
structural setup. The additivity property is motivated by the example
of the previous section. For example, for two spacelike separated
double cones $W_{1},W_{2}$ the vN algebra $\mathcal{A}\left(W_{1}\right)\vee\mathcal{A}\left(W_{2}\right)$
is the smallest vN algebra that we can assign to the region $W_{1}\lor W_{2}$
satisfying isotony, i.e. $\mathcal{A}\left(W_{1}\right),\mathcal{A}\left(W_{2}\right)\subset\mathcal{A}\left(W_{1}\vee W_{2}\right)$. Any other choice has to contain more operators, which are not locally
generated. The aim here is to clarify which are the other possible
choices. The duality condition is a kind of maximality condition for
the algebras of complementary regions. In this case, we assume that
such a condition holds for double cones, which are the only topological
trivial (connected and simply connected) causally complete regions.
The physical meaning is that, for topological trivial regions, any
operator can be defined locally. This assumption has been proved
to hold in many known concrete models of AQFT.

\subsubsection{Localized endomorphisms}

Now we discuss the consequences of definition \ref{c6-cri-dhr}.
The aim of this subsection is to obtain a clearer way to describe the DHR superselection
sectors.

First of all, we see that the vacuum representation $\pi_{0}\left(\mathcal{A}\right)$
has been identified with the defining representation $\mathcal{A}$.
In other words, $\pi_{0}\left(\mathcal{A}\left(\mathcal{O}\right)\right)\equiv\mathcal{A}\left(\mathcal{O}\right)$.
According to \eqref{c6-dhr-uni}, given a DHR representation $\pi$
and a double cone $W$, there exists a unitary operator $V_{W}:\mathcal{H}_{0}\rightarrow\mathcal{H}_{\pi}$
such that
\begin{equation}
\pi\left(A\right)=V_{W}AV_{W}^{\dagger}\,,\quad\forall A\in\mathcal{A}\left(W'\right)\,.\label{c6-dhr-unit}
\end{equation}
Now we define $\rho_{W}:\mathcal{A}\rightarrow\mathcal{B}\left(\mathcal{H}_{0}\right)$
by\footnote{Notice that $\rho_{W}$ not only depends on $W$ but also on the representation $\pi$.}
\begin{equation}
\rho_{W}\left(A\right):=V_{W}^{\dagger}\pi\left(A\right)V_{W}\,,\quad\forall A\in\mathcal{A}\, ,\label{c6-def-local-endo}
\end{equation}
which is also a representation of the algebra $\mathcal{A}$ (globally)
unitarily equivalent to $\pi$, an hence, also satisfies \eqref{c6-dhr-uni}.
Moreover, it can be shown that $\rho_{W}\in\mathrm{End}\left(\mathcal{A}\right)$,
which, in particular, implies that $\rho_{W}\left(A\right)\in\mathcal{A}$ for all $A\in\mathcal{A}$.
According to \eqref{c6-dhr-unit} and \eqref{c6-def-local-endo}, we have that
\begin{equation}
\rho_{W}\left(A\right)=A\,,\quad\forall A\in\mathcal{A}\left(W'\right)\,.\label{c6-local-endo}
\end{equation}
Any endomorphism satisfying \eqref{c6-local-endo} is called a \textit{localized
endomorphism} (with support in $W$).

Since the unitary equivalence \eqref{c6-dhr-unit} holds not for
one but for all double cones, we also have that for any other double
cone $\tilde{W}$ there exists a localized endomorphism $\rho_{\tilde{W}}\left(\cdot\right):=V_{\tilde{W}}^{\dagger}\pi\left(\cdot\right)V_{\tilde{W}}$
with support in $\tilde{W}$ and a unitary operator $\mathcal{I}_{W\tilde{W}}:=V_{\tilde{W}}^{\dagger}V_{W}\in\mathcal{A}$
such that 
\begin{equation}
\rho_{\tilde{W}}\left(\cdot\right)=\mathcal{I}_{W\tilde{W}}\rho_{W}\left(\cdot\right)\mathcal{I}_{W\tilde{W}}^{\dagger}\,.\label{c6-def-trans}
\end{equation}
Any localized endomorphism satisfying \eqref{c6-def-trans} is called
a \textit{localized transportable endomorphism}. The set of localized
transportable endomorphism of $\mathcal{A}$ is denoted by $\Gamma$,
and the subset of $\Gamma$ containing only irreducible endomorphisms
is denoted by $\Gamma_{\mathrm{irr}}$.
\begin{lem}
\label{c6-lem-corr1}For any DHR representation $\pi$ of the observable
algebra $\mathcal{A}$, there exists a transportable localized endomorphism
$\rho$ such that $\rho\cong\pi$. And certainly, any localized transportable
endomorphism can be considered as a representation of $\mathcal{A}$
satisfying \eqref{c6-local-endo}.
\end{lem}

In other words, the study of DHR representations can be carried out
considering localized transportable endomorphisms. The advantage of
localized transportable endomorphisms is that they are representations
implemented in the vacuum Hilbert space. An endomorphism $\rho\in\Gamma$
may be considered as the physical operation of adding a localized
charge over the vacuum state. In fact, the state $\omega_{\rho}:=\omega\circ\rho$
has the property $\omega_{\rho}\left(A\right)=\omega\left(A\right)$
for all $A\in\mathcal{A}\left(W'\right)$. Such a charge could be
also transported according to \eqref{c6-def-trans}. Another advantage
of considering transportable localized endomorphisms instead of general DHR representations,
is that they could be composed. In fact, if $\rho_{1},\rho_{2}\in\Gamma$
then $\rho_{1}\circ\rho_{2}\in\Gamma$. This has to be viewed as the
physical operation of adding charges on different spacetime regions.

Of course, the correspondence $\pi\mapsto\rho$ of lemma \ref{c6-lem-corr1}
is one to many. Moreover, since (globally) unitarily equivalent representations
describe the same physics, we must ``identify'' elements of $\Gamma$
coming from unitarily equivalent DHR representations. To
do that, we define the \textit{inner} or \textit{trivial} elements
of $\Gamma$ as
\begin{equation}
\mathcal{J}:=\left\{ \sigma\in\Gamma\,:\,\sigma\left(A\right)=UAU^{\dagger}\textrm{ with }U\in\mathcal{A}\left(W\right)\textrm{ unitary and }W\textrm{ a double cone}\right\} \, ,
\end{equation}
which is a proper subset of the inner automorphisms of $\mathcal{A}$.\footnote{An automorphism $\sigma \in \mathrm{Aut}(\mathcal{A})$ is called \textit{inner}, if there exists a unitary operator $U \in \mathcal{A}$ such that $\sigma(A)=UAU^{\dagger}$ for all $A \in \mathcal{A}$.} Two endomorphisms $\rho_{1},\rho_{2}\in\Gamma$ are defined to be
\textit{equivalent} $\rho_{1}\sim\rho_{2}$ iff there exists $\sigma\in\mathcal{J}$
such that $\rho_{2}=\sigma\circ\rho_{1}$. Then, we have the following
lemma.
\begin{lem}
Let $\pi_{1}$ and $\pi_{2}$ be (globally) unitarily equivalent DHR
representations, and let $\rho_{1}$ and $\rho_{2}\in\Gamma$ be its corresponding
transportable localized endomorphisms. Then $\rho_{1}\sim\rho_{2}$.
\end{lem}
\begin{proof}
See \cite{Doplicher:1969kp}.
\end{proof}
According to the above lemma, it is useful to consider the quotient
space $\Gamma/\mathcal{J}$. The elements of $\Gamma/\mathcal{J}$
are equivalence classes containing transportable localized endomorphism
which are related to each other by inner transportable localized endomorphism.
Then, it is not difficult to see that $\Gamma/\mathcal{J}$ is in
one-to-one correspondence with the set of all unitarily inequivalent
DHR representations. In other words, the set $\Gamma/\mathcal{J}$
contains exactly one copy of any unitarily inequivalent DHR representation.
Furthermore, according to definition \ref{c6-cri-dhr}, $\Gamma_{\mathrm{irr}}/\mathcal{J}$ is in one-to-one correspondence
with the DHR superselection sectors.

Many operations could be performed inside the set $\Gamma$. For example,
given any two endomorphisms $\rho_{1},\rho_{2}\in\Gamma$ there exists
$\rho\in\Gamma$ such that $\rho_{1}\oplus\rho_{2}$ is unitarily
equivalent (as a representation) to $\rho$.\footnote{Notice that $\rho_{1}\oplus\rho_{2}$ is a representation in $\mathcal{H}_{0}\oplus\mathcal{H}_{0}$,
whereas $\rho$ is a representation in $\mathcal{H}_{0}$.} We also have that, if $\mathcal{S}\subset\mathcal{H}_{0}$ is a proper
invariant subspace of $\rho\in\Gamma$, then there exists $\tilde{\rho}\in\Gamma$
such that $\left.\rho\right|_{\mathcal{S}}$ is unitarily equivalent
to $\tilde{\rho}$. These operations are the usual operations of taking
direct sum or restriction (to invariant subspaces) of representations.
Furthermore, as we have claimed above, $\Gamma$ naturally has a (semi)
product operation given by the composition of endomorphisms. In fact,
$\Gamma\subset\mathrm{End}\left(\mathcal{A}\right)$. From the mathematical
standpoint, the set $\Gamma$, equipped with the three operations above, defines a \textit{$C^{*}$-tensor category} \cite{Halvorson06}. All the operations are also well-defined in the quotient space $\Gamma/\mathcal{J}$.

Since DHR sectors are in one-to-one correspondence with $\Gamma_{\mathrm{irr}}/\mathcal{J}$,
any element of $\Gamma_{\mathrm{irr}}$ is called a DHR sector. Because of this, any element of $\Gamma_{\mathrm{irr}}$ is called an (irreducible) DHR sector. Moreover, according to the discussion above, any non-irreducible element of
$\Gamma$ could be formed by taking direct sums of irreducible ones. It is for this reason that, making an abuse of
language, any (generally reducible) transportable localized endomorphism is also called a DHR sector. Trivial transportable localized
endomorphisms correspond to the trivial (vacuum) sector.

\subsubsection{Intertwiners and failure of duality\label{c6-sec-Intertwiners-and-failure}}

For our purpose, we are interested in two endomorphism $\rho_{1},\rho_{2}\in\Gamma$
with supports in strictly spacelike separated double cones $W_{1}$ and $W_{2}$.
Any operator $\mathcal{I}_{12}\in\mathcal{B}\left(\mathcal{H}_{0}\right)$
satisfying 
\begin{equation}
\mathcal{I}_{12}\rho_{1}\left(A\right)=\rho_{2}\left(A\right)\mathcal{I}_{12}\,,\quad\forall A\in\mathcal{A}\,,\label{c6-def-int}
\end{equation}
is called an \textit{intertwiner}. If $\rho_{1}\sim\rho_{2}$, the
intertwiner is a unitary operator, but, in general, $\mathcal{I}_{12}$
is simply a partial isometry. When $\rho_{1}$ and $\rho_{2}$ are disjoint
(in particular, when they are unitarily inequivalent irreducible endomorphisms),
we have that $\mathcal{I}_{12}=0$. From \eqref{c6-def-int} we have that
\begin{equation}
\mathcal{I}_{12}A=A\mathcal{I}_{12}\,,\quad\forall A\in\mathcal{A}\left(W_{3}'\right)\,,\label{c6-int-alg}
\end{equation}
where $W_{3}$ is any double cone containing $W_{1}$ and $W_{2}$.
Then, by the duality condition for double cones, $\mathcal{I}_{12}\in\mathcal{A}\left(W_{3}'\right)'=\mathcal{A}\left(W_{3}\right)\subset\mathcal{A}$,
which means that the intertwiner belongs to the observable algebra. Furthermore, relation \eqref{c6-int-alg} could be strengthened
to $\mathcal{I}_{12}A=A\mathcal{I}_{12}$ for all $A\in\mathcal{A}\left(W_{1}'\right)\cap\mathcal{A}\left(W_{2}'\right)$,
which means 
\begin{equation}
\mathcal{I}_{12}\in\left(\mathcal{A}\left(W_{1}'\right)\cap\mathcal{A}\left(W_{2}'\right)\right)'\,.\label{c6-int-reg}
\end{equation}
Moreover, because of isotony, we have that $\mathcal{A}\left(\left(W_{1}\vee W_{2}\right)'\right)\subset\mathcal{A}\left(W_{1}'\right),\mathcal{A}\left(W_{2}'\right)$
which implies $\mathcal{A}\left(\left(W_{1}\vee W_{2}\right)'\right)\subset\mathcal{A}\left(W_{1}'\right)\cap\mathcal{A}\left(W_{2}'\right)$.
Hence, equation \eqref{c6-int-reg} implies
\begin{equation}
\mathcal{I}_{12}\in\mathcal{A}\left(\left(W_{1}\vee W_{2}\right)'\right)'\,.
\end{equation}
Formula \eqref{c6-int-reg} suggests that the existence of intertwiners
may contribute to the failure of the duality condition for two double cones. In fact, we have shown that
\begin{equation}
\mathcal{A}\left(W_{1}\vee W_{2}\right)\subset\mathcal{A}\left(W_{1}\vee W_{2}\right)\lor\left\{ \mathcal{I}_{12}\right\} \subset\mathcal{A}\left(\left(W_{1}\vee W_{2}\right)'\right)'\,,
\end{equation}
where $\left\{ \mathcal{I}_{12}\right\}$ denotes the set of all
intertwiners between $W_{1}$ and $W_{2}$. It is important to remark
that $\left\{ \mathcal{I}_{12}\right\}$ includes all the intertwiners
between all DHR representations, despite they are reducible or irreducible.
However, because we are interested in the algebra generated by such
intertwiners plus the additive algebra $\mathcal{A}_{W_{1}\vee W_{2}}$,
we need only a ``basis'' of intertwiners. We will show in section
\ref{c6-sec-int-twi} that such a basis could be constructed using
only the irreducible sectors $\rho\in\Gamma_{\mathrm{irr}}$. A precise
knowledge of the general structure of the set $\left\{ \mathcal{I}_{12}\right\} $
will be remarkably useful to compute entanglement measures on the
algebra $\mathcal{A}\left(\left(W_{1}\vee W_{2}\right)'\right)'$.

Unitary intertwiners can be used to construct localized endomorphisms.
Let $\rho\in\Gamma$ with support in $W$ and take a sequence of equivalent
morphisms $\rho_{\lambda}\in\Gamma$ ($\lambda\in\mathbb{R}$) whose
respective localization regions move to infinity, so as to become
eventually spacelike to any given double cone. In other words, we
can consider that $\rho_{\lambda}$ has support in $W_{\lambda}:=W+\lambda a$,
where $a\in\mathbb{R}^{d}$ is a fixed spacelike vector. Then, there exists
a corresponding sequence of unitary intertwiners $\mathcal{I}_{\lambda}\in\mathcal{A}\left(\left(W\vee\left(W+\lambda a\right)\right)'\right)'$
such that
\begin{equation}
A=\mathcal{I}_{\lambda}\rho\left(A\right)\mathcal{I}_{\lambda}^{\dagger}\,,\quad\forall A\in\mathcal{A}\left((W+\lambda a)'\right)\,.\label{c6-end-inf}
\end{equation}
In the limit when $\lambda\rightarrow+\infty$ we have that \eqref{c6-end-inf}
holds for all $A\in\mathcal{A}\left(D\right)$ where $D\in\mathcal{K}$
is any bounded region. Since any element of global algebra $\mathcal{A}$
is the uniform limit of elements belonging to algebras of bounded
regions, we expect that \eqref{c6-end-inf} holds for all $A\in\mathcal{A}$
in the limit when $\lambda\rightarrow+\infty$. More precisely, it
can be shown that \cite{Doplicher:1971wk}
\begin{equation}
\rho\left(A\right)=\lim_{\lambda\rightarrow+\infty}\mathcal{I}_{\lambda}^{\dagger}A\mathcal{I}_{\lambda}\,,\label{c6-end-lim}
\end{equation}
where this limit has to be understood in the strong operator topology.
In other words, the intertwiner operator could be used to create the
localized endomorphism. Essentially, the intertwiner changes the position
of the charge, and then, \eqref{c6-end-lim} represents the physical act of bringing the charge from infinity.

Let $\sigma_{i}\in\mathcal{J}$ with $\sigma_{i}\left(A\right):=U_{i}AU_{i}^{\dagger}$
with unitaries $U_{i}\in\mathcal{A}\left(W_{i}\right)$. In this case,
we have that the unitary operator $\mathcal{I}_{12}:=U_{2}U_{1}^{\dagger}$
satisfies
\begin{equation}
\mathcal{I}_{12}\sigma_{1}\left(A\right)=\sigma_{2}\left(A\right)\mathcal{I}_{12}\,,\quad\forall A\in\mathcal{A}\,,
\end{equation}
i.e. it is an intertwiner. In this case, we have that $\mathcal{I}_{12}\in\mathcal{A}\left(W_{1}\vee W_{2}\right)$
and we say that $\mathcal{I}_{12}$ is a \textit{trivial} intertwiner.
This happens because $\mathcal{I}_{12}$ intertwines between inner
endomorphisms both representing the vacuum sector. However, we expect
that intertwiners corresponding to non-trivial sectors do not belong
to the additive algebra $\mathcal{A}\left(\left(W_{1}\vee W_{2}\right)\right)$,
and they produce an explicit breakdown of the duality condition for
two double cones. In fact, if we assume that an intertwiner $\mathcal{I}_{12}\in\mathcal{A}\left(W_{1}\vee W_{2}\right)$
can be written as $\mathcal{I}_{12}:=U_{2}U_{1}^{\dagger}$ with unitaries
$U_{j}\in\mathcal{A}\left(W_{j}\right)$, then it is not difficult
to show that the endomorphisms $\rho_{j}$ in \eqref{c6-def-int}
are trivial, $\rho_{j}\left(A\right)=U_{j}AU_{j}^{\dagger}$. In other
words, we expect that any non-trivial DHR sector gives place to a non-additive
intertwiner.

Conversely, we expect that any operator $\mathcal{I}_{12}\in\mathcal{A}\left(\left(W_{1}\vee W_{2}\right)'\right)'$
but $\mathcal{I}_{12}\notin\mathcal{A}\left(\left(W_{1}\vee W_{2}\right)\right)$
should correspond to an intertwiner of a non-trivial DHR sector. Despite
there is no general proof of this statement, we can argue in favor
of that as follows. Starting with such an operator $\mathcal{I}_{12}$,
we may expect that we can use it to construct a family of operators
$\mathcal{I}_{\lambda}\in\mathcal{A}\left(\left(W_{1}\vee\left(W_{2}+\lambda a\right)\right)'\right)'$
which translates the charge to $W_{2}+\lambda a$, such that in the
limit $\lambda\rightarrow+\infty$ gives place to localized endomorphism
according to \eqref{c6-end-lim}. Furthermore, it is expected that
such a localized endomorphism is non-trivial whenever $\mathcal{I}_{12}$
is non-trivial, i.e. $\mathcal{I}_{12}\notin\mathcal{A}\left(\left(W_{1}\vee W_{2}\right)\right)$.

In other words, it is expected that
\begin{equation}
\mathcal{A}\left(\left(W_{1}\vee W_{2}\right)'\right)'=\mathcal{A}\left(W_{1}\vee W_{2}\right)\lor\left\{ \textrm{DHR intertwiners between }W_{1}\textrm{ and }W_{2}\right\} \,,
\end{equation}
and intertwiners corresponding to non-trivial DHR sectors does not
belong to the additive algebra $\mathcal{A}\left(\left(W_{1}\vee W_{2}\right)\right)$.

\paragraph{Example: bosonic subnet of the free fermion field\protect \\
}

Now, we continue with the example introduced above. The Fock Hilbert space of
the free fermion is decomposed as $\mathcal{H}=\mathcal{H}_{0}\oplus\mathcal{H}_{1}$,
where $\mathcal{H}_{0}$ (resp. $\mathcal{H}_{1}$) is formed by all
the vectors with an even (resp. odd) number of fermions. In particular,
we recall that the vacuum vector $\left|0\right\rangle $ belongs
$\mathcal{H}_{0}$. Moreover, the observable algebra $\mathcal{A}$
leaves both subspaces $\mathcal{H}_{0},\,\mathcal{H}_{1}$ invariant.
In this way, we can identify two representations
\begin{eqnarray}
\left(\pi_{0},\mathcal{H}_{0}\right): &  & \pi_{0}\left(A\right)\left|\psi_{0}\right\rangle :=A\left|\psi_{0}\right\rangle \,,\quad A\in\mathcal{A}\textrm{ and }\left|\psi_{0}\right\rangle \in\mathcal{H}_{0}\,,\\
\left(\pi_{1},\mathcal{H}_{1}\right): &  & \pi_{1}\left(A\right)\left|\psi_{1}\right\rangle :=A\left|\psi_{1}\right\rangle \,,\quad A\in\mathcal{A}\textrm{ and }\left|\psi_{1}\right\rangle \in\mathcal{H}_{1}\,.
\end{eqnarray}
Despite both representations has the same assignation rule $\pi_{k}\left(A\right)=A$
($k=1,2$), they are not the same representation since they act in
different Hilbert spaces. Moreover, both representations are not unitarily
equivalent. This is because both are covariant respect to unitarily
inequivalent representations of the Poincaré group. In fact, the one
acting on $\mathcal{H}_{0}$ is a vacuum representation, but the one
acting on $\mathcal{H}_{1}$ cannot be a vacuum representation since
it has not a Poincaré invariant vector.

Let $W$ be a double cone, and consider the fermionic operator
\begin{equation}
V_{W}:=\psi^{\dagger}\left(f\right)+\psi\left(f\right)\,,\label{siste}
\end{equation}
where $\sup\left(f\right)\subset W$. This operator does not belong
to the observable algebra $\mathcal{A}$, but it belongs to the fermionic
algebra $\mathfrak{F}$. Moreover, $V_{W}\in\mathfrak{F}\left(W\right)$.
In addition, we choose $f\left(x\right):=\alpha\left(\bar{x}\right)\delta\left(x^{0}\right)$
with $\int d^{d-1}x\,\alpha(\bar{x})^{\dagger}\alpha(\bar{x}):=1$.\footnote{Here we are smearing in space only. This can be done for a free field (see footnote \ref{fn_fer_t}). We could have also chosen spacetime smearing functions as well,
at the expense of replacing $\int d^{d-1}x\,\alpha(\bar{x})^{\dagger}\alpha(\bar{x})=1$
by an integral in $d$ dimensions weighted with the anticommutator
distribution \label{fer_anti}.} Using the anticommutation relations $\{\psi(0,\bar{x}),\psi^{\dagger}(0,\bar{y})\}=\delta(\bar{x}-\bar{y})$, it can be shown that $V_{W}^{\dagger}=V_{W}=V_{W}^{-1}$. We also have
that $V_{W}\mathcal{H}_{0}=\mathcal{H}_{1}$ and $V_{W}\mathcal{H}_{1}=\mathcal{H}_{0}$.
Then, we can consider the operator $V_{W}$ as acting\footnote{In other words, we take the restriction of $V_{W}$ to the vacuum
subspace $\mathcal{H}_{0}$.}
\begin{equation}
V_{W}:\mathcal{H}_{0}\rightarrow\mathcal{H}_{1}\,.
\end{equation}
In this way, $V_{W}$ is a unitary operator from $\mathcal{H}_{0}$
onto $\mathcal{H}_{1}$, and $V_{W}^{\dagger}:\mathcal{H}_{1}\rightarrow\mathcal{H}_{0}$.
It is not hard to see that
\begin{equation}
V_{W}^{\dagger}\pi_{1}\left(A\right)V_{W}=\pi_{0}\left(A\right)\,,\quad\forall A\in\mathcal{A}\left(W'\right)\,.\label{c6-ferm-dhr}
\end{equation}
Moreover, since the choice of the double cone $W$ in \eqref{siste}
is arbitrary, we also have that relation \eqref{c6-ferm-dhr} holds for
any double cone. This shows that $\pi_{1}$ is a DHR representation.
The localized endomorphism corresponding to $\pi_{1}$ is defined as
\begin{equation}
\rho_{W}\left(A\right):=V_{W}^{\dagger}\pi_{1}\left(A\right)V_{W}\,,\label{c6-ferm-endo}
\end{equation}
We have that $\rho_{W}\left(A\right)\mathcal{H}_{0}\subset\mathcal{H}_{0}$
for all $A\in\mathcal{A}$, and moreover,
\begin{equation}
\rho_{W}\left(A\right)\subset\pi_{0}\left(\mathcal{A}\right)\subset\mathcal{B}\left(\mathcal{H}_{0}\right)\,,\quad\forall A\in\mathcal{A}\,.
\end{equation}
Identifying the observable algebra $\mathcal{A}$ with the vacuum
representation $\pi_{0}\left(\mathcal{A}\right)$, expressions \eqref{c6-ferm-dhr}
and \eqref{c6-ferm-endo} imply that
\begin{equation}
\rho_{W}\left(A\right)=A\,,\quad\forall A\in\mathcal{A}\left(W'\right)\,.
\end{equation}
Now, we can do the same construction for two strictly spacelike separated double
cones $W_{1}$ and $W_{2}$. Then, we have fermionic unitary operators
$V_{W_{1}},V_{W_{2}}$ and localized endomorphisms $\rho_{1},\rho_{2}$,
according to \eqref{c6-ferm-endo}. The operator $\mathcal{I}_{12}:=V_{W_{2}}^{\dagger}V_{W_{1}}$
satisfies
\begin{equation}
\mathcal{I}_{12}\rho_{1}\left(\cdot\right)=\rho_{2}\left(\cdot\right)\mathcal{I}_{12}\,,
\end{equation}
i.e. it is an intertwiner between both endomorphisms. This intertwiner
belongs to the observable algebra $\mathcal{A}$, it is made out of
products of fermionic operators, one belonging to $\mathfrak{F}\left(W_{1}\right)$
and the other one to $\mathfrak{F}\left(W_{2}\right)$, but the product
$\mathcal{I}_{12}$ does not belong to the additive algebra $\mathcal{A}\left(W_{1}\vee W_{2}\right)$.
This is in accordance with what we have anticipated above. In this
way, we have shown explicitly, in this example, that the DHR intertwiners
are the operators which break down the duality condition for two double cones.

In this model, there is no another non-equivalent DHR representation beyond $\pi_{0}$
and $\pi_{1}$. There is no non-zero intertwiner between $\rho_{W}$ and the
``vacuum'' endomorphism, which is defined as $\iota\left(A\right):=A$ for
all $A\in\mathcal{A}$. This is because such representations are disjoint, i.e.,
they correspond to non-equivalent irreducible sectors. Non-irreducible
sectors can be formed taking direct sums of irreducible ones, and
their corresponding intertwiners could be generated taking products
and sums of irreducible intertwiners and operators of the additive
algebra $\mathcal{A}\left(W_{1}\vee W_{2}\right)$. In other words,
for the purpose of the algebra $\mathcal{A}\left(W_{1}\vee W_{2}\right)\vee\left\{ \mathcal{I}_{12}\right\} $,
we lose nothing by only considering irreducible intertwiners. The
operator $V_{W}$ is called a \textit{charge creating operator}. It
transforms vectors belonging to different sectors, but it does not belong to the
algebra $\mathcal{A}$.

We notice in this example, that there exists a local QFT $\mathfrak{F}\left(\mathcal{O}\right)$ containing exactly
one copy of each irreducible sector. In fact, the fermionic net $\mathfrak{F}\left(\mathcal{O}\right)$
is an AQFT satisfying all the axioms of the definition \ref{c1_def_fields}.
This net is covariant respect to the universal covering of the Poincaré
group. The observable net is the subnet formed by the bosonic elements.
The operator $\Gamma \in \mathcal{B}(\mathcal{H})$ defined by\footnote{Do not confuse this operator with the set of localized transportable endomorphisms of $\mathcal{A}$.}
\begin{eqnarray}
\Gamma\left|\psi_{0}\right\rangle :=\left|\psi_{0}\right\rangle \,, &  & \left|\psi_{0}\right\rangle \in\mathcal{H}_{0}\,,\\
\Gamma\left|\psi_{1}\right\rangle :=-\left|\psi_{1}\right\rangle \,, &  & \left|\psi_{1}\right\rangle \in\mathcal{H}_{1}\,,
\end{eqnarray}
forms a group of unitaries $G:=\left\{ \mathbf{1},\Gamma\right\} $ isomorphic to $\mathbb{Z}_{2}$.\footnote{In this case, $\Gamma$ is also the grading of the field algebra (see section \ref{c1-sec-fermions}).}
An operator $A\in\mathfrak{F}$ belongs to the observable algebra
if and only if it is $G$-invariant, i.e. $A\in G'.$ This enveloping
AQFT is named the ``field algebra'', in contrast to $\mathcal{A}$, which is called the observable algebra.\footnote{As we have anticipated in section \ref{c1-sec-fermions}, the field algebra may contain non-observable quantities which they do not commute at spacelike distance.}
Remarkably, this structure is quite general. Given any observable
AQFT $\mathcal{A}$, it always exists a field algebra $\mathfrak{F}$
containing copies of all DHR sectors, and a group of unitaries $G$
such that the observable algebra is formed by the $G$-invariant elements
of $\mathfrak{F}$. Moreover, the field algebra contains no further
SS.

\subsection{Field algebra\label{c6-secc-field}}

A \textit{field algebra} consists in a net of $C^{*}$-algebras $\mathcal{O}\in\mathcal{K} \mapsto \mathfrak{F}\left(\mathcal{O}\right)\subset\mathcal{B}\left(\mathcal{H}\right)$
satisfying all the axioms of definition \ref{c1_def_fields}, and
a compact group $G$ of unitaries, called the \textit{symmetry group},\footnote{In the literature of AQFT, $G$ is called the gauge group regardless it represents a group of global symmetries.} such that the observable algebra is defined as the $G$-invariant
part of $\mathfrak{F}$, i.e
\begin{equation}
\mathcal{A}:=\mathfrak{F}\cap G'\,.\label{c6-field-obs}
\end{equation}
Here, we again define the algebras of bounded regions as vN algebras,
whereas the algebras of the unbounded regions are just $C^{*}$ but
not vN algebras. The field algebra is assumed to satisfy twisted duality
for double cones. i.e ${\mathfrak{F}\left(W'\right)^{t}}\sp{\prime}=\mathfrak{F}\left(W\right)$.

The symmetry group must commute with the Poincaré unitaries\footnote{From now on, $U$ denotes the representation of the symmetry group
whereas $U_{P}$ denotes the representation of the Poincaré group.}
\begin{equation}
\left[U_{G}\left(g\right),U_{P}\left(l\right)\right]=0\,,\quad\forall g\in G,\,\forall l\in\tilde{\mathcal{P}}_{+}^{\uparrow}\,.
\end{equation}
Because of this, the elements of $G$ preserve the local algebras
\begin{equation}
\alpha_{g}\left(\mathfrak{F}\left(\mathcal{O}\right)\right):=U\left(g\right)\mathfrak{F}\left(\mathcal{O}\right)U\left(g\right)^{\dagger}=\mathfrak{F}\left(\mathcal{O}\right)\,,\quad\forall\mathcal{O}\in\mathcal{K}\,,\forall g\in G\,,\label{c6-acc-G}
\end{equation}
i.e., they represent \textit{internal symmetries}. Moreover, the Poincaré unitary $U_{P}\left(z\right)$ corresponding
to the element $z:=\left(0,-\mathbf{1}\right)\in\mathbb{R}^{d}\ltimes\tilde{\mathcal{L}}_{+}^{\uparrow}$,
which is the grading of the field algebra, must belong also to the symmetry group $G$. Then, the observable algebra $\mathcal{A}$ does not contain fermionic operators.

The local observable algebras are defined as
\begin{equation}
\mathcal{A}\left(W\right):=\mathfrak{F}\left(W\right)\cap G'\,,\quad\textrm{ if }W\textrm{ is a double cone ,}
\end{equation}
and the algebras for non-double cones are defined in order to satisfy
additivity. It can be shown that $\mathfrak{F}\left(\mathcal{O}\right)\subset\mathcal{A}\left(\mathcal{O}'\right)'$.
The observable net $\mathcal{O}\in\mathcal{K} \mapsto \mathcal{A}\left(\mathcal{O}\right)\subset\mathcal{B}\left(\mathcal{H}\right)$
satisfies all the axioms of definition \ref{c1_def_aqft}.

It is important to emphasize that the representation of the observable
algebra $\mathcal{A}$ on the Hilbert space $\mathcal{H}$ is highly
reducible. In fact, it can be shown that the Hilbert space $\mathcal{H}$
decomposes as
\begin{equation}
\mathcal{H}:=\bigoplus_{\sigma\in\hat{G}}\mathbb{C}^{d_{\sigma}}\otimes\mathcal{H}_{\sigma}\,,\label{c6-h-dec}
\end{equation}
where the direct sum runs over the spectrum $\hat{G}$ of $G$, i.e.
the set of all non-equivalent unitary irreducible representations.
All group representations are finite dimensional since $G$ is compact,
and $d_{\sigma}<\infty$ denotes their dimensions. The observable
algebra $\mathcal{A}$ and the symmetry group reduce according to
the decomposition \eqref{c6-h-dec} as
\begin{eqnarray}
U\left(g\right)\!\!\! & = & \!\!\!\bigoplus_{\sigma\in\hat{G}}U_{\sigma}\left(g\right)\otimes\mathbf{1}_{\mathcal{H}_{\sigma}}\,,\\
\mathcal{A}\ni A\!\!\! & = & \!\!\!\bigoplus_{\sigma\in\hat{G}}\mathbf{1}_{d_{\sigma}}\otimes\pi_{\sigma}\left(A\right)\,.
\end{eqnarray}
$U_{\sigma}$ is the unitary irreducible representation of $G$ of
type $\sigma\in\hat{G}$. $\pi_{\sigma}:\mathcal{A}\rightarrow\mathcal{B}\left(\mathcal{H}_{\sigma}\right)$
are inequivalent irreducible representations of the observable algebra
$\mathcal{A}$ \cite{Doplicher:1969tk}. Each of them gives place
to different AQFTs
\begin{eqnarray}
\mathcal{O}\in\mathcal{K} & \mapsto & \mathcal{A_{\sigma}}\left(\mathcal{O}\right):=\pi_{\sigma}\left(\mathcal{A_{\sigma}}\left(\mathcal{O}\right)\right)\subset\mathcal{B}\left(\mathcal{H}_{\sigma}\right)\,,\label{c6-falg-rep}
\end{eqnarray}
satisfying all the axioms of definition \ref{c1_def_aqft}. Moreover,
the vacuum vector $\left|0\right\rangle \in\mathcal{H}$ belongs to
the Hilbert space $\mathcal{H}_{0}:=\mathcal{H}_{\mathrm{triv}}$
corresponding to the trivial representation of $G$, and its corresponding
observable algebra representation $\mathcal{A}_{0}\left(\mathcal{O}\right):=\mathcal{A}_{\mathrm{triv}}\left(\mathcal{O}\right)$
is a vacuum representation, i.e. it satisfies also all the axioms
of definition \ref{c1_def_vacuum}.

It is important to remark that all the representations in \eqref{c6-falg-rep}
satisfy the DHR criteria (definition \ref{c6-dhr-uni}) respect to
the vacuum representation $\pi_{0}:=\pi_{\mathrm{triv}}$. Moreover,
the localized endomorphism $\rho_{W,\sigma}\in\mathrm{End}\left(\mathcal{A}\right)$
corresponding to a one-dimensional representation can be always constructed
as in \eqref{c6-ferm-endo}
\begin{equation}
\rho_{W,\sigma}\left(A\right):=V_{W,\sigma}^{\dagger}AV_{W,\sigma}\,,\label{c6-field-endo}
\end{equation}
where the charge creating operators $V_{W,\sigma}\in\mathfrak{F}$
but $V_{W,\sigma}\notin\mathcal{A}$. For a general irreducible representation,
a slightly modification of relation \eqref{c6-field-endo} holds (see
subsection \ref{c6-sec-int-twi} below). Non-irreducible representations
could be constructed in similar fashion invoking property B (see \cite{araki,haag,Halvorson06}).\footnote{Property B is a fundamental property concerning the local algebras
of QFT which is expected to hold in general because it follows from
the physically motivated postulates of causality, spectrum condition
and weak additivity \cite{borchers1967,DAntoni:1989aih,Halvorson06}.} 

In conclusion, starting with a field algebra $\mathfrak{F}$ with
a compact symmetry group $G$, and defining the observable algebra
$\mathcal{A}$ as the $G$-invariant part of $\mathfrak{F}$, then,
the observable algebra decomposes into irreducible DHR representations.
The vacuum representation appears only once in the above decomposition.
As we did above, it is useful to identify the observable algebra with
the vacuum representation, which by the way, they are $C^{*}$-isomorphic.
In fact, for the purpose of computing information/entanglement measures
in the vacuum state (or in any another normal state of $\mathcal{A}_{0}$), there is no difference in using $\mathcal{A}$
or $\mathcal{A}_{0}$. The GNS-representation of $\mathcal{A}$ and
the vacuum state $\omega\left(\cdot\right)=\left\langle 0\right|\cdot\left|0\right\rangle $
is unitarily equivalent to $\left(\mathcal{A}_{0},\mathcal{H}_{0}\right)$.

The most remarkable part of this story is that the above procedure 
can be reversed. If we start with an observable algebra $\mathcal{A}_{0}$
in the vacuum representation, it always exists a unique (up to net isomorphisms) field algebra $\mathfrak{F}$
and a compact symmetry group $G$, such that the $G$-invariant part
of $\mathfrak{F}$ contains copies of all DHR representations of the
observable algebra $\mathcal{A}_{0}$. This is the famous Doplicher-Roberts
reconstruction theorem \cite{Doplicher:1969kp,Doplicher:1990pn}.\footnote{This could be done for spacetime dimensions greater than $2$. For
$d=1,2$ the DHR SS may not come from a compact group, and a slight
modification is required.} In this way, we do not lose any generality assuming that the observable
algebra is embedded in bigger algebra containing copies of all DHR SS, as we
did for the free fermion field. According to our discussion in section
\ref{c6-sec-Intertwiners-and-failure}, we expect that the field algebra
satisfies (twisted) duality for two double cones, since it has no further
DHR SS.

\subsection{Intertwiners and twists\label{c6-sec-int-twi}}

Once we have recognized that the DHR intertwiners are the responsible
of the failure of the duality for two double cones, we need to better
understand their algebraic properties in order to be able to use them for
entanglement measures computations. In the end, we want to construct
an entropic order parameter which ``measures'' the difference between
the algebras $\mathcal{A}\left(W_{1}\vee W_{2}\right)$ and $\mathcal{A}\left(\left(W_{1}\vee W_{2}\right)'\right)'=\mathcal{A}\left(\left(W_{1}\vee W_{2}\right)\right)\vee\left\{ \mathcal{I}_{12}\right\} $.

With this in mind, we start the structural assumptions of the previous section.
The observable algebra $\mathcal{A}$ is defined as the $G$-invariant
part of a theory without superselection sectors. This later is field
algebra $\mathfrak{F}$. The neutral charge sector (vacuum representation) $\mathcal{A}_{0}$
is embedded in $\mathcal{A}$. $\mathcal{A}$ and $\mathcal{A}_{0}$
are $C^{*}$-isomorphic, and they are frequently identified.

In similar way as in relation \eqref{c6-field-endo}, a localized endomorphism
associated with a generic (not necessarily irreducible) representation
$\xi$ of $G$ localized in a double cone $W$ can be written by means of the
\textit{charge creating operators} $V_{W,\xi}\in\mathfrak{F}\left(W\right)$.
It is possible to choose a family of these operators $\left\{ V_{\xi}^{j}\,:,\,j=1,\ldots,d_{\xi}\right\} \subset\mathfrak{F}\left(W\right)$
such that they transform according the unitary representation
$\xi$ of the symmetry group\footnote{We do not write the subscript $W$ in order to lighten the notation.}
\begin{equation}
U(g)^{\dagger}\,V_{\xi}^{j}\,U(g)=D_{\xi}(g)_{k}^{j}V_{\xi}^{k}\,,\quad g\in G\,,\label{beca}
\end{equation}
where $D_{\xi}(g)_{k}^{j}$ is the representation matrix. To obtain
an endomorphism of ${\cal A}$ associated with the representation
$\xi$, we define 
\begin{equation}
A\in\mathcal{A}\mapsto\rho_{\xi}(A):=\sum_{j=1}^{d_{\xi}}V_{\xi}^{j}\,A\,V_{\xi}^{j\,\dagger}\,.\label{c6-endo-charge}
\end{equation}
Then, we have that $\rho_{\xi}\left(A\right)\in\mathcal{A}$ because of \eqref{beca}.
In order to \eqref{c6-endo-charge} defines an endomorphism (respects the product of operators 
and maps the identity into itself), we need additionally\footnote{In particular $\rho_{\xi}$ is a completely positive map.}
\begin{eqnarray}
V_{\xi}^{j\,\dagger}V_{\xi}^{k}\!\!\! & = & \!\!\!\delta_{jk}\mathbf{1}_{\mathcal{H}}\,,\label{s}\\
\sum_{j=1}^{d_{\xi}}V_{\xi}^{j}V_{\xi}^{j\,\dagger}\!\!\! & = & \!\!\!\mathbf{1}_{\mathcal{H}}\,,\label{t}
\end{eqnarray}
such that $V_{\xi}^{j}$ are partial isometries. For the case
of one dimensional (irreducible) representations, $V_{\xi}^{j}$ is
unitary, like in the example of the previous sections.

Acting with these operators in a vector $|\psi_{0}\rangle\in\mathcal{H}_{0}$,
we obtain vectors $|\psi_{\xi}^{j}\rangle=V_{\xi}^{j\,\dagger}|\psi_{0}\rangle$
transforming according to the representation $\xi$, i.e. $U(g)|\psi_{\xi}^{j}\rangle=D_{\xi}(g)_{k}^{j}|\psi_{\xi}^{k}\rangle$.
In this way, acting with these operators in the vacuum sector $\mathcal{H}_{0}$
we are able to construct any vector in the Hilbert space $\mathcal{H}$,
whenever $\xi$ contains copies of all irreducible representations.
Furthermore, any operator $F\in\mathfrak{F}$ can be written as
\begin{equation}
F:=\sum_{\xi}\sum_{j=1}^{d_{\xi}}A_{\xi,j}\,V_{\xi}^{j}\,,\label{esis}
\end{equation}
with $A_{\xi,j}\in{\cal A}$, where the representation $\xi$ and
the charge creating operators $V_{\xi}^{j}$ have to be appropriately
chosen depending on $F$. To write \eqref{esis}, we have assumes that every
irreducible representation appears at least one representation
in the above decomposition. 

This endomorphism arises when considering the GNS-representation of
the vector state
\begin{equation}
|\psi_{\sigma}^{j}\rangle:=\sqrt{d_{\sigma}}V_{\sigma}^{j\,\dagger}|0\rangle\,,\label{dietrich}
\end{equation}
corresponding to the irreducible representation $\sigma\in\hat{G}$.
The factor $\sqrt{d_{\sigma}}$ is necessary to have a properly normalized
state, since\footnote{In the case of infinite compact group, we may replace $\frac{1}{|G|}\sum_{g\in G}$
by $\int_{G}d\mu\left(g\right)$ where $\mu\left(g\right)$ is the
normalized Haar measure of $G$.} 
\begin{eqnarray}
\langle0|V_{\sigma}^{j}V_{\sigma}^{j\,\dagger}|0\rangle\!\!\! & = & \!\!\!\frac{1}{|G|}\sum_{g\in G}\langle0|U(g)V_{\sigma}^{j}U(g)^{\dagger}U(g)V_{\sigma}^{j\,\dagger}U(g)^{\dagger}|0\rangle\nonumber \\
 & = & \!\!\!\frac{1}{|G|}\sum_{g\in G}\sum_{j,l=1}^{d_{\sigma}}D_{\sigma}(g)_{k}^{j}D_{\sigma}(g)_{l}^{j\,*}\langle0|V_{\sigma}^{k}V_{\sigma}^{l\,\dagger}|0\rangle=\frac{1}{d_{\sigma}}\langle0|\sum_{k=1}^{d_{\sigma}}V_{\sigma}^{k}V_{\sigma}^{k\,\dagger}|0\rangle=\frac{1}{d_{\sigma}}\,, \hspace{1cm}
\end{eqnarray}
where we have used the orthogonality relation between irreducible representations,
\begin{equation}
\sum_{g\in G}D_{\sigma_{1}}(g)_{k_{1}}^{j_{1}}D_{\sigma_{2}}(g)_{k_{2}}^{j_{2}\,*}(g)=\frac{|G|}{d_{\sigma_{1}}}\delta_{\sigma_{1}\sigma_{2}}\delta_{j_{1}j_{2}}\delta_{k_{1}k_{2}}\,.
\end{equation}
A same type of algebraic manipulations shows that for any element $A\in\mathcal{A}$ 
\begin{equation}
\langle\psi_{\sigma}^{j}|A|\psi_{\sigma}^{j}\rangle=\langle0|\rho_{\sigma}(A)|0\rangle=\omega\circ\rho_{\sigma}(A)\,,
\end{equation}
where $\omega$ is the vacuum state and $\rho_{\sigma}$ is as in
\eqref{c6-endo-charge}. In other words, the GNS-representation corresponding
to the vector state \eqref{dietrich} is unitarily equivalent to $\rho_{\sigma}$.

The operators $V_{\xi}^{i}$ equipped with the relations \eqref{s} and \eqref{t}
generates what is called a \textit{Cuntz algebra}. This cannot be represented
in finite dimensions. However, the algebra of operators of the form
\cite{Longo:1989tt} 
\begin{equation}
A:=\sum_{j,k=1}^{d_{\xi}}A_{jk}V_{\xi}^{j}V_{\xi}^{k\,\dagger}\,,\quad A_{jk}\in\mathbb{C}\,,
\end{equation}
closes with respect to the matrix multiplication of the coefficients
\begin{equation}
A\cdot B=\sum_{j,k=1}^{d_{\xi}}\left(\sum_{l=1}^{d_{\xi}}A_{jl}B_{lk}\right)V_{\xi}^{j}V_{\xi}^{k\,\dagger}\,.
\end{equation}
Hence, it is a finite subalgebra of the Cuntz isomorphic to $M_{d_{\sigma}}\left(\mathbb{C}\right)$.

If $V_{1}^{j},V_{2}^{j}\in\mathfrak{F}$ are charge creating operators
localized in two double cones $W_{1}\text{\Large\ensuremath{\times\!\negmedspace\!\times}}W_{2}$
associated with the same representation $\xi$ of $G$, an intertwiner
corresponding to this sector can be written as
\begin{equation}
{\cal I}_{12}^{\xi}:=\sum_{j=1}^{d_{\xi}}V_{\xi,W_{1}}^{j}V_{\xi,W_{2}}^{j\,\dagger}\,.\label{c6-uni-int}
\end{equation}
It follows from \eqref{s} and \eqref{t} that ${\cal I}_{12}$ is
unitary. This belongs to observable algebra ${\cal A}$ because it
is invariant under the symmetry group. Moreover, it commutes will all the operators
localized outside the two double cones, but it is not generated by ${\cal A}_{W_{1}}$
and ${\cal A}_{W_{2}}$. Therefore, duality for two double cones does not hold in ${\cal A}$.

The unitary intertwiners \eqref{c6-uni-int} corresponding to the irreducible representations $\sigma\in\hat{G}$ could be chosen in order they satisfy the character algebra or fusion algebra, i.e.
\begin{equation}
{\cal I}_{12}^{\sigma_{1}}{\cal I}_{12}^{\sigma_{2}}=\sum_{\sigma_{3}\in\hat{G}}\eta_{\sigma_{1}\sigma_{2}}^{\sigma_{3}}{\cal I}_{12}^{\sigma_{3}}\,,
\end{equation}
where $\eta_{\sigma_{1}\sigma_{2}}^{\sigma_{3}}\in\mathbb{N}_{0}$ is the times that the representation $\sigma_{3}$ is included in the decomposition of $\sigma_{1}\otimes\sigma_{2}$.\footnote{$\eta_{\sigma_{1}\sigma_{2}}^{\sigma_{3}}=\langle\chi_{\sigma_{3}}\mid\chi_{\sigma_{1}\otimes\sigma_{2}}\rangle_{G}$ where $\chi_{\sigma}$ is the character of the representation $\sigma$ and $\langle\cdot\mid\cdot\rangle_{G}$ denotes the invariant and normalized scalar product on the group algebra $G$.} We further can chose that
\begin{equation}
{\cal I}_{12}^{\sigma\,\dagger}={\cal I}_{12}^{\bar{\sigma}} \;\;\; \mathrm{and} \;\;\; {\cal I}_{12}^{0}= \mathbf{1}_{\mathcal{H}} \, ,
\end{equation}
where $\bar{\sigma}$ is the representation conjugate to $\sigma$.

The \textit{twist} operators appear in the commutant of the algebra
of the two double cones ${\cal A}'_{W_{1}\vee W_{2}}$ and they are
labeled by elements of the group $g\in G$. They commute with the
algebras ${\cal A}_{W_{1}}$ and ${\cal A}_{W_{2}}$, but they do not
commute with $\mathfrak{F}_{W_{1}}$ (or equivalently $\mathfrak{F}_{W_{2}}$).
In fact, they can be chosen in a way that they implement the symmetry
group in the region $W_{1}$. More precisely, given any double cone
$\tilde{W}_{1}\supset\overline{W}_{1}$, there exists $\tau_{g}\in\mathfrak{F}_{\tilde{W}_{1}}$ such that
\begin{align}
\tau_{g}F\tau_{g}^{\dagger} & =U\left(g\right)FU\left(g\right)^{\dagger}\,, &  & \negthickspace\negthickspace\negthickspace\negthickspace\negthickspace\negthickspace\negthickspace\negthickspace\negthickspace\negthickspace\negthickspace\negthickspace\negthickspace\negthickspace\negthickspace\negthickspace\negthickspace\negthickspace\negthickspace\negthickspace\negthickspace\negthickspace\negthickspace\forall F\in\mathfrak{F}_{W_{1}}\,, \hspace{2cm} \label{c6-twsit-action}\\
\tau_{g}F'\tau_{g}^{\dagger} & =F'\,, &  & \negthickspace\negthickspace\negthickspace\negthickspace\negthickspace\negthickspace\negthickspace\negthickspace\negthickspace\negthickspace\negthickspace\negthickspace\negthickspace\negthickspace\negthickspace\negthickspace\negthickspace\negthickspace\negthickspace\negthickspace\negthickspace\negthickspace\negthickspace\forall F'\in\mathfrak{F}_{\tilde{W}'_{1}}\,.\nonumber 
\end{align}
In particular, on the charge creating operators $V_{\xi}^{j}\in\mathfrak{F}_{W_{1}}$ we have
\begin{equation}
\tau_{g}^{\dagger}\,V_{\xi}^{j}\,\tau_{g}=\sum_{j=1}^{d_{\xi}}R_{\xi}(g)_{k}^{j}V_{\xi}^{k}\,.\label{cinco}
\end{equation}
Moreover, the twist can be chosen such that they satisfy the group
operation \cite{Doplicher:1984zz}\footnote{This is not the case of the simple twist operator \eqref{c6-twist}.
To construct twists with these special properties one needs to invoke
the split property (definition \ref{def_split_property}) \cite{Doplicher:1984zz,araki}.} 
\begin{eqnarray}
\tau_{g}\tau_{h}\!\!\! & = & \!\!\!\tau_{gh}\\
U(g)\tau_{h}U(g)^{\dagger}\!\!\! & = & \!\!\!\tau_{gh\,g^{-1}}\,.\label{seis}
\end{eqnarray}
In other words, the twist operators forms the group algebra of $G$, which will be denoted by $\mathfrak{A}_{G}$.

Twist operators for $W'_{1}$ (the complement of a double cone) can
be easily defined as $\tau_{g}:=U\left(g\right)\tilde{\tau}_{g}^{\dagger}$,
where $U\left(g\right)$ are the global symmetry group transformations
and $\tilde{\tau}_{g}$ are the twist operators of double cone $W_{2}$
strictly included in $W_{1}$, i.e. $\bar{W}_{2}\subset W_{1}$.

The twist operators are not in general elements of ${\cal A}$, in
the case of non-Abelian symmetry groups. We can form (generally non-unitary)
elements of ${\cal A}$ by taking linear combinations $\tau_{c}=\sum_{g\in G}c_{g}\tau_{g}$,
$c_{g}\in\mathbb{C}$, and imposing 
\begin{equation}
U(g)\tau_{c}U(g)^{\dagger}=\tau_{c}\,.
\end{equation}
Then, the invariant twist operators are naturally associated with
the center of the group algebra $\mathcal{Z}\left(\mathfrak{A}_{G}\right)$
where the coefficients $c_{g}=c_{hgh^{-1}}$ are invariant under conjugation.
The dimension of this center $d_{G}$ is equal to the number of irreducible
representations or the number of different conjugacy classes. The
group algebra is equivalent to a direct sum of full matrix algebras
$\mathfrak{A}_{G}:=\bigoplus_{\sigma\in\hat{G}}M_{d_{\sigma}}\left(\mathbb{C}\right)$,
where the group elements are represented by with matrices 
\begin{eqnarray}
g\in G & \mapsto & \bigoplus_{\sigma\in\hat{G}}D_{\sigma}(g)\in\mathfrak{A}_{G}\,.
\end{eqnarray}
The center of the group algebra is then clearly spanned by all diagonal
matrices, which are linear combinations of the projectors on each irreducible
representation. These projectors are precisely\footnote{That these operators are projectors follows from the convolution property
of the characters 
\begin{equation}
\sum_{g\in G}\chi_{\sigma_{1}}(g)\chi_{\sigma_{2}}(hg^{-1})=\frac{|G|}{d_{\sigma_{1}}}\delta_{\sigma_{1}\sigma_{2}}\,\chi_{\sigma_{1}}(h)\,.
\end{equation}
} 
\begin{equation}
P_{\sigma}:=\frac{d_{\sigma}}{|G|}\sum_{g\in G}\chi_{\sigma}(g)^{*}\tau_{g}\,,\label{projec}
\end{equation}
where $\chi_{\sigma}(g)$ is the character of the irreducible representation
$\sigma\in\hat{G}$. Then, the invariant twists can be written  as
\begin{equation}
\tau_{c}=\sum_{\sigma\in\hat{G}}c_{\sigma}P_{\sigma}\in\mathcal{A}\,,\quad c_{\sigma}\in\mathbb{C}\,.\label{center}
\end{equation}

\begin{rem}
In $d\leq2$, there is a difference with respect to higher dimensions.
For example, for the chiral CFT as in the previous chapter, we can
divide the compactified line into four intervals. Let $A_{1}$, $A_{2}$
be two disjoint intervals and $A_{3}$, $A_{4}$ the two disjoint
intervals forming the complement of $A_{1}\vee A_{2}$ (see section
\ref{failure}). The intertwiner between $A_{1}$ and $A_{2}$ belongs
to ${\cal A}_{A_{3}\vee A_{4}}'$, but the twist operator crossing
$A_{3}$ (or $A_{4}$) also belongs to this algebra. Therefore, the
number of additional elements in the algebra of ${\cal A}_{A_{3}\vee A_{4}}'$
respect to the additive one ${\cal A}_{A_{1}\vee A_{2}}$ is larger.
The difference with respect to higher dimensions is because the topology
of the two intervals and its complement is the same in this case. In
higher dimensions, the intertwiner between two double cones $W_{3}$
and $W_{4}$, placed inside $(W_{1}\vee W_{2})'$ is trivially included
in the additive algebra $\mathcal{A}_{(W_{1}\vee W_{2})'}$ because
$W_{3},W_{4}$ can be deformed to a coinciding position without crossing
$W_{1},W_{2}$. In addition, in $d\leq2$ the DHR sectors do not necessarily
come from a group symmetry, and we have a more general theory of sectors
determined by their fusion rules under composition, which replace the
decomposition of tensor product of group representations as a sum
of irreducible representations. The reasons for this difference with
higher dimensions are also related to the complications that appear when
analyzing the spin-statistics theorem, since charged operators cannot
smoothly interchange its positions without crossing each other (see for example \cite{Halvorson06}).
\end{rem}

\section{Entropy and DHR sectors\label{entropyDHR}}

We are interested in the mutual information $I(W_{1},W_{2})$ between two double cones $W_{1}$ and $W_{2}$ for the vacuum state $\omega$. We have seen that in the model ${\cal A}$, there are two possible different
choices of algebras for the region $W_{1}\vee W_{2}$: one with and one without the intertwiners. The algebra ${\cal A}_{W_{1}\vee W_{2}}$ without the intertwiners is additive, and hence, it is the appropriate one to produce the mutual information 
\begin{equation}
I_{{\cal A}}(W_{1},W_{2}):=I({\cal A}_{W_{1}},{\cal A}_{W_{2}})\label{c6-mi-obs}
\end{equation}
in the observable algebra ${\cal A}$.\footnote{The above mutual information can be computed indistinctly in the observable
algebra ${\cal A}$ or in the its vacuum sector ${\cal A}_{0}$. Moreover,
if for example, we use the Araki formula to compute the above expression,
we have to use the vacuum GNS-representation of the algebra ${\cal A}(W_{1})\lor{\cal A}(W_{2})$
which is unitarily equivalent to ${\cal A}_{0}(W_{1})\lor{\cal A}_{0}(W_{2})$.} However, it is of obvious interest to look for an information theoretic quantity that senses the contributions of the intertwiners in the
algebra of the union. To start with, the simplest thing to do is to
focus on the MI corresponding to the field algebra
\begin{equation}
I_{\mathfrak{F}}(W_{1},W_{2}):=I(\mathfrak{F}_{W_{1}},\mathfrak{F}_{W_{1}})\,.\label{c6-mi-fields}
\end{equation}
The algebra of ${\cal F}(W_{1}\vee W_{2})$ naturally contains the
intertwiners while retaining additivity in $\mathfrak{F}$. Hence,
as an order parameter indicative of the presence of DHR SS in ${\cal A}$
we can compute 
\begin{equation}
\Delta I:=I_{\mathfrak{F}}(W_{1},W_{2})-I_{{\cal A}}(W_{1},W_{2})\,.\label{sity}
\end{equation}
This is always positive by the monotonicity property of the MI (see
proposition \ref{c2_mi_prop}). We emphasize that, even if $I_{\mathfrak{F}}(W_{1},W_{2})$
and $\Delta I$ look like quantities that depend on $\mathfrak{F}$,
they are in fact properties of ${\cal A}$ itself for $G$-invariant
states, from which $\mathfrak{F}$ can be reconstructed. The ultimate
physical reason is that the only non-zero vacuum expectation values
in $\mathfrak{F}$ are equal to expectation values in ${\cal A}$.
In fact, we will later show, in detail, how both quantities are directly
written in terms of the model ${\cal A}$.

To put this in a sound ground, we will follow some ideas presented
in \cite{Longo_xu}. We first need to review some quantum information
tools that will also be useful in the rest of the chapter. This is
done in the next section \ref{tools}. Next, in section \ref{lower},
we describe the entropic order parameter in terms of a RE, which
is determined by intertwiners' expectation values. This allows us
to put useful lower bounds. The description of the order parameter
in terms of twist expectation values is done in section \ref{upper}.
This gives us a tool for computing upper bounds on $\Delta I$. We
show that the difference of mutual informations saturates to $\log|G|$,
where $|G|$ is the number of elements in the symmetry group, in the
limit when the two regions touch each other. Twist and intertwiners
do not commute and satisfy entropic certainty and uncertainty relations.
This is described in section \ref{certainty}. After that, we make
different computations using the main ideas developed in sections
\ref{lower} and \ref{upper}. We study the case of Lie group
symmetries in section \ref{U1}, the case of regions with different
topologies in section \ref{OT}, states with excitations of non-Abelian
sectors in section \ref{EX}, and the case of spontaneous symmetry
breaking in section \ref{SSB}. Finally, in section \ref{dosd}, we
make some remarks about the special case of $d=2$.

\subsection{Entropic order parameter\label{tools}}

In order to better understand the difference \eqref{sity}, we
make use of the material developed in section \ref{c2_sec_ce}. For
the present case, the natural conditional expectation from the field
algebra to the observable algebra $\varepsilon:\mathfrak{F}\rightarrow\mathcal{A}$
is
\begin{equation}
\varepsilon(F):=\int_{G}d\mu\left(g\right)\,\alpha_{g}\left(F\right)\,,\label{cee}
\end{equation}
where $d\mu\left(g\right)$ is the normalized Haar measure of the compact group
$G$ acting unitarily in $\mathfrak{F}$, and $\alpha_{g}$ is defined as
in \eqref{c6-acc-G}.\footnote{For a finite symmetry group $G$, \eqref{cee} can be rewritten as
\begin{equation}
\varepsilon(A)=\frac{1}{|G|}\sum_{g\in G}U\left(g\right)\,A\,U\left(g\right)^{\dagger}\,.
\end{equation}
} All the properties stated in definition \ref{c2-def-ce} about conditional
expectations are easily seen to hold for $\varepsilon$. Essentially,
$\varepsilon$ takes the $G$-invariant part of an element of $\mathfrak{F}$. For the case of the even part of the fermion algebra,
a general element $F\in\mathfrak{F}$ is of the form $F=A_{0}+A_{1}\psi$
with $A_{0},A_{1}\in\mathcal{A}$ and $\psi\in\mathfrak{F}$ is a
smeared fermion field appropriately chosen depending on $F$. Then, $\varepsilon(F)=A_{0}$.
It is important to notice that the global vacuum state $\omega\in\mathfrak{S}\left(\mathfrak{F}\right)$
is $\varepsilon$-invariant
\begin{equation}
\omega=\omega\circ\varepsilon\,,\label{c6-ce-vacuum}
\end{equation}
since the vacuum vector $\left|0\right\rangle $ is $G$-invariant.

We are particularly interested in applying property \eqref{c2_ce_prop}
to \eqref{cee}, when we restrict the action of the conditional expectation to local subalgebras of $\mathfrak{F}$.
For a double cone $W$, we have that
\begin{equation}
\varepsilon(\mathfrak{F}_{W})=\mathfrak{F}_{W}\cap G'=\mathcal{A}_{W}\,.
\end{equation}
Let us now consider the states $\omega_{\mathfrak{F}}:=\left.\omega\right|_{\mathfrak{F}_{W}}$
and $\omega_{\mathcal{A}}:=\left.\omega\right|_{\mathcal{A}_{W}}$, which are the restriction of the vacuum state to the algebras $\mathfrak{F}_{W}$ and $\mathcal{A}_{W}$.
Then, if we apply \eqref{c2_ce_prop} to the the states $\omega\rightarrow\omega_{\mathfrak{F}}$
and $\phi\rightarrow\omega_{\mathcal{A}}\circ\varepsilon$, we obtain\footnote{To lighten the notation and when there is no place for confusion,
we frequently use the subscripts $\mathfrak{F}$ and $\mathcal{A}$,
in the relative entropies, instead of $\mathfrak{F}\left(W\right)$
and $\mathcal{A}\left(W\right)$.}
\begin{equation}
S_{\mathfrak{F}}\left(\omega_{\mathfrak{F}}\mid\omega_{\mathcal{A}}\circ\varepsilon\right)-S_{\mathcal{A}}\left(\left.\omega_{\mathfrak{F}}\right|_{\mathcal{A}_{W}}\mid\left.\omega_{\mathcal{A}}\circ\varepsilon\right|_{\mathcal{A}_{W}}\right)=S_{\mathfrak{F}}\left(\omega_{\mathfrak{F}}\mid\omega_{\mathfrak{F}}\circ\varepsilon\right)\,.\label{c6-ce-1dc}
\end{equation}
Equation \eqref{c6-ce-1dc} is trivial since every term is zero independently.
This is an immediate consequence of \eqref{c6-ce-vacuum}, which implies
$\omega_{\mathfrak{F}}=\omega_{\mathcal{A}}\circ\varepsilon=\omega_{\mathfrak{F}}\circ\varepsilon$
and $\left.\omega_{\mathfrak{F}}\right|_{\mathcal{A}_{W}}=\left.\omega_{\mathcal{A}}\circ\varepsilon\right|_{\mathcal{A}_{W}}$.

Now we consider the union of strictly spacelike separated double cones $W_{1}$ and $W_{2}$.
In this case, we have that
\begin{equation}
\varepsilon(\mathfrak{F}_{W_{1}\lor W_{2}})=\mathfrak{F}_{W_{1}\lor W_{2}}\cap G'={\mathfrak{F}^{t}}\sp{\prime}_{\left(W_{1}\lor W_{2}\right)'}\cap G'=\mathcal{A}'_{\left(W_{1}\lor W_{2}\right)'}=\mathcal{A}{}_{W_{1}\lor W_{2}}\vee\left\{ \mathcal{I}_{12}\right\} \,,\label{c6-ce-2dc}
\end{equation}
where we have used that the field algebra satisfies (twisted) duality
for two double cones since it has no DHR SS. From now on, we denote
$\mathfrak{F}_{12}:=\mathfrak{F}\left(W_{1}\vee W_{2}\right)$ and
$\mathfrak{F}_{j}:=\mathfrak{F}\left(W_{j}\right)$ (idem for $\mathcal{A}$). We apply \eqref{c2_ce_prop}
to the case of the algebras $\mathfrak{F}_{12}\cong\mathfrak{F}_{1}\hat{\otimes}\mathfrak{F}_{2}$
and its subalgebra $\mathcal{A}_{12}\cong\mathcal{A}_{1}\otimes\mathcal{A}_{2}$,
in order to gain information about the differences of the MIs.
Note that $\mathfrak{F}_{12}$ contains the intertwiners that belong
to $\mathcal{A}$ on top of the elements of $\mathcal{A}_{12}$. This
last algebra does not contain the intertwiners, that belong to the
global neutral algebra but not to the one formed additively in $W_{1}\vee W_{2}$.
To exploit this fact, we use the conditional expectation $\varepsilon_{12}=\varepsilon_{1}\otimes\varepsilon_{2}:\mathfrak{F}_{1}\hat{\otimes}\mathfrak{F}_{2}\rightarrow\mathcal{A}_{1}\otimes\mathcal{A}_{2}$
\begin{equation}
\varepsilon_{12}\left(F_{1}\hat{\otimes}F_{2}\right)=\varepsilon_{1}\left(F_{1}\right)\otimes\varepsilon_{2}\left(F_{2}\right)\,,\quad F_{j}\in\mathfrak{F}_{j}\,,\label{c6-ce-prod}
\end{equation}
where $\varepsilon_{j}:\mathfrak{F}_{j}\rightarrow\mathcal{A}_{j}$
are just the restrictions of \eqref{cee} to $\mathfrak{F}_{j}$,
i.e. $\varepsilon_{j}\left(F_{j}\right)=\varepsilon\left(F_{j}\right)$
for all $F_{j}\in\mathfrak{F}_{j}$. The equation \eqref{c6-ce-prod}
is extended to more general elements by linearity. Notice that in
$\varepsilon_{12}=\varepsilon_{1}\otimes\varepsilon_{2}$ the group
average \eqref{cee} is done on each factor independently. To do that,
we can use the twist operators rather than the global group transformations
(see eq. \eqref{c6-twsit-action}).

Let us define the restricted states
\begin{eqnarray}
\omega_{12}:=\left.\omega\right|_{\mathfrak{F}_{12}}\,, &  & \omega_{j}:=\left.\omega\right|_{\mathfrak{F}_{j}}\,,\label{c6-states1}\\
\phi_{12}:=\left.\omega\right|_{\mathcal{A}_{12}}\,, &  & \phi_{j}:=\left.\omega\right|_{\mathcal{A}_{j}}\,.\label{c6-states2}
\end{eqnarray}
The states we choose for using in \eqref{c2_ce_prop} are $\omega_{12}$
and $(\phi_{1}\otimes\phi_{2})\circ\varepsilon_{12}$. We have $(\phi_{1}\otimes\phi_{2})\circ\varepsilon_{12}=\omega_{1}\otimes\omega_{2}$
because both states are invariant under the group transformations
on each region separately, and hence they give the same expectation
values for any operator. We also have trivially $\omega_{12}|_{\mathcal{A}_{12}}=\phi_{12}$
and $\left.(\phi_{1}\otimes\phi_{2})\circ\varepsilon_{12}\right|_{\mathcal{A}_{12}}=\phi_{1}\otimes\phi_{2}$.
Hence, from \eqref{c2_ce_prop} follows
\begin{equation}
S_{\mathfrak{F}}\left(\omega_{12}\mid\omega_{1}\otimes\omega_{2}\right)-S_{\mathcal{A}}\left(\phi_{12}\mid\phi_{1}\otimes\phi_{2}\right)=S_{\mathfrak{F}}\left(\omega_{12}\mid\omega_{12}\circ\varepsilon_{12}\right)\,.
\end{equation}
The terms on the l.h.s of the above equation are exactly the MIs \eqref{c6-mi-obs} and
\eqref{c6-mi-fields}. Then, we have that
\begin{equation}
I_{\mathfrak{F}}(W_{1},W_{2})-I_{{\cal A}}(W_{1},W_{2})=S_{{\cal \mathfrak{F}}}(\omega_{12}|\omega_{12}\circ\varepsilon_{12})\,.\label{labe}
\end{equation}
The MI $I_{\mathfrak{F}}(W_{1},W_{2})$ measures the vacuum correlations
between regions $W_{1}$ and $W_{2}$ including the intertwiners,
while $I_{{\cal A}}(W_{1},W_{2})$ does not include correlations coming
from them. When the regions $W_{1}$ and $W_{2}$ are near
to each other, and the set of intertwiners is finite, there will be
plenty of correlations but these will be essentially the same in $\mathfrak{F}$
and ${\cal A}$. The leading divergent terms of the MIs will cancel out and
only the effect of the intertwiners will make a difference. On the other
hand, $S_{\mathfrak{F}}(\omega_{12}|\omega_{12}\circ\varepsilon_{12})$
is not a MI. This measures the difference between two states on
the algebra $\mathfrak{F}_{1}\hat{\otimes}\mathfrak{F}_{2}$: one
is the vacuum state $\omega_{12}$ and the other is essentially the
same state but where the intertwiners have been projected to the neutral
algebras on each region independently. Intuitively, the conditional expectation
$\varepsilon_{12}$ kills the intertwiners by destroying their vacuum
correlations. The term $S_{\mathfrak{F}}(\omega_{12}|\omega_{12}\circ\varepsilon_{12})$
is called the \textit{entropic order parameter}.

Now we have all the necessary tools to show that the entropic order parameter
$S_{{\cal \mathfrak{F}}}(\omega_{12}|\omega_{12}\circ\varepsilon_{12})$
and the field MI $I_{\mathfrak{F}}(W_{1},W_{2})$
are indeed objects that pertain directly to the observable algebra $\mathcal{A}$.
Concerning the entropic order parameter, we first notice that the two
states appearing in the RE, namely $\omega_{12}$ and
$\omega_{12}\circ\varepsilon_{12}$, are invariant under the action
of the global symmetry group. Secondly, following expression \eqref{c6-ce-2dc},
we have that $\varepsilon(\mathfrak{F}_{12})={\cal A}'_{(12)'}$. Therefore,
according to \eqref{c2-ce-igual}, we conclude that the order parameter
can be computed equivalently as
\begin{equation}
S_{\mathfrak{F}}(\omega_{12}|\omega_{12}\circ \varepsilon_{12})=S_{{\cal A}'_{(12)'}}(\omega_{12}|\omega_{12}\circ\varepsilon_{12})\,.
\end{equation}
This formula shows transparently that the order parameter is an
intrinsic quantity of the model ${\cal A}$ itself. Even more surprisingly,
the same is true for the MI $I_{{\cal \mathfrak{F}}}(W_{1},W_{2})$.
This can be written in the theory ${\cal A}$ more succinctly as 
\begin{equation}
I_{{\cal \mathfrak{F}}}(W_{1},W_{2})=S_{{\cal A}'_{(12)'}}(\omega_{12}|(\omega_{1}\otimes\omega_{2})\circ\varepsilon_{12})\,.\label{IO}
\end{equation}
Such a relation follows by applying again the formula \eqref{c2_ce_prop}
to the present scenario, which leads to
\begin{eqnarray}
S_{{\cal A}'_{(12)'}}\left(\omega_{12}|(\omega_{1}\otimes\omega_{2})\circ\varepsilon_{12}\right)\!\!\! & = & \!\!\!S_{{\cal A}_{12}}(\omega_{12}|\omega_{1}\otimes\omega_{2})+S_{{\cal A}'_{(12)'}}(\omega_{12}|\omega_{12}\circ\varepsilon_{12})\nonumber \\
 & = & \!\!\!I_{{\cal A}}(W_{1},W_{2})+S_{\mathfrak{F}}(\omega_{12}|\omega_{12}\circ\varepsilon_{12})=I_{\mathfrak{F}}(W_{1},W_{2})\,. \hspace{1cm}
\end{eqnarray}
Although we have defined the conditional expectation by means of the
field algebra $\mathfrak{F}$, the conditional expectation $\varepsilon_{12}$
can be defined directly in the algebra $\mathcal{A}$ as well. It
is the restriction of the above conditional expectation to the smaller algebra
${\cal A}'_{(12)'}$. More quantitatively, the action of $\varepsilon_{12}$
in ${\cal A}$ can be expressed in the following way. A generic element
of ${\cal A}_{12}$ can be written $A=\sum_{\sigma\in\hat{G}}A_{\sigma}\,{\cal I}_{12}^{\sigma}$
as an expansion in intertwiners of different irreducible representations
and where operators $A_{\sigma}$ commute with the (invariant) twists in ${\cal A}_{12}$). Then, $\varepsilon_{12}(A)=A_{\mathrm{triv}}=:A_{0}$.
Furthermore, from a conceptual perspective, it is unnecessary to know
the structure of SS to construct the conditional expectation $\varepsilon_{12}$.
In fact, it can be shown that it always exists a unique conditional expectation
for the inclusion of algebras ${\cal A}{}_{12}\subset{\cal A}'_{(12)'}$.\footnote{This follows because both algebras ${\cal A}{}_{12}$ and ${\cal A}'_{(12)'}$
are factors, and the relative commutant is trivial ${\cal A}'_{12}\cap{\cal A}'_{(12)'}=\left\{ \lambda\cdot\mathbf{1}\right\}$ (see \cite{Longo:1994xe}).}

Having shown that even if one computes REs in the field algebra $\mathfrak{F}$ one actually ends
up with REs of the invariant algebra $\mathcal{A}$,
it turns out to be technically and conceptually simpler to work with
the field algebra $\mathfrak{F}$, and we will do this in what follows.

\paragraph{Remarks on lattice QFT and EE\protect \\
}

We usually are interested to work in the lattice QFT because we work
with matrix algebras (or more in general type I vN algebras), and
hence, we can use the expressions in terms of density matrices developed
in section \ref{c2_sec_finite}. We first notice that lattice models with global
symmetries are easily constructed. We have a Hilbert space ${\cal H}_{n}$
for each vertex $n\in\mathbb{Z}^{d-1}$ of the lattice on which there
is a faithful representation $U_{n}$ of the group $G$. The global
Hilbert space is the tensor product ${\cal H}:=\bigotimes_{n}{\cal H}_{n}$
and the group acts with the tensor product representation $U:=\bigotimes_{n}U_{n}$.
The algebra $\mathfrak{F}:=\mathcal{B}\left(\mathcal{H}\right)$ is
the full algebra of operators in ${\cal H}$, and ${\cal A}\subset\mathfrak{F}$
is the subalgebra of invariant operators.

One interest in looking at finite dimensional algebras is the following.
One may entertain the idea that, even if the entropies have ambiguities
in continuum limit in QFT, the particular difference $S_{\mathfrak{F}}(W)-S_{{\cal A}}(W)$,
of the complete and the neutral models in the same region and
for the vacuum state, could be well-defined in the continuum limit, independently
of the chosen lattice regularization and the exact definition of the algebras. In such
a case, we could use just this difference as an order parameter. With
a focus in investigating this question, we collect some formulas for
matrix algebras that will be very useful along this chapter.

As we have explained in section \ref{c2_sec_ce}, given an inclusion
of finite dimensional algebras $\mathfrak{B}\subset\mathfrak{A}$,
there always exists a unique conditional expectation $\varepsilon:\mathfrak{A}\rightarrow\mathfrak{B}$
that preserves the trace. This is not the general case for conditional
expectations, but the conditional expectation \eqref{cee} restricted
to any local algebra $\mathfrak{F}\left(\mathcal{O}\right)$ preserves
the trace since it is an average over automorphisms. For any state
$\phi$ on $\mathfrak{A}$,  we have that
\begin{equation}
S_{\mathfrak{A}}(\phi|\phi\circ\varepsilon)=S_{\mathfrak{A}}(\phi\circ\varepsilon)-S_{\mathfrak{A}}(\phi)\,.\label{estasi}
\end{equation}
To show this we write this RE as $\Delta\langle K\rangle-\Delta S$,
where $\Delta S$ is the difference of the entropies of the two
states, $\Delta\langle K\rangle$ is the difference of the expectation
values of the "inner" modular Hamiltonian $K:=-\log(\rho_{\phi\circ\varepsilon})$, and $\rho_{\phi\circ\varepsilon}\in\mathfrak{A}$
is the statistical operator corresponding to the state $\phi\circ\varepsilon$.
Then, we have that
\begin{equation}
\textrm{Tr}_{\mathfrak{A}}(\rho_{\phi\circ E}\,a)=\textrm{Tr}_{{\cal \mathfrak{A}}}(\rho_{\phi}\varepsilon(A))=\textrm{Tr}_{{\cal \mathfrak{A}}}(\varepsilon(\rho_{\phi})\varepsilon(A))=\textrm{Tr}_{{\cal \mathfrak{A}}}(\varepsilon(\rho_{\phi})A)\,,\quad\forall A\in\mathfrak{A}\,,
\end{equation}
and we get $\rho_{\phi\circ E}=\varepsilon(\rho_{\phi})$ as an operator
in $\mathfrak{A}$. We then have that $\varepsilon(K)=K$ and hence
\begin{equation}
\textrm{Tr}_{\mathfrak{A}}(\rho_{\phi}K)=\textrm{Tr}_{\mathfrak{A}}(\varepsilon(\rho_{\phi}K))=\textrm{Tr}_{\mathfrak{A}}(\varepsilon(\rho_{\phi})K)=\textrm{Tr}_{\mathfrak{A}}(\rho_{\phi\circ\varepsilon}K)\,.\label{xilo1}
\end{equation}
Then, it follows that $\Delta\langle K\rangle=0$ and equation \eqref{estasi}
holds.

Eq. \eqref{estasi} is the difference of entropies on the same algebra
$\mathfrak{A}$, between the state $\phi$ and the corresponding invariant
one $\phi\circ\varepsilon$. In contrast, the quantity $S_{\mathfrak{F}}(W)-S_{{\cal A}}(W)$
refers to an entropy difference between an invariant state in two
different algebras. Let $\phi_{1}$ and $\phi_{2}$ be two states
on $\mathfrak{A}$, then we have that\footnote{In equation \eqref{c6-re-finite}, the restricted state $\left.\phi_{j}\right|_{\mathfrak{B}}$
is also denoted by $\phi_{j}$. }
\begin{eqnarray}
S_{\mathfrak{A}}(\phi_{1})\!\!\! & = & \!\!\!-S_{\mathfrak{A}}(\phi_{1}|\phi_{2})-\textrm{Tr}_{\mathfrak{A}}(\rho_{\phi_{1}}^{\mathfrak{A}}\log\rho_{\phi_{2}}^{\mathfrak{A}})\,,\\
S_{\mathfrak{B}}(\phi_{1})\!\!\! & = & \!\!\!-S_{\mathfrak{B}}(\phi_{1}|\phi_{2})-\textrm{Tr}_{\mathfrak{B}}(\rho_{\phi_{1}}^{\mathfrak{B}}\log\rho_{\phi_{2}}^{\mathfrak{B}})\,.\label{c6-re-finite}
\end{eqnarray}
Let us now assume that $\phi_{1}$ and $\phi_{2}$ are invariant states under some conditional expectation $\varepsilon:\mathfrak{A}\rightarrow\mathfrak{B}$, i.e. $\phi_{j}=\phi_{j}\circ\varepsilon$. Then we have that $S_{\mathfrak{A}}(\phi_{1}|\phi_{2})=S_{\mathfrak{B}}(\phi_{1}|\phi_{2})$,
and subtracting the above equations we get
\begin{equation}
S_{\mathfrak{A}}(\phi_{1})-S_{{\cal B}}(\phi_{1})=\textrm{Tr}_{\mathfrak{B}}(\rho_{\phi_{1}}^{\mathfrak{B}}\log\rho_{\phi_{2}}^{\mathfrak{B}})-\textrm{Tr}_{\mathfrak{A}}(\rho_{\phi_{1}}^{\mathfrak{A}}\log\rho_{\phi_{2}}^{\mathfrak{A}})=\langle\log\rho_{\phi_{2}}^{\mathfrak{B}}\rangle_{\phi_{1}}-\langle\log\rho_{\phi_{2}}^{\mathfrak{A}}\rangle_{\phi_{2}}\,.\label{tirania}
\end{equation}
This holds independently of the invariant state $\phi_{2}$ we have chosen,
and it is linear in the state $\phi_{1}$.

Equation \eqref{tirania} shows that expecting $S_{\mathfrak{F}}(W)-S_{{\cal A}}(W)$
to be well-defined is incorrect. In particular, it is not ordered
by inclusion as in the case of \eqref{estasi}. The following example exhibits
the problems that may occur. 
\begin{example}
Consider the fermion algebra at one site. It is given by the matrix
algebra $\mathfrak{A}:=M_{2}\left(\mathbb{C}\right)$ with basis elements
given by the operators $\left\{ \mathbf{1},c,c^{\dagger},c^{\dagger}c\right\} $,
where $c,c^{\dagger}$ are the creation and annihilation operators, and the
$\mathbb{Z}_{2}$ fermionic symmetry is the symmetry group. For an even
state $\phi$, such as the vacuum, the entropy in this algebra $\mathfrak{A}$
is equal to the one in the neutral algebra $\mathfrak{B}$ generated
by $\left\{ \mathbf{1},c^{\dagger}c\right\}$, i.e. $S_{\mathfrak{A}}\left(\phi\right)-S_{\mathfrak{B}}\left(\phi\right)=0$. However, if we choose the algebra $\mathfrak{\tilde{A}}$
generated by $\left\{ \mathbf{1},c+c^{\dagger}\right\} $, the entropy
will be $S_{\tilde{\mathfrak{A}}}\left(\phi\right)=\log(2)$ for any
even state, while the entropy in the even subalgebra $\mathfrak{\tilde{B}:=}\{\lambda\cdot1\}$
is zero. Hence, we have that $S_{\mathfrak{\tilde{A}}}\left(\phi\right)-S_{\mathfrak{\tilde{B}}}\left(\phi\right)=\log\left(2\right)$
for any even state. 

The above example illustrates what we expect in a lattice QFT. As
we enlarge the algebras to obtain the continuum limit, $S_{\mathfrak{F}}\left(\omega\right)-S_{\mathcal{A}}\left(\omega\right)$
can be fluctuating depending on the precise details on which the algebras
are chosen. This highlights the necessity of using the MI difference
\eqref{sity} to get unambiguous results.
\end{example}

\subsection{Intertwiner version: lower bound\label{lower}}

The consequence of expressing the difference of MIs \eqref{sity}
as a RE is that we can use monotonicity of the RE to put
lower bounds. In particular, to produce a lower bound, we can restrict
the states to a subalgebra ${\cal B}_{12}$ of $\mathfrak{F}_{12}$,
\begin{equation}
I_{{\cal \mathfrak{F}}}(W_{1},W_{2})-I_{{\cal A}}(W_{1},W_{2})\ge S_{{\cal B}_{12}}(\omega_{12}|\omega_{12}\circ\varepsilon_{12})\,.\label{left}
\end{equation}
Moreover, since the expectation values of the intertwiners are the main
difference between these states, we have to find a useful subalgebra ${\cal B}_{12}$
that contains the relevant information about the intertwiners.

If we have a finite dimensional ${\cal B}_{12}$ (or more generally
a type I vN algebra) the l.h.s. of \eqref{left} can be written
in terms of EEs if we further require that the conditional
expectation maps the subalgebra into itself, i.e. $\varepsilon_{12}({\cal B}_{12})\subseteq{\cal B}_{12}$.
Using \eqref{estasi}, we get a lower bound given by the difference of entropies
\begin{equation}
I_{{\cal \mathfrak{F}}}(W_{1},W_{2})-I_{{\cal A}}(W_{1},W_{2})\ge S_{{\cal B}_{12}}(\omega_{12}|\omega_{12}\circ\varepsilon_{12})=S_{{\cal B}_{12}}(\omega\circ\varepsilon_{12})-S_{{\cal B}_{12}}(\omega)\,.\label{left1}
\end{equation}
The fact that this difference is positive is because the charged operators
on $W_{1}$ and $W_{2}$ can have entanglement in the vacuum state, what
is reflected in the expectation values of the intertwiners. This entanglement
will count for the entropy of the first state on the r.h.s.
of \eqref{left1} but not for the second one.

To improve the lower bound, we can try to maximize the entropy difference
over all choices of intertwiner operators, i.e. over all choices of
the subalgebra ${\cal B}_{12}$. To see what we can do, let us suppose we have
a unitary intertwiner ${\cal I}_{12}=V_{1}V_{2}^{\dagger}$ for an
Abelian (one-dimensional) sector with unitaries $V_{j}\in\mathfrak{F}_{j}$.
A good idea is trying to maximize the expectation value\footnote{As we did before, $\langle F\rangle:=\omega\left(F\right)$ denotes the vacuum expectation value of an operator $F\in\mathfrak{F}$.}
\begin{equation}
\langle\mathcal{I}_{12}\rangle=\langle V_{1}V_{2}^{\dagger}\rangle\rightarrow1\,.\label{11}
\end{equation}
In this case, it could be said that both $V_{1}$ and $V_{2}$ create
essentially the same vector state acting on the vacuum. If $V_{2}=V_{1}^{-1}$
we would get exactly $1$, but this is not possible since they belong
to local algebras of spacelike separated regions.

Of special interest is the case where the region $W_{2}$ is essentially
$W_{1}'$, and hence both regions cover the full space. This is what we
have when we compute the universal regularized EE as we have explained
in section \ref{c3-sec-univee}. In this case, we will be able to
get the maximum value in \eqref{11}. Then, let us think directly
in the case $W_{2}=W'_{1}$.\footnote{In this case, the algebras of interest are the vN algebra $\mathfrak{F}(W_{1})$ and the vN algebra $\mathfrak{F}(W'_{1})''\equiv{\mathfrak{F}(W_{1})^{t}}\sp{\prime}$
generated by $\mathfrak{F}(W'_{1})$.} By using the modular conjugation $J$ associated with the algebra
$\mathfrak{F}(W_{1})$ and the vacuum state, we are able to convert\footnote{In the case of fermionic algebras, the modular theory requires a slightly
modification in order to take into account the twisted duality (see
for example \cite{Guido:1995fy}). However, here we use the usual
approach explained in section \ref{c2_sec-mod_th} because, in the
end, we obtain the same result.} 
\begin{equation}
\langle V_{1}V_{2}^{\dagger}\rangle=\langle V_{1}J\tilde{V}_{2}J\rangle\,,
\end{equation}
with $\tilde{V}_{2}:=JV_{2}^{\dagger}J$ belonging to the algebra $\mathfrak{F}\left(W_{1}\right)$. By modular theory, this is the same as 
\begin{equation}
\langle V_{1}V_{2}^{\dagger}\rangle=\langle V_{1}\Delta^{\frac{1}{2}}\tilde{V}_{2}^{\dagger}\rangle\,,
\end{equation}
where $\Delta\geq0$ is the modular operator. Using Schwarz inequality,
we get
\begin{equation}
|\langle V_{1}V_{2}^{\dagger}\rangle|^{2}=|\langle V_{1}\Delta^{\frac{1}{2}}\tilde{V}_{2}^{\dagger}\rangle|^{2}\le\langle V_{1}\Delta^{\frac{1}{2}}V_{1}^{\dagger}\rangle\,\langle\tilde{V}_{2}\Delta^{\frac{1}{2}}\tilde{V}_{2}^{\dagger}\rangle\,.
\end{equation}
Therefore, to maximize the expectation value we can choose either $V_{1}JV_{1}J$
or $\tilde{V}_{2}J\tilde{V}_{2}J$ as intertwiners. Without loss of
generality, we take 
\begin{equation}
V_{2}^{\dagger}:=JV_{1}J\,.
\end{equation}
Note that $V_{1}$ and $V_{2}^{\dagger}$ will be formed by representations
of opposite charge because of the action of $J$, and this is exactly
what we need to produce an intertwiner.

Therefore, we need to maximize $\langle V_{1}\Delta^{\frac{1}{2}}V_{1}^{\dagger}\rangle$
along all possible charge creating operators belonging to $\mathfrak{F}(W_{1})$.
If we could choose $V_{1}$ such as it commutes with the modular operator $\Delta$,
then we would get the desired result $\langle V_{1}V_{2}^{\dagger}\rangle=\langle V_{1}V_{1}^{\dagger}\rangle=1$
because of $\Delta|0\rangle=|0\rangle$. Intuitively, this commutation
can be achieved by writing $V_{1}$ in the basis that diagonalizes
the modular Hamiltonian. We can always write a unitary operator that
commutes with the density matrix by choosing phases in this basis.
However, this unitary will have zero charge because, roughly speaking,
the vacuum density matrix commutes with the charge operator. Hence,
we can solve the problem only in an approximate way, choosing charge
creating operators corresponding to modes of the modular Hamiltonian
with modular energy tending to zero, or as much invariant under the
modular flow as possible. In QFT, we can always approach $\langle V_{1}V_{2}^{\dagger}\rangle\rightarrow1$
as much as we want for complementary regions (in many different ways)
since $0\in\mathbb{R}$ is included in the spectrum of the modular
Hamiltonian, which is continuous in $\mathbb{R}$. In section \ref{bounds},
we will explore the physical content of this requirement developing some explicit
examples.

Notice that this cannot be done if $W_{1}$ and $W_{2}$ are at a finite
distance, since, in that case, $J$ would map the algebra $\mathfrak{F}(W_{1})$
into the algebra of the region $W'_{1}\supsetneq W_{2}$. Then, in general, the maximal correlator
cannot be achieved exactly for non-zero distance. However,
if the regions touch each to the other along some part of their entanglement surfaces, no matter how small is this part, we can think in putting highly localized
excitations very near this region of the boundary, where the modular
energy is small. In this region, we can think the states are similar
to the case where the full region $W_{1}'$ is covered by $W_{2}$.
For any such a case, we expect that the maximal correlation can be
achieved for a convenient choice of excitations approaching the entanglement
surface.

Now, coming back to the bound \eqref{left1} on the MI difference, we can have a
universal bound for a finite group $G$ when the spacelike separated
double cones $W_{1}$ and $W_{2}$ touch each to the other along a $d-2$
dimensional piece of their entanglement surfaces. In this case, we
expect that we can maximize the value of the intertwiner expectation values.
We will see this bound depends only on the number of elements $|G|$
of the symmetry group $G$. To see this, let us think we have finite subalgebras of operators
on each region $\mathcal{B}_{j}\subset\mathfrak{F}(W_{j})$, which
are isomorphic to the matrix algebra $\mathcal{B}_{1},\mathcal{B}_{2}\cong M_{N}\left(\mathbb{C}\right)$.
We further require that these algebras are transformed into itself by the global group
transformations. Let us call $P_{ij}^{1}$ and $P_{ij}^{2}$
the operators forming the matrix canonical basis of the algebras $\mathcal{B}_{1}$
and $\mathcal{B}_{2}$, i.e.
\begin{equation}
P_{ij}^{1}P_{kl}^{1}=\delta_{jk}\,P_{il}^{1}\,,\qquad P_{ij}^{1\,\dagger}=P_{ji}^{1}\,,\qquad\sum_{i=1}^{N}P_{ii}^{1}=\mathbf{1}\,,\label{416}
\end{equation}
and analogously for $W_{2}$. One way to generate these finite algebras
is to use the charge generating operators $V_{\xi}^{i}$ for some
representation (not necessarily irreducible). In general, they generate an infinite
dimensional algebra. However, the finite dimensional algebra
(discussed in section \ref{c6-sec-int-twi}) formed by the operators
\begin{equation}
A=\sum_{i,j=1}^{d_{\xi}}A_{ij}V_{\xi}^{i}V_{\xi}^{j\,\dagger}\,,\quad A_{ij}\in\mathbb{C}\,,\:V_{\xi}^{i}\in\mathfrak{F}(W_{1})\,,\label{esa1}
\end{equation}
form a matrix algebra isomorphic to $M_{d_{\xi}}\left(\mathbb{C}\right)$. However,
one can produce a subalgebra without worrying about the partial isometries
$V_{\xi}^{i}$. We will give some examples in the next section.

Now we want to maximize the entanglement between these two algebras. We choose $P_{ij}^{1}:=JP_{ij}^{2}J$ and we think that these
operators approximately commute with the modular operator. Under this
choice, we notice that if $\mathbf{D}_{1}\left(g\right)$ is the unitary
matrix representation of the group $G$ in the algebra $\mathcal{B}_{1}$,
then $\mathbf{D}_{2}\left(g\right):=\mathbf{D}_{1}\left(g\right){}^{*}$
is the representation of $G$ in the algebra $\mathcal{B}_{2}$. The
statistical operator (density matrix) of the vacuum state $\omega$ on the algebra $\mathcal{B}_{12}:=\mathcal{B}_{1}\otimes\mathcal{B}_{2}$ is 
\begin{equation}
\rho_{jl,ik}^{\omega}:=\langle P_{ij}^{1}P_{kl}^{2}\rangle=\langle P_{ij}^{1}JP_{kl}^{1}J\rangle=\langle P_{ij}^{1}\Delta^{\frac{1}{2}}P_{lk}^{1}\rangle\simeq\langle P_{ij}^{1}P_{lk}^{1}\rangle=\delta_{jl}\langle P_{ik}^{1}\rangle\,.
\end{equation}
Under these assumptions for the state, hermiticity of $\rho_{jl,ik}^{\omega}$ implies that
\begin{eqnarray}
\langle P_{ik}^{1}\rangle\!\!\! & = & \!\!\!\frac{1}{N}\delta_{ik}\,,\\
\rho_{jl,ik}^{\omega}\!\!\! & = & \!\!\!\frac{1}{N}\delta_{ik}\delta_{jl}\,.\label{md}
\end{eqnarray}
This state is invariant under the adjoint action by any unitary
matrix of the form $\mathbf{U}\otimes\mathbf{U}^{*}$, and
in particular, it is invariant under global group transformations that
have this form given our choice of algebras. Then, $\left.\omega\right|_{\mathcal{B}_{12}}$
is a pure state on $\mathcal{B}_{12}$, and hence we get
\begin{equation}
S_{\mathcal{B}_{12}}(\omega)=0\,.\label{twisted11}
\end{equation}
This means that $\omega$ is maximally entangled between $\mathcal{B}_{1}$ and
$\mathcal{B}_{2}$ as expected.

In order to compute the other state $\phi:=\omega\circ\varepsilon_{12}$,
we need to know how the group acts on each of the algebras. Let us
decompose the action of the group on each algebra $\mathcal{B}_{1}$
and $\mathcal{B}_{2}$ into irreducible representations. Suppose we
have irreducible representations $\sigma\in\hat{G}$ of dimension
$d_{\sigma}$ with multiplicity $n_{\sigma}$. Then, we have that 
\begin{equation}
\sum_{\sigma\in\hat{G}}n_{\sigma}d_{\sigma}=N\,.
\end{equation}
Without loss of generality we take the basis vectors that decompose
the group representation into irreducible ones and rename the indices
of the basis elements as $i\rightarrow(\sigma,s,l)$, where $s=1\,\cdots,n_{\sigma}$ and
$l=1,\cdots,d_{\sigma}$. The state $\phi=\omega\circ\varepsilon_{12}$
is represented by the density matrix 
\begin{eqnarray}
 &  &  \hspace{-.9cm} \rho_{(\sigma_{1}s_{1}l_{1})(\sigma_{2}s_{2}l_{2}),(\sigma_{3}s_{3}l_{3})(\sigma_{4}s_{4}l_{4})}^{\phi}\nonumber \\
 & & \hspace{.5cm} =   \frac{1}{|G|^{2}}\!\!\sum_{g_{1},g_{1}\in G}D_{\sigma_{1}}(g_{1})_{l'_{1}}^{l_{1}}D_{\sigma_{2}}(g_{2})_{l'_{2}}^{l_{2}\,*}\rho_{(\sigma_{1},s_{1},l_{1}')(\sigma_{2},s_{2},l_{2}'),(\sigma_{3},s_{3},l_{3}')(\sigma_{4},s_{4},l_{4}')}^{\omega}D_{\sigma_{3}}(g_{1})_{l'_{3}}^{l_{3}\,*}D_{\sigma_{4}}(g_{2})_{l'_{4}}^{l_{4}}\nonumber  \\
&  & \hspace{.5cm}= \frac{1}{d_{\sigma_{1}}\,N}\delta_{\sigma_{1}\sigma_{2}}\delta_{\sigma_{2}\sigma_{3}}\delta_{\sigma_{3}\sigma_{4}}\delta_{s_{1}s_{2}}\delta_{s_{3}s_{4}}\delta_{l_{1}l_{3}}\delta_{l_{2}l_{4}}\,.
\end{eqnarray}
In the last equation, we have used the orthogonality relations for irreducible representations
\begin{equation}
\sum_{g\in G}D_{\sigma_{1}}(g)_{l_{2}}^{l_{1}}D_{\sigma_{2}}(g)_{l_{4}}^{l_{3}\,*}=\frac{|G|}{d_{\sigma_{1}}}\delta_{\sigma_{1},\sigma_{2}}\delta_{l_{1}l_{3}}\delta_{l_{2}l_{4}}\,,\label{orthogonality}
\end{equation}
and the formula \eqref{md}. Therefore, the non-zero part of the density matrix has the structure of a direct sum of blocks labelled by the irreducible representations. The density matrix is then
\begin{equation}
\rho^{\phi}=\bigoplus_{\sigma\in\hat{G}}\left[\frac{n_{\sigma}d_{\sigma}}{N}\left(\frac{1}{n_{\sigma}}\mathbf{J}{}_{n_{\sigma}}\oplus\mathbf{0}{}_{n_{\sigma}^{2}-n_{\sigma}}\right)\otimes\left(\frac{1}{d_{\sigma}^{2}}\mathbf{1}_{d_{\sigma}^{2}}\right)\right]\,,\label{irredi}
\end{equation}
where $\mathbf{J}{}_{n}$ (resp. $\mathbf{0}{}_{n}$) denotes the
$n$-dimensional square matrix with all entries equal to $1$ (resp.
$0$). The second factor of \eqref{irredi} is proportional to the
identity matrix. Both factors in \eqref{irredi} are normalized to
have unit trace. Hence, writing the fraction of basis vectors with
representation $\sigma\in\hat{G}$ as 
\begin{equation}
q_{\sigma}:=\frac{n_{\sigma}d_{\sigma}}{N}\,,\quad\sum_{\sigma\in\hat{G}}q_{\sigma}=1\,,\label{con}
\end{equation}
the vN entropy is 
\begin{equation}
S_{\mathcal{B}_{12}}(\phi)=-\sum_{\sigma\in\hat{G}}q_{\sigma}\log q_{\sigma}+\sum_{\sigma\in\hat{G}}q_{\sigma}\log d_{\sigma}^{2}\,.\label{twisted}
\end{equation}
We can vary the frequency $q_{\sigma}$ of the representation $\sigma$
in order to achieve maximal entropy difference $S_{\mathcal{B}_{12}}(\phi)-S_{\mathcal{B}_{12}}(\omega) = S_{\mathcal{B}_{12}}(\phi)$,
taking into account the constraint \eqref{con}. We get that the maximum is achieved for 
\begin{equation}
q_{\sigma}:=\frac{d_{\sigma}^{2}}{|G|}\,,\label{proji}
\end{equation}
where we have used the relation $|G|=\sum_{\sigma\in\hat{G}}d_{\sigma}^{2}$
valid for finite groups. This implies that
\begin{equation}
n_{\sigma}=d_{\sigma}\frac{N}{|G|}\,,
\end{equation}
and from \eqref{twisted} we arrive to
\begin{equation}
S_{\mathcal{B}_{12}}(\omega\circ\varepsilon_{12})-S_{\mathcal{B}_{12}}(\omega)=\log|G|\,.
\end{equation}
Therefore, the optimal multiplicity of a representation is proportional
to the dimension of the representation. This is exactly the case of
the regular representation of the group. Then, the optimal representation consists of any number of copies of the regular one. Other representations
will give weaker constraints. Notice that there is no increase in
the entropy by arbitrarily multiplying the representations and enlarging
the algebra. The conditional expectation will take into account
that redundant copies are not measuring any new difference between the models since they are produced by the neutral algebra.\footnote{It is interesting to consider differences of Rényi entropies of the
state \eqref{irredi} of the intertwiner algebra and the vacuum state.
These differences of Rényi entropies are all equal to the same constant
$\log|G|$ when taking the regular representation and in this limit
of maximal entanglement. This feature of a state is named ``flat
spectrum'' in the literature. Presumably, this leads to a flat spectrum
of the difference of Rényi MIs between the two models in the limit
of touching regions. }

For the regular representation we have the best lower bound for complementary
regions
\begin{equation}
I_{\mathfrak{F}}(W_{1},W_{2})-I_{{\cal A}}(W_{1},W_{2})\ge\log|G|\,.\label{c6.-lower}
\end{equation}
As we will see in the next subsection, $\log|G|$ is also an upper
bound for the difference of the mutual informations.

In appendix \ref{regular}, we show formally that the regular representation
can always be achieved using the charge generators $V_{\sigma}^{i}$
of all irreducible representations. But, from a physical standpoint, we remark that the regular representation is naturally
constructed with high frequency by fusion. We will use this idea in
the example in section \ref{freexamples}. The reason is that the
character of the regular representation is $\chi_{\mathrm{reg}}(g)=|G|\delta_{g,1}$
and then, the regular representation itself is stable under fusion, i.e $\mathrm{reg}\otimes\mathrm{reg}\cong G\,\mathrm{reg}$.
The tensor product of a regular representation with another representation
of dimension $d_{\xi}$ has character $\chi_{\mathrm{reg}\otimes\xi}(g)=d_{\xi}|G|\delta_{g,1}$,
and then it decomposes into exactly $d_{\xi}$ copies of the regular
representation. This is not the case of other representations. For
any representation $\xi$ of dimension $d_{\xi}>1$, its character
satisfies
\begin{equation}
\frac{\chi_{\xi}(1)}{d_{\xi}}=1\,,\quad\left|\frac{\chi_{\xi}(g)}{d_{\xi}}\right|<1\,,
\end{equation}
and then, for the product $r=r_{1}\otimes r_{2}$ of two representations
\begin{equation}
\frac{\chi_{r}(1)}{d_{1}d_{2}}=1\,,\quad\left|\frac{\chi_{12}(g)}{d_{1}d_{2}}\right|=\left|\frac{\chi_{1}(g)}{d_{1}}\right|\left|\frac{\chi_{2}(g)}{d_{2}}\right|\,,
\end{equation}
i.e. the normalized character always approaches to the one of the regular
representation.

Another way to see this is to realize that the tensor product of arbitrary
representations $R$ with some fix representation $R_{0}$ can be
thought of as a stochastic process in the space of the probabilities
$q_{\sigma}$ with $\sigma\in\hat{G}$. In fact, the new representation
$R':=R_{0}\otimes R$ will have
\begin{equation}
q_{\sigma}^{R'}=\sum_{\nu\in\hat{G}}M_{\sigma\nu}^{R_{0}}\,q_{\nu}^{R}\,,
\end{equation}
where
\begin{equation}
M_{\sigma\nu}^{R_{0}}=\sum_{\xi\in\hat{G}}\frac{N_{\xi\nu}^{\sigma}}{d_{\xi}d_{\nu}}d_{\sigma}\,q_{\xi}^{R_{0}}\,,
\end{equation}
and $N_{\xi\nu}^{\sigma}$ is the fusion matrix giving the number
of irreducible representations of type $\sigma$ that appear in the
tensor product of representations $\xi$ and $\nu$. The matrix $M^{R_{0}}$
is stochastic and it represents a stochastic process since it has positive
entries and $\sum_{\sigma\in\hat{G}}M_{\sigma\nu}^{R_{0}}=1$. Since
for any fixed $\xi$ we have $\sum_{\nu\in\hat{G}}N_{\xi\nu}^{\sigma}d_{\nu}\sim d_{\sigma}$. \footnote{This follows from the fact that the tensor product of the regular representation with any other representation is proportional to the regular representation.} It follows that the probability vector $q_{\sigma}=\frac{d_{\sigma}^{2}}{|G|}$
is the fixed point of the stochastic process, i.e., an eigenvector of $M^{R_{0}}$
of eigenvalue $1$. As for any stochastic process, applying it repeatedly
will approach the fixed point exponentially fast.

Roughly speaking, the infinite algebra of QFT in a region is formed
by infinitely many products of subalgebras and the group representation
is closed under fusion. Hence, the frequency of each irreducible representation
must be that of the regular representation. In the regular representation,
the basis elements are treated on equal footing by the group transformations,
and the subspace of the irreducible representation $\sigma$ has dimension
$d_{\sigma}^{2}$. Then, the probability of each irreducible sector
in vacuum must be given by \eqref{proji}.

\subsection{Twist version: upper bound\label{upper}}

The simplest upper bound for $\Delta I$ uses the convexity property
of the RE (\ref{c2-e-re-co} in proposition \ref{c2_re_prop}). To use
this property in the present context, we note that
\begin{equation}
\omega_{12}\circ\varepsilon_{12}=\frac{1}{|G|^{2}}\sum_{g_{1}\in G_{1},g'_{2}\in G_{2}}\omega_{g_{1}g'_{2}}=\frac{1}{|G|}\sum_{g_{1}\in G_{1}}\omega_{g_{1}}\,,\label{accor}
\end{equation}
where $\omega_{g_{1}g'_{2}}\left(\cdot\right):=\omega\left(\tau_{g}^{1\,\dagger}\tau_{g'}^{2\,\dagger}\cdot\tau_{g}^{1}\tau_{g'}^{2}\right)$
and $\tau_{g}^{j}$ are the twist operators of the regions $W_{j}$
($j=1,2$), which they commute between each other. In expression \eqref{accor},
the group transformations act on the two regions $W_{1}$ and $W_{2}$
independently. In the second equality we have used the invariance
of $\omega$ under global group transformations, which implies that
$\omega_{g_{1}g_{2}}=\omega_{g_{1}}$. Using the convexity property
we get
\begin{equation}
\sum_{g_{1}\in G_{1}}\frac{1}{|G|}S_{\mathfrak{F}}(\omega_{g_{1}}|\omega_{12}\circ\varepsilon_{12})-S_{\mathfrak{F}}\Bigl(\frac{1}{|G|}\sum_{g_{1}\in G_{1}}\omega_{g_{1}}|\omega_{12}\circ\varepsilon_{12}\Bigr)\leq-\sum_{g\in G}\frac{1}{|G|}\log\left(\frac{1}{|G|}\right)=\log|G|.\label{tirsos}
\end{equation}
The second RE in \eqref{tirsos} vanishes while the
REs
\begin{equation}
S_{\mathfrak{F}}(\omega_{g_{1}}|\omega_{12}\circ\varepsilon_{12})=S_{\mathfrak{F}}\Bigl(\omega_{g_{1}}|\frac{1}{|G|}\sum_{g_{1}\in G_{1}}\omega_{g_{1}}\Bigr)
\end{equation}
for different $g_{1}\in G_{1}$, are all equal because we can transform
any one into any other by a group automorphism. Therefore, we get the
upper bound\footnote{This upper bound might be considered as an intertwiner or twist upper
bound, depending on the focus one is taking. But this bound is not
tight in general. The tightest upper bound, which we will derive
below, comes from analyzing the problem from a twist perspective.} 
\begin{equation}
I_{\mathfrak{F}}(W_{1},W_{2})-I_{{\cal A}}(W_{1},W_{2})=S_{\mathfrak{F}}(\omega_{12}|\omega_{12}\circ\varepsilon_{12})\le\log|G|\,,\label{supp}
\end{equation}
which together with the lower bound of the previous section, allows
us to conclude that, as the two entanglement surfaces touch each other,
the bound becomes saturated for a finite group $G$
\begin{equation}
I_{\mathfrak{F}}(W_{1},W_{2})-I_{{\cal A}}(W_{1},W_{2})=\log|G|\,.\label{fini}
\end{equation}
Defining the \textit{quantum dimension} ${\cal D}$ by ${\cal D}^{2}:=\sum_{\sigma\in\hat{G}}d_{\sigma}^{2}=|G|$,
we can also write this result in the form 
\begin{equation}
\Delta I=\log({\cal D}^{2})\,,
\end{equation}
and for the regularized entropy
\begin{equation}
\Delta S=\frac{\Delta I}{2}=\log({\cal D})\,.
\end{equation}
Written in this way, the contribution coincides with the formula for
the topological entanglement entropy \cite{Kitaev:2005dm,Levin:2006zz}.

It is interesting to note that \eqref{fini} is a purely topological
contribution and does not depend on the interactions or whether the
models are massive or massless. Of course, the size $\epsilon:=\mathrm{dist}(\gamma_{W_{1}},\gamma_{W_{2}})$
between the entanglement surfaces of $W_{1}$ and $W_{2}$ where saturation
is achieved, depends on the typical size where the intertwiners have
appreciable expectation values. For a conformal theory and two double
cones, $\Delta I$ will be a function of the cross-ratio determining
the geometry, while for a massive theory we need to cross the scale
of the energy gap to see some difference between the MIs
to arise, independently of the size of the regions $W_{1},W_{2}$. 

Moreover, as we have explained above, the order parameter $S_{\mathfrak{F}}(\omega_{12}|\omega_{12}\circ\varepsilon_{12})$
could be in principle computed in the observable algebra $\mathcal{A}$
itself without further necessity of the field algebra nor of the knowledge
of the structure of the DHR SS. At the end, if for a given model we
were able to compute it in the limit when both regions touch each
other, we would know if the underlying model contains DHR SS and which
is the size $|G|$ of the symmetry group $G$. This gives a novel
connection between SS and EE.

According to the derivation of \eqref{tirsos}, saturation is only
possible if the supports for the states $\omega_{g}$ become disjoint
for different $g\in G$. This requires the vacuum expectation values
for the squeezed twists that implement group operations in $W_{1}$
and not in $W_{2}$ to go to zero in this limit. We will see later
this is also implied by uncertainty relations between twist and intertwiners
that do not commute with each other.

An improved upper bound can be obtained by considering the dual version
of \eqref{labe} where the RE is based on the complementary algebra
of the two regions $S:=\left(W_{1}\vee W_{2}\right)'$, which is called the
\textit{shell}.\footnote{The term shell is motivated in the case when we want to compute the
regularized entropy for a double cone as we have explained in section
\ref{c3-sec-univee}. } This requires a more specific property that we could not find in
the mathematical literature. We are proving this property in the lattice
and taking the continuum limit afterward.

We again consider the algebra $\mathfrak{F}_{12}\cong\mathfrak{F}_{1}\hat{\otimes}\mathfrak{F}_{2}$
and we call $\mathfrak{F}_{S}:=\mathfrak{F''}_{12'}={\mathfrak{F}^{t}}\sp{\prime}_{12}$.\footnote{For the purpose of computing REs, we need that all
algebras be vN algebras. Hence the local algebras of unbounded regions
must be replaced by its double commutant. For example, the field algebra
of the shell $\mathfrak{F}_{S}=\mathfrak{F}_{12'}$ must be replaced
by $\mathfrak{F}''_{12'}$ and this later is equal to ${\mathfrak{F}^{t}}\sp{\prime}_{12}$
because of the twist duality.} The local algebras $\mathfrak{F}_{1}$, $\mathfrak{F}_{2}$ and $\mathfrak{F}_{S}$
commute with each other, and since $W_{1}\lor W_{2}\vee S$ fulfills
the whole spacetime $\mathbb{R}^{d}$, then we have that 
\begin{equation}
\mathfrak{F}_{1}\vee\mathfrak{F}_{2}\vee\mathfrak{F}_{S}=\mathcal{B}\left(\mathcal{H}\right)\,.
\end{equation}
The twist operators on the region $W_{1}$, which form the group
algebra, are denoted by $G_{\tau}$. The invariant part of $\mathfrak{F}_{12}$
under $G_{\tau}$ is ${\cal A}_{1}\vee\mathfrak{F}_{2}\cong{\cal A}_{1}\otimes\mathfrak{F}_{2}$.
The commutant of this algebra is $({\cal A}_{1}\vee\mathfrak{F}_{2})'=\mathfrak{F}_{S}\vee G_{\tau}$.
We have two conditional expectations. The first one is 
\begin{equation}
\varepsilon_{1}:\mathfrak{F}_{1}\hat{\otimes}\mathfrak{F}_{2}\rightarrow{\cal A}_{1}\otimes\mathfrak{F}_{2}
\end{equation}
 which follows as \eqref{cee} but where now we act with the twists
$\tau_{g}\in G_{\tau}$ in region $W_{1}$. The ``dual'' conditional
expectation maps 
\begin{equation}
\varepsilon_{\tau}:\mathfrak{F}_{S}\vee G_{\tau}\rightarrow\mathfrak{F}_{S}\,.
\end{equation}
To describe the action of $\varepsilon_{\tau}$ we note that any element
$A\in\mathfrak{F}_{S}\vee G_{\tau}$ can be uniquely written as $A=\sum_{g\in G}A_{g}\,\tau_{g}$,
where the $A_{g}\in\mathfrak{F}_{S}$. Then, we take 
\begin{equation}
\varepsilon_{\tau}(A):=A_{1}\,.\label{c6-ce-dual}
\end{equation}
The conditional expectation $\varepsilon_{\tau}$
can be obtained directly with the help of the charge creating operators
$V_{\mathrm{reg}}^{h}\in\mathfrak{F}(W_{1})$, corresponding to the
regular representation, in the following way
\begin{eqnarray}
\frac{1}{|G|}\sum_{h\in G}V_{\mathrm{reg}}^{h\,\dagger}AV_{\mathrm{reg}}^{h} \!\!\! &=& \!\!\!\frac{1}{|G|}\sum_{h\in G}\sum_{g\in G}A_{g}\,V_{\mathrm{reg}}^{h\,\dagger}\tau_{g}V_{\mathrm{reg}}^{h} \nonumber \\
&=& \!\!\! \frac{1}{|G|}\sum_{h\in G}\sum_{g\in G}A_{g}\,V_{\mathrm{reg}}^{h\,\dagger}V_{\mathrm{reg}}^{gh}\tau_{g}=A_{1}=\varepsilon_{\tau}(A)\,.
\end{eqnarray}
Furthermore, the definition of ${\cal F}_{S}\vee G_{\tau}$ and $\varepsilon_{\tau}$
does not depend on the precise form of the twists chosen. Moreover,
in the continuum QFT, it can be shown that \eqref{c6-ce-dual} is
the unique conditional expectation for the inclusion of algebras $\mathfrak{F}_{S}\subset\mathfrak{F}_{S}\vee G_{\tau}$.\footnote{This is non-longer true in the lattice QFT.}

For simplicity we take $W_{1}$ and $W_{2}$ to be two disjoint sets
of vertices on the lattice and take $\mathfrak{F}_{1}$ and $\mathfrak{F}_{2}$ as the algebras of all lattice operators at these vertices. These
algebras are in tensor product with the rest of the lattice operators,
which means 
\begin{eqnarray}
\mathcal{H}\!\!\! & := & \!\!\!\mathcal{H}_{1}\otimes\mathcal{H}_{2}\otimes\mathcal{H}_{S}\,,\\
\mathcal{B}\left(\mathcal{H}\right)\!\!\! & := & \!\!\!\mathfrak{F}_{1}\hat{\otimes}\mathfrak{F}_{2}\hat{\otimes}\mathfrak{F}_{S}\,,\qquad\quad\mathfrak{F}_{j}:=\mathcal{B}\left(\mathcal{H}_{j}\right),\:j=1,2,S\,.
\end{eqnarray}
In this setting, we can choose $G_{\tau}$ as the elements of
the group acting on the vertices of $W_{1}$, such that $G_{\tau}$
commutes with $\mathfrak{F}_{S}$. In this way, we define
\begin{equation}
\tau_{g}:=\bigotimes_{n\in W_{1}}U_{n}\left(g\right)\,.
\end{equation}
Then, we have that $G_{\tau}\subset\mathfrak{F}_{1}$ and $\mathfrak{F}_{S}\vee G_{\tau}=\mathfrak{F}_{S}\otimes G_{\tau}$. 

Because of the invariance of the global vacuum state, we have as in
\eqref{accor} 
\begin{equation}
S_{\mathfrak{F}_{12}}(\omega|\omega\circ\varepsilon_{12})=S_{\mathfrak{F}_{12}}(\omega|\omega\circ\varepsilon_{1})\,.
\end{equation}
In this lattice setting, using \eqref{estasi} we get
\begin{equation}
S_{\mathfrak{F}_{12}}(\omega|\omega\circ\varepsilon_{1})=S_{\mathfrak{F}_{12}}(\omega\circ\varepsilon_{1})-S_{\mathfrak{F}_{12}}(\omega)\,.\label{c6-rel-as-ee}
\end{equation}
Since the global vacuum state $\omega$ is pure, we can use lemma
\ref{c2_vn_dual} twice and transform \eqref{c6-rel-as-ee} successively
as
\begin{eqnarray}
\hspace{-1cm} S_{\mathfrak{F}_{12}}(\omega|\omega\circ\varepsilon_{1})\!\!\! & = & \!\!\!S_{\mathfrak{F}_{12}}(\omega\circ\varepsilon_{1})-S_{\mathfrak{F}_{S}}(\omega)\nonumber \\
 & = & \!\!\!S_{{\cal A}_{1}\otimes\mathfrak{F}_{2}}(\omega)-S_{\mathfrak{F}_{S}}(\omega)+\left(S_{\mathfrak{F}_{12}}(\omega\circ\varepsilon_{1})-S_{{\cal A}_{1}\otimes\mathfrak{F}_{2}}(\omega)\right)\nonumber \\
 & = & \!\!\!S_{\mathfrak{F}_{S}\vee G_{\tau}}(\omega)-S_{\mathfrak{F}_{S}}(\omega)+\left(S_{\mathfrak{F}_{12}}(\omega\circ\varepsilon_{1})-S_{{\cal A}_{1}\otimes\mathfrak{F}_{2}}(\omega)\right)\nonumber \\
 & = & \!\!\!S_{\mathfrak{F}_{S}\vee G_{\tau}}(\omega)-S_{\mathfrak{F}_{S}\vee G_{\tau}}(\omega\circ\varepsilon_{\tau})\nonumber \\
 &  & \!\!\!+\left(S_{\mathfrak{F}_{12}}(\omega\circ\varepsilon_{1})-S_{{\cal A}_{1}\otimes\mathfrak{F}_{2}}(\omega)\right)+\left(S_{\mathfrak{F}_{S}\vee G_{\tau}}(\omega\circ\varepsilon_{\tau})-S_{\mathfrak{F}_{S}}(\omega)\right)\,.\label{c6-s12p}
\end{eqnarray}
Since the conditional expectation $\varepsilon_{\tau}$ does not preserve
the trace unless the group is Abelian, we cannot convert the first
two terms into a RE using \eqref{estasi}. However, we can use the fact that $\omega\circ\varepsilon_{\tau}$ is
a product state on $\mathfrak{F}_{S}\otimes G_{\tau}$. In fact, this
state is equal to $\omega_{S}\otimes\varphi$, where $\omega_{S}:=\left.\omega\right|_{\mathfrak{F}_{S}}$
and $\varphi$ is the state on $G_{\tau}$ defined by $\varphi(\tau_{g}):=\delta_{g,1}$.
Then, we write 
\begin{equation}
S_{\mathfrak{F}_{S}\vee G_{\tau}}(\omega)-S_{\mathfrak{F}_{S}\vee G_{\tau}}(\omega\circ\varepsilon_{\tau})=-S_{\mathfrak{F}_{S}\vee G_{\tau}}(\omega|\omega\circ\varepsilon_{\tau})+\Delta\langle K_{\varphi}\rangle\,,
\end{equation}
where $\Delta\langle K_{\varphi}\rangle:=-\omega(\log\rho_{\varphi}^{G_{\tau}})+\varphi(\log\rho_{\varphi}^{G_{\tau}})$.
Replacing this last equation into \eqref{c6-s12p}
\begin{eqnarray}
 &  & S_{\mathfrak{F}_{12}}(\omega|\omega\circ\varepsilon_{12})+S_{{\cal \mathfrak{F}}_{S}\vee G_{\tau}}(\omega\circ\varepsilon_{\tau})\nonumber \\
 &  & \hspace{1cm}=\Delta\langle K_{\varphi}\rangle+\left(S_{\mathfrak{F}_{12}}(\omega\circ\varepsilon_{1})-S_{{\cal A}_{1}\otimes\mathfrak{F}_{2}}(\omega)\right)+\left(S_{\mathfrak{F}_{S}\vee G_{\tau}}(\omega\circ\varepsilon_{\tau})-S_{\mathfrak{F}_{S}}(\omega)\right). \hspace{2cm} \label{c6-cer-rel-antes}
\end{eqnarray}
Both two last terms within parenthesis on the r.h.s. are formed by differences of entropies between states that are invariant under the
conditional expectations but computed in the algebra and its fix-point
subalgebra. The last term within brackets
gives the entropy of the state $\varphi$ on the group algebra $G_{\tau}$,
since the state $\omega\circ\varepsilon_{\tau}$ is a tensor product
state on $\mathfrak{F}_{S}\otimes G_{\tau}$. To compute it we note that
the group algebra is a sum of full matrix algebras
\begin{equation}
G_{\tau}\simeq\bigoplus_{\sigma\in\hat{G}}M_{d_{\sigma}}\left(\mathbb{C}\right)=\mathfrak{A}_{G}\,,
\end{equation}
which dimensions $d_{\sigma}$ corresponding to the irreducible representations
$\sigma\in\hat{G}$. The projectors onto the different blocks are the
$P_{\sigma}$ in \eqref{projec}, which have expectation values $\varphi\left(P_{\sigma}\right)=d_{\sigma}^{2}/|G|$.
Then, the density matrix is block diagonal with elements $d_{\sigma}/|G|$
on the diagonal in each block. The entropy is then
\begin{equation}
S_{\mathfrak{F}_{S}\vee G_{\tau}}(\omega\circ\varepsilon_{\tau})-S_{\mathfrak{F}_{S}}(\omega)=S_{G_{\tau}}(\varphi)=\log|G|-\sum_{\sigma\in\hat{G}}\frac{d_{\sigma}^{2}}{|G|}\log d_{\sigma}\,.\label{360}
\end{equation}
To evaluate the first parenthesis on the r.h.s. of \eqref{c6-cer-rel-antes},
we must notice that ${\cal A}_{1},G_{\tau}\subset\mathfrak{F}_{1}$ and they are respective commutants inside inside $\mathfrak{F}_{1}$, i.e.
\begin{equation}
G_{\tau}={\cal A}'_{1}\cap\mathfrak{F}_{1}\quad\textrm{and}\quad{\cal A}{}_{1}=G'_{\tau}\cap\mathfrak{F}_{1}\,,
\end{equation}
and its common center $\mathcal{Z}\left(\mathcal{A}_{1}\right)=\mathcal{Z}\left(G_{\tau}\right)$
is again formed by the algebra of projectors $P_{\sigma}$ spanning the center of the group algebra. Then, diagonalizing these projectors,
we have a representation ${\cal A}_{1}\lor G_{\tau}\cong\oplus_{\sigma\in\hat{G}}M_{d_{\sigma}}\left(\mathbb{C}\right)\otimes M_{n_{\sigma}}\left(\mathbb{C}\right)$,
where the group acts as $D_{\sigma}(g)$ in each block in the first
factor and $M_{n_{\sigma}}\left(\mathbb{C}\right)$ represents matrix
algebras of invariant operators of $\mathcal{A}_{1}$. An invariant
state like $\omega\circ\varepsilon_{1}$ has a density matrix
\begin{equation}
\rho_{\omega\circ\varepsilon_{1}}^{\mathfrak{F}_{1}}=\bigoplus_{\sigma\in\hat{G}}\left(q_{\sigma}\,\frac{\mathbf{1}_{d_{\sigma}}}{d_{\sigma}}\otimes\rho_{\sigma}\right),
\end{equation}
where $q_{\sigma}:=\omega(P_{\sigma})$ are the frequencies with which
each irreducible sector appears in the algebra $\mathfrak{F}_{1}$,
and $\rho_{\sigma}$ are density matrices in $M_{n_{\sigma}}\left(\mathbb{C}\right)$.
Then, we get 
\begin{equation}
S_{\mathfrak{F}_{12}}(\omega\circ\varepsilon_{1})-S_{{\cal A}_{1}\otimes\mathfrak{F}_{2}}(\omega)=\sum_{\sigma\in\hat{G}}q_{\sigma}\log d_{\sigma}\,.\label{c6-S1}
\end{equation}
Moreover, taking into account that the vacuum state is invariant under global
group symmetries, we have that the statistical operator corresponding
to the state $\left.\omega\right|_{G_{\tau}}$ must be
\begin{equation}
\rho_{\left.\omega\right|_{G_{\tau}}}=\bigoplus_{\sigma\in\hat{G}}q_{\sigma}\frac{\mathbf{1}_{d_{\sigma}}}{d_{\sigma}}\,.
\end{equation}
Then, we can write
\begin{equation}
\Delta\langle K_{\varphi}\rangle=-\sum_{\sigma\in\hat{G}}q_{\sigma}\log d_{\sigma}+\sum_{\sigma\in\hat{G}}\frac{d_{\sigma}^{2}}{|G|}\log d_{\sigma}\,.\label{c6-deltaK}
\end{equation}
Therefore, replacing \eqref{360}, \eqref{c6-S1}, and \eqref{c6-deltaK}
into \eqref{c6-cer-rel-antes}, we finally arrive to
\begin{equation}
S_{\mathfrak{F}_{12}}(\omega|\omega\circ\varepsilon_{12})=\log|G|-S_{{\cal \mathfrak{F}}_{S}\vee G_{\tau}}(\omega|\omega\circ\varepsilon_{\tau})\,.\label{sedien}
\end{equation}
Since this relation holds in any lattice discretization, it should
also hold in the continuum limit. This is because the terms in the
equation are all well-defined in such a limit. Finally, collecting
all the results together, we arrive to
\begin{equation}
I_{\mathfrak{F}}(W_{1},W_{2})-I_{{\cal A}}(W_{1},W_{2})=\log|G|-S_{{\cal \mathfrak{F}}_{S}\vee G_{\tau}}(\omega|\omega\circ\varepsilon_{\tau})\,.\label{sedaa}
\end{equation}
If $W_{1}$ and $W_{2}$ become larger, the l.h.s. of \eqref{sedaa}
increases and the RE on the shell must decreases. This is why the
RE on the algebra ${\cal \mathfrak{F}}_{S}\vee G_{\tau}$ appears with a minus
sign in \eqref{sedaa}.

Equation \eqref{sedaa} again expresses an upper bound of $\log|G|$
to $\Delta I$, but this is a finer upper bound since it is improved
by the RE on the r.h.s. of \eqref{sedaa}. As in the case of the intertwiners,
we can take any subalgebra $\mathcal{B}_{S\tau}\subset{\cal F}_{S}\vee G_{\tau}$
to get a convenient upper bound 
\begin{equation}
I_{\mathfrak{F}}(W_{1},W_{2})-I_{{\cal A}}(W_{1},W_{2})\le\log|G|-S_{\mathcal{B}_{S\tau}}(\omega|\omega\circ\varepsilon_{\tau})\,.\label{sedien1}
\end{equation}
Any set of twists $\tau_{g}$ that close a representation of the group
form a linear basis for some choice of the algebra $G_{\tau}$.
Then, we can restrict to this algebra.\footnote{Note that these are smeared twists, as spread as possible, to increase their expectation
values, in contrast to the sharp twists we have used in the derivation
above.} In this case, recalling that in the twist algebra the states $\left.\omega\right|_{G_{\tau}}$
and $\varphi$ are represented by the statistical operators
\begin{equation}
\rho_{\left.\omega\right|_{G_{\tau}}}:=\bigoplus_{\sigma\in\hat{G}}q_{\sigma}\frac{\mathbf{1}_{d_{\sigma}}}{d_{\sigma}}\quad\mathrm{and}\quad\rho_{\varphi}:=\bigoplus_{\sigma\in\hat{G}}\frac{d_{\sigma}}{|G|}\,\mathbf{1}_{d_{\sigma}}\,,
\end{equation}
we get
\begin{equation}
I_{\mathfrak{F}}(W_{1},W_{2})-I_{{\cal A}}(W_{1},W_{2})\le-\sum_{\sigma\in\hat{G}}q_{\sigma}\log q_{\sigma}+\sum_{\sigma\in\hat{G}}q_{\sigma}\log(d_{\sigma}^{2})\,.\label{sed}
\end{equation}
Equation \eqref{sed} is the same as expression \eqref{twisted}, which is positive and bounded from above by $\log|G|$. It is a function
of the twist expectation values through (see \eqref{projec}) 
\begin{equation}
q_{\sigma}=\langle P_{\sigma}\rangle=\frac{d_{\sigma}}{|G|}\sum_{g\in G}\chi_{\sigma}^{*}(g)\langle\tau_{g}\rangle\,.\label{370}
\end{equation}
We get $\log|G|$ on the r.h.s. of \eqref{sedien1} for ``sharp'' twists
satisfying $\langle\tau_{g}\rangle=\delta_{g,1}$. These expectation
values imply the regular representation probabilities through the
previous relation. In a realistic scenario, the smallest upper bound
will be for the most spread out twists, where the expectation values
of the twists are bigger and the RE on the twist algebra is larger.
On the other side of the story, the upper bound goes to zero when
$\langle\tau_{g}\rangle=1$ for all $\tau_{g}$. This is the case
for the vacuum and the global group tra\textcolor{black}{nsformations,
which satisfy $q_{\sigma}=\delta_{\sigma,\mathrm{triv}}$}. Finally, we notice that for Abelian
groups \eqref{sed} is just the entropy on the twist algebra since
the second term vanishes. This is not the case of a non-Abelian group
where the entropy in the twist algebra is $-\sum_{\sigma\in\hat{G}}q_{\sigma}\log q_{\sigma}+\sum_{\sigma\in\hat{G}}q_{\sigma}\log(d_{\sigma})$
rather than \eqref{sed}. Hence, there is an additional contribution
in \eqref{sed}. This is necessary to match the intertwiner RE in the special cases where the upper and the lower bounds coincide.

We want to remark that the expression \eqref{sed} for an upper bound
should remain valid for continuous groups as far as the group is compact
and the statistics of the sectors give a finite result. In fact, it
will turn out this expression is generally finite for Lie group symmetries
in QFT.

\subsection{Entropic certainty relation\label{certainty}}

We recall equation \eqref{sedien} 
\begin{equation}
S_{\mathfrak{F}_{12}}(\omega|\omega\circ\varepsilon_{12})+S_{{\cal \mathfrak{F}}_{S}\vee G_{\tau}}(\omega|\omega\circ\varepsilon_{\tau})=\log|G|\,.\label{c6-cer}
\end{equation}
For large intertwiner expectation values (small separation between
the regions $W_{1}$ and $W_{2}$), the first relative entropy will
approach $\log|G|$, implying the twist one goes to zero, while the
opposite is true for large separation between the regions, where there
are some twists with large expectation values. Equation \eqref{c6-cer}
is an \textit{entropic certainty relation}. We remark a close relation
between the algebras involved in such a relation, that it could be
represented in the following diagram
\begin{equation}
\begin{array}{ccc}
\mathfrak{F}_{12} & \overset{\varepsilon_{12}}{\longrightarrow} & \!\!\!\mathcal{A}{}_{12}\\
\:\updownarrow\prime &  & \updownarrow\prime\\
{\cal \mathfrak{F}}_{S} & \overset{\varepsilon_{\tau}}{\longleftarrow} & {\cal \mathfrak{F}}_{S}\vee G_{\tau}\,.
\end{array}\label{c6-diag-cer}
\end{equation}
The horizontal arrows indicates the action of the conditional expectations,
whereas the vertical arrows indicates commutants. In other words,
the initial space of the conditional expectation $\varepsilon_{\tau}$
coincides with the commutant of the target space of $\varepsilon_{12}$
and viceversa. We also have that the underlying state is pure in the
global algebra $\mathfrak{F}$. Moreover, there is a strong relation
between the number $\left|G\right|$ and the algebras $\mathfrak{F}_{12}$
and $\mathcal{A}{}_{12}$. Given any inclusion of vN algebras and
a conditional expectation
\begin{equation}
\mathcal{N}\subset\mathcal{M}\subset\mathcal{B}\left(\mathcal{H}\right)\,,\quad\varepsilon:\mathcal{M}\rightarrow\mathcal{N}\,,\label{c6-incl-vn}
\end{equation}
it can be defined an algebraic index $\left[\mathcal{M}:\mathcal{N}\right]_{\varepsilon}\geq1$
which measures, in some sense, the ``size'' of the portion of $\mathcal{M}$
that is killed by the conditional expectation $\varepsilon$ \cite{Longo:1989tt,Longo:1990zp,Longo:1994xe}.
In our case, we have exactly $[\mathfrak{F}_{12}:\mathcal{A}{}_{12}]_{\varepsilon_{12}}=\left|G\right|$.
Moreover, we believe that the structure \eqref{c6-diag-cer} can be extended to more general scenarios as follows. Given the general inclusion of vN algebras and the conditional expectation \eqref{c6-incl-vn}, it should exists a ``dual'' conditional
expectation $\varepsilon':\mathcal{N}'\rightarrow\mathcal{M}'$, with the same index $\left[\mathcal{M}:\mathcal{N}\right]_{\varepsilon}=\left[\mathcal{N}':\mathcal{M}'\right]_{\varepsilon'}$, such that the following relation holds
\begin{equation}
S_{\mathcal{M}}(\omega|\omega\circ\varepsilon)+S_{\mathcal{N}}(\omega|\omega\circ\varepsilon')=\log\left[\mathcal{M}:\mathcal{N}\right]_{\varepsilon}\,,\label{c6-cer-gen}
\end{equation}
for any pure state on $\mathcal{B}\left(\mathcal{H}\right)$. We are not sure if this is true for any given conditional expectation $\varepsilon:\mathcal{M}\rightarrow\mathcal{N}$, but perhaps for the one that minimizes the index, which gives the lower possible value on the r.h.s. of \eqref{c6-cer-gen}. In the case of \eqref{c6-diag-cer}, the conditional expectations are the unique possible ones for such an inclusion of algebras. In fact, this relation have been proved for an arbitrary inclusion of finite dimensional algebras \cite{ecr2020}. The case involving general (infinite dimensional) algebras is left for future work.

If in \eqref{c6-cer} we restrict to subalgebras $\mathcal{B}_{12}$ and $\mathcal{B}_{G\tau}$, we obtain instead the \textit{entropic uncertainty
relation} 
\begin{equation}
S_{\mathcal{B}_{12}}(\omega|\omega\circ\varepsilon_{12})+S_{\mathcal{B}_{G\tau}}(\omega|\omega\circ\varepsilon_{\tau})\le\log|G|\,.
\end{equation}
In order to both relative entropies  being non-zero,
$\mathcal{B}_{12}$ (resp. $\mathcal{B}_{G\tau}$) must contain, at
least, some non-trivial subalgebra of intertwiners (resp. twists).

Similar entropic uncertainty relations occur for generalized measurements
\cite{coles2017entropic,berta2016entropic}. Notice that the maximal
RE for each term needs minimal uncertainty: expectation values of
the twist operators or the intertwiners equal to maximal ones. In
the case of minimal uncertainty, each RE can achieve $\log|G|$. Therefore, minimal uncertainty cannot be achieved at the same time
for intertwiners and twists. The non-trivial commutation relations
between twists and intertwiners is what prevents the left-hand side
of this inequality to reach $2\log|G|$, while $\log|G|$ is the maximum
that can be achieved for each of the two terms separately.

In the same way, if we have an impure global state that is invariant
under the group  transformations (i.e. a thermal state), we can purify it in a larger
Hilbert space (GNS representation) and upon reduction we get 
\begin{equation}
S_{\mathfrak{F}_{12}}(\omega|\omega\circ\varepsilon_{12})+S_{{\cal \mathfrak{F}}_{S}\vee G_{\tau}}(\omega|\omega\circ\varepsilon_{\tau})\le\log|G|\,.
\end{equation}

Uncertainty relations may be derived for operator expectation values
rather than entropies using the commutations relations between twists
and intertwiners. For example, in the case of the bosonic subnet of
the fermion net described above, we have just one twist and one intertwiner
satisfying
\begin{equation}
\tau\,{\cal I}_{12}=-{\cal I}_{12}\,\tau\,.
\end{equation}
The usual uncertainty relation for non-commuting operators gives 
\begin{equation}
|\langle\tau\rangle|^{2}+|\langle{\cal I}_{12}\rangle|^{2}\le1\,.
\end{equation}
Then, when the twist has maximal expectation value $|\langle\tau\rangle|=1$,
the expectation value of the intertwiner is zero, and viceversa.

It is important to emphasize, that relation \eqref{c6-cer} holds
not only for double cones but also for any pair of strictly spacelike
separated topological trivial regions. The better way to understand
such a kind of regions is fixing a Cauchy surface $\Sigma\subset\mathbb{R}^{d}$
and taking regions $\mathcal{O}_{j}:=D\left(\mathcal{C}_{j}\right)$
with spacelike regions $\mathcal{C}_{j}\subset\Sigma$. Then, \eqref{c6-cer}
holds for regions $\mathcal{C}_{j}$ which are topologically equivalent to a sphere
or to the complement of a sphere.\footnote{In algebraic topology, all the homotopy groups of the topological trivial region $\mathcal{C}_{j}$
are trivial, i.e. $\pi_{n}\left(\mathcal{C}_{j}\right)=0$ for all
$n=1,\ldots,d-1.$}

In fact, DHR intertwiners are non-trivial objects for disconnected
regions. Let us assume that the regions $\mathcal{O}_{1},\mathcal{O}_{2}\in\mathcal{K}$
are connected and strictly spacelike separated. The set of DHR intertwiners
between $\mathcal{O}_{1}$ and $\mathcal{O}_{2}$ would be always
the same regardless of the topology of $\mathcal{O}_{1}$ and $\mathcal{O}_{2}$
whenever both $\mathcal{O}_{1}$ and $\mathcal{O}_{2}$ remain connected.
However, for other topologies for the connected regions $\mathcal{O}_{1}$
and $\mathcal{O}_{2}$, a new different type of intertwiners may arise.
However, these intertwiners do not correspond to sectors localized
in double cones \cite{Buchholz:1981fj}. To summarize, relation \eqref{c6-cer} holds
always for topological trivial regions, and it also holds just for
connected regions whenever the theory has no other localized sectors
rather than the DHR ones. Of course the same happens for relations
\eqref{c6.-lower} and \eqref{sedaa}. In section \ref{OT}, we study
how the order parameter $\Delta I$ behaves for regions having different non-trivial
topologies.

\subsection{Lie groups\label{U1}}

When the symmetry group is not finite, $\Delta I(W_{1},W_{2})$ will
be divergent in the limit when the regions touch each other. The interest
lies in understanding how this quantity depends on the distance $\epsilon$
between the two regions $W_{1}$ and $W_{2}$, when $\epsilon\rightarrow0^{+}$.
This is important, for example, when we want to compute the universal
regularized EE explained in section \ref{c3-sec-univee}. With that
idea in mind and for simplicity, we consider regions $W_{1}:=D\left(\mathcal{C}_{R}\right)$
and $W_{2}:=D\left(\mathcal{C}'_{R+\epsilon}\right)$ ($\epsilon>0$).
Let us first analyze the case of a group $G:=U(1)$. We have a continuum
of twists
\begin{equation}
\tau_{k}:=\mathrm{e}^{ikQ_{1}}\,,\quad k\in[-\pi,\pi)\,,\label{c6-lie-twist}
\end{equation}
where $Q_{1}$ is the selfadjoint generator of the twist algebra crossing
the region $W_{1}$. Since the group $G$ is compact, the group representation
\eqref{c6-lie-twist} decomposes into irreducible representations
\begin{equation}
\mathcal{H}=\bigoplus_{q\in\mathbb{Z}}\mathcal{H}_{q}\,,\quad\tau_{k}=\sum_{q\in\mathbb{Z}}\mathrm{e}^{ikq}P_{q}\,,\label{c6-lie-deco}
\end{equation}
where $P_{q}$ is the projector in the subspace $\mathcal{H}_{q}$.

In general, computing the exact operators $Q_{1}$ and the expectation
values of the twists on a specific theory will be a problem depending
on the dynamics. However, we are interested in the limit $\epsilon\rightarrow0^{+}$, and we will argue that the leading divergent term is universal.
We know that inside the double cone $W_{1}$, the operator $Q_1$ may be expressed as
\begin{equation}
Q_{1}\sim\int d^{d}x\,J^{0}(x)\alpha(x)\,,
\end{equation}
where $J^{0}\left(x\right)$ is the charge density and $\alpha(x)$
is a convenient smearing function. This functions integrates to $1$
in time, it is spatially constant inside $W_{1}$ and vanish along
$W_{2}$. On the shell $S:=(W_{1}\vee W_{2})'$ the operator content
and the smearing changes, such as to give $\tau_k$ the desired group
properties.

For small $\epsilon$, the leading term of the total charge fluctuation
inside the ball will come from short distance charge fluctuations
distributed all along the entanglement surface, with a particle-antiparticle
on each side the corridor separating the two regions, see figure \ref{fig6}.
We can then picture the fluctuations of the total charge contributing
to $Q_{1}$ as given by a large sum of independent random variables
$Q_{1,n}$ ($n=1,\ldots,N$ and $N\gg1$), since short distance fluctuations
that are separated by a macroscopic distance along the surface of
the sphere will not see each other.\footnote{We will come back to this point in section \ref{twistcontinuo}, where we elaborate a little more about general properties of twists expectation values.}  Then, due to the central limit theorem, the total charge distribution
$Q_{1}:=\sum_{n=1}^{N}Q_{1,n}$ can be very well approximated by
a Gaussian distribution

\begin{equation}
p_{q}:=\left\langle P_{q}\right\rangle \simeq\frac{c}{\sqrt{2\pi\langle Q_{1}^{2}\rangle}}\mathrm{e}^{-\frac{q^{2}}{2\langle Q_{1}^{2}\rangle}}\,,\quad c>0\,,\label{prubi}
\end{equation}
where $\langle Q_{1}^{2}\rangle=\sum_{n=1}^{N}\langle Q_{1,n}^{2}\rangle\gg1$
for small $\epsilon$. Therefore, replacing these probabilities into \eqref{c6-lie-deco}, we get
\begin{equation}
\langle\tau_{k}\rangle=\sum_{q\in\mathbb{Z}}p_{q}\,\mathrm{e}^{iqk}\simeq\mathrm{e}^{-\frac{1}{2}k^{2}\langle Q_{1}^{2}\rangle}\,,\label{proyi}
\end{equation}
where we have used that $\langle Q_{1}^{2}\rangle\gg1$ in order to
approximate the above expression using a continuous Fourier transform.
This is why the result is not periodic in $k$, but it will hold
very approximately in the limit we are studying.

An upper bound for $\Delta I$ is then easily computed from \eqref{sed},
to be the entropy of this distribution\footnote{More precisely, the central limit theorem asserts that
\begin{equation}
\sum_{q\leq q_{0}}p_{q}\simeq\int_{-\infty}^{q_{0}}dq\,\frac{1}{\sqrt{2\pi\langle Q_{1}^{2}\rangle}}\mathrm{e}^{-\frac{q^{2}}{2\langle Q_{1}^{2}\rangle}}\,,\quad\langle Q_{1}^{2}\rangle\gg1\,,\label{c6-cl}
\end{equation}
and moreover, the characteristic function \eqref{proyi} converges
pointwise to
\begin{equation}
\sum_{q\in\mathbb{Z}}p_{q}\,\mathrm{e}^{iqk}\simeq\int_{-\infty}^{q_{0}}dq\,\frac{1}{\sqrt{2\pi\langle Q_{1}^{2}\rangle}}\mathrm{e}^{-\frac{q^{2}}{2\langle Q_{1}^{2}\rangle}}\mathrm{e}^{ikq}=\mathrm{e}^{-\frac{1}{2}k^{2}\langle Q_{1}^{2}\rangle}\,,\quad\langle Q_{1}^{2}\rangle\gg1,\,k\ll1\,.
\end{equation}} 
\begin{equation}
-\sum_{q\in\mathbb{Z}}p_{q}\log(p_{q})=1/2\log\langle Q_{1}^{2}\rangle+\textrm{constant}\,.\label{thesame}
\end{equation}
Notice that even if the twist algebra has a continuum of operators,
the upper bound is well-defined because it is the entropy of a classical
discrete set of charges, or equivalently, because the group is compact.
We expect that the difference of MIs is divergent in the
non-compact case. There should be no problem with the MI on ${\cal A}$, but however, the one on $\mathfrak{F}$ is the one not
well-defined in this case. We think that the problem is that $\mathfrak{F}$
contains too many sectors that would make fail the splitting property
that guarantees we can take the algebra of two regions as a tensor
product. This splitting property is related to the finiteness of a
nuclearity index \cite{haag}, which in turn is related to the partition
function. Similar observations have been made recently using other
arguments \cite{Harlow:2018tng}.

The best upper bound corresponds to the lowest $\langle Q_{1}^{2}\rangle$.
This corresponds to the most spread out twists. As the smearing function
on the shell becomes wider, the probability of charge fluctuations
on each side of the shell decreases, and the charge fluctuations inside
the smearing region are averaged to zero. We give a more direct calculation
of $\langle Q_{1}^{2}\rangle$ in section \ref{twistcontinuo} below.
Here we notice that the result must be proportional to the area
since bulk virtual fluctuations of the charge are suppressed because
they will appear with both signs and the total charge average zero.
For a current that is conformal in the UV, the area $A$ must be compensated
by powers of the cutoff, giving $\langle Q_{1}^{2}\rangle\sim A/\epsilon{}^{d-2}$. Then, we have that
\begin{equation}
\Delta I\le\frac{1}{2}\log\frac{A}{\epsilon^{d-2}}+\textrm{constant}\sim\frac{(d-2)}{2}\log\frac{R}{\epsilon}+\textrm{constant}\,.
\end{equation}

A lower bound can be given by using the intertwiners. There
is one intertwiner $\mathcal{I}_{q}$ for each number $q\in\mathbb{Z}$
representing the charge, which labels the irreducible representations
of the group. This Abelian intertwiner algebra $\mathcal{B}_{12}$
is $C^{*}$-isomorphic to the algebra of continuous periodic functions
in the interval in $\mathbb{R}$, or equivalently, to the algebra
of continuous functions in the unit circle $S^{1}$.\footnote{For any Abelian group $G$ the intertwiners are labeled by its representations,
and we can represent the Abelian algebra of the intertwiners with
the algebra of functions on the group. This coincides with the algebra
of the characters, which has the product law $\chi_{r_{1}}(g)\chi_{r_{2}}(g)=\chi_{r_{1}\otimes r_{2}}(g)$.} Each intertwiner is represented by the function
\begin{equation}
\mathcal{I}_{q}\mapsto f_{q}\left(k\right):=\mathrm{e}^{ikq}\,,\quad k\in[-\pi,\pi)\,.
\end{equation}
In fact, given any state $\psi$ on the Abelian algebra $\mathcal{B}_{12}$,
$(\mathcal{B}_{12},\psi)$ corresponds to the classical probability
space $\left(S^{1},\mathcal{B}(S^{1}),p_{\psi}\left(k\right)dk\right)$
where the probability measure $p_{\psi}(k)dk$ is given by the probability
density\footnote{$\left(S^{1},\mathcal{B}(S^{1})\right)$ corresponds to the measurable
space over $S^{1}$ generated by its Borel sets $\mathcal{B}\left(S^{1}\right)$. }
\begin{equation}
p_{\psi}\left(k\right):=\frac{1}{2\pi}\sum_{q\in\mathbb{Z}}\mathrm{e}^{ikq}\,\langle\mathcal{I}_{q}\rangle_{\psi}\,,\quad k\in[-\pi,\pi)\,.\label{c6-class-prob}
\end{equation}
Then, this Abelian algebra corresponds to a continuous classical random
variable on the space $[-\pi,\pi)$. Furthermore, the (quantum) relative
entropy between two states $\psi$ and $\phi$ on $\mathcal{B}_{12}$
is equivalent to the classical relative entropy (Kullback–Leibler
divergence) between the classical probabilities densities $p_{\psi}(k)$ and $p_{\phi}(k)$,
\begin{equation}
S_{\mathcal{B}_{12}}(\psi\mid\phi)=H_{KL}(p_{\psi}\mid p_{\phi}):=\int_{-\pi}^{\pi}p_{\psi}\left(k\right)\log\left(\frac{p_{\psi}\left(k\right)}{p_{\phi}\left(k\right)}\right).\label{c6-kld}
\end{equation}

In the present scenario, the state $\omega\circ\varepsilon_{12}$
is represented by the constant probability density $p_{\omega\circ\varepsilon_{12}}\left(k\right):=1/(2\pi)$
since we have that $\omega\circ\varepsilon_{12}\left(\mathcal{I}_{q}\right)=0$
for all $q\neq0$. The probability density $p_{\omega}\left(k\right)$
corresponding to the state $\omega$ depends on the specific algebra
of intertwiners we choose. We have to select it in order to maximize
the relative entropy $S_{\mathcal{B}_{12}}(\omega\mid\omega\circ\varepsilon_{12})$.
This can be achieved by concentrating the probability $p_{\omega}\left(k\right)$
around $k=0$ as much as possible. This means that the probability of the
different charges is as much flat as possible. In particular, to sense
the probability distribution \eqref{prubi} of charge vacuum fluctuations,
our intertwiners will have to be spread out on the surface of the
sphere. Any smaller localization will lead to a less flat distribution
of probabilities of charges. Heuristically, the intertwiner $\mathcal{I}_{q}$
will then carry a state $\sqrt{p_{q_{0}}}\,|q_{0}\rangle_{1}\otimes|-q_{0}\rangle_{2}$
to $\sqrt{p_{q_{0}}}\,|q_{0}+q\rangle\otimes|-q_{0}-q\rangle_{2}$.
The expectation value will be
\begin{equation}
\langle{\cal I}_{q}\rangle\simeq\sum_{q_{0}\in\mathbb{Z}}\sqrt{p_{q_{0}}p_{q_{0}+q}}=\mathrm{e}^{-\frac{q^{2}}{8\langle Q_{1}^{2}\rangle}}\,.\label{distic}
\end{equation}
According to \eqref{c6-class-prob}, the classical probability $p_{\omega}\left(k\right)$
is therefore
\begin{equation}
p_{\omega}\left(k\right)\simeq\sqrt{\frac{2\langle Q_{1}^{2}\rangle}{\pi}}\mathrm{e}^{-2\langle Q_{1}^{2}\rangle k^{2}}\,,\quad\langle Q_{1}^{2}\rangle\gg1\,.
\end{equation}
and following \eqref{c6-kld} the relative entropy $S_{\mathcal{B}_{12}}(\omega\mid\omega\circ\varepsilon_{12})$
gives the same leading order calculation \eqref{thesame}. We then
get that the asymptotic behavior is
\begin{equation}
\Delta I\simeq\frac{1}{2}\log\frac{A}{\epsilon^{d-2}}\sim\frac{(d-2)}{2}\log\frac{R}{\epsilon}\,,\quad\epsilon\rightarrow0^{+}\,.\label{epifa}
\end{equation}
This term should be attributed to the model $\mathcal{A}$ as a contribution
$-\frac{(d-2)}{2}\log\frac{R}{\epsilon}$ to the MI. This logarithmic term is ``topological'' in the sense that it appearsin any dimension $d>2$, and it does not depend
on the curvature of the boundary as the usual logarithmic anomaly terms. 

In $d=2$ we have to replace $(R/\epsilon)^{(d-2)}\rightarrow\log(R/\epsilon)$, and the leading term is
\begin{equation}
\Delta I\simeq\frac{1}{2}\log(\log(R/\epsilon))\,,\quad\epsilon\rightarrow0^{+}\,.\label{cft}
\end{equation}
However, in $d=2$ this is correct for two intervals that touch each
other, while in the case of nearly complementary regions the shell
consists of two intervals and the coefficient gets duplicated for
massive fields, while it is still \eqref{cft} for CFTs (see section
\ref{dosd}).

For a non-Abelian compact Lie group $G$, we have different twist
generators $L_{i}$, $i=1\,\cdots,{\cal G}$, where ${\cal G}$ is
the dimension of its Lie algebra. For each of these charges we expect
to have a Gaussian probability of charges as in \eqref{prubi} for
the same reasons as above. The group is non-commutative though. However,
the typical expectation values of the charges are very large in the
limit of small $\epsilon$, and therefore we are in the regime of
``large numbers'' where the non-commutativity is not relevant. Then, the intertwiner version gives us a picture of ${\cal G}$ independent
charges with 
\begin{equation}
\Delta I\simeq\frac{1}{2}\,(d-2)\,{\cal G}\,\log\frac{R}{\epsilon}\,,\quad\epsilon\rightarrow0^{+}\,.\label{nona}
\end{equation}
The twist version matches this expectation but there is a subtlety.
A twist $\mathrm{e}^{ik_{i}L_{i}}$ has appreciable expectation value
only for small parameters $k_{i}$ as in \eqref{proyi}. This means
only the neighborhood of the identity is probed in the group. Therefore
we might expect to have effectively the case of ${\cal G}$ Abelian
generators. This is correct, but the conditional expectation knows
that these different directions in the Lie algebra can be connected
by group transformations and cannot be considered independent. Hence,
the vN entropy in the group algebra is in fact smaller than what
is expected for the case of ${\cal G}$ Abelian generators. However,
the formula \eqref{sed} contains an additional piece on top of the
twist entropy in the non-Abelian case, and taking into account this
contribution, the calculation with the twists matches the expectation
\eqref{nona} from the intertwiners.

Let us see how this work in a concrete example. Consider the case
of $G:=SO\left(3\right)$. According to the discussion above, for
small $\epsilon$, the expectation values of the twist $\tau_{g}$
are non-zero only for those corresponding to group elements near the
identity element. In this situation, as in the Abelian
case, it is useful to parametrized the twist operators with a 3-vector
$\bar{k}$ according to 
\begin{equation}
\tau_{\bar{k}}=\mathrm{e}^{i\bar{k}\cdot\bar{L}}\,,\;\label{twist_so3}
\end{equation}
where $\bar{L}:=(L_{1},L_{2},L_{3})$ are like angular momentum operators
with commutation relations $\left[L_{j},L_{k}\right]=i\epsilon_{jkl}L_{l}$
. As argued above, the vacuum expectation values of such twist operators
in the small $\epsilon$ limit is Gaussian, and has to be rotationally
invariant 
\begin{equation}
\left\langle \tau_{\bar{k}}\right\rangle \simeq\mathrm{e}^{-\frac{1}{2}\left|\bar{k}\right|^{2}\left\langle \bar{L}^{2}\right\rangle }\,.\label{exp_so3}
\end{equation}
Then, it behaves as if they were the twist operators associated with
three independent generators of the Abelian group $U(1)^{3}$. The
computation using these expectation values is straightforward. First,
we have that the irreducible representations of $SO\left(3\right)$
are labeled by a non-negative integer $l\in\mathbb{Z}_{\geq0}$. The
$l$-representation has dimension $d_{l}:=\left(2l+1\right)$ and
its character $\chi_{l}$ is given by \cite{hamermesh2012group} 
\begin{equation}
\chi_{l}\left(\theta\right):=\frac{\sin\left(\left(l+\frac{1}{2}\right)\left|\theta\right|\right)}{\sin\left(\frac{1}{2}\left|\theta\right|\right)}\,,\label{char_so3}
\end{equation}
where $\theta$ is the angle of rotation from the identity. This coincides
with $\theta\simeq|\vec{k}|$ for small $\theta$. To compute the
desired upper bound using equation \eqref{sed}, we need first to
calculate the probabilities $q_{l}$ attached to the $l$-representation.
For that we use the Lie group continuum version of \eqref{370} 
\begin{equation}
q_{l}:=\left(2l+1\right)\frac{1}{\pi}\int_{0}^{\pi}dk\,(1-\cos(\theta))\chi_{l}\left(\theta\right)\,\mathrm{e}^{-\frac{1}{2}\theta^{2}\left\langle \bar{L}^{2}\right\rangle }\,,\label{ql_so3}
\end{equation}
where the finite sum was replaced by the integral over the full group
$SO\left(3\right)$ using the normalized Haar measure (see \cite{hamermesh2012group}),
and we are assuming $\langle\bar{L}^{2}\rangle\gg1$. Replacing \eqref{char_so3}
into \eqref{ql_so3}, we can compute analytically the probabilities,
which are given in terms of $\textrm{Erf}$ functions. At the end,
replacing such probabilities into \eqref{sed} we can check 
\begin{equation}
I_{\mathcal{F}}\left(1,2\right)-I_{\mathcal{\mathcal{O}}}\left(1,2\right)\leq-\sum_{l=0}^{\infty}q_{l}\log\left(q_{l}\right)+\sum_{l=0}^{\infty}q_{l}\log\left(d_{l}^{2}\right)\sim\frac{3}{2}\log\left\langle \bar{L}^{2}\right\rangle +\mathrm{constant}\,,\label{bound_so3}
\end{equation}
as we have claimed above.

We notice that each term in \eqref{bound_so3}, for large $\left\langle \bar{L}^{2}\right\rangle $
(small $\epsilon$), reads
\begin{eqnarray}
-\sum_{l=0}^{\infty}q_{l}\log\left(q_{l}\right)\!\!\! & \sim & \!\!\!\frac{1}{2}\log\left\langle \bar{L}^{2}\right\rangle \,,\\
2\sum_{l=0}^{\infty}q_{l}\log\left(d_{l}\right)\!\!\! & \sim & \!\!\!\log\left\langle \bar{L}^{2}\right\rangle \,.
\end{eqnarray}
For an invariant state $\omega$, the density matrix for the twist algebra $G_{\tau}$ decomposes according to the irreducible representations as 
\begin{equation}
\rho_{\omega}:=\bigoplus_{l\in\mathbb{Z}_{\geq0}}q_{l}\cdot\frac{\mathbf{1}_{d_{l}}}{d_{l}}\,,\label{dm_so3}
\end{equation}
where $\mathbf{1}_{d_{l}}$ is the identity matrix in the full matrix
algebra $M_{d_{l}}\left(\mathbb{C}\right)$. The vN entropy on this
algebra is then 
\begin{equation}
S_{G_{\tau}}\left(\omega\right)=-\sum_{l=0}^{\infty}q_{l}\log\left(q_{l}\right)+\sum_{l=0}^{\infty}q_{l}\log\left(d_{l}\right)\,.\label{asa}
\end{equation}
Then, this entropy contributes only with a $\log\langle\bar{L}^{2}\rangle$
and the missing $1/2\log\langle\bar{L}^{2}\rangle$ comes from the
fact that the last term in \eqref{asa} has a factor $2$ in the correct
formula \eqref{bound_so3}. This is in contrast with the Abelian case,
where \eqref{bound_so3} gives the entropy in the twist algebra.

\subsection{Other topologies\label{OT}}

\begin{figure}[t]
\centering
\begin{centering}
\includegraphics[width=12cm]{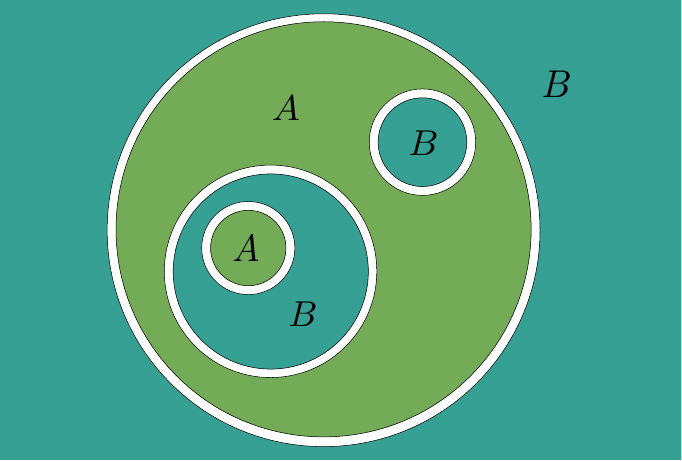} 
\par\end{centering}
\centering{}\captionsetup{width=0.9\textwidth} \caption{\foreignlanguage{english}{\label{topofig}Two almost complementary regions $A$ and $B$ with
non trivial topology. There is an independent set of intertwiners
and twists for each connected component of the common boundary between
$A$ and $B$ (white region). In this case, the number of connected
components of the boundary is $n_{\partial}=4$.}}
 \end{figure}

The same type of ideas can be used to try to understand the difference
of MIs between the models $\mathfrak{F}$ and ${\cal A}$ for regions
with different topologies, such as the one shown in figure \ref{topofig}.
We remember that when we derived formula \eqref{labe} using the conditional
expectation property \eqref{c2_ce_prop} for the two double cones
$W_{1}$ and $W_{2}$, we used the relation
\begin{equation}
\left(\phi_{1}\otimes\phi_{2}\right)\circ\left(\varepsilon_{1}\otimes\varepsilon_{2}\right)=\omega_{1}\otimes\omega_{2}\,,
\end{equation}
where $\omega_{j}$ and $\phi_{j}$ are the restriction of the vacuum
state to the algebras $\mathfrak{F}(W_{j})$ and $\mathcal{A}(W_{j})$
respectively as in \eqref{c6-states1} and \eqref{c6-states2}. However,
this is no longer true if any of the regions are formed by different
connected components. For example, let us suppose that $W_{1}$ is replaced
for region $\mathcal{O}_{1}:=W_{1,1}\vee W_{1,2}$ where $W_{1,k}$
are strictly separated double cones. Now, in this case, we may use
of the conditional expectation $\varepsilon_{1}:\mathfrak{F}(W_{1,1}\vee W_{1,2})\rightarrow\mathcal{A}(W_{1,1}\vee W_{1,2})$
\begin{equation}
\varepsilon_{1}\left(F_{1}\hat{\otimes}F_{2}\right)=\varepsilon_{1,1}\left(F_{1}\right)\otimes\varepsilon_{1,1}\left(F_{2}\right)\,,
\end{equation}
where $\varepsilon_{1,j}:\mathfrak{F}(W_{1,j})\rightarrow\mathcal{A}(W_{1,j})$
are the usual conditional expectations of single double cones giving
by the group average in each region $W_{1,j}$. Then, we have that
$\phi_{1}\circ\varepsilon_{1}\neq\omega_{1}$ as states in $\mathfrak{F}(W_{1})$,
since $\omega_{1}(\mathcal{I})\neq0$ for intertwiners between the
regions $W_{1,1}$ and $W_{1,2}$, whereas $\varepsilon_{1}(\mathcal{I})=0$
because it projects to the additive algebra $\mathcal{A}\left(\mathcal{O}_{1}\right)=\mathcal{A}(W_{1,1}\vee W_{1,2})$.

However, we can obtain a generalization of \eqref{labe} complementing
\eqref{c2_ce_prop} with the property \ref{c2-e-re-tp} of proposition \ref{c2_re_prop}.
Let $\mathcal{O}_{1},\mathcal{O}_{2}\in\mathcal{K}$ be strictly spacelike
separated regions, $\varepsilon_{j}:\mathfrak{F}(\mathcal{O}_{j})\rightarrow\mathcal{A}(\mathcal{O}_{j})$
be conditional expectations, $\varepsilon_{12}:=\varepsilon_{1}\otimes\varepsilon_{2}$
and $\omega_{j}$, $\phi_{j}$, $\omega_{12}$ and $\phi_{12}$ the
restricted vacuum states as in \eqref{c6-states1} and \eqref{c6-states2}.
Then, we have that
\begin{eqnarray}
I_{\mathfrak{F}}\left(\mathcal{O}_{1},\mathcal{O}_{2}\right)-I_{\mathcal{A}}\left(\mathcal{O}_{1},\mathcal{O}_{2}\right)\!\!\! & = & \!\!\!S_{\mathfrak{F}}\left(\omega_{12}|\omega_{1}\otimes\omega_{2}\right)-S_{\mathcal{A}}\left(\phi|\phi_{1}\otimes\phi_{2}\right)\nonumber \\
 & = & \!\!\!S_{\mathfrak{F}}\left(\omega_{12}|\omega_{1}\otimes\omega_{2}\right)-S_{\mathfrak{F}}\left(\omega|\left(\omega_{1}\otimes\omega_{2}\right)\circ\varepsilon_{12}\right)\nonumber \\
 &  & \!\!\!+\left[S_{\mathfrak{F}}\left(\omega|\left(\omega_{1}\otimes\omega_{2}\right)\circ\varepsilon_{12}\right)-S_{\mathcal{A}}\left(\phi|\phi_{1}\otimes\phi_{2}\right)\right], \hspace{1cm}
\end{eqnarray}
where for the term between brackets we have used the conditional expectation property
\eqref{c2_ce_prop} and for the second term we use $(\omega_{1}\otimes\omega_{2})\circ\varepsilon_{12}=(\omega_{1}\circ\varepsilon_{1})\otimes(\omega_{2}\circ\varepsilon_{2})$,
\begin{equation}
I_{\mathfrak{F}}\left(\mathcal{O}_{1},\mathcal{O}_{2}\right)-I_{\mathcal{A}}\left(\mathcal{O}_{1},\mathcal{O}_{2}\right)=S_{\mathfrak{F}}\left(\omega_{12}|\omega_{1}\otimes\omega_{2}\right)-S_{\mathfrak{F}}\left(\omega|(\omega_{1}\circ\varepsilon_{1})\otimes(\omega_{2}\circ\varepsilon_{2})\right)+S_{\mathfrak{F}}\left(\omega|\omega\circ\varepsilon_{12}\right)\,.
\end{equation}
Using property \ref{c2-e-re-tp} of proposition \ref{c2_re_prop} in the second term
above, we finally arrive to
\begin{equation}
I_{\mathfrak{F}}\left(\mathcal{O}_{1},\mathcal{O}_{2}\right)-I_{\mathcal{A}}\left(\mathcal{O}_{1},\mathcal{O}_{2}\right)=S_{\mathfrak{F}}\left(\omega_{12}|\omega\circ\varepsilon_{12}\right)-S_{\mathfrak{F}}\left(\omega_{1}|\omega_{1}\circ\varepsilon_{1}\right)-S_{\mathfrak{F}}\left(\omega_{2}|\omega_{2}\circ\varepsilon_{2}\right)\,.\label{ecui}
\end{equation}
We remark that the first term on the l.h.s. of \eqref{ecui} is a
RE on $\mathfrak{F}(\mathcal{O}_{1}\vee\mathcal{O}_{2})$, whereas
the last two terms are REs in $\mathfrak{F}(\mathcal{O}_{1})$ and
$\mathfrak{F}(\mathcal{O}_{2})$ respectively. When $\mathcal{O}_{1}$
and $\mathcal{O}_{2}$ are double cones, equation \eqref{ecui} coincides
with \eqref{labe} because the last terms two on the r.h.s of \eqref{ecui}
vanish. However, \eqref{ecui} remains valid for any non-globally-invariant
state $\omega$. 

Now we take a Cauchy surface $\Sigma\subset\mathbb{R}^{d}$ and two
strictly spacelike separated regions $\mathcal{O}_{A}:=D\left(A\right)$
and $\mathcal{O}_{B}:=D\left(B\right)$, such that $A,B\subset\Sigma$
have $n_{A}$ and $n_{B}$ connected components respectively. Using
\eqref{ecui} we get 
\begin{eqnarray}
& &\hspace{-1cm} I_{\mathfrak{F}}(\mathcal{O}_{A},\mathcal{O}_{B})-I_{{\cal A}}(\mathcal{O}_{A},\mathcal{O}_{B}) \nonumber \\
&& =S_{\mathfrak{F}}(\omega_{AB}|\omega_{AB}\circ\otimes_{i}\varepsilon_{A_{i}}\otimes_{j}\varepsilon_{B_{j}})-S_{\mathfrak{F}}(\omega_{A}|\omega_{A}\circ\otimes_{i}\varepsilon_{A_{i}})-S_{\mathfrak{F}}(\omega_{B}|\omega_{B}\circ\otimes_{j}\varepsilon_{B_{j}}) \, , \hspace{1cm}\label{rhs}
\end{eqnarray}
where $i=1,\ldots,n_{A}$ and $j=1,\ldots,n_{B}$. In contrast to
the single component case, the two last terms do not vanish for the
vacuum state when $n_{A},n_{B}>1$. To obtain \eqref{rhs}, we use that any
connected region $\mathcal{O}_{A_{i}}$, $\mathcal{O}_{B_{j}}$ has
a corresponding set of twist operators.

Let us try to understand the value of this MI difference. We are mainly interested in the limit case of two regions $A$ and
$B$ that are nearly complementary to each other (see figure \ref{topofig}). Let us focus first on the first term on the right-hand side of \eqref{rhs}.
In general, there will be a complicated pattern of interference between
the intertwiners crossing pairs of the $n_{A}+n_{B}$ regions, but
under the current assumptions, this will be dominated by the intertwiners
crossing between adjacent boundaries of $A$ and $B$. In the particular
limit we are considering here, each connected component of the boundary
of $A$ meets with a connected component of the boundary of $B$ (see
figure \ref{topofig}). Therefore, the number of connected components
of the boundary of $A$ and the ones of $B$ agree. Let us call this
number $n_{\partial}$. Since each connected boundary divides the
space $\Sigma$ in two, the different boundaries form a tree under
inclusion. This leads to the fact that the sets of intertwiners crossing
each of these boundaries are algebraically independent to each other.
They are also statistically independent because they are well localized
in different boundaries. The total number of independent set of intertwiners
is given by $n_{A}+n_{B}-1$, because it is given by the total number
of independent charge creating operator algebras, minus one to account
for the neutrality of the operator. This coincides with the number
of boundaries, $n_{\partial}=n_{A}+n_{B}-1$, since in going from
the interior to the exterior, each time we cross a boundary we can
add a unique connected component of $A$ or $B$. Then, the maximization
of intertwiner entropy will indeed select the ones crossing each boundary
with no free choices left, and we have
\begin{equation}
S_{\mathfrak{F}}(\omega_{AB}|\omega_{AB}\circ\otimes_{i}\varepsilon_{A_{i}}\otimes_{j}\varepsilon_{B_{j}})=n_{\partial}\,\log|G|.\label{instance}
\end{equation}
This last entropy has the form of a topological contribution.

There is a parallel twist version of this story. The difference between
the entropies of the involved states in this relative entropy must
come from the small region $C:=(A\lor B)'$. This has exactly $n_{\partial}$
connected components which are thin regions at the interfaces between
$A$ and $B$. Each of these connected surfaces divides the space
in two and carries one set of independent twist operators that make
the difference between the algebras of the two models in $C$. In
the limit of small separation, independent sets of twists have all
vanishing expectation value and are statistically independent. Hence again, each boundary contributes $\log|G|$.

Thinking in the mutual information difference in the generic case,
the first term in \eqref{rhs} is bounded above by \eqref{instance}.
From \eqref{rhs} we then have the general bound 
\begin{equation}
0\le I_{\mathfrak{F}}(\mathcal{O}_{A},\mathcal{O}_{B})-I_{{\cal A}}(\mathcal{O}_{A},\mathcal{O}_{B})\le n_{\partial}\,\log|G|.\label{instance1}
\end{equation}

The second term on the right-hand side of \eqref{rhs} depends on
the intertwiners crossing the different connected components of $A$,
which are in the local algebra ${\cal A}'_{\Sigma}\left(A'\right)$,
but not in the algebra ${\cal A}_{\Sigma}\left(A\right)$. This
term is in fact equal to the difference of generalized MIs\footnote{In finite dimensional algebras $S(\omega_{A}|\omega_{A_{1}}\otimes\cdots\otimes\omega_{A_{n}})=S(\omega_{A_{1}})+\cdots+S(\omega_{A_{n}})-S(\omega_{A})$. }
\begin{equation}
S_{\mathfrak{F}}(\omega_{A}|\omega_{A}\circ\otimes_{i}\varepsilon_{A_{i}})=S_{\mathfrak{F}}(\omega_{A}|\omega_{A_{1}}\otimes\cdots\otimes\omega_{A_{n}})-S_{{\cal \mathcal{A}}}(\phi_{A}|\phi_{A_{1}}\otimes\cdots\otimes\phi_{A_{n}})\,.
\end{equation}
This is in general difficult to compute but we can simplify this contribution
if we focus in the case where the theory is gapped, and we are in
the infrared regime with regions much larger than the gap scale.
The expectation values of the intertwiners crossing components of
$A$ are exponentially small in the regime of large mass because the
typical distances between components are large compared to the mass
scale. In consequence, this contribution vanishes in this approximation.
Equivalently, the same holds for the third term in \eqref{rhs}, which
concerns the multicomponent region $B$. Then, in this limit and for
$\epsilon$ smaller than the gap scale, the bound is saturated and
we obtain 
\begin{equation}
I_{\mathfrak{F}}(\mathcal{O}_{A},\mathcal{O}_{B})-I_{{\cal A}}(\mathcal{O}_{A},\mathcal{O}_{B})=n_{\partial}\,\log|G|.\label{instance2}
\end{equation}

\subsection{Excitations\label{EX}}

We want to investigate how the entropy on the algebra of a double cone $W$ changes
when we insert a well localized charged excitation. Evidently, in
the case that the charged excitation corresponds to a sector of dimension
$d_{\sigma}=1$ (such as any excitation of an Abelian group symmetry),
there will not be any change in the entropy because this is represented
by an automorphism.

Let we take the state $\psi_{\sigma}^{j}$ on $\mathfrak{F}$ given
by the vector
\begin{equation}
|\psi_{\sigma}^{j}\rangle:=\sqrt{d_{\sigma}}V_{\sigma}^{j\,\dagger}|0\rangle\,,\quad V_{\sigma}^{j}\in\mathfrak{F}\left(W\right)\,, \label{exc_state_1}
\end{equation}
corresponding to the irreducible representation $\sigma\in\hat{G}$ and
already introduced in \eqref{dietrich}. For any element $A\in\mathcal{A}$, we have that
\begin{equation}
\langle\psi_{\sigma}^{j}|A|\psi_{\sigma}^{j}\rangle=\langle0|\rho_{\sigma}(A)|0\rangle=\omega\circ\rho_{\sigma}(A)\,,
\end{equation}
where 
\begin{equation}
\rho_{\sigma}(A):=\sum_{j=1}^{d_{\sigma}}V_{\sigma}^{j}\,A\,V_{\sigma}^{j\,\dagger}\,
\end{equation}
is the corresponding localized endomorphism. Notice that $\psi_{\sigma}^{j}$ is pure
in $\mathfrak{F}$ but not in ${\cal A}$.

To measure the entropy of this impurity we can compute
\begin{equation}
S_{\mathfrak{F}_{W}}(\psi_{\sigma}^{j}|\psi_{\sigma}^{j}\circ\varepsilon)=S_{\mathfrak{F}_{W}}(\psi_{\sigma}^{j}|\omega\circ\rho\circ\varepsilon)\,.
\end{equation}
To get an upper bound to this quantity we can use \ref{c2-e-re-i} in proposition \ref{c2_re_prop}, and we get
\begin{equation}
S_{\mathfrak{F}_{W}}(\psi_{\sigma}^{j}|\psi_{\sigma}^{j}\circ\varepsilon)=S_{\mathfrak{F}_{W}}(\psi_{\sigma}^{j}\circ\alpha_{g}|\psi_{\sigma}^{j}\circ\varepsilon\circ\alpha_{g})=S_{\mathfrak{F}_{W}}(\psi_{\sigma}^{j}\circ\alpha_{g}|\psi_{\sigma}\circ\varepsilon)\,,
\end{equation}
where $\alpha_{g}$ is the group automorphism \eqref{c6-acc-G}. In
the last equality we used that $\varepsilon=\varepsilon\circ\alpha_{g}$.
Then, we can average over $g\in G$ the first state to obtain the second
one. Even better, it is possible to average only over $G_{\sigma}:=\left\{ g_{\sigma,k}\,:\,k=1,\ldots,d_{\sigma}\right\} \subset G$
group elements, which just change the basis elements in the representation $\sigma$, to get the second state. Using again the convexity of the RE (\ref{c2-e-re-co} in proposition \ref{c2_re_prop}), we get the upper bound
\begin{eqnarray}
S_{\mathfrak{F}_{W}}(\psi_{\sigma}^{j}|\psi_{\sigma}^{j}\circ\varepsilon)\!\!\! & = & \!\!\!\frac{1}{d_{\sigma}}\sum_{g_{\sigma\in G}}S_{\mathfrak{F}_{W}}(\psi_{\sigma}^{j}\circ\alpha_{g_{\sigma}}|\psi_{\sigma}\circ\varepsilon)\\
 & \le & \!\!\!S_{\mathfrak{F}_{W}}\Bigl(\frac{1}{d_{\sigma}}\sum_{g_{\sigma\in G}}\psi_{\sigma}^{j}\circ\alpha_{g_{\sigma}}|\psi_{\sigma}\circ\varepsilon\Bigr)+\log(d_{\sigma})=\log(d_{\sigma})\,.\label{c6-exc-upp}
\end{eqnarray}
The vectors $|\psi_{\sigma}^{j}\rangle$ are orthonormalized
because they belong to different superselection sectors, i.e. $\langle\psi_{\sigma}^{j}|\psi_{\sigma'}^{k}\rangle=\delta_{\sigma\sigma'}\delta_{jk}$. As long as we make the excitation support smaller, or the size of the double
cone bigger, the reduced states for each $g_{\sigma}\in G_{\sigma}$
in the sum \eqref{c6-exc-upp} become disjoint. This is the condition
for the bound to become saturated
\begin{equation}
S_{\mathfrak{F}_{W}}(\psi_{\sigma}^{j}|\psi_{\sigma}^{j}\circ\varepsilon)=\log d_{\sigma}\,.
\end{equation}
A completely localized state $\tilde{\psi}_{\sigma}^{j}$ on $\mathfrak{F}$
is produced by the operators $V_{\sigma}^{j}$ acting over the vacuum
vector, i.e.
\begin{equation}
|\tilde{\psi}_{\sigma}^{j}\rangle:=V_{\sigma}^{j}|0\rangle\,,\quad V_{\sigma}^{j}\in\mathfrak{F}\left(W\right)\,.\label{vector}
\end{equation}
Contrary to \eqref{exc_state_1}, there is no factor $\sqrt{d_{r}}$ since $\langle\tilde{\psi}_{\sigma}^{j}|\tilde{\psi}_{\sigma}^{k}\rangle=\delta_{jk}$
because of $V_{\sigma}^{j\,\dagger}V_{\sigma}^{k}=\delta_{jk}$. The
vector \eqref{vector} corresponds to an excitation in the conjugate representation
$\bar{\sigma}$. The state $\tilde{\psi}_{\sigma}^{j}$ on ${\cal \mathcal{A}}$ is equivalent to the state 
\begin{equation}
\tilde{\psi}_{\sigma}:=\frac{1}{d_{\sigma}}\sum_{\sigma=1}^{d_{\sigma}}\tilde{\psi}_{\sigma}^{j}\,.\label{tirosami}
\end{equation}
In the same way as above, we have an upper bound $\log d_{\sigma}$
for the RE. The lower bound can be obtained by reducing to the finite
dimensional subalgebra \eqref{esa1}, generated by the operators $\{V_{\sigma}^{j}V_{\sigma}^{k\,\dagger}\,:\,j,k=1,\ldots,d_{\sigma}\}$. We have for the two states on this subalgebra
\begin{eqnarray}
\langle\tilde{\psi}_{\sigma}^{j}|V_{\sigma}^{k}V_{\sigma}^{l\,\dagger}|\tilde{\psi}_{\sigma}^{j}\rangle\!\!\! & = & \!\!\!\delta_{jk}\delta_{jl}\,,\label{tirso}\\
\frac{1}{d_{\sigma}}\sum_{j=1}^{d_{\sigma}}\langle\tilde{\psi}_{\sigma}^{j}|V_{\sigma}^{k}V_{\sigma}^{l\,\dagger}|\tilde{\psi}_{\sigma}^{j}\rangle\!\!\! & = & \!\!\!\frac{1}{d_{\sigma}}\delta_{kl}\,.\label{tarso}
\end{eqnarray}
The RE on this subalgebra is $\log(d_{r})$. Then, we have 
\begin{equation}
S_{\mathfrak{F}_{W}}(\tilde{\psi}_{\sigma}^{j}|\tilde{\psi}_{\sigma}^{j}\circ\varepsilon)=\log d_{\sigma}\,.
\end{equation}

Instead of using this relative entropy we can try to compute the change
in the mutual information for two touching double cones $W_{1}$ and $W_{2}$.
The excitation does not change correlations of operators in $\mathfrak{F}$
outside the support of $V_{\sigma}^{j}$. Then, if this support is\textcolor{black}{{}
small and well-inside of $W_{1}$}, we expect that
\begin{equation}
I_{\mathfrak{F}}(W_{1},W_{2};\tilde{\psi}_{\sigma}^{j})-I_{\mathfrak{F}}(W_{1},W_{2};\omega)\simeq0\,.
\end{equation}
To compute the change in the model ${\cal A}$ we use the formula, which gives
\eqref{ecui}
\begin{equation}
I_{\mathfrak{F}}(W_{1},W_{2};\tilde{\psi}_{\sigma}^{j})-I_{\mathcal{A}}(W_{1},W_{2};\tilde{\psi}_{\sigma}^{j})=S_{\mathfrak{F}_{12}}(\tilde{\psi}_{\sigma}^{j}|\tilde{\psi}_{\sigma}^{j}\circ\varepsilon_{12})-S_{\mathfrak{F}_{1}}(\tilde{\psi}_{\sigma}^{j}|\tilde{\psi}_{\sigma}^{j}\circ\varepsilon_{1})-S_{\mathfrak{F}_{2}}(\tilde{\psi}_{\sigma}^{j}|\tilde{\psi}_{\sigma}^{j}\circ\varepsilon_{2})\,.\label{dista}
\end{equation}
The last term is zero because the two states are equal in $W_{2}$.
The second term is $\log(d_{\sigma})$ as we have seen above. The first
term is upper bounded by $\log|G|+\log d_{\sigma}$ since it is the
minimal number of transformed states $\tilde{\psi}_{\sigma}^{j}$
we need to mix to get $\tilde{\psi}_{\sigma}^{j}\circ\varepsilon_{12}$.
But it is also lower bounded by the same number since we can use a
subalgebra formed by the one in (\ref{tirso}-\ref{tarso}) plus some
intertwiner algebra near the touching boundary of $W_{1}$ and $W_{2}$.
Expectation values for these algebras are uncorrelated and the effect
of the conditional expectation can also be decoupled. Then, we conclude
\begin{equation}
I_{\mathfrak{F}}(W_{1},W_{2};\tilde{\psi}_{\sigma}^{j})-I_{\mathcal{A}}(W_{1},W_{2};\tilde{\psi}_{\sigma}^{j})=\log|G|\,.
\end{equation}
Thus, the MI on ${\cal A}$ does not change with respect to the vacuum, as happens on $\mathfrak{F}$. The excitation
is impure in the model ${\cal A}$, but its impurity is due to a transformation
of the vacuum well-inside the region $W_{1}$, that it does not modify
correlations with $\mathcal{A}(W_{2})$. Therefore, it will not change
the MI.

If we instead create the state $\varphi_{\sigma\bar{\sigma}}^{j}$, which
represents a particle in $W_{1}$ and an antiparticle in $W_{2}$
\begin{equation}
|\varphi_{\sigma\bar{\sigma}}^{j}\rangle:=V_{\sigma,1}^{j}V_{\bar{\sigma},2}^{j}|0\rangle\,,\quad V_{\sigma,1}^{j}\in\mathfrak{F}\left(W_{1}\right),\,V_{\bar{\sigma},2}^{j}\in\mathfrak{F}\left(W_{2}\right)\,,
\end{equation}
the MI on $\mathfrak{F}$ will not change with respect
to the vacuum one, because of the same reasons as above. For the model
${\cal A}$ we can use again \eqref{dista}. Now, the last two REs are $\log d_{\sigma}$. The first one is again upper bounded
and lower bounded by $\log|G|+\log d_{\sigma}$. Then, we conclude that
\begin{eqnarray}
I_{\mathfrak{F}}(W_{1},W_{2};\varphi_{\sigma\bar{\sigma}}^{j})-I_{\mathcal{A}}(W_{1},W_{2};\varphi_{\sigma\bar{\sigma}}^{j})\!\!\! & = & \!\!\!\log|G|-\log d_{\sigma}\,,\\
I_{{\cal A}}(W_{1},W_{2};\varphi_{\sigma\bar{\sigma}}^{j})-I_{\mathcal{A}}(W_{1},W_{2};\omega)\!\!\! & = & \!\!\!\log d_{\sigma}\,.
\end{eqnarray}
Remarkably this does not depend on how far separated are the excitations.
We would have expected $2\log d_{\sigma}$ for the MI of a maximally
entangled state on a Hilbert space of dimension $d_{\sigma}$. But
here, the effect is rather the MI of classical random variables with perfect
correlation and maximal uncertainty (the effective state is analogous
to \eqref{tirosami} with $\tilde{\psi}_{\sigma}^{j}\rightarrow\varphi_{\sigma\bar{\sigma}}^{j}$),
which gives half this number, and it is not produced by entanglement. 

For the pure state $\varphi_{\sigma\bar{\sigma}}$ on $\mathfrak{F}$
given by
\begin{equation}
|\varphi_{\sigma\bar{\sigma}}\rangle:=\frac{1}{\sqrt{d_{\sigma}}}\sum_{j=1}^{d_{\sigma}}V_{\sigma,1}^{j}V_{\bar{\sigma},2}^{j}|0\rangle\,,\quad V_{\sigma,1}^{j}\in\mathfrak{F}\left(W_{1}\right),\,V_{\bar{\sigma},2}^{j}\in\mathfrak{F}\left(W_{2}\right)\,,\label{c6-state3}
\end{equation}
we get, along the same lines,
\begin{equation}
I_{\mathfrak{F}}(W_{1},W_{2};\varphi_{\sigma\bar{\sigma}})-I_{\mathcal{A}}(W_{1},W_{2};\varphi_{\sigma\bar{\sigma}})=\log|G|\,.
\end{equation}
This is because the state \eqref{c6-state3} is invariant under group
automorphisms and hencd, we have $S_{\mathfrak{F}_{12}}(\varphi_{\sigma\bar{\sigma}}|\varphi_{\sigma\bar{\sigma}}\circ\varepsilon_{12})=\log|G|$
as for the vacuum state, and the two last terms in \eqref{dista}
vanish. Then, we expect 
\begin{equation}
I_{{\cal A}}(W_{1},W_{2};\varphi_{\sigma\bar{\sigma}})-I_{\mathcal{A}}(W_{1},W_{2};\omega)=I_{\mathfrak{F}}(W_{1},W_{2};\varphi_{\sigma\bar{\sigma}})-I_{\mathfrak{F}}(W_{1},W_{2};\omega)=2\log d_{\sigma}\,,
\end{equation}
as it corresponds to a pure state in both models.

Several results about the entropy of charged states, analogous to the
ones studied in this section, have been previously appeared in the literature for
specific models. See, for example, \cite{Lewkowycz:2013laa,Dong:2008ft,Caputa:2014eta,Nozaki:2014hna,Alcaraz:2011tn,Longo:2018obd,Hollands17}.

\subsection{Spontaneous symmetry breaking\label{SSB}}

If the observable algebra $\mathcal{A}$ satisfies all the structural
assumptions given in assumption \ref{c6-ass-dhr}, we can uniquely reconstruct
the field algebra $\mathfrak{F}$ and the compact symmetry group $G$,
such that $\mathcal{A}$ is made of the $G$-invariant elements of $\mathfrak{F}$.
Moreover, the symmetry group acts unitarily in $\mathcal{H}$ leaving
the vacuum vector invariant. Then, there are no broken symmetries.
In order to leave the possibility of the appearance of spontaneous
symmetry breaking, we must relax some of the assumptions listed there.
The manner this phenomenon is described in the algebraic setting is to
impose that local algebras attached to doubles cones are no longer maximal
in the vacuum sector, i.e. they do not satisfy the duality condition
\ref{c6-ass-dua} of assumption \ref{c6-ass-dhr}. The local operators
which are then missing are, roughly speaking, fields that transform
non-trivially under the symmetry group but do not leave out of
the vacuum sector \cite{Roberts:1974mg,Doplicher:1990qa,Buchholz:1992bg}.

A symmetry is said to be spontaneously broken if there is no unitary
operator implementing it in the vacuum representation of $\mathfrak{F}$,
or what amount to the same thing, if the vacuum state is not stable
under the action of such a symmetry.

Despite, the local algebras attached to double cones do not satisfy
the duality condition, we assume they satisfy essential duality
(see section \ref{c1_subsec_lattice}). Then, we can define the dual
net $\mathcal{O}\in\mathcal{K}\rightarrow\mathcal{A}^{d}\left(\mathcal{O}\right)\subset\mathcal{B}\left(\mathcal{H}_{0}\right)$, acting over the same Hilbert space $\mathcal{H}_{0}$ where $\mathcal{A}$
acts, as
\begin{itemize}
\item $\mathcal{A}^{d}\left(W\right):=\mathcal{A}^{d}\left(W'\right)'$
for all double cones $W$,
\item and imposing additivity (condition \ref{c6-ass-add} in assumption
\ref{c6-ass-dhr}) for more general causally complete regions.
\end{itemize}
Then, we have that $\mathcal{A}\subset\mathcal{A}^{d}$ and, it can
be shown, that the dual net satisfies all the axioms of definitions
\ref{c1_def_aqft} and \ref{c1_def_vacuum}. In fact, the unitary
representation of the Poincaré group and the vacuum vector of $\mathcal{A}$
and $\mathcal{A}^{d}$ are the same. 

Any representation of $\mathcal{A}$ fulfilling \eqref{c6-dhr-uni}
extends canonically to representations of $\mathcal{A}^{d}$ of the
same type on the same Hilbert space \cite{Roberts:1978kt}. Then,
as we have explained in section \ref{c6-secc-field}, we can associate
to $\mathcal{A}^{d}$ a unique compact symmetry group $H$ and a complete
field system $\mathfrak{F}$ on the Hilbert space $\mathcal{H}$.
Then, $\mathcal{A}^{d}$ is made of the $H$-invariant elements of
$\mathfrak{F}$. We have then the inclusion of nets
\begin{equation}
\mathcal{A}\left(\mathcal{O}\right)\subset\mathcal{A}^{d}\left(\mathcal{O}\right)\subset\mathfrak{F}\left(\mathcal{O}\right)\,.
\end{equation}
Then, we can define the full symmetry group $G$ as the group of all
automorphisms of $\mathfrak{F}$ leaving $\mathcal{A}$ pointwise
fixed. Of course, we have that $H\subset G$. Contrary to $H$, $G$
may be non-compact. An element of $G$ lies actually in $H$ if and
only if it corresponds to a non-spontaneously broken symmetry (i.e. it is unitarily
implementable) if and only if it leaves $\mathcal{A}^{d}$ pointwise
fixed. Similarly, the elements of $G$ that do not belong to $H$
are the \textit{spontaneously broken symmetries}. It can be shown
that the elements of $G$ leave all the local field algebras $\mathfrak{F}\left(\mathcal{O}\right)$
globally stable, i.e.
\begin{equation}
\alpha\left(\mathfrak{F}(\mathcal{O})\right)=\mathfrak{F}(\mathcal{O})\,,\quad\forall\alpha\in G\,,\:\forall\mathcal{O}\in\mathcal{K}\,,
\end{equation}
and $H$ is the stabilizer in $G$ of the vacuum state on $\mathfrak{F}$. 

Now, let us assume that $G$ commutes with the Poincaré unitaries in
$\mathfrak{F}$ and leaves $\mathcal{A}^{d}$ globally fixed.\footnote{If it does not happen, we can consider the subgroup of $G$ satisfying the above properties.} Then, $G$ leaves all the local algebras $\mathcal{A}^{d}\left(\mathcal{O}\right)$
globally stable. We also have that $H$ is a normal subgroup of $G$
and $G$ is an extension of $H$ by $G/H$. The action of an element
$g\in G$ takes the vacuum state, induced by the vector $\left|0\right\rangle \in\mathcal{H}_{0}$,
to the state $\omega_{g}$. Such a state coincides with the vacuum
state over $\mathcal{A}$, but they may be different states over $\mathcal{A}^{d}$.
Moreover, such a state is Poincaré-invariant in $\mathfrak{F}$. The
set $\mathcal{V}:=\{\omega_{g}\,:\,g\in G\}$ is the set of \textit{degenerate
vacua} and it is in one-to-one correspondence with $G/H$.

For simplicity, we assume that the group $G$ is compact. Then, we can define
two different conditional expectations
\begin{equation}
\varepsilon:\mathfrak{F}\rightarrow\mathcal{A}\quad\textrm{and}\quad\tilde{\varepsilon}:\mathfrak{F}\rightarrow\mathcal{A}^{d}\,,
\end{equation}
given by the group averages on $G$ and $H$ respectively. 

When a symmetry is broken it is not longer true that the RE for the vacuum on the two models is zero for a double cone
$W$. This is because the vacuum expectation values do not generally
vanish for the charged operators. In fact, the RE
\begin{equation}
S_{\mathfrak{F}_{W}}(\omega_{g}|\omega_{g}\circ\varepsilon)\,,\label{c6-ssb-eor-one}
\end{equation}
is an interesting quantity to compute in this case and serves as an
entropic order parameter for symmetry breaking. Using the invariance
of the RE under automorphisms and the invariance of
the conditional expectation $\varepsilon$ under $G$, it can be shown
that the above RE does not depend on the degenerate vacua chosen
$\omega_{g}$. Moreover, $S_{\mathfrak{F}_{W}}(\omega_{g}|\omega_{g'}\circ\varepsilon)$
is independent of $g,g'\in G$. From the rest of this section, $\omega$
denotes any vacua of $\mathcal{V}$.\footnote{We do not consider a mixed state of vacua, since the model $\mathfrak{F}$
does not satisfy clustering in such a state.} Using \eqref{c2_ce_prop} for $\tilde{\varepsilon}:\mathfrak{F}(W)\rightarrow\mathcal{A}^{d}(W)$
and the states $\omega$ and $\omega\circ\varepsilon$, we get
$S_{\mathfrak{F}_{W}}(\omega|\omega\circ\varepsilon)=S_{\mathcal{A}_{W}^{d}}(\omega|\omega\circ\varepsilon)$.

Let us first discuss the case of a finite group $G$. As in section
\ref{upper}, the RE is upper bounded because of the convexity property,
\begin{equation}
S_{\mathfrak{F}_{W}}(\omega|\omega\circ\varepsilon)\le\log\left(\frac{|G|}{|H|}\right)\,.\label{c6-ssb-cono}
\end{equation}
We have that $\omega=\omega\circ\tilde{\varepsilon}$, and to convert
such a state into the second one on the l.h.s. of \eqref{c6-ssb-cono},
we need to average only with $|G|/|H|$ independent elements of $G/H$.
As a curiosity, we note that this is a RE for a single region, which is invariant under Poincar\'e transformations of the region. This
rare luxury is possible precisely because of the existence of degenerate vacua.

Let $W:=D\left(\mathcal{C}_{R}\right)$ with $\mathcal{C}_{R}$ as
in \eqref{c6-reg1}. The RE increases with the size $R$ of the region
$W$. We can take the entropy difference in any finite subalgebra stable
under $\varepsilon$ to get a lower bound. We expect that, for small $R$ with respect to the scale of the symmetry breaking, the symmetry
is effectively restored and the RE approaches to zero.
In other words, there are no operators inside the region that are
able to efficiently distinguish the two states. On the other hand, for regions larger
than the symmetry breaking scale, we expect saturation of the bound
\eqref{c6-ssb-cono}. As an example, we can take a theory with a $\mathbb{Z}_{2}$
broken symmetry, where the order parameter is given by the non-zero expectation value of a scalar field $\phi\left(x\right)$. As an order parameter
we can use a smeared mode $\phi_{f}:=\int d^{d}x\,f(x)\phi(x)$ such
that $\int d^{d}x\,f(x)=1$. We have that $\langle\phi_{f}\rangle=:\mu$
in the state $\omega$, and $\langle\phi_{f}\rangle=0$ in the state
$\omega\circ\varepsilon$. But the fluctuations of this mode for small
support of the test function $f$ will be much bigger than $\mu$
and of order of $R^{-1}$. Hence, we cannot efficiently distinguish
the states in a small region.\footnote{To get a rough estimate of this behavior, we may assume Gaussian fluctuations.
The RE, between a classical continuous variable with Gaussian
distribution of width $R^{-1}$ centered around the origin and another
Gaussian distribution centered in $\mu$, is $\sim(R\mu)^{2}$.}

To understand the behavior of the MI difference $\Delta I:=I_{\mathfrak{F}}(W_{1},W_{2})-I_{{\cal A}}(W_{1},W_{2})$
we use formula \eqref{ecui} 
\begin{equation}
\Delta I=S_{{\cal \mathfrak{F}}_{12}}(\omega_{12}|\omega_{12}\circ\varepsilon_{1}\otimes\varepsilon_{2})-S_{{\cal \mathfrak{F}}_{1}}(\omega_{1}|\omega_{1}\circ\varepsilon_{1})-S_{{\cal \mathfrak{F}}_{2}}(\omega_{2}|\omega_{2}\circ\varepsilon_{2})\,,\label{ecui4}
\end{equation}
which is valid for general states. We are mainly interested in the case of nearly complementary regions $W_{2}\rightarrow W'_{1}$. Then, let us consider $W_{1}:=D\left(\mathcal{C}_{R}\right)$
and $W_{2}:=D\left(\mathcal{C}'_{R+\epsilon}\right)$ ($\epsilon>0$).
In this case, the last term, for an unbounded region $W_{2}$, saturates
to $\log\left(|G|/|H|\right)$. Using convexity we can show that
\begin{equation}
S_{\mathfrak{F}_{12}}(\omega_{12}|\omega_{12}\circ\varepsilon_{1}\otimes\varepsilon_{2})\leq2\log|G|-\log|H|\,.\label{c5-ssb-2reg}
\end{equation}
In this case, to transform the state $\omega_{12}$ into $\omega_{12}\circ\varepsilon_{1}\otimes\varepsilon_{2}$
we need to average with the elements of the group $G$ in both regions
independently, but we can dispense on average on $|H|$ elements in
one of the regions, because the vacuum state is still globally
invariant under the unbroken subgroup $H$.\footnote{Despite the broken symmetries are not related to superselection sectors,
there exist local implementations of the elements of $G$ for any
local algebra $\mathfrak{F}\left(\mathcal{O}\right)$ for which the
vacuum vector $\left|0\right\rangle $ is standard \cite{Buchholz:1992bg}.} In the present limit, we can argue that this term always saturates the
bound \eqref{c5-ssb-2reg}. This is because we can use, as a lower
bound, the RE on a subalgebra formed by charged operators
localized far away from $W_{1}$ (e.g. the same subalgebra that one
can use to show $S_{\mathfrak{F}_{1}}(\omega_{2}|\omega_{2}\circ\varepsilon_{2})=\log\left(|G|/|H|\right)$)
and an intertwiner subalgebra around the common boundary of the
regions $W_{1}$ and $W_{2}$. Expectation values are independent
for these subalgebras because the charged operators in the intertwiner subalgebra
have large fluctuations, and we get the saturation of the bound \eqref{c5-ssb-2reg}
independently of the size $R$. Therefore, we expect 
\begin{equation}
I_{\mathfrak{F}}(W_{1},W_{2})-I_{{\cal A}}(W_{1},W_{2})=\log|G|-S_{{\cal \mathfrak{F}}_{1}}(\omega_{1}|\omega_{1}\circ\varepsilon_{1})\,.
\end{equation}
Hence, this is controlled by the same entropic order parameter discussed
above. This approaches the result
\begin{equation}
\Delta I\rightarrow\begin{cases}
\log|G|\,, & R\mu\ll1\,,\\
\log|H|\,, & R\mu\gg1\,,
\end{cases}\qquad\epsilon\ll R,\mu^{-1}\,,\label{c6-ssb-deltaI-ee}
\end{equation}
where $\mu$ is the symmetry breaking scale, and $\epsilon$ is the separation between the regions $W_{1}$ and $W_{2}$. The first case corresponds to the case when the broken symmetry is restored in the
region $W_{1}$, whereas the second case corresponds to the IR behavior
where the symmetry is completely broken inside $W_{1}$.

To understand the regime $\mu^{-1}\ll\epsilon,R$, when the separation
between the regions is larger than the breaking scale, we have to split
\eqref{ecui4} into two contributions using the intermediate algebra
$\mathcal{A}^{d}$
\begin{eqnarray}
I_{\mathfrak{F}}(W_{1},W_{2})-I_{{\cal A}^{d}}(W_{1},W_{2})\!\!\! & = & \!\!\!S_{{\cal \mathfrak{F}}_{12}}(\omega_{12}|\omega_{12}\circ\tilde{\varepsilon}_{1}\otimes\tilde{\varepsilon}_{2})\,,\label{c6-ssb-delta1}\\
I_{{\cal A}^{d}}(W_{1},W_{2})-I_{{\cal A}}(W_{1},W_{2})\!\!\! & = & \!\!\!S_{{\cal A}_{12}^{d}}(\omega_{12}|\omega_{12}\circ\varepsilon_{1}\otimes\varepsilon_{2})\nonumber \\
 &  & \!\!\!-S_{{\cal A}_{1}^{d}}(\omega_{1}|\omega_{1}\circ\varepsilon_{1})-S_{{\cal A}_{2}^{d}}(\omega_{2}|\omega_{2}\circ\varepsilon_{2})\,.\label{c6-ssb-delta2}
\end{eqnarray}
The first term is just the case when there are no broken symmetries,
and it behaves as $I_{\mathfrak{F}}(W_{1},W_{2})-I_{{\cal A}^{d}}(W_{1},W_{2})=\log|H|$
for $\epsilon\ll R$, and $I_{\mathfrak{F}}(W_{1},W_{2})-I_{{\cal A}^{d}}(W_{1},W_{2})=0$
for $\epsilon\gg R$. The second term is the case where all symmetries
are broken. In this case, the group of these symmetries is the quotient
group $G/H$.\footnote{The conditional expectation $\left.\varepsilon\right|_{\mathcal{A}^{d}}$
restricted to $\mathcal{A}^{d}$ involves the average over $|G|/|H|$
independents elements of $G$ because $\mathcal{A}^{d}$ is $H$-invariant.} For the same reasons as above, the first term in \eqref{c6-ssb-delta2}
is bounded above by $2\log(|G|/|H|)$, whereas the second and third
terms are each one bounded above by $\log(|G|/|H|)$. For $\mu^{-1}\ll R$,
we have that both last two terms saturate their respective bounds, pushing the
first term in \eqref{c6-ssb-delta2} to saturates since the r.h.s.
of the same expression is positive due to monotonicity. Then,
combining these results for \eqref{c6-ssb-delta1} and \eqref{c6-ssb-delta2},
we obtain
\begin{equation}
\Delta I\rightarrow\begin{cases}
\log|H|\,, & \epsilon\ll R\,,\\
0\,, & \epsilon\gg R\,,
\end{cases}\qquad\mu^{-1}\ll\epsilon,R\,.
\end{equation}
We remark, that in the case we want to interpret $\Delta I$ as a
regularized entropy (see section \ref{c3-sec-univee}), we must consider
the regime \eqref{c6-ssb-deltaI-ee} since we must take $\epsilon$
smaller than any other scale of the theory.

According to our analysis of the entropic certainty relation in section
\ref{certainty}, we conclude that the following relation applies
\begin{equation}
S_{\mathfrak{F}_{12}}(\omega|\omega\circ\varepsilon_{12})+S_{{\cal \mathfrak{F}}_{S}\vee G_{\tau}}(\omega|\omega\circ\varepsilon_{\tau})=2\log|G|-\log|H|\,.\label{c6-ssb-cer}
\end{equation}
The conditional expectation $\varepsilon_{\tau}:{\cal \mathfrak{F}}_{S}\vee G_{\tau}\rightarrow{\cal \mathfrak{F}}_{S}$
projects the local operators that implement (broken and unbroken) the symmetries in the region $W_{1}$. It is the dual conditional
expectation of $\varepsilon_{12}$ for the inclusion of algebras $\mathcal{A}{}_{12}\subset\mathfrak{F}_{12}$,
which has algebraic index $|G|^{2}/|H|$.\footnote{This is because, in this case, to pass from $\mathfrak{F}_{12}$ to
$\mathcal{A}_{12}$ we can use first the conditional expectation $\tilde{\varepsilon}_{12}:\mathfrak{F}_{12}\rightarrow\mathcal{A}_{12}^{d}$
which has index $|H|$ as in the unbroken case, and then the conditional
expectation $\left.\varepsilon_{1}\otimes\varepsilon_{2}\right|_{\mathcal{A}_{12}^{d}}:\mathcal{A}_{1}^{d}\otimes\mathcal{A}_{2}^{d}\rightarrow\mathcal{A}_{1}\otimes\mathcal{A}_{2}$
which has index $(|G|$/$|H|)^{2}$, because the average in each factor
is independent and is given by the group $G/H$. The total index for the
composition of the conditional expectations is the product of the
indices and gives $|G|^{2}/|H|$.}

In this context, it is also interesting to consider the certainty relation
for the SSB order parameter
\begin{equation}
S_{\mathfrak{F}_{1}}(\omega|\omega\circ\varepsilon_{1})+S_{\mathfrak{F}_{1'}\vee G_{\tau}}(\omega|\omega\circ\varepsilon_{\tau_{1}})=\log|G|\,,\label{c6-ssb-cer-1}
\end{equation}
which is represented in the diagram
\begin{equation}
\begin{array}{ccc}
\mathfrak{F}(W_{1}) & \overset{\varepsilon_{1}}{\longrightarrow} & \quad\mathcal{A}(W_{1})\\
\quad\updownarrow\prime &  & \quad\;\:\,\updownarrow\prime\\
{\cal \mathfrak{F}}(W'_{1}) & \overset{\varepsilon_{\tau_{1}}}{\longleftarrow} & {\cal \mathfrak{F}}(W'_{1})\vee G_{\tau}\,.
\end{array}
\end{equation}
However, using the intermediate algebra $\mathcal{A}^{d}$, we have a different certainty relation
\begin{equation}
S_{\mathcal{A}_{1}^{d}}(\omega|\omega\circ\varepsilon_{1})+S_{\mathcal{A}_{1'}^{d}}(\omega|\omega\circ\varepsilon_{\tau_{1}})=\log|G|-\log|H|\,,\label{c6-ssb-cer-2}
\end{equation}
according to the diagram
\begin{equation}
\begin{array}{ccc}
\mathcal{A}^{d}(W_{1})\!\! & \overset{\varepsilon_{1}}{\longrightarrow} & \!\!\!\!\mathcal{A}(W_{1})\\
\quad\updownarrow\prime &  & \updownarrow\prime\\
\mathcal{A}^{d}(W_{1})'=\mathcal{A}(W'_{1}) & \overset{\varepsilon_{\tau_{1}}}{\longleftarrow} & \mathcal{A}(W_{1})'=\mathcal{A}^{d}(W'_{1})\,.
\end{array}
\end{equation}
We now argue why both certainty relations could be simultaneously
valid despite we have that $S_{\mathfrak{F}_{1}}(\omega|\omega\circ\varepsilon_{1})=S_{\mathcal{A}_{1}^{d}}(\omega|\omega\circ\varepsilon_{1})$.
This is because in \eqref{c6-ssb-cer-1} the commutants are taken
inside the Hilbert space $\mathcal{H}$ since all the field algebras
involved are subalgebras $\mathcal{B}\left(\mathcal{H}\right)$, whereas
in \eqref{c6-ssb-cer-2} we used that $\mathcal{A}\subset\mathcal{A}^{d}\subset\mathcal{B}\left(\mathcal{H}_{0}\right)$
as long as we take commutants inside the vacuum Hilbert space $\mathcal{H}_{0}$.
For this reason, the commutant $\mathcal{A}^{d}(W'_{1})$ of $\mathcal{A}(W_{1})$
inside $\mathcal{B}\left(\mathcal{H}_{0}\right)$ does not include the
DHR twists, but it includes the operators which implement the broken
symmetries locally. On the other hand, the algebra ${\cal \mathfrak{F}}(W'_{1})\vee G_{\tau}$
include all the broken and unbroken twists. In this way, we have always
$S_{\mathfrak{F}_{1'}\vee G_{\tau}}(\omega|\omega\circ\varepsilon_{\tau_{1}})\geq\log|H|$
because there is a group subalgebra $H_{\tau}\subset G_{\tau}$ of
$|H|$ independent twists of the unbroken part of $G$ such that
$S_{H_{\tau}}(\omega|\omega\circ\varepsilon_{\tau_{1}})=\log|H|$
independently of the value of $R\mu$.

All the above discussion suits perfectly for a finite group $G$.
In order to understand the case of SSB of a Lie group, we
first study the simple model of a compactified free scalar.

\subsubsection{Free compactified scalar\label{scalar}}

Let us take the global algebra $\mathfrak{F}$ of the free massless
scalar real field $\phi\left(x\right)$ for $d>2$.\footnote{For $d=2$, this theory is not well-defined since it does not satisfies
positivity. Then, we have to consider the subalgebra of the derivatives
$\partial_{\mu}\phi$ of $\phi$. But, this is unsatisfactory for the
purpose of this subsection because we want to study the SSB $\phi\rightarrow\partial_{\mu}\phi$.
This is consistent with the fact that SSB with continuous symmetry
does not exists in $d=2$. As we saw in chapter \ref{CURRENT}, in
$d=2$, the algebra of $\partial_{\mu}\phi$ is the invariant subalgebra
of the free fermion field under the unbroken $U(1)$-symmetry.} The way to construct this algebra is similar to what we have explained
in chapter \ref{RE_CS}, with some small differences coming that, in this case, the
mass is zero. The global algebra $\mathfrak{F}$ is defined as a concrete
$C^{*}$-algebra of operators acting on the vacuum Fock Hilbert
space $\mathcal{H}_{0}$. This Hilbert space is the symmetric tensor
product of the Hilbert space of one-particle states of zero mass and zero
spin. The local algebras are defined using the Weyl unitaries $\mathrm{e}^{i\phi\left(f\right)}$
, $f\in\mathcal{S}\left(\mathbb{R}^{d},\mathbb{R}\right)$, as
\begin{equation}
\mathfrak{F}\left(B\right):=\left\{ \mathrm{e}^{i\phi\left(f\right)}\,:\,\mathrm{supp}\left(f\right)\subset B\right\} ''\subset\mathcal{B}(\mathcal{H}_{0})\,,\quad B\in\mathcal{K}\textrm{ bounded.}\label{c6-massless}
\end{equation}
For an unbounded region $\mathcal{O}$, we define $\mathfrak{F}\left(\mathcal{O}\right)$
as the $C^{*}$-algebra generated by the unitaries $W\left(f\right)$
with bounded $\mathrm{supp}\left(f\right)\subset\mathcal{O}$. This algebra
satisfies duality for double cones (\ref{c6-ass-dua} in assumption
\ref{c6-ass-dhr}). Moreover, $\mathfrak{F}$ has no superselection
sectors and it also satisfies duality for any pair of double cones $W_{1}\text{\Large\ensuremath{\times\negmedspace\!\!\times}}W_{2}$. 

We define the subalgebra $\mathcal{A}\subset\mathfrak{F}$ of the
derivatives $\partial_{\mu}\phi\left(x\right)$ of the massless scalar
field. This is the pointwise fixed subalgebra under the automorphisms
\begin{equation}
\alpha_{s}(\phi):=\phi+s\,,\quad s\in\mathbb{R}\,.\label{c6-der-sca}
\end{equation}
In this case, the group of such automorphisms is $\mathbb{R}$. They
can be locally implemented for any standard region $\mathcal{O}\in\mathcal{K}$
\cite{Buchholz:1992bg}. Moreover, for a bounded region $B:=D\left(\mathcal{C}\right)$
with $\mathcal{C}\subset\Sigma_{0}:=\left\{ x^{0}=0\right\} \subset\mathbb{R}^{d}$,
we have that
\begin{eqnarray}
 &  & \alpha_{s}(F)=\mathrm{e}^{isQ_{B}}F\mathrm{e}^{-isQ_{B}}\,,\quad F\in\mathfrak{F}\left(B\right)\,,\label{c6-loc-imp-ssb}\\
 &  & Q_{B}:=\int_{\mathbb{R}^{d-1}}J\left(\bar{x}\right)f_{B}\left(\bar{x}\right)\,d^{d-1}x\,,\label{c6-loc-imp-ssb_2}
\end{eqnarray}
where $J\left(\bar{x}\right):=\left.\partial_{0}\phi\left(x\right)\right|_{x^{0}=0}$
and $f_{B}\in\mathcal{S}\left(\mathbb{R}^{d-1},\mathbb{R}\right)$
is any compactly supported function such that $f_{B}(\bar{x})=1$
for all $\bar{x}\in\mathcal{C}$.\footnote{Expressions (\ref{c6-loc-imp-ssb}-\ref{c6-loc-imp-ssb_2}) do not
hold for unbounded regions nor for the whole spacetime, because in
these cases, the integral \eqref{c6-loc-imp-ssb_2} does not converge
to a well-defined self-adjoint operator. However, for unbounded standard
regions we can still construct local implementations of \eqref{c6-der-sca} \cite{Buchholz:1992bg}. This is no longer possible for the whole spacetime, i.e.
there is no unitary global implementation of \eqref{c6-der-sca} \cite{Bogolyubov}.} In term of the Weyl unitaries, \eqref{c6-der-sca} reads
\begin{equation}
\alpha_{s}\left(\mathrm{e}^{i\phi\left(f\right)}\right)=\mathrm{e}^{is\int_{\mathbb{R}^{d}}f\left(x\right)\,d^{d}x}\mathrm{e}^{i\phi\left(f\right)}\,,\quad s\in\mathbb{R}\,.\label{c6-ssb-weyl}
\end{equation}
Hence, the $C^{*}$-algebra $\mathcal{A}$ is generated by the collection
of all Weyl unitaries satisfying the subsidiary condition
\begin{equation}
\int_{\mathbb{R}^{d}}f\left(x\right)\,d^{d}x=0\,.\label{c6-smear-der}
\end{equation}
Thereby, the local algebras of $\mathcal{A}$ attached to double
cones are defined as in \eqref{c6-massless} but restricting the smearing
functions to the ones satisfying \eqref{c6-smear-der}, whereas the local
algebras for non-double cones are constructed from the double cone
ones imposing additivity. The algebra $\mathcal{A}$ acts irreducible
in $\mathcal{H}_{0}$. However, the local algebras attached to doubles
cones are not maximal in the vacuum representation, i.e. they do not
satisfy the duality condition \ref{c6-ass-dua} of assumption \ref{c6-ass-dhr}.
As we discussed in the previous section, we shall consider the dual
net $\mathcal{A}^{d}\supset\mathcal{A}$. In this example, it results
that $\mathcal{A}^{d}\equiv\mathfrak{F}$. Then, we have that $\mathcal{A}$
has no non-trivial superselection sectors and all the symmetries of
are spontaneously broken.

In this case, the symmetry group is non-compact. In order to not deal
with issues coming from non-compact groups,\footnote{For non-compacts groups the group average \eqref{cee} is, in principle,
not well-defined. However, there is a subclass of non-compact groups,
called \textit{amenable} \textit{groups}, where a sort of group average,
called \textit{mean}, could already be constructed. However, such
a mean is far to be unique as in the compact case. We remark that
any abelian group, and in particular $\mathbb{R}$, is amenable.} we consider the following intermediate algebra
\begin{equation}
\mathcal{A}\subsetneq\mathfrak{F}_{\lambda}\subsetneq\mathfrak{F}\,,\;\lambda>0\,,
\end{equation}
where $\mathfrak{F}_{\lambda}$ is the pointwise fixed subalgebra
under the (restricted set of) automorphisms
\begin{equation}
\phi\rightarrow\phi+2\pi n\lambda\,,\quad n\in\mathbb{Z}\,.
\end{equation}
Thus, the algebra $\mathfrak{F}_{\lambda}$ describes a free compactified
real scalar field with compactification radius $\lambda$. The parameter
$\lambda$ has dimension $(d-2)/2$. We have that $\mathfrak{F}_{\lambda} \subset \mathfrak{F}_{\lambda'}$ if and only if $\lambda'$ is an integer multiple of $\lambda$. According to \eqref{c6-ssb-weyl}, $\mathfrak{F}_{\lambda}$ is generated
by the Weyl operators $\mathrm{e}^{i\phi\left(f\right)}$ satisfying the constrain
\begin{equation}
\int_{\mathbb{R}^{d}}f\left(x\right)\,d^{d}x=\frac{k}{\lambda}\,,\quad\textrm{for some }k\in\mathbb{Z}\,.\label{c6-smear-der-1}
\end{equation}
The local algebras of $\mathfrak{F}_{\lambda}$ can be defined in
the same way that we did above for the algebra $\mathcal{A}$, but restricting the
smearing functions to the ones satisfying \eqref{c6-smear-der-1} instead of \eqref{c6-smear-der}. The inclusion $\mathfrak{F}_{\lambda}\subsetneq\mathfrak{F}$ behaves
in similar way as the one $\mathcal{A}\subsetneq\mathfrak{F}$ above.
Here, we have again that the local algebras attached to doubles cones
are not maximal in the vacuum representation, where $\mathfrak{F}_{\lambda}$
acts irreducible. The dual net corresponding to $\mathfrak{F}_{\lambda}$
is also $\mathfrak{F}$. However, in this case the symmetry group
$\mathfrak{F}\rightarrow\mathfrak{F}_{\lambda}$ is $\mathbb{Z}$,
which is also spontaneously broken.

Here we are interested in the inclusion $\mathcal{A}\subsetneq\mathfrak{F}_{\lambda}$.
The algebra $\mathcal{A}$ is the pointwise fixed subalgebra under the automorphisms
of $\mathfrak{F}_{\lambda}$
\begin{equation}
\phi\rightarrow\phi+k\lambda\,,\quad k\in[-\pi,\pi)\,.\label{c6-ssb-compact-neu}
\end{equation}
Therefore, there is a $G:=U(1)$ spontaneously broken symmetry between
$\mathfrak{F}_{\lambda}$ and $\mathcal{A}$. Let $W$ be a double
cone and define
\begin{equation}
V:=\mathrm{e}^{i\lambda^{-1}\phi\left(f\right)}\,,\quad\mathrm{supp}\left(f\right)\subset W\,,\quad\int_{\mathbb{R}^{d}}d^{d}x\,f(x)=1\,,\label{c6-ssb-cco}
\end{equation}
which belongs to $\mathfrak{F}_{\lambda}(W)$ but not to $\mathcal{A}(W)$.
We have that $\mathfrak{F}_{\lambda}\left(W\right)=\mathcal{A}\left(W\right)\vee\{V_{W}\}$.\footnote{We also have that $\mathfrak{F}_{\lambda}=\mathcal{A}\vee\{V_{W}\}$
for any operator $V_{W}$ as in \eqref{c6-ssb-cco} and any double cone
$W$.} Thereby, $\left\{ V^{n}\,:\,n\in\mathbb{Z}\right\} $ is an Abelian
subalgebra of $\mathfrak{F}_{\lambda}\left(W\right)$ which is not
included in $\mathcal{A}(W)$. In fact, such an Abelian subalgebra is
the main difference between $\mathcal{A}(W)$ and $\mathfrak{F}_{\lambda}(W)$.
The expectation values of charged operators in $\mathfrak{F}_{\lambda}$,
with respect to the $U(1)$-symmetry, have non-zero expectation values in the vacuum state
\begin{equation}
\langle V^{n}\rangle=\mathrm{e}^{-\frac{n^{2}}{2\lambda^{2}}(f\cdot C\cdot f)}\,,\label{vn}
\end{equation}
where we have defined
\begin{eqnarray}
\hspace{-1cm} f_{1}\cdot C\cdot f_{2}\!\!\! & := & \!\!\!\int_{\mathbb{R}^{2d}}d^{d}x\,d^{d}y\,f_{1}(x)C(x-y)f_{2}(y)\,,\nonumber \\
C(x)\!\!\! & := & \!\!\!\langle\phi(x)\phi(0)\rangle=\frac{\Gamma(\Delta)}{4\pi^{\frac{d}{2}}}\frac{1}{\left(x-i0^{+}e_{0}\right)^{2\Delta}}\,,\quad\Delta:=\frac{d-2}{2},\,e_{0}:=\left(1,\bar{0}\right)\,.\label{c6-corr-cft}
\end{eqnarray}
Now we consider two double cones $W_{1}\text{\Large\ensuremath{\times\negmedspace\!\!\times}}W_{2}$
and we define ${\cal I}_{12}:=V_{1}V_{2}^{\dagger}$, where $V_{j}\in\mathfrak{F}_{\lambda}(W_{j})$
are as in \eqref{c6-ssb-cco}. The operator ${\cal I}_{12}$ belongs
to the algebra ${\cal A}$, commutes with all operators in $\mathfrak{F}_{\lambda}\left((W_{1}\vee W_{2})'\right)$,
but it does not belong to the additive algebra $\mathcal{A}(W_{1}\lor W_{2})$.
Hence, the operator ${\cal I}_{12}$ is a kind of intertwiner, but
not related to any SS.

Let us now study the entropic order parameter \eqref{c6-ssb-eor-one}
for a single region. For that, we consider the standard double cone $W:=D\left(\mathcal{C}_{R}\right)$.
We can estimate $S_{\mathfrak{F}_{\lambda}\left(W\right)}(\omega|\omega\circ\varepsilon)$
computing the RE on the subalgebra $\left\{ V^{n}\,:\,n\in\mathbb{Z}\right\} $,
maximizing over all smearing functions satisfying \eqref{c6-ssb-cco}.
As happens with the intertwiner algebra for a $U(1)$-symmetry in
section \ref{U1}, this subalgebra corresponds to the classical probability
space of probability distributions over the unit circle $S^{1}$.
The state $\omega\circ\varepsilon$ corresponds to the the constant probability
density $p_{\omega\circ\varepsilon}(k):=(2\pi)^{-1}$. The other
state depends on the vacuum expectation values of $V_{W}^{n}$ (see equation \eqref{vn})
\begin{equation}
p_{\omega}\left(k\right):=\frac{1}{2\pi}\sum_{n\in\mathbb{Z}}\mathrm{e}^{ikn}\,\langle V^{n}\rangle\,,\quad k\in[-\pi,\pi)\,.\label{c6-ssb-fourier}
\end{equation}
We have to take wide smearing functions to get the maximal RE. In
analogy with the computation in section \ref{U1}, if the coefficient $(f\cdot C\cdot f)/(2\lambda^{2})$
multiplying $n^{2}$ in the exponent of \eqref{vn} is small, we get a RE 
\begin{equation}
S_{\mathfrak{F}_{\lambda}\left(W\right)}(\omega|\omega\circ\varepsilon)\simeq-1/2\log((f\cdot C\cdot f)/\lambda^{2})\,.
\end{equation}
This is the case when $R\mu\gg1$, where we have defined $\mu:=\lambda^{\frac{2}{d-2}}$
to the energy scale of $\lambda$. On the other hand, for small $R$,
the coefficient of the exponent in \eqref{vn} is large and the probability
is concentrated around $n=0$ as for $\omega\circ\varepsilon$. The RE
has a change of regime at $R\mu\sim1$ and goes to zero for $R\mu\rightarrow0^{+}$
as happens for finite groups. Then, we get
\begin{equation}
S_{\mathfrak{F}_{\lambda}\left(W\right)}(\omega|\omega\circ\varepsilon)\simeq\begin{cases}
\frac{(d-2)}{2}\log(R\mu)\,, & R\mu\gg1\,,\\
0\,, & R\mu\ll1\,.
\end{cases}\label{c6-ssb-com-1}
\end{equation}

We now study the MI difference between $\mathfrak{F}_{\lambda}$
and ${\cal A}$, which can be investigated using the same tools developed
so far. Then, we consider the regions $W_{1}:=D\left(\mathcal{C}_{R}\right)$
and $W_{2}:=D\left(\mathcal{C}'_{R+\epsilon}\right)$ ($\epsilon>0$).
For $W_{2}$ unbounded, the first and third terms on the r.h.s. of
\eqref{ecui4} independently diverge, but it difference keeps finite.
For the calculations below, it is useful to take $W_{2}$ large but
bounded, to maintain both terms independently finite, and take the
limit of $W_{2}\rightarrow D\left(\mathcal{C}'_{R+\epsilon}\right)$
in the end.

To evaluate the MI difference let us first investigate the contribution
of the intertwiners. This model has the nice feature that we can explicitly
compute their expectation values. We form a subalgebra of intertwiners
using the integer powers of the one mode $(V_{1}V_{2}^{\dagger})^{n}$.
This Abelian subalgebra $\mathcal{B}_{12}$ corresponds to the classical
probability space of probability distributions over the unit circle
$S^{1}$. The expectation values are 
\begin{equation}
\langle V_{1}^{n}V_{2}^{-n}\rangle=\mathrm{e}^{-\frac{n^{2}}{2}\sigma^{2}}\,,\label{ene-1}
\end{equation}
with
\begin{equation}
\sigma^{2}:=\frac{1}{\lambda^{2}}(f_{1}\cdot C\cdot f_{1}+f_{2}\cdot C\cdot f_{2}-2\,f_{1}\cdot C\cdot f_{2})\,.\label{minsig}
\end{equation}
This has the general Gaussian form of \eqref{distic}, but here the
expression is exact. Again, this gives, for small $\sigma^{2}\ll1$
and through a Fourier transform, a Gaussian classical probability
distribution $p_{\omega}\left(k\right)$. The state $\omega\circ\varepsilon_{12}$
corresponds to a uniform classical probability density $p_{\omega\circ\varepsilon_{12}}\left(k\right):=(2\pi)^{-1}$.
The RE is given by the classical Kullback–Leibler divergence
\begin{equation}
S_{{\cal B}_{12}}(\omega|\omega\circ\varepsilon_{12})=H_{KL}(p_{\omega}|p_{\omega\circ\varepsilon_{12}})=-\log(\sigma)\,.
\end{equation}
In order to minimize $\sigma$ in \eqref{minsig}, $\alpha_{1}$ and
$\alpha_{2}$ have to be near to each other lying along the boundary.
The minimization depends on the geometry, more precisely, on the total area
available $A\sim R^{d-2}$ and the separation distance $\epsilon$,
but is independent of the compactification radius $\lambda$. Then,
we can use symmetric test functions approximately translational invariant
along the boundary surface to get 
\begin{equation}
\sigma^{2}\simeq(\lambda^{2}R^{d-2}\,f(R/\epsilon))^{-1}\,.
\end{equation}
The area factor within the brackets is dictated by dimensional reasons
and the extensivity of the problem along the entanglement surface.
The factor $f(R/\epsilon)$ should be a slowly varying function resulting
from the minimization in the shape of the test functions in the direction
perpendicular to the boundary. This factor should ensure $\sigma\rightarrow0$
whenever $\epsilon\rightarrow0^{+}$, though at a slow pace. What we want
to emphasize is that this cannot be further improved to be of the
order $(\epsilon/R)^{d-2}$ as in the case of a the $U(1)$-symmetry
with a conformal current in the UV studied in section \ref{U1}. It
can also be checked that it cannot be improved by taking a larger algebra
formed by charge creating operators for different modes along the
surface.

The twist version tells a parallel story but it is easier to compute
the dependence on $\epsilon$. The twists can be constructed with
integrals of the current $J^{0}=\partial_{0}\phi$ as in (\ref{c6-loc-imp-ssb}-\ref{c6-loc-imp-ssb_2}).
Its expectation values are 
\begin{equation}
\langle\tau_{k}\rangle=\mathrm{e}^{-\frac{1}{2}\lambda^{2}\,k^{2}\,(f\cdot C_{j}\cdot f)}\,,
\end{equation}
where $C_{J}(x):=-\partial_{0}^{2}C\left(x\right)$. As in section
\ref{U1}, these expectation values (and the twists \eqref{c6-loc-imp-ssb_2} itself)
are very good approximations for an exponent which is large around
$|k|\gtrsim\pi$, because the result is not actually periodic $k\rightarrow k+2\pi$. However, it is important that these expectation values are interpreted
in terms of discrete probabilities of a classical random variable
taking values $q\in\mathbb{Z}$, and the twist expectation values are
represented as $\langle\tau_{k}\rangle=\sum_{q\in\mathbb{Z}}\mathrm{e}^{iqk}p_{q}$.
Following the same reasoning as in section \ref{U1}, the twist subalgebra
entropy is
\begin{equation}
S_{G_{\tau}}\left(\omega\right)\simeq\frac{1}{2}\log(\lambda^{2}(f\cdot C_{J}\cdot f))\,.\label{484}
\end{equation}
To estimate the argument of the logarithm and avoid integrals over
coinciding points we can use the fact that the twist belongs to the
neutral algebra $\mathcal{A}$ and the vacuum state on the neutral
algebra is invariant under the symmetry transformations \eqref{c6-ssb-compact-neu}.
A similar computation as the one performed in subsection \ref{twistcontinuo}
gives
\begin{equation}
f\cdot C_{J}\cdot f\simeq R^{d-2}\log(R/\epsilon)\,,\quad\frac{R}{\epsilon}\gg1\,.
\end{equation}
Hence, we expect
\begin{equation}
S_{G_{\tau}}\left(\omega\right)\simeq\frac{d-2}{2}\log(\mu R)+\frac{1}{2}\log(\log(R/\epsilon))\,,\quad\frac{R}{\epsilon}\gg1,\,R\mu\gg1\,,\label{tyu}
\end{equation}
independently of the value of $\epsilon\mu$. This is similar to the
case of a general $U(1)$-symmetry with a conformal current in the
UV given by \eqref{epifa}, but the cutoff has been replaced by the
scale of compactification. The dependence on the cutoff is subleading
but still divergent as it must be, since the size of the group is
infinite. The reason of the difference with \eqref{epifa} is clearly
that the conserved current $J_{\mu}(x)$ is not conformal, it has
dimension $d/2$ instead of $d-1$, giving a much smaller charge fluctuation
rate for short distances.
The result \eqref{tyu} also holds for $R$ smaller than the compactification
scale provided the argument in the logarithm in \eqref{484} is still
large, or equivalently, \eqref{tyu} is still positive. This curiously
seems to require very small $\epsilon$ as we decrease the radius.
For small $R$ and fix $\epsilon$, the intertwiner expectation value
is very concentrated around the identity, i.e. around $n=0$. We cannot use a continuous
charge approximation to get the probability distribution $p_{\omega}\left(k\right)$,
and these probabilities are given by a Fourier series with coefficients
proportional to \eqref{ene-1} as in \eqref{c6-ssb-fourier}. Then, the RE
goes to zero fast with $R\mu\rightarrow0^{+}$.

Coming back to the MI difference, we can follow the same reasoning
as above for finite groups. For $W_{2}$ bounded the result is finite
and then we take the limit of a large region $W_{2}$ with the rest
of the geometry fixed. Then, the last term in (\ref{ecui4}) should
be canceled by similar contribution coming from the first term, which is given by the same
charged fields as the ones contributing to the last term. After this
cancellation, there is a competition between the intertwiner and a
charged operator in $W_{1}$ to the first term. For $\epsilon$ small
enough, as corresponds to a cutoff, the intertwiner dominates and
we should have 
\begin{equation}
\Delta I\simeq\frac{d-2}{2}\log(\mu R)+\frac{1}{2}\log(\log(R/\epsilon))-S_{\mathfrak{F}_{\lambda}(W_{1})}(\omega|\omega\circ\varepsilon)\,,\quad\epsilon\ll R,\mu^{-1}\,.\label{c6-ssb-com-DI-casi}
\end{equation}
Then, replacing \eqref{c6-ssb-com-1} into \eqref{c6-ssb-com-DI-casi},
we get, for $\epsilon\ll R,\mu^{-1}$,
\begin{equation}
\Delta I\simeq\begin{cases}
\frac{d-2}{2}\log(\mu R)+\frac{1}{2}\log(\log(R/\epsilon))\,, & R\mu\ll1\,,\\
\frac{1}{2}\log(\log(R/\epsilon))\,, & R\mu\gg1\,.
\end{cases}\label{123}
\end{equation}
For small $R\mu\ll1$, we note that the negative sign of the first term for
small $R\mu$ must be supported by a compensating sign of the second term,
and we need an exponentially small cutoff, as already remarked
previously. For large $R\mu\gg1$, $\Delta I$ does not vanish, in
contrast to the case of a finite group. It has the same dependence
on $\epsilon$ as in the UV. On the other hand,
for large $\epsilon$, we expect $\Delta I$ to vanish in the regime
of small $R\mu$.

\subsubsection{SSB of a compact Lie group}

Now we consider the case of a compact Lie symmetry group spontaneously
broken with a conformal current in the UV. Let us first discuss
the order parameter $S_{\mathfrak{F}\left(W\right)}(\omega|\omega\circ\varepsilon)$
for the double cone $W:=D\left(\mathcal{C}_{R}\right)$. For large
$R$, we can compute this quantity with the RE on the algebra of compactified
scalars that are the Goldstone modes. As in the previous discussion,
we get 
\begin{equation}
S_{\mathfrak{F}\left(W\right)}(\omega|\omega\circ\varepsilon)\simeq\frac{{\cal G}(d-2)}{2}\log(R\mu)\,,\label{3121}
\end{equation}
where $\mu$ is the SSB scale that is taken of the same order as the
compactification radius, and ${\cal G}$ is the number of Goldstone
bosons. For smaller radius, we cannot use the Goldstone boson approximation
any more, but we expect that the RE has a change of
regime at $R\mu\sim1$ and goes to zero for $R\mu\rightarrow0^{+}$,
as happens for finite groups and compact scalars.

For the MI difference, we can follow the same reasoning as above for
the compact scalar and finite groups. We take spacetime regions $W_{1}:=D\left(\mathcal{C}_{R}\right)$
and $W_{2}:=D\left(\mathcal{C}'_{R+\epsilon}\right)$ ($\epsilon>0$).
The intertwiners dominate the rest of the contribution to the first
term of \eqref{ecui4} and, using the results on section \ref{U1},
we have 
\begin{equation}
\Delta I\simeq\frac{{\cal G}(d-2)}{2}\log(R/\epsilon)-S_{\mathfrak{F}\left(W_{1}\right)}(\omega|\omega\circ\varepsilon)\,.\label{67}
\end{equation}
Then, according to the \textcolor{black}{preceding discussion, we
get, for $\epsilon\ll R,\mu^{-1}$,}
\begin{equation}
\Delta I\simeq\begin{cases}
\frac{{\cal G}(d-2)}{2}\log(R/\epsilon)\,, & R\mu\ll1\,,\\
-\frac{{\cal G}(d-2)}{2}\log(\epsilon\mu)\,, & R\mu\gg1\,.
\end{cases}
\end{equation}
This does not vanish for $R\mu\gg1$, in contrast to the case of a
finite group. It has the same dependence on $\epsilon$ as in the
UV, but the dependence on the radius has been replaced by the SSB scale.
Note that both models contain the massless scalar contribution of
the Goldstone modes in the IR on top of this difference.

\subsubsection{Remarks on previous results in the literature}

We make some comments on previous results in the literature. In \cite{Casini:2014aia}
there is a numerical study of the EE of a free Maxwell field in $d=3$.
This model is equivalent to the algebra of derivatives of a free massless
real scalar through the identification $\varepsilon_{\mu\nu\delta}F^{\nu\delta}=\partial_{\mu}\phi$.
Then, the relation between the scalar and the Maxwell models is the
same as between the scalar and the algebra of its derivatives. The
symmetry $\phi\rightarrow\phi+s$ is uncompactified and the model
does not contain any scales. This is equivalent to the model $\mathfrak{F}_{\lambda}$
in the decompactification limit $\lambda\rightarrow+\infty$. Because
of that, the RE $S_{\mathfrak{F}\left(W\right)}(\omega|\omega\circ\varepsilon)$ is not finite and we get the divergent quantity $\frac{1}{2}\log(R\mu)$
as $\mu=\lambda^{\frac{2}{d-2}}\rightarrow+\infty$. In the presence
of a short distance cutoff $\delta>0$ and with the naive lattice
interpretation of the difference in entropies between the two models
in place of the RE (see the discussion at the end of
section \ref{tools}), the compactification radius should get trade
off by the cutoff, and we get, up to lattice ambiguities,
\begin{equation}
S_{\textrm{Maxwell}}(R)-S_{\textrm{scalar}}(R)\simeq\frac{1}{2}\log(R/\delta)\,.\label{dife}
\end{equation}
This is what was found numerically in \cite{Casini:2014aia}. For the
MI difference, according to \eqref{123}, in the
limit of small $\epsilon$ and $\mu\rightarrow+\infty$, we get
\begin{equation}
I_{\textrm{scalar}}(W_{1},W_{2})-I_{\textrm{Maxwell}}(W_{1},W_{2})\simeq\frac{1}{2}\log(\log(R/\epsilon))\,.\label{mild}
\end{equation}
This does not contain a $\log(R/\epsilon)$ term, and hence,
it does not reproduce the difference in lattice entropies \eqref{dife},
which would have given the contradictory result that the MI of the
smaller model would have been bigger than the one of the larger model.
We checked numerically in the lattice, following the methods in
\cite{Casini:2014aia}, this dependence of the MI difference in the short $\epsilon$ limit, and we found agreement with
\eqref{mild}. This term should be attributed as a negative contribution $-1/2\log(\log(R\epsilon))$
to the Maxwell field MI rather than a positive one to the the scalar, which
has a finite constant term. An analogous result is expected between
the scalar and its derivatives (dual to higher form gauge fields)
in any dimensions. Notice that the MI of the free scalar
$\mathfrak{F}$ is finite as well as the one of ${\cal A}$, even
if there is a non-compact SSB relating it to ${\cal A}$. This is
different from what we expect for a non-compact symmetry that is not
spontaneously broken.

In \cite{Agon:2013iva} the authors study the change in entropy, between
a free real scalar field $\mathfrak{F}$ and a free real compact scalar
$\mathfrak{F}_{\lambda}$, using the replica trick. For the MI, this
is given by the subtraction of results in \eqref{123} with \eqref{mild}.
Then, we get
\begin{equation}
I_{\mathfrak{F}_{\lambda}}(W_{1},W_{2})-I_{\mathfrak{F}}(W_{1},W_{2})=\begin{cases}
\frac{d-2}{2}\log(\mu R)\,, & R\mu\ll1\,,\\
0\,, & R\mu\gg1\,.
\end{cases}
\end{equation}
This coincides with the result of \cite{Agon:2013iva} for the difference
in entropies. Then, using $\Delta I/2$ as a proper renormalized entropy,
our result differs from the one in \cite{Agon:2013iva} by a factor
$1/2$. This factor is typical of the difference between EE and $\Delta I/2$
for classically correlated variables.

The EE in models with SSB of continuous symmetries is also discussed
in \cite{Metlitski:2011pr}. The authors compute the EE in the case
of SSB for a finite volume space. At finite size, symmetry is restored
and the physics is the one of $\mathfrak{F}$, but with the symmetric
state, which does not satisfy clustering in the large volume limit.
They show that the EE for a large region of size $R$ contains precisely
a term of the form \eqref{3121} as long as $R\mu\gg1$. This explains previous
lattice simulations \cite{kallin2011anomalies}. This also coincides
with the results from our calculations, since the lattice expression
for the RE in \eqref{3121} is
\begin{equation}
S_{\mathfrak{F}}(\omega\circ\varepsilon)-S_{{\cal \mathfrak{F}}}(\omega)\sim\frac{{\cal G}(d-2)}{2}\log(R\mu)\,,\quad R\mu\gg1\,.
\end{equation}
The symmetric state has a new term that shows up in the calculations
of \cite{Metlitski:2011pr}. Note that the MI difference has a completely
different behavior.

\subsection{The case of $d=2$ and chiral CFTs\label{dosd}}

For the purpose of this section let us consider the algebras attached
to spacelike open regions over the Cauchy surface $\Sigma_{0}\cong\mathbb{R}$
at time $x^{0}=0$. For the case of a chiral CFT, the algebras are
attached to open regions over the null line $\mathcal{N}\cong\mathbb{R}$.
In both cases, the topology of the space is $\mathbb{R}$. The double
cones corresponds to bounded open intervals $A\subset\mathbb{R}$.

The main reason of why the theory of DHR SS has significant differences
in this case with respect to the case $d\geq3$ is that the complement
of double cone is a disconnected region formed by two semi-infinite
regions. This gives a non-trivial statistics for the charged sectors.
This implies that the SS structure does not necessarily come from
a compact symmetry group, and it has to be described more generally
by their statistical dimensions and fusion rules. 

Apart from that, a second difference arises when we compute the MI between a double cone $W_{1}$ and a nearly complementary
region $W_{2}$. In this case, the shell $S:=(W_{1}\lor W_{2})'$
is a disconnected region formed by two bounded intervals of length
$\epsilon\apprge0$, whereas for $d\geq3$ this region is connected.
Let still consider the case where ${\cal A}$ is the pointwise fixed
subalgebra of $\mathfrak{F}$ under a finite symmetry group $G$.
For a massive theory, we get twice the value corresponding to higher
dimensions 
\begin{equation}
\Delta I=2\log|G|\,.
\end{equation}
This can be thought as an instance of \eqref{instance} since the
shell and $W_{2}$ are disconnected regions, and hence, there are
two sets of independent intertwiners connecting $W_{1}$ with the
two connected components of $W_{2}$. However, if the theory is conformal,
$W_{2}$ can be thought of as a single interval because the theory
is defined over the unit circle $S^{1}\cong\overline{\mathbb{R}}$.
In this case, we obtain 
\begin{equation}
\Delta I=\log|G|\,.
\end{equation}
If the two regions $W_{1}$ and $W_{2}$ touch each other on one side
and they stay enough separated from the other side, we get $\Delta I=\log|G|$
in both cases, conformal and massive.

A case which can be computed exactly is the free chiral current that
we have studied in chapter \ref{CURRENT}. By bosonization, this is
the same model as the one obtained by restricting the algebra $\mathfrak{F}$,
of a free chiral fermion field, to the subalgebra ${\cal \mathcal{A}}$
generated by the fermionic current $:\psi^{\dagger}(x)\psi(x):$.
The group symmetry is $G=U(1)$. As both theories are conformally invariant,
their MIs on the vacuum state are only functions of the cross-ratio
\begin{equation}
\eta:=\frac{(b_{1}-a_{1})(b_{2}-a_{2})}{(a_{2}-a_{1})(b_{2}-b_{1})}\in(0,1)\,,
\end{equation}
which depends on the endpoints of the intervals $A_{1}:=(a_{1},b_{1})$ and $A_{2}:=(a_{2},b_{2})$.
Using the results of chapter \ref{CURRENT}, we write\footnote{For the MI of the chiral fermion see also \cite{Casini:2005rm,reduced_density,Longo_xu}}
\begin{eqnarray}
I_{\mathfrak{F}}(\eta)\!\!\! & = & \!\!\!-\frac{1}{6}\log(1-\eta)\,,\label{ffermion}\\
I_{{\cal A}}(\eta)\!\!\! & = & \!\!\!-\frac{1}{6}\log(1-\eta)-\tilde{U}(\eta)\,,
\end{eqnarray}
where $\tilde{U}(\eta)\geq0$.\footnote{$\tilde{U}(\eta)$ is called $-U(\eta)$ in \eqref{mic}. }
Using the results of \eqref{c5-subsec:Mutual-information} for the present
set up ($b_{1}-a_{1}=:R$, $a_{2}=b_{1}+\epsilon$ and $b_{2}=a_{1}-\epsilon$),
we obtain that the MI difference $\Delta I\left(\eta\right)=\tilde{U}(\eta)$
behaves as
\begin{equation}
\Delta I\left(\eta\right)\sim\frac{1}{2}\log(-\log(1-\eta))\sim\frac{1}{2}\log(\log(R/\epsilon))\,,\quad\epsilon\rightarrow0^{+}\,,\label{sistea}
\end{equation}
where $\eta\sim1-(\epsilon/R)^{2}$ and $R$ is the size of the interval
corresponding to the region $A_{1}$.

Assuming the usual relation that the entropy for complementary
regions is the same in a pure global state, we would have 
\begin{equation}
S(I_{1}\cup I_{3})=S(I_{2}\cup I_{4})\;,\label{ruin}
\end{equation}
where we are thinking in a compactified real line $S^{1}\cong\overline{\mathbb{R}}$
divided in four intervals. Completing this relation with the entropies
of the single intervals to get MIs,\footnote{The single interval entropies are $S(r)=(c/6)\log(r/\epsilon)+\mathrm{const.}$,
where $r$ is the size of the interval.} for a CFT in $d=2$, \eqref{ruin} translates into the symmetry relation \cite{Casini:2004bw}
\begin{equation}
I(\eta)=I(1-\eta)-\frac{c}{6}\log\left(\frac{1-\eta}{\eta}\right)\,,\label{fds}
\end{equation}
where $c$ is the central charge (summed over both chiralities). This
symmetry property does not hold when there are superselection sectors
which ruin \eqref{ruin}, and in particular is badly broken for the
chiral current (not for the chiral fermion) as we have shown in the
previous chapter. This symmetry leads to $\tilde{U}(\eta)=\tilde{U}(1-\eta)$,
while for the current we have $\tilde{U}(0)=0$, because large distance
mutual information vanishes, and $\tilde{U}(0)=+\infty$. For the
case of a finite symmetry group, $\Delta I\left(\eta=0\right)=0$ while
$\Delta I\left(\eta=1\right)=\log|G|$.

Relation \eqref{fds} can be shown from the replica trick using modular
invariance for non-chiral models \cite{Cardy:2017qhl}. In connection
to this, it has been shown more generally that modular invariant models
are complete (duality holds for two intervals) and the symmetry property
\eqref{fds} holds \cite{Xu:2018fsv}.

It is to be noted that in $d=2$ it does not hold any more $\Delta I=S_{{\cal A}'_{34}}(\omega|\omega\circ\varepsilon)$,
where $\varepsilon:{\cal A}'_{34}\rightarrow{\cal A}{}_{12}$. This
is because, in $d=2$, the algebra ${\cal A}'_{34}$, in addition to
the intertwiners, contains the twists, which are not in $\mathfrak{F}_{12}$.
It is expected that $S_{{\cal A}'_{34}}(\omega|\omega\circ\varepsilon)$
has limit $2\log|G|$ at the point of contact, instead of $\log|G|$
as happens for $\Delta I$.

The case of general SS not necessarily coming from a symmetry group
was treated in \cite{Longo_xu,Xu:2018uxc,Xu:2018fsv}. Each sector
has a statistical dimension $d_{r}\geq1$, which can be a non-integer
and it can only take some specific values for $d_{r}<2$ \cite{Longo:1989tt}.
Generalizing the result we got for a group, we have, for the limit of contact
between the complementary intervals, in a $2d$ CFT 
\begin{equation}
S_{{\cal A}'_{34}}(\omega|\omega\circ\varepsilon)=\log{\cal D}^{2}=\log\sum_{r}d_{r}^{2}\,,\label{redh}
\end{equation}
where the sum runs over the irreducible sectors, and ${\cal D}^{2}:=\sum_{r}d_{r}^{2}$
is called the \textit{quantum dimension} of the model $\mathcal{A}$.
In terms of the quantum dimensions of the SS, this formula is the
same in any spacetime dimension. It is also the index of inclusion
of algebras ${\cal A}_{12}\subset{\cal A}'_{34}$ \cite{Longo:1994xe,Longo:1989tt,Kawahigashi:1999jz}.
The result \eqref{redh} holds for finite index.

\section{Examples of intertwiner and twist bounds\label{bounds}}

In this section, we compute in some concrete examples, the intertwiner
and twists expectation values. The objective is to build up intuition
on how they generally behave in order to provide the best bounds
available. We describe how intertwiners are assimilated to edge modes
localized near the boundary and how they tend to minimize the modular
energy for nearly complementary regions, while they spread out for distant regions. We also describe how the charge creating operators
in the same region try to minimize their mutual entanglement, and
hence, to repel each other. Finally, concerning the squeezed twists, we show
that, in the short $\epsilon$ limit, they are exponentially suppressed
by the area and have Gaussian expectation values for Lie groups.

\subsection{Intertwiners at short distance\label{c6-subsec:Intertwiners-at-short}}

In this subsection, we explain, for the fermion field, how the unitary
intertwiner $\mathcal{I}_{12}$ of two complementary regions can be
suitable chosen in order to maximize its vacuum expectation value,
i.e $|\left\langle \mathcal{I}_{12}\right\rangle | \simeq1$.\footnote{We always have that $|\left\langle \mathcal{I}\right\rangle |\leq1$, since
$\mathcal{I}$ is unitary.} We consider the theory of a free fermion field and the $\mathbb{Z}_{2}$
bosonic symmetry discussed in section \ref{DHR}. An intertwiner
between a region $W_{1}$ and its complement $W_{2}:=W_{1}^{'}$ can
be written as 
\begin{equation}
\mathcal{I}_{12}:=V_{1}V_{2}^{\dagger}\,,\label{int_5}
\end{equation}
where $V_{i}\in\mathfrak{F}\left(W_{i}\right)$ are fermionic unitary
operators made out of the fermion fields as in \eqref{c6-ferm-uni}
and \eqref{siste}
\begin{equation}
V_{j}:=\psi\left(f_{j}\right)+\psi^{\dagger}\left(f_{j}\right)\,,\label{eq: ferm_int}
\end{equation}
where $f_{j}$ are spinor valued functions supported in the regions
$W_{j}$. Here we use the description at fixed time $x^{0}$. The
spacetime regions $W_{j}:=D\left(\mathcal{C}_{j}\right)$ are the Cauchy
development of the complementary space regions $\mathcal{C}_{1}\subset\mathbb{R}^{d-1}$
and $\mathcal{C}_{2}=\mathbb{R}^{d-1}-\mathcal{C}_{1}$. Then, we
write the smearing functions as $f_{j}\left(x\right):=\delta\left(x^{0}\right)\alpha_{j}\left(\bar{x}\right)$.
Expression \eqref{eq: ferm_int} automatically gives $V_{j}=V_{j}^{\dagger}$.
In order to have $V_{j}^{-1}=V_{j}^{\dagger}$ we must also impose
\begin{equation}
\int_{\mathcal{C}_{j}}d^{d-1}x\,\alpha_{j}^{\dagger}\left(\bar{x}\right)\alpha_{j}\left(\bar{x}\right)=1\,.\label{norm_u}
\end{equation}
Using that the local field algebra $\mathfrak{F}\left(W_{1}\right)$
satisfies duality,\footnote{This for sure happens for double cones because our assumption. According
to Bisognano-Wichmann theorem it also happens for Wedges. Moreover,
according to our discussion at the end of section \ref{c6-secc-field},
duality should also hold for two double cones, since the field algebra
has no DHR SS. } we can choose the unitary $V_{2}\in\mathcal{\mathfrak{F}}\left(W_{2}\right)$
as the vacuum modular conjugated of some unitary operator $\tilde{V}_{1}\in\mathcal{\mathfrak{F}}\left(W_{1}\right)$
\begin{equation}
V_{2}=:-i\,(ZJ)^{\dagger}\tilde{V}_{1}^{\dagger}ZJ\,,
\end{equation}
where $Z$ is the operator defined in definition \ref{c1-def-Z},
constructed from the grading of the algebra and it takes account
that the algebras of $W_{1}$ and $W_{2}$ graded commute.\footnote{For fermionic nets, the modular conjugation $J$ must be replaced
by the twisted modular conjugation $ZJ$ \cite{carpi2008structure}.
Regardless this technicality, the outcome of the argument below holds. } Since modular conjugation respects statistics, $V_{2}$ is a fermionic
operator iff $\tilde{V}_{1}$ is fermionic.\footnote{Moreover, for free fields, $V_{2}$ is of the form \eqref{eq: ferm_int}
if $\tilde{V}_{1}$ is of that form too.} Using the relations for the modular operator and modular conjugation
for a fermionic model \cite{carpi2008structure}, we can rewrite the
vacuum expectation value of the intertwiner \eqref{int_5} as 
\begin{equation}
\left\langle V_{1}V_{2}^{\dagger}\right\rangle =\left\langle V_{1}\Delta^{\frac{1}{2}}\tilde{V}_{1}^{\dagger}\right\rangle =\left\langle V_{1}\mathrm{e}^{-\frac{1}{2}K}\tilde{V}_{1}^{\dagger}\right\rangle \,,\label{exp_int_mod}
\end{equation}
where $K$ is the ``full'' vacuum modular Hamiltonian of the algebra $\mathfrak{F}(W_{1})$.
We can write it in terms of the inner modular Hamiltonians $K:=K_{W_{1}}-K_{W'_{1}}$.
Moreover, for the purpose of the expression \eqref{exp_int_mod},
we need only to known $K_{W_{1}}$ since $V_{1},\tilde{V}_{1}\in\mathfrak{F}(W_{1})$.
To lighten the notation, we make an abuse of notation and denote $K_{1}:=K_{W_{1}}$.

Expression \eqref{exp_int_mod} indicates that we can search for a
maximum in the expectation value within the choices $\tilde{V}_{1}=V_{1}$.
Here we invoke all the relations developed in sections \ref{subsec-CAR-algebra-(fermions)}
and \ref{c3-sec_cont}. The ``inner'' modular Hamiltonian is quadratic
in the field operator and the eigenfunctions $u_{s,k}\left(\bar{x}\right)$
of the modular Hamiltonian kernel are the same as those of the correlator
kernel (see equations from \eqref{c3-fer-corr-ker} to \eqref{c3-mh-ker-diag}).
To compute \eqref{exp_int_mod} is useful to use the operators defined
in equation \eqref{c3-field-s}, namely
\begin{equation}
\tilde{\psi}\left(s,k\right):=\int_{\mathcal{C}_{1}}d^{d-1}x\,u_{s,k}^{\dagger}\left(\bar{x}\right)\psi\left(\bar{x}\right)\,,\quad\psi\left(x\right)=\sum_{k}\int_{-\infty}^{+\infty}ds\,u_{s,k}\left(\bar{x}\right)\,\tilde{\psi}\left(s,k\right)\,,\label{new_modes}
\end{equation}
which satisfy the CAR relations \eqref{c3-car-fer-mod}. Replacing
\eqref{new_modes} into \eqref{c3-mh-ker}, we show that the modular
Hamiltonian is diagonal in these new modes 
\begin{equation}
K_{1}=\sum_{k\in\Upsilon}\int_{-\infty}^{+\infty}ds\,\tilde{\psi}\left(s,k\right)^{\dagger}\,2\pi s\,\tilde{\psi}\left(s,k\right)\,.\label{new_H}
\end{equation}
Using \eqref{new_modes}, we can rewrite \eqref{eq: ferm_int} as\footnote{From now on, we omit the subscripts $j$ in $V_{j}$, $\alpha_{j}$
and $\mathcal{C}_{j}$.} 
\begin{equation}
V:=\sum_{k\in\Upsilon}\int_{-\infty}^{+\infty}ds\left[\tilde{\psi}\left(s,k\right)\tilde{\alpha}\left(s,k\right)^{*}+\tilde{\psi}^{\dagger}\left(s,k\right)\tilde{\alpha}\left(s,k\right)\right]\,,\label{new_v}
\end{equation}
where
\begin{equation}
\tilde{\alpha}\left(s,k\right):=\int_{\mathcal{C}}d^{d-1}x\,u_{s,k}^{\dagger}\left(\bar{x}\right)\alpha_{1}\left(\bar{x}\right)\,,\label{mod_ft}
\end{equation}
and the normalization condition \eqref{norm_u} implies
\begin{equation}
\sum_{k\in\Upsilon}\int_{-\infty}^{+\infty}ds\,\left|\tilde{\alpha}\left(s,k\right)\right|^{2}=1\,.
\end{equation}
The vacuum correlators for these new modes can be easily obtained
from \eqref{c3-ferm-con-eig} and \eqref{new_modes} 
\begin{equation}
\left\langle \tilde{\psi}^{\dagger}\left(s,k\right)\tilde{\psi}\left(s',k'\right)\right\rangle =\frac{1}{1+\mathrm{e}^{2\pi s}}\delta_{kk'}\delta\left(s-s'\right)\,.\label{new_corr1}
\end{equation}
Replacing \eqref{new_H} and \eqref{new_v} into \eqref{exp_int_mod},
and using \eqref{new_corr1} and the fact that 
\begin{eqnarray}
\left[K,V_{1}\right]=0 & \textrm{and} & K\left|0\right\rangle =0\,,
\end{eqnarray}
a straightforward computations gives 
\begin{equation}
\left\langle V_{1}V_{2}^{\dagger}\right\rangle =\left\langle V_{1}\Delta^{\frac{1}{2}}V_{1}^{\dagger}\right\rangle =\sum_{k\in\Upsilon}\int_{-\infty}^{+\infty}\frac{\left|\tilde{\alpha}\left(s,k\right)\right|^{2}}{\cosh\left(\pi s\right)}ds\,.
\end{equation}
It is not possible to choose a charge creating operator that commutes
with the modular Hamiltonian, which would imply $\left\langle V_{1}V_{2}^{\dagger}\right\rangle =1$
exactly, because it is charged and the modular Hamiltonian is neutral.\footnote{More precisely, the modular group $\Delta_{\Omega}^{it}$, of any
invariant vector $\left|\Omega\right\rangle \in\mathcal{H}_{0}$ and
any subalgebra $\mathfrak{F\left(\mathcal{O}\right)}$, belongs to
the neutral algebra $G'$. Then, the unbounded modular Hamiltonian
$K_{\Omega}$ is \textit{neutral} in the sense that all its bounded
spectral projectors belongs to $G'$. In particular, we have that
$\left\langle K_{\Omega}\Psi_{1}\right|U\left(g\right)\left|\Psi_{2}\right\rangle =\left\langle \Psi_{1}\right|U\left(g\right)K_{\Omega}\left|\Psi_{2}\right\rangle $
for all $\left|\Psi_{1}\right\rangle ,\left|\Psi_{2}\right\rangle \in\mathrm{Dom}\left(K_{\Omega}\right)$
and all $g\in G$.} But we can choose a charged mode with a very small modular energy.
As this formula clearly displays, in order to maximize the expectation
value of the intertwiner we have to construct a wave packet with small
modular energy by localizing $\tilde{\alpha}\left(s,k\right)$ sharply
around $s=0$. There is a lot of freedom in approaching this limit.
For example, we can choose Gaussian wave packets 
\begin{equation}
\tilde{\alpha}\left(s,k\right):=\frac{\sqrt{p_{k}}}{\left(2\pi\right)^{\frac{1}{4}}\sqrt{\sigma}}\mathrm{e}^{-\frac{s^{2}}{4\sigma^{2}}-i\lambda_{k}s}\,,\label{gausi}
\end{equation}
with $\lambda_{k}\in\mathbb{R}$, $\sigma>0$, $0\leq p_{k}\leq1$
and $\sum_{k\in\Upsilon}p_{k}=1$. Under this choice, we have 
\begin{eqnarray}
\left\langle V_{1}V_{2}^{\dagger}\right\rangle  & \underset{\sigma\rightarrow0^{+}}{\longrightarrow} & 1\,.\label{int_to_one}
\end{eqnarray}
As an example, we have $\left\langle V_{1}V_{2}^{\dagger}\right\rangle >0.99$
for $\sigma=\frac{1}{22}$. Notice that we still have the freedom to choose
different phases and probabilities for the different values of the degeneracy parameter
$k\in\Upsilon$. The values of the degeneracy parameters are in correspondence
with the variables describing the boundary of the region \cite{Arias17}.

After we have shown that the expectation value of the free fermion
intertwiner can be (asymptotically) maximized, we want to see how these ``maximized'' wave packets are localized in position space.
We expect that such wave packets are more and more supported around
the boundary $\partial\mathcal{C}{}_{1}$ as long as the expectation
value \eqref{int_to_one} approximates to $1$. Certainly, the modular
conjugated operator $V_{2}$ will be also located around $\partial\mathcal{C}{}_{1}$.
In the following, we will show this behavior in some examples.

\subsubsection{Free chiral fermion and Rindler wedge\label{subsec:1+1-chiral-fermion}}

\begin{figure}
\centering
\begin{subfigure}{.5\textwidth}
  \centering
  \includegraphics[width=.9\linewidth]{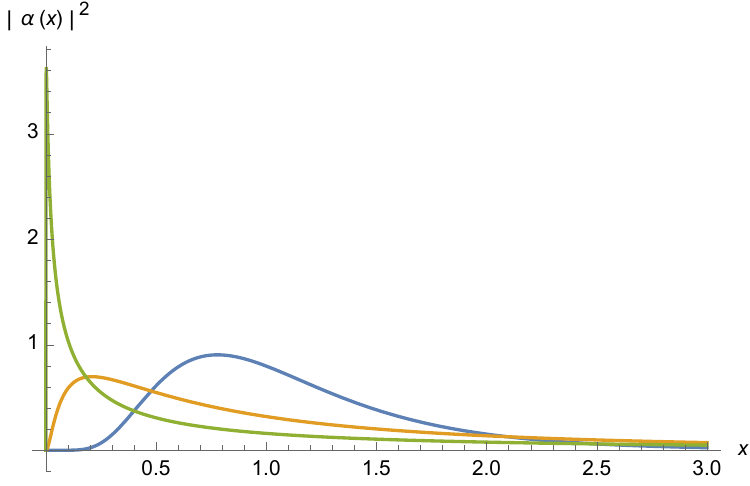}
  \caption{$\lambda=0$ and different values of $\sigma=\frac{1}{5},\frac{2}{5},1$.}
  \label{fig:2d_rw_var}
\end{subfigure}%
\begin{subfigure}{.5\textwidth}
  \centering
  \includegraphics[width=.9\linewidth]{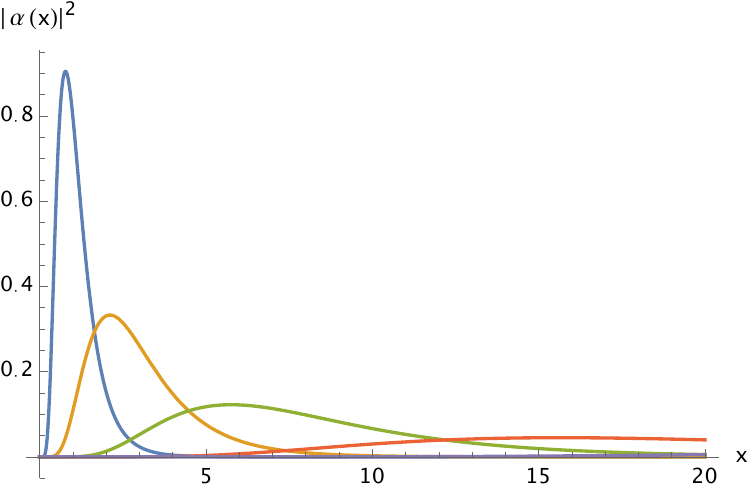}
  \caption{$\sigma=1$ and different values of $\lambda=0,1,2,3$.}
  \label{fig:2d_rw_disp}
\end{subfigure}
\caption{Localization of the wave packet $\left|\alpha\left(x\right)\right|^{2}$ (eq. \eqref{wp_2d_rw}) for the free chiral fermion and the right Rindler wedge.}
\end{figure}

We consider the right Rindler wedge $\mathcal{C}_{1}:=\left\{ x>0\right\} $.
The normalized eigenfunctions are \cite{scalar_nuestro} 
\begin{equation}
u_{s}\left(x\right):=\frac{\mathrm{e}^{is\log\left(x\right)}}{\sqrt{2\pi x}}\,,\quad x>0\,,
\end{equation}
and the ``modular'' Fourier transform \eqref{mod_ft} can be analytically
done 
\begin{equation}
\alpha\left(x\right)=\left(\frac{2}{\pi}\right)^{\frac{1}{4}}\sqrt{\frac{\sigma}{x}}\,\mathrm{e}^{-\sigma^{2}\left(\log\left(x\right)-\lambda\right)^{2}}\,.\label{wp_2d_rw}
\end{equation}
The probability density $\left|\alpha\left(x\right)\right|^{2}$ is
an ordinary Gaussian wave packet but in the logarithmic variable $z:=\log\left(x\right)$.
In figure \ref{fig:2d_rw_var}, we plot the the wave packet \eqref{wp_2d_rw}
for a fixed $\lambda$ and different values of $\sigma$. Similarly,
in figure \ref{fig:2d_rw_disp}, we plot the same wave packet for a
fixed $\sigma$ and different values of $\lambda$. As we can see
from figure \ref{fig:2d_rw_var}, as long as $\sigma\rightarrow0$,
the wave packet $\left|\alpha\left(x\right)\right|^{2}$ concentrates
around $x=0$. The job of the parameter $\lambda$ is to move the
wave packet center of mass inside the region $x>0$. For a given $\lambda$,
the wave packet concentrates $\frac{1}{2}$ of the probability between
$0<x<\mathrm{e}^{\lambda}$.

\subsubsection{Free chiral fermion and one interval}

\begin{figure}
\centering
\begin{subfigure}{.5\textwidth}
  \centering
  \includegraphics[width=.9\linewidth]{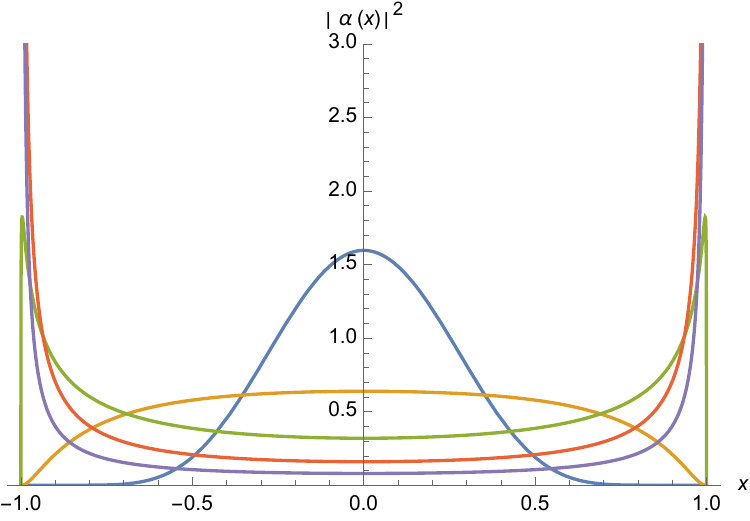}
  \caption{$\lambda=0$ and different values of $\sigma=\frac{1}{20},\frac{1}{10},\frac{1}{5},\frac{2}{5},1$.}
 \label{fig:2d_1_var}
\end{subfigure}%
\begin{subfigure}{.5\textwidth}
  \centering
  \includegraphics[width=.9\linewidth]{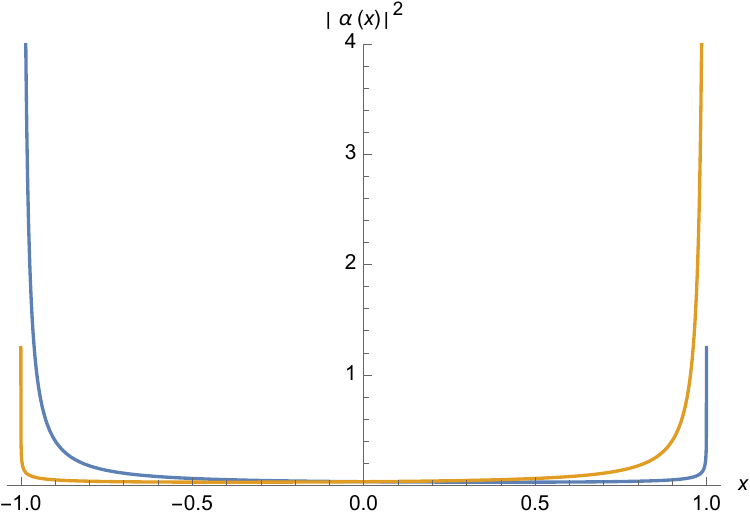}
  \caption{$\sigma=\frac{1}{10}$ and $\lambda=-10,10$.}
  \label{fig:2d_1_disp}
\end{subfigure}
\caption{Localization of the wave packet $\left|\alpha\left(x\right)\right|^{2}$ (eq. \eqref{wp_2d_1}) for the free chiral fermion and a single interval. We set the interval
endpoints to $a=-1$ and $b=1$. The apparent divergencies when $x\rightarrow a^{+},b^{-}$ are a matter of the plot resolution. The wave packet $\left|\alpha\left(x\right)\right|^{2}$ goes continuously to zero at the endpoints of the intervals.}
\end{figure}

We set $\mathcal{C}{}_{1}:=\left(a,b\right)$ an interval. In this
case, the normalized eigenfunctions are the one computed in chapter
\ref{CURRENT}
\begin{equation}
u_{s}\left(x\right)=\sqrt{\frac{z'\left(x\right)}{2\pi}}\mathrm{e}^{isz\left(x\right)}\,,\quad a<x<b\,,
\end{equation}
where $z\left(x\right):=\log\left(\frac{x-a}{b-x}\right)$.\footnote{In chapter \ref{CURRENT}, we have used the notation $\omega(x)$ for
the function $z(x)$.} Then, the integral \eqref{mod_ft} can be analytically done 
\begin{equation}
\alpha\left(x\right)=\left(\frac{2}{\pi}\right)^{\frac{1}{4}}\sqrt{z'\left(x\right)}\sqrt{\sigma}\mathrm{e}{}^{-\sigma^{2}\left(\lambda-z\left(x\right)\right)^{2}}\,.\label{wp_2d_1}
\end{equation}
The probability density $\left|\alpha\left(x\right)\right|^{2}$ is
a Gaussian wave packet in the variable $z$. In figure \ref{fig:2d_1_var}
we show the wave packet \eqref{wp_2d_1} for a fixed $\lambda$ and
different values of $\sigma$. As we can appreciate, the support of the
function $\left|\alpha\left(x\right)\right|^{2}$ concentrates around
the endpoints $a,b$ of the interval as long as $\sigma\rightarrow0^{+}$.

For example, a straightforward computation, shows that, for $\lambda=0$,
the region $\left(a,a+\epsilon\right)\cup\left(b-\epsilon,\epsilon\right)$
with $\epsilon=\left(b-a\right)\left(1+\mathrm{e}^{\frac{1}{200\sigma}}\right)^{-1}$
concentrates $\apprge0.99$ of the probability of $\left|\alpha\left(x\right)\right|^{2}$.
In figure \ref{fig:2d_1_disp}, we plot the wave packet \eqref{wp_2d_1}
for a fixed $\sigma$ and two different values of $\lambda$. As we
can see, the job of the parameter $\lambda$ is to distribute the probability
of the wave packet asymmetrically between the endpoints of the interval.
In other words, we can freely choose the ``maximizer'' intertwiner
to be located around $x=a$ or around $x=b$, or simultaneously around
both endpoints with some relative probability that we can choose at
our own will. To be more precise, we first redefine the real parameter
$\lambda$ as $\mu:=\sqrt{2}\lambda\sigma$. Now, the probability distribution
\begin{equation}
\left|\alpha\left(x\right)\right|^{2}=\left(\frac{2}{\pi}\right)^{\frac{1}{2}}z'\left(x\right)\sigma\,\mathrm{e}{}^{-\left(\mu-\sqrt{2}\sigma z\left(x\right)\right)^{2}}\,,
\end{equation}
has the limit 
\begin{equation}
\lim_{\sigma\rightarrow0^{+}}\left|\alpha\left(x\right)\right|^{2}=q\cdot\delta\left(x-a\right)+\left(1-q\right)\cdot\delta\left(x-b\right)\,,
\end{equation}
where $q=q\left(\mu\right):=\frac{1-\text{erf}(\mu)}{2}\in\left[0,1\right]$
and $\text{erf}$ is the usual Gaussian distribution error function.
If we combine different wave functions with different phases, we see
that, as long as we are not interested in the precise form of the packet
concentrated in the extremes of the interval, we have the freedom
of a quantum mechanical wave function on a two dimensional Hilbert
space seated at the two endpoints.

\subsubsection{Free massive fermion in $d=2$ and the Rindler wedge}

We consider the right Rindler region $\mathcal{C}_{1}:=\left\{ x>0\right\} $.
In this case, the theory does not decouples in chiralities and hence
the eigenfunction are two dimensional spinor functions. To compute
them, we follow the same strategy of \cite{Arias17}\footnote{We must first solve the Euclidean Dirac equation $\left(\gamma_{\mu}\partial_{\mu}+m\right)S_{s}\left(x,y\right)=0$
for $\left(x,y\right)\in\mathbb{R}^{2}$ with a multiplicative boundary
condition $S_{s}\left(x,0^{+}\right)=-\mathrm{e}^{2\pi s}S_{s}\left(x,0^{-}\right)$
for $x>0$ (the Rindler wedge region). Then, the eigenfunctions are
$u_{s}\left(x\right):=c_{s}S_{s}\left(x,0^{+}\right)$ with $c_{s}\in\mathbb{C}$
a normalization constant.} and we obtain
\begin{equation}
u_{s}\left(x\right):=\left(\begin{array}{c}
u_{s,+}\left(x\right)\\
u_{s,-}\left(x\right)
\end{array}\right)=\frac{1}{\pi}\sqrt{m\,\cosh\left(\pi s\right)}\left(\begin{array}{c}
K_{\frac{1}{2}-is}\left(mx\right)\\
-i\,K_{\frac{1}{2}+is}\left(mx\right)
\end{array}\right)\,,\label{ef_2d_m}
\end{equation}
where $K_{\nu}\left(z\right)$ is the modified Bessel function of
2nd kind. The probability of the a wave packet
\begin{equation}
\alpha\left(x\right)=\int_{-\infty}^{+\infty}ds\,\tilde{\alpha}\left(s\right)u_{s}\left(x\right)\,,
\end{equation}
to be localized around $x=0$ is given by the behavior of the eigenfunctions
in the limit $x\rightarrow0$. We remember the limiting formulas
\begin{eqnarray}
K_{\frac{1}{2}-is}\left(mx\right)\!\!\! & \simeq & \!\!\!\frac{1}{2}\Gamma\left(\frac{1}{2}-is\right)\left(\frac{mx}{2}\right)^{-\frac{1}{2}+is}\textrm{ ,}\\
K_{\frac{1}{2}+is}\left(mx\right)\!\!\! & \simeq & \!\!\!\frac{1}{2}\Gamma\left(\frac{1}{2}+is\right)\left(\frac{mx}{2}\right)^{-\frac{1}{2}-is}\textrm{ .}
\end{eqnarray}
Then, for $x\simeq0$, the eigenfunctions \eqref{ef_2d_m} become
\begin{eqnarray}
u_{s,+}\left(x\right)\!\!\! & \simeq & \!\!\!\frac{1}{\sqrt{2\pi x}}\mathrm{e}^{is\log\left(\frac{mx}{2}\right)}\textrm{ ,}\\
u_{s,-}\left(x\right)\!\!\! & \simeq & \!\!\!\frac{\left(-i\right)}{\sqrt{2\pi x}}\mathrm{e}^{is\log\left(\frac{mx}{2}\right)}.
\end{eqnarray}
Then, for a Gaussian wave packet \eqref{gausi} we get
\begin{eqnarray}
\alpha_{+}\left(x\right)\!\!\! & = & \!\!\!\int_{-\infty}^{+\infty}ds\,\tilde{\alpha}\left(s\right)u_{s,+}\left(x\right)\simeq\left(\frac{2}{\pi}\right)^{\frac{1}{4}}\sqrt{\frac{\sigma}{x}}\,\mathrm{e}^{-\sigma^{2}\left(\log\left(\frac{mx}{2}\right)-\lambda\right)^{2}}\,,\label{2d_wp_1}\\
\alpha_{-}\left(x\right)\!\!\! & = & \!\!\!\int_{-\infty}^{+\infty}ds\,\tilde{\alpha}\left(s\right)u_{s,-}\left(x\right)\simeq\left(-i\right)\left(\frac{2}{\pi}\right)^{\frac{1}{4}}\sqrt{\frac{\sigma}{x}}\,\mathrm{e}^{-\sigma^{2}\left(\log\left(\frac{mx}{2}\right)+\lambda\right)^{2}}\,.\label{2d_wp_2}
\end{eqnarray}
In other words, the probability density $\left|\alpha\left(x\right)\right|^{2}=\left|\alpha_{+}\left(x\right)\right|^{2}+\left|\alpha_{-}\left(x\right)\right|^{2}$
behaves, for wave packets localized near $x\simeq0$, as the sum of
two gaussian distributions in the variable $z:=\log\left(\frac{mx}{2}\right)$.
The analysis of the localization of such a wave packet can be followed
from the massless case (subsection \ref{subsec:1+1-chiral-fermion}).
We have to emphasize that expressions \eqref{2d_wp_1} and \eqref{2d_wp_2}
are only valid for $x\simeq0$. Then, only for $x_{*}\apprge0$, the
probability$\int_{0}^{x_{*}}\left(\left|\alpha_{+}\left(x\right)\right|^{2}+\left|\alpha_{-}\left(x\right)\right|^{2}\right)dx$
can be calculated using expressions \eqref{2d_wp_1} and \eqref{2d_wp_2}. 
\begin{itemize}
\item For $\lambda\ll0$, we have that $x_{*}=\frac{2}{m}\mathrm{e}^{\lambda}\apprge0$,
and hence
\begin{eqnarray}
p_{+}\left(0,x_{*}\right)\!\!\! & = & \!\!\!\int_{0}^{x_{*}}dx\left|\alpha_{+}\left(x\right)\right|^{2}=\frac{1}{2}\,,\\
p_{-}\left(0,x_{*}\right)\!\!\! & = & \!\!\!\int_{0}^{x_{*}}dx\left|\alpha_{-}\left(x\right)\right|^{2}\simeq0\,.
\end{eqnarray}
\item Conversely, for $\lambda\gg0$, we have that $x_{*}=\frac{2}{m}\mathrm{e}^{-\lambda}\apprge0$,
and then
\begin{eqnarray}
p_{+}\!\!\! & = & \!\!\!\int_{0}^{x_{*}}dx\left|\alpha_{+}\left(x\right)\right|^{2}\simeq0\,,\\
p_{-}\!\!\! & = & \!\!\!\int_{0}^{x_{*}}dx\left|\alpha_{-}\left(x\right)\right|^{2}=\frac{1}{2}\,.
\end{eqnarray}
\end{itemize}
In other words, the free parameter $\lambda$ determines which of
the chiralities of the spinor wave function $\alpha\left(x\right)$
is localized around $x\simeq0$. Since the full wave packet is normalized
to $1$, the other $\frac{1}{2}$ of the probability is  distributed
between both chiralities in order to have 
\begin{equation}
p\left(x_{*},+\infty\right)=\int_{x_{*}}^{+\infty}dx\left(\left|\alpha_{+}\left(x\right)\right|^{2}+\left|\alpha_{-}\left(x\right)\right|^{2}\right)\simeq\frac{1}{2}\,.
\end{equation}
Furthermore, the localization becomes sharper if we take into account
the dependance of the probability with the dispersion $\sigma$. We
recall that the above approximate wave packets \eqref{2d_wp_1} and
\eqref{2d_wp_2} are only valid for a small dispersion $\sigma\rightarrow0^{+}$
since they were calculated from the approximated eigenfunctions for
$s\simeq0$. 
\begin{itemize}
\item For $\sigma\simeq0$ and $\lambda\ll-\frac{1}{\sigma}$, we have
that $x_{*}=\frac{2}{m}\mathrm{e}^{\lambda+\frac{1}{\sigma}}\apprge0$
and hence
\begin{equation}
p_{+}\left(0,x_{*}\right)=\int_{0}^{x_{*}}\left|\alpha_{+}\left(x\right)\right|^{2}\apprge0.977\,,
\end{equation}
\item For $\sigma\simeq0$ and $\lambda\gg\frac{1}{\sigma}$, we now have
that $x_{*}=\frac{2}{m}\mathrm{e}^{-\lambda+\frac{1}{\sigma}}\apprge0$,
and then
\begin{equation}
p_{-}\left(0,x_{*}\right)=\int_{0}^{x_{*}}\left|\alpha_{-}\left(x\right)\right|^{2}\apprge0.977\,.
\end{equation}
\end{itemize}
In other words, in the small limit $\sigma\rightarrow0^{+}$ and
for large positive (resp. negative) mean value $\lambda$, the positive
(resp. negative) chirality of the wave packet $\alpha\left(x\right)$
is localized around the origin with probability very close to $1$,
while the negative (resp. positive) chirality has probability very
close to $0$ in the full region $x>0$.

\subsubsection{Free chiral fermion and $n$ intervals}

The multi-interval region is denoted as $\mathcal{C}:=\bigcup_{j=1}^{n}\left(a_{j},b_{j}\right)$,
where $a_{j}<b_{j}<a_{j+1}$. According to our results of chapter
\ref{CURRENT}, the eigenfunctions of the modular Hamiltonian are
\begin{equation}
u_{s}\left(x\right):=\frac{\left(-1\right)^{l+1}}{\sqrt{2\pi}}\frac{P\left(x\right)}{\sqrt{-\prod_{j=1}^{n}\left(x-a_{j}\right)\left(x-b_{j}\right)}}\mathrm{e}^{isz\left(x\right)}\,,\quad x\in\left(a_{l},b_{l}\right)\,,
\end{equation}
where 
\begin{equation}
z\left(x\right):=\log\left(-\frac{\prod_{j=1}^{n}\left(x-a_{j}\right)}{\prod_{j=1}^{n}\left(x-b_{j}\right)}\right)\,,
\end{equation}
and $P\left(x\right)$ is a polynomial of degree $n-1$. Moreover, we
can choose $n$ linearly independent polynomials $P_{k}$ in order
to form an orthonormal basis. As happened for the one interval case,
for a wave packet in modular coordinates $\tilde{\alpha}\left(s,k\right)$
sharply localized around $s=0$, the above eigenfunctions make the
wave packet in position space to be highly localized around the intervals
endpoints $a_{j},b_{j}$. To be more precise, let us take a single normalized
eigenfunction. Then, in analogy with the single interval case, we
know that the wave function 
\begin{equation}
\varphi\left(x\right):=\sqrt{z'\left(x\right)}\int_{-\infty}^{+\infty}ds\,\tilde{\alpha}\left(s\right)\mathrm{e}^{isz\left(x\right)}
\end{equation}
is localized in such a way that 
\begin{equation}
\left|\varphi\left(x\right)\right|^{2}=\sum_{l=1}^{n}\left[q\,\delta\left(x-a_{l}\right)+\left(1-q\right)\delta\left(x-b_{l}\right)\right]\,,\label{deltas_n}
\end{equation}
in the limit when the support of $\tilde{\alpha}\left(s\right)$ shrinks
to $s=0$.\footnote{In expression \eqref{deltas_n} we assume that $\int_{-\infty}^{+\infty}ds\,\left|\tilde{\alpha}\left(s,k\right)\right|^{2}=1$. }
The probability $q\in\left[0,1\right]$ above can be freely chosen.
Then, the wave packet becomes 
\begin{eqnarray}
\alpha\left(x\right)\!\!\! & = & \!\!\!\int_{-\infty}^{+\infty}ds\,\tilde{\alpha}\left(s\right)u_{s}\left(x\right)=\frac{r\left(x\right)}{\sqrt{z'\left(x\right)}}\varphi\left(x\right)\nonumber \\
 & \simeq & \!\!\!\sum_{l=1}^{n}\left[\sqrt{q}\frac{r\left(a_{l}\right)}{\sqrt{z'\left(a_{l}\right)}}\,\eta_{a_{l}}\left(x\right)+\sqrt{1-q}\frac{r\left(b_{l}\right)}{\sqrt{z'\left(b_{l}\right)}}\eta_{b_{l}}\left(x\right)\right]\,,\label{wp_n_1e}
\end{eqnarray}
where we have written 
\begin{equation}
r\left(x\right):=\frac{\left(-1\right)^{l+1}}{\sqrt{2\pi}}\frac{P\left(x\right)}{\sqrt{-\prod_{j=1}^{n}\left(x-a_{j}\right)\left(x-b_{j}\right)}}\,,
\end{equation}
and $\eta_{c}\left(x\right)$ is a wave packet sharply concentrated
around $x=c$ and normalized according to $\int\left|\eta_{c}\left(x\right)\right|^{2}dx=1$.

For example, if we choose conveniently the normalized eigenfunction
given by 
\begin{equation}
P_{k}^{\left(a\right)}\left(x\right):=\sqrt{\frac{\prod_{j=1}^{n}\left(b_{j}-a_{k}\right)}{\prod_{j\neq k}\left(a_{j}-a_{k}\right)}}\frac{\prod_{j\neq k}\left(x-a_{j}\right)}{\sqrt{-\prod_{j=1}^{n}\left(x-a_{j}\right)\left(x-b_{j}\right)}}\,,\label{eig_n_a}
\end{equation}
we have that $\frac{r_{k}\left(a_{l}\right)}{\sqrt{z'\left(a_{l}\right)}}=\delta_{kl}$.
Choosing $q_{k}=1$, eq. \eqref{wp_n_1e} simplifies to 
\begin{equation}
\alpha\left(x\right)\simeq\eta_{a_{k}}\left(x\right)\,.
\end{equation}

Similarly, we can choose the normalized eigenfunction given by\footnote{We remark that the eigenfunctions \eqref{eig_n_a} and \eqref{eig_n_b}
are not orthogonal.}
\begin{equation}
P_{k}^{\left(b\right)}\left(x\right):=\sqrt{\frac{\prod_{j}\left(b_{k}-a_{j}\right)}{\prod_{j\neq k}\left(b_{k}-b_{j}\right)}}\frac{\prod_{j\neq k}\left(x-b_{j}\right)}{\sqrt{-\prod_{j}\left(x-a_{j}\right)\left(x-b_{j}\right)}}\,.\label{eig_n_b}
\end{equation}
In this case, we have that $\frac{r_{k}\left(b_{l}\right)}{\sqrt{z'\left(b_{l}\right)}}=\delta_{kl}$, and choosing $q_{k}=0$, \eqref{wp_n_1e} simplifies to
\begin{equation}
\alpha\left(x\right)\simeq\eta_{b_{k}}\left(x\right)\,.
\end{equation}

In other words, in the limit of small modular parameter $s\simeq0$,
there is always a wave packet for the fermion intertwiner localized
around any chosen endpoint. A general intertwiner can be constructed
as a superposition of such endpoints localized wave packets 
\begin{equation}
\alpha\left(x\right)\simeq\sum_{l=1}^{n}\left(\mathrm{e}^{i\phi_{a_{l}}}\,\sqrt{p_{a_{l}}}\,\eta_{a_{l}}\left(x\right)+\mathrm{e}^{i\phi_{b_{l}}}\sqrt{p_{b_{l}}}\eta_{b_{l}}\left(x\right)\right)\,.
\end{equation}
According to the normalization relation \eqref{norm_u} for the wave
packet and the localization properties of the functions $\eta_{a_{l}}\left(x\right)$
and $\eta_{b_{l}}\left(x\right)$, the probabilities $p_{a_{l}}$, $p_{b_{l}}$ and the phases $\phi_{a_{l}}$, $\phi_{b_{l}}$
can be freely chosen with the exception that they must satisfy 
\begin{equation}
\sum_{l=1}^{n}p_{a_{l}}+\sum_{l=1}^{n}p_{b_{l}}=1\,.
\end{equation}
Thus, we can picture these wave functions as a quantum mechanical
degree of freedom in a Hilbert space with one basis vector for each
endpoint.

\subsubsection{Free massive fermion in $d$ dimensions and the Rindler wedge}

We consider the right Rindler region $\mathcal{C}:=\left\{ \bar{x}\in\mathbb{R}^{d-1}\,:\,x^{1}>0\right\} $.
In this case, by dimensional reduction, the eigenfunctions must be
of the form
\begin{equation}
u_{s,k}\left(x^{1},\bar{x}_{\Vert}\right):=w_{s}\left(\bar{k},m,x^{1}\sqrt{\bar{k}^{2}+m^{2}}\right)\mathrm{e}^{i\bar{k}\cdot\bar{x}_{\Vert}}\,,
\end{equation}
where $\bar{x}_{\Vert}:=\left(x^{2},\cdots,x^{d-1}\right)$ are the
transverse directions to the wedge, and $\bar{k}:=\left(k^{2},\cdots,d^{d-1}\right)$
are the momentum variables in that directions (degeneracy parameters for the modular
eigenfunctions). Each of the components\footnote{For $d$ dimensions, the spinor space has dimension $2^{\left\lfloor \frac{d}{2}\right\rfloor }$.}
$w_{s,j}$ ($j=1,\ldots,n$) of the spinor function, $w_{s}$ satisfy
the Euclidean Klein-Gordon equation $\left(-\nabla^{2}+\sqrt{m^{2}+\bar{k}^{2}}\right)w_{s,j}=0$.
Hence, $u_{s,k}$ behaves for the parallel direction $x^{1}$ as the
massive $1+1$ eigenfunction with reduced mass $\sqrt{m^{2}+\bar{k}^{2}}$.
The relative phases between each spinor components must to be adjusted
in order to the full spinor solution $v_{s,k}$ satisfies the Euclidean
Dirac equation. For any $d$ dimensions, this is a cumbersome problem
involving higher dimensional gamma matrices. In order to pleasantly 
display the main features of the problem, we study the case $d=3$,
where the spinor space is the same as in $d=2$. Following the arguments
above, we obtain the eigenfunctions
\begin{equation}
u_{s,k}\left(x^{1},x^{2}\right)=\frac{1}{\pi}\frac{\sqrt{\cosh\left(\pi s\right)}}{\sqrt[4]{m^{2}+k^{2}}}\left(\begin{array}{c}
\sqrt{m^{2}+k^{2}}\,K_{\frac{1}{2}-is}\left(x^{1}\sqrt{m^{2}+k^{2}}\right)\\
\left(k-im\right)\,K_{\frac{1}{2}+is}\left(x^{1}\sqrt{m^{2}+k^{2}}\right)
\end{array}\right)\frac{\mathrm{e}^{ikx^{2}}}{\sqrt{2\pi}}\,.
\end{equation}

To maximize the intertwiner correlation, we must choose a modular
wave $\tilde{\alpha}\left(s,k\right)$ highly supported around $s\simeq0$.
This implies that the shape of $\tilde{\alpha}\left(s,k\right)$
in the $k$ variable can be freely chosen. To better understand this,
we can start with a wave packet of the form $\tilde{\alpha}\left(s,k\right):=\tilde{\alpha}_{1}\left(s\right)\tilde{\alpha}_{2}\left(k\right)$ and define
\begin{equation}
\beta\left(x^{1},k\right):=\int_{-\infty}^{+\infty}ds\,\tilde{\alpha}_{1}\left(s\right)\frac{1}{\pi}\frac{\sqrt{\cosh\left(\pi s\right)}}{\sqrt[4]{m^{2}+k^{2}}}\left(\begin{array}{c}
\sqrt{m^{2}+k^{2}}\,K_{\frac{1}{2}-is}\left(x^{1}\sqrt{m^{2}+k^{2}}\right)\\
\left(k-im\right)\,K_{\frac{1}{2}+is}\left(x^{1}\sqrt{m^{2}+k^{2}}\right)
\end{array}\right)\,,
\end{equation}
which implies that
\begin{equation}
\alpha\left(x^{1},x^{2}\right)=\int_{-\infty}^{+\infty}dk\,\tilde{\alpha}_{2}\left(k\right)\beta\left(x^{1},k\right)\,\frac{\mathrm{e}^{ikx^{2}}}{\sqrt{2\pi}}\,.
\end{equation}
Now, we do a similar analysis as in the previous section. For any given
fixed $k$, the function $\beta\left(x^{1},k\right)$ is sharply localized
around $x^{1}\simeq0$. Being more specific, given any $s_{max}\apprge0$
and $\epsilon\apprge0$, we can always choose a modular wave packet
$\tilde{\alpha}_{1}\left(s\right)$ with support (essentially) contained
in $\left[-s_{max},s_{max}\right]$ such that the function $\beta\left(x^{1},k=0\right)$
is localized with probability almost $1$ in the interval $\left[0,\frac{\epsilon}{m}\right]$.\footnote{For example, as we have done in the previous section, we can take a Gaussian wave packet $\tilde{\alpha}\left(s\right)$ with dispersion
$\sigma=s_{max}\gtrsim0$ and mean value $\left|\lambda\right|\gg\frac{1}{\sigma}$.
Then, $\epsilon=\mathrm{e}^{-\left|\lambda\right|+\frac{1}{\sigma}}$.} Then, the supports of $\beta\left(x^{1},k\right)$ for the other $k$ modes, will be (essentially) localized in the intervals $\left[0,\frac{\epsilon}{\sqrt{m^{2}+k^{2}}}\right]\subset\left[0,\frac{\epsilon}{m}\right]$,
or in other words, the whole function $\beta\left(x^{1},k\right)$
will be localized in the interval $x^{1}\in\left[0,\frac{\epsilon}{m}\right]$
for all $k$. We can still freely choose the shape of the wave packet
$\tilde{\alpha}_{2}\left(k\right)$ and get any shape to the wave packet
$\alpha\left(x^{1},x^{2}\right)$ in the transverse $x^{2}$-direction, without affecting the relation $\left\langle V_{1}V_{2}^{\dagger}\right\rangle \simeq1$ for the intertwiner. In other words, the wave packet $\alpha\left(x^{1},x^{2}\right)$ can be chosen such that
\begin{eqnarray}
 & \left\langle V_{1}V_{2}^{\dagger}\right\rangle \simeq1\Rightarrow\alpha\left(x^{1},x^{2}\right)\simeq0\textrm{ for }x^{1}>\frac{\epsilon}{m}\,,\\
 & \textrm{and}\nonumber \\
 & \gamma\left(x^{2}\right):=\int_{0}^{\infty}dx^{1}\,\left|\alpha\left(x^{1},x^{2}\right)\right|^{2}\textrm{ with any shape.}
\end{eqnarray}
The above discussion can be extended to any region $\mathcal{C}:=A\times\mathbb{R}^{d-2}$
with $A\subset\mathbb{R}$. In particular, when $A:=\bigcup_{l=1}^{n}\left(a_{l},b_{l}\right)$
is any multi-interval, the fermion intertwiner $\alpha\left(x^{1},\bar{x}_{\Vert}\right)$
wave function will behave as
\begin{equation}
\alpha\left(x^{1},\bar{x}_{\Vert}\right)\simeq\sum_{l=1}^{n}f_{a_{l}}\left(\bar{x}_{\Vert}\right)\,\eta_{a_{l}}\left(x^{1}\right)+f_{b_{l}}\left(\bar{x}_{\Vert}\right)\eta_{b_{l}}\left(x^{1}\right)\,,
\end{equation}
where $\eta_{c}\left(x\right)$ is a wave packet highly localized around
$x=c$ with $\int\left|\eta_{c}\left(x\right)\right|^{2}dx=1$, and
$f_{c}\left(\bar{x}_{\Vert}\right)$ are arbitrary functions in the
transverse directions satisfying
\begin{equation}
\sum_{l=1}^{n}\int_{\mathbb{R}^{d-2}}dx_{\Vert}\,\left(\left|f_{a_{l}}\left(\bar{x}_{\Vert}\right)\right|^{2}+\left|f_{b_{l}}\left(\bar{x}_{\Vert}\right)\right|^{2}\right)=1\,.
\end{equation}

\subsubsection{Double cones in CFTs}

Suppose we have a sphere in a CFT and we have an Abelian sector with
unitary charge creating operator inside a double cone $W_{1}$, which for simplicity we take $W_{1}:=D\left(\mathcal{C}_{R}\right)$.
The operator has to be chosen such as to have almost zero modular energy.
The double cone can be conformally mapped to a hyperbolic space,
with curvature scale $R$ and temperature $(2\pi R)^{-1}$ \cite{chm}. In this space, the
modular Hamiltonian is just $K=2\pi RH$, where $H$ is the ordinary
Hamiltonian in the hyperboloid. To produce a unitary operator $V$ with small modular
energy, the excitation has to carry low momentum. This requires it
to be spread on regions much bigger than the curvature radius. On
the other hand, it can be placed anywhere in the translational invariant
hyperbolic space. However, once mapped back to the sphere, it will
be highly concentrated along the boundary of the sphere in the Minkowski
spacetime, but can be spread in the angular coordinates.

\subsection{Free examples for finite groups\label{freexamples}}

In this section, we study a simple example of intertwiner lower bounds
for finite groups. Let us think we have independent free fermion fields
$\psi_{j}(x)$ ($j=1,\cdots,n$) and consider symmetries that interchange
the different fields. The field algebra $\mathfrak{F}$ acts on the
Hilbert space
\begin{equation}
\mathcal{H}:=\bigotimes_{j=1}^{n}\mathcal{H}_{j}\,,
\end{equation}
where all $\mathcal{H}_{j}$ are equal to the Fock Hilbert
space of a single fermionic operator. The vacuum vector is given by
the tensor product of the vacuum vector in each factor, i.e. $\left|0\right\rangle :=\bigotimes_{j=1}^{n}\left|0_{j}\right\rangle $.
To define the action of the fermion field $\psi_{j}(x)$ in $\mathcal{H}$
we have to invoke the $\mathbb{Z}_{2}$-grading operators $\Gamma_{j}$
which act as
\begin{equation}
\Gamma_{j}\bigotimes_{k=1}^{n}\left|v_{k}\right\rangle =\left(-1\right)^{\left|v_{j}\right|}\bigotimes_{k=1}^{n}\left|v_{k}\right\rangle \,,
\end{equation}
where $\left|v_{j}\right|$ is the fermionic number of the vector
$\left|v_{j}\right\rangle \in\mathcal{H}_{j}$. Then, the field operator
$\psi_{j}(x)$ is represented in $\mathcal{H}$ as
\begin{equation}
\mathcal{\psi}_{j}\left(x\right)\mapsto\Gamma_{1}\otimes\cdots\otimes\Gamma_{j-1}\otimes\mathcal{\psi}_{j}\left(x\right)\otimes\mathbf{1}\otimes\cdots\otimes\mathbf{1}\,,
\end{equation}
in order to have the right anticommutation relations, i.e. $\left\{ \mathcal{\psi}_{j}\left(x\right),\mathcal{\psi}_{k}\left(y\right)\right\} =0$
for all $j\neq k$ and all $x,y\in\mathbb{R}^{d}$. The local algebra
of a bounded region $\mathcal{O}\in\mathcal{K}$ is defined as
\begin{equation}
\mathfrak{F}(\mathcal{O}):=\left\{ \sum_{j=1}^{n}\left[\mathcal{\psi}_{j}\left(f_{j}\right)+\mathcal{\psi}_{j}^{\dagger}\left(g_{j}\right)\right]\,:\,\mathrm{supp}(f_{j}),\,\mathrm{supp}(g_{j})\subset\mathcal{O}\right\} '',
\end{equation}
and for an unbounded region $B\in\mathcal{K}$, $\mathfrak{F}(B)$
is defined imposing additivity (condition \ref{c6-ass-add} in assumption
\ref{c6-ass-dhr}). As in the case of a single fermion field, this
net is ``complete'', i.e. it satisfies the axioms \ref{c1-conj-irr},
\ref{c1-conj-add-bis}, and \ref{c1-conj-inter}-\ref{c1-conj-cov} of
definition \ref{c1_conj}.

Here we consider two different symmetry groups: $S_{n}$ the permutation
group of $n$ elements, and $\mathbb{Z}_{n}$ the cyclic permutation
group of $n$ elements. These groups acts naturally on $\mathfrak{F}$
by permuting the field operators $\psi_{j}(x)$ and giving place to
two different observable algebras.\footnote{The observable algebra defined in this way contains operators which
anticommute at spacelike distance. This issue is unimportant for the
outcome of this section.}

As in the previous section, we consider two nearly complementary regions
$W_{1}$ and $W_{2}$. We can build charge generating operators using
the same type of operator $V_{j}$ of equation \eqref{siste} used in the
preceding section. To simplify calculations, and since we are not
interested in the fermion character of the fields but on the permutation
symmetries between different fields, we are going to use bosonic operators
$B_{j}$ for each field. These can be constructed out of the product
of two of the $V_{j}$ operators corresponding to non-overlapping
test functions in the same region $W_{1}$, 
\begin{equation}
B_{j}:=iV_{j,x}V_{j,y}\,,
\end{equation}
such that $B_{j}^{2}=\mathbf{1}$ and $B_{j}^{\dagger}=B_{j}$. We
also want these operators to have very small vacuum expectation value
$\langle B_{j}\rangle\simeq0$, what can be done by taking modes with
small correlation.\footnote{We can take $V_{j,x}$, $V_{j,y}$ as in \eqref{siste} with test
functions $f_{j,x}$, $f_{j,y}$ having support inside small spheres
of radius $R_{j,x},R_{j,y}$ centered at spacelike separated points
$x,y\in W_{1}$. Then, it is enough to take the $\mathrm{dist}\left(x,y\right)\gg R_{j,x},R_{j,y}$.} These operators commute for different fields. As discussed in the
previous section, we can choose $B_{j}^{1}\in\mathfrak{F}(W_{1})$
and $B_{j}^{2}\in\mathfrak{F}(W_{2})$ such that $\langle B_{j}^{1}B_{j}^{2}\rangle\simeq1$.
It is also clear that the expectation values $\langle B_{j}^{1}B_{k}^{2}\rangle=0$
for any $j\ne k$.

To start, let us take $n=3$ and the group $G:=\mathbb{Z}_{3}$ of
cyclic permutations of the fields ($|G|=3$). Let us take the finite
dimensional subalgebra $\mathcal{\mathcal{B}}_{W_{1}}$ generated
by the unitaries $B_{1},B_{2},B_{3}$ of the three fields inside $W_{1}$.
These unitaries are chosen in such a way the group elements transforms
one into the others. We have that $\dim(\mathcal{\mathcal{B}}_{W_{1}})=2^{3}=8$.
This algebra is generated by the projectors 
\begin{equation}
P_{j,\pm}:=\frac{\mathbf{1}_{\mathcal{H}}\pm B_{j}}{2}\,,\hspace{0.5cm}P_{j,\pm}{}^{2}=P_{j,\pm}=P_{j,\pm}^{\dagger}\,,\hspace{0.5cm}P_{j,+}\,P_{j,-}=0\,,
\end{equation}
or alternatively, it is also generated by the following set of eight
orthogonal projectors
\begin{equation}
P_{\pm\pm\pm}:=P_{1,\pm}P_{2,\pm}P_{3,\pm}\,.
\end{equation}
We call simply $P_{\beta}$ to these projectors, where $\beta=1,\ldots,8$.
There is an analogous algebra $\mathcal{\mathcal{B}}_{W_{2}}$ in
region $W_{2}$, and define $\mathcal{\mathcal{B}}_{12}:=\mathcal{\mathcal{B}}_{W_{1}}\vee\mathcal{\mathcal{B}}_{W_{2}}$.
The vacuum state just gives non-zero expectation value to the same
projector in $W_{1}$ and $W_{2}$, and we have 
\begin{equation}
\omega(P_{\beta}^{1}P_{\beta'}^{2})=\frac{1}{8}\delta_{\beta,\beta'}.
\end{equation}
This state has vN entropy $S_{\mathcal{\mathcal{B}}_{12}}(\omega)=\log(8)$.

Under the action of the group $G=\mathbb{Z}_{3}$, the above eight
projectors in $W_{1}$ are interchanged in the following form. There
are two regular representations (of three elements) spanned by the
projectors with $\beta$ having two plus signs or with two minus signs,
and two trivial representations due to the projectors with\textcolor{black}{{}
all signs equal. Each representation matches with one corresponding
representation in $W_{2}$. }All the projectors $P_{\beta}^{1}P_{\beta'}^{2}$
of the same regular representation lead to the same expectation value
\begin{equation}
\omega\circ\varepsilon(P_{\beta}^{1}P_{\beta'}^{2})=\frac{1}{8}\times\frac{1}{3}\,,
\end{equation}
because the conditional expectation mixes $\beta,\beta'$ on all the
possible values of the representation. Then, each regular representation
contributes with $-9\times1/8\times1/3\log(1/8\cdot1/3)$ to $S_{\mathcal{\mathcal{B}}_{12}}(\omega\circ\varepsilon)$,
while the trivial representations contribute with the same amount
as for the entropy of $\omega$, i.e. $-1/8\log(1/8)$ each. Then,
we get the lower the bound\footnote{Here we use that the conditional expectation preserves the trace,
and hence $S_{\mathcal{\mathcal{B}}_{12}}(\omega\mid\omega\circ\varepsilon)=-\Delta S_{\mathcal{\mathcal{B}}_{12}}$.} 
\begin{equation}
\Delta I\ge S_{\mathcal{\mathcal{B}}_{12}}(\omega\circ\varepsilon)-S_{\mathcal{\mathcal{B}}_{12}}(\omega)=\frac{3}{4}\log(3)\,.
\end{equation}
This coincides with the general result \eqref{twisted}.\footnote{In section \ref{lower}, we got $S_{\mathcal{\mathcal{B}}_{12}}(\omega)=0$
because we choose a bigger non-commutative algebra containing the
projectors to the diagonal elements we are using here. Even, if the
two entropies change when we enlarge the algebra in this way, the RE given by the difference $S_{\mathcal{\mathcal{B}}_{12}}(\omega\circ\varepsilon)-S_{\mathcal{\mathcal{B}}_{12}}(\omega)$
does not change.} In the factor $3/4$, we recognize the total probability of the regular
representations, which equals $6/8$.

To improve this bound we add a new site on each region, i.e. we take
two modes for each field and we call then $B_{j}^{\alpha}$, where
$\alpha=a,b$. Let us assume that the modes are decoupled, i.e. $\langle B_{j}^{a}B_{j}^{b}\rangle\simeq0$.
In other words, there is no entanglement between the two modes. We
expect that this automatic holds if the modes commute with each other,
e.g. if they are spatially separated, since by monogamy of entanglement,
they cannot have correlations between them if they are maximally entangled
with modes in the complementary region $W_{2}$. Now, the algebra
$\mathcal{\mathcal{B}}_{W_{1}}$ is spanned by the orthogonal projectors
\begin{equation}
P_{\beta}:=P_{\pm\pm,\pm\pm,\pm\pm}:=P_{1,\pm}^{a}P_{1,\pm}^{b}P_{2,\pm}^{a}P_{2,\pm}^{b}P_{3,\pm}^{a}P_{3,\pm}^{b}\,,\quad\beta=1,\ldots,2^{6}=64\,.
\end{equation}
Now, when we apply the group transformations we will have a larger
proportion of regular representations because there are four possibilities
$(\pm\pm)$ for each field, and the three fields have to have equal
this index in order not to form a regular representation. In general,
taking $N$ independent sites we get that the probability of the regular
representation is $\left(1-\frac{1}{2^{2N}}\right)$, and following
the same calculation as above we arrive at 
\begin{equation}
\Delta I\ge\left(1-\frac{1}{2^{2N}}\right)\log(3)\,.
\end{equation}
This mens that our lower bound can approach $\log|G|$ as much as we want.

Different groups can be treated similarly. Let us take the non-Abelian
group $G:=S_{3}$ which has $|G|=6$. Using $N$ sites, we again get
for each field $2^{N}$ labels for the projectors. Then, starting with one of the projectors, in order to the permutations
of the fields do not generate $3!=6$ different projectors (and hence
the regular representation), it must be that at least two of the labels
for the different fields are equal. Then, the probability of the regular
representation is the same as the probability of having the three
labels different. As it was shown in section \ref{lower}, the regular representation
always contribute a $\log|G|$. This gives
\begin{equation}
\Delta I\ge\frac{(2^{N}-1)(2^{N}-2)}{2^{2N}}\log6\,.
\end{equation}
We need more an more sites for better precision, but the approach
is exponentially fast.

It is evident that an example for the permutation group $S_{n}$ can
be constructed in the same way by using $n$ fields. Since each finite
group is a subgroup of a permutation group, and the regular representation
of the permutation group decomposes into regular representations of
the subgroups, an example can be devised in the same lines for any
finite group.

\subsection{Intertwiners at a finite distance}

In subsection \ref{c6-subsec:Intertwiners-at-short} we have seen
that the intertwiners are concentrated around the boundary for complementary
regions. Here we want to show that they will spread out in the coordinates
orthogonal to the boundary if the two regions are separated. We cannot
use now the vacuum modular conjugation to obtain a good charge creating
operator partner. Then, we simply minimize the expectation value to
obtain the optimal intertwiner. 

Here we still deal with the simple case of the $\mathbb{Z}_{2}$-symmetry
of the free fermion. Then, we consider spacetime regions $W_{j}:=D\left(\mathcal{C}_{j}\right)$,
where $\mathcal{C}_{1},\mathcal{C}_{2}\subset\Sigma_{0}$ are spacelike
regions over the Cauchy surface $\Sigma_{0}:=\left\{ x^{0}=0\right\} \subset\mathbb{R}^{d}$.
Moreover, $\overline{\mathcal{C}}_{1}\subset\mathcal{C}'_{2}$ which
means $W_{1}\text{\Large\ensuremath{\times\negmedspace\!\!\times}}W_{2}$.
We construct an intertwiner $\mathcal{I}_{12}:=V_{1}V_{2}^{\dagger}$
between $W_{1}$ and $W_{2}$ as in \eqref{int_5} and \eqref{eq: ferm_int}.
The vacuum expectation of such a intertwiner is
\begin{eqnarray}
\langle\mathcal{I}\rangle\!\!\! & = & \!\!\!\int_{\mathcal{C}_{1}\times\mathcal{C}_{2}}\negthickspace\negthickspace d^{d-1}x\,d^{d-1}y\,\left[\alpha_{1}(\bar{x})^{\dagger}C(\bar{x}-\bar{y})\alpha_{2}(\bar{y})-\alpha_{2}(\bar{x})^{\dagger}C(\bar{x}-\bar{y})\alpha_{1}(\bar{y})\right]\nonumber \\
 & = & \!\!\!\int_{\mathcal{C}_{1}\times\mathcal{C}_{2}}\negthickspace\negthickspace d^{d-1}x\,d^{d-1}y\,\alpha_{1}(\bar{x})^{T}\left[C(\bar{x}-\bar{y})-C(\bar{x}-\bar{y})^{*}\right]\alpha_{2}(\bar{y})\,,\label{var_ifd}
\end{eqnarray}
where $C(\bar{x}-\bar{y}):=\langle\psi(\bar{x})\psi^{\dagger}(\bar{y})\rangle$
is the fermion correlator, and we have used that the support of the
two functions are disjoint. In the last equality in \eqref{var_ifd},
we have assume that the smearing functions $\alpha_{j}\left(\bar{x}\right)$
are real. When $\mathcal{C}_{1}$ and $\mathcal{C}_{2}$ are two spheres
of size $R$ separated by a distance $L\gg R$, we have that this
expectation value falls as $(R/L)^{d-1}$ in the massless case and
exponentially in the massive one.

Taking variations of \eqref{var_ifd} with respect to $\alpha_{1}$
and $\alpha_{2}$ with the constraints $\int_{\mathcal{C}_{j}}\alpha_{j}(\bar{x})^{T}\alpha_{j}(\bar{x})=1$,
we get 
\begin{equation}
\int_{\mathcal{C}_{1}}d^{d-1}x\,\alpha_{1}^{T}(\bar{x})\left[C(\bar{x}-\bar{y})-C(\bar{x}-\bar{y})^{*}\right]=\lambda_{2}\,\alpha_{2}(\bar{y})\,,
\end{equation}
where $\lambda_{2}$ is a constant, the Lagrange multiplier. We have
an analogous equation for $\alpha_{1}$. The solutions of these integral
equations are generally not easy to obtain, but we can think for example
in the easy case of very separated regions. In that case, the correlator
function is almost constant when $\bar{x},\bar{y}$ belong to each
of the regions. Then, it follows that the optimal distribution is given
by constant functions $\alpha_{1},\alpha_{2}$. Hence, the charged
modes have spread as much as possible. One can easily compute the
contribution to the entropy of this intertwiner and check that it
is less than the MI for the fermion at large distances, while it has
the same falling rate $L^{d-1}$ (in the massless case), where $L$ is the distance
between the regions.

As an example, for a chiral fermion, we can compute the RE for one
intertwiner mode. Here we consider two intervals $A_{j}:=\left(a_{j},b_{j}\right)$
of length $R_{j}:=b_{j}-a_{j}$, separated by a distance $L:=a_{2}-b_{1}$.
In this case, we have a two dimensional Abelian algebra $\mathcal{B}_{12}\cong\mathbb{C}\oplus\mathbb{C}$
generated by the elements $\left\{ \mathbf{1},U\right\} $, and the unitary
intertwiner $U$ is defined as\footnote{We introduce a $i$ pre-factor in the definition \eqref{int_uni}
in order to $U$ be a unitary operator.} 
\begin{equation}
U_{12}:=i\mathcal{I}_{12}=iV_{1}V_{2}^{\dagger}=i\int_{A_{1}\times A_{2}} \!\!\!\!\! dx\,dy\,\alpha_{1}\left(x\right)\alpha_{2}\left(y\right)\left[\psi\left(x\right)+\psi^{\dagger}\left(x\right)\right]\left[\psi\left(y\right)+\psi^{\dagger}\left(y\right)\right].\label{int_uni}
\end{equation}
For simplicity, we have used real functions $\alpha_{j}(x)$ with support
in $A_{j}$. The vacuum expectation value of such operator is 
\begin{eqnarray}
\langle U_{12}\rangle_{\omega}\!\!\! & = & \!\!\!i\int dx\,dy\,\alpha_{1}\left(x\right)\alpha_{2}\left(x\right)\left[C\left(x-y\right)-C\left(y-x\right)\right]\nonumber \\
 & \simeq & \!\!\!-\frac{1}{\pi L}\left[\int_{A_{1}}dx\,\alpha_{1}\left(x\right)\right]\left[\int_{A_{2}}dy\,\alpha_{2}\left(y\right)\right]\,.
\end{eqnarray}
In the above expression we have used $C\left(x-y\right)=\left\langle \psi\left(x\right)\psi^{\dagger}\left(y\right)\right\rangle =\delta\left(x-y\right)+\frac{i}{2\pi}\frac{1}{x-y}\simeq\frac{i}{2\pi L}$
when $L\gg R_{1},R_{2}$. For constants functions $\alpha_{j}(x)$
normalized according to $\int_{A_{j}}dx\,\alpha_{j}\left(x\right)^{2}=1$,
we have $\int_{A_{j}}dx\,\alpha_{j}\left(x\right)=\sqrt{R_{j}}$,
and hence 
\begin{equation}
\langle U_{12}\rangle_{\omega}=-\frac{\sqrt{R_{1}R_{2}}}{\pi L} \,.
\end{equation}
On the other hand, the expectation value on the state $\omega\circ\varepsilon_{12}$
is $\langle U_{12}\rangle_{\omega\circ\varepsilon_{12}}=1$.

The classical probability distribution of any state $\psi$ on the
abelian algebra $\mathcal{B}_{12}$ is $\left(p_{1},p_{2}\right)$,
where $p_{1}-p_{2}=\left\langle U_{12}\right\rangle _{\psi}$ and
$p_{1}+p_{2}=\left\langle \mathbf{1}\right\rangle _{\psi}=1$. Then,
we have the following two probabilities distributions
\begin{eqnarray}
\omega\!\!\! & \rightarrow & \!\!\!\left(\frac{1}{2}-\frac{\sqrt{R_{1}R_{2}}}{2\pi L},\frac{1}{2}+\frac{\sqrt{R_{1}R_{2}}}{2\pi L}\right)\,,\\
\omega\circ\varepsilon_{12}\!\!\! & \rightarrow & \!\!\!\left(\frac{1}{2},\frac{1}{2}\right)\,.
\end{eqnarray}
Then, the RE restricted to the algebra $\mathcal{B}_{12}$ between
such states is 
\begin{equation}
S_{\mathcal{B}_{12}}\left(\omega\left|\omega\circ\varepsilon_{12}\right.\right)=\frac{R_{1}R_{2}}{2\pi^{2}L^{2}}\,,\label{srel_int_alg}
\end{equation}
which is strictly smaller than the mutual information in the field
algebra $I_{\mathfrak{F}}\left(A_{1},A_{2}\right)=\frac{R_{1}R_{2}}{6L^{2}}$
(see \eqref{ffermion}). The model ${\cal A}$ does not contain the
fermion and its mutual information at large distances falls with a
larger power than the one of the fermion. 

We can speculate on two reasons why \eqref{srel_int_alg} does not
coincide with the fermion MI at large distances. First,
the two dimensional algebra $\mathcal{B}_{12}$ may be too small, and
second, our election of the intertwiner is not good enough (for example,
we can still make different elections multiplying the intertwiner
$U_{12}$ by any unitaries in $\mathcal{A}\left(A_{1}\vee A_{2}\right)$).

As a commentary to the previous calculations in section \ref{freexamples},
if we choose the different charged modes on each region such that
they are not independent to each other, we clearly get a less optimal
result. In the limit when these modes are maximally entangled with
the complementary region, this just means we have to take non-overlapping
modes for the different sites. However, if the two regions $W_{1}$,
$W_{2}$ do not touch each other, we cannot produce maximally entangled
modes, and the charged modes will have a finite width in the direction
perpendicular to the boundary. In this case, even if we have several
sites on $W_{1}$ that are spatially separated, in general, the correlation
of these modes will not vanish. To improve the result we need to diminish
these correlations as much as possible since these correlations between
charged modes in $W_{1}$ are not intertwiner correlations. This means
the modes tend to repel each other in the direction parallel to the
boundary in order to maximize the bound.

\subsection{Sharp twists have Gaussian correlations with area law\label{twistcontinuo}}

In general it is difficult to obtain an exact explicit expression
for the twist operators having all the desired properties listed in
section \ref{c6-sec-int-twi}. For example, let consider a $U(1)$
symmetry with a conserved current $J_{\mu}$ and regions $W_{1}:=D\left(\mathcal{C}_{R}\right)$
and $W_{2}:=D\left(\mathcal{C}'_{R+\epsilon}\right)$ with $\epsilon>0$.
Then, we can construct twist operators $\tau_{k}$ for the region
$W_{1}$ acting trivially along $W_{2}$ as
\begin{eqnarray}
\tau_{k}\!\!\! & := & \!\!\!\mathrm{e}^{i\,k\,Q_{1}}\,,\quad k\in\mathbb{R}\,,\\
Q_{1}\!\!\! & := & \!\!\!\int_{S^{d-2}}\negthickspace d\Omega\int_{\mathbb{R}_{\geq0}}\negthickspace dr\,r^{d-2}\int_{\mathbb{R}}dt\,\alpha(t)\,\gamma(r)\,J_{0}(x)\,,\label{esp}
\end{eqnarray}
where $\gamma$ and $\alpha$ are any smooth smearing functions such that 
\begin{equation}
\gamma(x):=\begin{cases}
0\,, & r\geq R+\frac{\epsilon}{2}\,,\\
1\,, & r\leq R\,.
\end{cases}
\end{equation}
and $\alpha(t)=0$ for $|t|\geq\epsilon/2$, while $\int_{\mathbb{R}}dt\,\alpha(t)=1$.
These twists form a one-parameter group of unitaries ($\tau_{k_{1}}\tau_{k_{2}}=\tau_{k_{1}+k_{2}}$).
They act as conjugation on the operators of $\mathfrak{F}\left(W_{1}\right)$
in the same way as the global symmetry group, and act trivially on
the elements of $\mathfrak{F}\left(W_{2}\right)$. However, $\tau_{k}$
is not $2\pi$-periodic. To obtain this periodicity one should deform
the twist inside the shell $S:=(W_{1}\vee W_{2})$. This can be accomplished
using the split property (see \cite{Doplicher:1984zz}), but the result
would have a less transparent expression. For the $U(1)$ case, as
we have discussed in subsection \ref{U1}, it turns out that the expectation
values will fall fast with $|k|$ for small $\epsilon\apprge0$. Then,
in the the limit $\epsilon\rightarrow0^{\text{+}}$, we do not expect
difference in the leading divergent term in $\epsilon$ of the MI
if we use the twist operators of expression \eqref{esp} rather than
any twist operators fulfilling the group composition law of $U(1)$.

Because of the CPT symmetry, expectation values of odd powers of $Q_{1}$
vanish. For computing $\langle Q_{1}^{2}\rangle$ we use that, because
of conservation, the correlation function of the currents writes 
\begin{equation}
\langle J_{\mu}(x)J_{\nu}(0)\rangle=(g_{\mu\nu}\nabla^{2}-\partial_{\mu}\partial_{\nu})\,C(x)\,.
\end{equation}
For a general CFT, $C(x)$ is as in \eqref{c6-corr-cft}. Integrating
by parts we get 
\begin{equation}
\langle Q_{1}^{2}\rangle=\int d^{d}x\,d^{d}x'\,\alpha(t)\,\alpha(t')\,\beta(r)\,\beta(r')\,C(x-x')\,,
\end{equation}
where $\beta(r):=\gamma'(r)$ is a smooth function with support in
the shell. Keeping one point fixed and moving the other on the shell,
the result is seen to be proportional to the area times the remaining
integral. Because the result is dimensionless, it is universally
given, in a CFT, by 
\begin{equation}
\langle Q_{1}^{2}\rangle=c\,\frac{R^{d-2}}{\epsilon^{d-2}}\,,
\end{equation}
where the dimensionless constant $c$ depends on the precise shape
of the smearing functions. If there are mass scales in the theory
nothing changes for the leading term as far as $\epsilon\apprge0$
is in the UV regime.

To compute $\langle Q_{1}^{4}\rangle$ exactly we should know the
four-point functions of the current and these functions depend on
the specific details of the theory. However, if we want to compute
the leading term in $\epsilon\rightarrow0^{+}$ we can argue as follows.
Because of conservation and translation invariance, the four-point
function of the charge density $J_{0}$ can be written as a combination
of spatial derivatives of some functions $H$ of the coordinate differences.
The bulk integral can be then integrated out to get integrals on the
shell. One way to convince oneself of this, is that each of the four
operators $Q_{1}$ do not depend on the smearing inside the ball and
the flux of the current can be written in a different Cauchy surface,
as long as the shell part is not changed. Then, as above, in the
thin shell, the leading contribution comes from points of coincidence
of the correlator functions $H$. But the behavior of $H$ at coincidence
points can be read off from the points of coincidence of the correlators
of $J_{\mu}$, which satisfy clustering properties. Then, the leading
term comes from two pairs of coincidence points, and for each coincidence
points we have the same contribution as for the two-point function.
There are also three and four-point coincidences, but these give subleading
terms since we lose powers of the area. Since we have three possible
pairings between the four points, the leading term should read 
\begin{equation}
\langle Q_{1}^{4}\rangle\simeq3\,c^{2}\left(\frac{R^{d-2}}{\epsilon^{d-2}}\right)^{2}\,.
\end{equation}
With the same reasoning, we see that, for the purpose of computing
the leading term for small $\epsilon\apprge0$ in $\langle Q_{1}^{n}\rangle$,
we can assume that $Q_{1}$ is a free operator with Gaussian statistics
and Wick's theorem holds. The same conclusion arises from thinking that
the charge fluctuations are a sum over a large number of independent
fluctuations along the surface, and then using the central limit theorem.
We then arrive to a Gaussian distribution
\begin{equation}
\langle\tau_{k}\rangle=\langle\mathrm{e}^{ikQ_{1}}\rangle\simeq\mathrm{e}^{-k^{2}\frac{\langle Q_{1}^{2}\rangle}{2}}\,.
\end{equation}
For small enough $\epsilon$, only twist operators $\tau_{k}$ with
small $k$ have non-zero expectation values. As explained in section
\ref{U1}, this leads to 
\begin{equation}
\Delta I\simeq\frac{1}{2}\log\langle Q_{1}^{2}\rangle\simeq\frac{d-2}{2}\log(R/\epsilon)\,.
\end{equation}

More generally, we expect that in the $\epsilon\rightarrow0^{+}$
limit, twists for any finite group symmetry that affects the UV fix
point should also have an area law 
\begin{equation}
\langle\tau\rangle\simeq\mathrm{e}^{-c\frac{R^{d-2}}{\epsilon^{d-2}}}\,.\label{leat}
\end{equation}
We can argue that this has to be the case in the following way. For a sharp
twist operator, the charges that they measure are formed by the tensor
product of a large number $\sim R^{d-2}/\epsilon^{d-2}$ of independent
charge fluctuations (representations) along the surface. These form
a large representation of the group, and because of the arguments
in section \ref{lower}, this representation is mainly formed by copies
of the regular representation except for a fraction of the Hilbert
space that is exponentially small in the number of fused representations.
The expectation value of the twist, for any element of the group except
the identity, is zero for the regular representation. Then, we get
the leading behavior \eqref{leat}.

A well-known example is to take a QFT and replicate it $N$ times.
We can then take the subalgebra of pointwise fixed elements under
cyclic permutations of operators between copies. The twist operator
for this symmetry is the Rényi twist operator \cite{Calabrese:2009qy},
with expectation value for small $\epsilon$ 
\begin{equation}
\langle\tau_{n}\rangle=\textrm{tr}\rho_{1}^{n}=\mathrm{e}^{-(n-1)S_{n}(W_{1})}\simeq e^{-c\frac{R^{d-2}}{\epsilon^{d-2}}}\,,
\end{equation}
where $S_{n}$ is the Rényi EE of the region on the original model.
Thus, the expectation value of the twist is exponentially small with
an area law for the exponent. This coincides with the area law for
EE.\footnote{The results of our work, for this particular scenario, give $NI_{QFT}-I_{\textrm{Renyi orbifold}}=\log(N)$,
when $\epsilon\rightarrow0^{+}$. }

\section{Conclusions of the chapter\label{c6-conclu}}

In the context of QFT, the definition and computation of meaningful
information theoretic quantities can become extremely complicated.
The reason is simple. The most basic building block, the EE, is infinite and therefore ill-defined. To overcome this obstacle,
two natural avenues have been pursued in the past. The first and most
natural thing to do is to regularize the QFT with a lattice, which
makes EE finite. The problem is that we should only
trust aspects of such EE that do not depend on the
regularization scheme. Unfortunately, in several examples, it turns
out that to obtain the expected universal results one needs to fine-tune
the UV lattice definition, for example, by ad hoc choices of boundary
operators/algebras. The second and most rigorous avenue is to consider MI or related quantities, which can be considered
either directly in the continuum QFT or as limits of lattice quantities
\cite{chmy}. The advantage of this approach is that it is, in principle,
free from ambiguities, but the surprise is that in some cases it apparently
turns out not to provide the expected universal results. The questions
are thus clear: How do we extract the universal terms in the expansion
of the EE correctly and unambiguously? What are
the new physical features involved?

The main objective of this chapter has been to study these problems
for the case of theories with global symmetries. These symmetries
have the property that charged operators can be constructed locally.
In the context of algebraic QFT, these charged superselection sectors
are called DHR (because of Haag, Doplicher, and Roberts \cite{Doplicher:1969tk,Doplicher:1969kp,Doplicher:1973at}).

The solution to the problem stated above starts with the key observation
that theories with DHR sectors have certain ambiguities in the assignation
of algebras to regions. These ambiguities have been known for a long
time (see \cite{haag} for example), and we have described them in section \ref{algebra-regions}.
The main important message in this regard is that in theories with
DHR sectors it is not possible to assign algebras to regions in a
satisfactory way, where this means, a way satisfying the properties
of isotony, duality, additivity and intersection.\footnote{Duality for two intervals in CFT in $d=2$ is related to modular invariance.
Then, duality in higher dimensions and different regions can also be
thought as requirements generalizing the ones of modular invariance
for $d=2$ to other QFT and dimensions.} More concretely, for global symmetries, there is a clash between
duality and additivity for certain topologically non-trivial regions.
In particular, for two disconnected regions, such as the ones used
to define EE through MI, the additive
algebra of regions $W_{1}$ and $W_{2}$, defined as usual as $\mathcal{A}_{W_{1}}\vee\mathcal{A}_{W_{2}}$,
is not equal to the commutant algebra of the complementary region.
In fact, we have a violation of the duality property
\begin{equation}
\mathcal{A}_{W_{1}}\vee\mathcal{A}_{W_{2}}\subsetneq\mathcal{A}'_{(W_{1}\vee W_{2})'}\,.
\end{equation}
The reason for such a proper inclusion is that one can find neutral
operators $\mathcal{I}_{\sigma}$, which are called intertwiners for
group theoretic reasons, which do not belong to the additive algebra
$\mathcal{A}_{W_{1}}\vee\mathcal{A}_{W_{2}}$ but commute with the
algebra of the complementary region $\mathcal{A}{}_{(W_{1}\vee W_{2})'}$.
Basically, for localized charge creating operators $V^{j}_\sigma$, transforming
according certain representation $\sigma$ with dimension $d_{\sigma}$ ($j=1,\ldots,d_{\sigma}$)
of the symmetry group, one can form the neutral operator
\begin{equation}
\mathcal{I}_{\sigma}:=\sum_{j=1}^{d_{\sigma}}V_{W_{1}}^{j}V_{W_{1}}^{j\,\dagger}\,,\label{c6-conclu-int}
\end{equation}
where the subscript indicates the localization properties of the operator.
From this expression, it is transparent that $\mathcal{I}_{\sigma}\in\mathcal{A}'_{(W_{1}\vee W_{2})'}$,
but $I_{r}\notin\mathcal{A}_{W_{1}}\vee\mathcal{A}_{W_{2}}$.

In turn, this is due to the existence of twist operators $\tau_{[g]}$,
labeled by the conjugacy classes of the global group, which basically
implement the symmetry transformation just in one of the connected
components of $W_{1}\vee W_{2}$, but they belong to the neutral algebra
as well, even in the non-abelian case. These twists do not belong
to the additive algebra of the complementary region $(W_{1}\vee W_{2})'$,
but since it is a symmetry transformation, it commutes with all $\mathcal{A}_{W_{1}}\vee\mathcal{A}_{W_{2}}$,
which is composed of products of neutral operators. Most importantly,
as it has been described in this chapter, these twists operators do
not commute with the intertwiners \eqref{c6-conclu-int}.

We want to remark that these observations, the appearance of these
intertwiners and twists when considering topologically non-trivial
regions, do not depend on the regularization scheme. In particular,
it does not depend on algebra choices in a lattice regularization.
It is a true physical feature of the continuum QFT, a macroscopic
manifestation of the underlying global symmetry group. It is also
important to notice that these observations are purely made within
the vacuum sector of the theory, no charge creating operator is needed,
since both $\mathcal{I}_{\sigma}$ and $\tau_{[g]}$ are neutral operators
that indeed belong to the additive algebra of a sufficiently big spacetime
region.

The solution to the problem stated above is rooted in the implications
of the existence of such operators for the MI. From both,
a technical and physical perspective, the section \ref{entropyDHR}
has been devoted to analyzing the modifications to the MI
due to this enlarged operator algebras. The main tool that has been
used is the key formula \eqref{c2_ce_prop} described in section
\ref{c2_sec_ce}. When applied to QFT for the inclusion of algebras
${\cal A}\subset\mathfrak{F}$ and choosing the conditional expectation
and the states appropriately, \eqref{c2_ce_prop} gives the difference
of the MI on the two models between the regions $W_{1}$ and $W_{2}$
\begin{equation}
I_{\mathfrak{F}}(W_{1},W_{2})-I_{\mathcal{A}}(W_{1},W_{2})=S_{\mathfrak{F}}(\omega\mid\omega\circ\varepsilon_{12})\,,
\end{equation}
implying that such a RE difference can be computed solely
on the neutral algebra in the vacuum sector of the theory. Even more,
$I_{\mathfrak{F}}$ has a natural and direct definition in $\mathcal{A}$
(see \eqref{IO}).

The fact that the difference of MIs is itself a RE greatly simplifies the analysis of such an object since one
can resort to monotonicity and convexity to constraint it in several
ways. More concretely, we have found two dual ways to study the problem.
In the first approach, we compute a lower bound to such a RE by restricting to a certain finite algebra of intertwiners,
constructed basically from \eqref{esa1}. The challenge is to find
the best finite intertwiner subalgebra, i.e a finite subalgebra providing
the best lower bound to the RE. Interestingly, this
maximization procedure requires two concrete physical ingredients.
First, from a group theory point of view, we need to choose the intertwiner
subalgebra associated with the regular representation of the group.
Second, from the point of view of QFT, once such a regular representation
is chosen, we have to make sure that we maximize the correlation functions
in the vacuum state. This forces us to choose the intertwiners so
as to commute as much as possible with the vacuum modular Hamiltonian.
Explicit examples of this maximization of correlation functions and
of how the regular representation is inherently present in the vacuum,
have been described in section \ref{bounds}. The identification
of these two physical features, the regular representation and choosing
intertwiners that commute with the modular Hamiltonian, are two of
the most important physical messages of our analysis.

The second line of study uses the equality of entanglement entropies
for complementary algebras to relate the previous RE
to another relative entropy on the complementary algebra, which includes
the additive algebra and the twists operators $\tau_{[g]}$. From
this perspective, the problem is similar: we need to find the best
subalgebra that provides the best upper bound. The connection with
the intertwiner version is rooted in the fact that the group algebra
has the same dimension as the regular representation. While the intertwiners
are labeled by the irreducible representations of the group, the invariant twists can
be labeled by conjugacy classes, and both labels run over the same
number of elements.

Moreover, the twist/intertwiner duality is best described by both,
the entropic certainty and uncertainty relations, which were derived in section
\ref{certainty} and nicely codify the non-commuting character
of the twist/intertwiner algebras in an information theoretic manner.
These uncertainty relations are also among the most important
physical outcomes of our analysis.

Using these features, we have been able to compute the modifications
to universal contributions to the MI associated with
finite and continuous Lie groups. We have also computed the universal
contributions when considering different topologies, excited states,
scenarios with spontaneous symmetry breaking, and analyze the particularities
of two dimensional QFTs. All these results have been described
in section \ref{entropyDHR}. Some of these results were found previously
in the literature, and some of them are new. But we want to stress
that all of them arise from the same basic physical principles discussed
above. Hence, in this sense, the present approach provides a unification
of all these seemingly disconnected results.

We want to end with some important remarks. It is sometimes said that
the problems we have been considering in the present work arise in
theories with gauge symmetries, and are due to a certain arbitrariness
in the choice of algebras in lattice regularizations. Our first important
remark is that this is wrong. The problems only appear when the operator
algebra considered is incomplete and the theory has a structure of SS. To sense the difference, we could have a ``gauge''
theory with charges in all representations. This last theory has no
problems of assignations of algebras to regions in any meaningful
sense, where meaning is always related to properties of the continuum
QFT. Indeed, the converse is also true, we can have theories with
no gauge symmetry which actually show macroscopic ambiguities in the
definition of the MI. All the cases considered in
this paper are examples of such a scenario. The second important remark
is that whenever we have a structure of SS, their
contribution to the MI can be obtained only by focusing
on the vacuum sector. This is pretty impressive and it can
be related to the fact that the neutral algebra is an example of a
sufficient algebra, whenever the state considered is invariant under
the symmetry (see \cite{ohya-petz,petz_easy} for the definition of
a sufficient algebra).


\renewcommand\chaptername{Chapter}
\selectlanguage{english}

\chapter{Conclusions\label{CONCL}}

This thesis is advocated to the study of aspects of EE in QFT. We
followed an algebraic perspective. As we argued in the introduction
and chapters \ref{AQFT} and \ref{INFO}, the algebraic approach to
QFT fits naturally with this purpose. On one hand, the local degrees
of freedom of QFT organize themselves in algebras (subalgebras of
the global algebra more precisely), leading to a map
\begin{eqnarray}
\mathcal{O} & \mapsto & \mathfrak{A}\left(\mathcal{O}\right)
\end{eqnarray}
from the set of spacetime regions to the set of subalgebras of the
global algebra of QFT \cite{haag,horuzhy,Halvorson06}. On the other
hand, the natural way to define statistical/information measures is
in the context of vN algebras. In fact, any vN algebra represents
a statistical ``quantum'' system. These are the mathematical non-commutative
generalizations of classical probability spaces, in the sense of Kolmogorov.
In the later case, the measurable subsets (usually taken as the Borel
subsets) of the probability space form a distributive lattice, giving
place to usual classical rules between probabilities called \textit{classical
logic}. In the non-commutative case, the set of projectors of a vN
algebra (which by the way generate all the algebra itself) form an
orthomodular, but not distributive, lattice. This reflects the fact
that there are non-commuting projectors, or equivalently, that the
system has incompatible observables. Physical experiments involving
systems described under this approach display a richer structure of
outcomes called \textit{quantum logic} \cite{gleason,beltrametti,Mittelstaedt}.

This richer structure is still more significant when one considers
bipartite quantum systems. Some states on such systems show a particular
kind of correlations, between the subsystems, which are present only
due to  the quantum nature of the systems involved. This pure quantum
phenomenon, known as entanglement, has already brought out several
problems to philosophers of quantum theory, giving place to a collection
of anti-intuitive gedanken and real experiments. Moreover, when used
in our favor, it allows to construct skillful quantum devices (simulators
or computers) to solve problems more efficiently than devices that
are governed by laws of classical physics.

In QFT, this phenomenon is present on fundamental grounds. The Reeh-Schlieder
(RS) theorem asserts that the vacuum state, of any QFT, is entangled
at any distance. The key point that makes this theorem true is the
relativistic covariance of the theory and the spectrum condition.
However, the RS theorem does not hold for general non-relativistic
QFTs. This suggests a deep connection between geometry and entanglement.
This connection has been subsequently enhanced in the context of quantum
gravity due to the interpretation of the entropy of a black hole as
EE, and its generalization with the holographic prescription for EE
in AdS/CFT. From the pure QFT perspective, the RS theorem allows us
to define, for any region of the spacetime, a state-dependent dynamics
which leaves invariant the observables of the corresponding region.
This evolution is called modular flow and is represented by a one-parameter
group of unitaries whose Hermitian generator is called modular Hamiltonian.
The modular Hamiltonian appears in many conceptual aspects of the
foundations of AQFT. In particular, from our perspective, we remark
the following two. On one hand, it is used to define statistical/entanglement
measures for general quantum systems, e.g. the relative entropy (RE).
RE is a central concept in the study of entanglement in QFT, because,
in contrast to vN entropy, it is a well-defined quantity in the continuum
QFT and it also allows us to derive many other quantum information
measures, such as the mutual information (MI). On the other hand,
it is deeply related to spacetime symmetries according to Bisognano-Wichmann
(BW) theorem. In fact, this relation is so strong that, under general
assumptions, we have that the upshot of BW theorem is equivalent to
Poincaré covariance of the theory \cite{Davidson95}.

All these insights point out the relevance of the entanglement in
QFT, and the benefits of the algebraic approach to study its consequences.
Since AQFT is a perfect framework to study QFT rigorously, we proposed
to compute entanglement measures and modular Hamiltonians from this
perspective. The advantage of this approach is that it allows us to
compute quantities, from precise mathematical grounds, in an unambiguous
way, without the need to deal with cutoffs and cumbersome regularization
prescriptions. In this way, we can argue that the relation between
EE and AQFT is very strong. On one hand, EE entropy, or any entanglement
measure, needs the algebraic approach to be properly well-defined.
On the other hand, we may entertain the expectation that QFT may admit
a complete precise formulation in terms of entropic quantities. In
the Wightman approach to QFT, the knowledge of all vacuum expectation
values of product of field operators (Wightman functions) is enough
to reconstruct uniquely the theory. In the algebraic framework, the
local algebras are the central objects in the description of the theory,
and in this setting, the equivalent to the vacuum correlators are
the statistical/entanglement measures attached to such algebras. In
particular, the mutual information of the vacuum state universally
quantifies such correlations for commuting algebras attached to spacelike
separated regions. We can expect that the knowledge of all the mutual
informations for the vacuum state and any pair of regions, would be
enough to reconstruct uniquely the AQFT model. One can also hope that
it would help us in the search of a “dynamical principle” which allows
us to construct non-trivial models satisfying the axioms of AQFT discussed
along this thesis.

With this perspective in mind, in chapter \ref{RE_CS}, we computed
the relative entropy for coherent states using Araki formula. The
advantage of using this formula is that we could unambiguously compute
the RE in the continuum QFT, without the need of invoking non-rigorous
methods such as lattice or replica trick. We showed that using such
non-rigorous methods, ambiguities can arise in the final result of
such RE. These ambiguities come when one uses the formula
\begin{equation}
S(\phi\mid\omega)=\Delta\langle K\rangle-\Delta S\,,\label{conc-ds-dh}
\end{equation}
which holds for finite quantum systems, but not for the local algebras
of QFT. More precisely, both terms on the r.h.s. of \eqref{conc-ds-dh}
are mathematically ill-defined. The second one is ill-defined because
the EE entropy is UV-divergent in QFT, and despite the difference
of two EE could be finite, its value, in general, depends on the regularization
prescription. The first term is trickier. It involves the ``half''
modular Hamiltonian, which, despite not being a mathematical well-defined
operator, is a sesquilinear form, and hence, expectation values of
this operator are generally finite. However, to define this operator,
we need to cut the ``full'' modular Hamiltonian into two pieces,
and in this act, ambiguities may appear in the boundary of the region.
In the case of the free scalar field and the Rindler wedge, these
ambiguities come from possible improving terms of the stress-tensor.
Using the Araki formula, we solved these ambiguities in the case of
coherent states. Moreover, the general structure of modular theory
allowed us to prove general statements about the RE without the need
of specifying any particular model. More concretely, we showed that,
for free theories, the RE between coherent states is symmetric.

Following in the same direction, in chapter \ref{CURRENT}, we computed
the modular Hamiltonian and the MI for a free chiral scalar field.
A similar computation has been done in the past for the free chiral
fermion \cite{reduced_density}. We rederives such a result, in a more
transparent way, using a novel method that relates the eigenfunctions
of the correlator kernel with solutions of the wave equation in the
Euclidean plane \cite{Arias17}. In this case, since the theory is
massless, the wave equation is equivalent to finding holomorphic functions
in the complex plane with suitable boundary conditions, which is nothing
else than a Riemann-Hilbert problem. For the free chiral fermion,
the modular Hamiltonian is non-local whenever the regions contain
more than one interval. However, in that case, the non-locality has
a very particular structure: the non-local terms are quadratic in
the fermion operator pairing each point in any given interval, with
exactly one point in all others. The reason for this special structure
could be, perhaps, the multi-local symmetries described by Rehren
and Tedesco \cite{Rehren:2012wa}. Using the same techniques, we computed
the modular Hamiltonian and the MI for the free chiral scalar for
two intervals. However, in contrast to the fermion field, the non-local
part is given by a smooth kernel, which convolutes any point in any
of the two intervals with all the other points in both intervals.
In both cases, the local part is given by an integral of the stress
tensor smeared by a \textit{local temperature} function. This function
is universal for free fields \cite{Arias17}, and we expect that it
should be universal for any $d=2$ CFT. The local term gives the leading
contribution to the RE between the vacuum state and a highly-energetic
and well-localized excitation inside the region \cite{Arias17}. Remarkably,
the MI for the scalar is upper bounded by the one of the fermion field.
This should not be surprising since the free chiral scalar net is
a subnet of the free chiral fermion net. To our surprise, the scalar
shows a failure of the duality condition (assumption \ref{c1-conj-duality}
in definition \ref{c1_conj}) for the algebras assigned to two intervals. This
fact is translated into a loss of a symmetry property for the mutual
information usually associated with modular invariance \cite{Cardy:2017qhl}.
In the analysis of this curious phenomenon, we identified which operators
are the responsible for the failure of the duality, and how it can be restored
if we merge the two chiralities in a $d=2$ CFT in Minkowski spacetime.
On the other hand, the free fermion satisfies the duality conditions
and its MI satisfies the mentioned symmetry property.

This last observation motivated the study of our last chapter \ref{EE_SS}.
This led us to investigate the origin and scope, among general QFTs
in any dimensions, of the failure of the duality behind the failure
of the symmetry property in the EE. There we studied entanglement
aspects of theories having non-trivial superselection sectors (SS)
on the EE entropy. The structure of SS has direct consequences in
the considered model, which is accessible from the vacuum representation
itself. In fact, it leaves a definite imprint in the relations between
the different local subalgebras of operators assigned to the different
regions of the spacetime. The superselection structure affects the
relations between algebras and regions, either violating the property
of duality and/or additivity for some topologically non-trivial regions.
In this new scenario, where models with SS sectors are considered,
there is more than one choice for the algebra of topologically non-trivial
regions, being this directly sensitive to the EE. We constructed an
entropic order parameter given by the difference of two MIs between
two disconnected regions. These MIs can be reinterpreted as corresponding
to different models, with and without SS. We found that such a difference
is bounded above by the dimension of the global symmetry group and
saturates whenever the regions touch each other. Moreover, such an
entropic parameter leads to a certainty entropic relation involving
REs in two non-commutative algebras. The non-commutativity of such
algebras is due to the operators that break the duality condition
for non-connected regions, and are directly related to the SS of the
model. Furthermore, we argued that such a certainty relation has a
very general structure that should hold for any inclusion of algebras
and relates entropic quantities with the index theory of vN subfactors.
The subtle certainty relation is stronger than usual uncertainty relations
between non-commuting operators. There is no prior example of this
relation in the literature. We hope it should have wider applications
than the ones discussed in this thesis. Once the entropic order parameter
was recognized and fully characterized, we proceeded to apply it to
different situations: continuous Lie symmetry groups, regions with
other topologies, excited states, and scenarios with spontaneous symmetry
breaking. We successfully obtained the relevant universal contribution
in all of these scenarios, unifying, generalizing, and clarifying
some disperse specific results that have appeared in the literature,
as well as, obtaining some previously unknown ones.

We expect that the contributions we have made in this field will be
useful to better understand the structure of entanglement in QFT. In particular, with the motivation of finding a possible quantum
information version of QFT, we think that we have made great progress
in understanding the connection of EE with internal symmetries, which
are very important in particle physics. On the other hand, this connection
could be extended to more general scenarios. For example, in the case
of gauge symmetries, the mismatch between the A-anomaly
coefficient and the coefficient of the logarithmic term in the EE
of a Maxwell field could be solved if we compute the MI for the complete
model that includes the charged fields, instead of the “incomplete”
model of the pure gauge Maxwell field \cite{Casini:2019nmu}. In fact, the difference in
the mismatch should be computable using the techniques and the entropic
order parameter we have developed along chapter \ref{EE_SS}.\footnote{Though this investigation is well-advanced and shortly to be published,
it is not included in this thesis.} Another important application of the results of that chapter is in
the context of AdS/CFT. In \cite{Casini:2019kex}, we also argue that
the features of holographic EE correspond to a picture of a sub-theory
with a large number of superselection sectors.\footnote{This investigation is more speculative to fit in the design of this
thesis, and for this reason, has not been included in the present
volume.} In fact, in the holographic context, the formula \eqref{labe} could
be interpreted as the Ryu-Takayanagi formula plus quantum corrections.
In that case, when it is applied to connected regions on the boundary
theory, it gives the known result of “relative entropy equals bulk
relative entropy” \cite{Jafferis:2015del}. This interpretation of
the features of holographic EE in terms of sub-theory and conditional
expectations would open the way to interpret the bit-thread picture
of holographic entanglement in terms of more concrete and physical
objects, the intertwiners, in the boundary theory.


\appendix
\renewcommand\appendixname{Appendix}
\selectlanguage{english}

\chapter{Lattices\label{APP_LATTICE}}

\begin{defn}
Let $\left(L,\leq\right)$ be a partially ordered set and $S\subset L$
a subset. An element $u\in L$ is said to be an \textit{upper bound}
of $S$ if $s\leq u$ for all $s\in S$, and an element $l\in L$
is said to be a \textit{lower bound} of $S$ if $l\leq s$ for all
$s\in S$.
\end{defn}

\begin{rem}
A set $S\subset L$ may have many upper bounds (or lower bounds),
or none at all.
\end{rem}

\begin{defn}
Let $\left(L,\leq\right)$ be a partially ordered set, and $S\subset L$
an arbitrary subset. An upper bound $u$ of $S$ is said to be a \textit{supremum}
if $u\leq\tilde{u}$ for all upper bounds $\tilde{u}$ of $S$. Dually,
a lower bound $l$ of $S$ is said to be an \textit{infimum} if $\tilde{l}\leq l$
for all lower bounds $\tilde{l}$ of $S$.
\end{defn}

\begin{rem}
A set $S\subset L$ may have many supremums (or infimums), or none
at all.
\end{rem}

\begin{defn}
A \textit{lattice} is a partially ordered set $\left(L,\leq\right)$
such that any pair of elements $\{a,b\}\subset L$ has a unique supremum,
which is denoted by $a\vee b$, and a unique infimum, which is denoted by $a\wedge b$. 
\end{defn}

\begin{defn}
A lattice $L$ is called \textit{bounded} if there exists a \textit{greatest
element} $1\in L$ such that $l\leq1$ for all $l\in L$, and a \textit{least
element} $0\in L$ such that $0\leq l$ for all $l\in L$. 
\end{defn}

\begin{defn}
A bounded lattice $L$ is called \textit{complemented} if every element
$a\in L$ has a \textit{complement}, i.e. an element $b\in L$ such that
$a\wedge b=0$ and $a\vee b=1$. 
\end{defn}

\begin{rem}
An element $a\in L$ of a complemented lattice may have many complements.
A bounded lattice in which every element has exactly one complement
is called a\textit{ uniquely complemented} lattice.
\end{rem}

\begin{defn}
A \textit{complementation} on a complemented lattice $L$ is a function
that maps each element $a\in L$ to a complement $a'\in L$ in such
a way that the following axioms are satisfied,
\end{defn}

\begin{enumerate}
\item (complement laws) $a\wedge a'=0$ and $a\vee a'=1$,
\item (involution) $(a')'=a$,
\item (order-reversing) $a\leq b\Rightarrow b'\leq a'$,
\end{enumerate}
for all $a,b\in L$. A complemented lattice equipped with a complementation
is called an \textit{orthocomplemented lattice}.
\begin{example}
The prototypical example of an orthocomplemented lattice is the power
set of any given set. In this case, the order is given by inclusion,
the supremum (resp. infimum) of any two sets is the union (resp. the
intersection) of such sets, and the complementation is given by the
complement of sets.
\end{example}

\begin{lem}
Let $L$ be an orthocomplemented lattice. Then, we have that 
\begin{equation}
(a\vee b)'=a'\land b'\quad\textrm{and}\quad(a\land b)'=a'\vee b'\,,\label{c1-fn-dm-1}
\end{equation}
for all $a,b\in L$. The relations \eqref{c1-fn-dm-1} are known as
De Morgan laws.
\end{lem}

\begin{defn}
Let $(L,\leq)$ be a lattice and $S\subset L$ a subset. We say that
$S$ is a \textit{sublattice} if $a\lor b\in S$ and $a\wedge b\in S$
whenever $a,b\in S$. In the case of orthocomplemented lattices, we
also require that $a'\in S$ whenever $a\in S$. Automatically, $(S,\leq)$
is an (orthocomplemented) lattice.
\end{defn}

\begin{defn}
A lattice $L$ is said to be \textit{distributive} if it satisfies
any of the following two equivalent properties.
\end{defn}

\begin{itemize}
\item $a\lor(b\land c)=(a\lor b)\land(a\lor c)$, for all $a,b,c\in L$.
\item $a\land(b\lor c)=(a\land b)\lor(a\land c)$, for all $a,b,c\in L$.
\end{itemize}
Both relations above are usually called \textit{distributive laws}.
\begin{prop}
Any complemented and distributive lattice is uniquely complemented.
\end{prop}

\begin{defn}
A distributive and orthocomplemented lattice is called \textit{Boolean}
lattice.
\end{defn}

\begin{defn}
\label{fn_comm_sets-1}Let $L$ be an orthocomplemented lattice and
$a,b\in L$. We say that $a$ \textit{commutes} with $b$ if $a=(a\wedge b)\vee(a\wedge b')$.
It can be shown that $a$ commutes with $b$ if and only if $b$ commutes
with $a$. In other words, commutation is a symmetric relation.
\end{defn}

\begin{prop}
Let $L$ be an orthocomplemented lattice. Then, $L$ is distributive
(and hence Boolean) if and only if $a$ commutes with $b$ for all
$a,b\in L$.
\end{prop}

\begin{defn}
An orthocomplemented lattice $L$ is said to be \textit{orthomodular}
if 
\begin{equation}
a=(a\wedge b')\vee b,\quad\textrm{for all}\;b\leq a\,.
\end{equation}
\end{defn}

\begin{rem}
Any Boolean lattice is orthomodular, but the converse is false.
\end{rem}

\begin{defn}
A \textit{lattice homomorphism} is a map $\Phi:L_{1}\rightarrow L_{2}$
between lattices $L_{1},L_{2}$, such that $\Phi$
preserves all the lattice operations, i.e.
\begin{equation}
\Phi\left(a\vee b\right)=\Phi\left(a\right)\vee\Phi\left(b\right)\quad\textrm{and}\quad\Phi\left(a\land b\right)=\Phi\left(a\right)\land\Phi\left(b\right)\,,
\end{equation}
for all $a,b\in L_{1}$. In the case of orthocomplemented lattices,
we also require that $\Phi\left(a'\right)=\Phi\left(a\right)'$ for
all $a\in L_{1}$.
\end{defn}

\begin{defn}
A lattice homomorphism $\Phi:L_{1}\rightarrow L_{2}$ is a \textit{lattice
isomorphism} if $\Phi$ is invertible and $\Phi^{-1}:L_{2}\rightarrow L_{1}$
is also a lattice homomorphism. When $L_{1}=L_{2}=L$, any lattice
isomorphism $\Phi:L\rightarrow L$ is called a \textit{lattice automorphism}.
The set of all lattice automorphisms is denoted by $\mathrm{Aut}(L)$
and it forms a group under composition.
\end{defn}


\renewcommand\appendixname{Appendix}
\selectlanguage{english}

\chapter{Sobolev spaces\label{AP-sec-sobo}}

For the definition and properties of Sobolev spaces, we follow \cite{Evans}.
However, here we adapt the notation to our convenience.

Consider the test function space $\mathcal{D}\left(\mathbb{R}^{n}\right):=C_{c}^{\infty}\left(\mathbb{R}^{n}\right)\varsubsetneq\mathcal{S}\left(\mathbb{R}^{n}\right)$
of smooth and compactly supported functions with its usual topology.
The $n$-dimensional complex \textit{Sobolev space} of order $\alpha\in\mathbb{R}$
is defined as 
\begin{equation}
H^{\alpha}\left(\mathbb{R}^{n}\right):=\left\{ f\in\mathcal{D}'\left(\mathbb{R}^{n}\right)\,:\,\hat{f}\left(\bar{p}\right)\omega_{\bar{p}}^{\alpha}\in L^{2}\left(\mathbb{R}^{n}\right)\right\} \,,
\end{equation}
where $\omega_{\bar{p}}=\sqrt{\bar{p}^{2}+1}$ and $\hat{f}\left(\bar{p}\right):=\left(2\pi\right)^{-\frac{n}{2}}\int_{\mathbb{R}^{n}}f\left(\bar{x}\right)\mathrm{e}^{-i\bar{p}\cdot\bar{x}}d^{n}x$
is the usual Fourier transform. From the definition follows that $H^{0}\left(\mathbb{R}^{n}\right)=L^{2}\left(\mathbb{R}^{n}\right)$
and $H^{\alpha}\left(\mathbb{R}^{n}\right)\subset H^{\alpha'}\left(\mathbb{R}^{n}\right)$
if $\alpha>\alpha'$.

The Sobolev space $H^{\alpha}\left(\mathbb{R}^{n}\right)$ is a Hilbert
space under the inner product 
\begin{equation}
\left\langle f,g\right\rangle _{H^{\alpha}}:=\langle\hat{f}\omega_{\bar{p}}^{\alpha}\mid\hat{g}\omega_{\bar{p}}^{\alpha}\rangle_{L^{2}}=\int_{\mathbb{R}^{n}}d^{n}p\,\hat{f}\left(\bar{p}\right)^{*}\hat{g}\left(\bar{p}\right)\omega_{\bar{p}}^{2\alpha}\,.
\end{equation}
Furthermore, for $f\in H^{\alpha}\left(\mathbb{R}^{n}\right)$ we
have that $\left\Vert f\right\Vert _{H^{\alpha'}}\leq\left\Vert f\right\Vert _{H^{\alpha}}$
if $\alpha>\alpha'$, and hence the natural injections $H^{\alpha}\left(\mathbb{R}^{n}\right)\hookrightarrow H^{\alpha'}\left(\mathbb{R}^{n}\right)$
for $\alpha>\alpha'$ are continuous. We also have that the set $C^{\infty}\left(\mathbb{R}^{n}\right)\subset\mathcal{S}\left(\mathbb{R}^{n}\right)$
is dense in $H^{\alpha}\left(\mathbb{R}^{n}\right)$.

When $\alpha=k\in\mathbb{N}_{0}$, there is also another useful equivalent
characterization of the Sobolev spaces in term of weak derivatives\footnote{The weak derivative of an element of $\mathcal{D}'\left(\mathbb{R}^{n}\right)$
is its usual derivative in the distributional sense.}
\begin{equation}
H^{k}\left(\mathbb{R}^{n}\right)=\left\{ f\in\mathcal{D}'\left(\mathbb{R}^{n}\right)\,:\,D^{\mu}f\in L^{2}\left(\mathbb{R}^{n}\right)\,,\textrm{ for all }\left|\mu\right|\leq k\right\} \,.\label{eq: sobolev_2}
\end{equation}
It is useful to introduce a new norm in $H^{k}\left(\mathbb{R}^{n}\right)$ as 
\begin{equation}
\left\Vert f\right\Vert '_{H^{k}}:=\left(\sum_{\left|\mu\right|\leq k}\int_{\mathbb{R}^{n}}d^{n}x\left|D^{\mu}f\left(x\right)\right|^{2}\right)^{\frac{1}{2}}\,,
\end{equation}
which is equivalent to the former norm $\left\Vert \cdot\right\Vert {}_{H^{k}}$.

The real Sobolev spaces $H^{\alpha}\left(\mathbb{R}^{n},\mathbb{R}\right)$
are defined in a similar manner as above, but restricting to real
valued functions.\\

In general, it is easier to calculate the usual pointwise derivatives
rather than the weak derivatives. Then, the following lemma states
sufficient conditions for both notions of derivatives coincide. Before
we formulate it, we need to introduce the notions of \textit{$C^{k}$}-piecewise
function. 
\begin{defn}
Let $U\subset\mathbb{R}^{n}$ open, $f\in L_{loc}^{1}\left(U\right)$
and $k\in\mathbb{N}_{0}$. We say that $f$ is a \textit{$C^{k}$-piecewise}
function iff there exists a finite family of pairwise disjoint open
sets $\left\{ \Omega_{j}\right\} _{j=1,\ldots,J}\subset U$ such that
\end{defn}
\begin{enumerate}
\item $\bigcup_{j=1}^{J}\overline{\Omega}_{j}=\overline{U}$. 
\item $f\in C^{k}\left(\Omega_{j}\right)$ for all $j=1,\ldots,J$. 
\item For all $j=1,\ldots,J$, $\forall x_{0}\in\partial\Omega_{j}$ and
for all multi-index $\left|\alpha\right|\leq k$, the $\lim_{x\rightarrow x_{0}}\left.D^{\alpha}f\left(x\right)\right|_{\Omega_{j}}$
exist and are finite, where $D^{\alpha}$ is the usual multi-order
pointwise derivative. 
\end{enumerate}
We denote $C_{t}^{k}\left(U\right)$ the set of \textit{$C^{k}$}-piecewise
functions on $U$.\\

Now, we formulate the lemma that ensures that weak and pointwise derivatives coincide. 
\begin{lem}
\label{lem:derivatives}

\textup{Let $U\subset\mathbb{R}^{n}$ be open and $f\in C^{0}\left(U\right)\cap C_{t}^{1}\left(U\right)$.
Then, the (first order) weak derivatives of $f$ coincides with the usual pointwise derivatives.} 
\end{lem}

\begin{proof}
Since $f\in C^{0}\left(U\right)\cap C_{t}^{1}\left(U\right)$ we have
that $f$ is locally Lipschitz continuous on $U$ (see Corollary 4.1.1
on \cite{scholtes}). Then, we have that $f$ is locally absolute continuous
on $U$, and hence, $f\in L_{loc}^{1}\left(U\right)$. Then $f$
is weakly differentiable and the (first order) weak and pointwise
derivatives of $f$ coincide a.e. 
\end{proof}
Now, using the above lemma and the alternative definition (eq. (\ref{eq: sobolev_2}))
for the Sobolev space $H^{1}\left(\mathbb{R}^{n}\right)$, the proof
of lemma \ref{par:lemma-1} is trivial.


\renewcommand\appendixname{Appendix}
\selectlanguage{english}

\chapter{Computations of chapter \ref{RE_CS}}

\section{Calculation of $\left\langle f_{R}\right|k_{1}\left|f_{R}\right\rangle _{\mathrm{1p}}$\label{b2-1particle}}

Take $f_{R}\in\mathcal{S}\left(\mathbb{R}^{d},\mathbb{R}\right)$
and for simplicity call $f:=f_{R}$. Then,
\begin{eqnarray}
\left\langle f\right|k_{1}\left|f\right\rangle _{\mathrm{1p}}\!\!\! & = & \!\!\!\mathrm{Re}\left\langle f\right|k_{1}\left|f\right\rangle _{\mathrm{1p}}=\mathrm{Re}\left(-i\left.\frac{\mathrm{d}}{\mathrm{d}s}\right|_{s=0}\left\langle f\right|\mathrm{e}^{ik_{1}s}\left|f\right\rangle _{\mathrm{1p}}\right)=\nonumber \\
 & = & \!\!\!\left.\frac{\mathrm{d}}{\mathrm{d}s}\right|_{s=0}\mathrm{Im}\left\langle f\right|u\left(\Lambda_{1}^{s},0\right)\left|f\right\rangle _{\mathrm{1p}}=\left.\frac{\mathrm{d}}{\mathrm{d}s}\right|_{s=0}\mathrm{Im}\langle f\mid f^{s}\rangle_{\mathrm{1p}}\,,\label{eq:evqm_1}
\end{eqnarray}
where we have defined $f^{s}:=f_{\left(\Lambda_{1}^{s},0\right)}$.
As we have explained in section \ref{subsec:Relation-between-algebras},
there exist functions $f_{\varphi},\,f_{\pi},\,f_{\varphi}^{s},\,f_{\pi}^{s}\in\mathcal{S}\left(\mathbb{R}^{d-1},\mathbb{R}\right)$
such that $E\left(f\right)=E_{\varphi}\left(f_{\varphi}\right)+E_{\pi}\left(f_{\pi}\right)$
and $E\left(f^{s}\right)=E_{\varphi}\left(f_{\varphi}^{s}\right)+E_{\pi}\left(f_{\pi}^{s}\right)$.
Replacing these into \eqref{eq:evqm_1}, we get
\begin{eqnarray}
\left\langle f\right|k_{1}\left|f\right\rangle _{\mathrm{1p}}\!\!\! & = & \!\!\!\left.\frac{\mathrm{d}}{\mathrm{d}s}\right|_{s=0}\mathrm{Im}\langle f_{\varphi}+f_{\pi}\mid f_{\varphi}^{s}+f_{\pi}^{s}\rangle_{\mathrm{1p}}=\left.\frac{\mathrm{d}}{\mathrm{d}s}\right|_{s=0}\left(\mathrm{Im}\langle f_{\varphi}\mid f_{\pi}^{s}\rangle_{\mathrm{1p}}+\mathrm{Im}\langle f_{\pi}\mid f_{\varphi}^{s}\rangle_{\mathrm{1p}}\right)\nonumber \\
 & = & \!\!\!\left.\frac{\mathrm{d}}{\mathrm{d}s}\right|_{s=0}\left(\frac{1}{2}\int_{\mathbb{R}^{d-1}}f_{\varphi}\left(\bar{x}\right)\,f_{\pi}^{s}\left(\bar{x}\right)d^{d-1}x-\mathrm{\frac{1}{2}}\int_{\mathbb{R}^{d-1}}f_{\varphi}^{s}\left(\bar{x}\right)\,f_{\pi}\left(\bar{x}\right)d^{d}x\right)\,,\label{eq:evqm_2}
\end{eqnarray}
where have used the relations \eqref{eq:rel_prod_int} in the second line
and \eqref{eq:img_pe} in the last line. From the Poincaré invariance
of the distribution $\Delta\left(x\right)$, we have that
\begin{equation}
F^{s}\left(x\right)=\int_{\mathbb{R}^{d}}\Delta\left(x-y\right)\,f^{s}\left(x\right)d^{d}y\,\textrm{,}
\end{equation}
where we have defined $F^{s}:=F_{\left(\Lambda_{1}^{s},0\right)}$.
Then, according to \eqref{eq:f_fixed_time}, the following relations
hold
\begin{eqnarray}
f_{\varphi}\left(\bar{x}\right)\!\!\! & = & \!\!\!-\frac{\partial F}{\partial x^{0}}\left(0,\bar{x}\right)\,,\\
f_{\pi}\left(\bar{x}\right)\!\!\! & = & \!\!\!F\left(0,\bar{x}\right)\,,\\
f_{\varphi}^{s}\left(\bar{x}\right)\!\!\! & = & \!\!\!-\cosh\left(s\right)\frac{\partial F}{\partial x^{0}}\left(\bar{x}^{s}\right)+\sinh\left(s\right)\frac{\partial F}{\partial x^{1}}\left(\bar{x}^{s}\right)\,,\\
f_{\pi}^{s}\left(\bar{x}\right)\!\!\! & = & \!\!\!F\left(\bar{x}^{s}\right)\,,
\end{eqnarray}
where $\bar{x}^{s}:=\left(-x^{1}\sinh\left(s\right),x^{1}\cosh\left(s\right),x_{\bot}\right)$.
And hence, 
\begin{eqnarray}
\left.\frac{\mathrm{d}}{\mathrm{d}s}\right|_{s=0}f_{\varphi}^{s}\left(\bar{x}\right)\!\!\! & = & \!\!\!x^{1}\frac{\partial^{2}F}{\left(\partial x^{0}\right)^{2}}\left(0,\bar{x}\right)+\frac{\partial F}{\partial x^{1}}\left(0,\bar{x}\right)\,,\\
\left.\frac{\mathrm{d}}{\mathrm{d}s}\right|_{s=0}f_{\pi}^{s}\left(\bar{x}\right)\!\!\! & = & \!\!\!-x^{1}\frac{\partial F}{\partial x^{0}}\left(0,\bar{x}\right)\,\textrm{.}
\end{eqnarray}
Replacing such expressions into \eqref{eq:evqm_2}, using the equation
of motion for $F$ and performing an integration by parts, we finally get
\begin{equation}
\left\langle f\right|k_{1}\left|f\right\rangle _{\mathrm{1p}}=\int_{\mathbb{R}^{d-1}}d^{d-1}x\,x^{1}\,\frac{1}{2}\left.\left(\left(\frac{\partial F}{\partial x^{0}}\right)^{2}+\left|\nabla F\right|^{2}+m^{2}F^{2}\right)\right|_{x^{0}=0}\textrm{.}\label{apend_sr_easy}
\end{equation}

\section{Explicit computations of section \ref{subsec:determ_alpha}\label{subsec:appendix_3}}

We first define $g_{R}^{s}:=E_{\varphi}\left(g_{\varphi,R}^{s}\right)+E_{\pi}\left(g_{\pi,R}^{s}\right)\in\mathfrak{H}$. Then, we compute
\begin{eqnarray}
 &  & \!\!\!\left\langle 0\right|\Delta_{\Phi,0}^{it_{1}}\Delta_{\Phi,0}^{it_{2}}\left|0\right\rangle \nonumber \\
 & = & \!\!\!\left\langle 0\right|\mathrm{e}^{i\alpha\left(s_{1}\right)}W_{\varphi}\left(g_{\varphi,R}^{s_{1}}\right)W_{\pi}\left(g_{\pi,R}^{s_{1}}\right)\Delta_{0}^{it_{1}}\mathrm{e}^{i\alpha\left(s_{2}\right)}W_{\varphi}\left(g_{\varphi,R}^{s_{2}}\right)W_{\pi}\left(g_{\pi,R}^{s_{2}}\right)\Delta_{0}^{it_{1}}\left|0\right\rangle \nonumber \\
 & = & \!\!\!\mathrm{e}^{i\alpha\left(s_{1}\right)+i\alpha\left(s_{2}\right)}\left\langle 0\right|W_{\varphi}\left(g_{\varphi,R}^{s_{1}}\right)W_{\pi}\left(g_{\pi,R}^{s_{1}}\right)\mathrm{e}^{is_{1}K_{1}}W_{\varphi}\left(g_{\varphi,R}^{s_{2}}\right)W_{\pi}\left(g_{\pi,R}^{s_{2}}\right)\left|0\right\rangle \nonumber \\
 & = & \!\!\!\mathrm{e}^{i\alpha\left(s_{1}\right)+i\alpha\left(s_{2}\right)-i\mathrm{Im}\langle g_{\varphi,R}^{s_{1}}\mid g_{\pi,R}^{s_{1}}\rangle_{\mathrm{1p}}-i\mathrm{Im}\langle g_{\varphi,R}^{s_{2}}\mid g_{\pi,R}^{s_{2}}\rangle_{\mathrm{1p}}}\left\langle 0\right|W\left(g_{R}^{s_{1}}\right)\mathrm{e}^{is_{1}K_{1}}W\left(g_{R}^{s_{2}}\right)\left|0\right\rangle \nonumber \\
 & = & \!\!\!\mathrm{e}^{i\alpha\left(s_{1}\right)+i\alpha\left(s_{2}\right)-i\mathrm{Im}\langle g_{\varphi,R}^{s_{1}}\mid g_{\pi,R}^{s_{1}}\rangle_{\mathrm{1p}}-i\mathrm{Im}\langle g_{\varphi,R}^{s_{2}}\mid g_{\pi,R}^{s_{2}}\rangle_{\mathrm{1p}}}\left\langle 0\right|W\left(g_{R}^{s_{1}}\right)\mathrm{e}^{is_{1}K_{1}}W\left(g_{R}^{s_{2}}\right)\mathrm{e}^{-is_{1}K_{1}}\left|0\right\rangle \nonumber \\
 & = & \!\!\!\mathrm{e}^{i\alpha\left(s_{1}\right)+i\alpha\left(s_{2}\right)-i\mathrm{Im}\langle g_{\varphi,R}^{s_{1}}\mid g_{\pi,R}^{s_{1}}\rangle_{\mathrm{1p}}-i\mathrm{Im}\langle g_{\varphi,R}^{s_{2}}\mid g_{\pi,R}^{s_{2}}\rangle_{\mathrm{1p}}}\left\langle 0\right|W\left(g_{R}^{s_{1}}\right)W\left(u\left(\Lambda_{1}^{s_{1}}\right)g_{R}^{s_{2}}\right)\left|0\right\rangle \nonumber \\
 & = & \!\!\!\mathrm{e}^{i\alpha\left(s_{1}\right)+i\alpha\left(s_{2}\right)-i\mathrm{Im}\left[\langle g_{\varphi,R}^{s_{1}}\mid g_{\pi,R}^{s_{1}}\rangle_{\mathrm{1p}}+\langle g_{\varphi,R}^{s_{2}}\mid g_{\pi,R}^{s_{2}}\rangle_{\mathrm{1p}}+\langle g_{R}^{s_{1}}\mid u(\Lambda_{1}^{s_{1}})g_{R}^{s_{2}}\rangle_{\mathrm{1p}}\right]}\left\langle 0\right|W\left(g_{R}^{s_{1}}+u(\Lambda_{1}^{s_{1}})g_{R}^{s_{2}}\right)\left|0\right\rangle \nonumber \\
 & = & \!\!\!\mathrm{e}^{i\alpha\left(s_{1}\right)+i\alpha\left(s_{2}\right)-i\mathrm{Im}\left[\langle g_{\varphi,R}^{s_{1}}\mid g_{\pi,R}^{s_{1}}\rangle_{\mathrm{1p}}+\langle g_{\varphi,R}^{s_{2}}\mid g_{\pi,R}^{s_{2}}\rangle_{\mathrm{1p}}+\langle g_{R}^{s_{1}}\mid u(\Lambda_{1}^{s_{1}})g_{R}^{s_{2}}\rangle_{\mathrm{1p}}\right]-\frac{1}{2}\left\Vert g_{R}^{s_{1}}+u\left(\Lambda_{1}^{s_{1}}\right)g_{R}^{s_{2}}\right\Vert _{\mathrm{1p}}^{2}}\,,\label{apen_conca_1}
\end{eqnarray}
and
\begin{eqnarray}
\left\langle 0\right|\Delta_{\Phi,0}^{i\left(t_{1}+t_{2}\right)}\left|0\right\rangle \!\!\! & = & \!\!\!\left\langle 0\right|\mathrm{e}^{i\alpha\left(s_{1}+s_{2}\right)}W_{\varphi}\left(g_{\varphi,R}^{s_{1}+s_{2}}\right)W_{\pi}\left(g_{\pi,R}^{s_{1}+s_{2}}\right)\Delta_{0}^{i\left(t_{1}+t_{2}\right)}\left|0\right\rangle \nonumber \\
 & = & \!\!\!\mathrm{e}^{i\alpha\left(s_{1}+s_{2}\right)}\left\langle 0\right|W_{\varphi}\left(g_{\varphi,R}^{s_{1}+s_{2}}\right)W_{\pi}\left(g_{\pi,R}^{s_{1}+s_{2}}\right)\left|0\right\rangle \nonumber \\
 & = & \!\!\!\mathrm{e}^{i\alpha\left(s_{1}+s_{2}\right)-i\mathrm{Im}\langle g_{\varphi,R}^{s_{1}+s_{2}}\mid g_{\pi,R}^{s_{1}+s_{2}}\rangle_{\mathrm{1p}}}\left\langle 0\right|W\left(g_{R}^{s_{1}+s_{2}}\right)\left|0\right\rangle \nonumber \\
 & = & \!\!\!\mathrm{e}^{i\alpha\left(s_{1}+s_{2}\right)-i\mathrm{Im}\langle g_{\varphi,R}^{s_{1}+s_{2}}\mid g_{\pi,R}^{s_{1}+s_{2}}\rangle_{\mathrm{1p}}-\frac{1}{2}\left\Vert g_{R}^{s_{1}+s_{2}}\right\Vert _{\mathrm{1p}}^{2}}\,.\label{apen_conca_2}
\end{eqnarray}
Taking $\left.\frac{\mathrm{d}}{\mathrm{d}s_{1}}\right|_{s_{1}=0}$
on both expressions above, we obtain
\begin{eqnarray}
\left.\frac{\mathrm{d}}{\mathrm{d}s_{1}}\right|_{s_{1}=0}\!\!\!\!\!\!\!\!\left\langle 0\right|\Delta_{\Phi,0}^{it_{1}}\Delta_{\Phi,0}^{it_{2}}\left|0\right\rangle \!\!\! & = & \!\!\!i\alpha'\left(0\right)-i\underset{=0}{\underbrace{\left.\frac{\mathrm{d}}{\mathrm{d}s_{1}}\right|_{s_{1}=0}\!\!\!\!\!\mathrm{Im}\langle g_{\varphi,R}^{s_{1}}\mid g_{\pi,R}^{s_{1}}\rangle_{\mathrm{1p}}}}\nonumber \\
 &  & \!\!\!\!-i\left.\frac{\mathrm{d}}{\mathrm{d}s_{1}}\right|_{s_{1}=0}\!\!\!\!\!\mathrm{Im}\langle g_{R}^{s_{1}}\mid u\left(\Lambda_{1}^{s_{1}}\right)g_{R}^{s_{2}}\rangle_{\mathrm{1p}}-\frac{1}{2}\left.\frac{\mathrm{d}}{\mathrm{d}s_{1}}\right|_{s_{1}=0}\!\!\!\!\left\Vert g_{R}^{s_{1}}+u\left(\Lambda_{1}^{s_{1}}\right)g_{R}^{s_{2}}\right\Vert _{\mathrm{1p}}^{2}\nonumber \hspace{5mm} \\
 & = & \!\!\!i\alpha'\left(0\right)\!-\!\left.\frac{\mathrm{d}}{\mathrm{d}s_{1}}\right|_{s_{1}=0}\!\!\!\left(\!i\mathrm{Im}\langle g_{R}^{s_{1}}\mid g_{R}^{s_{2}}\rangle_{\mathrm{1p}}+\frac{1}{2}\left\Vert g_{R}^{s_{1}}+u\left(\Lambda_{1}^{s_{1}}\right)g_{R}^{s_{2}}\right\Vert _{\mathrm{1p}}^{2}\right)\!,\label{apen_der_1}
\end{eqnarray}
and
\begin{eqnarray}
\left.\frac{\mathrm{d}}{\mathrm{d}s_{1}}\right|_{s_{1}=0}\!\left\langle 0\right|\Delta_{\Phi,0}^{i\left(t_{1}+t_{2}\right)}\left|0\right\rangle \!\!\! & = & \!\!\!i\alpha'\left(s_{2}\right)-i\left.\frac{\mathrm{d}}{\mathrm{d}s_{1}}\right|_{s_{1}=0}\!\mathrm{Im}\langle g_{\varphi,R}^{s_{1}+s_{2}}\mid g_{\pi,R}^{s_{1}+s_{2}}\rangle_{\mathrm{1p}}-\frac{1}{2}\left.\frac{\mathrm{d}}{\mathrm{d}s_{1}}\right|_{s_{1}=0}\!\left\Vert g_{R}^{s_{1}+s_{2}}\right\Vert _{\mathrm{1p}}^{2}\nonumber \\
 & = & \!\!\!i\alpha'\left(s_{2}\right)-i\frac{\mathrm{d}}{\mathrm{d}s_{2}}\mathrm{Im}\langle g_{\varphi,R}^{s_{2}}\mid g_{\pi,R}^{s_{2}}\rangle_{\mathrm{1p}}-\frac{1}{2}\left.\frac{\mathrm{d}}{\mathrm{d}s_{1}}\right|_{s_{1}=0}\!\left\Vert g_{R}^{s_{1}+s_{2}}\right\Vert _{\mathrm{1p}}^{2}\,\textrm{.}\label{apen_der_2}
\end{eqnarray}
Matching separately the real and imaginary parts of these last two expressions, we
arrive to formulas (\ref{eq:ec_dif_alpha}-\ref{eq:obvia}).

Expressions (\ref{c5-gamma}-\ref{eq:QyR}) follow from
\begin{eqnarray}
2\,\mathrm{Im}\langle g_{R}^{s_{1}}\mid g_{R}^{s_{2}}\rangle_{\mathrm{1p}}\!\!\! & = & \!\!\!2\mathrm{Im}\langle g_{\varphi,R}^{s_{1}}+g_{\pi,R}^{s_{1}}\mid g_{\varphi,R}^{s_{2}}+g_{\pi,R}^{s_{2}}\rangle_{\mathrm{1p}}\nonumber \\
 & = & \!\!\!\int_{\Sigma}\!d^{d-1}x\,g_{\varphi}^{s_{1}}\left(\bar{x}\right)g_{\pi}^{s_{2}}\left(\bar{x}\right)-\int_{\Sigma}\!d^{d-1}x\,g_{\varphi}^{s_{2}}\left(\bar{x}\right)g_{\pi}^{s_{1}}\left(\bar{x}\right)\nonumber \\
 & = & \!\!\!\int_{\Sigma}\!d^{d-1}x\left(f_{\varphi}^{s_{1}}\left(\bar{x}\right)-f_{\varphi}\left(\bar{x}\right)\right)\left(f_{\pi}^{s_{2}}\left(\bar{x}\right)-f_{\pi}\left(\bar{x}\right)\right)\nonumber \\
 &  & \!\!\!-\int_{\Sigma}\!d^{d-1}x\left(f_{\varphi}^{s_{2}}\left(\bar{x}\right)-f_{\varphi}\left(\bar{x}\right)\right)\left(f_{\pi}^{s_{1}}\left(\bar{x}\right)-f_{\pi}\left(\bar{x}\right)\right)\nonumber \\
 & = & \!\!\!\underset{:=P\left(s_{1}\right)}{\underbrace{\int_{\Sigma_{R}}\!d^{d-1}x\,f_{\varphi}\left(\bar{x}\right)f_{\pi}^{s_{1}}\left(\bar{x}\right)-\int_{\Sigma}\!d^{d-1}x\,f_{\varphi}^{s_{1}}\left(\bar{x}\right)f_{\pi}\left(\bar{x}\right)}}\nonumber \\
 &  & \!\!\!+\underset{:=Q\left(s_{1},s_{2}\right)}{\underbrace{\int_{\Sigma_{R}}\!d^{d-1}x\,f_{\varphi}^{s_{1}}\left(\bar{x}\right)f_{\pi}^{s_{2}}\left(\bar{x}\right)}}-\underset{:=R\left(s_{1},s_{2}\right)}{\underbrace{\int_{\Sigma_{R}}\!d^{d-1}x\,f_{\varphi}^{s_{2}}\left(\bar{x}\right)f_{\pi}^{s_{1}}\left(\bar{x}\right)}}+\gamma\left(s_{2}\right), \hspace{7mm}
\end{eqnarray}
where the function $\gamma$ includes all the $s_{1}$-independent
terms.

The function $P\left(s_{1}\right)$ is essentially the same as \eqref{eq:evqm_2}
in appendix \ref{b2-1particle}, with the difference that now the
integration is over the region $\Sigma_{R}=\left\{ \bar{x}\in\mathbb{R}^{d-1}\,:\,x^{1}>0\right\} $
instead of the whole space $\mathbb{R}^{d-1}$. Despite this, the final
result is the same, and hence we get\footnote{In contrast with the computation of \eqref{eq:evqm_2} in section \ref{b2-1particle},
now it appears a boundary term after the integration by parts. Fortunately,
this term vanishes since the integrand is zero at the boundary of
$\Sigma_R$. }
\begin{equation}
\left.\frac{\mathrm{d}P}{\mathrm{d}s_{1}}\right|_{s_{1}=0}=\int_{\Sigma_{R}}\!d^{d-1}x\,x^{1}\left.\left(\left(\frac{\partial F}{\partial x^{0}}\right)^{2}+\left(\nabla F\right)^{2}+m^{2}F^{2}\right)\right|_{x^{0}=0}=:\boldsymbol{S}\,.
\end{equation}
Now, we explicitly obtain the relations \eqref{eq:cool_relation_2}. Then, we compute
\begin{eqnarray}
\left.\frac{\mathrm{d}R}{\mathrm{d}s_{1}}\right|_{s_{1}=0}\!\!\! & = & \!\!\!\left.\frac{\mathrm{d}}{\mathrm{d}s_{1}}\right|_{s_{1}=0}\int_{\Sigma_{R}}\!d^{d-1}x\,\left(-\cosh\left(s_{2}\right)\frac{\partial F}{\partial x^{0}}\left(\bar{x}^{s_{2}}\right)+\sinh\left(s_{2}\right)\frac{\partial F}{\partial x^{1}}\left(\bar{x}^{s_{2}}\right)\right)F\left(\bar{x}^{s_{1}}\right)\nonumber \\
 & = & \!\!\!\int_{\Sigma_{R}}\!d^{d-1}x\,\left(-\cosh\left(s_{2}\right)\frac{\partial F}{\partial x^{0}}\left(\bar{x}^{s_{2}}\right)+\sinh\left(s_{2}\right)\frac{\partial F}{\partial x^{1}}\left(\bar{x}^{s_{2}}\right)\right)\left(-x^{1}\frac{\partial F}{\partial x^{0}}\left(\bar{x}\right)\right)\nonumber \\
 & = & \!\!\!\int_{\Sigma_{R}}\!d^{d-1}x\,\left(-\frac{\partial F}{\partial x^{0}}\left(\bar{x}\right)\right)\left(-x^{1}\cosh\left(s_{2}\right)\frac{\partial F}{\partial x^{0}}\left(\bar{x}^{s_{2}}\right)+x^{1}\sinh\left(s_{2}\right)\frac{\partial F}{\partial x^{1}}\left(\bar{x}^{s_{2}}\right)\right)\nonumber \\
 & = & \!\!\!\frac{\mathrm{d}}{\mathrm{d}s_{2}}\int_{\Sigma_{R}}\!d^{d-1}x\,\left(-\frac{\partial F}{\partial x^{0}}\left(\bar{x}\right)\right)F\left(\bar{x}^{s_{2}}\right)\nonumber \\
 & = & \!\!\!\left.\frac{\mathrm{d}}{\mathrm{d}s_{2}}\right|_{s_{1}=0}\int_{\Sigma_{R}}\!d^{d-1}x\,\left(-\cosh\left(s_{1}\right)\frac{\partial F}{\partial x^{0}}\left(\bar{x}^{s_{1}}\right)+\sinh\left(s_{1}\right)\frac{\partial F}{\partial x^{1}}\left(\bar{x}^{s_{1}}\right)\right)F\left(\bar{x}^{s_{2}}\right)\nonumber \\
 & = & \!\!\!\left.\frac{\mathrm{d}}{\mathrm{d}s_{2}}\right|_{s_{1}=0}\int_{\Sigma_{R}}\!d^{d-1}x\,f_{\varphi}^{s_{1}}\left(\bar{x}\right)f_{\pi}^{s_{2}}\left(\bar{x}\right)=\left.\frac{\mathrm{d}Q}{\mathrm{d}s_{2}}\right|_{s_{1}=0}.\label{eq:RaQ}
\end{eqnarray}
Similarly, we start with
\begin{eqnarray}
\left.\frac{\mathrm{d}Q}{\mathrm{d}s_{1}}\right|_{s_{1}=0}\!\!\! & = & \!\!\!\left.\frac{\mathrm{d}}{\mathrm{d}s_{1}}\right|_{s_{1}=0}\int_{\Sigma_{R}}\!d^{d-1}x\,f_{\varphi}^{s_{1}}\left(\bar{x}\right)f_{\pi}^{s_{2}}\left(\bar{x}\right)\nonumber \\
 & = & \!\!\!\left.\frac{\mathrm{d}}{\mathrm{d}s_{1}}\right|_{s_{1}=0}\int_{\Sigma_{R}}\!d^{d-1}x\,\left(-\cosh\left(s_{1}\right)\frac{\partial F}{\partial x^{0}}\left(\bar{x}^{s_{1}}\right)+\sinh\left(s_{1}\right)\frac{\partial F}{\partial x^{1}}\left(\bar{x}^{s_{1}}\right)\right)F\left(\bar{x}^{s_{2}}\right)\nonumber \\
 & = & \!\!\!\int_{\Sigma_{R}}\!d^{d-1}x\,\left(x^{1}\frac{\partial^{2}F}{\left(\partial x^{0}\right)^{2}}\left(\bar{x}\right)+\frac{\partial F}{\partial x^{1}}\left(\bar{x}\right)\right)F\left(\bar{x}^{s_{2}}\right)\nonumber \\
 & = & \!\!\!\int_{\Sigma_{R}}\!d^{d-1}x\,\left(x^{1}\left(\nabla^{2}-m^{2}\right)F\left(\bar{x}\right)+\frac{\partial F}{\partial x^{1}}\left(\bar{x}\right)\right)F\left(\bar{x}^{s_{2}}\right)\,.
\end{eqnarray}
Then, we integrate the Laplacian term by parts
\begin{eqnarray}
\left.\frac{\mathrm{d}Q}{\mathrm{d}s_{1}}\right|_{s_{1}=0}\!\!\! & = & \!\!\!-\int_{\Sigma_{R}}\!d^{d-1}x\,x^{1}m^{2}F\left(\bar{x}\right)F\left(\bar{x}^{s_{2}}\right)-\int_{\Sigma}\!d^{d-1}x\,x^{1}\nabla_{\bot}F\left(\bar{x}\right)\cdot\nabla_{\bot}F\left(\bar{x}^{s_{2}}\right)\nonumber \\
 &  & \!\!\!-\int_{\Sigma_{R}}\!d^{d-1}x\,x^{1}\frac{\partial F}{\partial x^{1}}\left(\bar{x}\right)\left(-\sinh\left(s_{2}\right)\frac{\partial F}{\partial x^{0}}\left(\bar{x}^{s_{2}}\right)+\cosh\left(s_{2}\right)\frac{\partial F}{\partial x^{1}}\left(\bar{x}^{s_{2}}\right)\right). \hspace{1cm}
\end{eqnarray}
After a second integration by parts, we get
\begin{eqnarray}
\left.\frac{\mathrm{d}Q}{\mathrm{d}s_{1}}\right|_{s_{1}=0}\!\!\! & = & \!\!\!\int_{\Sigma_{R}}\!d^{d-1}x\,x^{1}F\left(\bar{x}\right)\left(\nabla_{\bot}^{2}-m^{2}\right)F\left(\bar{x}^{s_{2}}\right)\nonumber \\
 &  & \!\!\!+\int_{\Sigma_{R}}\!d^{d-1}x\,F\left(\bar{x}\right)\left(-\sinh\left(s_{2}\right)\frac{\partial F}{\partial x^{0}}\left(\bar{x}^{s_{2}}\right)+\cosh\left(s_{2}\right)\frac{\partial F}{\partial x^{1}}\left(\bar{x}^{s_{2}}\right)\right)\nonumber \\
 &  & \!\!\!+\int_{\Sigma_{R}}\!d^{d-1}x\,x^{1}F\left(\bar{x}\right)\left(\sinh^{2}\left(s_{2}\right)\frac{\partial^{2}F}{\left(\partial x^{0}\right)^{2}}\left(\bar{x}^{s_{2}}\right)\right.\nonumber \\
 &  & \!\!\!\left.-2\sinh\left(s_{2}\right)\cosh\left(s_{2}\right)\frac{\partial^{2}F}{\partial x^{0}\partial x^{1}}\left(\bar{x}^{s_{2}}\right)+\cosh^{2}\left(s_{2}\right)\frac{\partial^{2}F}{\left(\partial x^{1}\right)^{2}}\left(\bar{x}^{s_{2}}\right)\right)\,.
\end{eqnarray}
Now, we form a Laplacian term in the first line and we use the equation of motion for $F$,
\begin{eqnarray}
\left.\frac{\mathrm{d}Q}{\mathrm{d}s_{1}}\right|_{s_{1}=0}\!\!\! & = & \!\!\!\int_{\Sigma_{R}}\!d^{d-1}x\,x^{1}F\left(\bar{x}\right)\frac{\partial^{2}F}{\left(\partial x^{0}\right)^{2}}\left(\bar{x}^{s_{2}}\right)\nonumber \\
 &  & \!\!\!+\int_{\Sigma_{R}}\!d^{d-1}x\,F\left(\bar{x}\right)\left(-\sinh\left(s_{2}\right)\frac{\partial F}{\partial x^{0}}\left(\bar{x}^{s_{2}}\right)+\cosh\left(s_{2}\right)\frac{\partial F}{\partial x^{1}}\left(\bar{x}^{s_{2}}\right)\right)\nonumber \\
 &  & \!\!\!+\int_{\Sigma_{R}}\!d^{d-1}x\,x^{1}F\left(\bar{x}\right)\left(\sinh^{2}\left(s_{2}\right)\frac{\partial^{2}F}{\left(\partial x^{0}\right)^{2}}\left(\bar{x}^{s_{2}}\right)\right.\nonumber \\
 &  & \!\!\!\left.-2\sinh\left(s_{2}\right)\cosh\left(s_{2}\right)\frac{\partial^{2}F}{\partial x^{0}\partial x^{1}}\left(\bar{x}^{s_{2}}\right)+\sinh^{2}\left(s_{2}\right)\frac{\partial^{2}F}{\left(\partial x^{1}\right)^{2}}\left(\bar{x}^{s_{2}}\right)\right)\,.
\end{eqnarray}
Finally, a straightforward computation shows that
\begin{eqnarray}
\left.\frac{\mathrm{d}Q}{\mathrm{d}s_{1}}\right|_{s_{1}=0}\!\!\! & = & \!\!\!\int_{\Sigma_{R}}\!d^{d-1}x\,\frac{\mathrm{d}}{\mathrm{d}s_{2}}\left(-\cosh\left(s_{2}\right)\frac{\partial F}{\partial x^{0}}\left(\bar{x}^{s_{2}}\right)+\sinh\left(s_{2}\right)\frac{\partial F}{\partial x^{1}}\left(\bar{x}^{s_{2}}\right)\right)F\left(\bar{x}\right)\nonumber \\
 & = & \!\!\!\left.\frac{\mathrm{d}}{\mathrm{d}s_{2}}\right|_{s_{1}=0}\int_{\Sigma_{R}}\!d^{d-1}x\,\left(-\cosh\left(s_{2}\right)\frac{\partial F}{\partial x^{0}}\left(\bar{x}^{s_{2}}\right)+\sinh\left(s_{2}\right)\frac{\partial F}{\partial x^{1}}\left(\bar{x}^{s_{2}}\right)\right)F\left(\bar{x}^{s_{1}}\right)\nonumber \hspace{1cm} \\
 & = & \!\!\!\left.\frac{\mathrm{d}}{\mathrm{d}s_{2}}\right|_{s_{1}=0}\int_{\Sigma_{R}}\!d^{d-1}x\,f_{\varphi}^{s_{2}}\left(\bar{x}\right)f_{\pi}^{s_{1}}\left(\bar{x}\right)=\left.\frac{\mathrm{d}R}{\mathrm{d}s_{2}}\right|_{s_{1}=0}.\label{eq:QaR}
\end{eqnarray}
Using \eqref{eq:RaQ} and \eqref{eq:QaR}, we arrive to \eqref{eq:cool_relation_2}.

\section{Analytic continuation for $N\left(s\right)$\label{subsec:Analytic-continuation}}

In order to show that formulas (\ref{eq: is_zero_dos}-\ref{eq: is_zero}) hold, we need
to explicitly obtain the analytic continuation of the function 
\begin{equation}
N\left(s\right)=\frac{i}{2}\left(Q\left(0,s\right)-R\left(0,s\right)\right)-\frac{1}{2}\left\Vert g_{R}^{s}\right\Vert _{\mathrm{1p}}^{2}\textrm{ ,}\label{eq:N(s)}
\end{equation}
or more specifically, we need to show that there exists a continuous
function $\tilde{N}:\mathbb{R}+i\left[0,2\pi\right]\rightarrow\mathbb{C}$,
analytic on $\mathbb{R}+i\left(0,2\pi\right)$, such that
\begin{equation}
\tilde{N}\left(s+i0\right)=N\left(s\right)\textrm{ .}
\end{equation}
To begin with, we notice that
\begin{eqnarray}
\frac{i}{2}Q\left(0,s\right)\!\!\! & = & \!\!\!\frac{i}{2}\int_{x^{1}>0}d^{d-1}x\,f_{\varphi}\left(\bar{x}\right)f_{\pi}^{s}\left(\bar{x}\right)=i\,\mathrm{Im}\langle f_{\varphi,R}\mid f_{\pi,R}^{s}\rangle_{\mathrm{1p}}\,,\\
\frac{i}{2}R\left(0,s\right)\!\!\! & = & \!\!\!\frac{1}{2}\int_{x^{1}>0}d^{d-1}x\,f_{\varphi}^{s}\left(\bar{x}\right)f_{\pi}\left(\bar{x}\right)=i\,\mathrm{Im}\langle f_{\varphi,R}^{s}\mid f_{\pi,R}\rangle_{\mathrm{1p}}\,,
\end{eqnarray}
where the above expressions make sense regardless $f_{\pi,R}^{s}\notin\mathfrak{H}$.
This is because 
\begin{equation}
\langle f_{\varphi,R}\mid f_{\pi,R}^{s}\rangle_{\mathrm{1p}}=\int_{\mathbb{R}^{d-1}}\frac{d^{d-1}p}{2\omega_{\bar{p}}}\hat{f}_{\varphi,R}\left(\bar{p}\right)^{*}i\omega_{\bar{p}}\hat{f}_{\pi,R}^{s}\left(\bar{p}\right)=\frac{i}{2}\langle\hat{f}_{\varphi,R}\mid\hat{f}_{\pi,R}^{s}\rangle_{L^{2}}\,,
\end{equation}
which is convergent. The problem involving scalar products of split
functions $f_{\varphi,R}^{s}$ and $f_{\pi,R}^{s}$ happens only when
we try to compute the scalar product of two sharply cut test functions
of the momentum operator, e.g. 
\begin{equation}
\langle f_{\text{\ensuremath{\pi}},R}\mid f_{\pi,R}^{s}\rangle_{\mathrm{1p}}=\int_{\mathbb{R}^{d-1}}\frac{d^{d-1}p}{2\omega_{\bar{p}}}\left(i\omega_{\bar{p}}\hat{f}_{\pi,R}\left(\bar{p}\right)\right)^{*}i\omega_{\bar{p}}\hat{f}_{\pi,R}^{s}\left(\bar{p}\right)=\frac{1}{2}\left\langle \hat{f}_{\pi,R},\hat{f}_{\pi,R}^{s}\right\rangle _{H^{\frac{1}{2}}}\,,\label{eq: scalar_pr_bad}
\end{equation}
which is in general divergent. Such divergence comes from the non-continuity of the function $f_{\pi,R}\left(\bar{x}\right)=f_{\pi}\left(\bar{x}\right)\Theta\left(x^{1}\right)$
at $x^{1}=0$. To overcome this difficulty, we introduce a family of
smooth functions (for $\epsilon>0$) 
\begin{equation}
f_{\varphi,R}^{\epsilon}\left(\bar{x}\right):=f_{\varphi}\left(\bar{x}\right)\Theta_{\epsilon}\left(x^{1}\right)\quad\mathrm{and}\quad f_{\pi,R}^{\epsilon}\left(\bar{x}\right):=f_{\pi}\left(\bar{x}\right)\Theta_{\epsilon}\left(x^{1}\right)\,,
\end{equation}
where $\Theta_{\varepsilon}\in C^{\infty}\left(\mathbb{R}\right)$
is a regularized Heaviside function such that
\begin{equation}
\Theta_{\epsilon}\left(t\right)=\begin{cases}
0 & \textrm{if }t\leq\frac{\epsilon}{2}\\
1 & \textrm{if }t\geq\epsilon
\end{cases}\textrm{ .}\label{eq:step_smooth}
\end{equation}
Then, we have that
\begin{eqnarray}
f_{\varphi,R}^{\epsilon}\left(\bar{x}\right)\underset{\epsilon\rightarrow0^{+}}{\longrightarrow}f_{\varphi,R}\left(\bar{x}\right) & \textrm{ and } & f_{\pi,R}^{\epsilon}\left(\bar{x}\right)\underset{\epsilon\rightarrow0^{+}}{\longrightarrow}f_{\pi,R}\left(\bar{x}\right)\,,\label{eq: conv_funcs}
\end{eqnarray}
where the above convergence must be in a sense that we will specify
opportunely below. Before we get into such convergence issues, we
notice that $f_{\varphi,R}^{\epsilon},f_{\pi,R}^{\epsilon}\in\mathcal{S}\left(\mathbb{R}^{d-1},\mathbb{R}\right)$
and hence, the scalar product \eqref{eq: scalar_pr_bad} is now well-defined. Then, we define the function
\begin{equation}
N^{\epsilon}\left(s\right):=i\,\mathrm{Im}\langle f_{\varphi,R}^{\epsilon}\mid f_{\pi,R}^{s,\epsilon}\rangle_{\mathrm{1p}}-i\,\mathrm{Im}\langle f_{\varphi,R}^{s,\epsilon}\mid f_{\pi,R}^{\epsilon}\rangle_{\mathrm{1p}}-\frac{1}{2}\left\Vert g_{R}^{s,\epsilon}\right\Vert _{\mathrm{1p}}^{2}\,,\label{eq:Ne(s)}
\end{equation}
which is just the regularized version of \eqref{eq:N(s)}. In the
next subsection, we will show that $N^{\epsilon}\left(s\right)\rightarrow N\left(s\right)$
when $\epsilon\rightarrow0^{+}$. Expression \eqref{eq:Ne(s)} can be rewritten as
\begin{eqnarray}
N^{\epsilon}\left(s\right)\!\!\! & = & \!\!\!i\,\mathrm{Im}\langle f_{\varphi,R}^{\epsilon}\mid f_{\pi,R}^{s,\epsilon}\rangle_{\mathrm{1p}}+i\,\mathrm{Im}\langle f_{\pi,R}^{\epsilon}\mid f_{\varphi,R}^{s,\epsilon}\rangle_{\mathrm{1p}}-\frac{1}{2}\langle f_{R}^{\epsilon}-f_{R}^{s,\epsilon}\mid f_{R}^{\epsilon}-f_{R}^{s,\epsilon}\rangle_{\mathrm{1p}}\nonumber \\
 & = & \!\!\!\langle f_{R}^{\epsilon}\mid f_{R}^{s,\epsilon}\rangle_{\mathrm{1p}}-\frac{1}{2}\langle f_{R}^{\epsilon}\mid f_{R}^{\epsilon}\rangle_{\mathrm{1p}}-\frac{1}{2}\langle f_{R}^{s,\epsilon}\mid f_{R}^{s,\epsilon}\rangle_{\mathrm{1p}}=\langle f_{R}^{-\frac{s}{2},\epsilon}\mid f_{R}^{\frac{s}{2},\epsilon}\rangle_{\mathrm{1p}}-\langle f_{R}^{\epsilon}\mid f_{R}^{\epsilon}\rangle_{\mathrm{1p}}\nonumber \\
 & = & \!\!\!\int_{\mathbb{R}^{d-1}}\!\!\frac{d^{d-1}p}{2\omega_{\bar{p}}}\left[\left(\hat{f}_{\varphi,R}^{-\frac{s}{2},\epsilon}+i\omega_{\bar{p}}\hat{f}_{\pi,R}^{-\frac{s}{2},\epsilon}\right)^{*}\left(\hat{f}_{\varphi,R}^{\frac{s}{2},\epsilon}+i\omega_{\bar{p}}\hat{f}_{\pi,R}^{\frac{s}{2},\epsilon}\right)\right.\nonumber \\
 &  & \!\!\!\left.-\left(\hat{f}_{\varphi,R}^{\epsilon}+i\omega_{\bar{p}}\hat{f}_{\pi,R}^{\epsilon}\right)^{*}\left(\hat{f}_{\varphi,R}^{\epsilon}+i\omega_{\bar{p}}\hat{f}_{\pi,R}^{\epsilon}\right)\right],\label{eq:estudiar_conver}
\end{eqnarray}
where in the penultimate line we have used that $f_{R}^{s_{1}+s_{2},\epsilon}=u\left(\Lambda_{1}^{s_{2}}\right)f_{R}^{s_{1},\epsilon}$
for all $s_{1},s_{2}\in\mathbb{R}$. For a moment, let assume that
this last expression converges to
\begin{eqnarray}
N\left(s\right)\!\!\! & = & \!\!\!\int_{\mathbb{R}^{d-1}}\!\!\frac{d^{d-1}p}{2\omega_{\bar{p}}}\left[\left(\hat{f}_{\varphi,R}^{-\frac{s}{2}}+i\omega_{\bar{p}}\hat{f}_{\pi,R}^{-\frac{s}{2}}\right)^{*}\left(\hat{f}_{\varphi,R}^{\frac{s}{2}}+i\omega_{\bar{p}}\hat{f}_{\pi,R}^{\frac{s}{2}}\right)\right.\nonumber \\
 &  & \!\!\!\left.-\left(\hat{f}_{\varphi,R}+i\omega_{\bar{p}}\hat{f}_{\pi,R}\right)^{*}\left(\hat{f}_{\varphi,R}+i\omega_{\bar{p}}\hat{f}_{\pi,R}\right)\right]\,,\label{eq: estudiar_anal}
\end{eqnarray}
when $\epsilon\rightarrow0^{+}$. We will prove this in the next subsection.
The second term of the above integrand is independent on $s$ and
hence its analytic continuation is trivial. Let us then focus in the
first term. Using the Poincaré covariance and causality of the Klein-Gordon
equation, it is not difficult to show that
\begin{equation}
\hat{f}_{\varphi,R}^{s}\left(\bar{p}\right)+i\omega_{\bar{p}}\hat{f}_{\pi,R}^{s}\left(\bar{p}\right)=\hat{f}_{\varphi,R}\left(\Lambda_{1}^{s}\bar{p}\right)+i\,\Lambda_{1}^{s}\omega_{\bar{p}}\,\hat{f}_{\pi,R}\left(\Lambda_{1}^{s}\bar{p}\right)\,,\label{eq:cortar_boostear}
\end{equation}
where $\Lambda_{1}^{s}\bar{p}=\left(p^{1}\cosh\left(s\right)-\omega_{\bar{p}}\sinh\left(s\right),\bar{p}_{\bot}\right)$
and $\Lambda_{1}^{s}\omega_{\bar{p}}=\omega_{\bar{p}}\cosh\left(s\right)-p^{1}\sinh\left(s\right)$.
Then, the first integrand term of \eqref{eq: estudiar_anal} becomes
\begin{eqnarray}
\hspace{-1.3 cm} &  & \!\!\!\left(\hat{f}_{\varphi,R}^{-\frac{s}{2},\epsilon}\left(\bar{p}\right)+i\omega_{\bar{p}}\hat{f}_{\pi,R}^{-\frac{s}{2},\epsilon}\left(\bar{p}\right)\right)^{*}\left(\hat{f}_{\varphi,R}^{\frac{s}{2},\epsilon}\left(\bar{p}\right)+i\omega_{\bar{p}}\hat{f}_{\pi,R}^{\frac{s}{2},\epsilon}\left(\bar{p}\right)\right)\nonumber \\
\hspace{-1.3 cm} & = & \!\!\!\!\int_{\mathbb{R}^{2\left(d-1\right)}}\!\!\!\!\!\! d^{d-1}x\,d^{d-1}y\left(f_{\varphi,R}\left(\bar{x}\right)-i\omega_{\bar{p}}f_{\pi,R}\left(\bar{x}\right)\right)\left(f_{\varphi,R}\left(\bar{y}\right)+i\omega_{\bar{p}}f_{\pi,R}\left(\bar{y}\right)\right)\mathrm{e}^{i\Lambda^{-\frac{s}{2}}\left(\bar{p}\right)\cdot\bar{x}-i\Lambda^{\frac{s}{2}}\left(\bar{p}\right)\cdot\bar{y}},\label{eq:pi_ext}
\end{eqnarray}
where $-i\Lambda^{\frac{s}{2}}\left(\bar{p}\right)\cdot\bar{y}=-i\left(-\sinh\left(\frac{s}{2}\right)\omega_{\bar{p}}+\cosh\left(\frac{s}{2}\right)p^{1}\right)y^{1}-i\bar{p}_{\bot}\cdot\bar{y}_{\bot}$,
and equivalently for $i\Lambda^{-\frac{s}{2}}\left(\bar{p}\right)\cdot\bar{x}$.
Then, we have that
\begin{eqnarray}
 -i\!\left(-\sinh\!\left(\frac{s}{2}\right)\omega_{\bar{p}}+\cosh\!\left(\frac{s}{2}\right)p^{1}\right)\!y^{1} \!\!\!\! & \underset{s\rightarrow s+i\sigma}{\longrightarrow} & \!\!\!\! -i\!\left(\!-\sinh\left(\frac{s+i\sigma}{2}\right)\omega_{\bar{p}}+\cosh\left(\frac{s+i\sigma}{2}\right)p^{1}\!\right)y^{1}\nonumber \\
 & = & \!\!\!\! -i\left(-\sinh\left(\frac{s}{2}\right)\omega_{\bar{p}}+\cosh\left(\frac{s}{2}\right)p^{1}\right)y^{1}\cos\left(\frac{\sigma}{2}\right)\nonumber \\
 &  & \!\!\!\! -\underset{\geq m}{\underbrace{\left(\cosh\!\left(\frac{s}{2}\right)\omega_{\bar{p}}-\sinh\!\left(\frac{s}{2}\right)p^{1}\right)}}y^{1}\sin\!\left(\frac{\sigma}{2}\right)\!,
\end{eqnarray}
where the second term provides an exponential dumping in equation
\eqref{eq:pi_ext} when $\sigma\in\left(0,2\pi\right)$ because $\mathrm{supp}\left(f_{\varphi,R}\right),$
$\mathrm{supp}\left(f_{\text{\ensuremath{\pi}},R}\right)\subset\Sigma_{R}$.
Equivalently, it can be shown that $i\Lambda^{-\frac{s}{2}}\left(\bar{p}\right)\cdot\bar{x}$
also provides an exponential dumping for $\sigma\in\left(0,2\pi\right)$.
Hence, we have that
\begin{equation}
\tilde{N}\left(s+i\sigma\right)\textrm{ is an analytic function for }s+i\sigma\in\mathbb{R}+i\left(0,2\pi\right)\textrm{. }
\end{equation}
Looking at expressions \eqref{eq: estudiar_anal} and \eqref{eq:pi_ext},
it is easy to determine that 
\begin{equation}
\lim_{\sigma\rightarrow2\pi^{-},s=0}\tilde{N}\left(s+i\sigma\right)=0\,.
\end{equation}

\subsection{Convergence of $N^\epsilon(s)$}

In order to show that expression \eqref{eq: estudiar_anal} holds,
we need to prove the following two limits
\begin{eqnarray}
N^{\epsilon}\left(s\right)\!\!\! & \underset{\epsilon\rightarrow0^{+}}{\longrightarrow} & \!\!\!N\left(s\right)=\frac{i}{2}\left(Q\left(0,s\right)-R\left(0,s\right)\right)-\frac{1}{2}\left\Vert g_{R}^{s}\right\Vert _{\mathfrak{H}}^{2}\,,\label{eq:limite1}\\
N^{\epsilon}\left(s\right)\!\!\! & \underset{\epsilon\rightarrow0^{+}}{\longrightarrow} & \!\!\!\int_{\mathbb{R}^{d-1}}\!\!\frac{d^{d-1}p}{2\omega_{\bar{p}}}\left[\left(\hat{f}_{\varphi,R}^{-\frac{s}{2}}+i\omega_{\bar{p}}\hat{f}_{\pi,R}^{-\frac{s}{2}}\right)^{*}\left(\hat{f}_{\varphi,R}^{\frac{s}{2}}+i\omega_{\bar{p}}\hat{f}_{\pi,R}^{\frac{s}{2}}\right)\right.\nonumber \\
 &  & \!\!\!\left.-\left(\hat{f}_{\varphi,R}+i\omega_{\bar{p}}\hat{f}_{\pi,R}\right)^{*}\left(\hat{f}_{\varphi,R}+i\omega_{\bar{p}}\hat{f}_{\pi,R}\right)\right]\,.\label{eq: limite2}
\end{eqnarray}
To do this, we must precise in which sense the functions $f_{\varphi,R}^{s,\epsilon},f_{\pi,R}^{s,\epsilon}$
converge in \eqref{eq: conv_funcs}. To start, we choose the following
smooth step function \eqref{eq:step_smooth}
\begin{equation}
\Theta_{\varepsilon}\left(t\right)=\begin{cases}
0 & \textrm{if }t\leq\frac{\epsilon}{2}\,,\\
\left[\exp\left(\frac{\epsilon\left(t-\frac{3\epsilon}{4}\right)}{\left(t-\frac{3\epsilon}{4}\right)^{2}-\left(\frac{\epsilon}{4}\right)^{2}}\right)+1\right]^{-1} & \textrm{if }\frac{\epsilon}{2}<t<\epsilon\,,\\
1 & \textrm{if }t\geq\epsilon\textrm{ .}
\end{cases}\label{eq: theta_reg}
\end{equation}
Let us first focus in the limit \eqref{eq: limite2}. Looking back
to \eqref{eq:estudiar_conver}, we can rewrite the r.h.s. of that
expression as
\begin{eqnarray}
N^{\epsilon}\left(s\right)\!\!\! & = & \!\!\!\langle f_{R}^{-\frac{s}{2},\epsilon}\mid f_{R}^{\frac{s}{2},\epsilon}\rangle_{\mathrm{1p}}-\langle f_{R}^{\epsilon}\mid f_{R}^{\epsilon}\rangle_{\mathrm{1p}}=\langle f_{R}^{-\frac{s}{2},\epsilon}\mid f_{R}^{\frac{s}{2},\epsilon}-f_{R}^{\epsilon}+f_{R}^{\epsilon}\rangle_{\mathrm{1p}}-\langle f_{R}^{\epsilon}\mid f_{R}^{\epsilon}\rangle_{\mathrm{1p}}\nonumber \\
 & = & \!\!\!\langle f_{\varphi,R}^{-\frac{s}{2},\epsilon}\mid f_{\varphi,R}^{\frac{s}{2},\epsilon}-f_{\varphi,R}^{\epsilon}\rangle_{\mathrm{1p}}+\langle f_{\varphi,R}^{-\frac{s}{2},\epsilon}\mid f_{\pi,R}^{\frac{s}{2},\epsilon}-f_{\pi,R}^{\epsilon}\rangle_{\mathrm{1p}}+\langle f_{\varphi,R}^{-\frac{s}{2},\epsilon}-f_{\varphi,R}^{\epsilon}\mid f_{\varphi,R}^{\epsilon}\rangle_{\mathrm{1p}}\nonumber \\
 &  & \!\!\!+\langle f_{\pi,R}^{-\frac{s}{2},\epsilon}-f_{\pi,R}^{\epsilon}\mid f_{\varphi,R}^{\epsilon}\rangle_{\mathrm{1p}}+\langle f_{\pi,R}^{-\frac{s}{2},\epsilon}\mid f_{\varphi,R}^{\frac{s}{2},\epsilon}-f_{\varphi,R}^{\epsilon}\rangle_{\mathrm{1p}}+\langle f_{\varphi,R}^{-\frac{s}{2},\epsilon}-f_{\varphi,R}^{\epsilon}\mid f_{\pi,R}^{\epsilon}\rangle_{\mathrm{1p}}\nonumber \hspace{1cm} \\
 &  & \!\!\!+\underset{\varoast}{\underbrace{\langle f_{\pi,R}^{-\frac{s}{2},\epsilon}\mid f_{\pi,R}^{\frac{s}{2},\epsilon}-f_{\pi,R}^{\epsilon}\rangle_{\mathrm{1p}}}}+\underset{\varoast}{\underbrace{\langle f_{\pi,R}^{-\frac{s}{2},\epsilon}-f_{\pi,R}^{\epsilon}\mid f_{\pi,R}^{\epsilon}\rangle_{\mathrm{1p}}}}\,.\label{eq:analizar_conver}
\end{eqnarray}
It is not difficult to see that
\begin{eqnarray}
f_{\varphi,R}^{s,\epsilon}\underset{\epsilon\rightarrow0^{+}}{\longrightarrow}f_{\varphi,R}^{s} & \textrm{ and } & f_{\pi,R}^{s,\epsilon}\underset{\epsilon\rightarrow0^{+}}{\longrightarrow}f_{\pi,R}^{s}\,,\quad\textrm{in }L^{2}\left(\mathbb{R}^{d-1}\right)\,,\label{eq:conv_l2}
\end{eqnarray}
which implies that all terms in \eqref{eq:analizar_conver} are convergent,
except, perhaps, those pointed by $\varoast$. Now, we concentrate in
those remaining terms, for example
\begin{equation}
\langle f_{\pi,R}^{-\frac{s}{2},\epsilon}\mid f_{\pi,R}^{\frac{s}{2},\epsilon}-f_{\pi,R}^{\epsilon}\rangle_{\mathrm{1p}}=\frac{1}{2}\int_{\mathbb{R}^{d-1}}d^{d-1}p\,\hat{f}_{\pi,R}^{-\frac{s}{2},\epsilon}\left(\bar{p}\right)\left(\hat{f}_{\pi,R}^{-\frac{s}{2},\epsilon}\left(\bar{p}\right)-\hat{f}_{\pi,R}^{\epsilon}\left(\bar{p}\right)\right)\omega_{\bar{p}}\,.\label{eq:conv_casi_final}
\end{equation}
The convergence of \eqref{eq:conv_casi_final} is guarantee by the
fact that 
\begin{eqnarray}
f_{\pi,R}^{\epsilon}-f_{\pi,R}^{-\frac{s}{2},\epsilon} & \underset{\epsilon\rightarrow0^{+}}{\longrightarrow} & f_{\pi,R}-f_{\pi,R}^{-\frac{s}{2}}\quad\textrm{in }H^{1}\left(\mathbb{R}^{d-1}\right)\,,\label{eq:conv_final}\\
 & \Downarrow\nonumber \\
\left(\hat{f}_{\pi,R}^{\epsilon}-\hat{f}_{\pi,R}^{-\frac{s}{2},\epsilon}\right)\omega_{\bar{p}} & \underset{\epsilon\rightarrow0^{+}}{\longrightarrow} & \left(\hat{f}_{\pi,R}-\hat{f}_{\pi,R}^{-\frac{s}{2}}\right)\omega_{\bar{p}}\quad\textrm{in }L^{2}\left(\mathbb{R}^{d-1}\right)\,.\label{eq:conv_final_bis}
\end{eqnarray}
In order to probe \eqref{eq:conv_final}, we remember that $f_{\pi,R}\left(\bar{x}\right)-f_{\pi,R}^{s}\left(\bar{x}\right)=g_{\pi}^{s}\left(\bar{x}\right)\Theta\left(x^{1}\right)$,
with $g_{\pi}^{s}\in\mathcal{S}\left(\mathbb{R}^{d-1},\mathbb{R}\right)$
and $\left.g_{\pi}^{s}\right|_{x^{1}=0}=0$. Then, the following lemma
ensures \eqref{eq:conv_final}. 
\begin{lem}
\textup{Let $g\in\mathcal{S}\left(\mathbb{R}^{n}\right)$ with $\left.g\right|_{x^{1}=0}=0$,
$g_{R}\left(\bar{x}\right)=g\left(\bar{x}\right)\Theta\left(x^{1}\right)$
and $g_{R}^{\epsilon}\left(\bar{x}\right)=g\left(\bar{x}\right)\Theta_{\epsilon}\left(x^{1}\right)$
with $\Theta_{\epsilon}$ as \eqref{eq: theta_reg}. Then, $g_{R}\in H^{1}\left(\mathbb{R}^{n}\right)$
and $g_{R}^{\epsilon}\underset{\epsilon\rightarrow0^{+}}{\longrightarrow}g_{R}$
in $H^{1}\left(\mathbb{R}^{n}\right)$.} 
\end{lem}

\begin{proof}
The fact that $g_{R}\in H^{1}\left(\mathbb{R}^{n}\right)$ is guaranteed
by lemma \ref{par:lemma-1}. Here, we prove the convergence for $n=1$.
The generalization to $n>1$ is straightforward. Since $g_{R}$ and $g_{R}^{\epsilon}$
satisfy the hypothesis of the lemma \ref{lem:derivatives}, their
weak derivatives coincide with their pointwise derivatives, and hence
\begin{eqnarray}
\left\Vert g_{R}^{\epsilon}-g_{R}\right\Vert '^{2}{}_{H^{1}}\!\!\!\! & = & \!\!\!\!\int_{-\infty}^{+\infty}\!\!\!\!\!dx\left|g\left(x\right)\Theta_{\epsilon}\left(x\right)-g\left(x\right)\Theta\left(x\right)\right|^{2}+\!\int_{-\infty}^{+\infty}\!\!\!\!\! dx\left|\partial_{x}\left[g\left(x\right)\Theta_{\epsilon}\left(x\right)-g\left(x\right)\Theta\left(x\right)\right]\right|^{2}\nonumber \\
 & \leq & \!\!\!\!\int_{-\infty}^{+\infty}\!\!\!\!dx\left|g\left(x\right)\right|^{2}\left|\Theta_{\epsilon}\left(x\right)-\Theta\left(x\right)\right|^{2}+\!\int_{-\infty}^{+\infty}\!\!\!\!dx\left|g'\left(x\right)\right|^{2}\left|\Theta_{\epsilon}\left(x\right)-\Theta\left(x\right)\right|^{2}\nonumber \\
 &  & \!\!\!\!+\int_{-\infty}^{+\infty}\!\!\!\!dx\left|g\left(x\right)\right|^{2}\left|\Theta'_{\epsilon}\left(x\right)\right|^{2}+2\int_{-\infty}^{+\infty}\!\!\!\!dx\left|g\left(x\right)\right|\left|g'\left(x\right)\right|\left|\Theta_{\epsilon}\left(x\right)-\Theta\left(x\right)\right|\left|\Theta'_{\epsilon}\left(x\right)\right|\nonumber \\
 & \leq & \!\!\!\!\int_{\frac{\epsilon}{2}}^{\epsilon}dx\left(\left|g\left(x\right)\right|^{2}+\left|g'\left(x\right)\right|^{2}\right)+\int_{\frac{\epsilon}{2}}^{\epsilon}dx\left|g\left(x\right)\right|^{2}\left|\Theta'_{\epsilon}\left(x\right)\right|^{2}\nonumber \\
 &  & \!\!\!\!+2\int_{\frac{\epsilon}{2}}^{\epsilon}dx\left|g\left(x\right)\right|\left|g'\left(x\right)\right|\left|\Theta'_{\epsilon}\left(x\right)\right|\,.
\end{eqnarray}
We notice that since $g\in C^{\infty}\left(\mathbb{R}\right)$ and
$g\left(0\right)=0$, by Taylor's theorem, we have that $g\left(x\right)=g'\left(0\right)x+r\left(x\right)x$,
with $r\left(x\right)\underset{x\rightarrow0}{\longrightarrow}0$
and $r\in C^{\infty}\left(\mathbb{R}\right)$. We also have that $\max_{x\in\mathbb{R}}\left|\Theta'_{\epsilon}\left(x\right)\right|=\frac{4}{\epsilon}$,
which follows from the definition of that function . Then, using the
above properties and assuming $0<\varepsilon\leq1$, we have
\begin{eqnarray}
\left\Vert g_{R}^{\epsilon}-g_{R}\right\Vert '^{2}{}_{H^{1}}\!\!\!\! & \leq & \!\!\!\!\underset{x\in\left[0,1\right]}{\max}\left(\left|g\left(x\right)\right|^{2}+\left|g'\left(x\right)\right|^{2}\right)\int_{\frac{\epsilon}{2}}^{\epsilon}dx+\underset{x\in\left[0,1\right]}{\max}\left|g'\left(0\right)+r\left(x\right)\right|^{2}\frac{16}{\epsilon^{2}}\int_{\frac{\epsilon}{2}}^{\epsilon}dx\,x^{2}\nonumber \\
 &  & \!\!\!\!+\underset{x\in\left[0,1\right]}{\max}\left|g'\left(x\right)\right|\underset{x\in\left[0,1\right]}{\max}\left|g'\left(0\right)+r\left(x\right)\right|\frac{8}{\epsilon}\int_{\frac{\epsilon}{2}}^{\epsilon}dx\,x\nonumber \\
 & \leq & \!\!\!\!\underset{x\in\left[0,1\right]}{\max}\left(\left|g\left(x\right)\right|^{2}+\left|g'\left(x\right)\right|^{2}\right)\frac{\epsilon}{2}+\underset{x\in\left[0,1\right]}{\max}\left|g'\left(0\right)+r\left(x\right)\right|^{2}\frac{14}{3}\epsilon\nonumber \\
 &  & \!\!\!\!+\underset{x\in\left[0,1\right]}{\max}\left|g'\left(x\right)\right|\underset{x\in\left[0,1\right]}{\max}\left|g'\left(0\right)+r\left(x\right)\right|3\epsilon\underset{\epsilon\rightarrow0^{+}}{\longrightarrow}0\,.
\end{eqnarray}
\end{proof}
Then, we have that all terms in \eqref{eq:analizar_conver} converge.
By continuity of the inner product, the limit of \eqref{eq:analizar_conver}
is just this same expression but evaluated at $\epsilon=0$, which
coincides with the l.h.s of \eqref{eq:conv_l2}.

We can use the same arguments to prove the limit \eqref{eq:limite1}.
The first two terms of \eqref{eq:Ne(s)} are convergent due to \eqref{eq:conv_l2},
and the remaining term is also convergent due to \eqref{eq:conv_final}
and \eqref{eq:conv_final_bis}. Then, by the continuity of the inner product, we have that
\begin{equation}
N^{\epsilon}\left(s\right)\underset{\epsilon\rightarrow0^{+}}{\longrightarrow}N\left(s\right)=\frac{i}{2}\left(Q\left(0,s\right)-R\left(0,s\right)\right)-\frac{1}{2}\left\Vert g_{R}^{s}\right\Vert _{\mathrm{1p}}^{2}\,.
\end{equation}
Finally, expression \eqref{eq: estudiar_anal} holds.


\renewcommand\appendixname{Appendix}
\selectlanguage{english}

\chapter{Constructing the regular representation\label{regular}}

In this appendix we show how to construct the endomorphism corresponding
to the regular representation of the group. We follow the constructions developed
in \cite{Longo:1994xe}. The regular representation is defined as
the direct sum of all its irreducible representations, each one appearing a number
of times equal to its dimension
\begin{equation}
\rho_{reg}=\bigoplus_{\sigma\in\hat{G}}d_{\sigma}\,\rho_{\sigma}\,.\label{regend}
\end{equation}
By constructing such a reducible representation, we mean to provide
a set of charged intertwiners $\mathcal{I}_{\sigma}^{j}$ ($j=1,\ldots,d_{\sigma}$) satisfying\footnote{An intertwiner from one representation $\rho$ to another $\sigma$ is an operator $T\in\mathcal{A}$
satisfying $T\,\rho=\sigma\,T$, and it is usually denoted by $T:\rho\rightarrow\sigma$.}
\begin{eqnarray}
\mathcal{I}_{\sigma}^{j}:\iota\!\!\! & \rightarrow & \!\!\!\rho_{reg}\,,\\
\mathcal{I}_{\sigma}^{j\,\dagger}\mathcal{I}_{\nu}^{k} & = & \!\!\!\delta_{jk}\delta_{\sigma\nu}\,,\\
\sum_{\sigma\in\hat{G}}\sum_{j=1}^{d_{\sigma}}\mathcal{I}_{\sigma}^{j}\mathcal{I}_{\sigma}^{j\,\dagger}\!\!\! & = & \!\!\!\mathbf{1}\,.\label{regv}
\end{eqnarray}
The first relation says that all $\mathcal{I}_{\sigma}^{j}$ intertwine the vacuum representation
$\iota$ to the regular one $\rho_{reg}$. Equivalently, $\mathcal{I}_{\sigma}^{j}\vert0\rangle$
is a state that transforms under the regular representation of the
group. The second and third relations ensure that the regular endomorphism
can be explicitly written as 
\begin{equation}
\rho_{reg}(A):=\sum_{\sigma\in\hat{G}}\sum_{j=1}^{d_{\sigma}}\mathcal{I}_{\sigma}^{j}\,A\,\mathcal{I}_{\sigma}^{j\,\dagger}\,,\quad A\in\mathcal{A}\,.
\end{equation}
More importantly for us is that such relations allow to construct the finite
dimensional algebra spanned by the elements
\begin{equation}
A=\sum_{\sigma,\nu\in\hat{G}}\sum_{j,k=1}^{d_{\sigma}}a_{\sigma j,\nu k}\mathcal{I}_{\sigma}^{j}\mathcal{I}_{\nu}^{k\,\dagger}\,,\quad a_{\sigma j,\nu k}\in\mathbb{C}\,.\label{closedVr}
\end{equation}
This algebra is used in chapter \ref{EE_SS} to find lower bounds for the entropic order parameter.

There are two possible avenues to construct such a space of intertwiners.
The first is to use the method described in the text for constructing
the irreducible sectors. This approach requires to have some operator
$O_{reg}$ that takes us from the vacuum to the given charged sector,
in this case the, regular one. This approach is quite sensible and
physical when such operators can be found easily. For example, in gauge
theories, it is simple to consider Wilson lines in any given representation.

There is also a complementary approach, that only requires knowledge
of the charged intertwiners associated with the irreducible sectors
$\sigma\in\hat{G}$. These are the operators $V_{\sigma}^{j}$  ($j=1,\cdots,d_{\sigma}$)
that are described in chapter \ref{EE_SS}. They satisfy 
\begin{eqnarray}
V_{\sigma}^{j}:\iota\!\!\! & \rightarrow & \!\!\!\rho_{\sigma}\,,\\
V_{\sigma}^{j\,\dagger}V_{\sigma}^{k}\!\!\! & = & \!\!\!\delta_{jk}\,,\\
\sum_{j=1}^{d_{\sigma}}V_{\sigma}^{j}V_{\sigma}^{j\,\dagger}\!\!\! & = & \!\!\!\mathbf{1}\,,\\
\rho_{\sigma}(A)\!\!\! & = & \!\!\!\sum_{j=1}^{d_{\sigma}}V_{\sigma}^{j}\,A\,V_{\sigma}^{j\,\dagger}\,,\quad A\in\mathcal{A}\,.\label{vrel}
\end{eqnarray}
Now, the regular representation as defined above in \eqref{regend},
implies the existence of partial isometries $\omega_{\sigma}^{j}$ ($j=1,\cdots,d_{\sigma}$), satisfying the following properties 
\begin{eqnarray}
\omega_{\sigma}^{j}:\rho_{\sigma}\!\!\! & \rightarrow & \!\!\!\rho_{reg}\,,\\
\omega_{\sigma}^{j\,\dagger}\omega_{\nu}^{k}\!\!\! & = & \!\!\!\delta_{jk}\delta_{\sigma\nu}\,,\\
\sum_{j=1}^{d_{\sigma}}\omega_{\sigma}^{j}\omega_{\sigma}^{j\,\dagger}\!\!\! & = & \!\!\!\mathbf{1}\,,\\
\sum_{j=1}^{d_{\sigma}}\omega_{\sigma}^{j}\,\rho_{s}(A)\,\omega_{\sigma}^{j\,\dagger}(\omega_{s}^{i})^{\dagger}\!\!\! & = & \!\!\!\rho_{reg}\left(A\right)\,,\quad A\in\mathcal{A}\,.\label{omegrel}
\end{eqnarray}
Furthermore, these operators were explicitly constructed in \cite{Longo:1994xe},
with the help of particular charged intertwiner to the regular endomorphism
$v:\iota\rightarrow\rho_{reg}$. Its explicit construction might be
cumbersome, but its existence is guaranteed for finite groups \cite{Longo:1994xe}.
Assuming we have such an operator, in \cite{Longo:1994xe} it was shown
that there is an anti-isomorphism, such that
\begin{equation}
\omega_{\sigma}^{j}=\vert G\vert\,\varepsilon(vV_{\sigma}^{j\,\dagger})\,,\quad \textrm{for all }\sigma\in\hat{G}\textrm{ and }j=1,\cdots,d_{\sigma} \,,
\end{equation}
where $\varepsilon$ is the conditional expectation and $\vert G\vert$
is the order of the group. Then, given $V_{\sigma}^{j}$ and $v$, we can construct all the operators $\omega_{\sigma}^{j}$.
It is simple now to find the charged operators of the regular representation.
Since $V_{\sigma}^{j}:\iota\rightarrow\rho_{\sigma}$ and $\omega_{\sigma}^{j}:\rho_{\sigma}\rightarrow\rho_{reg}$,
it is clear that
\begin{equation}
\omega_{\sigma}^{j}V_{\sigma}^{j}:\iota\rightarrow\rho_{reg}\,.
\end{equation}
Moreover, given \eqref{vrel} and \eqref{omegrel}, it is simple to
verify that relations \eqref{regv} hold for $\mathcal{I}_{\sigma}^{j}:=\omega_{\sigma}^{j}V_{\sigma}^{j}$.


\begin{biblio}
\bibliography{mibib}
\end{biblio}

\begin{postliminary}

\begin{seccion}{Publications}

\noindent The thesis is based on the following publications:

\begin{enumerate}
\item \textit{Entanglement entropy and superselection sectors. Part I. Global symmetries}, H. Casini, M. Huerta, J. Mag\'an, D. Pontello, JHEP, 2020 (2), 14 (2020).
\item \textit{Relative entropy for coherent states from Araki formula}, H. Casini, S. Grillo, D. Pontello, Phys. Rev. D 99, 125020 (2019).
\item \textit{Entropy and modular Hamiltonian for a free chiral scalar in two intervals }, R. Arias, H. Casini, M. Huerta, D. Pontello,  Phys. Rev. D 98, 125008 (2018).
\end{enumerate}
  
\noindent Other publications not discussed in this thesis: 

\begin{enumerate}[start=4]
\item \textit{Quantum Complementarity through Entropic Certainty Principles}, J. Mag\'an, D. Pontello, Phys. Rev. A, 103 (1), 012211, (2021).
\item \textit{Logarithmic coefficient of the entanglement entropy of a Maxwell field}, H. Casini, M. Huerta, J. Mag\'an, D. Pontello, Phys. Rev. D, 101 (6), 065020 (2020).
\item \textit{Anisotropic Unruh temperatures}, Arias, R., Casini, H., Huerta, M., Pontello, D., Phys. Rev., D96 (10), 105019 (2017).
\end{enumerate}
  
\end{seccion}


\end{postliminary}

\end{document}